\documentclass[titlepage]{tcibook}
\usepackage{fancyhea}
\usepackage{work}
\usepackage{bm}       
\usepackage{graphicx}
\usepackage{hyperref}      
\usepackage[usenames,dvipsnames]{color} 
\usepackage{colortbl} 
\usepackage{amssymb}
\usepackage{amsmath}
\usepackage{siunitx}
\let\counterwithout\relax

\usepackage{chngcntr}
\usepackage{booktabs}
\usepackage{multirow}
\usepackage{multicol}
\usepackage[numbers]{natbib}
\counterwithout{figure}{chapter}
\usepackage{titlesec}
\usepackage{enumitem}
\usepackage{wrapfig}
\usepackage{changepage}
\usepackage{framed}

\usepackage[utf8]{inputenc}

\newcommand{\nc}{\newcommand}  

\def\beq{\begin{equation}}
\def\eeq#1{\label{#1}\end{equation}}
\def\eeqn{\end{equation}}

\newenvironment{Eqnarray}%
   {\arraycolsep 0.14em\begin{eqnarray}}{\end{eqnarray}}
\def\beqa{\begin{Eqnarray}}
\def\eeqa#1{\label{#1}\end{Eqnarray}}
\def\eeqan{\end{Eqnarray}}

\nc{\ra}{\rightarrow}  
\nc{\slsh}{\slash\hspace*{-0.22cm}}
\def\Re{{\cal R \mskip-4mu \lower.1ex \hbox{\it e}\,}}
\def\Im{{\cal I \mskip-5mu \lower.1ex \hbox{\it m}\,}}

\nc{\vev}[1]{ \left\langle {#1} \right\rangle }
\nc{\bra}[1]{ \langle {#1} | }
\nc{\ket}[1]{ | {#1} \rangle }
\nc{\fb}{\,{\rm fb}^{-1}}
\nc{\ev}{{\rm eV}}
\nc{\kev}{{\rm keV}}
\nc{\Mev}{{\rm MeV}}
\nc{\gev}{{\rm GeV}}
\nc{\tev}{{\rm TeV}}
\nc{\mev}{{\rm MeV}}

\def\del{\partial}
\def\Dslash{\not{\hbox{\kern-4pt $D$}}}
\def\dslash{\not{\hbox{\kern-2pt $\del$}}}
\def\pslash{\not{\hbox{\kern-2pt $p$}}}
\def\ETmiss{ \not{\hbox{\kern-4pt $E$}}_T }

\def\msb{{\bar{\ssstyle M \kern -1pt S}}}

\newcommand{\cmbexp}{{CMB-S4}}
\def\umux{$\mu$mux}
\def\fmux{fmux}
\newcommand{\wmap}{{\sl WMAP}}
\newcommand{\planck}{{\sl Planck}}

\newcommand{\Neff}{\ensuremath{N_\mathrm{eff}}}

\definecolor{orange}{rgb}{1,0.3,0}

\newcommand{\ic}{I\&C}
\newcommand{\commentout}[1]{{}}

\newcommand{\Mpl}{M_{\rm P}}

\def\lsim{\raise-.75ex\hbox{$\buildrel<\over\sim$}}

\DeclareUrlCommand\email{\urlstyle{rm}}

\newcommand{\prelim}[1]{#1}
\renewcommand{\prelim}[1]{} 
\usepackage{appendix}
\usepackage{changepage}

\def \nsatcryochile {0}
\def \nsattubeschile  {0}
\def \nsatcryosp {6}
\def \nsattubessp {18}
\def \nlatchile {2}
\def \nlatsp {1}

\begin{document}

\def\bibname{References}

\bibliographystyle{utphys}  

\raggedbottom

\pagenumbering{roman}

\parindent=0pt
\parskip=8pt
\setlength{\evensidemargin}{0pt}
\setlength{\oddsidemargin}{0pt}
\setlength{\marginparsep}{0.0in}
\setlength{\marginparwidth}{0.0in}
\marginparpush=0pt


\renewcommand{\chapname}{chap:science_}
\renewcommand{\chapterdir}{.}
\renewcommand{\arraystretch}{1.25}
\addtolength{\arraycolsep}{-3pt}

\pagenumbering{roman} 
\chapter*{CMB-S4 Science Case, Reference Design, and Project Plan}
\vskip -9.5pt
\hbox to\headwidth{%
       \leaders\hrule height1.5pt\hfil}
\vskip-6.5pt
\hbox to\headwidth{%
       \leaders\hrule height3.5pt\hfil}

  \begin{center}
   {\Large\bf
      CMB-S4 Collaboration\\
      \bigskip
      July 9, 2019\\
   }
 \end{center}
\eject

\setcounter{page}{1}

\begin{center}
  {\Large \bf Preface}
\end{center}
\bigskip

This document on the CMB-S4 Science Case, Reference Design, and Project Plan is the product of a global community of scientists who are united in support of advancing CMB-S4 to cross key thresholds in our understanding of the fundamental nature of space and time and the evolution of the Universe.  CMB-S4 is planned to be a joint National Science Foundation (NSF) and Department of Energy (DOE) project, with the construction phase to be funded as an NSF Major Research Equipment and Facilities Construction (MREFC) project and a DOE High Energy Physics (HEP) Major Item of Equipment (MIE) project. At the time of this writing, an interim project office has been constituted and tasked with 
advancing the CMB-S4 project in the NSF MREFC Preliminary Design Phase and toward DOE Critical Decision CD-1.  DOE CD-0 is expected imminently.

CMB-S4 has been in development for six years. 
Through the Snowmass Cosmic Frontier planning process, experimental groups in the cosmic microwave background (CMB) and broader cosmology communities came together to produce two influential CMB planning papers, endorsed by over 90 scientists, that outlined the science case as well as the CMB-S4 instrumental concept \cite{Abazajian:2013vfg,Abazajian:2013oma}.  It immediately became clear that an enormous increase in the scale of ground-based CMB experiments would be needed to achieve the exciting threshold-crossing scientific goals, necessitating a phase change in the ground-based CMB experimental program. To realize CMB-S4, a partnership of the university-based CMB groups, the broader cosmology community, and the national laboratories would be needed.

The community proposed CMB-S4 to the 2014 Particle Physics Project Prioritization Process (P5) as a single, community-wide experiment, jointly supported by DOE and NSF. Following P5's recommendation of CMB-S4 under all budget scenarios,  the CMB community started in early 2015 to hold biannual workshops -- open to CMB scientists from around the world -- to develop and refine the concept.  Nine workshops have been held to date, typically with 150 to 200 participants. The workshops have focused on developing the unique and vital role of the future ground-based CMB program. 

This growing CMB-S4 community produced a detailed and influential CMB-S4 Science Book \cite{Abazajian:2016yjj} and a CMB-S4 Technology Book \cite{TechBookarXiv170602464A}. Over 200 scientists contributed to these documents. These and numerous other reports, workshop and working group wiki pages, email lists, and much more may be found at the website \url{http://CMB-S4.org}.

Soon after the CMB-S4 Science Book was completed in August 2016, DOE and NSF requested the Astronomy and Astrophysics Advisory Committee (AAAC) to convene a Concept Definition Taskforce (CDT) to conduct a CMB-S4 concept study. The resulting report was unanimously accepted in late 2017.\footnote{See \url{https://www.nsf.gov/mps/ast/aaac/cmb\_s4/report/CMBS4\_final\_report\_NL.pdf}\,.}

One recommendation of the CDT report was that the community should organize itself into a formal collaboration. An Interim Collaboration Coordination Committee was elected by the community to coordinate this process. The resulting draft bylaws were refined at the Spring 2018 CMB-S4 community workshop, and overwhelmingly ratified on March 19th 2018, bringing the CMB-S4 Science Collaboration into being, and the first elections for the  officers of the collaboration were completed by the end of April 2018.  As of summer 2019 the collaboration has 198 members, 71 of whom hold positions within the organizational structure. These members represent 11 countries on 4 continents, and 76 institutions comprising 16 national laboratories and 60 universities. In parallel, the maturation of CMB-S4 as a project, strongly supported by project expertise at the national laboratories,
is manifest in the organization of a preProject Development Group and now the Interim Project Office.

\clearpage

\begin{center}
 {\Large \bf List of Contributors and Endorsers} 
\end{center}
\tolerance=4000

\newcounter{affilcount}
\newcommand{\affil}[1]{\refstepcounter{affilcount}\label{#1}}
\affil{UCIrvine}
\affil{JohnsHopkinsUniversity}
\affil{UniversityofIllinoisatUrbana-Champaign}
\affil{SLAC}
\affil{StanfordUniversity}
\affil{OxfordUniversity}
\affil{UCBerkeley}
\affil{LawrenceBerkeleyNationalLaboratory}
\affil{Fermilab}
\affil{UCSanDiego}
\affil{SISSA}
\affil{ArgonneNationalLaboratory}
\affil{HarvardUniversity}
\affil{UniversityofNewMexico}
\affil{UniversityofChicago}
\affil{AstroParticleandCosmologyLaboratory}
\affil{Caltech}
\affil{CornellUniversity}
\affil{UniversityofPennsylvania}
\affil{CenterforAstrophysicsHarvardandSmithsonian}
\affil{YaleUniversity}
\affil{UniversityofCincinnati}
\affil{LMUMunich}
\affil{INAF}
\affil{ItalianALMARegionalCentre}
\affil{CITA}
\affil{InstitutdAstrophysiquedeParis}
\affil{UniversityofManchester}
\affil{ArizonaStateUniversity}
\affil{CenterforComputationalAstrophysicsFlatironInstitute}
\affil{RutgersUniversity}
\affil{CardiffUniversity}
\affil{DartmouthCollege}
\affil{UniversityofSussex}
\affil{InstituteofAstronomyandDAMTPUniversityofCambridge}
\affil{McGillUniversity}
\affil{NIST}
\affil{StockholmUniversity}
\affil{NASAGoddardSpaceFlightCenter}
\affil{PrincetonUniversity}
\affil{SimonFraserUniversity}
\affil{UniversityofSouthernCalifornia}
\affil{HaverfordCollege}
\affil{UniversityofBritishColumbia}
\affil{UniversityofColoradoBoulder}
\affil{UniversityofMinnesota}
\affil{UniversityofMichigan}
\affil{KEK}
\affil{PennsylvaniaStateUniversity}
\affil{InstituteforAdvancedStudy}
\affil{UniversityofToronto}
\affil{FloridaStateUniversity}
\affil{ColumbiaUniversity}
\affil{NationalResearchCouncilCanada}
\affil{UniversityofVictoria}
\affil{KavliIPMU}
\affil{FitBit}
\affil{UCDavis}
\affil{UniversityofPittsburgh}
\affil{AaltoUniversity}
\affil{JPL}
\affil{StonyBrookUniversity}
\affil{UniversityofTokyo}
\affil{UniversityofWashington}
\affil{UniversityofGroningen}
\affil{CEASaclay}
\affil{SouthernMethodistUniversity}
\affil{InstitutLagrangedeParis}
\affil{EuropeanSouthernObservatory}
\affil{PerimeterInstitute}
\affil{DunlapInstitute}
\affil{NationalTaiwanUniversity}
\affil{UniversityofMilano-Bicocca}
\affil{BrookhavenNationalLaboratory}
\affil{UCLA}
\affil{UniversityofMelbourne}
\affil{CaseWesternReserveUniversity}
\affil{UniversityofCambridge}
\affil{DescartesLab}
\affil{KyotoUniversity}
\affil{UniversityofWisconsinMadison}
\affil{UniversitadegliStudidiMilan}
\affil{LAL}
\affil{BrownUniversity}
\affil{MassachusettsInstituteofTechnology}
\affil{SyracuseUniversity}
Kevork Abazajian,\textsuperscript{\ref{UCIrvine}}
Graeme Addison,\textsuperscript{\ref{JohnsHopkinsUniversity}}
Peter Adshead,\textsuperscript{\ref{UniversityofIllinoisatUrbana-Champaign}}
Zeeshan Ahmed,\textsuperscript{\ref{SLAC}}
Steven W.~Allen,\textsuperscript{\ref{StanfordUniversity}}
David Alonso,\textsuperscript{\ref{OxfordUniversity}}
Marcelo Alvarez,\textsuperscript{\ref{UCBerkeley},\ref{LawrenceBerkeleyNationalLaboratory}}
Adam Anderson,\textsuperscript{\ref{Fermilab}}
Kam~S.\ Arnold,\textsuperscript{\ref{UCSanDiego}}
Carlo Baccigalupi,\textsuperscript{\ref{SISSA}}
Kathy Bailey,\textsuperscript{\ref{ArgonneNationalLaboratory}}
Denis Barkats,\textsuperscript{\ref{HarvardUniversity}}
Darcy Barron,\textsuperscript{\ref{UniversityofNewMexico}}
Peter S.~Barry,\textsuperscript{\ref{UniversityofChicago}}
James~G.\ Bartlett,\textsuperscript{\ref{AstroParticleandCosmologyLaboratory}}
Ritoban Basu Thakur,\textsuperscript{\ref{Caltech}}
Nicholas Battaglia,\textsuperscript{\ref{CornellUniversity}}
Eric Baxter,\textsuperscript{\ref{UniversityofPennsylvania}}
Rachel Bean,\textsuperscript{\ref{CornellUniversity}}
Chris Bebek,\textsuperscript{\ref{LawrenceBerkeleyNationalLaboratory}}
Amy~N.\ Bender,\textsuperscript{\ref{ArgonneNationalLaboratory}}
Bradford A.~Benson,\textsuperscript{\ref{Fermilab}}
Edo Berger,\textsuperscript{\ref{CenterforAstrophysicsHarvardandSmithsonian}}
Sanah Bhimani,\textsuperscript{\ref{YaleUniversity}}
Colin A.~Bischoff,\textsuperscript{\ref{UniversityofCincinnati}}
Lindsey Bleem,\textsuperscript{\ref{ArgonneNationalLaboratory}}
Sebastian Bocquet,\textsuperscript{\ref{LMUMunich}}
Kimberly Boddy,\textsuperscript{\ref{JohnsHopkinsUniversity}}
Matteo Bonato,\textsuperscript{\ref{INAF},\ref{ItalianALMARegionalCentre}}
J.~Richard Bond,\textsuperscript{\ref{CITA}}
Julian Borrill,\textsuperscript{\ref{LawrenceBerkeleyNationalLaboratory},\ref{UCBerkeley}}
François R.~Bouchet,\textsuperscript{\ref{InstitutdAstrophysiquedeParis}}
Michael L.~Brown,\textsuperscript{\ref{UniversityofManchester}}
Sean Bryan,\textsuperscript{\ref{ArizonaStateUniversity}}
Blakesley Burkhart,\textsuperscript{\ref{CenterforComputationalAstrophysicsFlatironInstitute},\ref{RutgersUniversity}}
Victor Buza,\textsuperscript{\ref{HarvardUniversity}}
Karen Byrum,\textsuperscript{\ref{ArgonneNationalLaboratory}}
Erminia Calabrese,\textsuperscript{\ref{CardiffUniversity}}
Victoria  Calafut,\textsuperscript{\ref{CornellUniversity}}
Robert Caldwell,\textsuperscript{\ref{DartmouthCollege}}
John E.~Carlstrom,\textsuperscript{\ref{UniversityofChicago}}
Julien Carron,\textsuperscript{\ref{UniversityofSussex}}
Thomas Cecil,\textsuperscript{\ref{ArgonneNationalLaboratory}}
Anthony Challinor,\textsuperscript{\ref{InstituteofAstronomyandDAMTPUniversityofCambridge}}
Clarence L.~Chang,\textsuperscript{\ref{ArgonneNationalLaboratory}}
Yuji Chinone,\textsuperscript{\ref{UCBerkeley}}
Hsiao-Mei Sherry Cho,\textsuperscript{\ref{SLAC}}
Asantha Cooray,\textsuperscript{\ref{UCIrvine}}
Thomas M.~Crawford,\textsuperscript{\ref{UniversityofChicago}}
Abigail Crites,\textsuperscript{\ref{Caltech}}
Ari  Cukierman,\textsuperscript{\ref{SLAC}}
Francis-Yan Cyr-Racine,\textsuperscript{\ref{HarvardUniversity}}
Tijmen de Haan,\textsuperscript{\ref{LawrenceBerkeleyNationalLaboratory}}
Gianfranco de Zotti,\textsuperscript{\ref{INAF}}
Jacques Delabrouille,\textsuperscript{\ref{AstroParticleandCosmologyLaboratory}}
Marcel Demarteau,\textsuperscript{\ref{ArgonneNationalLaboratory}}
Mark Devlin,\textsuperscript{\ref{UniversityofPennsylvania}}
Eleonora Di Valentino,\textsuperscript{\ref{UniversityofManchester}}
Matt Dobbs,\textsuperscript{\ref{McGillUniversity}}
Shannon Duff,\textsuperscript{\ref{NIST}}
Adriaan Duivenvoorden,\textsuperscript{\ref{StockholmUniversity}}
Cora Dvorkin,\textsuperscript{\ref{HarvardUniversity}}
William Edwards,\textsuperscript{\ref{LawrenceBerkeleyNationalLaboratory}}
Joseph Eimer,\textsuperscript{\ref{JohnsHopkinsUniversity}}
Josquin Errard,\textsuperscript{\ref{AstroParticleandCosmologyLaboratory}}
Thomas Essinger-Hileman,\textsuperscript{\ref{NASAGoddardSpaceFlightCenter}}
Giulio Fabbian,\textsuperscript{\ref{UniversityofSussex}}
Chang Feng,\textsuperscript{\ref{UniversityofIllinoisatUrbana-Champaign}}
Simone Ferraro,\textsuperscript{\ref{LawrenceBerkeleyNationalLaboratory}}
Jeffrey P.~Filippini,\textsuperscript{\ref{UniversityofIllinoisatUrbana-Champaign}}
Raphael Flauger,\textsuperscript{\ref{UCSanDiego}}
Brenna Flaugher,\textsuperscript{\ref{Fermilab}}
Aurelien A.~Fraisse,\textsuperscript{\ref{PrincetonUniversity}}
Andrei Frolov,\textsuperscript{\ref{SimonFraserUniversity}}
Nicholas Galitzki,\textsuperscript{\ref{UCSanDiego}}
Silvia Galli,\textsuperscript{\ref{InstitutdAstrophysiquedeParis}}
Ken Ganga,\textsuperscript{\ref{AstroParticleandCosmologyLaboratory}}
Martina Gerbino,\textsuperscript{\ref{ArgonneNationalLaboratory}}
Murdock Gilchriese,\textsuperscript{\ref{LawrenceBerkeleyNationalLaboratory}}
Vera Gluscevic,\textsuperscript{\ref{UniversityofSouthernCalifornia}}
Daniel Green,\textsuperscript{\ref{UCSanDiego}}
Daniel Grin,\textsuperscript{\ref{HaverfordCollege}}
Evan Grohs,\textsuperscript{\ref{UCBerkeley}}
Riccardo Gualtieri,\textsuperscript{\ref{UniversityofIllinoisatUrbana-Champaign}}
Victor Guarino,\textsuperscript{\ref{ArgonneNationalLaboratory}}
Jon E.~Gudmundsson,\textsuperscript{\ref{StockholmUniversity}}
Salman Habib,\textsuperscript{\ref{ArgonneNationalLaboratory}}
Gunther Haller,\textsuperscript{\ref{SLAC}}
Mark Halpern,\textsuperscript{\ref{UniversityofBritishColumbia}}
Nils W.~Halverson,\textsuperscript{\ref{UniversityofColoradoBoulder}}
Shaul Hanany,\textsuperscript{\ref{UniversityofMinnesota}}
Kathleen Harrington,\textsuperscript{\ref{UniversityofMichigan}}
Masaya Hasegawa,\textsuperscript{\ref{KEK}}
Matthew Hasselfield,\textsuperscript{\ref{PennsylvaniaStateUniversity}}
Masashi Hazumi,\textsuperscript{\ref{KEK}}
Katrin Heitmann,\textsuperscript{\ref{ArgonneNationalLaboratory}}
Shawn Henderson,\textsuperscript{\ref{SLAC}}
Jason W.~Henning,\textsuperscript{\ref{UniversityofChicago}}
J. Colin Hill,\textsuperscript{\ref{InstituteforAdvancedStudy}}
Ren\'{e}e Hlo\v{z}ek,\textsuperscript{\ref{UniversityofToronto}}
Gil Holder,\textsuperscript{\ref{UniversityofIllinoisatUrbana-Champaign}}
William Holzapfel,\textsuperscript{\ref{UCBerkeley}}
Johannes Hubmayr,\textsuperscript{\ref{NIST}}
Kevin M.~Huffenberger,\textsuperscript{\ref{FloridaStateUniversity}}
Michael Huffer,\textsuperscript{\ref{SLAC}}
Howard Hui,\textsuperscript{\ref{Caltech}}
Kent Irwin,\textsuperscript{\ref{StanfordUniversity}}
Bradley R.~Johnson,\textsuperscript{\ref{ColumbiaUniversity}}
Doug Johnstone,\textsuperscript{\ref{NationalResearchCouncilCanada},\ref{UniversityofVictoria}}
William C.~Jones,\textsuperscript{\ref{PrincetonUniversity}}
Kirit Karkare,\textsuperscript{\ref{UniversityofChicago}}
Nobuhiko Katayama,\textsuperscript{\ref{KavliIPMU}}
James Kerby,\textsuperscript{\ref{ArgonneNationalLaboratory}}
Sarah Kernovsky,\textsuperscript{\ref{FitBit}}
Reijo Keskitalo,\textsuperscript{\ref{LawrenceBerkeleyNationalLaboratory},\ref{UCBerkeley}}
Theodore Kisner,\textsuperscript{\ref{LawrenceBerkeleyNationalLaboratory},\ref{UCBerkeley}}
Lloyd Knox,\textsuperscript{\ref{UCDavis}}
Arthur Kosowsky,\textsuperscript{\ref{UniversityofPittsburgh}}
John Kovac,\textsuperscript{\ref{HarvardUniversity}}
Ely D.~Kovetz,\textsuperscript{\ref{JohnsHopkinsUniversity}}
Steve Kuhlmann,\textsuperscript{\ref{ArgonneNationalLaboratory}}
Chao-lin Kuo,\textsuperscript{\ref{StanfordUniversity}}
Nadine Kurita,\textsuperscript{\ref{SLAC}}
Akito Kusaka,\textsuperscript{\ref{LawrenceBerkeleyNationalLaboratory}}
Anne Lahteenmaki,\textsuperscript{\ref{AaltoUniversity}}
Charles R.~Lawrence,\textsuperscript{\ref{JPL}}
Adrian T.~Lee,\textsuperscript{\ref{UCBerkeley},\ref{LawrenceBerkeleyNationalLaboratory}}
Antony Lewis,\textsuperscript{\ref{UniversityofSussex}}
Dale Li,\textsuperscript{\ref{SLAC}}
Eric Linder,\textsuperscript{\ref{LawrenceBerkeleyNationalLaboratory}}
Marilena Loverde,\textsuperscript{\ref{StonyBrookUniversity}}
Amy Lowitz,\textsuperscript{\ref{UniversityofChicago}}
Mathew S. Madhavacheril,\textsuperscript{\ref{PrincetonUniversity}}
Adam Mantz,\textsuperscript{\ref{StanfordUniversity}}
Frederick Matsuda,\textsuperscript{\ref{UniversityofTokyo}}
Philip Mauskopf,\textsuperscript{\ref{ArizonaStateUniversity}}
Jeff McMahon,\textsuperscript{\ref{UniversityofMichigan}}
Matthew McQuinn,\textsuperscript{\ref{UniversityofWashington}}
P.~Daniel Meerburg,\textsuperscript{\ref{UniversityofGroningen}}
Jean-Baptiste Melin,\textsuperscript{\ref{CEASaclay}}
Joel Meyers,\textsuperscript{\ref{SouthernMethodistUniversity}}
Marius Millea,\textsuperscript{\ref{InstitutLagrangedeParis}}
Joseph Mohr,\textsuperscript{\ref{LMUMunich}}
Lorenzo Moncelsi,\textsuperscript{\ref{Caltech}}
Tony Mroczkowski,\textsuperscript{\ref{EuropeanSouthernObservatory}}
Suvodip Mukherjee,\textsuperscript{\ref{InstitutdAstrophysiquedeParis}}
Moritz M\"{u}nchmeyer,\textsuperscript{\ref{PerimeterInstitute}}
Daisuke Nagai,\textsuperscript{\ref{YaleUniversity}}
Johanna Nagy,\textsuperscript{\ref{DunlapInstitute},\ref{UniversityofToronto}}
Toshiya  Namikawa,\textsuperscript{\ref{NationalTaiwanUniversity}}
Federico Nati,\textsuperscript{\ref{UniversityofMilano-Bicocca}}
Tyler Natoli,\textsuperscript{\ref{DunlapInstitute}}
Mattia Negrello,\textsuperscript{\ref{CardiffUniversity}}
Laura Newburgh,\textsuperscript{\ref{YaleUniversity}}
Michael D.~Niemack,\textsuperscript{\ref{CornellUniversity}}
Haruki Nishino,\textsuperscript{\ref{KEK}}
Martin Nordby,\textsuperscript{\ref{SLAC}}
Valentine Novosad,\textsuperscript{\ref{ArgonneNationalLaboratory}}
Paul O'Connor,\textsuperscript{\ref{BrookhavenNationalLaboratory}}
Georges Obied,\textsuperscript{\ref{HarvardUniversity}}
Stephen Padin,\textsuperscript{\ref{UniversityofChicago}}
Shivam Pandey,\textsuperscript{\ref{UniversityofPennsylvania}}
Bruce Partridge,\textsuperscript{\ref{HaverfordCollege}}
Elena Pierpaoli,\textsuperscript{\ref{UniversityofSouthernCalifornia}}
Levon Pogosian,\textsuperscript{\ref{SimonFraserUniversity}}
Clement Pryke,\textsuperscript{\ref{UniversityofMinnesota}}
Giuseppe Puglisi,\textsuperscript{\ref{StanfordUniversity}}
Benjamin Racine,\textsuperscript{\ref{HarvardUniversity}}
Srinivasan Raghunathan,\textsuperscript{\ref{UCLA}}
Alexandra Rahlin,\textsuperscript{\ref{Fermilab}}
Srini Rajagopalan,\textsuperscript{\ref{BrookhavenNationalLaboratory}}
Marco Raveri,\textsuperscript{\ref{UniversityofChicago}}
Mark Reichanadter,\textsuperscript{\ref{SLAC}}
Christian L.~Reichardt,\textsuperscript{\ref{UniversityofMelbourne}}
Mathieu Remazeilles,\textsuperscript{\ref{UniversityofManchester}}
Graca Rocha,\textsuperscript{\ref{JPL}}
Natalie A.~Roe,\textsuperscript{\ref{LawrenceBerkeleyNationalLaboratory}}
Anirban Roy,\textsuperscript{\ref{SISSA}}
John Ruhl,\textsuperscript{\ref{CaseWesternReserveUniversity}}
Maria Salatino,\textsuperscript{\ref{AstroParticleandCosmologyLaboratory}}
Benjamin Saliwanchik,\textsuperscript{\ref{YaleUniversity}}
Emmanuel Schaan,\textsuperscript{\ref{LawrenceBerkeleyNationalLaboratory}}
Alessandro Schillaci,\textsuperscript{\ref{Caltech}}
Marcel M.~Schmittfull,\textsuperscript{\ref{InstituteforAdvancedStudy}}
Douglas Scott,\textsuperscript{\ref{UniversityofBritishColumbia}}
Neelima Sehgal,\textsuperscript{\ref{StonyBrookUniversity}}
Sarah Shandera,\textsuperscript{\ref{PennsylvaniaStateUniversity}}
Christopher Sheehy,\textsuperscript{\ref{BrookhavenNationalLaboratory}}
Blake~D.\ Sherwin,\textsuperscript{\ref{UniversityofCambridge}}
Erik Shirokoff,\textsuperscript{\ref{UniversityofChicago}}
Sara M.~Simon,\textsuperscript{\ref{UniversityofMichigan}}
An\v{z}e Slosar,\textsuperscript{\ref{BrookhavenNationalLaboratory}}
Rachel Somerville,\textsuperscript{\ref{CenterforComputationalAstrophysicsFlatironInstitute},\ref{RutgersUniversity}}
David Spergel,\textsuperscript{\ref{PrincetonUniversity},\ref{CenterforComputationalAstrophysicsFlatironInstitute}}
Suzanne T.~Staggs,\textsuperscript{\ref{PrincetonUniversity}}
Antony Stark,\textsuperscript{\ref{HarvardUniversity}}
Radek Stompor,\textsuperscript{\ref{AstroParticleandCosmologyLaboratory}}
Kyle T.~Story,\textsuperscript{\ref{DescartesLab}}
Chris Stoughton,\textsuperscript{\ref{Fermilab}}
Aritoki Suzuki,\textsuperscript{\ref{LawrenceBerkeleyNationalLaboratory}}
Osamu Tajima,\textsuperscript{\ref{KyotoUniversity}}
Grant P.~Teply,\textsuperscript{\ref{UCSanDiego}}
Keith Thompson,\textsuperscript{\ref{StanfordUniversity}}
Peter Timbie,\textsuperscript{\ref{UniversityofWisconsinMadison}}
Maurizio Tomasi,\textsuperscript{\ref{UniversitadegliStudidiMilan}}
Jesse I.~Treu,\textsuperscript{\ref{PrincetonUniversity}}
Matthieu Tristram,\textsuperscript{\ref{LAL}}
Gregory Tucker,\textsuperscript{\ref{BrownUniversity}}
Caterina Umiltà,\textsuperscript{\ref{UniversityofCincinnati}}
Alexander van Engelen,\textsuperscript{\ref{CITA}}
Joaquin D.~Vieira,\textsuperscript{\ref{UniversityofIllinoisatUrbana-Champaign}}
Abigail G.~Vieregg,\textsuperscript{\ref{UniversityofChicago}}
Mark Vogelsberger,\textsuperscript{\ref{MassachusettsInstituteofTechnology}}
Gensheng Wang,\textsuperscript{\ref{ArgonneNationalLaboratory}}
Scott Watson,\textsuperscript{\ref{SyracuseUniversity}}
Martin White,\textsuperscript{\ref{LawrenceBerkeleyNationalLaboratory},\ref{UCBerkeley}}
Nathan Whitehorn,\textsuperscript{\ref{UCLA}}
Edward J.\ Wollack,\textsuperscript{\ref{NASAGoddardSpaceFlightCenter}}
W.~L.~Kimmy Wu,\textsuperscript{\ref{UniversityofChicago}}
Zhilei Xu,\textsuperscript{\ref{UniversityofPennsylvania}}
Siavash Yasini,\textsuperscript{\ref{UniversityofSouthernCalifornia}}
James Yeck,\textsuperscript{\ref{UniversityofWisconsinMadison}}
Ki Won Yoon,\textsuperscript{\ref{StanfordUniversity}}
Edward Young,\textsuperscript{\ref{SLAC}}
Andrea Zonca\textsuperscript{\ref{UCSanDiego}}

\begin{multicols}{2}
  \scriptsize
  \setlength{\parskip}{2pt}

\noindent\textsuperscript{\ref{UCIrvine}}UC Irvine

\noindent\textsuperscript{\ref{JohnsHopkinsUniversity}}Johns Hopkins University

\noindent\textsuperscript{\ref{UniversityofIllinoisatUrbana-Champaign}}University of Illinois at Urbana-Champaign

\noindent\textsuperscript{\ref{SLAC}}SLAC

\noindent\textsuperscript{\ref{StanfordUniversity}}Stanford University

\noindent\textsuperscript{\ref{OxfordUniversity}}Oxford University

\noindent\textsuperscript{\ref{UCBerkeley}}UC Berkeley

\noindent\textsuperscript{\ref{LawrenceBerkeleyNationalLaboratory}}Lawrence Berkeley National Laboratory

\noindent\textsuperscript{\ref{Fermilab}}Fermilab

\noindent\textsuperscript{\ref{UCSanDiego}}UC San Diego

\noindent\textsuperscript{\ref{SISSA}}SISSA

\noindent\textsuperscript{\ref{ArgonneNationalLaboratory}}Argonne National Laboratory

\noindent\textsuperscript{\ref{HarvardUniversity}}Harvard University

\noindent\textsuperscript{\ref{UniversityofNewMexico}}University of New Mexico

\noindent\textsuperscript{\ref{UniversityofChicago}}University of Chicago

\noindent\textsuperscript{\ref{AstroParticleandCosmologyLaboratory}}AstroParticle \& Cosmology Laboratory

\noindent\textsuperscript{\ref{Caltech}}Caltech

\noindent\textsuperscript{\ref{CornellUniversity}}Cornell University

\noindent\textsuperscript{\ref{UniversityofPennsylvania}}University of Pennsylvania

\noindent\textsuperscript{\ref{CenterforAstrophysicsHarvardandSmithsonian}}Center for Astrophysics, Harvard \& Smithsonian

\noindent\textsuperscript{\ref{YaleUniversity}}Yale University

\noindent\textsuperscript{\ref{UniversityofCincinnati}}University of Cincinnati

\noindent\textsuperscript{\ref{LMUMunich}}LMU Munich

\noindent\textsuperscript{\ref{INAF}}INAF

\noindent\textsuperscript{\ref{ItalianALMARegionalCentre}}Italian ALMA Regional Centre

\noindent\textsuperscript{\ref{CITA}}CITA

\noindent\textsuperscript{\ref{InstitutdAstrophysiquedeParis}}Institut d'Astrophysique de Paris

\noindent\textsuperscript{\ref{UniversityofManchester}}University of Manchester

\noindent\textsuperscript{\ref{ArizonaStateUniversity}}Arizona State University

\noindent\textsuperscript{\ref{CenterforComputationalAstrophysicsFlatironInstitute}}Center for Computational Astrophysics, Flatiron Institute

\noindent\textsuperscript{\ref{RutgersUniversity}}Rutgers University

\noindent\textsuperscript{\ref{CardiffUniversity}}Cardiff University

\noindent\textsuperscript{\ref{DartmouthCollege}}Dartmouth College

\noindent\textsuperscript{\ref{UniversityofSussex}}University of Sussex

\noindent\textsuperscript{\ref{InstituteofAstronomyandDAMTPUniversityofCambridge}}Institute of Astronomy and DAMTP, University of Cambridge

\noindent\textsuperscript{\ref{McGillUniversity}}McGill University

\noindent\textsuperscript{\ref{NIST}}NIST

\noindent\textsuperscript{\ref{StockholmUniversity}}Stockholm University

\noindent\textsuperscript{\ref{PrincetonUniversity}}Princeton University

\noindent\textsuperscript{\ref{NASAGoddardSpaceFlightCenter}}NASA Goddard Space Flight Center

\noindent\textsuperscript{\ref{SimonFraserUniversity}}Simon Fraser University

\noindent\textsuperscript{\ref{UniversityofSouthernCalifornia}}University of Southern California

\noindent\textsuperscript{\ref{HaverfordCollege}}Haverford College

\noindent\textsuperscript{\ref{UniversityofBritishColumbia}}University of British Columbia

\noindent\textsuperscript{\ref{UniversityofColoradoBoulder}}University of Colorado Boulder

\noindent\textsuperscript{\ref{UniversityofMinnesota}}University of Minnesota

\noindent\textsuperscript{\ref{UniversityofMichigan}}University of Michigan

\noindent\textsuperscript{\ref{KEK}}KEK

\noindent\textsuperscript{\ref{PennsylvaniaStateUniversity}}Pennsylvania State University

\noindent\textsuperscript{\ref{InstituteforAdvancedStudy}}Institute for Advanced Study

\noindent\textsuperscript{\ref{UniversityofToronto}}University of Toronto

\noindent\textsuperscript{\ref{FloridaStateUniversity}}Florida State University

\noindent\textsuperscript{\ref{ColumbiaUniversity}}Columbia University

\noindent\textsuperscript{\ref{NationalResearchCouncilCanada}}National Research Council Canada

\noindent\textsuperscript{\ref{UniversityofVictoria}}University of Victoria

\noindent\textsuperscript{\ref{KavliIPMU}}Kavli IPMU

\noindent\textsuperscript{\ref{FitBit}}FitBit

\noindent\textsuperscript{\ref{UCDavis}}UC Davis

\noindent\textsuperscript{\ref{UniversityofPittsburgh}}University of Pittsburgh

\noindent\textsuperscript{\ref{AaltoUniversity}}Aalto University

\noindent\textsuperscript{\ref{JPL}}JPL

\noindent\textsuperscript{\ref{StonyBrookUniversity}}Stony Brook University

\noindent\textsuperscript{\ref{UniversityofTokyo}}University of Tokyo

\noindent\textsuperscript{\ref{UniversityofWashington}}University of Washington

\noindent\textsuperscript{\ref{UniversityofGroningen}}University of Groningen

\noindent\textsuperscript{\ref{CEASaclay}}CEA Saclay

\noindent\textsuperscript{\ref{SouthernMethodistUniversity}}Southern Methodist University

\noindent\textsuperscript{\ref{InstitutLagrangedeParis}}Institut Lagrange de Paris

\noindent\textsuperscript{\ref{EuropeanSouthernObservatory}}European Southern Observatory

\noindent\textsuperscript{\ref{PerimeterInstitute}}Perimeter Institute

\noindent\textsuperscript{\ref{DunlapInstitute}}Dunlap Institute

\noindent\textsuperscript{\ref{NationalTaiwanUniversity}}National Taiwan University

\noindent\textsuperscript{\ref{UniversityofMilano-Bicocca}}University of Milano-Bicocca

\noindent\textsuperscript{\ref{BrookhavenNationalLaboratory}}Brookhaven National Laboratory

\noindent\textsuperscript{\ref{UCLA}}UCLA

\noindent\textsuperscript{\ref{UniversityofMelbourne}}University of Melbourne

\noindent\textsuperscript{\ref{CaseWesternReserveUniversity}}Case Western Reserve University

\noindent\textsuperscript{\ref{UniversityofCambridge}}University of Cambridge

\noindent\textsuperscript{\ref{DescartesLab}}Descartes Lab

\noindent\textsuperscript{\ref{KyotoUniversity}}Kyoto University

\noindent\textsuperscript{\ref{UniversityofWisconsinMadison}}University of Wisconsin--Madison

\noindent\textsuperscript{\ref{UniversitadegliStudidiMilan}}Università degli Studi di Milan

\noindent\textsuperscript{\ref{LAL}}LAL

\noindent\textsuperscript{\ref{BrownUniversity}}Brown University

\noindent\textsuperscript{\ref{MassachusettsInstituteofTechnology}}Massachusetts Institute of Technology

\noindent\textsuperscript{\ref{SyracuseUniversity}}Syracuse University

\end{multicols}
\clearpage

\begin{center}
  {\Large \bf Executive Summary}
\end{center}

\label{sec:ExecSum}


This document presents the science case, Reference Design, and project plan for the Stage-4 ground-based cosmic microwave background experiment CMB-S4. CMB-S4 was conceived by the community during the 2013 Snowmass physics planning activity, as the path forward to realizing the enormous potential of CMB measurements for understanding the origin and evolution of the Universe, from the highest energies at the dawn of time through the growth of structure to the present day.  The science case is spectacular, including the search for primordial gravitational waves as predicted from inflation, constraints on relic particles, setting the neutrino mass scale, unique and complementary insights into dark energy and tests of gravity on large scales, elucidating the role of baryonic feedback on galaxy formation and evolution, opening up a window on the transient Universe at millimeter wavelengths, and even the exploration of the outer solar system.  
The CMB-S4 Legacy Survey covering over half the sky with unprecedented sensitivity through the millimeter-wave band 
will have profound and lasting impact on Astronomy and Astrophysics and provide a powerful complement to surveys  at other wavelengths, such as the Large Synoptic Survey Telescope (LSST) and the {\it Wide Field Infrared Survey Telescope (WFIRST)}, and others yet to be imagined.

CMB-S4 was recommended by the 2014 Particle Physics Project Prioritization Panel (P5) report {\it Building for Discovery: Strategic Plan for U.S. Particle Physics in the Global Context\/} and by the 2015 National Academies report {\it A Strategic Vision for NSF Investments in Antarctic and Southern Ocean Research}. The community further developed the science case in the 2016 {\it CMB-S4 Science Book\/} and surveyed the status of the technology in the 2017 {\it CMB-S4 Technology Book}.  This work formed the foundation for the joint NSF-DOE Concept Definition Task Force (CDT), a subpanel of the Astronomy and Astrophysics Advisory Committee (AAAC), an FACA committee advising DOE, NASA, and NSF. The CDT report was enthusiastically accepted by the AAAC in October 2017.  The CDT report provides clear guidance on the science goals and measurement requirements for CMB-S4, along with a strawperson instrument design, schedule, and cost.  

CMB-S4 will be a joint agency program with roughly comparable support from NSF and DOE. The construction phase is expected to be funded as an NSF MREFC project and a  DOE HEP MIE project. The CMB-S4 Collaboration and the DOE laboratory-based CMB-S4 pre-Project Development Group were established shortly after the CDT report was accepted. They are working together as a team to advance the CMB-S4 project. This team brings together the CMB community, as well as the considerable expertise of the national laboratories, and it is this team has produced this report. 

This report builds on the work done by the CDT and the strawperson design concept in the CDT report. We fully support the two key elements of the CDT concept for CMB-S4: {\it (1) it requires multiple cameras and telescopes distributed across two sites,} 
{\it and (2) the experiment will be undertaken by a single collaboration and run as one project.}  
The latter point is essential given the magnitude of the increase in science reach and complexity over existing CMB projects.

The CMB-S4 Reference Design presented here uses proven existing technology that has been developed and demonstrated by CMB experimental groups over the last decade, scaled up to unprecedented levels. The design and implementation plan addresses the considerable technical challenges presented by the required scaling-up of the instrumentation and by the scope and complexity of the data analysis and interpretation.  Features of the design and plan include: superconducting detector arrays with well-understood and robust material properties and processing techniques; high-throughput mm-wave telescopes and optics with unprecedented precision and rejection of systematic contamination; full internal characterization of astronomical foreground emission; large cosmological simulations and improved theoretical modeling; and computational methods for extracting minute correlations in massive, multi-frequency data sets that include noise and a host of known and unknown signals. 

\subsection*{Science Goals and Measurement Capabilities}

Following the CMB-S4 Science Book and the CDT report we set three transformative science goals for CMB-S4.

\begin{itemize}

\item The first goal is to measure the imprint of primordial gravitational waves on the 
CMB polarization anisotropy, quantified by the tensor-to-scalar ratio $r$. The simplest models that naturally explain the observed departure from scale invariance of the density perturbations predict $r > 0.001$, and a particularly well-motivated subclass of these models predicts $r > 0.003$. CMB-S4 will be able to detect primordial gravitational waves for $r > 0.003$ at greater than 5$\sigma$. Such a detection would yield the first evidence of the quantization of gravity and point to inflationary physics near the energy scale associated with grand unified theories, probing energy scales far beyond the reach of the LHC or any conceivable collider experiment, and providing additional evidence in favor of the idea of the unification of forces. The measurement of the energy scale of inflation would have broad implications for many other aspects of fundamental physics, including key aspects of string theory. In the absence of a detection, the upper limit of $r < 0.001$ at 95\% CL achievable by CMB-S4 would significantly advance our understanding of inflation. It would rule out large classes of inflationary models and dramatically impact how we think about the theory. To some, the remaining class of models would be contrived enough to give up on inflation altogether.

\item The second goal for CMB-S4 is to detect or strongly constrain departures from the thermal history of the Universe predicted by the Standard Model of particle physics. Many well-motivated extensions of the Standard Model to higher energies predict low-mass relic particles. Departures from the standard history are conveniently quantified by the contribution of light relic particles to the effective number of relativistic species in the early Universe, $N_{\rm eff}$. CMB measurements are sensitive to the contribution of relic particles to the energy density in the early Universe and therefore only depend on the interaction cross-sections of the relics with Standard-Model particles through the temperature at which the relics decouple.
CMB-S4 will constrain $\Delta N_{\rm eff}\leq 0.06$ at the $95\%$ confidence level allowing detection of, or constraints on, a wide range of light relic particles even if they are too weakly interacting to be detected by lab-based experiments. CMB-S4 will be the most robust and precise probe of the thermal history of our Universe and will improve bounds on the decoupling temperature compared to Stage-3 CMB experiments or planned large-scale structure surveys by an order of magnitude or more, depending on the spin of the particle.

\item The third science goal for CMB-S4  is to provide a unique and powerful survey of 
a large fraction of
the sky at centimeter (cm) to millimeter (mm) wavelengths at unprecedented depth and angular resolution.
Such a data set would provide 
enormous legacy value to the broader Astronomy and Astrophysics communities 
and would complement and enhance the LSST optical survey of the same region, as well as other planned and yet-to-be-imagined surveys and data from both ground- and space-based instruments. 
Using the signature of gravitational and electromagnetic interactions between matter and the CMB
as it traverses the expanse of the Universe, mm-wave maps of sufficient depth and resolution would
provide highly complementary data for investigations of dark energy, modifications to general relativity, 
and neutrino properties.  For example, a sufficiently deep, wide, and high-resolution data set would enable two independent and competitive determinations of the sum of the neutrino masses, using weak gravitational lensing and the evolution of the number density of galaxy clusters.
These data would also provide a unique and powerful probe of the influence of baryonic feedback on the formation of galaxies and clusters of galaxies. With sufficient depth and observing cadence, such a cm-to-mm-wave survey
would also open an entirely new window on the transient and dynamic Universe, including mm-wave searches for orphan GRB afterglows and dwarf planets.
\end{itemize}

This third goal---the legacy cm-to-mm-wave dataset---will have the broadest benefit to both the cosmological and astronomical communities, 
and while it does not 
provide sharp measurement thresholds that drive the design 
of CMB-S4, it is crucial that the instrument be designed to deliver the full promise of the legacy science.
For instance, the sensitivity and sky coverage required to meet the goal of $\Delta N_{\rm eff}\leq 0.06$ at the $95\%$ confidence level are a good match to the requirements for the legacy data, but we must also keep in mind
parameters unique to this science goal such as observing cadence.

The science goals lead to the following measurement requirements.

\begin{itemize}

\item Low-resolution ultra-deep measurements (noise levels $< 1\,\mu$K-arcmin) 
over an exceptionally low-foreground region covering 3\% of the sky are required to meet the primordial gravitational wave goals.  These measurements must have high fidelity and low contamination over a wide range of angular scales and frequencies. Large-angular-scale measurements with resolution of around 30 arcminutes and well-determined beam properties and excellent control of systematic contamination are needed to image the $B$-mode polarization signature of the primordial gravitational waves.  Small-angular-scale measurements with resolution of order 1.5 arcminutes are needed for removing the contamination of the degree-scale $B$ modes caused by gravitational lensing of the much stronger CMB $E$-mode polarization, a process referred to as ``delensing.''

\item High-resolution ($\leq 1.5$ arcminutes) deep and wide measurements at a noise level of $1\,\mu$K-arcmin over approximately 70\% of the sky (60\% of the sky after applying a Galactic cut) are required to meet the light relic and legacy data goals.

\item Multifrequency coverage is required for foreground mitigation. As current measurements  
have shown, even in the cleanest regions of the sky, the rms fluctuation in Galactic foreground emission is an order of magnitude larger than the predicted $B$-mode fluctuations for $r = 0.001$.  Simulations based on the current best knowledge of the dust and synchrotron foreground emission indicate CMB-S4 can meet its primordial gravitational wave goals using nine frequency bands spanning 20 to 270\,GHz for the degree-scale measurements.  Fewer bands are needed for the high-angular-resolution delensing and deep/wide surveys.

\end{itemize}

The CMB-S4 science goals are therefore met with two nested, highly complementary surveys, the deep and wide survey covering 70\% of the sky and the ultra-deep survey focused on the detection of degree-angular-scale $B$-mode polarization generated by primordial gravitational waves, such as those predicted by inflationary models.  
In the context of their legacy value to the wider community, we will refer to the deep/wide and ultra-deep high-resolution surveys together as the CMB-S4 Legacy Survey.

\subsection*{Design Overview}

This document presents the Reference Design for CMB-S4. This design meets the measurement requirements and therefore can deliver the CMB-S4 science goals. 
This report provides the details of the design, including the technology choices, as well as design or technology options that, if developed in time, could lead to improved performance, lower cost, or reduced risk.
Building on the work done for the CDT report strawperson design, the Reference Design is supported by the extensive use of simulations based on our current understanding of the expected level and complexity of foreground emission and on noise levels scaled from existing experiments, as summarized in the appendices. 

The major components of the Reference Design are as follows.

\begin{itemize}

\item An ultra-deep survey covering 3\% of the sky, more if a primordial gravitational wave signal is detected, to be conducted over seven years
using: fourteen 0.55-m refractor SATs (at 155\,GHz and below) and four 0.44-m SATs (at 220/280\,GHz), 
with dichroic, horn-coupled transition-edge sensor (TES) detectors in each SAT measuring two of the eight targeted frequency bands between 
30 and 270\,GHz; and one 6-m class 
crossed-Dragone 
``delensing'' LAT, 
equipped with detectors distributed over seven bands from 20 to 270\,GHz.
Measurements at degree angular scales and larger made using refractor telescopes with roughly 0.5-m apertures have been demonstrated to deliver high-fidelity, low-contamination polarization measurements at these scales.  The combination of the SATs with the 6-m LAT therefore provides low-resolution $B$-mode measurements with excellent control of systematic contamination, as well as the high-resolution measurements required for delensing.  The ultra-deep survey SATs and 6-m LAT are to be located at the South Pole to allow targeted observations of the single deep narrow field, with provisions to relocate a fraction of the SATs in Chile if, for example, a high level of $r$ is detected or unforeseen systematic issues are encountered.

The total detector count for the eighteen SATs is 
153,232
with the majority of the detectors allocated to the 85- to 155-GHz bands.  There are 
four
pixel designs. The total number of science-grade 150-mm detector wafers required for eighteen SATs is 
204.

The delensing LAT will have a total TES detector count of
114,432,
with the majority of the detectors allocated to the 95- to 150-GHz bands. There are four pixel designs. The total number of science grade 150-mm diameter detector wafers required is 76.

\item The deep and wide survey covering approximately 70\% of the sky to be conducted over seven years using two 6-m 
crossed-Dragone 
LATs located in Chile, each equipped with 
121,760
TES detectors distributed over eight frequency bands spanning 30 to 270\,GHz. The total number of science grade 150-mm diameter detector wafers required is 152.
\end{itemize}

The total detector count for CMB-S4 is 
511,184, 
and it requires 
432
science grade wafers. This is an enormous increase over the detector count of all Stage-3 experiments combined
and is required to meet the CMB-S4 science goals. 

\subsection*{Project Plan Overview}  

The CMB-S4 Collaboration and the pre-Project Development Group 
of experienced project leaders drawn primarily from the national laboratories,  
jointly contributed to the development of a Work Breakdown Structure (WBS), Organization, Cost Book, Resource Loaded Schedule, and Risk Registry.  The Reference Design and project baseline prepared for this document is the basis for subsequent design and project development work that is now being led by the Interim Project Office.  A permanent Integrated Project Office will be established in 2020 to manage the construction phase which starts in 2021.  

The CMB-S4 Project is a collaborative project, with the scientific Collaboration members serving in technical leadership roles in the project.  This is similar to many successful projects including IceCube, ATLAS, and CMS.  
The project office is responsible for forming partnerships with key stakeholder institutions including DOE National Laboratories, universities, and potential collaborating observatories/projects such as the Simons Observatory, South Pole Observatory, and the CCAT-prime project.  Partnerships are also expected to include foreign institutions participating in the CMB-S4 Collaboration and contributing to the CMB-S4 Project.

The CMB-S4 project total estimated cost is currently \$591.6M (fully loaded and escalated to the year of expenditure) including a 35\% contingency budget. The cost estimate is the full cost, i.e., it does not take credit for use of any legacy infrastructure or for contributions from collaborating institutions supported by private and international partners, e.g., large-aperture telescopes currently under construction in Chile as part of the Simons Observatory, and large- and small-aperture telescopes proposed by international collaborators.  In-kind contributions delivered by private and international partners are expected and would reduce the total cost to NSF and DOE.  It is estimated that the value of in-kind contributions could reduce the total cost of the CMB-S4 project by 20--25\%.

The cost contingency estimate was constructed using input from subject matter experts with previous experience in previous CMB experiments and similar NSF MREFC projects and DOE MIE projects.  As the design, cost estimates, and schedules mature the contingency as a percentage of the base cost estimate is expected to decrease to 30\% or less.  

The project has developed a task based detailed resource loaded schedule which was reviewed by an external panel of experts in December 2018.  The schedule has 1110 activities, 1928 relationships, 5 Level 1, 20 Level 2 and 299 Level 3 Milestones. The estimates follow the guidance in the NSF Large Facilities Manual, NSF 17-066 and the Project Management for the Acquisition of Capital Assets, DOE 413.3b.

A CMB-S4 Risk and Opportunity Management Plan describes the continuous risk and opportunity management process implemented by the project, consistent with DOE O413.3B, ``Project Management for the Acquisition of Capital Assets,'' and the NSF 17-066, ``NSF Large Facilities Manual.''  The plan establishes the methods of assessing CMB-S4 project risk and opportunities for all subsystems as well as the system as a whole. The project is working on mitigations to ensure that the highest risks identified in the risk registry are lowered to reasonable levels on a time scale consistent with the overall project timeline.

The basic operations model for CMB-S4 will be observations with multiple telescopes and cameras distributed across two sites, with observing priorities and specifications optimized for the CMB-S4 science goals, and data from all instruments shared throughout the entire CMB-S4 collaboration.  The operations cost is based on a preliminary bottom-up estimate that includes management, site staff, utilities, instrument maintenance, data transmission, data products, pipeline upgrades, collaboration management, and science analysis. The annual operations cost is \$32M in 2019 dollars, excluding 20 FTE/year of scientist effort supported by DOE research funds. 

\subsection*{Data Analysis and Data Release Plan Overview}  

Many of the key data analysis tools that will be needed for the scientific exploitation of the CMB-S4 data are in fact needed to optimize the final design of the instrument---simulations are particularly important. These tools are included in the Data Management section of the Reference Design and are therefore included as part of the construction phase.  We also outline an operations plan that includes developing the science analysis tools that will be needed for the scientific exploitation of the data. 

The production and release of the CMB-S4 program data deliverables is an integral part of operations, and the plan is for this effort to be supported by NSF and DOE.  These deliverables include the data products from the powerful CMB-S4 Legacy Survey that covers 70\% of the sky.  The data products will include maps of the  temperature and polarization in each of the nine CMB-S4 observing bands, the projected mass map reconstructed from CMB lensing, and source catalogs (e.g., galaxy clusters, active and dusty galaxies, and transient events). These products will be made available through a series of releases to the entire Astronomy and Astrophysics community and will provide a powerful complement to the surveys that will be available or planned at other wavelengths, such as  LSST and {\it WFIRST}.  The CMB-S4 maps and catalogs will also provide exciting targets for detailed follow-up study by the Atacama Large Millimeter/submillimeter Array (ALMA), the {\it James Webb Space Telescope (JWST)}, and other facilities.

The key science analyses, such as the searches for primordial gravitational waves and light relics, will be pursued by the CMB-S4 collaboration with support shared by NSF and DOE. Non-key science analyses, for example studies of the Legacy Survey maps and catalogs, will be carried out by the wider community of laboratory and university scientists with support expected to be provided by individual NSF and DOE awards.

\eject

\tableofcontents


\def\as#1{[{\bf AS:} {\it #1}] }


\eject
\pagenumbering{arabic} 
\setcounter{page}{1}
\chapter{Science Case} 
\label{chap:science}

\definecolor{shadecolor}{rgb}{0.8,0.90,0.95}


\vskip\baselineskip

CMB-S4 will probe fundamental physics and astrophysics with a millimeter-wavelength survey of unprecedented depth over a large sky area.  These measurements will represent a leap forward in the study of the cosmic microwave background (CMB), which in the half century since its first discovery, has again and again transformed our understanding of the early Universe.

CMB experiments on the ground, on balloons, and in space have provided conclusive evidence that our Universe evolved from an early hot, dense state.  They have determined the age and the composition of our Universe with percent-level precision.  They have provided the strongest evidence that dark matter cannot consist of non-luminous baryonic matter. They have measured the polarization of the CMB, and shown us in a model-independent way that the fluctuations we see in CMB intensity were already present at the time of recombination.  To be consistent with Einstein's theory of general relativity,  some process  must have generated these fluctuations long before the moment when our Universe became filled with a hot and dense plasma; i.e., long before the ``hot Big Bang'' began.

Measurements of the CMB continue to provide us with a remarkable opportunity to study the Universe over its history.  Although CMB observations have already harvested nearly all the information accessible in the primary temperature anisotropies,
 current experiments have only begun to make precise measurements of the CMB polarization.  The secondary anisotropies, such as the weak gravitational lensing of the CMB by large-scale structure and scattering by the thermal and kinematic Sunyaev-Zeldovich (tSZ and kSZ) effects, also hold tremendous promise to improve our understanding of astrophysics and cosmology.

CMB-S4 will push beyond these new frontiers in CMB science.  It will exploit the enormous potential of CMB measurements to once again transform our understanding of the early Universe and of particle physics.  In doing so, it will fulfill the goals set out in the 2010 Astronomy and Astrophysics Decadal Survey and the 2014 report of the Particle Physics Project Prioritization Panel.  Specifically, it will search for primordial gravitational waves and for light relic particles.  CMB-S4 will also constrain neutrino properties and provide critical measurements of the evolution of cosmic structure from the early Universe to the present day, advancing a key goal of the 2010 Decadal Survey.

CMB-S4 will also be built as an astronomical survey machine.  To date, CMB experiments have produced the only wide-area mm-wave
sky surveys, providing new insights
in astrophysics.  CMB-S4 will image a large fraction of the sky, including
much of the Galaxy, with high-fidelity measurements of intensity and linear polarization, on scales from arcminutes to many degrees.

These wavelengths reveal a wide variety of astrophysical effects.
Compton scattering of mm-wave light on electrons in the late Universe provides
key information about the ionization history of the Universe and thermal evolution of diffuse gas.
Individual galaxy clusters can be detected through this Compton scattering, and large samples of
galaxies can be detected through either thermal emission from their dust or by synchrotron 
emission from active galactic nuclei (AGN). By surveying large areas of sky, these surveys are
able to probe large volumes and detect large samples of objects that are too rare to be found in smaller-area surveys. 

CMB-S4 will provide measurements on a wide variety
of timescales.  Typical CMB experiments survey the same patch of
sky repeatedly to build up the signal-to-noise ratio. This leads to sampling
on timescales as fast as multiple times per second, with multiple observations extending over
several years. CMB-S4's unprecendented depth means that it should observe many objects in the mm-wave transient and variable sky, including gammay-ray bursts and blazars.  It should also observe the thermal emission from
moving Solar System objects such as asteroids, dwarf planets, and unknown bodies in the outer
Solar System. 

CMB-S4 will generate measurements with broad and diverse scientific applications.  The following sections will give a tour
of some of these applications and touch on many ideas in particle physics, cosmology, astrophysics, and astronomy.


\section{Summary of science goals}

We have organized the rich and diverse set of CMB-S4 scientific goals into four themes: 

\begin{enumerate}
\begin{shaded*}
\setlength{\itemsep}{0mm} 
\setlength{\parsep}{0mm}
\item \textit{primordial gravitational waves and inflation;}
\item \textit{the dark Universe;}
\item \textit{mapping matter in the cosmos;} 
\item \textit{the time-variable millimeter-wave sky.} 
\end{shaded*}
\end{enumerate}
The first two science themes relate to fundamental physics and have been primary drivers of the CMB-S4 concept from the beginning. The final two themes have emerged through engagement with the wider astronomical community and studies of broader scientific opportunities made possible by a millimeter-wave survey of unprecedented depth and breadth, and together we refer to them as our Legacy Survey themes. 
We summarize the four science themes in this short section and present the full science case in the remainder of the chapter.
 
\paragraph{Primordial gravitational waves and inflation.}

We have a historic opportunity to open up a window to the primordial Universe. If the predictions of some of the leading models for the origin of the hot big bang are borne out, CMB-S4 will detect primordial gravitational waves. This detection would provide the first evidence for the quantization of gravity, reveal new physics at the energy scale of grand unified theories, and yield insight into the symmetries of nature and possibly into the properties of quantum gravity. Conversely, a null result would rule out large classes of models and put significant strain on the leading paradigm for early-Universe cosmology, the theory of cosmic inflation.

Cosmic inflation refers to 
a period of accelerated expansion prior to the hot big bang. During this epoch, quantum fluctuations were imprinted on all spatial scales in the cosmos. These fluctuations seeded the density perturbations that developed into all the structure in the Universe today. While we cannot yet claim with high confidence that the Universe underwent cosmic inflation, the simplest models of inflation are exceptionally successful at matching the data. Specifically, these predictions include small mean spatial curvature and initial density perturbations drawn from a nearly Gaussian distribution with a variance that is slightly larger on large scales than on small scales. Each of these predictions has been verified to high precision. 

Tantalizingly, the observed (weak) scale-dependence of the amplitude of density perturbations has quantitative implications for the detection of primordial gravitational waves. In the simplest class of models, the amplitude of primordial gravitational waves is comparable to the deviation from scale invariance, quantified by $n_{\rm s}-1$. However, {\em all\/} inflation models that naturally explain the observed $n_{\rm s}-1$ value, and that also have a characteristic scale larger than the Planck mass, generate primordial gravitational waves above the 95\% confidence upper limit that CMB-S4 can set (see Fig.~\ref{fig:nsrp01} for example models). A well-motivated sub-class within this set of models is detectable by CMB-S4 at 5$\sigma$. 

\begin{figure}[!th]
\begin{center}
\includegraphics[width=6in]{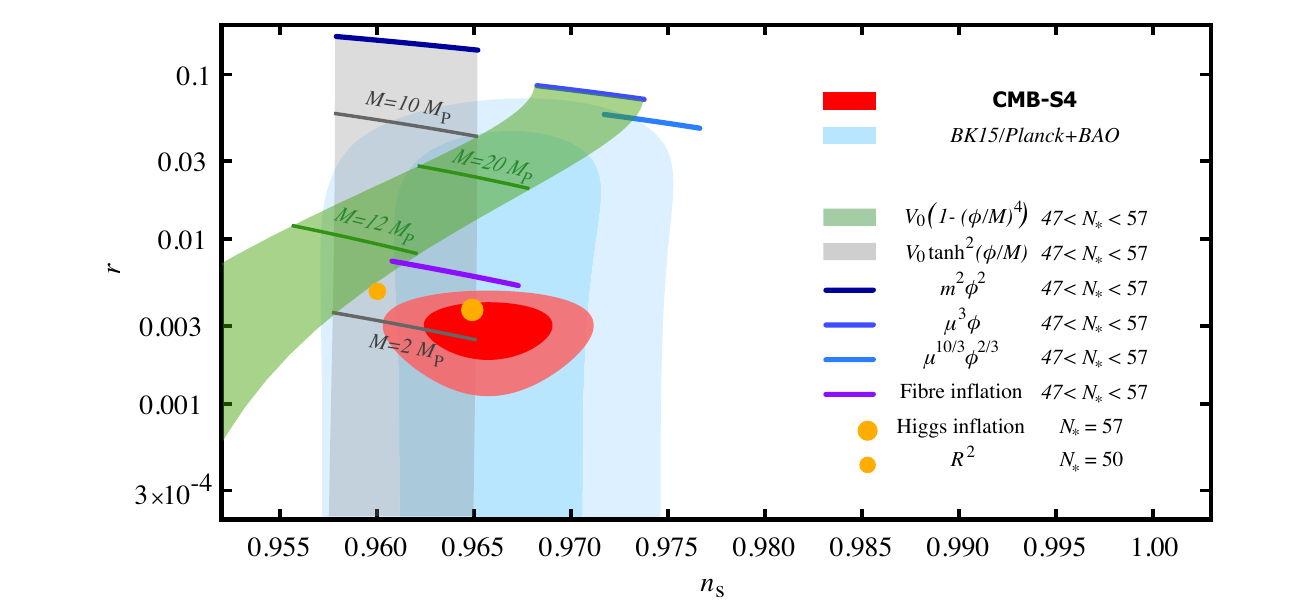}
\end{center}
\caption{Forecast of CMB-S4 constraints in the $n_{\rm s}$--$r$ plane for a fiducial model with $r=0.003$. 
 Also shown are the current best constraints from a combination of the BICEP2/{\em Keck Array\/} experiments and \planck\ \cite{Array:2015xqh}. Models that naturally explain the observed departure from scale invariance separate into two viable classes: monomial and plateau. The monomial models ($V(\phi)=\mu^{4-p}\phi^{\,p}$) are shown for three values of $p$ as blue lines for $47<N_\ast<57$ (with the spread in $N_\ast$ reflecting uncertainties in reheating, and smaller $N_\ast$ predicting lower values of $n_{\rm s}$). This class is not completely ruled out by the data, but is disfavored. The plateau models divide into those with plateaus near the scalar field origin, for which we include the quartic hilltop (green band) as an example, and those with plateaus away from the origin, for which we include the $tanh$ form (gray band) as an example, as this form arises in  a sub-class of $\alpha$-attractor models~\cite{Kallosh:2013hoa}. Some particular realizations of physical models in the plateau class are also shown: the Starobinsky model~\cite{Starobinsky:1980te} and Higgs inflation~\cite{Bezrukov:2007ep} (small and large orange filled circles, respectively) and fibre inflation~\cite{Cicoli:2008gp} (purple line). The differing choices of $N_\ast$ for Higgs and Starobinsky reflect differing expectations for reheating efficiency. 
}
\label{fig:nsrp01}
\end{figure}

Because the Universe has expanded by a tremendous amount since the time when primordial perturbations were imprinted, CMB observations can probe physics at extraordinarily small length scales, {\em up to $10^{10}$ times smaller than those probed in terrestrial particle colliders}. The CMB provides a unique window to test new phenomena at these length scales.  The observational requirement is also clear: we must measure the polarization to high precision on angular scales from several arcminutes to degrees.

Primordial gravitational waves source an odd-parity fluctuation pattern in the polarization across the sky, called ``$B$ modes'' by analogy with electromagnetism (while scalar density perturbations source only an even parity polarization pattern, called ``$E$ modes'').  To measure primordial $B$ modes, we must observe at multiple frequencies to remove Galactic foreground contamination and also measure small angular scales to remove the (non-primordial) gravitational lensing-generated polarization $B$ modes.  The CMB-S4 reference design has sufficient sensitivity to detect or tightly constrain the degree-scale $B$ modes generated by gravitational waves in many models, and to measure the  amount of gravitational waves (tensor perturbations), detecting or setting an upper limit on the tensor-to-scalar ratio $r$.  With an order of magnitude more detectors than precursor observations, and exquisite control of systematic errors, we will improve upon limits from previous observations by a factor of 5, allowing us to either detect primordial gravitational waves or rule out a broad class of models with a super-Planckian characteristic scale.

Complementary to the search for gravitational waves, CMB-S4 will provide exquisite measurements of primordial {\em density\/} fluctuations via $E$ modes. The polarization sensitivity will surpass current measurements of $E$-mode polarization, which are far from being sample-variance-limited.  Because polarization has lower Galactic foregrounds than temperature, we will improve measurements across the angular scales already observed in temperature, and push to yet smaller angular scales. These polarization measurements will significantly extend and enhance searches for non-power-law features in the primordial power spectrum, small variations in the equation of state, and small departures from Gaussianity.
The CMB is the most robust observable for non-Gaussianities to date and CMB-S4 will provide the tightest constraints on the most compelling signatures, improving the constraints from the {\it Planck\/} satellite. Non-Gaussianities can also arise in models with undetectably small gravitational wave production, and provide an independent handle on the early Universe. 
Non-Gaussianity can also be measured via cross correlation of the CMB-S4 mass map with galaxy surveys, a measurement that has the potential to rule out a large class of inflationary models.

\paragraph{The dark Universe.}
In the standard cosmological model, about ninety five percent of the mass--energy density of the Universe is in dark matter and dark energy.  With CMB-S4 we can address numerous questions about these dark ingredients, such as: How is the dark (plus baryonic) matter distributed on large scales? Does the dark matter have non-gravitational interactions with baryons? And are there additional unseen components beyond dark matter and dark energy? 

\begin{figure}[!th]
\begin{center}
\includegraphics[width=6in]{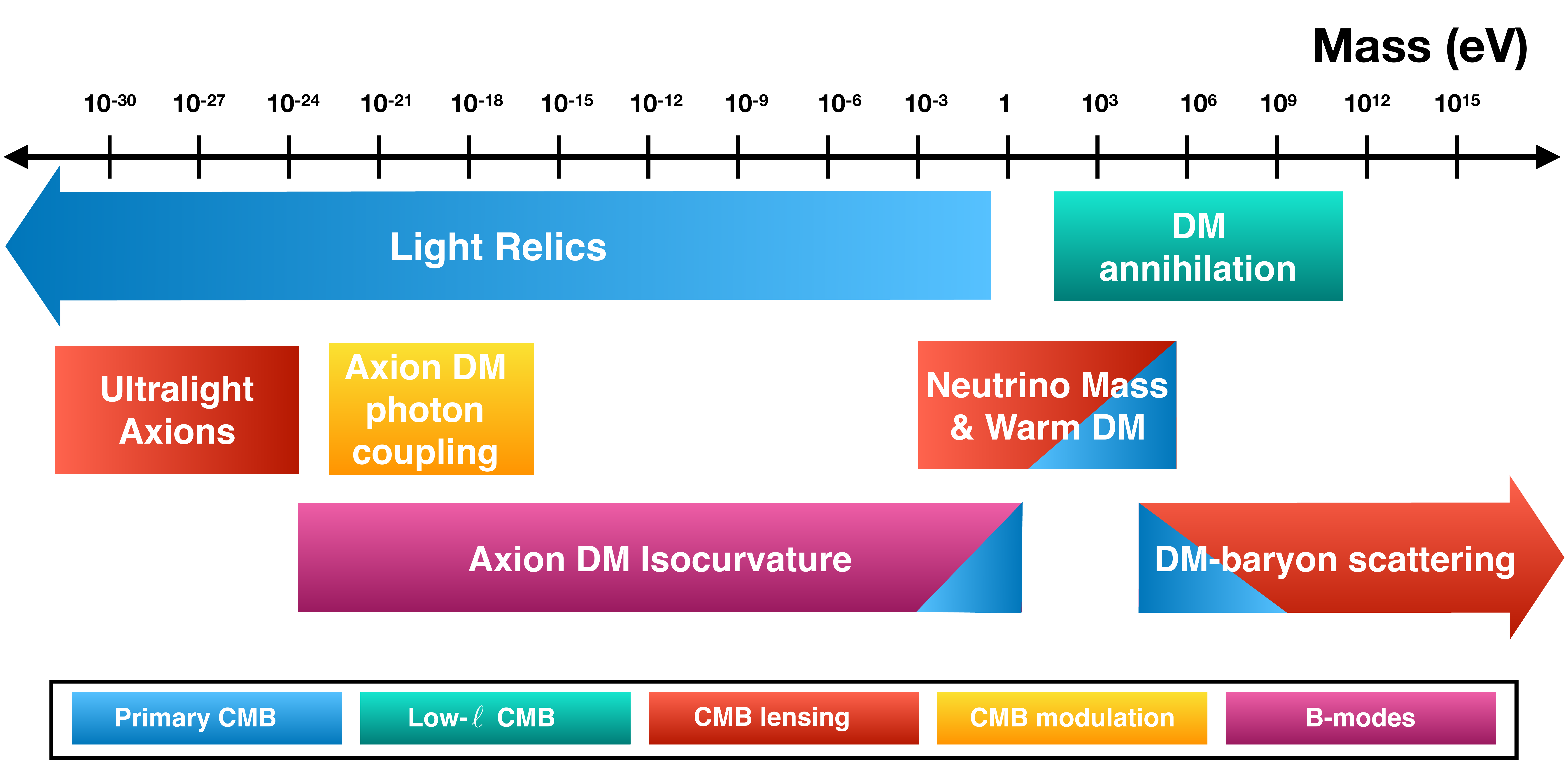}
\end{center}
\caption{CMB-S4-enabled exploration of light relics, axions, neutrino mass, and dark matter properties.    In each case, there is a window in the mass of the relevant particle where the CMB is particularly sensitive.  Each such region is shown in a color (or colors) representing the observable(s) that drives the constraint.   The primary CMB anisotropies at high-$\ell$ (blue) is particularly sensitive to light relics, as discussed in Section~\ref{sec:lr}, and properties of the dark matter discussed in Section~\ref{sec:dm}.  Low-$\ell$ modes (green) most directly impact constraints on dark matter annihilation.   CMB lensing reconstruction (red) is a sensitive probe of matter in the late Universe, particularly effects that suppress clustering power.  Axion dark matter can create additional modulations of the CMB polarization angles (yellow) through their coupling to photons.  A detection of primordial gravitational waves (pink) would severely constrain the QCD axion because of the implied high scale of inflation.   }
\label{fig:mass_summary}
\end{figure}

Light relic particles are one well-motivated possibility.  These additional components to the dark sector are light particles produced in the early Universe, and are sometimes referred to as ``dark radiation.''  Many extensions of the Standard Model~(SM) predict such light relics, including axion-like particles and sterile neutrinos~\cite{Abazajian:2012ys, Essig:2013lka, Alexander:2016aln}. For large regions of the unexplored parameter space, these light particles are thermalized in the early Universe. To date, CMB observations by the {\it Planck\/} satellite can probe light particles that departed from equilibrium (``froze out'') as early as the first 50~micro-seconds of the Universe. With CMB-S4 we can push back this frontier by over a factor of 10,000, to the first fractions of a nanosecond. 

CMB-S4 will achieve sensitivity to relics that froze out well before the quark-hadron phase transition (the epoch when the Universe cooled sufficiently that quarks became locked into hadrons like neutrons and protons).  The contribution of light relics to the energy density leads to observable consequences in the CMB temperature and polarization anisotropy. This is often parameterized with the ``effective number of neutrino species,'' $\Neff$.  The collective influence of the three already-known light relics (the three families of neutrinos) has already been detected at high significance. Current data are only sensitive enough to detect additional relics that froze out after the quark-hadron transition, and Stage-3 CMB experiments can only push somewhat into that epoch, so CMB-S4's ability to probe times well before that transition is a major advance.

In addition to precise constraints on $\Neff$, CMB-S4 will give an {\it independent\/} high-precision measurement of the primordial helium abundance, $Y_\mathrm{p}$. This is particularly useful since $Y_\mathrm{p}$ is sensitive to $\Neff$ a few minutes after reheating, while the CMB power spectrum is affected by $\Neff$ prior to recombination, about 370,000 years later.  Measuring the radiation content at these well-separated times provides a window onto any non-trivial evolution in the energy density of radiation in the early Universe.  Furthermore, $\Neff$ and $Y_\mathrm{p}$ are sensitive to neutrino physics and  physics beyond the Standard Model in related, but different ways, allowing even finer probes of particle physics, especially in the neutrino and dark sectors.

CMB-S4 will also enable a broader exploration of the dark Universe in combination with other probes, often significantly enhancing them by breaking their intrinsic degeneracies.  It will improve or detect various possibilities for the dark matter properties beyond the simplest cold-dark-matter scenario, as described in Fig.~\ref{fig:mass_summary}. It will add to dark energy constraints through precision measurements of the primordial power spectrum (where dark energy physics enters through projection effects), through precision measurements of the lensing convergence power spectrum, through the CMB-lensing-derived mass calibration of galaxy clusters, and through CMB lensing tomography.

\commentout{
\subsection{(Old version) Light Relics and Other Dark Sector Ingredients}

The abundances of baryons, photons, neutrinos, and (possibly) dark matter were determined during the hot thermal phase that dominated the early Universe. It is the abundances of these particles and the forces between them that determine the conditions of the Universe that we see today. 
There is strong motivation to determine if there is more to this list of ingredients, in particular more light relics: light particles produced in thermal equilibrium. 
Many extensions of the Standard Model~(SM) predict new light particles, including axions, axion-like particles, and sterile neutrinos~\cite{Abazajian:2012ys, Essig:2013lka, Alexander:2016aln}. For large regions of unexplored parameter space these light particles are thermalized in the early Universe. To date, using CMB observations by the {\it Planck\/} satellite, we can probe the creation of light particles back to the first $\simeq 50$ micro-seconds. With CMB-S4 we can push back this frontier by over a factor of 10,000 to the first fractions of a nanosecond. 

Light particles are ubiquitous in models of the late Universe as well.  They may form the dark matter (e.g., axions), be an essential ingredient of a more complicated dark sector as the force carrier between dark matter and the Standard Model (or itself), or provide a source of dark radiation for a dark thermal history.  Furthermore, they could explain discrepancies in the measurements of the Hubble constant~$H_0$, the amplitude of large-scale matter fluctuations~$\sigma_8$, and the properties of clustering on small scales.  

Their contribution to the energy density, often parameterized with the ``effective number of neutrino species,'' $\Neff$, leads to observable consequences in the CMB temperature and polarization anisotropy. The collective influence of 3 light relics (the 3 families of neutrinos) has already been detected at high significance. 
Current data are sufficiently sensitive to detect relics that froze out after the quark-hadron phase transition. Stage 3 experiments push somewhat into the epoch of the transition, and CMB-S4 will achieve sensitivity to relics that froze out well before the transition, significantly increasing discovery potential.

In addition to precise constraints on $\Neff$, another advance of CMB-S4 will be an {\it independent\/} high-precision measurement of the primordial helium abundance, $Y_\mathrm{p}$. This is particularly useful since $Y_\mathrm{p}$ is sensitive to $\Neff$ a few minutes after reheating, while the CMB power spectrum is affected by $\Neff$ prior to recombination, about 370,000 years later.  Measuring the radiation content at these well-separated times provides a window onto non-trivial evolution in the energy density of radiation in the early Universe.  Furthermore, $\Neff$ and $Y_\mathrm{p}$ are sensitive to neutrino and BSM physics in related, but different ways, allowing even finer probes of physics beyond the particle physics standard model, especially in the neutrino and dark sectors.

The search for light relics exists in the larger context of the exploration of the dark sector enabled by CMB-S4; see Fig.~\ref{fig:mass_summary}. [Expand]
}

\paragraph{Mapping matter in the cosmos.}
Matter in the Universe can be sorted into two categories, ``normal'' or ``baryonic''
matter that is described by the Standard Model of particle physics
and ``dark" matter that has only been observed to interact gravitationally.
Observations indicate there is
more than five times more dark matter than baryonic matter,
and most of the  baryonic matter is in the form of hot ionized gas rather
than cold gas or stars.
CMB-S4 will be able to map out
normal and dark
matter separately by measuring the fluctuations in the
total mass density (using gravitational lensing) and the ionized gas density (using Compton scattering).

\begin{figure}[!th]
\begin{center}
\hspace{-0.4in}
\begin{tabular}{m{0.1in}m{1.9in}m{1.9in}m{1.9in}}
      &  \multicolumn{1}{c}{Planck fidelity} & \multicolumn{1}{c}{Simulated sky} & \multicolumn{1}{c}{CMB-S4 fidelity} \\
    \rotatebox[origin=c]{90}{Lensing} &
    \includegraphics[width=2in]{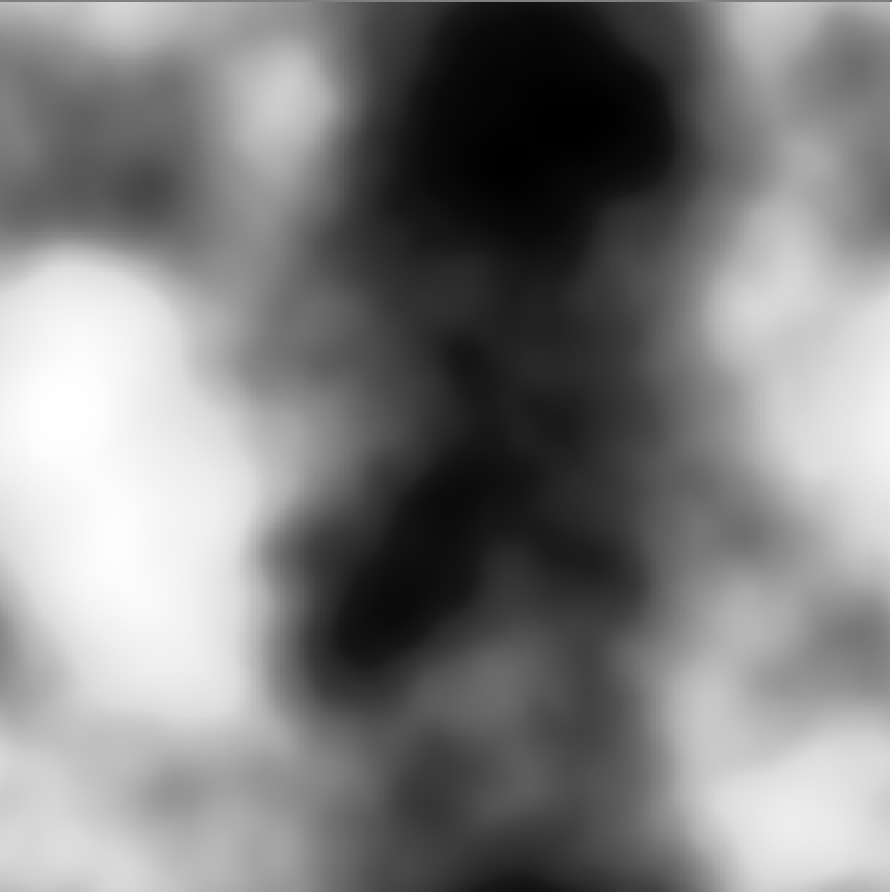} &
    \includegraphics[width=2in]{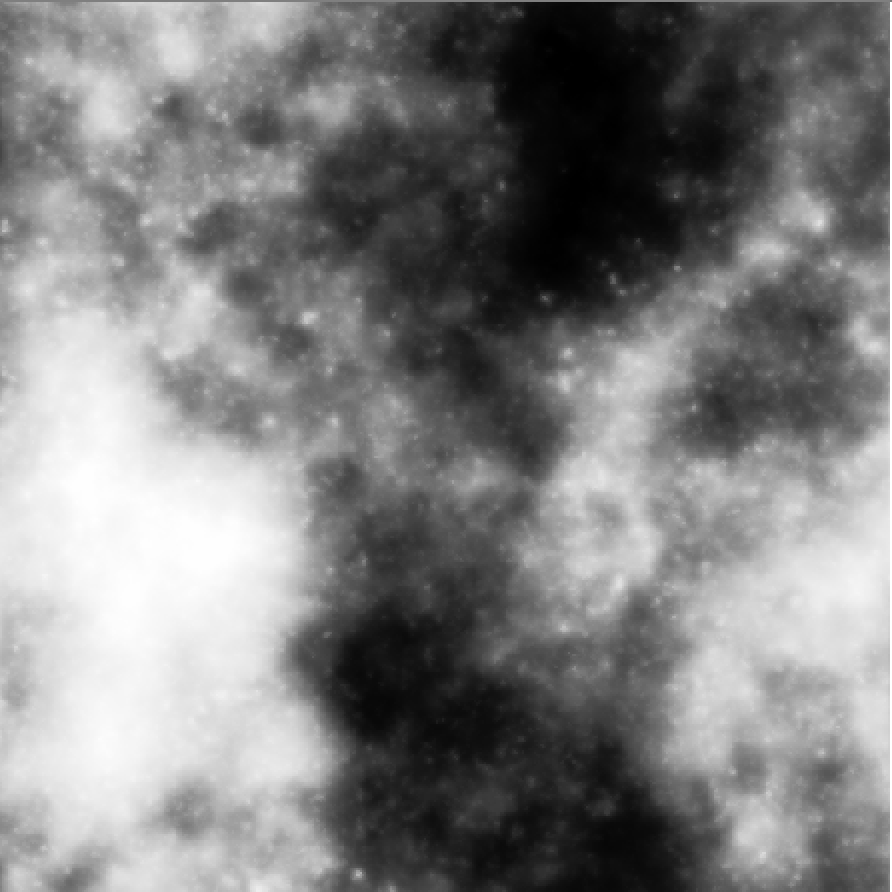} &
    \includegraphics[width=2in]{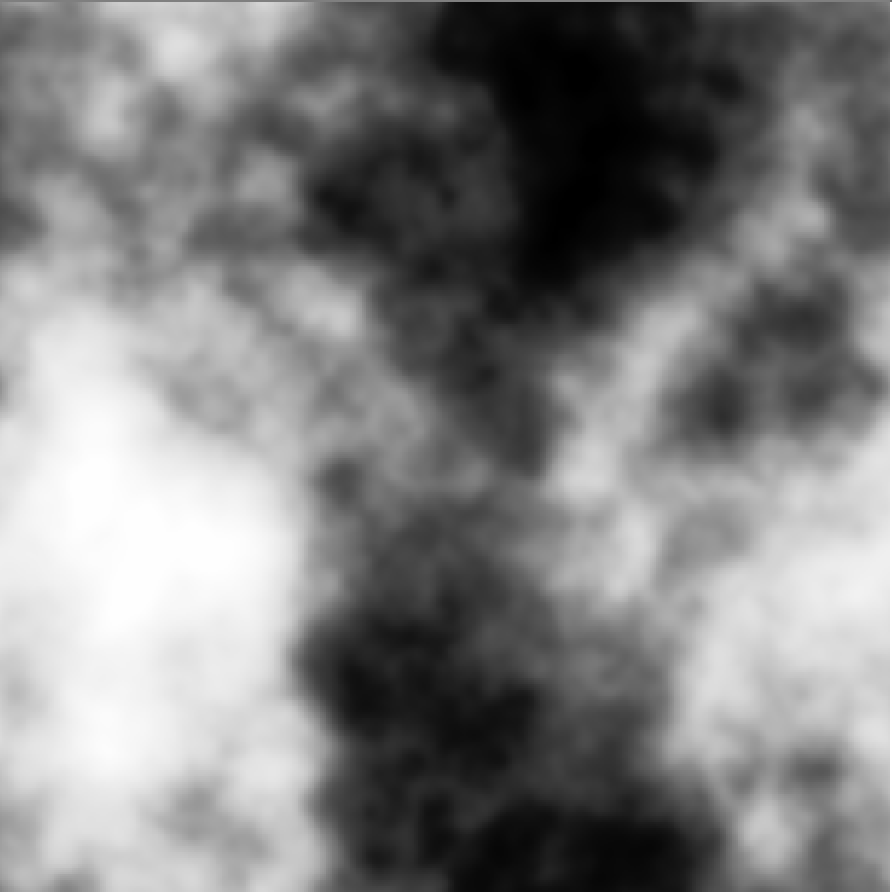} \\
        \rotatebox[origin=c]{90}{Compton-$y$} &
   	\includegraphics[width=2in]{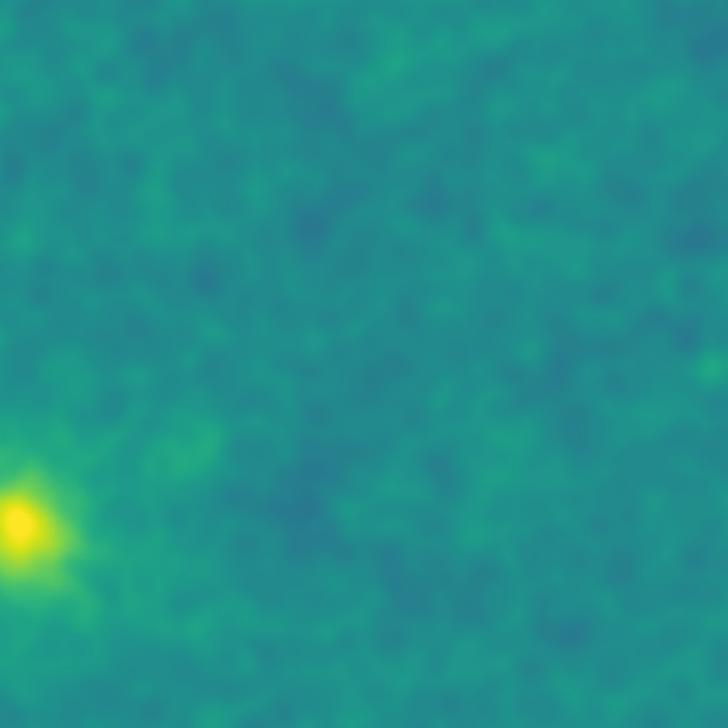} &
	\includegraphics[width=2in]{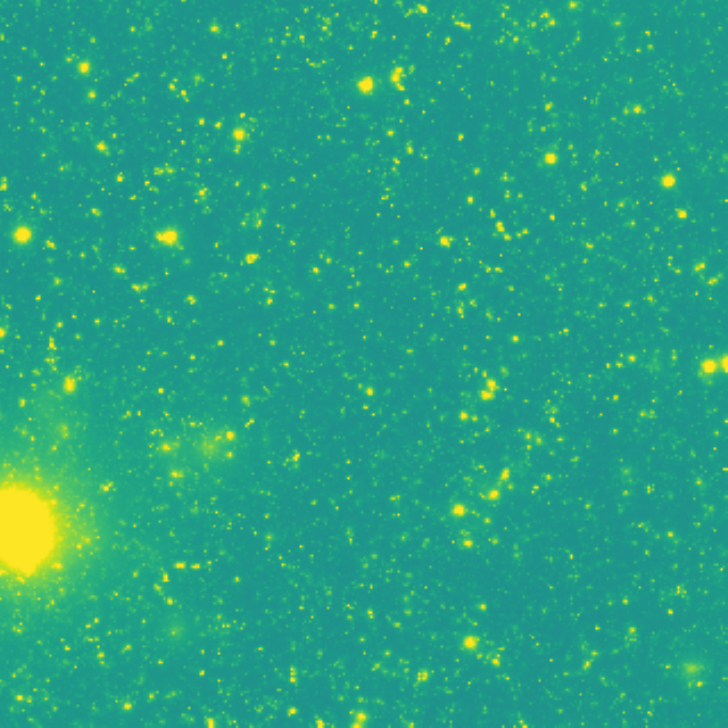} &
	\includegraphics[width=2in]{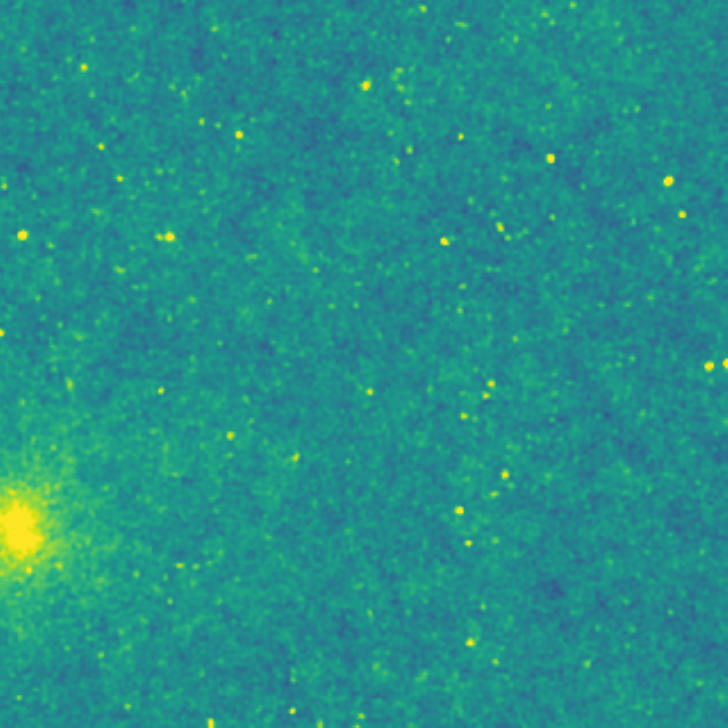} 
 \end{tabular}
\end{center}

\caption{Example lensing-deflection maps (top) and thermal SZ (Compton $y$-maps, bottom)
reconstructed with {\it Planck\/} (left) and CMB-S4 data (right).  
The center panels show a 25 deg${}^2$ patch of the all-sky lensing-deflection field in the WebSky
simulations (top) and Compton $y$ (bottom).
The left panels show Wiener-filtered maps of the signal after adding (Gaussian) noise
and residual foregrounds with levels corresponding to the {\it Planck\/} 2018 
lensing-deflection map (top) and the {\it Planck\/} 2015 ``Needlet Internal Linear
Combination'' tSZ map (bottom).  The right panel shows analogous Wiener-filtered maps with
noise expected for CMB-S4 (top) and residual foregrounds determined by the CMB-S4 + {\it Planck\/} tSZ noise power
spectrum in Fig.~\ref{fig:s4_yy}.
The significantly higher fidelity of the CMB-S4
reconstruction is evident.}
\label{fig:mass_y_map}
\end{figure}

Gravitational lensing of background sources by intervening gravitational potentials leads
to detectable distortions that can be used to reconstruct fluctuations in the
mass density. Virtually all of the the density 
fluctuations within the observable Universe leave an imprint when the CMB is the background source.
The map resulting from CMB lensing reconstruction will be wide-area, highly sensitive,
and extremely well-calibrated.

On its own, we can use this map to precisely measure the amplitude of large-scale
structure at intermediate redshifts, with important applications to dark energy, 
modified gravity, and studies of neutrino masses.
In concert with catalogs of objects, we can use this map to weigh samples (of e.g., galaxies and galaxy clusters) to as high a redshift as such sources can be found.
 The technique of CMB lensing tomography, enabled by CMB-S4 and galaxy catalogs from---for example---the Large Synoptic Survey Telescope (LSST), will allow for the creation of mass maps in broad redshift slices out to redshifts as high as 5,  making possible new precision tests of cosmology. 
Such results explore the connection between
visible baryons and the underlying dark-matter scaffolding. 
In conjunction with cosmic-shear surveys (e.g., LSST) that measure the low-redshift
mass distribution, a map of the high-redshift mass distribution can be constructed, 
gaining new insight into the first galaxies.
By calibrating cluster masses
at high redshift, the abundance of galaxy clusters can be used as an additional
probe of dark energy and neutrino masses.

Low-redshift structure acts to lens both the CMB and the images of intermediate-$z$ galaxies.  Detailed comparisons  
provides a valuable cross-check on galaxy shear measurement calibration 
and enables geometric tests using the longest possible lever arm.

Most of the baryons in the late Universe are believed to be in a diffuse ionized plasma 
that is difficult to observe. This ionized plasma can leave imprints in the CMB
through Compton scattering, the so-called Sunyaev-Zeldovich effects. The two leading
variants are either a spectral distortion from hot electrons interacting with the
relatively cold CMB (thermal SZ or tSZ), or a general redshift or blueshift of the scattered
photons due to coherent bulk flows along the line of sight (kinematic SZ or kSZ). 

The nature
of the scattering makes the tSZ independent of redshift.
With the deep and wide field covering a large amount of volume and
the ultra-deep field imaging lower-mass clusters, CMB-S4 will be an
effective probe of
the crucial regime of $z \gtrsim 2$,
when galaxy clusters were vigorously accreting new hot gas while at the same time
forming the bulk of their stars.
The CMB-S4 catalog
will be more than an order of magnitude
larger than current catalogs based on tSZ or X-ray measurements, and 
will contain an order of magnitude more clusters at $z > 2$ than will be discovered with Stage 3 CMB
experiments.
CMB-S4 will also measure the diffuse tSZ signal everywhere on the sky and make a
temperature-weighted map of
ionized gas
which can be used to measure the average
thermal pressure profiles
around
galaxies and groups of galaxies.

\begin{figure}[!t]
\begin{center}
  \includegraphics[width=6in]{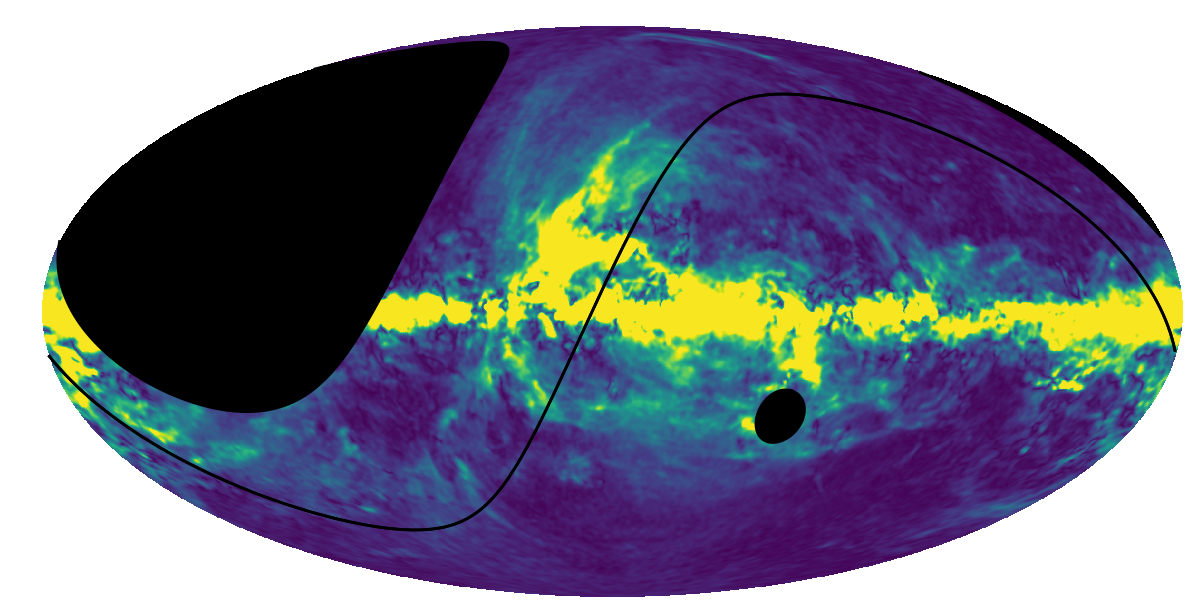}
\caption{CMB-S4 survey area, in Galactic coordinates, with the ecliptic plane marked as the solid line.}\label{fig:survey}
\end{center}
\end{figure}

CMB-S4 will measure the kSZ effect, which will be combined with data from
other surveys to make maps of the projected electron density
around samples of objects.
Applications of these maps include
measuring ionized gas as a function of radius, directly constraining
the impact of feedback from active galactic nuclei and supernovae on the
intergalactic medium  and
constraining theories of modified gravity with the bulk flow amplitude as a function of separation.

Even without overlapping galaxy catalogs, the kSZ signal adds extra small-scale
power that is significantly non-Gaussian.
Some of this excess power and non-Gaussianity will be coming from the relatively
local Universe where the galaxy catalogs overlap, but there should also be a substantial 
signal coming from the epoch of reionization. By directly probing the ionized
gas distribution, these measurements are completely complementary to the 
measurements of the neutral gas that can be obtained with redshifted Ly-$\alpha$
or redshifted 21-cm studies.

In the course of its survey, CMB-S4 will catalog the emission from galaxies in the mm-wave band, including AGN and dusty star-forming galaxies.  The matter in our own Galaxy will also be mapped in intensity and linear polarization over a
large fraction of the sky, with extracted images of synchrotron and dust emission
with high fidelity on scales ranging from arcminutes to
several degrees. 

\paragraph{The time-variable millimeter-wave sky.}

There have been relatively few studies of the variable sky at mm-wavelengths, with
only one systematic survey done to date (by a CMB experiment \cite{Whitehorn2016}).  
Known contributors to the time-varying sky are transient events, Solar System
objects, and variable AGN (especially blazars).

Targeted follow-up observations of gamma-ray bursts, core-collapse supernovae, tidal disruption events,
classical novae, X-ray binaries, and stellar flares have found that there are many transient events
with measured fluxes that would make them detectable by CMB-S4.  A systematic survey of the mm-wave sky with a
cadence of a day or two over a large fraction of the sky would be an excellent
complement to other transient surveys, filling a gap between radio and optical
searches. Gamma-ray burst afterglows can be detected within a few hours of the burst in many cases, and there is a possibility of capturing mm-wave afterglows that have no corresponding gamma-ray trigger either from the geometry of relativistic beaming and/or from sources being at very high redshift.  

\begin{figure}
\begin{center}  
\includegraphics[width=6in]{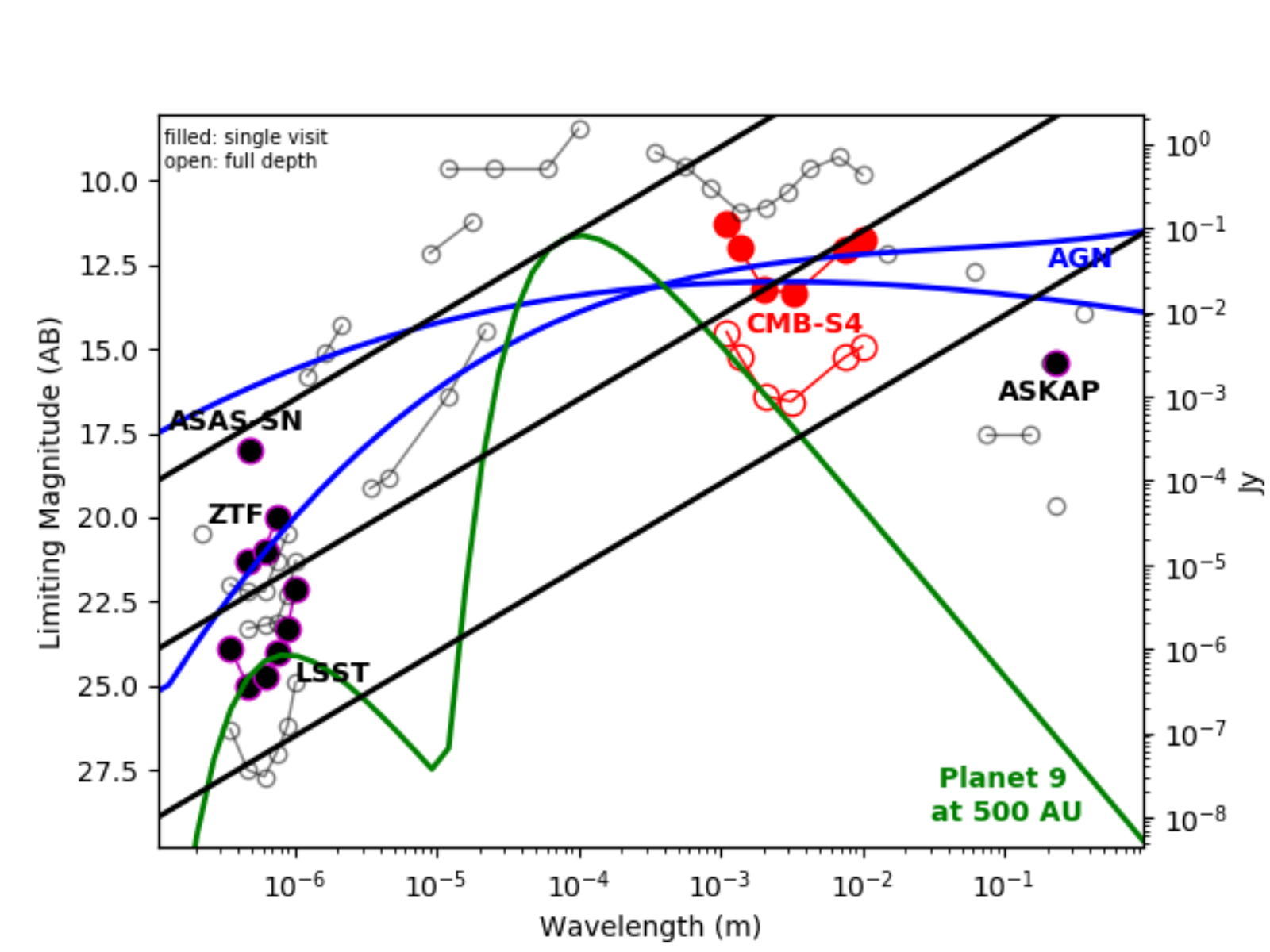}
\end{center}
\caption{Filled circles show 5$\sigma$ limiting magnitudes for 
transient surveys (ASAS-SN, Zwicky Transient Facility, Large Synoptic
Sky Survey, Australia SKA Pathfinder, all in black, CMB-S4 weekly flux limit in red) 
over a large fraction of the
sky. Diagonal lines indicate constant $\nu S_\nu$, lines separated by factors of 100: green
shows a Neptune-mass planet at 500 AU; and blue lines show SEDs corresponding to a quasar or blazar,
normalized to be representative in flux of the population measurable in daily CMB-S4 maps. Open
circles show the coadded $5\sigma$ depth for past and near-future surveys that cover large fractions of the sky, with CMB-S4 again shown in red. The ultra-deep CMB-S4 survey (not shown) is a factor of 2-6
deeper than the deep and wide survey, depending on the wavelength.}
\label{fig:transients}
\end{figure}

Thermal emission from planets, 
dwarf planets, and asteroids have been measured at these wavelengths, and since such  
sources have high proper motions, they should be easily differentiated from the 
relatively stationary extrasolar sky. Using the thermal emission rather than
reflected light has several complementary aspects: the fall-off with distance is less severe,
providing unique information on possible large objects in the distant reaches of
the Solar System; the physical information available is also very different, 
measuring long-wavelength emissivity rather than optical reflectivity; and with long-time
baselines for observation it will be possible to build up rotation curves for a large 
number of objects, enabling detailed comparison with the optical and infrared versions.

CMB-S4 will play an active role in multi-messenger astronomy.
Accreting black holes are known to be highly variable. A CMB survey can provide
a long baseline with high time sampling in both intensity and linear polarization.
This will create a mm-wave archive for multi-messenger astronomy, in particular for
future blazars that are discovered to be sources of high-energy neutrinos
(such as the blazar TXS 0506+056, thought to be associated with the IceCube event
IC170922A).
With a large catalog of time-variable blazars, it will be possible to derive
detailed variability statistics over several years with nearly daily monitoring for 
both the detected objects and the sources that are observed to {\it not\/} be neutrino sources.
Additionally, the natural wide-area nature of the survey will make it straightforward
to search for gravitational wave sources that happen to be poorly localized. 
Although the first binary neutron star merger, GW170817, was not detected at millimeter
wavelengths, this was likely due to the low density of the merger environment 
\cite{alexander2017}. There is reason to expect, based on observations of short gamma-ray bursts,
that at least some mergers can occur in denser environments, which will make them fainter in the optical, {\it but\/} enhance their mm
emission \cite{berger2014}.

\newpage
\section{Primordial gravitational waves and inflation}
\label{sec:PGWandI}

\begin{shaded}
\emph{The first science theme addresses physical processes in the very early Universe responsible for the origin of all structure.}  

CMB-S4 has the potential to detect a pristine relic from the primordial Universe, and to unravel the mechanism responsible for the generation of primordial perturbations.  Primordial gravitational waves at the time of recombination leave an imprint on the polarization pattern of the cosmic microwave background. 

Cosmologists widely regard inflation, a period of nearly exponential expansion, as the most compelling paradigm for the very early Universe. Many of the predictions of the simplest models of inflation have been verified. For example, we have detected at high significance the predicted small departure from scale invariance of the density perturbations. The sign and size of the departure from scale invariance implies that measurements of or constraints on the gravitational wave amplitudes by CMB-S4 will provide key information about the early Universe, as explained below.

Because gravitational waves are fluctuations in the spacetime metric rather than the density, such a detection would open a new window into the early Cosmos and transform our understanding of fundamental physics. For the foreseeable future, precise measurements of CMB polarization are our only way to detect primordial gravitational waves. Even an upper limit from CMB-S4 would provide invaluable insights into the first moment of our Universe.

In addition, CMB-S4 will provide unprecedented constraints on primordial fluctuations in general.  The simplest models of inflation predict that
the initial fluctuations in the number densities of the 
various components (dark matter, baryons, photons, and neutrinos) are completely
correlated.  Such fluctuations are adiabatic.
Other models predict fluctuations in some components that are anti-correlated with other components.  Such fluctuations are ``isocurvature'' fluctuations, and they could come in a wide variety of possible forms. To date, no experiment has detected isocurvature fluctuations, and CMB-S4 will substantially improve upon current constraints.

Inflation produces nearly Gaussian fluctuations, but many models make a definite prediction for a small amount of non-Gaussianity.  Discovery of that non-Gaussianity would present a true breakthrough, on par with a detection of primordial gravitational waves. 
{\it Planck} \cite{Ade:2013ydc,Ade:2015ava} showed that the
non-Gaussianity of the primordial fluctuations is indeed small; however, impressive as these constraints are, they do not yet reach the theoretically motivated thresholds that CMB-S4 can achieve. 
CMB-S4 will have multiple ways to measure non-Gaussianity, particularly in conjunction with large-scale structure measurements, opening an important range of parameter space. 
\end{shaded}

\subsection{Primordial gravitational waves}
\label{sec:gw}
Stars and galaxies formed through gravitational collapse from primordial density perturbations that were generated in the very early Universe. The detailed mechanism that created these density perturbations remains unknown, but one may expect it to not only produce density perturbations but also gravitational waves. The amplitude of the gravitational wave signal provides crucial information about the processes responsible for the origin of structure. By measuring or constraining the amplitude of primordial gravitational waves CMB-S4 will provide invaluable information about the origin of all structure in our Universe.

Our quantitative discussion of primordial graviational waves begins with the line element for a perturbed Friedmann-Lema\^{\i}tre-Robertson-Walker (FLRW) universe, which in the  ADM (Arnowitt-Deser-Misner) formalism \cite{Arnowitt:1962hi} is given by
\begin{eqnarray}
\label{eq:metric}
ds^2&=&-N^2dt^2 +h_{ij}(dx^i+N^idt)(dx^j+N^jdt)\,,\nonumber\\
h_{ij}&=&a^2(t)[e^{2\mathcal{R}}\delta_{ij}+\gamma_{ij}]\,.
\end{eqnarray}
The Hamiltonian and momentum constraints determine the lapse $N$ and the shift $N^i$ in terms of the dynamical scalar and tensor degrees of freedom $\mathcal{R}$
and $\gamma_{ij}$. In general $h_{ij}$ may also contain vector perturbations, but these rapidly decay and can be neglected unless they are actively sourced, e.g., by cosmic strings. 

Because the equations of motion for the fluctuations are invariant under translations, and the fluctuations are small enough to be treated perturbatively, it is convenient to work with the Fourier transforms
\begin{equation}
\label{eq:perts}
\mathcal{R}(t,\mathbf{x})=\int \frac{d^3 k}{(2\pi)^3}\mathcal{R}(t,\mathbf{k})e^{i \mathbf{k}\cdot\mathbf{x}}+{\rm h.c.}\qquad{\rm and}\qquad\gamma_{ij}(t,\mathbf{x})=\sum\limits_\lambda\int\frac{d^3k}{(2\pi)^3}\gamma_\lambda(t,\mathbf{k})e_{ij}(\mathbf{k},\lambda)e^{i \mathbf{k}\cdot\mathbf{x}}+{\rm h.c.}\,,
\end{equation}
where $e_{ij}(\mathbf{k},\lambda)$ is the transverse-traceless polarization tensor for the graviton, $\lambda$ labels the polarization states of the gravitational waves, and `h.c.' stands for the Hermitian conjugate. 

In a universe that is dominated by matter or radiation the expansion rate $H=\dot{a}/a$ decays more rapidly than the momentum $k/a$ redshifts. So at early times the modes are `outside the horizon,' $k\ll aH$, and are time-independent. As modes `enter the horizon,' $k\gg aH$, the modes oscillate. 

The statistical properties of the scalar and tensor fluctuations, $\mathcal{R}$ and $\gamma_\lambda$, at times when the modes are outside the horizon, provide the link between late-time observations and the primordial era. For a universe that is statistically homogeneous and isotropic, and in which the primordial fluctuations are adiabatic and Gaussian, the information about the statistical properties is contained entirely in the 2-point correlation functions
\begin{eqnarray}
\langle\mathcal{R}(\mathbf{k})\mathcal{R}(\mathbf{k}^{\prime})\rangle&=&(2\pi)^3\delta^3(\mathbf{k}+\mathbf{k}^{\prime})\frac{2\pi^2}{k^3}\Delta^2_{\mathcal{R}}(k),\nonumber\\
\langle\gamma_\lambda(\mathbf{k})\gamma_{\lambda^{\prime}}(\mathbf{k}^{\prime})\rangle&=&(2\pi)^3\delta_{\lambda\lambda^{\prime}}\delta^3(\mathbf{k}+\mathbf{k}^{\prime})\frac{2\pi^2}{k^3}\frac{1}{2}\Delta^2_{\rm \gamma}(k),
\end{eqnarray}
where the factor of $1/2$ in the last line accounts for the fact that the measured power includes contributions from each of the two graviton polarizations. 

\begin{figure}[h]
\begin{center}
\includegraphics[width=4in]{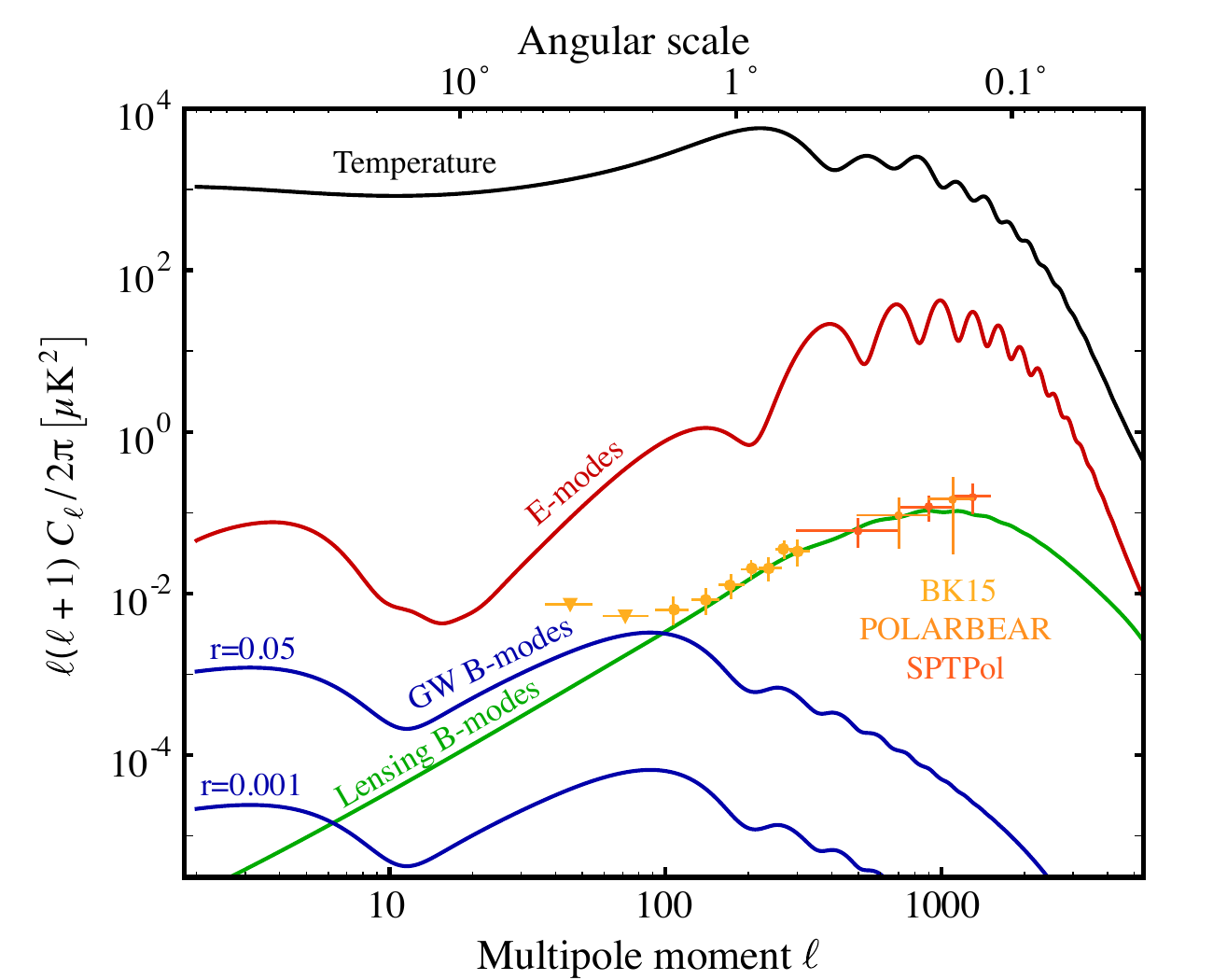}
\end{center}
\caption{Theoretical predictions for the temperature (black),
$E$-mode (red), and tensor $B$-mode (blue) power spectra. Primordial
$B$-mode spectra are shown for two representative values of the tensor-to-scalar
ratio: $r=0.001$ and $r=0.05.$
The contribution to tensor $B$ modes from scattering during the recombination epoch peaks at $\ell \approx 80$
and from reionization at $\ell < 10$.
Also shown are expected values for the contribution to $B$ modes from gravitationally lensed $E$ modes (green).
Current measurements of the $B$-mode spectrum are shown for BICEP2/{\em Keck Array} (light orange), POLARBEAR (orange), and SPTPol (dark orange).
The lensing contribution to the $B$-mode spectrum can be partially removed by measuring the
E-mode polarization and exploiting the non-Gaussian statistics of the lensing.
}
\label{fig:clall}
\end{figure}

All current observations are consistent with $\Delta^2_\gamma(k)=0$ and angular power spectra given by
\begin{equation}
C_{XX,\ell}=\int \frac{dk}{k}\Delta^2_\mathcal{R}(k)\left|\int\limits_0^{\tau_0} d\tau S_X(k,\tau)u_{X,\ell}(k(\tau_0-\tau))\right|^2\,,
\label{eq:clscalar}
\end{equation}
where $S_X(k,\tau)$ with $X=T,E$ are source functions that encode the evolution of the modes, and $u_{X,\ell}$ are functions that encode the geometry of the Universe. In a spatially flat universe $u_{T,\ell}$ is, for example, a spherical Bessel function, $u_{T,\ell}=j_\ell$. The spectrum of primordial perturbations is nearly scale-invariant, i.e.,
\begin{equation}\label{eq:power_spectra_power_law}
\Delta^2_{\mathcal{R}}(k)= A_{\rm s}\left(\frac{k}{k_\ast}\right)^{n_{\rm s}-1+\frac{1}{2}\left.\frac{dn_{\rm s}}{d\ln k}\right|_{k=k_\ast}\ln(k/k_\ast)+\dots}\qquad\text{with}\qquad n_{\rm s}\approx 1\qquad\text{and}\qquad \left.\frac{dn_{\rm s}}{d\ln k}\right|_{k=k_\ast}\approx 0\,.
\end{equation}
If primordial gravitational waves are present, they also contribute to the angular power spectra of temperature and polarization anisotropies. Given that the scalar perturbations are nearly scale invariant, we may expect the same for the tensor perturbations and hence parameterize their power spectrum as
\begin{equation}\label{eq:power_spectra_power_law_gw}
\Delta^2_{\gamma}(k)= A_{\rm r}\left(\frac{k}{k_\ast}\right)^{n_{\rm t}+\frac{1}{2}\left.\frac{dn_{\rm r}}{d\ln k}\right|_{k=k_\ast}\ln(k/k_\ast)+\dots}\qquad\text{with}\qquad n_{\rm t}\approx 0\qquad\text{and}\qquad \left.\frac{dn_{\rm t}}{d\ln k}\right|_{k=k_\ast}\approx 0\,.
\end{equation}
We can then define the tensor-to-scalar ratio as $r=A_{\rm t}/A_{\rm s}$. At linear order in perturbation theory only the tensor perturbations generate $B$-mode polarization, and a search for primordial gravitational waves can be cast as a search for $B$-mode polarization in the cosmic microwave background radiation.  Angular power spectra for anisotropies in the CMB temperature and polarization are shown in Fig.~\ref{fig:clall}. $B$-mode anisotropies created by gravitational waves are shown for two representative values of the tensor-to-scalar ratio and are significantly fainter than both temperature and $E$-mode anisotropies, highlighting the experimental precision necessary to detect them. 

Primordial gravitational waves are not the only source of degree-scale $B$ modes, since weak gravitational lensing of the cosmic microwave background by large-scale structure converts $E$- to $B$-mode polarization. This lensing contribution dominates over the primordial signal for $r\lesssim 0.01$. CMB-S4 will rely on precision measurements of $E$ and $B$ modes on smaller angular scales to remove 90\% of this lensing contribution.

 In addition, as we will discuss, emission from thermal dust and relativistic electrons in our galaxy lead to $B$-mode polarization in the microwave sky, and instrumental effects convert temperature or $E$-mode polarization to $B$-mode polarization. Both must be controlled to unprecedented levels.

\paragraph{Implications for inflation}
Inflation, a period of accelerated expansion of the early Universe, is the leading paradigm for explaining the origin of the primordial density perturbations that grew into the CMB anisotropies and eventually into the stars and galaxies we see around us. Accelerated expansion requires matter with an energy density that dilutes relatively slowly as the Universe expands. In inflationary models, such an energy density is usually obtained via the introduction of a new field $\phi$, called the inflaton. In the simplest scenarios its Lagrangian density is given by
\begin{equation}
{\cal L} = - \frac{1}{2} g^{\mu\nu}\partial_\mu \phi \partial_\nu \phi - V(\phi)\,,
\end{equation}
where $V(\phi)$ is the potential energy density. 

The overall evolution of the Universe is well-described by a FLRW line element
\begin{equation}
ds^2=-dt^2+a^2(t)\left[\frac{dr^2}{1-kr^2}+r^2d\Omega^2\right]\,,
\end{equation}
where $k=0$ for a flat spatial geometry, $k=\pm1$ allows for spatial curvature, and the time evolution is specified by the scale factor, $a(t)$. The Hubble parameter, $H=\dot{a}/a$, gives the rate of expansion of the Universe. 

A period of inflation will drive the spatial curvature close to zero, in good agreement with current observations. As a consequence, we will assume spatial flatness and set $k=0$ for most considerations; however, we discuss constraints on curvature obtainable with CMB-S4 in Sect.~\ref{sec:curvature}.

In addition to a flat universe, the simplest models of inflation predict that the primordial density perturbations are adiabatic, very nearly Gaussian, and nearly scale-invariant, in agreement with existing observations.

In single-field slow-roll inflation, the gauge-invariant combination of metric and scalar-field fluctuations that is conserved outside the horizon has the power spectrum
\begin{equation}
\label{eq:inf_PR}
\Delta^2_{\mathcal{R}}(k)=\frac{1}{2\epsilon M_{\rm P}^2}\left.\left(\frac{H}{2\pi}\right)^2\right|_{k=aH},
\end{equation}
where $\epsilon=-\dot{H}/H^2$ is the first slow-roll parameter, and $M_{\rm P}=1/\sqrt{8\pi G}$ is the reduced Planck mass.

\begin{figure}[!th]
\begin{center}
\includegraphics[width=6in]{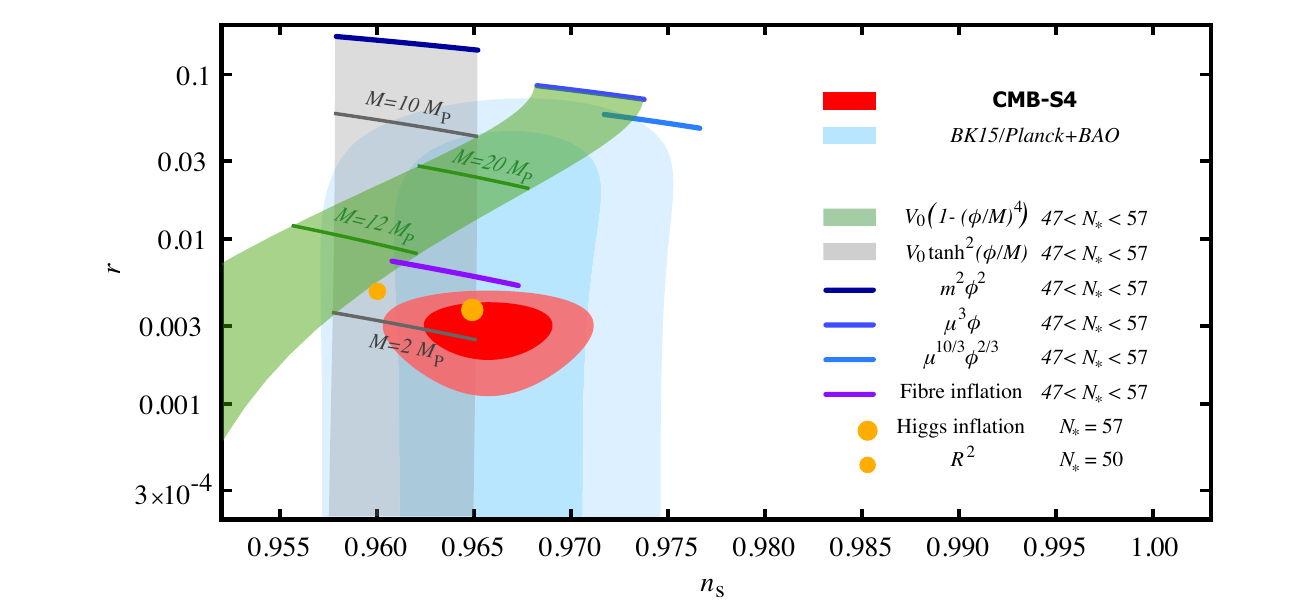}
\end{center}
\caption{Forecast of CMB-S4 constraints in the $n_{\rm s}$--$r$ plane for a fiducial model with $r=0.003$. 
 For comparison, we also show the current best constraints from a combination of the BICEP2/{\em Keck Array} experiments, \planck\ \cite{Array:2015xqh}, and BAO data. These are compared to several theoretical models. Chaotic inflation with $V(\phi)=\mu^{4-p}\phi^p$ for \mbox{$p=2/3,1,2$} are shown as blue lines for $47<N_\ast<57$ (with smaller $N_\ast$ predicting lower values of $n_{\rm s}$). The Starobinsky model and Higgs inflation are shown as small and large orange filled circles, respectively. The green band shows the predictions for a quartic hilltop model for which the potential throughout the inflationary period is described by $V(\phi)\approx V_0(1-(\phi/M)^4)$ before developing a minimum. The gray band shows the prediction of a sub-class of $\alpha$-attractor models~\cite{Kallosh:2013hoa}, and the purple line shows fibre inflation~\cite{Cicoli:2008gp}.
}
\label{fig:nsrp01}
\end{figure}
In addition to producing primordial density perturbations, the rapid expansion of spacetime creates primordial gravitational waves that imprint a characteristic polarization pattern onto the CMB. Their power spectrum is given by
\begin{equation}
\label{eq:inf_Pt}
\Delta^2_\gamma(k)=\frac{8}{M_{\rm P}^2}\left.\left(\frac{H}{2\pi}\right)^2\right|_{k=aH}\,,
\end{equation}
and is a measure of the expansion rate of the Universe during inflation. Together with the Friedmann equation a detection of primordial gravitational waves would then allow us to infer the inflationary energy scale.\footnote{In some models of inflation the relation between the tensor-to-scalar ratio and the energy scale of inflation are broken because there are additional sources of gravitational wave production~\cite{Namba:2015gja}. However, the signal in these models is highly non-Gaussian and would not be mistaken for quantum fluctuations.}  

CMB-S4 will be able to detect primordial gravitational waves for $r>0.003$, and a detection would point to inflationary physics near the energy scale associated with grand unified theories. As a consequence, a detection would provide additional evidence in favor of the idea of the unification of forces, and would probe energy scales far beyond the reach of the Large Hadron Collider (LHC) or any conceivable collider experiment. Selected models within reach of CMB-S4 are shown in Fig.~\ref{fig:nsrp01} together with current limits and constraints expected for CMB-S4. 

The knowledge of the energy scale of inflation would have broad implications for many other aspects of fundamental physics, including ubiquitous ingredients of string theory like axions and moduli, fields that control the shapes and sizes of the compact dimensions. 

Because the polarization pattern is due to quantum fluctuations in the metric of spacetime generated during inflation, a detection would also provide evidence that gravity is quantized. Moreover, because many of the models that predict a signal that is strong enough to be detected with CMB-S4 are based on symmetry principles and inflation occurs at high energy and large inflaton field range, a detection would also provide some insights into the nature of quantum gravity.

\begin{figure}[!th]
\begin{center}
\includegraphics[width=6in]{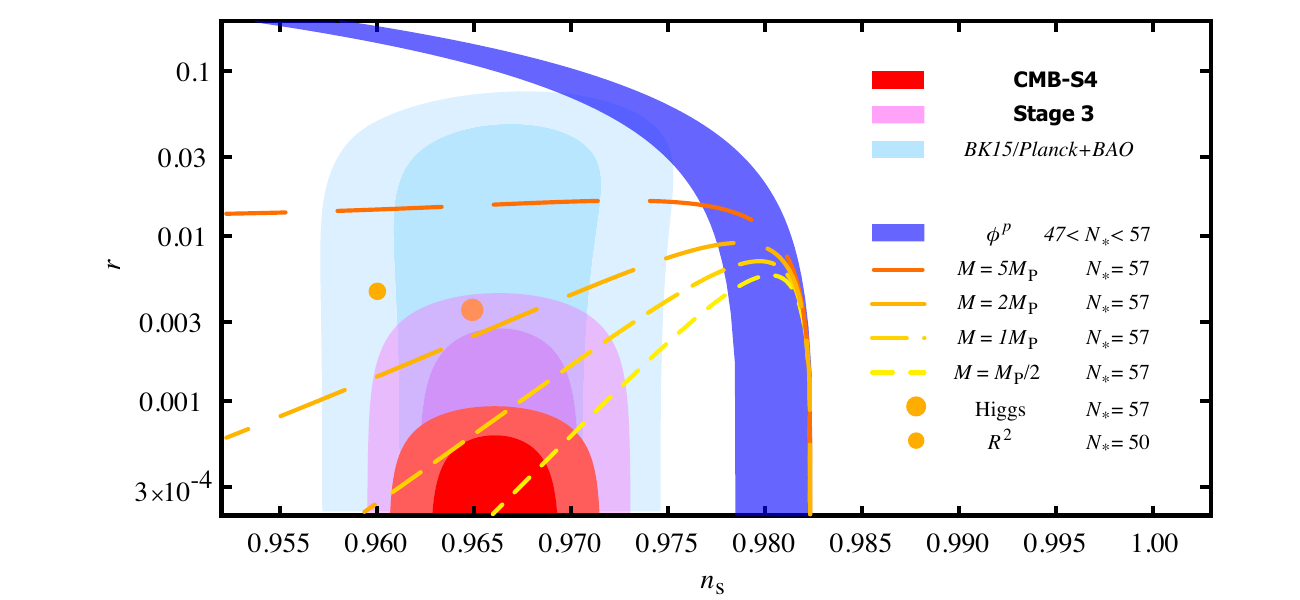}
\end{center}
\caption{Forecast of CMB-S4 constraints in the $n_{\rm s}$--$r$ plane for a fiducial model with $r=0$. 
Also shown are the current best constraints from a combination of the BICEP2/{\em Keck Array} experiments and \planck\ \cite{Array:2015xqh}.
The Starobinsky model and Higgs inflation are shown as small and large orange filled circles. The lines show the classes of model that naturally explain the observed value of $n_{\rm s}$. The corresponding potentials all either polynomially or exponentially approach a plateau. The scale in field space over which the potential approaches the plateau is referred to as the ``characteristic scale'' (see Ref.~\cite{Abazajian:2016yjj} for more details). We show different values, $M=M_{\rm P}/2$, $M=M_{\rm P}$, $M=2\,M_{\rm P}$, and $M=5\,M_{\rm P}$. Longer dashes correspond to larger values of the scale $M$. The Planck scale plays an important role because the gravitational scale and the characteristic scale share a common origin. The number of e-folds $N_\ast$ chosen for the figure corresponds to nearly instantaneous reheating, which leads to the smallest values for $r$ for a given model. Other reheating scenarios predict larger values of $r$ and are easier to detect or exclude.
}
\label{fig:nsr0}
\end{figure}

In the absence of a detection, the upper limit of $r < 0.001$ at 95\% CL achievable by CMB-S4 is nearly two orders of magnitude stronger than current limits and about a factor of 5 stronger than projected limits for Stage-3 experiments. As explained in more detail in Ref.~\cite{Abazajian:2016yjj}, such an upper limit would exclude large classes of inflationary models. In summary, there are only two classes of model that naturally explain the observed value of $n_{\rm s}$ and are consistent with current data. The first class consists of monomial models with potentials that during inflation are well approximated by $V(\phi)=\mu^{4-p}\phi^p$. Representatives in this class are shown in Fig.~\ref{fig:nsrp01}, and the entire class of models is shown in Fig.~\ref{fig:nsr0}. The second class consists of models in which the potential $V(\phi)$ approaches a plateau, either polynomially or exponentially. The potential for models in this class has a characteristic scale over which the potential varies~\cite{Abazajian:2016yjj}.\footnote{This characteristic scale was introduced in Ref.~\cite{Abazajian:2016yjj} and should not be confused with the field range or the energy scale of inflation. For a discussion see Refs.~\cite{Abazajian:2016yjj} and~\cite{Linde:2016hbb}.} The sensitivity of CMB-S4 is chosen to exclude all models in this class with a characteristic scale that exceeds the Planck scale. The Planck scale constitutes an important threshold because the scale of gravitational interactions and the characteristic scale may share a common origin and be linked to each other, such as in the Starobinsky model~\cite{Starobinsky:1980te}, in Higgs inflation~\cite{Bezrukov:2007ep}, or more general models involving non-minimally coupled scalar fields. As a consequence, even in the absence of a detection CMB-S4 would significantly advance our understanding of inflation, and would dramatically affect how we think about the theory. The classes of model that naturally predict the observed value of $n_{\rm s}$, together with current constraints and constraints expected for CMB-S4, are shown in Fig.~\ref{fig:nsr0}.

\subsection{Primordial density perturbations }
\label{subsec:PF}

CMB-S4 can also seek to characterize the primordial Universe by searching for well-motivated signatures in the scalar fluctuations, in the primordial power spectrum, and non-Gaussianities.

We will begin with discussions of the overall shape of the primordial power spectrum, and generalize to
investigate the possibility of measuring features in the primordial power spectrum.

Currently, constraints on non-Gaussianity rely on measurements of the CMB bispectrum, 
and CMB-S4 will improve on these measurements.
Below, we will summarize some well-motivated bispectrum shapes \cite{Babich:2004gb,2009astro2010S.158K}. In addition, the precise measurement of $B$-mode polarization opens up another possible window to explore the physics of the early Universe. Non-Gaussianities that measure interactions between tensors and scalars, can, for the first time, be constrained with sensitivity that has the potential to exceed that of scalar non-Gaussianities \cite{Meerburg:2016ecv}. Although the presence of these couplings is predicted to be small \cite{Maldacena:2011nz,Bordin:2016ruc} in the simplest inflation models, such a detection would present evidence of exciting new physics \cite{Agrawal:2017awz,Domenech:2017kno}. In addition, theoretical  consistency conditions between scalar and tensor non-Gaussianities provide a compelling observational target  \cite{Hayden:2016xxa,Lee:2016vti,Baumann:2017jvh}. 

In addition to the bispectrum, fluctuations in the galaxy density are also a powerful probe of 
primordial non-Gaussianity, since the galaxy power spectrum will be biased on large scales 
in the presence of primordial local non-Gaussianity. 
Recently, two independent techniques have been proposed to use CMB data to precisely measure
the scale-dependence of this galaxy bias on large scales. Precise measurements
require ``cosmic variance cancellation,'' which uses multiple tracers of the same field, to remove cosmic variance on this bias. We will present results of forecasts on the local type of non-Gaussianity using CMB-S4 data combined with data from the LSST. 

\subsubsection{Primordial Power Spectrum}

Currently, our best constraints on the shape of the primordial power spectrum on large scales come from measurements of the CMB temperature anisotropies by the {\it Planck\/} satellite.
The current value of the slope of the primordial curvature power spectrum is $n_{\rm s}=0.965 \pm 0.004$ \cite{Akrami:2018odb}. The forecasted uncertainty on the slope for a model of $\Lambda$CDM + running of the tilt using CMB-S4 noise, including the CMB temperature and polarization fields, is $\sigma_{n_{\rm s}}=0.002$, reducing the error from {\it Planck\/} by a factor of $2$.  
Slow-roll models of inflation predict a running of the tilt of order $(n_{\rm s}-1)^2\sim 10^{-3}$. With CMB-S4, the forecasted uncertainty on the running is $\sigma_{n_{\rm run}}=0.0029$ (see Eq.~\ref{eq:power_spectra_power_law}), which combined with large-scale structure constraints could be tightened even further. For comparison, we find $\sigma_{n_{\rm s}}=0.0027$ and $\sigma_{n_{\rm run}}=0.0038$ for Simons Observatory.

Features in the CMB temperature power spectrum \cite{Bennett:2003bz,Spergel:2003cb,Ade:2015lrj,Akrami:2018odb} have been claimed and interpreted as possible evidence for new physics during inflation. Evidence of a feature would provide us with yet another unique signature that can be used to understand the physics of the early Universe \cite{Slosar:2019gvt}. Specific forms of such features are typically handled on a case-by-case basis.  However, for model-independent analyses, it is desirable to have a simple and accurate prescription that relates CMB observables to the shape of the inflationary potential. Various approaches exist in the literature for reconstructing this primordial power spectrum (see e.g., \cite{2003MNRAS.342L..72B, 2011JCAP...08..031G, 2012ApJ...749...90H, 2013PhRvD..87h3526A, 2014JCAP...01..025H, 2014PhRvD..89j3502D, 2016PhRvD..93b3504M, 2016JCAP...09..009H})

\commentout{
In a series of papers \cite{Dvorkin:2009ne,Dvorkin:2010dn,Dvorkin:2011ui,Obied:2017tpd,Obied:2018qdr}, a formalism (known as ``generalized slow roll") has been developed to test the hypotheses of slow-roll and single-field inflation in a general and model-independent way. 
In this formalism, there is a single source function $G'$ that is responsible for the observable features and it is simply related to the local slope and curvature of the inflationary potential in the same way the tilt is for the case of standard slow roll:
\begin{equation}
G' \approx 3\left(\frac{V'}{V}\right)^2-2\frac{{V'}^{ \prime}}{ V},
\end{equation}
where primes denote derivatives with respect to the inflation field.
This framework can be used to map constraints from the CMB onto constraints on the shape of the inflationary potential, beyond any specific model of inflation.

Here we adopt the generalized slow-roll approximation (see Appendix \ref{subsec:forecasts_PDF} for more details on this formalism) to accommodate order unity variations in the power spectrum from slow-roll predictions. 
Under this formalism the primordial power spectrum $\Delta_\mathcal{R}^{2}(k)$ can be written in terms of a single unknown source function $G'$,  and the rest are known functions (the window functions, in Eq.~\ref{eq:window_function}).
}

In a series of papers \cite{Dvorkin:2009ne,Dvorkin:2010dn,Dvorkin:2011ui,Obied:2017tpd,Obied:2018qdr}, a formalism (known as ``generalized slow roll”) has been developed to test the hypotheses of slow-roll and single-field inflation in a general and model-independent way.
In this formalism, there is a single source function $G'$ that is responsible for the observable features and it is simply related to the local slope and curvature of the inflationary potential in the same way the tilt is for the case of standard slow roll:
\begin{equation}
G'\approx 3\left({V'\over V}\right)^2-2{V''\over V},
\end{equation}
where primes denote derivatives with respect to the inflation field.
This framework can be used to map constraints from the CMB onto constraints on the shape of the inflationary potential, without assuming any specific model of inflation.

Here we adopt the generalized slow-roll approximation to accommodate order unity variations in the power spectrum from slow-roll predictions.
Under this formalism the primordial power spectrum $\Delta_\mathcal{R}^{2}(k)$ can be written in terms of a single unknown source function $G’$,  and the rest are known functions. We perform a Fisher matrix analysis varying the six standard $\Lambda$CDM cosmological parameters ($A_{\rm s}$, $n_{\rm s}$, $\tau$, $\theta$, $\Omega_{\rm c}$, $\Omega_{\rm b}$) and $40$ coefficients that parametrize the source function.

\begin{figure}[t]
\begin{center}
\includegraphics[width=0.7\textwidth]{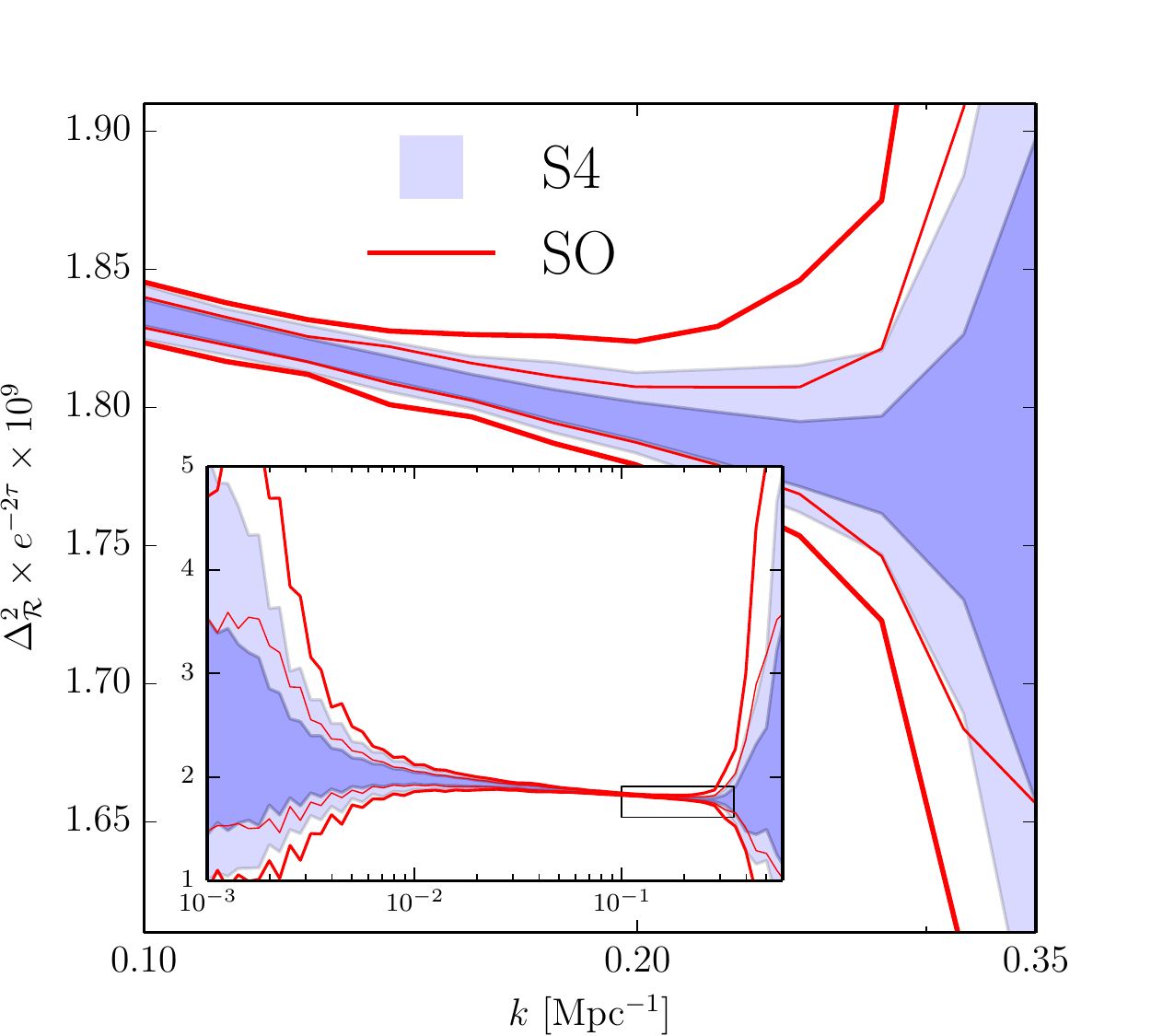}
\caption{Reconstructed primordial power spectrum with CMB-S4 noise curve including atmospheric noise co-added with Planck noise, noise from point sources, and lensing reconstruction. We parameterize our source function $\delta G'$ in the range $0.5 <s/{\rm Mpc}< 4870$ with a total of 40 parameters, equispaced in logarithmic scale. Also shown are the constraints coming from the Simons Observatory baseline, for the same fraction of the sky observed by CMB-S4 (as discussed in Ref. \cite{2018arXiv180807445T}).}\label{fig:Delta2}
\end{center}
\end{figure}

The ordinary slow-roll approximation corresponds to $G'(\ln s) = 1-n_{\rm s}$, and results in
a power-law curvature power spectrum.
We restrict our parameterization of fluctuations, $\delta G'$ (around the constant $G'=1-n_{\rm s}$) to $0.5 <s/{\rm Mpc}< 4870$. We sample $\delta G'$ for a total of 40 parameters, equispaced on a logarithmic scale. We then construct a smooth function using the natural spline of these points.
In Figure \ref{fig:Delta2} we show the forecasted reconstructed initial curvature power spectrum for CMB-S4. 

\subsubsection{Spatial Curvature}
\label{sec:curvature}

The theory of inflation predicts an almost flat universe with curvature $|\Omega_K|\sim10^{-4}$  (from contributions from large-scale modes). If departures from this value were measured at a significant level, this would give us information about the physics of inflation. For instance, it could indicate that the inflaton field was not slowly rolling when the largest scales exited the horizon.

Combining CMB-S4, {\it Planck\/} $TT$ on the part of the sky that is not observed by CMB-S4, and {\it Planck}'s low-$\ell$ ($\ell<30$) measurement of $TT$ over the full sky (along with a {\it Planck\/} prior on $\tau$ of $\sigma(\tau)=0.007$), the constraints on the curvature reach $\sigma(\Omega_K)=2.4\times10^{-3}$, which is a factor of 2.5 better than current constraints from the CMB, namely $\sigma(\Omega_K)=6.5\times10^{-3}$ using CMB temperature, polarization, and lensing data from the {\it Planck\/} satellite. This constraint could be sharpened even further by combining CMB and baryon acoustic oscillation (BAO) data.

\subsubsection{Isocurvature}
Measurements of CMB temperature and polarization power spectra indicate that the primordial initial conditions are primarily adiabatic; that is, if we define
\begin{align}
S_{i j}\equiv \frac{\delta n_{i}}{n_{i}}-\frac{\delta n_j}{n_j} \ ,
\end{align}
then the adiabatic condition is that $S_{i j}=0$.
Here the species labels $i,j$ can denote baryons, cold dark matter (CDM), photons ($\gamma$), or neutrinos. Number densities are denoted by $n_{i}$. Adiabatic perturbations are produced if the initial perturbations in all species are seeded by the inflaton. Otherwise, the initial conditions are a mixture of adiabatic and isocurvature perturbations, for which $S_{i\gamma}\neq 0$. These initial conditions determine the acoustic peak structure and large-scale amplitude of CMB anisotropies \cite{Bond:1984fp,Kodama:1986fg,Kodama:1986ud,Hu:1994jd,Moodley:2004nz,Bean:2006qz}, which can thus probe additional fields present during inflation. 
\label{sec:isosec}
Present data still allow subdominant CDM density isocurvature (CDI), baryon density isocurvature (BDI), or neutrino density isocurvature (NDI) initial conditions (described in Refs.
\cite{Enqvist:2000hp,MacTavish:2005yk,Dunkley:2008ie,Planck:2013jfk,Ade:2015lrj,Akrami:2018odb}) at the roughly 1\% level, depending on precise assumptions made about the spectral shape and multi-mode admixtures of isocurvature. CMB-S4 could shed light on inflationary physics by detecting isocurvature fluctuations. As specific examples of the physics that can be measured with CMB-S4,
we consider two well-studied scenarios: the curvaton; and compensated isocurvature perturbations (CIPs). Axion-type isocurvature is discussed in Section~\ref{sec:axioniso}.

The curvaton scenario is a model in which a sub-dominant field $\sigma$ acquires vacuum fluctuations during inflation, sources $\zeta$  (the total gauge-invariant curvature perturbation), and then decays \cite{Mollerach:1989hu,Mukhanov:1990me,Moroi:2001ct,Lyth:2001nq,Lyth:2002my}. Curvaton candidates include sneutrinos, among others \cite{Postma:2002et,Kasuya:2003va,Ikegami:2004ve,Mazumdar:2004qv,Allahverdi:2006dr,Papantonopoulos:2006xi,Mazumdar:2010sa,Mazumdar:2011xe}. Depending on whether a species $i$ (or its set of quantum numbers) is produced by, before, or after curvaton decay, perturbations in $i$ are offset from $\zeta$, with distinct density perturbation $\zeta_i$ in the species $i$ on surfaces of constant curvature. Here $i\in\left\{\nu, {\rm b}, {\rm c}\right\}$. The resulting isocurvature perturbations \cite{Lyth:2001nq,Lyth:2002my,Gordon:2002gv} have entropy fluctuation given by
\begin{eqnarray}
S_{i \gamma}=\left\{\begin{array}{ll}-3\zeta-3(\zeta_{\gamma}-\zeta),&\mbox{if $i$ is produced before $\sigma$ decay,}\\
3\left(\frac{1}{r_{\rm D}}-1\right)\zeta-3(\zeta_{\gamma}-\zeta),&\mbox{if $i$ is produced by $\sigma$ decay},\\
-3(\zeta_\gamma-\zeta),&\mbox{if $i$ is produced after $\sigma$ decay}.\end{array}\right.\label{eq:strew}
\end{eqnarray} The parameter $r_{\rm D}$ is the fractional energy density in the curvaton when it decays.

The mixture of isocurvature modes is determined by whether or not baryon number, lepton number, and CDM are produced before, by, or after curvaton decay. Curvaton-type isocurvature is distinct from axion isocurvature, because it is correlated (or anti-correlated) with $\zeta$. If lepton number is produced by curvaton decay, the lepton chemical potential $\xi_{\rm lep}$ sets the size of NDI modes \cite{Lyth:2002my,Gordon:2003hw,DiValentino:2011sv}:
\begin{equation}
S_{\nu \gamma}=
-\frac{135}{7}\left(\frac{\xi_{\rm lep}}{\pi}\right)^2\zeta_{\gamma}.\end{equation}
There are $27$ distinct curvaton scenarios, since baryon number, lepton number, and CDM could each be produced before, by, or after curvaton decay. Viable models are those in which one of baryon number or CDM is produced by curvaton decay, and those in which \textit{both} baryon number and CDM are produced after curvaton decay. For curvaton-decay scenarios, we use the notation ($b_{y_{\rm b}}$, $c_{y_{\rm c}}$, $L_{y_{\rm L}}$), where the subscripts run over $y_i\in \{{\rm before, by, after}\}$ and $L$ denotes lepton number. For example, $(b_{\rm before}, c_{\rm by}, L_{\rm by})$ is a model in which baryon number is produced before curvaton decay, CDM by curvaton decay, and lepton number by curvaton decay. Current isocurvature limits favor values of $r_{\rm D}\approx 1$, except for models in which baryon number is produced by curvaton decay and CDM before (or vice versa), which favor central values of $r_{\rm D} \approx 0.16$ ($r_{\rm D} \approx 0.84$). 

The current limits \cite{Smith:2015bln} on $r_{\rm D}$ are shown in Table~\ref{limits_rd}, along with a forecast of CMB-S4's sensitivity to $r_{\rm D}$ via isocurvature modes. There is dramatic improvement in constraints on the $(b_{\rm by},c_{\rm before},L_{\rm by})$ and $(b_{\rm before},c_{\rm by},L_{\rm by})$ scenarios because of the accompanying NDI perturbations. One unusual case is $(b_{\rm after},c_{\rm after}, L_{y_{\rm L}})$, where isocurvature just constrains the combination \cite{Smith:2015bln} $\chi_{\rm D} \equiv [1+\xi_{\rm lep}^2/(\pi^2) \left(1/r_D -1\right)]^{-1}$, while the independent constraint on $\xi_{\rm lep}^{2}$ is driven by the CMB limit on total relativistic energy density ($N_{\rm eff}$).

\begin{table}[htbp!]
\begin{center}
\begin{tabular}{| c || c | c |}
\hline
{\rm Isocurvature scenario} &  \planck & CMB-S4 \\ \hline \hline
  & $ \Delta r_D/r_{D}^{\rm adi}$ &$ \Delta r_D/r_{D}^{\rm adi}$\\\hline
$(b_{\rm by},c_{\rm before},L_{y_{\rm L}})$ & $0.03$&$0.005$\\
$(b_{\rm before},c_{\rm by},L_{y_{\rm L}})$ &  $0.01$ &$0.004$\\
$(b_{\rm by},c_{\rm after},L_{y_{\rm L}})$ &  $0.04$&$0.01$\\
$(b_{\rm after},c_{\rm by},L_{y_{\rm L}})$ & $0.008$&$0.002$\\
$(b_{\rm by},c_{\rm by},L_{y_{\rm L}})$ &  $0.007$&$0.002$\\ \hline\hline
& $\Delta \chi_{\rm D}/\chi_{\rm }^{\rm adi}$&$\Delta \chi_{\rm D}/\chi_{\rm }^{\rm adi}$ \\\hline
$(b_{\rm after},c_{\rm after},L_{y_{\rm L}})$ & $0.003$&$0.0004$ \\ \hline \hline
 &  $\Delta \xi^{2}_{\rm lep}$ &$\Delta \xi^{2}_{\rm lep}$\\\hline
$(b_{\rm by},c_{\rm before},L_{\rm by})$ &$0.02$ &$0.002$\\
$(b_{\rm before},c_{\rm by},L_{\rm by})$ &$0.4$  & $0.04$\\
$(b_{\rm by},c_{\rm after},L_{\rm by})$ &$0.3$  &$0.04$\\
$(b_{\rm after},c_{\rm by},L_{\rm by})$ & $0.3$&$0.04$\\
$(b_{\rm by},c_{\rm by},L_{\rm by})$ & $0.3$ & $0.04$\\
$(b_{\rm after},c_{\rm after},L_{\rm by})$ & $0.3$ & $0.04$\\
\hline
\end{tabular}
\end{center}
\caption{Isocurvature constraints on $r_{\rm D}$ and $\xi_{\rm lep}^{2}$, both at $95\%$ CL using \planck\ TT+BAO+LowP data \cite{Smith:2015bln} in viable curvaton decay-scenarios, and Fisher forecasts for CMB-S4 sensitivity. 
\label{limits_rd}}
\end{table}
If baryon number is produced before curvaton decay, while CDM is produced by curvaton decay,  CMB $3$-pt correlations (see Sec. \ref{sec:curv_ng}) are excited, with $f_{\rm NL}^{\rm local}\approx 5$, also within the reach of CMB-S4. As a result, CMB-S4 provides a useful set of consistency checks if certain curvaton signatures are observed \cite{Smith:2015bln}.

If baryon number is produced by curvaton decay, but CDM is produced before (or vice versa), a relatively large compensated isocurvature perturbation $S_{\rm bc}$ (CIP) is produced between the baryons and CDM.
Curvaton-generated CIPs are proportional to $\zeta$, $S_{\rm bc}=A\zeta$, where $A\approx 17$ in the $(b_{\rm by}, c_{\rm before}, L_{ y_{\rm L}})$ scenario and $A\approx -3$ for $(b_{\rm by}, c_{\rm before}, L_{ y_{\rm L}})$. CIPs would induce non-Gaussianities in the CMB \cite{Grin:2011nk,Grin:2011tf,Grin:2013uya,He:2015msa}. At CMB-S4 sensitivity \cite{He:2015msa}, the threshold for a $95\%$ CL detection is $A\approx 10$, and so a CIP test of the $(b_{\rm by}, c_{\rm before}, L_{ y_{\rm L}})$ scenario is within reach of CMB-S4, a substantial improvement over \planck\ sensitivity. For uncorrelated CIPs, we find that the sensitivity of CMB-S4 to a scale-invariant (SI) angular power spectrum of uncorrelated CIPs is $\Delta_{\rm cl}=0.003$ at the $95\%$ CL with current parameters \cite{He:2015msa}, assuming \planck~ cosmology \cite{Ade:2015lrj}. Here $\Delta_{\rm cl}$ is the rms CIP amplitude on cluster scales. This is a significant improvement over the upper limit of $\Delta_{\rm cl}\leq 0.077$ from the {\it Wilkinson Microwave Anisotropy Probe} (\wmap) \cite{Grin:2013uya}, or the \planck~upper limit $\Delta_{\rm cl}\leq 0.064$ obtained in Ref. \cite{Smith:2017ndr}.

\subsubsection{Primordial non-Gaussianity: the bispectrum}

Unlike the scalar and tensor power spectra, higher-order correlations of the scalar modes are directly sensitive to the dynamics and field content responsible for inflation (and its alternatives) \cite{Meerburg:2019qqi}.  While non-Gaussian correlations are small in conventional single-field slow-roll inflation, there exist many other possibilities for the nature of inflation that give strikingly different predictions when we move beyond the power spectrum.  The constraints on non-Gaussianity from the \wmap\ \cite{Senatore:2009gt,Bennett:2012fp} and \planck\ \cite{Ade:2013ydc,Ade:2015ava} satellites currently place the most stringent limits on a wide range of mechanisms for inflation; however, these measurements are not sufficiently sensitive to suggest that a particular mechanism is favored by the data.  

\label{sec:curv_ng}
The space of non-Gaussian signals from inflation can broadly be grouped into two conceptual categories that generate distinguishable features in the correlation functions. These are signals that: (1) indicate non-trivial self-interactions of the effective inflaton fluctuation; or (2) indicate interactions with degrees of freedom other than the inflaton. 
For the first category, self-interactions that respect the time-translation invariance during inflation lead to qualitatively different predictions from self-interactions that violate it. For the second category the signatures qualitatively depend on the mass of the additional degrees of freedom (fields). Additional light degrees of freedom can fluctuate significantly and may have large self-interactions. Heavy degrees of freedom do not fluctuate appreciably, but may come into existence by quantum fluctuations and decay into inflaton quanta, generating non-Gaussian correlations \cite{Arkani-Hamed:2015bza,Chen:2015lza,Lee:2016vti,Arkani-Hamed:2018kmz}. Alternatively, they may become excited by the dynamics of the inflaton and their backreaction on the inflationary dynamics may lead to non-Gaussian correlations \cite{Barnaby:2010vf,Barnaby:2011pe,Barnaby:2012xt,Namba:2013kia}. 

Constraints on non-Gaussianity are often expressed in terms of the correlator of three scalar modes, described by the bispectrum $B_{\mathcal{R}}(\mathbf{k}_1,\mathbf{k}_2,\mathbf{k}_3)$, defined through
\begin{equation}
\label{eq:Bsss}
\langle\mathcal{R}(\mathbf{k}_1)\mathcal{R}(\mathbf{k}_2)\mathcal{R}(\mathbf{k}_3)\rangle=(2\pi)^3\delta(\mathbf{k}_1+\mathbf{k}_2+\mathbf{k}_3)B_{\mathcal{R}}(\mathbf{k}_1,\mathbf{k}_2,\mathbf{k}_3)\,.
\end{equation}
The structure of particle interactions relevant for inflation provides both a general organizing principle for this functional space and several specific well-motivated forms of the bispectrum. 
We now briefly review the classification of scalar non-Gaussianity from inflation and comment on non-Gaussian signatures that are especially relevant for large-field inflation.

Our convention for extracting a normalization of the amplitude, $f_{\rm NL}$, follows Refs.~\cite{Babich:2004gb} and \cite{Ade:2013ydc}, i.e., 
\begin{equation}
B_{\mathcal{R}}(k_1,k_2,k_3) = \frac{3}{5} (4 \pi^4) 2 A_{\rm s}^2 f_{\rm NL} F(k_1,k_2,k_3) ,
\end{equation}
where $F$ is referred to as the shape of the bispectrum, which in a scale invariant universe $\propto k^{-6}$. Different physical effects during inflation lead to different shapes for the bispectrum.

All bispectra that come from fluctuations of the field that drives inflation (``single-clock'' scenarios) most strongly couple Fourier modes $k$ of similar amplitude. The ``squeezed limit'' of these bispectra (the coupling of modes $k_1\ll k_2\sim k_3$) is very restricted. 
A large fraction of the parameter space for scenarios involving interactions during inflation that respect the underlying shift symmetry (i.e., are approximately scale-invariant) are well captured by the equilateral~\cite{Babich:2004gb} and orthogonal shapes~\cite{Senatore:2009gt} shapes. 
This includes scenarios in which inflaton fluctuations have non-trivial self-interactions~\cite{Silverstein:2003hf,ArkaniHamed:2003uz,Alishahiha:2004eh,Chen:2006nt,Cheung:2007st,Senatore:2009gt} or couplings between the inflaton and other (potentially massive) degrees of freedom~\cite{Chen:2009zp,Tolley:2009fg, Cremonini:2010ua, Achucarro:2010da,Baumann:2011nk,Barnaby:2011pe,Arkani-Hamed:2015bza}. Single-field slow-roll inflation necessarily produces $f_{\rm NL}^{\rm equil} < 1$~\cite{Creminelli:2003iq} and therefore any detection of $f_{\rm NL}^{\rm equil} \geq 1$ would rule out this large class of models.  A detection would imply that inflation is a strongly coupled phenomenon and/or involved more than one field~\cite{Baumann:2014cja,Alvarez:2014vva,Baumann:2015nta}.  
In single-field inflation, the amplitude of the non-Gaussianity typically suggests a new energy scale, $M$, such that $f_{\rm NL}^{\rm equil} \propto A_{\rm s}^{-1/2} \, (H/M)^2$~\cite{Cheung:2007st,Baumann:2011su}; at this energy scale self-interactions become strongly coupled and current limits on non-Gaussianity \cite{Ade:2015ava} translate into $M>\mathcal{O}(10)H$.  In the presence of additional hidden sectors, the amplitude of non-Gaussanity scales with the strength of the coupling between the inflaton and these additional fields, usually suppressed by an energy scale $\Lambda$.  Current limits give $\Lambda  > (10{-}10^{5}) H$~\cite{Green:2013rd,Assassi:2013gxa}, with the variation depending mostly on the dimension of the operator coupling the two sectors.  For $r > 0.01$, these constraints require some of the interactions to be weaker than gravitational.  CMB-S4 would further tighten existing constraints on a wide variety of interactions of the inflaton with itself and any other fields that are excited during inflation.    

When light degrees of freedom other than the inflaton contribute to the observed scalar fluctuations, coupling between modes of very different wavelengths is allowed. Historically, the most well-studied signature of this type comes from the ``local'' shape, which couples short wavelength modes $k_2\sim k_3$ to a long wavelength mode $k_1$ ($\ll k_2, k_3$). A significant detection of this shape would rule out all models of single-clock inflation \cite{Creminelli:2004yq}. In addition, such a signal would open the door to significant cosmic variance on all scales, from coupling of fluctuations within our observed volume to any super-Hubble modes \cite{Nelson:2012sb,LoVerde:2013xka,Nurmi:2013xv}. Indeed, there would be room for a significant shift between the observed amplitude of scalar fluctuations (and so the observed $r$) and the mean value of fluctuations on much larger scales \cite{Bonga:2015urq}. Any scenario that predicts local non-Gaussianity, together with fluctuations on scales much larger than our observed volume, predicts a probability distribution for our observed $f_{\rm NL}^{\rm local}$, but many well-motivated scenarios also predict a small mean value; these include the simplest modulated reheating scenario \cite{Zaldarriaga:2003my} and ekpyrotic cosmology \cite{Lehners:2009ja}, both of which predict mean values of $f_{\rm NL}^{\rm local}\approx5$. 

Currently the strongest constraints on the local shape come from the \planck\ 2015 temperature and polarization analysis that finds $f_{\rm NL}^{\rm local} = 0.8 \pm 5.0$~\cite{Ade:2015ava}. CMB-S4 will improve these limits by a factor 2 for local and orthogonal shapes.\footnote{Note that constraints on orthogonal and local non-Gaussianities in the CMB scale almost as expected, i.e., as the square root of the number of observed modes. Unfortunately, some information is lost in projection, which particularly affects equilateral non-Gaussianity, scaling more like the fourth root of the number of modes.} Note that these forecasts assume that CMB-S4 can be combined with large-scale measurements from \planck,  the impact of secondaries \cite{Hill:2018ypf} can be reduced, and temperature and polarization can be sufficiently delensed to remove large contributions to the variance \cite{PhysRevD.70.083005}. Even under these assumptions, the constraints are insufficient to reach the most compelling theoretical threshold around $|f_{\rm NL}^{\rm local}|\lesssim 1$~\cite{Alvarez:2014vva}. Nonetheless, the landscape of inflationary models is large and a detection would instantly present a monumental discovery similar to that of gravitaional waves. CMB-S4 could, for example, provide hints for the mean level of non-Gaussianity expected from modulated reheating scenario or ekpyrotic cosmology at roughly the $2\,\sigma$ level. The simplest curvaton scenario, which predicts $f_{\rm NL} = -5/4$ \cite{Lyth:2001nq}, will be out of reach using just the bispectrum measured by CMB-S4, but we will show in the next subsection that combining CMB-S4 with LSST data is likely to reach this significant threshold. 

Table~\ref{tab:NGs} shows the forecasted constraints on the local, equilateral, and orthogonal shapes from CMB-S4 using bispectrum measurements. 
In fact we expect that CMB-S4 would likely provide the strongest constraints on equilateral and orthogonal templates that will be available for the foreseeable future. Since limits on non-Gaussianity provide a unique and fundamental insight into the nature of inflation, this increased sensitivity would provide a non-trivial improvement in our understanding of the Universe.

Perhaps of special interest for CMB-S4 are non-Gaussian signatures that would be expected in models of large-field inflation, which could lead to non-trivial features in the bispectrum. 
For example, in models in which the inflaton is an axion, there is an approximate discrete shift symmetry \cite{Silverstein:2008sg,McAllister:2008hb,Flauger:2009ab}
that naturally lead to periodic features in the bispectrum. 
Often these models predict counterparts in the power spectrum where the associated features are expected to be detected first~\cite{Behbahani:2011it}, but this need not be the case~\cite{Behbahani:2012be}. An attempt has been made~\cite{Ade:2015ava} to look for resonant and local features in the bispectrum; however, at present, these models have not yet been constrained systematically.

Correlators including at least one $B$ mode will benefit significantly from the improved sensitivity of CMB-S4. In particular, the three point correlation $\langle \mathcal{R}(\mathbf{k}_1)\mathcal{R}(\mathbf{k}_2)\gamma_\lambda(\mathbf{k}_3) \rangle$ can be constrained using $\langle BTT\rangle$, $\langle BTE\rangle $ and $\langle BEE\rangle$.

The details of the tensor-scalar-scalar correlator are contained in the bispectrum $B_{\mathcal{R}\mathcal{R}\gamma_\lambda}(\mathbf{k}_1,\mathbf{k}_2,\mathbf{k}_3)$, defined by pulling out the appropriate polarization structure associated with the tensor mode:
\begin{equation}
\label{eq:Bsst}
\qquad \langle \mathcal{R}(\mathbf{k}_1)\mathcal{R}(\mathbf{k}_2)\gamma_\lambda(\mathbf{k}_3) \rangle = (2\pi)^3 \delta(\mathbf{k}_1+\mathbf{k}_2+\mathbf{k}_3) B_{\mathcal{R}\mathcal{R}\gamma_\lambda}(k_1,k_2,k_3) e_{ij}(\mathbf{k}_3,\lambda)\hat{k}_1^i \hat{k}_2^j,
\end{equation}
where $e_{ij}(\mathbf{k},\lambda)$ is the transverse-traceless polarization tensor (see Eq.~\ref{eq:perts}). The amplitude and momentum dependence of the bispectrum $B_{\mathcal{R}\mathcal{R}\gamma_\lambda}$ can be parametrized by \cite{Meerburg:2016ecv}
\begin{equation}
\label{eq:Bzzg}
B_{\mathcal{R}\mathcal{R}\gamma_\lambda}(k_1,k_2,k_3)= 16 \pi^4 A_{\rm s}^2 \sqrt{r}f_\mathrm{NL}^{\mathcal{R}\mathcal{R}\gamma} F(k_1,k_2,k_3).
\end{equation} 

In the bottom half of Table~\ref{tab:NGs} we show the results of forecasts for $\sqrt{r}\tilde{f}_{\rm NL}$ using local-, equilateral-, and orthogonal-like templates. A detection of this correlation would be an immediate indication of some deviation from the simple inflationary paradigm \cite{Bordin:2016ruc,Dimastrogiovanni:2015pla}. There are known possibilities that would generate a scalar-scalar-tensor bispectrum with larger amplitude and/or different shape: different symmetry patterns (e.g., solid inflation \cite{Endlich:2012pz} or gauge-flation \cite{Maleknejad:2011jw, Adshead:2016iix}); gravitational waves not produced as vacuum fluctuations; or multiple tensors (e.g., bigravity \cite{Bordin:2016ruc}). Any non-zero tensor amplitude could also be sourced by a higher-order massive spin field that couples to two scalars and one graviton (see for example Ref.~\cite{Hayden:2016xxa} for a discussion of such signatures).

\begin{table}[t!]
\centering
\begin{tabular}{|c|cccc|}
\hline
 Shape: $\langle \mathcal{R}\mathcal{R}\mathcal{R}\rangle $ &Current & CMB-S4 goal & Conservative & CV-limited  \\
$\langle TTT \rangle $,$\langle TTE \rangle $,$\langle TEE \rangle $,$\langle EEE \rangle $ & & & &\\
$f_{\rm sky}$ &  75\%  &    43\% &  43\% & 100\%\\
\hline
$\sigma (f_{\rm NL}^{\rm local})$ & 5 &  1.9 ($5.3$) & 2.1& 0.8 ($7.1$) \\
$\sigma (f_{\rm NL}^{\rm equil})$ & 43 & 22.1 ($-0.4$) & 23.5& 11.8 ($-1.9$)  \\
$\sigma (f_{\rm NL}^{\rm ortho})$  & 21 &  9.0 ($-5.0$) & 10.6 & 4.4 ($-6.3$)  \\
\hline\hline
 Shape: $\langle \mathcal{R} \mathcal{R} \gamma \rangle $ & Current  &  CMB-S4 goal & Conservative &   CV-limited \\
 $\langle BTT \rangle $,$\langle BTE \rangle $,$\langle BEE \rangle$ & & & &\\
$f_{\rm sky}$  & 69\% & 3\% & 3\% &  100\% \\
\hline
$\sigma(\sqrt{r}\tilde{f}_{\rm NL}^{\rm local})$ & 28 &  0.79 & 1.2& 0.052  \\
$\sigma(\sqrt{r}\tilde{f}_{\rm NL}^{\rm equil})$ & \dots &  16 & 24 & 1.7 \\
$\sigma(\sqrt{r}\tilde{f}_{\rm NL}^{\rm ortho})$ & \dots &  4.4 & 7.4 & 0.41 \\
\hline
\hline
\end{tabular}
\caption[CMB-S4]{Forecast constraints on non-Gaussianity. $\langle \mathcal{R} \mathcal{R} \mathcal{R} \rangle$: Current constraints are from \planck\ . Forecasted constraints for CMB-S4 goal use \planck\ noise at low $\ell$ and  bispectra are computed using the \planck\ 2018 cosmology \cite{Aghanim:2018eyx}. Forecasts do not include foregrounds or lensing contributions to the covariance, but the signal bias from lensing is given in brackets ($T$ + $E$, see Ref.~\cite{Lewis:2011fk}); some secondaries are expected (see Ref. \cite{Hill:2018ypf}), in particular in the local and orthogonal shapes constrained using temperature data alone. We quote constraints without using $\langle TTT\rangle$ in the conservative column for reference, where $\sigma_{f_{\rm NL}} = ((\sigma^{\rm All}_{f_{\rm NL}})^{-2}-(\sigma^{T}_{f_{\rm NL}})^{-2})^{-1/2}$. Some of these can be cleaned using e.g., ILC techniques, while others would require simulations. Deprojection of tSZ was performed in Ref.~\cite{2018arXiv180807445T}, which suggested minimal impact, so we do not expect the constraints to be impacted as much as removing $\langle TTT\rangle$ entirely, but further study is required, especially for contributions that cannot be removed using multi-frequency information such as kSZ. For polarization, apart from the reionization-lensing bispectrum, leakage of secondaries is currently less well studied. Constraints can be combined over $35\%$ of the sky with Planck sensitivity, improving constraints by $\mathcal{O}(10\%)$. You can compare the results to the cosmic variance limited constrains with $\ell_{\rm max} = 5000$ in the last column. $ \langle \mathcal{R} \mathcal{R} \mathcal{\gamma} \rangle$: Current constraint are derived from temperature only and shape considered is not exactly $f_{\rm NL}^{\rm local}$. There are currently no constraints on equilateral and orthogonal non-Gaussianities $\langle \gamma \mathcal{R} \mathcal{R} \rangle $. Forecasts add the small-aperture telescopes over $3\%$ of the sky. We assume $13\%$ residual lensed $B$-mode power. Preliminary studies show that foregrounds could leak into $\langle BTE \rangle$ and $\langle BEE \rangle$, and we show how projected constraints are affected when these combinations are {\it not\/} included in the conservative column. Lensing contributions are expected to be present in these observables as well, but neither they nor foregrounds are included in these forecasts. \label{tab:NGs}}
\end{table}

\subsubsection{Primordial non-Gaussianity: large-scale structure}

CMB-S4 can also contribute to the measurement of local primordial non-Gaussianities in an indirect way, by complementing a galaxy survey to exploit so-called sample-variance cancellation. By combing CMB-S4 data with measurements of large-scale structure, for example with LSST \cite{LSSTScienceCollaboration2009}, impressive constraints could be achieved on $f^{\rm local}_{\rm NL}$. We briefly review this method and forecast constraints on the local parameter.

On linear scales, galaxy bias relates the amplitude of matter perturbations $\delta_{\rm m}$ and galaxy perturbations $\delta_{\rm g}$. Local non-Gaussianities lead to a scale-dependent correction to the galaxy bias $b_{\rm g}$ on large scales, proportional to $f^{\rm local}_{\rm NL} / k^2$ \cite{Dalal:2007cu}. If one can obtain a measurement of both the biased galaxy field and the unbiased matter field, one can cancel out the stochastic nature of the modes to measure $b_{\rm g} = \delta_{\rm g}/\delta_{\rm m}$ without sample variance \cite{Seljak:2008xr}. Of course the measurement of both modes will in practice be noisy and limit the effect of sample-variance cancellation. 

Two different methods have recently been proposed for how to use high-resolution CMB maps to measure the matter field $\delta_{\rm m}$ in conjunction with a large-scale galaxy survey
for sample-variance cancellation: the reconstructed
CMB lensing potential \cite{Schmittfull:2017ffw}; and the kSZ reconstruction \cite{Munchmeyer:2018eey}. 
Interestingly these two techniques trace different modes, since lensing probes transverse modes and kSZ velocities probe radial modes of the underlying matter distribution. We now discuss each technique in more detail.

\paragraph{Kinematic Sunyaev-Zeldovich effect}
The large-scale radial velocity field can be reconstructed by a process called kSZ tomography \cite{Deutsch:2017ybc, Smith:2018bpn}. This is possible because the kSZ temperature is proportional to the line-of-sight integral over the product of electron density $\rho_{\rm e}$ and radial velocity $v_{\rm r}$. If one uses galaxies as tracers for the electron distribution, one can construct a quadratic estimator that combines small-scale CMB temperature fluctuations and the small-scale galaxy distribution, to estimate the large-scale velocity field. 
The noise in this kSZ velocity reconstruction is independent of the scale of the perturbation. 
In linear theory the velocity field is related to the matter density field by a factor $f a H / k$, where $f$ is the growth rate, resulting in a density reconstruction noise proportional to $k^2$ that can 
be much smaller than galaxy shot noise (which is independent of $k$) on large scales. 

The kSZ-derived matter density can then be cross-correlated with a galaxy tracer of the
same large-scale density field \cite{Munchmeyer:2018eey}, as a function of scale, to determine the
scale dependence of the clustering bias. Since the same modes are being measured in each survey,
there is no sample variance in this comparison, greatly improving the precision on large
scales.

We forecast $\sigma( f^{\rm local}_{ \rm NL})  = 0.57$ for the combination of CMB-S4 kSZ and LSST, using the forecasting pipeline of Ref. \cite{Munchmeyer:2018eey}, which can be compared to $\sigma(f^{\rm local}_{\rm NL}) = 1.45$ from LSST alone (assuming no internal sample-variance cancellation within the LSST data). For comparison, for Simons Observatory plus LSST the sensitivity is forecasted to be $\sigma( f^{\rm local}_{ \rm NL})  = 1.0$~\cite{Ade:2018sbj}.

\paragraph{CMB Lensing}
An alternative method to measure $f^{\rm local}_{\rm NL}$ using sample-variance cancellation is to cross-correlate the CMB-S4 CMB lensing measurement with galaxy surveys such as LSST \cite{Jeong0910,Schmittfull:2017ffw}. 
Since the CMB lensing convergence field is determined by all the structure between the CMB last-scattering surface and the observer, we need to include galaxies at high redshift, $z>3$, to obtain a large cross-correlation coefficient between CMB lensing and galaxies, which is required for sample-variance cancellation to work efficiently. 
In addition to the LSST galaxies at $z\le 3$ specified in the last section, we therefore include additional LSST galaxies at $z=3$--7, based on extrapolating \cite{Schmittfull:2017ffw} recent Hyper Suprime-Cam observations \cite{Goldrush1,Goldrush2} of dropout galaxies \cite{Dunlop1205} in that redshift range (not including these high-redshift galaxies would degrade the forecasted $f^{\rm local}_{\rm NL}$ precision reported below by about a factor of 2).
We split the LSST galaxies into six tomographic redshift bins at $z=0$--0.5, 0.5--1, 1--2, 2--3, 3--4, and 4--7, and estimate the ability of CMB-S4 to measure local non-Gaussianity.

The result of this forecast depends strongly on the minimum multipole $\ell_{\rm min}$ of both the LSST galaxy density field and the CMB lensing convergence, since the signal scales as $1/k^2$. 
For $\ell_{\rm min}=2,8,15,20$, and $40$ we obtain $\sigma(f^{\rm local}_{\rm NL})=0.72,  0.99,  1.3,  1.5$, and $2.2$, respectively, assuming $f_{\rm sky}=0.3$ overlap with LSST.
Similarly to the kSZ forecast above, this method therefore allows us to probe the $\sigma(f^{\rm local}_{\rm NL})\approx 1$ regime.
With the Simons Observatory, the same forecast yields $\sigma(f^{\rm local}_{\rm NL})=1.2,  1.3,  1.5,  1.7$, and $2.6$, which is $60\%$ larger than CMB-S4 for $\ell_{\rm min}=2$ and about $15\%$ larger for $\ell_{\rm min}\ge 15$.

\newpage
\section{The dark Universe}

\begin{shaded}
\emph{The second science theme relates to the fundamental physics of invisible components of the Universe.}

CMB-S4 will probe the fundamental physics of components that are difficult or impossible for us to observe directly.
It will test for the presence of light, relativistic relic particles, beyond our Standard Model of particle physics, that were thermally produced in the early Universe.
To date, CMB observations by the {\it Planck\/} satellite can probe light particles that departed from equilibrium (``froze out'') as early as the first $\simeq 50$ micro-seconds of the Universe. With CMB-S4 we can push back this frontier by over a factor of 10,000, to the first fractions of a nanosecond.

CMB-S4 will achieve sensitivity to relics that froze out well before the quark-hadron phase transition (the epoch when the Universe cooled sufficiently that quarks became locked into hadrons like neutrons and protons).  The contribution of light relics to the energy density leads to observable consequences in the CMB temperature and polarization anisotropy. This is often parameterized with the ``effective number of neutrino species,'' $\Neff$.  The collective influence of the three already-known light relics (the three families of neutrinos) has already been detected at high significance. Current data are only sensitive enough to detect additional relics that froze out after the quark-hadron transition, and Stage-3 CMB experiments can only push somewhat into that epoch, so CMB-S4's ability to probe times well before the transition is a major advance.

CMB-S4 can measure the summed mass of the neutrino species. Together with terrestrial experiments this may shed light on the mechanism responsible for neutrino mass, one of the biggest mysteries of the Standard Model. The total mass unambiguously determines the absolute mass scale of neutrinos, and if the sum is $\lesssim 0.1$eV, CMB-S4 can furthermore disfavor the inverted mass hierarchy.  Together with neutrino-less double beta-decay experiments, CMB-S4 may provide evidence to help determine if neutrinos are Dirac or Majorana particles, or point to unknown physics in the neutrino sector.  The main neutrino observables are CMB lensing, lensing cross-correlation, and the abundances of galaxy clusters.

Current cosmological observations already require the existence of dark energy and dark matter, but these have not been observed in the laboratory.  CMB-S4 can target the various predictions that dark-energy models make for the clustering and growth of matter fluctuations at late times, examining the secondary anisotropies from gravitational lensing and interactions with the gas in galaxies and galaxy clusters.  Constraints on cosmic birefringence could probe the microphysics of the dark energy.

CMB-S4 can test dark-matter models and constrain parts of the parameter space that are inaccessible to laboratory experiments.  In particular, CMB observations directly probe the physics of dark matter throughout cosmic history, and do not rely on assumptions about the local dark-matter phase-space distribution within the Milky Way.  CMB-S4 can place constraints on a variety of scenarios, including dark matter that interacts with baryons or with dark radiation, or consists of ultra-light axion-like particles.
\end{shaded}

\subsection{Light relics}
\label{sec:lr}
\label{sec:science_light_relics} 

A natural and important question is whether there are additional, unknown forms of radiation in the Universe (i.e., additional relativistic species).
This radiation density leaves a measurable imprint on the CMB and can be determined with high precision by CMB-S4.  Radiation also changes the Universe at late times, by altering the amplitude of clustering, the scale and phase of the baryon acoustic oscillations (BAOs), and abundances of light elements.   Precision measurements of the radiation content of the Universe with the CMB is both a window into the dark sector and a necessary tool for calibrating measurements of the lower-redshift Universe.

The possibility of additional radiation is also compelling, in the contexts of both particle physics and cosmology~\cite{Essig:2013lka, Marsh:2015xka, Alexander:2016aln}. New light particles may arise in the form of axions and sterile neutrinos, or can appear as a byproduct of new symmteries that would explain the small mass of the Higgs boson \cite{Abazajian:2001nj, Strumia:2006db, Ackerman:2008gi, Boyarsky:2009ix, Arvanitaki:2009fg, Cadamuro:2010cz, Kaplan:2011yj, Abazajian:2012ys, CyrRacine:2012fz, Brust:2013xpv, Weinberg:2013kea, Salvio:2013iaa, Essig:2013lka, Kawasaki:2015ofa, Graham:2015cka, Marsh:2015xka, Baumann:2016wac, Alexander:2016aln, Arkani-Hamed:2016rle, Chacko:2016hvu, Craig:2016lyx, Chacko:2018vss}. Furthermore, light particles can thermalize in the early Universe for wide ranges of unexplored parameter space,  leading to an observable level of additional radiation. Light particles may form the dark matter (e.g., axions) or part of a dark sector, they can mediate forces in the dark and visible sectors, or they can result from the decay of new heavier particles. These possibilities may also play a role in explaining the discrepancies observed in the Hubble constant $H_0$ measurements~\cite{Archidiacono:2013fha, Bernal:2016gxb, Zhang:2017aqn, Addison:2017fdm, Aylor:2018drw}, the amplitude of fluctuations $\sigma_8$~\cite{MacCrann:2014wfa, Lesgourgues:2015wza, Kohlinger:2017sxk, Joudaki:2017zdt}, and clustering on small scales~\cite{Weinberg:2013aya,Aghanim:2018eyx}. 

CMB-S4 will provide a transformative measurement for the radiation density, particularly for particles that would have thermalized in the early Universe. Any such particle adds at least a percent-level contribution to the radiation energy density; the precise amount, per degree of freedom, is determined entirely by its decoupling temperature. CMB-S4 will be orders of magnitude more sensitive to decoupling temperatures than current experiments, and can reach targets that are not reachable by other means.

\paragraph{Cosmic neutrino background.}
The cosmic neutrino background is one of the remarkable predictions of the hot Big Bang scenario. In the very early Universe, neutrinos were kept in thermal equilibrium with the Standard Model plasma. As the Universe cooled, neutrinos decoupled from the plasma. A short time later, the relative number density and temperature in photons increased, primarily due to the transfer of entropy to photons from annihilating electron-positron pairs. The background of cosmic neutrinos persists today, with a temperature and number density similar to that of the CMB. Their energy density $\rho_\nu$ is most commonly expressed in terms of the effective number of neutrino species,\vspace{-7pt}
\begin{equation}
\Neff = \frac{8}{7}\left(\frac{11}{4}\right)^{\!4/3} \frac{\rho_\nu}{\rho_\gamma} \, ,\vspace{-5pt}
\end{equation}
where $\rho_\gamma$ is the energy density in photons. This definition is chosen so that $\Neff=3$ in the SM if neutrinos had decoupled instantaneously prior to electron-positron annihilation. The neutrino density $\rho_\nu$ receives a number of corrections from this simple picture of decoupling, and the best available calculations give $\Neff^\mathrm{SM} = 3.045$ in the SM~\cite{Mangano:2005cc, Grohs:2015tfy, deSalas:2016ztq}.

Cosmology is sensitive to the gravitational effects of neutrinos, both through their mean energy density~\cite{Peebles:1966zz, Dicus:1982bz, Hou:2011ec, Bashinsky:2003tk} and their fluctuations, which propagate at the speed of light in the early Universe due to the free-streaming nature of neutrinos~\cite{Bashinsky:2003tk, Baumann:2015rya, Baumann:2017lmt}. A radiation fluid whose fluctuations do not exceed the sound speed of the plasma~\cite{Bell:2005dr, Friedland:2007vv} could arise from large neutrino self-interactions~\cite{Cyr-Racine:2013jua,Oldengott:2014qra,Lancaster:2017ksf,oldengott17,Kreisch:2019yzn}, neutrino-dark sector interactions \cite{Wilkinson:2014ksa,Escudero:2015yka}, or dark radiation self-coupling \cite{Buen-Abad:2015ova,Chacko:2016kgg,Buen-Abad:2017gxg}. Such a radiation fluid can be observationally distinguished from free-streaming radiation, and can serve as both a foil for the cosmic neutrino background and a test of new physics in the neutrino and dark sectors~\cite{Baumann:2015rya, Brust:2017nmv, Choi:2018gho}.

\emph{Neutrinos are messengers from a few seconds after the Big Bang and provide a new window into our cosmological history.} While these relics have been detected in cosmological data, higher precision measurements would advance the use of neutrinos as a cosmological probe. Furthermore, the robust measurement of the neutrino abundance from the CMB is crucial for inferring cosmic parameters, including the expansion history using BAOs~\cite{Dodelson:2016wal}, the neutrino masses~\cite{Abazajian:2013oma}, and $H_0$~\cite{Aylor:2018drw}.

\paragraph{Other light relics.}
\emph{A measurement of the value of $\Neff$ provides vastly more information than just the energy density in cosmic neutrinos.} The parameter $\Neff$ is a probe of any particles that have the same gravitational influence as relativistic neutrinos, which is true of all kinds of (free-streaming) radiation. Furthermore, this radiation could have been created at much earlier times when the energy densities were even higher than in the cores of stars or supernovae, shedding light on the physics at new extremes of temperature and density, and on our early cosmic history.

New light particles that were thermally produced in the early Universe contribute to the neutrino density $\rho_\nu$ and increase $\Neff$ above the amount from neutrinos alone. The presence of any additional species can therefore be characterized by $\Delta\Neff \equiv \Neff - \Neff^\mathrm{SM}$. Since all such thermalized particles behave in the same way from a cosmological point of view, this parametrization captures a vast range of new physics, e.g., axions, sterile neutrinos, dark sectors, and beyond~\cite{Brust:2013xpv, Chacko:2015noa, Baumann:2016wac, Abazajian:2016yjj}.

\begin{figure}[htbp!]
\centering
\includegraphics[trim= 0 0 0 5]{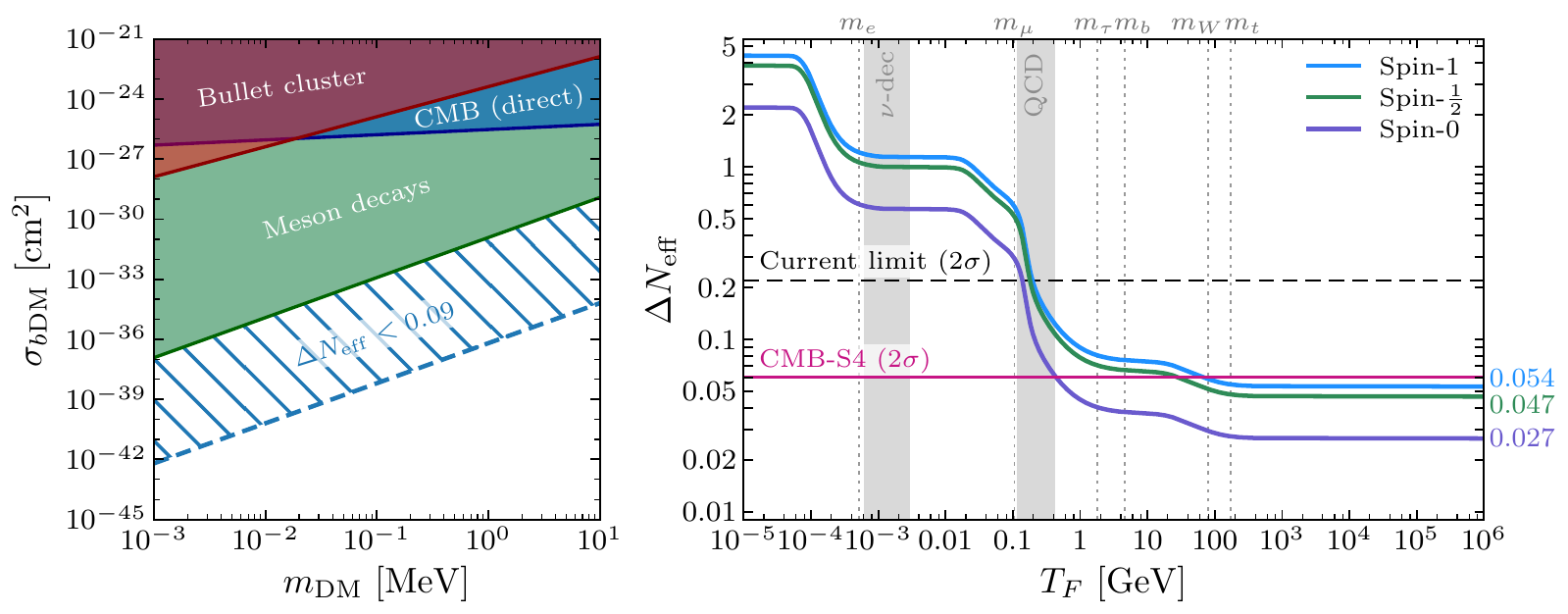}
\caption{\textit{Left:}~Limits on the dark-matter-baryon cross-section $\sigma_{b\mathrm{DM}}$ for a Yukawa potential. Future cosmological constraints will restrict $\Delta \Neff < 0.09$ and, therefore, exclude cross-sections large enough to thermalize the (\SI{200}{keV}-mass) particle mediating the force~\cite{Green:2017ybv}. This limit is compared to the direct bound on baryon-dark-matter scattering from the CMB~\cite{Gluscevic:2017ywp} and to the constraints on dark forces from the Bullet Cluster~\cite{Markevitch:2003at}. The strongest current constraint is from the absence of meson decays to the mediator~\cite{Essig:2010gu}. \textit{Right:}~Contributions of a single massless particle, which decoupled from the SM at temperature $T_F$, to the effective number of relativistic species, $\Neff = N_\mathrm{eff}^\mathrm{SM} + \Delta\Neff$, with the SM expectation $N_\mathrm{eff}^\mathrm{SM} = 3.045$ from neutrinos. The dashed line shows the 2$\sigma$ limit from a combination of current CMB, BAO, and Big Bang nucleosynthesis (BBN) observations~\cite{Aghanim:2018eyx}.  The purple line shows the projected sensitivity of CMB-S4 and illustrates its power to constrain light thermal relics. The displayed values on the right are the observational thresholds for particles with different spins and arbitrarily large decoupling temperatures.}
\label{fig:deltaNeff}
\end{figure}

Constraints on $\Neff$ are broadly useful and, most importantly, allow the exploration of new and interesting territory in a variety of well-motivated models. This can be seen with a simple example: dark-matter-baryon scattering. For low-mass (sub-\si{GeV}) dark matter, current data allow for relatively large scattering cross-sections~\cite{Battaglieri:2017aum}. If they scatter through a Yukawa potential, which is a force mediated by a scalar particle, this force is consistent with fifth-force experiments and stellar cooling if the mediator has a mass around \SI{200}{keV}. However, the particle that mediates the force necessarily\footnote{The mediator with a mass of \SI{200}{keV} is too heavy to contribute to $\Neff$, but it must decay to sub-\si{eV} mass particles, which will increase $\Neff$, in order to avoid more stringent constraints.} contributes $\Delta \Neff \geq 0.09$ when it comes into thermal equilibrium with the Standard Model~\cite{Green:2017ybv}. Excluding this value would require that the strength of the interactions is small enough to prevent the particle from reaching equilibrium at any point in the history of the Universe, which, consequently, limits the scattering cross-section, as shown in the left panel of Fig.~\ref{fig:deltaNeff}. This measurement is sensitive to 10--15 orders of magnitude in cross-section that are not probed by direct constraints from cosmology and astrophysics, and is five orders of magnitude stronger than meson decay searches. We see that \emph{cosmological measurements of $\Delta \Neff$ are an extremely sensitive probe of dark-sector physics that are complementary to more direct tests, both in the laboratory and with astrophysical observations}~\cite{Green:2017ybv, Knapen:2017xzo}.

More generally, the contribution to $\Neff$ from any thermalized new particle is easy to predict because its energy density in equilibrium is fixed by the temperature and the number of internal states (e.g., spin configurations). Under mild assumptions (see e.g., Ref.~\cite{Wallisch:2018rzj} for a detailed discussion), the contribution to $\Delta\Neff$ is determined by two numbers, the last temperature at which it was in equilibrium, $T_{\rm F}$, and the effective number of spin degrees of freedom, $g_{\rm s}$, according to\vspace{-8pt}
\begin{equation}
\Delta\Neff = g_{\rm s} \left(\frac{43/4}{g_{\star}(T_{\rm F})}\right)^{\!4/3} .\vspace{-7pt}
\end{equation}
The function $g_{\star}(T_{\rm F})$ is the number of effective degrees of freedom (defined as the number of independent states with an additional factor of $7/8$ for fermions) of the SM particle content at the temperature $T_{\rm F}$. This function appears in the formula for $\Delta\Neff$ because it determines how much the photons are heated relative to a new light particle due to the annihilation of the heavy SM particles as the Universe cooled (see the right panel of Fig.~\ref{fig:deltaNeff}). CMB-S4 is expected to reach a precision of $\sigma(\Neff) = 0.03$, which would extend our reach in $T_{\rm F}$ by several orders of magnitude for a particle with spin $s > 0$ and be the first measurement sensitive to a real scalar ($s=0$) that decouples prior to the QCD phase transition.

To understand the impact of such a measurement, recall that equilibrium at temperature $T$ arises when the production rate $\Gamma$ is much larger than the expansion rate $H(T)$. At high temperatures, production is usually fixed by dimensional analysis, $\Gamma \propto \lambda^2 T^{2n+1}$, where $\lambda$ is the coupling to the Standard Model with units of $[\mathrm{Energy}]^{-n}$. The particle is therefore in equilibrium if $\lambda^2 \gg \Mpl^{-1} T^{-2n+1}$. There are two important features of this formula: (i)~the appearance of the Planck scale $\Mpl$ implies that we are sensitive to very weak couplings ($\Mpl^{-2} = 8 \pi G_{\hskip-1ptN}$); and (ii)~for $n \geq 1$ it scales like an inverse power of $T$. As a result, sensitivity to increasingly large $T_{\rm F}$ implies that we are probing increasingly weak couplings (lower production rates) in proportion to the improvement in $T_{\rm F}$ (not $\Delta\Neff$). These two features explain why future measurements of $\Delta\Neff$ can be orders of magnitude more sensitive than terrestrial and astrophysical probes of the same physics~\cite{Baumann:2016wac, Abazajian:2016yjj}.

The impact of the current measurement of $\Neff$ and the projected sensitivity of CMB-S4 is illustrated in Fig.~\ref{fig:deltaNeff}. Anticipated improvement in measurements of $\Neff$ translate into orders of magnitude in sensitivity to the temperature $T_{\rm F}$. This temperature sets the reach in probing fundamental physics. Even in the absence of a detection, future cosmological probes would place constraints that can be orders of magnitude stronger than current probes of the same physics, including for axion-like particles~\cite{Baumann:2016wac} and dark sectors~\cite{Adshead:2016xxj, Craig:2016lyx, Green:2017ybv, Chacko:2018vss}. It is also worth noting that these contributions to $\Neff$ asymptote to specific values of $\Delta\Neff = 0.027$, 0.047, and 0.054 for a massless (real) spin-0 scalar, spin-1/2 (Weyl) fermion, and spin-1 vector boson, respectively (see Fig.~\ref{fig:deltaNeff}). 

Even without new light particles, $\Neff$ is a \emph{probe of new physics that changes our thermal history, including processes that result in a stochastic background of gravitational waves}~\cite{Boyle:2007zx, Stewart:2007fu, Meerburg:2015zua}. Violent phase transitions and other nonlinear dynamics in the primordial Universe could produce such a background, peaked at frequencies much larger than those accessible to $B$-mode polarization measurements of the CMB or, in many cases, direct detection experiments such as LIGO and LISA~\cite{Maggiore:1999vm, Easther:2006gt, Dufaux:2007pt, Amin:2014eta,Caprini:2018mtu}. For particularly violent sources, the energy density in gravitational waves can be large enough to make a measurable contribution to $\Neff$~\cite{Caprini:2018mtu, Adshead:2018doq, Amin:2019qrx}.

In addition to precise constraints on $\Neff$, cosmological probes will provide an \emph{independent high-precision measurement of the primordial helium abundance $Y_\mathrm{p}$}, due to the impact of helium on the free electron density prior to recombination. This is particularly useful since $Y_\mathrm{p}$ is sensitive to $\Neff$ a few minutes after the Big Bang, while the CMB and matter power spectra are affected by $\Neff$ prior to recombination, about \num{370,000}~years later. Measuring the radiation content at these well-separated times provides a window onto any nontrivial evolution in the energy density of radiation in the early Universe~\cite{Fischler:2010xz, Hasenkamp:2011em, Hooper:2011aj, Hasenkamp:2012ii}. Furthermore, $\Neff$ and $Y_\mathrm{p}$ are sensitive to neutrinos and physics beyond the Standard Model in related, but different ways, which allows for even finer probes of new physics, especially in the neutrino and dark sectors.

\subsubsection{Observational signatures}\label{sec:Neffsignatures}

The effect of the radiation density on the damping tail drives the constraint on $\Neff$ in the CMB in $\Lambda$CDM\,+\,$\Neff$ models.  The largest effect is from the impact on the mean free path of photons, which introduces an exponential suppression 
of short wavelength modes~\cite{Zaldarriaga:1995gi}.  In a more detailed analysis, the effect on the damping tail is subdominant to the change to the scale of matter-radiation equality and the location of the first acoustic peak~\cite{Hou:2011ec}, both of which are precisely measured.  As a result, the effect of neutrinos on the damping tail is more accurately represented by holding the first acoustic peak fixed.  This changes the sign of the effect on the damping, but the intuition for the origin of the effect 
remains the same.  At the noise level and resolution of CMB-S4, this effect is predominately measured through the $TE$ spectrum at $\ell >2500$.  

In addition to the effect on the expansion rate, perturbations in neutrinos (and dark radiation) affect the photon-baryon fluid through their gravitational influence.  The contributions from neutrino fluctuations are well described by a correction to the amplitude and the phase of the acoustic peaks in both temperature and polarization~\cite{Bashinsky:2003tk}.  The phase shift is a particularly compelling signature, since it is not degenerate with other cosmological parameters~\cite{Bashinsky:2003tk,Baumann:2015rya}.  This effect is the result of the free-streaming nature of neutrinos, which allows propagation speeds of effectively the speed of light (while the neutrinos are relativistic).  Any free-streaming light relics will also contribute to the amplitude and phase shift of the acoustic peaks.  Furthermore, light relics that do not freely stream can in principle be disinguished from those that do by measuring their differing impacts on the damping tail and phase shift~\cite{Baumann:2015rya,Choi:2018gho}.

\planck\ has provided a strong constraint on $\Neff = 2.92^{+0.18}_{-0.19}$ using temperature and polarization data~\cite{Aghanim:2018eyx}. 
Recently, the phase shift from neutrinos has also been established directly in the \planck\ temperature data~\cite{Follin:2015hya}.  This provides the most direct evidence for the presence of free-streaming radiation in the early Universe, consistent with the cosmic neutrino background.

\subsubsection{Target}

\begin{figure}[t!]
\begin{center}
\includegraphics[width=0.35\textwidth]{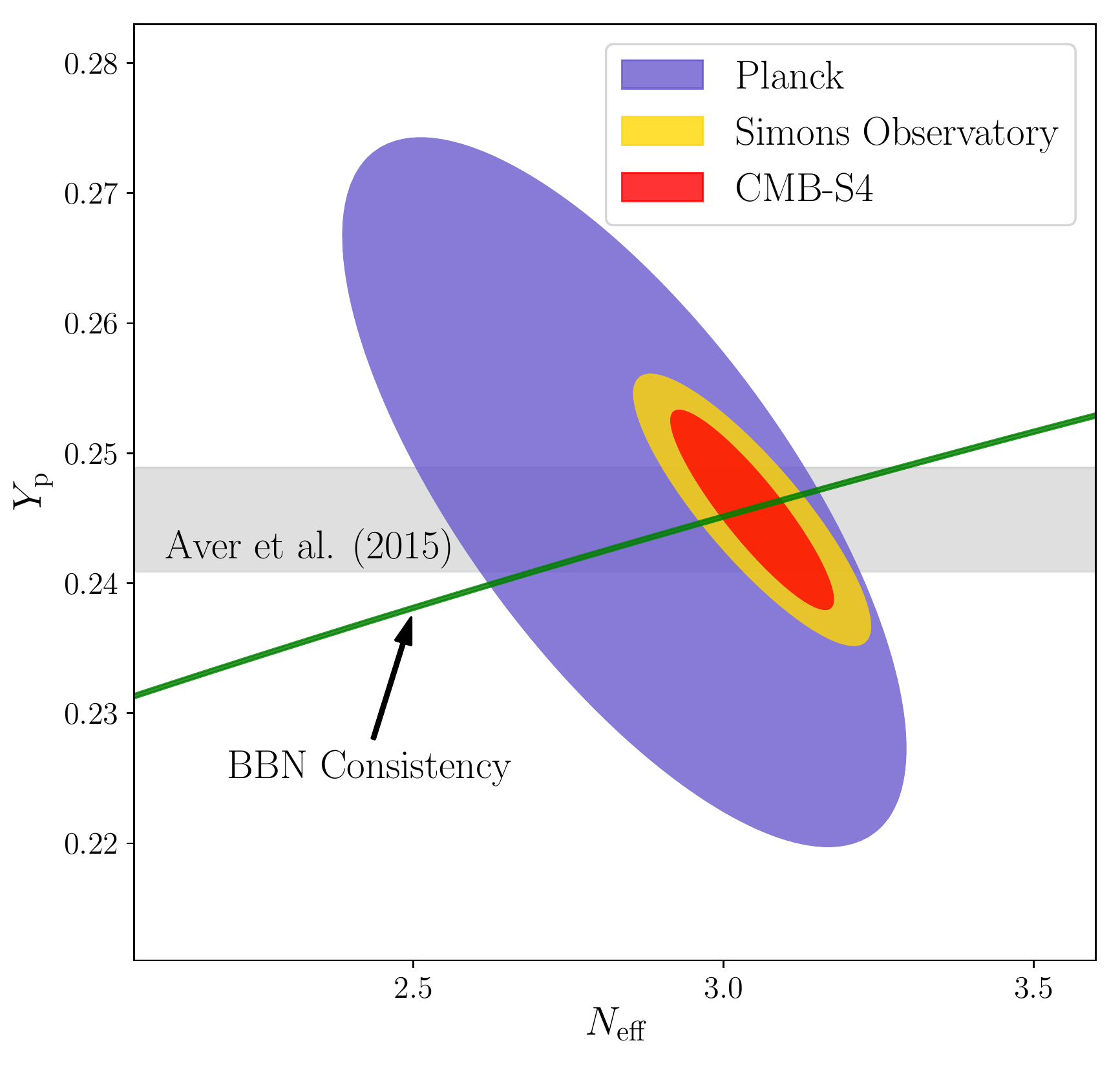}
\hskip 20pt
\includegraphics[width=0.45\textwidth]{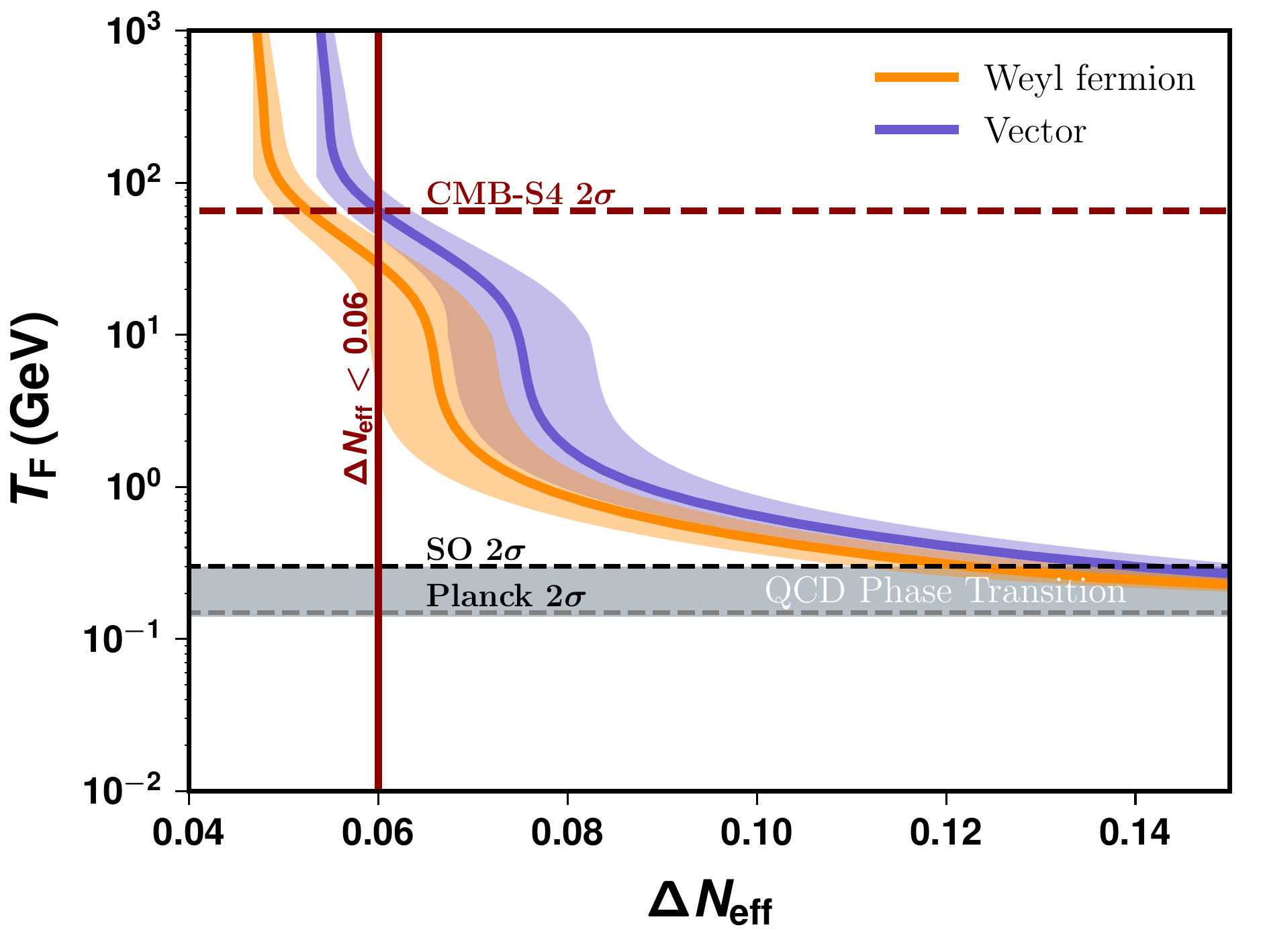}
\caption{{\it Left:}~Constraints in the $Y_\mathrm{p}$--$\Neff$ plane from current and future cosmological surveys, compared to the predictions of standard BBN and current astrophysical measurements of $Y_\mathrm{p}$~\cite{Aver:2015iza}.  {\it Right:}~Inferred reach in $T_\mathrm{F}$ for a given sensitivity to $\Neff$.  The vertical limits on $T_\mathrm{F}$ assume we are using a vector ($g=2$) to relate $2\sigma$ $\Neff$ limits to $T_\mathrm{F}$.  We see that at CMB-S4 sensitivity, constraints have a reach two orders of magnitude higher in $T_\mathrm{F}$ than either \planck\ or the Simons Observatory. Bands show error estimates based on Ref.~\cite{Saikawa:2018rcs}.}
\label{fig:Neff_thermal}
\end{center}
\end{figure}

One of the features that makes $\Neff$ a compelling theoretical target is the degree to which broad classes of models fall into two basic levels of $\Delta\Neff$.  As illustrated in the left panel of Fig.~\ref{fig:Neff_thermal}, any species that was in thermal equilibrium with the Standard Model degrees of freedom produces a characteristic correction to $\Neff$ that depends only on its spin and its freeze-out temperature.  For freeze-out after the QCD phase transition, one finds $\Delta \Neff \gtrsim 0.3$.  Freeze-out before the QCD phase transition instead produces $0.07>\Delta \Neff > 0.027$ per effective degree of freedom.  The first case has been tested by the data from the \planck\ satellite.  The second category, which is sensitive to freeze-out temperatures ranging from hundreds of MeV to the reheating temperature, falls into the level of sensitivity attainable by CMB-S4.

The right panel of Fig.~\ref{fig:Neff_thermal} shows how a measurement at CMB-S4 sensitivity will dramatically improve our knowledge of physics at higher temperatures (i.e., at earlier times, weaker coupling).  CMB-S4 at design sensitivity would place an upper limit of $\Delta \Neff < 0.060$ at 2$\sigma$, which translates into $T_\mathrm{F} =70$\,(30)\,GeV for a light vector (Weyl fermion).  
This dramatic improvement in sensitivity to $T_\mathrm{F}$ over existing limits from \planck\, $T_\mathrm{F} = 0.1$~(0.1)\,GeV, and projected constraints from Simons Observatory, $T_\mathrm{F} = 0.3$~(0.3)\,GeV, is a direct result of the particle content of the Standard Model, which changes dramatically from quarks and gluons to baryons and mesons around the QCD phase transition (0.1--1\,GeV) and thus significantly dilutes the energy densities of particles that decoupled at higher temperatures.  Reaching $\Delta \Neff < 0.06$--$0.07$ leads to dramatic improvements in the sensitivity to particles with spin. At this level of sensitivity, the zoo of particles present around the QCD phase transition is insufficient to dilute the energy below experimental sensitivity, and we see a very sharp corresponding rise in the $T_\mathrm{F}$ curves.  Instead we are limited only by the dilution due to a few heavy particles and thus it is easier to make significant gains.

\subsection{Neutrino mass}
\label{sec:mnu}

 The 2015 Nobel Prize in Physics recognized the discovery of neutrino oscillations, which demonstrated that neutrinos have mass. However, the overall scale of the neutrino masses and some mixing parameters of the full set that determine neutrino oscillations are still not measured.
Cosmology offers a unique view of neutrinos; they were produced in large numbers in the high temperatures of the early Universe and left a distinctive imprint in the cosmic microwave background and on the large-scale structure of the Universe. Therefore, CMB-S4 and the DESI and LSST surveys together will have the power to detect properties of neutrinos that complement those probed by large terrestrial experiments such as the Deep Underground Neutrino Experiment (DUNE), and those searching for beta-decay and neutrinoless double-beta decay.

Neutrino flavor oscillations---oscillations between electron, muon, and tau neutrinos---are described by a model where the three neutrino flavor states are a linear combination of three mass states. The matrix that relates the two is the Pontecorvo-Maki-Nakagawa-Sakata (PMNS) matrix. The PMNS matrix depends upon six real parameters: three mixing angles, $\theta_{12}$,  $\theta_{23}$, $\theta_{13}$; and three $CP$-violating phases, $\delta$, $\alpha_1$, $\alpha_2$.

Experiments have measured the three mixing angles of $U_{\rm PMNS}$ and the two mass-squared splittings, $\Delta m^2_{21}= 7.54 \times 10^{-5}\,$eV$^2$ and $|\Delta m_{13}^2|\approx 2.4 \times 10^{-3}\,$eV$^2$ \cite{Agashe:2014kda}. Despite this, fundamental questions about neutrino mass and mixing remain: (i) the absolute mass scale; (ii) the relative ordering, or {\em hierarchy} of the masses; (iii) whether neutrinos are Dirac or Majorana particles; and (iv)  and whether neutrinos violate charge and parity (measuring $\delta_{CP}$).
CMB-S4 will measure the sum of the neutrino masses,
\begin{equation}
    M_\nu \equiv \sum m_\nu
\end{equation} 
with sufficient sensitivity to be relevant to these open issues. The value of $M_\nu$ unambiguously determines the absolute mass scale of neutrinos, and if $M_\nu$ is determined to be $\lesssim 0.1\,$eV, CMB-S4 can furthermore rule out the inverted mass hierarchy at increasing statistical significance depending on $\sigma(M_\nu)$. There are also circumstances in which the combination of CMB-S4 constraints and neutrinoless double-beta decay measurements can point to neutrinos being Dirac particles. Finally, the observables available to CMB-S4 to probe neutrinos can also provide information about new physics in the neutrino sector: the existence of new sterile neutrino states, neutrino interactions and decays, or a nonstandard thermal history for the Universe.  

In the standard cosmology, neutrinos were in thermal equilibrium with photons, electrons, and positrons until the temperature of the Universe dropped below $T\sim1\,$MeV. At this time the weak interaction rate became smaller than the expansion rate of the Universe and neutrinos decoupled. Assuming the standard thermal history, the neutrino temperature can be related to the CMB temperature, and we can predict a relic abundance of $n_\nu \approx 56\,\mathrm{cm^{-3}}$ for each neutrino and antineutrino state. The CMB-inferred total radiation density in the early Universe is in excellent agreement with a contribution from neutrinos with sub-eV mass and this expected relic abundance. As discussed in \ref{sec:lr}, CMB-S4 constraints on the radiation density will put stringent tests on the thermal history of neutrinos. 

The neutrino mass-square splittings give lower limits on the masses of two neutrino states of $\geq \sqrt{\Delta m_{21}^2} \approx 0.01\,$eV and $\geq\sqrt{|\Delta m_{13}^2|} \approx 0.05\,$eV. These lower limits, in combination with the temperature $T_\nu$, imply that the energy of at least two species of relic neutrinos is today dominated by their rest mass, rather than their momentum. In combination with the inferred relic abundance, this predicts a neutrino contribution to the energy budget of $\Omega_\nu h^2 \gtrsim 0.0006$. This $\Omega_\nu$ contributes to the matter budget of the Universe at late times. 

Since neutrinos were relativistic for much of the history of the Universe and still have relatively large kinetic energy today, their gravitational clustering is qualitatively different from that of cold dark-matter particles or baryons. On large scales, neutrino clustering is identical to that of CDM and baryons. On smaller scales, where the neutrino velocity is important, neutrinos free stream out of gravitational potentials, leaving the CDM and baryons behind; since the free-streaming neutrinos' energy still contributes to the expansion, this causes a suppression to the growth of structure on small scales. The scale separating these two regimes is the neutrino free-streaming scale, defined by \cite{Bond:1983hb, Lesgourgues:2006nd} $k_{\rm fs}(a) \equiv \sqrt{\frac{3}{2}}aH(a)/v_\nu(a) \approx 0.04\, a^2 \sqrt{\Omega_{\rm m} a^{-3} + \Omega_\Lambda}\left(m_\nu/0.05\,{\rm eV}\right) h/{\rm Mpc}$ in comoving coordinates. For a fixed CDM and baryon density, massive neutrinos thus induce a suppression in the matter power spectrum given by \cite{Bond:1983hb, Ma:1996za, Hu:1997vi, Hu:1997mj}
\begin{equation}
\frac{P_{\rm mm}(k \gg k_{\rm fs} | \Omega_\nu)}{P_{\rm mm}( k \gg k_{\rm fs}| \Omega_\nu = 0)} \approx 1-6 \Omega_\nu/\Omega_{\rm m} .
\end{equation}
Cosmological data sets are primarily sensitive to the net suppression, and therefore the neutrino energy density $\Omega_\nu$. Since the neutrino number density is separately constrained and is the same for each mass eigenstate, these measurements of $\Omega_\nu$ determine $M_\nu$.

A central goal of the CMB-S4 experiment is to achieve a robust detection of the neutrino mass for the minimum mass sum implied by oscillation data, $M_\nu = 58 \, {\rm meV}$ (or $\Omega_\nu h^2 = 0.0006$).  In the next section, we will discuss methods through which CMB-S4 can directly constrain the neutrino mass.

\subsubsection{CMB-S4 observables for \textit{M}$_\nu$}
\label{ssec:Mnuobservables}
The main effect of massive neutrinos on large-scale structure is to suppress the matter power spectrum on scales smaller than the neutrino free streaming scale. By measuring the magnitude of this suppression, we can therefore determine the neutrino mass from cosmology. 

Typically, this suppression is measured by comparing the amplitude of structure below the free-streaming scale to the nearly unsupressed amplitude of structure seen in the CMB; however, a constraint on the optical-depth parameter $\tau$ is required to determine the primordial amplitude from the CMB, and the precision of this optical-depth constraint can therefore be a limiting factor. Beyond the typically limiting optical-depth degeneracy, another degeneracy arises with the matter density (on which structure growth also depends), though this is more easily overcome by combining CMB data with baryon acoustic oscillation measurements.

{\bf CMB Lensing:} One of the cleanest methods to probe the matter power spectrum, and hence the neutrino mass, is CMB lensing. Along their path to our telescopes, the photons of the CMB are gravitationally deflected by the entire mass distribution through which they pass, remapping the unlensed temperature and polarization as $X(\mathbf{\hat{n}})=X^{\rm un}(\mathbf{\hat{n}+\nabla }\phi(\mathbf{\hat{n}}))$ where the CMB $X={T,E,B}$ and the lensing potential $\phi$ is a direct gravitational probe of the projected mass distribution. The lensing potential power spectrum is related to the projected matter power spectrum by $C_\ell^{\phi \phi} = \int_0^{\eta^*} d \eta\, W^2(\eta)P(k=\ell/\eta)$,
where $W$ is a geometric projection kernel that depends mainly on distance ratios and $\eta^*$ is the distance to the CMB last scattering surface.
CMB lensing is a relatively clean probe because it arises from linear or mildly non-linear scales that are minimally affected by baryonic feedback and because the source redshift is perfectly known. While non-Gaussianity introduced by extragalactic and Galactic foregrounds must be taken into consideration, such systematics are typically small to begin with and can be further mitigated by 
improved lensing estimators and multifrequency cleaning.

{\bf Lensing cross-correlations:} A related probe that can also give insight into the neutrino mass is the measurement of cross-correlations of CMB-S4 lensing with upcoming galaxy surveys such as LSST or Euclid. Since a lensing-galaxy correlation $C_\ell^{\phi {\rm g}}$ is proportional to $b P_{\rm mm}$, where $b$ is galaxy bias, an unknown astrophysical parameter. The galaxy power spectrum scales as $C_\ell^{\rm g g} \propto b^2 P_{\rm mm}$, the combination of these two probes provides (to some extent) an independent measure of the amplitude of structure, and hence of the neutrino mass. 
While the optical-depth degeneracy remains, it is somewhat reduced by the galaxy shape constraint in $C_\ell^{\rm gg}$, as well as the sensitivity to low-redshift growth. Furthermore, since the cross-correlations can be broken up into tomographic redshift bins, cross-correlation measurements can help break degeneracies between dark-energy parameters and neutrino masses. While some systematic errors are nulled in cross-correlation, extragalactic foregrounds remain in temperature; in addition, the measurement of galaxy clustering can come with considerable challenges such as photometric redshift uncertainty.

{\bf Galaxy Clusters:} Via the thermal Sunyaev-Zeldovich effect, CMB-S4 will be able to find tens of thousands of galaxy clusters out to high redshifts. The masses of these galaxy clusters can be obtained via an SZ-mass scaling relation calibrated by both weak lensing (e.g., with LSST) and CMB lensing (internally). This will allow a determination of the mass function, which is a sensitive probe of the amplitude of structure on small scales, and hence will provide a precise probe of the neutrino mass. The redshift-dependence of the galaxy cluster abundance is expected to reduce
degeneracy between neutrino and dark-energy parameters. Although clusters are complex objects with rich astrophysics complicating precise cosmological measurements, great progress has been made in mass calibration and in the mitigation of other uncertainties. Clusters represent a physically very different probe to CMB gravitational lensing; the robustness of neutrino mass
measurements will be substantially improved by having 
independent measurement techniques with entirely different sources
of systematic errors.

\subsubsection{CMB-S4 forecasts for $M_\nu$}
\label{sec:mnuforecast}
\begin{figure}[h!]
\centering \includegraphics[height=0.31\textwidth]{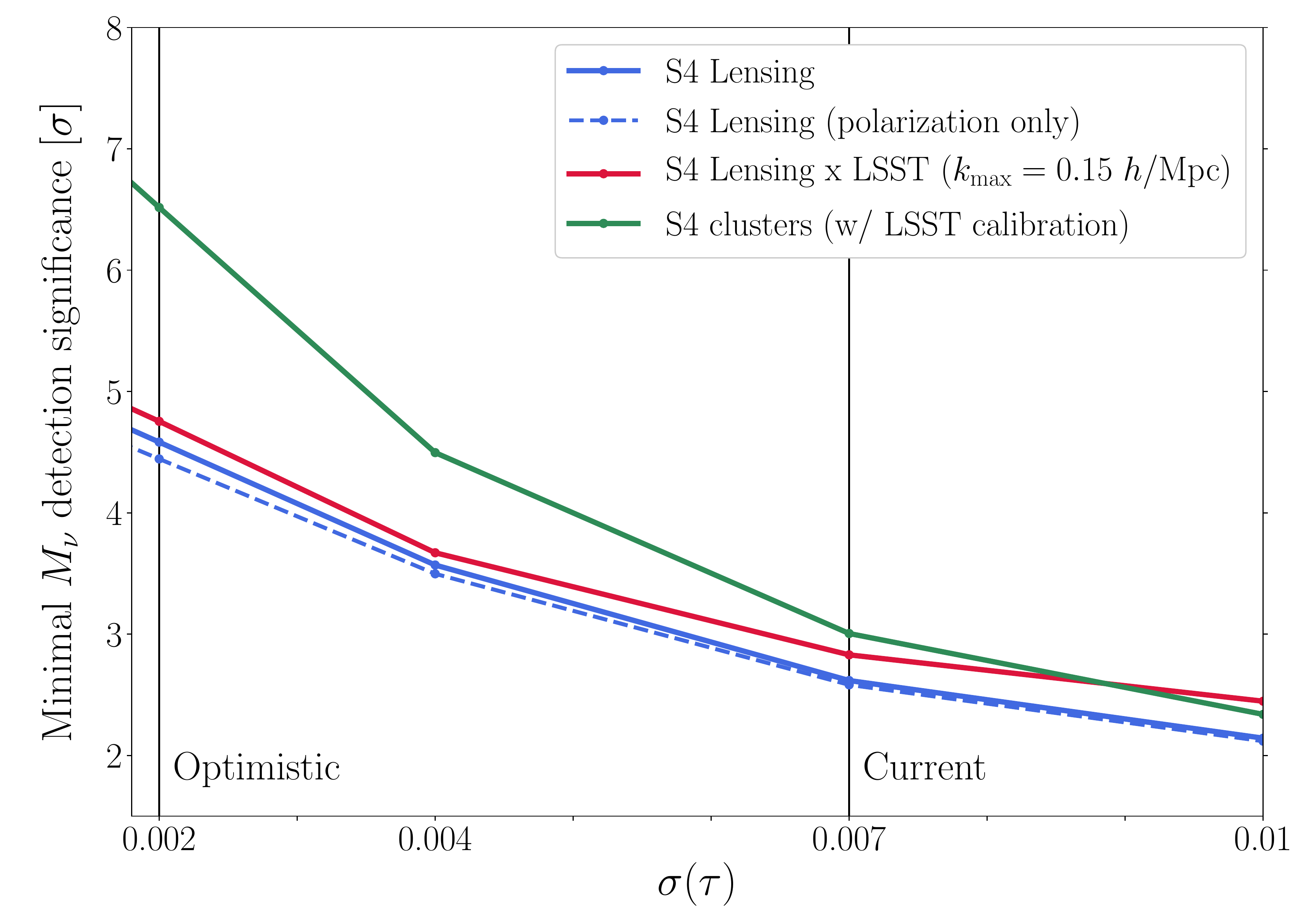} \hspace{0.16cm} \includegraphics[height=0.31\textwidth]{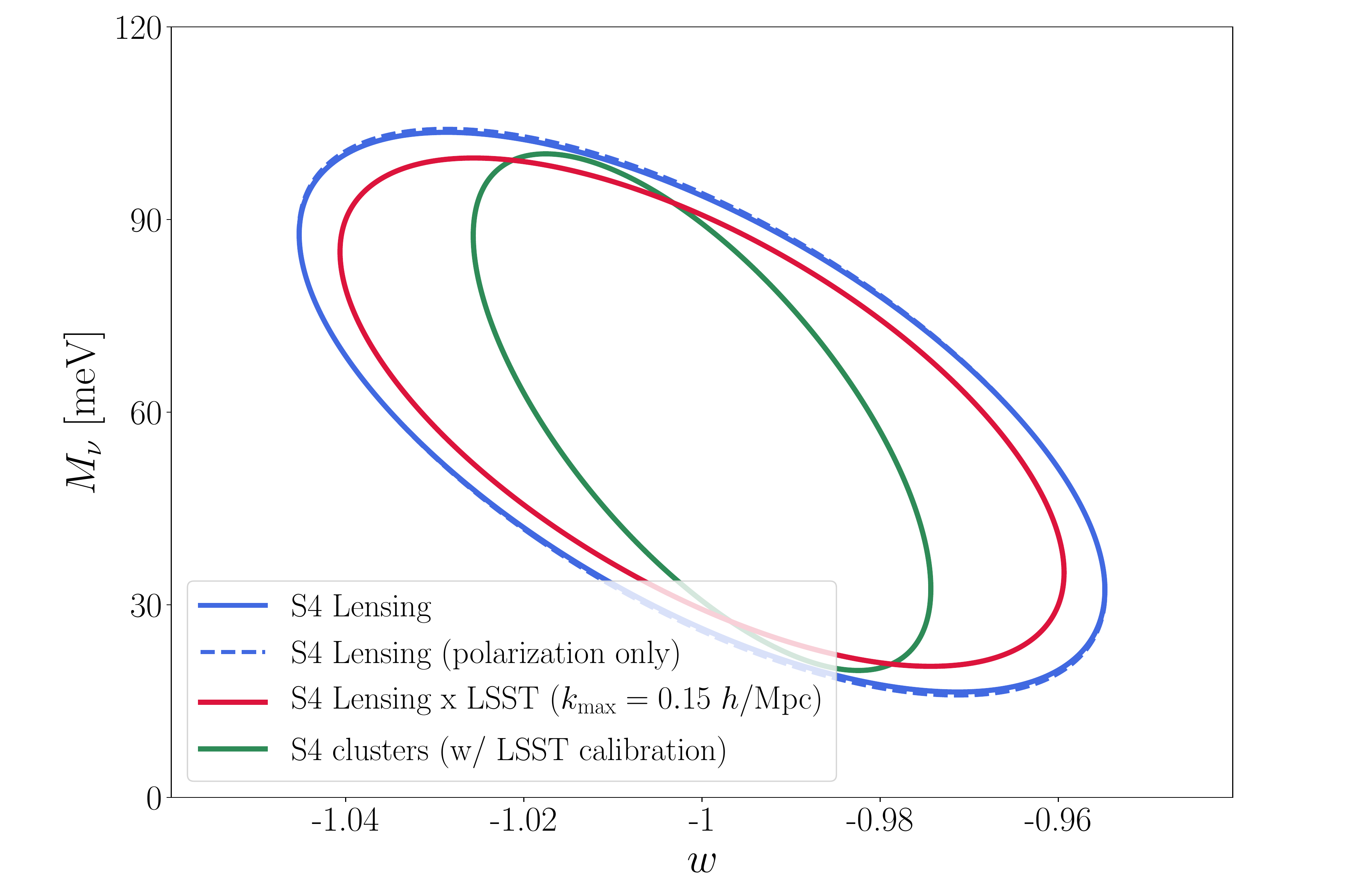}
\caption{Left panel: Forecasted constraints on the neutrino mass sum for several different CMB-S4-derived cosmological probes, written here in terms of the significance of a detection of the minimum value consistent with oscillation data $M_\nu = 58$meV. Neutrino mass constraints are degenerate with the optical depth to the CMB, $\tau$. Constraints here are shown as a function of the assumed $1-\sigma$ errors on $\tau$.  All probes give similar results for the current values $\sigma(\tau)\approx$0.007; constraints improve rapidly as the error on $\tau$ decreases. We have marked on the plot both the current $\tau$ constraint and an optimistic value derived from either kSZ analysis or an external large-scale CMB measurement.  Constraints are derived from: the CMB-S4 lensing power spectrum (blue solid line); the lensing power spectrum using only polarization reconstruction (blue dashed); the CMB-S4 lensing power spectrum with the addition of LSST gold sample galaxy cross-correlations and clustering (red); CMB-S4 Sunyaev-Zel'dovich selected clusters with LSST weak lensing mass calibration (green). In all cases, DESI BAO and CMB-S4 primary CMB constraints are also assumed. Right panel: Forecast errors for joint constraints on neutrino mass and the dark energy equation of state. Probes with redshift resolution generally show somewhat less degeneracy of neutrino constraints with equation of state values, as well as smaller errors in general.}
\label{fig:constraints}
\label{fig:MN}
\end{figure}

{\bf Forecasts within $\nu\Lambda$CDM:} In this section we will present the results of forecasts for the constraints on neutrino mass arising from multiple probes within a minimal model of $\Lambda$CDM + neutrino mass. 
The forecast errors on the sum of the neutino masses from different CMB-S4 derived probes are shown in Fig.~\ref{fig:constraints}, plotted as a function of a prior assumed on the CMB optical depth. It can be seen that all probes perform quite similarly in the absence of an
estimate of $\tau$ that is more precise than the current state of the art with {\it Planck\/} ($\sigma(\tau)=0.007$). All probes are limited by the incomplete knowledge of the CMB optical depth; this well-known result strongly motivates improvements in external determinations of the CMB optical depth. An alternative possibility, albeit one that has not yet been demonstrated in data, is to make use of measurements of the kSZ trispectrum to constrain reionization and obtain an estimate of the optical depth. Preliminary estimates indicate that tight constraints of order  $\sigma(\tau)=0.002$ (and powerful neutrino mass measurements) might be possible with CMB-S4 kSZ, though such constraints would be less model-independent than those arising from the large-scale CMB. Small improvements in neutrino mass constraints can also be seen by adding probes with redshift resolution or the ability to probe the galaxy-spectrum shape, because these provide information that is not degenerate with $\tau$. 

We note that using only polarization lensing-reconstruction preserves nearly all the signal-to-noise ratio in a neutrino mass measurement. 
While we adopt the CMB lensing constraints as our conservative baseline, we find that for low $\sigma(\tau)$, CMB-S4 cluster-cosmology forecasts (with an LSST weak-lensing mass calibration) appear even more powerful than CMB lensing, motivating more research in this area.

In all cases, even for present-day optical-depth constraints, an approximately 3$\sigma$ detection of the minimal mass sum is possible for CMB-S4. The fact that multiple different CMB-S4 probes should be able to achieve similar constraints will allow for cross-checks and thus increase the reliability of any claimed detection. 
Furthermore, unlike with prior experiments, such as Simons Observatory, we will 
make a near-equally significant detection using only lensing of CMB polarization; 
we do not need to rely on temperature-based reconstructions where extragalactic foreground systematics are a possible concern. CMB-S4 will thus provide definitive measurements of the neutrino mass sum that are more reliable than those from prior experiments. If the optimistic constraints on the optical depth of order $\sigma(\tau)\approx 0.002$ can be achieved, with either internal kSZ analyses or a large-scale cosmic variance limited, external CMB experiment, at least 5$\sigma$ detections of the minimal mass sum will become possible with CMB-S4.

{\bf Forecasts with more general dark-energy assumptions:} 
To examine the strength of the dependence of our constraints on the assumption of a particular cosmological model, we relax the assumption that the dark energy must be a cosmological constant and perform our forecasts again, this time freeing the dark-energy equation of state parameter $w$. The resulting constraint contours are shown in the right panel of Fig.~\ref{fig:constraints}.
While all probes show some degeneracy between the neutrino mass and the dark-energy equation of state, this degeneracy is, to some extent, reduced in the three probes that resolve growth of structure as a function of redshift, resulting in improved neutrino mass constraints. That is because
the different redshift dependences of the energy densities of dark energy and neutrinos allow the effects to be distinguished. We conclude that, especially with tomographic probes, powerful CMB-S4 measurements of neutrino mass are feasible even when allowing for somewhat more general dark-energy physics.

Before proceeding, it is useful to put a CMB-S4 detection of neutrino mass in context with anticipated constraints on $M_\nu$ from other cosmological data sets. At present, the tightest bounds on $M_\nu$ are achieved by combining information in the primary CMB with a measure of the amplitude of structure at late times (in effect determining $M_\nu$ by comparing $\sigma_8$ and $A_{\rm s}$). The amplitude of structure at late times can be determined from a variety of probes, including the CMB-S4 observables of CMB lensing and SZ cluster abundance discussed above, but also using data from galaxy surveys such as the abundance of galaxy clusters, the Lyman-$\alpha$ forest power spectrum, redshift-space distortions (RSDs), or weak-gravitational-lensing statistics. 
At present the tightest constraints on $M_\nu$ are from {\it Planck\/} CMB lensing, yielding $M_\nu < 120\,$meV at 95\% confidence \cite{Aghanim:2018eyx}), with the latest data from the Lyman-$\alpha$ forest giving competitive constrains of  $M_\nu < 140\,$meV at 95\% confidence. To date, other probes of the amplitude of structure give interesting but less stringent constraints, e.g., RSDs from the Baryon Oscillation Spectroscopic Survey (BOSS), $M_\nu < 230\,$meV \cite{Alam:2016hwk}, and $<260\,$meV from DES lensing and galaxy clustering \cite{Abbott:2017wau}. However, in the next decade a number of experiments are anticipated to constrain or detect $M_\nu$ at a level comparable to expectations for CMB-S4---for example, LSST, the Dark Energy Spectroscopic Instrument (DESI), {\it Euclid}, {\it WFIRST}, the Square Kilometer Array (SKA), and Simons Observatory are forecasted to reach $1\sigma$ limits on $M_\nu$ of order $10$--$40\,$meV when combined with current CMB data \cite{Ade:2018sbj, Aghamousa:2016zmz, Amendola:2016saw, Brinckmann:2018owf}). In the event of no improvement in our knowledge of $\tau$, each probe may only reach a 2--$3\sigma$ detection of $M_\nu = 58\,$meV.  Multiple independent approaches will therefore be critical for establishing the value of the neutrino mass from cosmology. The role of CMB-S4 is therefore as a particularly robust cross-check of any data on $M_\nu$ from other experiments and, in the event of improvements to our knowledge of $\tau$, as an experiment that provides a high-significance measure of $M_\nu$. 

\subsubsection{CMB-S4 measurements of \textit{M}$_\nu$: detection scenarios and implications}
\begin{table}[t!]
\begin{center}
\begin{tabular}
{| l | c c c c | p{5cm} | }\hline Scenario & $m_{\beta \beta}$ & $m_{\beta}$&  $M_\nu$ & $\Delta \Neff$ & Conclusion \\
\hline 
Normal hierarchy & $< 2\sigma$ & $< 2\sigma$  & $60 \, {\rm meV}$ & 0 & Normal neutrino physics; no evidence for BSM
\\[.2cm]
Dirac neutrinos & $< 2\sigma$ & $\dots$  & $350  \, {\rm meV}$ & 0 & Neutrino is a Dirac particle \\[.2cm]
Sterile neutrino & $< 2\sigma$ & $< 2\sigma$   & $350  \, {\rm meV}$ & $>0$ & Detection of sterile neutrino consistent with short-baseline data \\
\hline
Diluted neutrinos & $ 0.25 \, {\rm eV}$ & $ 0.25 \, {\rm eV}$  & $<150  \, {\rm meV}$ & $< 0$ & Modified thermal history (e.g., late decay) \\[.2cm]
Exotic Neutrinos & $ 0.25 \, {\rm eV}$ & $ 0.25 \, {\rm eV}$  & $<150  \, {\rm meV}$ & $0$ & Modified thermal history; (e.g., neutrino decay to new particle) \\[.2cm]
Excluded & $ 0.25 \, {\rm eV}$ & $ 0.25 \, {\rm eV}$  & $500  \, {\rm meV}$ & $0$ & Already excluded by cosmology \\
\hline
Dark radiation & $< 2\sigma$ & $< 2\sigma$  & $60  \, {\rm meV}$ & $>0$ & Evidence for new light particles; normal hierarchy for neutrinos
\\[.2cm]
Late decay & $< 2\sigma$ & $< 2\sigma$  & $60  \, {\rm meV}$ & $<0$ & Energy injection into photons at temperature $T \lesssim 1$ MeV \\
\hline 
\end{tabular}
\caption{Relation between neutrino experiments and cosmology.  We include the measurement of the Majorana mass via neutrinoless double-beta decay ($m_{\beta \beta}$) or a kinematic endpoint ($m_\beta$) compared to the cosmological measurement of the sum of the masses $M_\nu$ and the CMB measurement of $\Neff$.  Here ``$< 2 \sigma$'' indicates an upper limit from future observations. For observations on the timescale of CMB-S4, we use $\sigma(m_{\beta\beta}) \approx 0.075 \, {\rm eV} $ and $\sigma(m_\beta) \approx 0.1 \, {\rm eV} $.  For $\Delta\Neff$ the use of $\gtrless 0$ indicates a significant deviation from the Standard Model value.}
\label{table:neutrinoscenarios}
\end{center}
\end{table} 

CMB-S4 will make a cosmological measurement of $M_\nu$ that is complementary to terrestrial measurements of neutrino mass and other properties. In this section, we put the CMB-S4 measurement in context with laboratory neutrino experiments and present several detection scenarios and their implications. 

{\bf Determining the neutrino mass scale:} While CMB-S4 will determine $M_\nu$ through gravitational effects, terrestrial measurements of the neutrino mass using radioactive decay are kinematic and determine an effective electron (anti-)neutrino mass, which is an incoherent sum of mass eigenstates. Current measurements from Mainz~\cite{Kraus:2004zw} and Troitsk~\cite{Aseev:2011dq} limit this mass to $< 2.0\,$eV. The KArlsruhe TRItium Neutrino (KATRIN) experiment~\cite{Angrik:2005ep} is expected to improve this limit by a factor of 10 and Project~8 may reach a limit of $0.04\,$eV\cite{Esfahani:2017dmu}. Within the standard neutrino mass and cosmological paradigm, the kinematic and cosmological measurements of neutrino mass are connected through the PMNS neutrino mixing matrix. The combination of cosmological and terrestrial neutrino mass measurements can therefore test our cosmological neutrino model. A discrepancy could point to new physics, for example a modified thermal history through neutrino decay

{\bf Lepton number violation -- Majorana or Dirac neutrinos:} One of the most exciting connections between cosmological measurements of neutrino mass and terrestrial experiments is the complementarity between cosmological neutrino mass measurements and the search for neutrinoless double-beta decay (NLDBD). NLDBD is a hypothetical decay mode of certain nuclei, where two neutrons convert to two protons and two electrons with no emission of neutrinos. The observation of NLDBD would be transformational---it would demonstrate that neutrinos are Majorana particles and reveal a new lepton-number-violating mechanism for mass generation. This new physics could potentially explain both the smallness of neutrino masses and matter-antimatter asymmetry in the Universe.

\begin{figure}[t!]
\centering \includegraphics[width=0.49\textwidth]{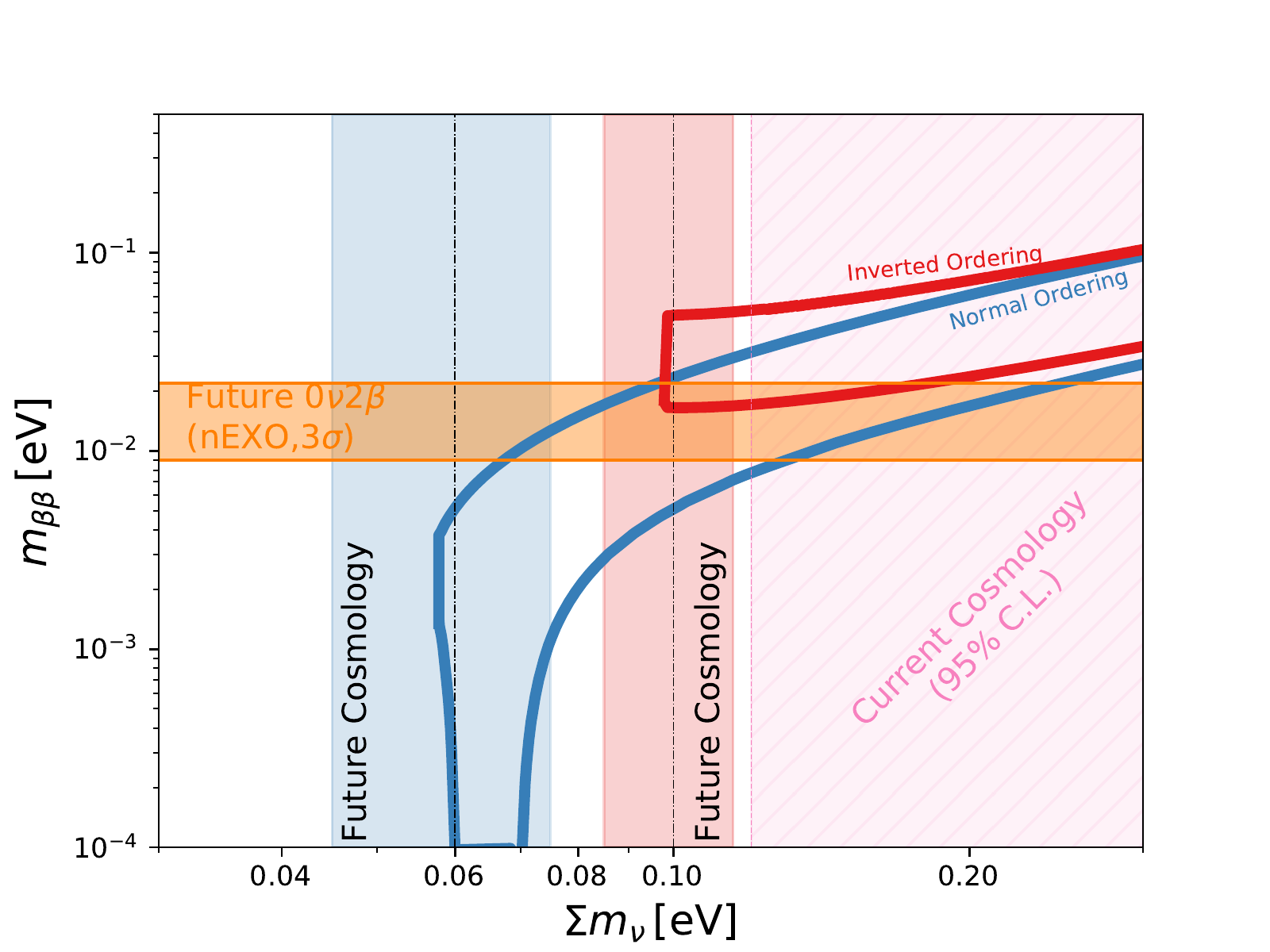} \includegraphics[width=0.49\textwidth]{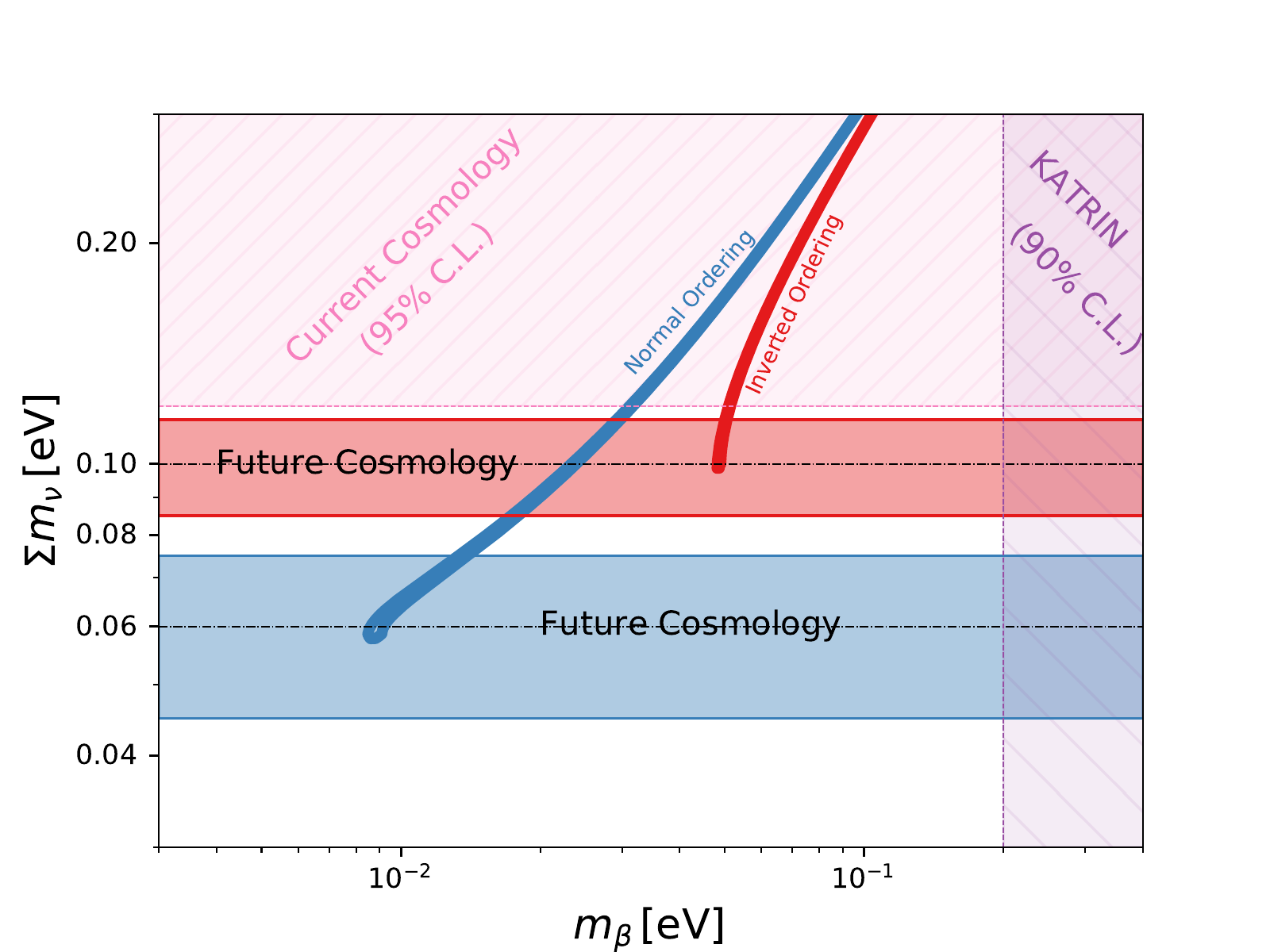}
\caption{Left: Majorana effective neutrino mass $m_{\beta\beta}$ versus $M_\nu$ in the scenario where NLDBD is mediated by light neutrino exchange. The area enclosed by the blue and red solid lines indicate the allowed 95\% ranges from neutrino oscillation experiments \cite{nufit} for normal ordering (NO) and inverted ordering (IO) assuming complete ignorance of the Majorana phases. The vertical blue and red bands show the forecasted $1\sigma$ constraints on $M_\nu$ from CMB-S4 for minimal mass NO and IO. The horizontal band shows the sensitivity of future NLDBD experiments. A CMB-S4 detection of $M_\nu$, in combination with a detection of $m_{\beta\beta}$ can constrain the Majorana phases. Right: Sum of individual the neutrino masses as a function of the electron-neutrino effective mass $m_\beta$ for the NO (blue) and the IO (red). Again, the area enclosed by the blue and red solid lines indicate the allowed 95\% ranges from neutrino oscillation experiments \cite{nufit} for NO and IO. The horizontal bands show the future cosmological constraints around each ordering (assuming the mass of the lightest neutrino state $m_\mathrm{lightest} =0$\,eV). Also shown are the anticipated limits on $m_\beta$ from the KATRIN experiment in the case of no detection.}
\label{fig:NLDBD}
\end{figure}

To illustrate the connection between NLDBD searches and $M_\nu$, consider the simplest case where NLDBD is mediated by exchange of light Majorana neutrinos. Within the context of this mechanism, these experiments determine an ``effective neutrino mass,'' $m_{\beta\beta}$, given in terms of the PMNS mixing matrix, including two unknown Majorana phases, and the individual neutrino masses. In this scenario, the absence of a signal in next generation NLDBD searches, combined with a cosmological measurement constraining $M_\nu > 100\,$meV (corresponding to either the inverted hierarchy or a minimum neutrino mass of 50\,meV), would strongly point to neutrinos being Dirac particles (see Fig.~\ref{fig:NLDBD}). On the other hand, if NLDBD is observed, cosmological measurements of $M_\nu$ are then sensitive to the Majorana phases. Perhaps the most interesting scenario would be if cosmological and NLDBD measurements cannot be jointly explained by the exchange of light Majorana neutrinos and new physics is required.

{\bf Neutrino mass ordering and CP violation:}
In the case of normal ordering with non-degenerate neutrino mass, the CMB-S4 measurement of $M_\nu$ will provide a 2--4$\sigma$ determination of the neutrino mass ordering. Fully characterizing neutrino mass ordering and CP violation are major goals of the terrestrial neutrino physics program ~\cite{Patterson:2015xja}. In the scenario where the neutrino mass spectrum is normally ordered and non-degenerate, CMB-S4 would be a strong complement to terrestrial experiments by providing a measurement of neutrino ordering that is independent of oscillation parameters and $\delta_{\rm CP}$. Under all circumstances, the combination of CMB-S4 with terrestrial determinations of neutrino ordering will provide a definitive measurement of the neutrino mass spectrum.

{\bf Sterile neutrinos:}
Mechanisms of introducing neutrino mass often include sterile
neutrinos, with both Majorana and Dirac terms potentially
contributing (e.g., Ref.~\cite{Langacker:2011bi}). A number of recent neutrino oscillation experiments have reported anomalies that are possible indications of four or more neutrino mass eigenstates with mass splittings $\mathcal{O}(1\,{\rm eV})$. To explain these anomalies, these neutrinos typically have relatively large mixing angles and would therefore thermalize in the early Universe, affecting primordial nucleosynthesis \cite{Abazajian:2002bj}, changing $\Neff$ by $\mathcal{O}(1)$, and $M_\nu$ by $\mathcal{O}(1\,{\rm eV})$. Introducing new neutrino interactions or a modified thermal history for neutrinos can accommodate additional light sterile neutrinos without violating current constraints on $\Neff$ and $M_\nu$. 
As discussed in the light relics section, CMB-S4 will provide additional constraints on neutrino-self interactions, potentially confirming or ruling out these scenarios.

Interestingly, there exist tensions in some combinations of CMB and LSS data sets that could be alleviated with the
presence of massive neutrinos, extra neutrinos, or both. CMB-S4 could shed light on the sterile neutrino mass and vacuum flavor-mixing parameters invoked to explain the experimental neutrino anomalies. Telltale signatures in $\Neff$, $M_{\nu}$, and $Y_\mathrm{p}$ can allow CMB-S4 to probe this larger parameter space.

\subsection{Dark energy}
\label{sec:de}
\label{subsec:DE}

\def\as#1{[{\color{red}{\bf AS:} {\it #1}}] }
\def\vg#1{[{\color{red}{\bf VG:} {\it #1}}] }

The discovery 20 years ago that the expansion of the Universe is
accelerating presented a profound puzzle to physics, one that we have
not yet solved.
Our current framework can explain
these observations only by invoking a new substance with unique
properties (dark energy) or by changing the century-old,
well-tested theory of general relativity developed by Einstein. 

As the current epoch of acceleration is much later than the epoch from
which the photons in the CMB originate, the behavior of dark energy or
modifications of gravity do not significantly influence the properties
of the primordial CMB. However, the CMB can still inform us about the
properties of dark energy through two fundamental pathways. First, the
projection from physical scales to the angular scales observed in a
CMB map involve distance, which is affected by the expansion history
along the line of sight to the last scattering surface.  It is
therefore sensitive to dark energy. The other pathway, involving secondary anisotropies, is even more important.
These include weak gravitational lensing, as
well as interactions with free electrons leading to the thermal and
kinematic Sunyaev-Zeldovich effects. Therefore, as discussed in the
dark-energy submission \cite{2019arXiv190312016S} to the 2020 Decadal Survey of Astronomy and
Astrophysics (and endorsed by all major dark energy experimental
collaborations), the CMB is a recognized probe of dark energy.

In the simplest model acceleration is driven by a cosmological
constant. Although theoretically unnatural, this model does satisfy
current constraints, so a clear target for CMB-S4 is to test the various
predictions this model makes at late times. Using  gravitational
lensing of the CMB, the abundance of galaxy clusters, and cosmic
velocities, CMB-S4 will measure both the expansion rate $H$ and the
amount of clustering, quantified by the parameter $\sigma_8$, as a
function of time. The constraints from CMB-S4 alone will be at the
0.1\% level on each and, when combined with other experiments,
will reach well below this level, particularly when the power of
CMB-S4 is also utilized to calibrate other probes. CMB-S4
constraints will be among the most powerful tests of the cosmological
constant---more crucially, the simultaneous sensitivity to the expansion
and the growth will allow us to distinguish the dark-energy paradigm from
modifications to general relativity. There are many models for acceleration in the
latter class, and CMB-S4 will be generically useful in  constraining them.

\subsubsection{Canonical probes: lensing-convergence power spectrum and galaxy clusters}

In the basic $\Lambda$CDM cosmological model, the dark energy is
assumed to be the energy of the vacuum, or equivalently an inert component
with equation of state $w=p/\rho=-1$. The most common phenomenological
extensions of this model are models with an expansion history
deviating from the $\Lambda$CDM model in a way that is parameterized by
the dark energy having a time-varying equation of state. 
A common parameterization is 
\begin{equation}
w(z) = w_0 + w_a \frac{z}{1+z}.
\end{equation}

Changing the energy content of the Universe has two main
consequences. The first is that the expansion history changes and
therefore distance measures as a function of redshift change. This is
how most of the low-redshift probes, such as baryon acoustic
oscillations (BAOs) and type Ia supernovae probe dark energy. For
CMB-S4 this effect affects the projection of the correlations at the
surface of last scattering into the anglar power spectrum. For
example, changing $w$ while keeping other parameters fixed will shift
the positions of acoustic peaks in the CMB power spectrum.  However,
the precise measurement of the basic cosmological parameters in the
presence of free $w$ and $w_a$ interacts in a non-trivial way with the
low-redshift constraints of dark energy, which leads to improved
constraints on the $w$--$w_a$ plane through changing both the shape and
amplitude of the observable CMB temperature and polarization power
spectra, as well as breaking degeneracies with other parameters. For
example, precise determination of parameters relevant for the
pre-recombination era affects the precision with which the BAO ruler
is known.

\begin{figure}[h!]
  \centering
  \includegraphics[width=0.8\textwidth]{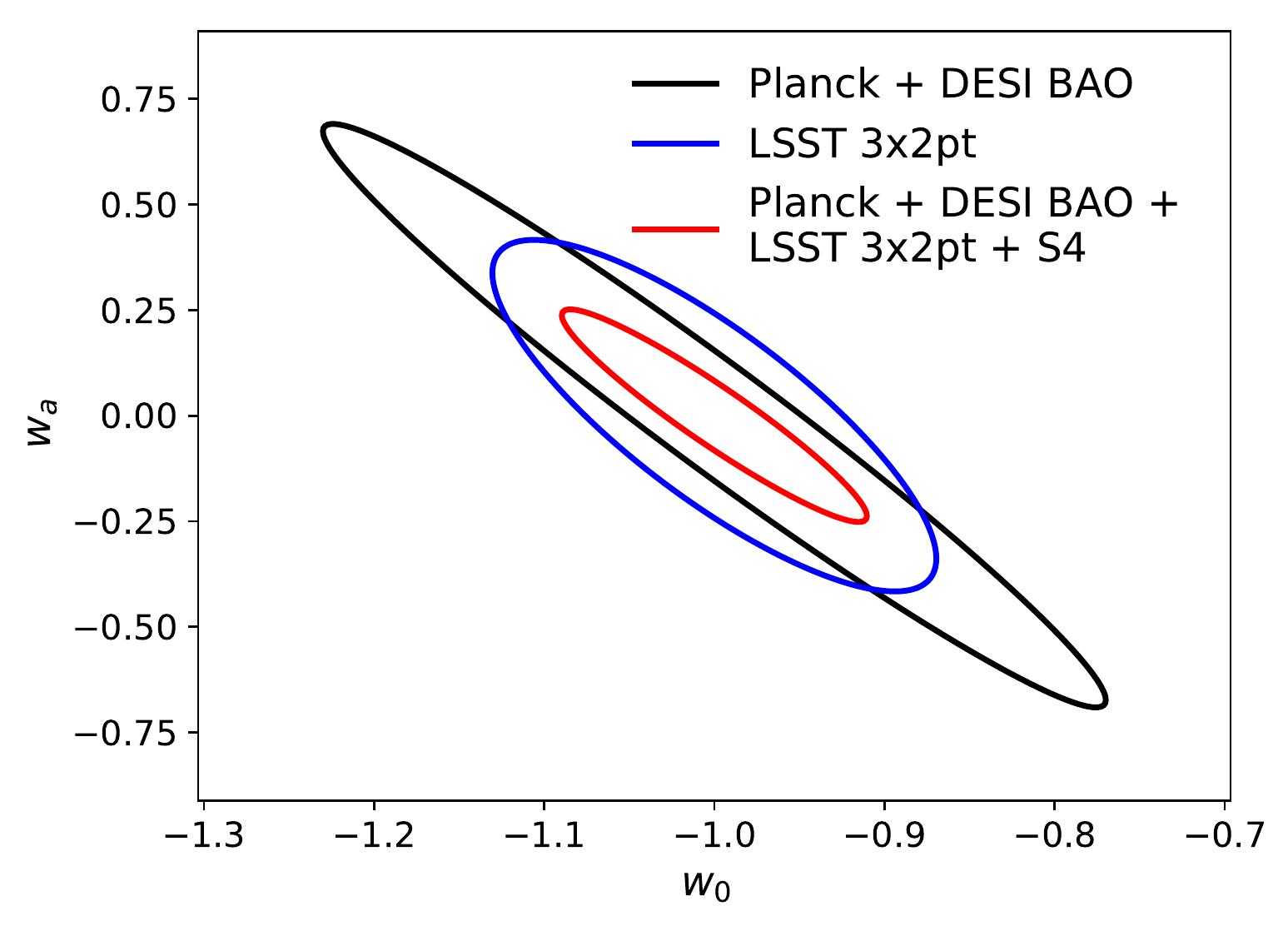}

  \caption{Improvement on the standard $w_0$--$w_a$
    parameters for the combination of: (i) {\it Planck\/} prior and expansion
    history measurements (DESI BAO) in black; (ii) LSST 3$\times$2-point
    function measurements (including auto- and cross-correlation of
    galaxy number density and shear field fluctuations) in blue; and
    (iii) the combination of the first two items, with CMB-S4 power-spectrum
    measurements (temperature, polarization, and weak lensing
    reconstruction, but importantly no galaxy cluster
    information) in red.}
\label{fig:w0wa}
\end{figure}

In Fig.~\ref{fig:w0wa} we show the improvement of constraints once
CMB-S4 spectra (both CMB and lensing reconstruction) are added to more
direct probes of dark energy. This specific forecast was made using the \texttt{GoFish}
package\footnote{\url{https://github.com/damonge/GoFish}} \cite{2018PhRvD..97l3544M} and we show three representative cases. The first is the determination of expansion history from the DESI experiment based on measurements of BAOs \cite{2016arXiv161100036D}, arguably the most robust probe of dark energy.   In blue we show the same contour based on ``3$\times$2-point'' analysis of the LSST data \cite{2009arXiv0912.0201L,2018arXiv180901669T}, i.e., the combination of the photometric galaxy clustering, galaxy-shear cross-correlations, and shear-shear auto-correlations. This probe has a very different set of systematic errors, but a comparable total constraining power. We note that both surveys will produce further constraints. For example DESI will perform analysis of the full shape of the redshift-space power spectrum while LSST will also use other probes, such as supernovae, strong lensing and galaxy clusters. Nevertheless, this choice presents a nominal set of constraints. Upon addition of the CMB-S4 temperature, polarization, and lensing reconstruction power spectra, these constraints tighten significantly, which we show as the red ellipse in Fig.~\ref{fig:w0wa}. We quantify this in terms of a figure of merit (FoM) that is defined to be inverse area under the ellipse ($1/\sqrt{|C|}$, where $C$ is the 2$\times$2 C matrix of $w$, $w_a$), which is very similar in spirit to the Dark Energy Task Force figure of merit \cite{2006astro.ph..9591A}. We additionally let the neutrino mass $\sum m_\nu$ be a free parameter (see also Sect.~\ref{sec:mnu}), since it is known to be degenerate with dark-energy parameters \cite{2005PhRvL..95v1301H}. We find that {\it Planck\/}+DESI BAO and LSST 3$\times$2 analyses have comparable FoMs of approximately 70. The combined FoM of these experiments is around 200, consistent with relatively modest degeneracy breaking. However, it increases to 400 with the addition of CMB-S4 data. CMB-S4 hence contributes to a doubling of the figure of merit.  We want to emphasize that part of this improvement is due to ability of CMB-S4 to break the degeneracy with the neutrino mass.

Besides the effect on the expansion history, dark energy also affects
the rate at which structures form in the Universe. Most importantly,
in standard general relativity, these are intrinsically linked---the
given expansion history determines the growth rate. By measuring the
growth over cosmic history we can in principle falsify this
picture, which would be a  strong indication of new physics in the
gravitational sector and a truly revolutionary discovery.

CMB-S4 will measure growth using the following two main methods.
\begin{itemize}
\item \textbf{CMB lensing reconstruction}: Change in the growth of
  structures will affect the amplitude (and to a lesser extent also
  the shape through geometric effects) of the CMB lensing convergence
  power spectrum.  This method is particularly powerful in combination
  with auto-spectra of low redshift tracers as we show in
 Fig.~\ref{fig:fs8-ksz};

\item  \textbf{Galaxy clusters}: The abundance and clustering of galaxy clusters are independent probes of growth. CMB-S4 will find clusters through the
tSZ effect using proven matched-filtering
techniques. The tSZ is a  powerful tool to find and count clusters for
CMB-S4 because the detection efficiency is nearly independent of
redshift for an instrument with arcminute-scale beams, and
the selection function is well behaved and simple to model. The utility
of cluster abundances as cosmological probes is limited by systematic
uncertainties in the the observable-to-mass scaling relations. CMB-S4
will use CMB halo lensing (or optical weak lensing) to calibrate
the observable-to-mass relation, aided by the well-understood selection
function. The tSZ selected clusters will require optical surveys to
confirm and provide redshifts for the low redshift clusters
($z \lesssim1.5$) and near-IR follow-up observations for the remaining
clusters.

\end{itemize}

We show how CMB lensing reconstruction can be converted into redshift-resolved measurements of the clustering amplitude
$\sigma_8$ in the left panel of Fig.~\ref{fig:fs8-ksz}.

\subsubsection{Kinematic Sunyaev-Zeldovich (kSZ) effect}

The kSZ effect is the Doppler shift of CMB
photons caused by scattering off the plasma in late-time galaxies and
clusters, which are moving with respect to the CMB.  This Doppler
shift causes a slight change in the measured CMB temperature, while
preserving the blackbody spectrum, and therefore cannot be isolated
by a multi-frequency analysis, unlike the tSZ effect.  However,
the broad frequency coverage of CMB-S4 is ideal for reducing
other foregrounds and thus the noise on the kSZ signal.

The kSZ-induced temperature fluctuation is proportional to the galaxy
\textit{radial\/} peculiar velocity $v_{\rm r}$, and its optical depth to
Thomson scattering $\tau_{\rm g}$ (proportional to the number of free
electrons in the galaxy itself):
$(\Delta T / T)_{\rm kSZ} = - \tau_{\rm g} \ v_{\rm r}$.

\begin{figure}[h!]
  \centering

  \includegraphics[width=0.49\textwidth]{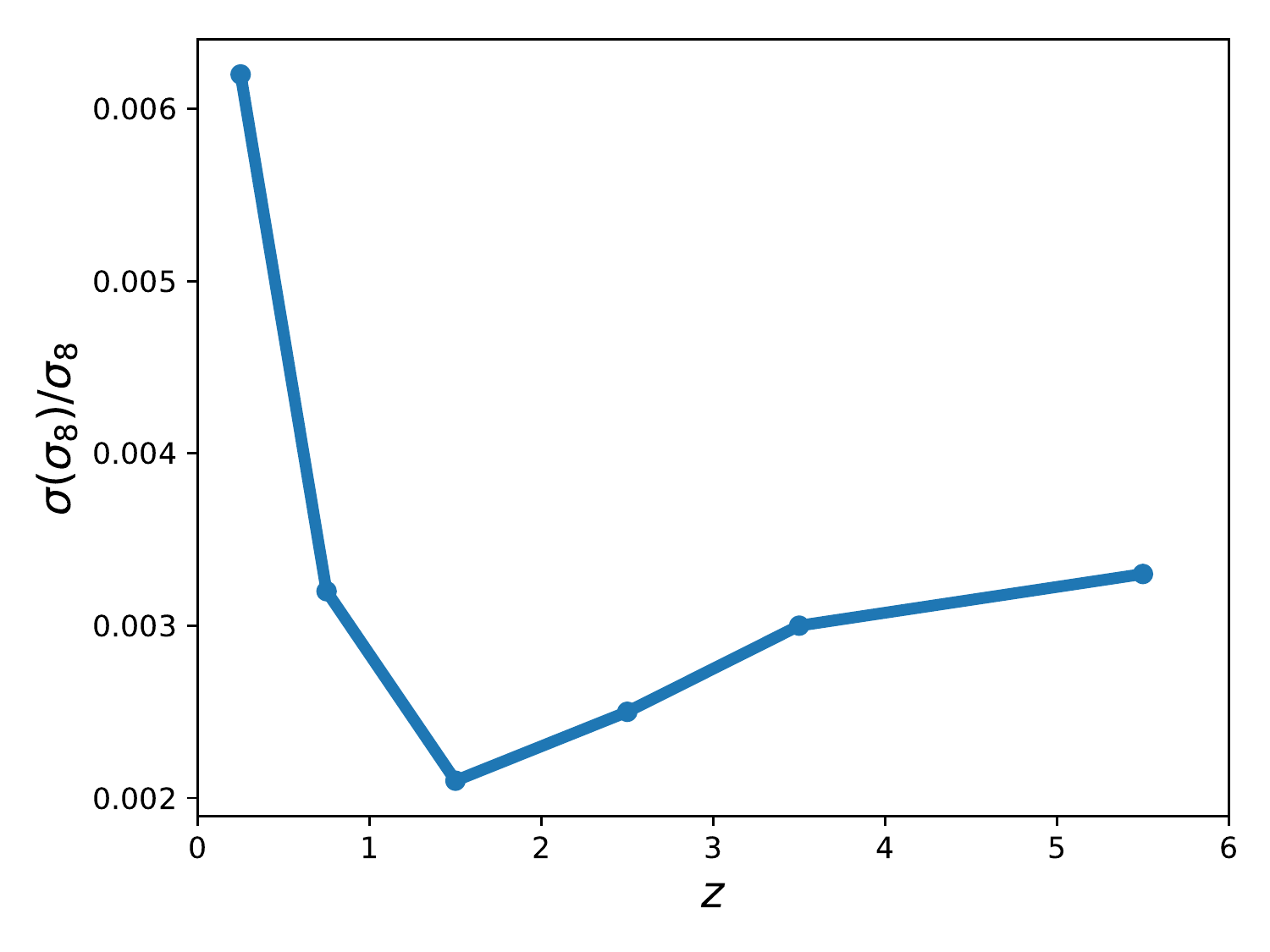}
  \includegraphics[width=0.49\textwidth]{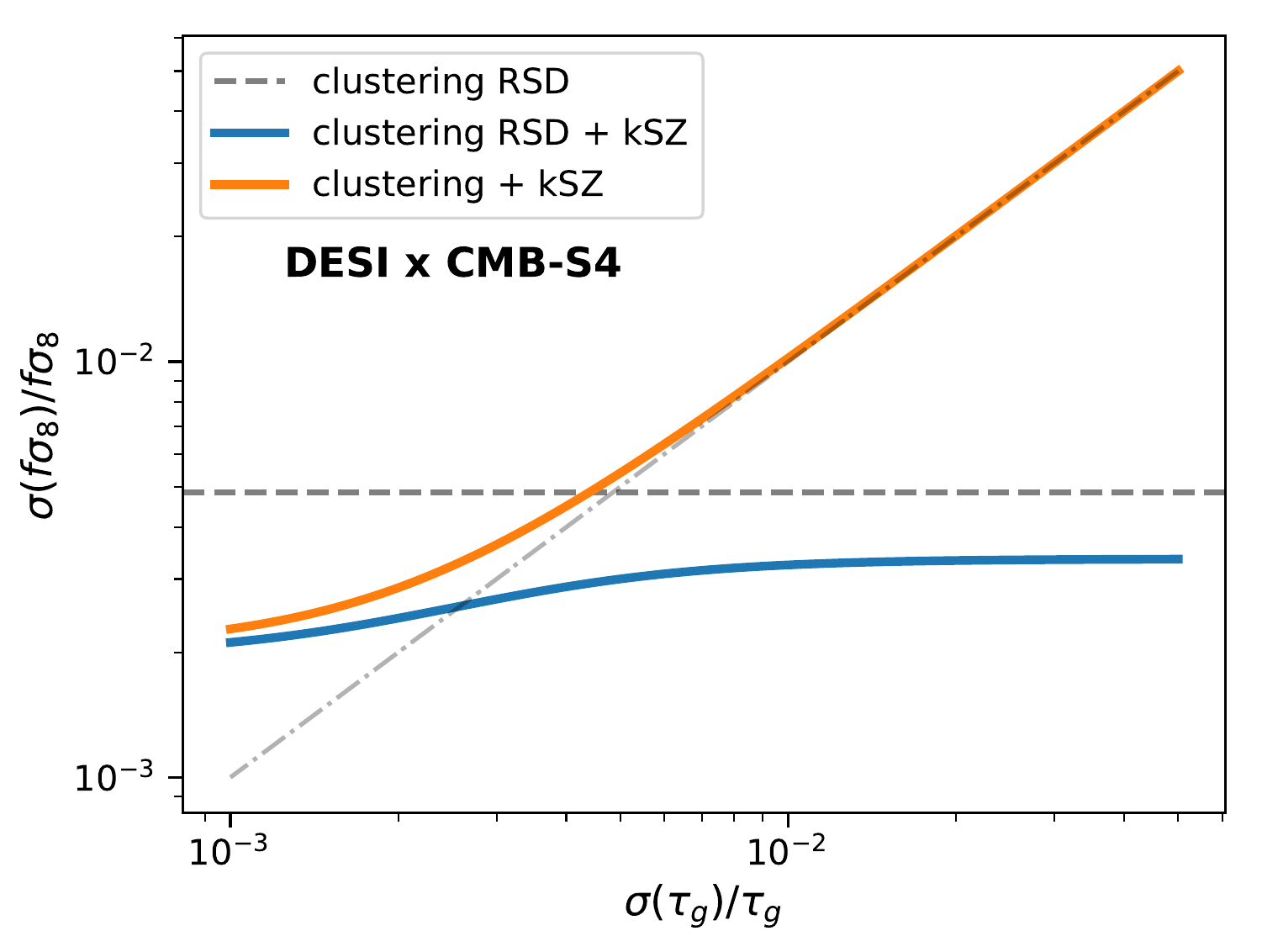}

  \caption{Constraints on the growth parameter from CMB-S4 from two
    independent sets of measurements. Left panel: Constraints on the
    matter amplitude $\sigma_8$ in tomographic redshift bins (indicated
    by the positions of points) from the combination of LSST galaxies
    and CMB-S4 lensing, assuming a fixed $\Lambda$CDM
    cosmology. Relaxing this assumption does not diminsh our ability
    to measure departures from the fiducial model. Right Panel:
    Constraints on $f\sigma_8$ from the kSZ effect as a function of the
    size of prior on $\tau_{\rm g}$ assuming a single redshift bin centered
    at $z=0.75$. The straight dashed line shows results from
    RSDs coming from the DESI experiment. The
    signal from kSZ in galaxy clustering alone is plotted in orange
    and the combination of everything in blue.  We see that even in
    the absence of an informative prior, the constraining power is
    twice that of DESI alone. For a sufficiently tight prior on
    $\tau_{\rm g}$, the CMB-S4 data alone can surpass DESI RSD
    measurements.}
\label{fig:fs8-ksz}
\end{figure}

Thanks to its high resolution, CMB-S4 will allow unprecedented
measurements of velocity fields through the kSZ effect. A measurement
of the peculiar velocity can in turn constrain the amplitude of matter
density fluctuations $\sigma_8(z)$ at the effective redshift of the
sample, as well as the linear growth factor $f$, which is sensitive to
dark-energy or modified-gravity models
\cite{Bhattacharya:2007sk,Kosowsky:2009nc,Mueller:2014nsa,Mueller:2014dba}.
These measurements will be able to distinguish dark energy from
interesting models of modified gravity and will provide complementary
constraints to redshift-space distortions and weak lensing
measurements, probing larger physical scales.  Figure~\ref{fig:fs8-ksz} shows
constraints on $f \sigma_8$ expected from CMB-S4, together with a
galaxy sample from the upcoming DESI survey. We assume an overlap $f_{\mathrm{sky}}$ of
0.2, resulting in a total survey volume of $116\,\mathrm{Gpc}^3$, containing 19.6 million
galaxies from all DESI galaxy samples. 

One caveat is that the amplitude of the signal is proportional to the optical depth of the galaxy sample
used in the analysis, $\tau_{\rm g}$. This is not known a priori and is in principle 
degenerate with $f \sigma_8$, as can be seen in the ``clustering+kSZ'' curve in Fig.~\ref{fig:fs8-ksz}.
There are several proposed ideas for breaking the optical-depth degeneracy, which range from tSZ or
X-ray measurements of the galaxies/clusters themselves \cite{2018PhRvD..97f3514A,Flender:2016cjy}
to measuring RSDs in conjunction with the
kSZ effect \cite{2017JCAP...01..057S}. The effect of
degeneracy breaking of the optical depth can be seen in the blue solid
curve in Fig.~\ref{fig:fs8-ksz}.
Scale-dependence in growth, either caused by dark-energy perturbations 
(for example from a non-standard speed of sound), or by screening mechanisms in modified gravity,
is not subject to the optical-depth degeneracy and can be constrained directly.

In addition, kSZ measurements can be important for dark-energy studies
in a more indirect way.  The high resolution of CMB-S4 allows for a direct
measurement of the optical-depth profile $\tau_{\rm g}(\theta)$, 
since the large-scale velocity field is
constant over the size of a galaxy.  In turn, $\tau_{\rm g}(\theta)$ is
proportional to the gas profile of the galaxy sample,
allowing for calibration of baryon effects on the power spectrum, a
leading systematic for weak lensing surveys aimed at measuring dark
energy \cite{vanDaalen:2011xb, Mohammed:2014mba}.

\subsubsection{Cosmic birefringence}
\label{sec-biref}

\begin{figure}[h!]
\centering \includegraphics[width=0.7\textwidth]{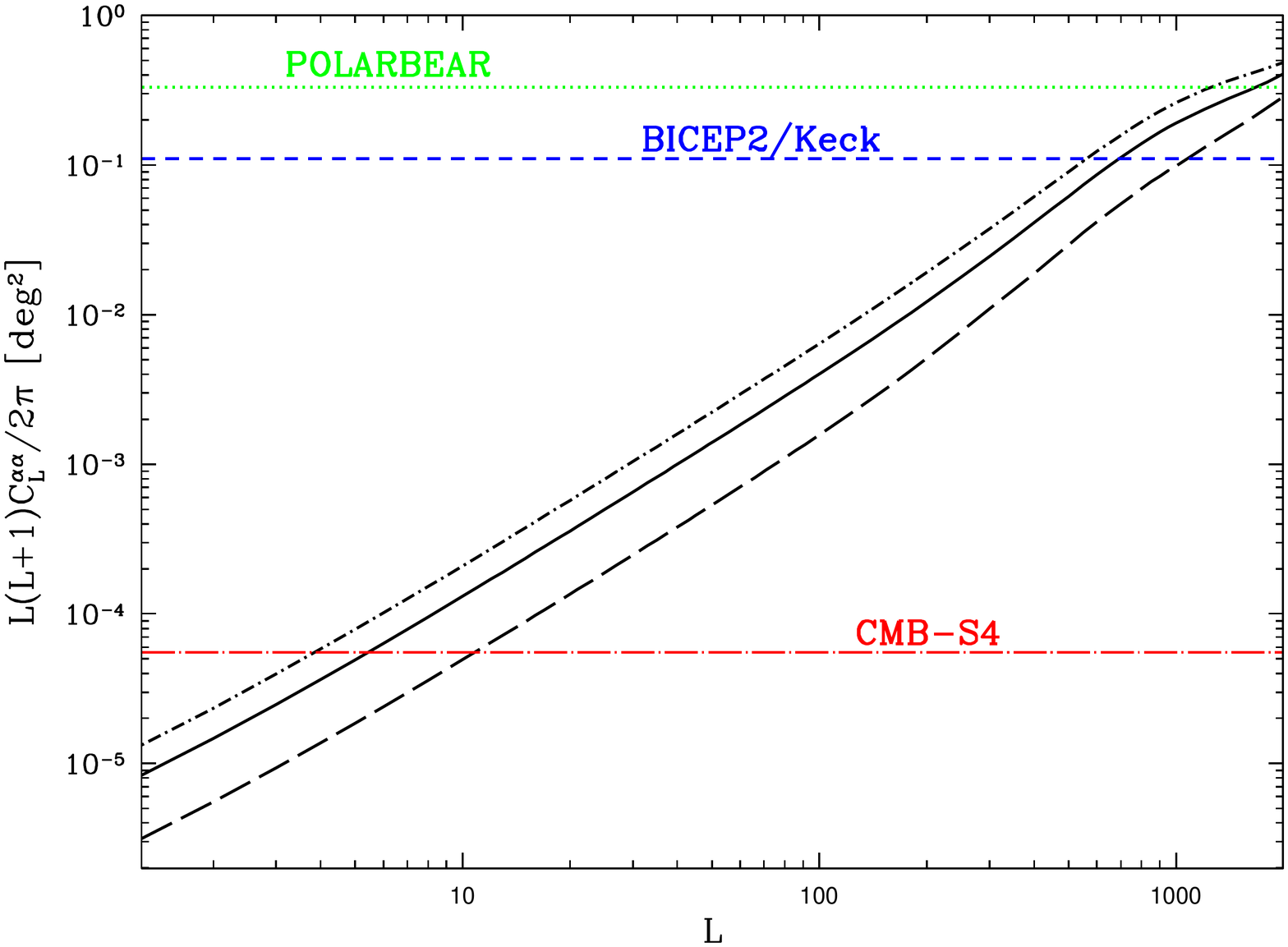}
\caption{Black lines represent model-independent projected noise for the cosmic-birefringence rotation-angle spectrum for CMB-S4. The noise assumes no rotation signal and is calculated in three different ways: (a) assuming no delensing and using the forecasted noise in the ILC (dot-short dash line); (b) assuming 80\% delensing and with forecasted noise remaining after the foreground subtraction (solid line); and (c) assuming 100\% delensing and perfect foreground subtraction (long dash line). The colored lines show an example of a specific model for birefringent rotation---the scale-invariant power spectrum. The amplitude of this rotation power spectrum is set to the current 95\% CL bound from POLARBEAR \cite{Ade:2015cao} (green dot) and BICEP2/Keck \cite{Array:2017rlf} (blue short dash), and to the projected bound for CMB-S4 with the noise calculated under assumption (b) (red dot-long dash line).} 
\label{fig:CB-forecast}
\end{figure}

In addition to probing the global dynamics of dark energy, as discussed in
previous sections, CMB-S4 will also be sensitive to the imprint of new
parity-violating physics within the dark-energy sector: $TB$ and $EB$ correlations
that arise from the effect of cosmic birefringence.  Detection of
these parity-violating correlations would have paradigm-changing
implications for cosmological scale physics, and may present a unique handle
to probing microphysics of dark energy.

The simplest dynamical way to model the accelerated expansion of the
Universe is to invoke a new slowly evolving scalar field that
dominates its energy budget (the quintessence models for dark energy). Such a
field could couple to photons through the Chern-Simons term in
the electromagnetic Lagrangian, causing rotation of the linear
polarization of photons propagating cosmological distances---an
effect known as ``cosmic birefringence'' \cite{Carroll:1998zi}. In the
case of the CMB, such rotation converts some of the primordial $E$ mode into $B$
modes, producing characteristic $TB$ and $EB$ cross-correlations in the CMB
maps \cite{Kamionkowski:2008fp,Gluscevic:2009mm}. Even though there is
no firm theoretical prediction for the size of this effect,\footnote{One exception, for example, are the ``Axiverse'' models that arise in the context of string theory and feature scalar fields that produce sky-averaged rotation angles on the order of $1/137$ radians (see Ref.~\cite{Arvanitaki:2009fg}).}
if observed, it would be a clear ``smoking-gun'' evidence for physics
beyond the Standard Model, in the form of a new scalar field. 

Previous studies have constrained both a uniform rotation angle $\alpha$, as well as anisotropic rotation described by  a power spectrum. The current tightest bound on the uniform rotation angle is $\alpha < 0.5^\circ$ at 68\% CL derived from {\it Planck\/} data \cite{Aghanim:2016fhp}. 
The effect of a uniform cosmological polarization rotation is degenerate with systematic uncertainty on the overall 
instrument polarization-angle calibration. Assuming this overall angle can be calibrated to arbitrary precision, CMB-S4
would improve the constraint on uniform cosmological rotation to
$\alpha < 0.2'$. 

A promising way to pursue the search for cosmic birefringence is via the quadratic estimator formalism \cite{Gluscevic:2012me}, which explores the off-diagonal (mode-coupling) $EB$ correlations on small angular scales. The existing constraints were derived under the assumption of a scale-invariant rotation spectrum, for which $A \equiv  L (L+1) C_L^{\alpha  \alpha}/2 \pi$ is independent of $L$. Such spectra could originate from fluctuations in a spectator scalar field present during inflation \cite{Pospelov:2008gg}. The best current bound, obtained from sub-degree scale polarization measurements with BICEP2/Keck \cite{Array:2017rlf}, is $A<0.11$ deg$^2$ (and essentially the same limit from {\it Planck\/} data \cite{Contreras2017}). More accurate measurements of polarization anisotropy at higher resolution and over a wider range of scales will significantly improve on this.
 
Figre~\ref{fig:CB-forecast} shows the forecasted noise (assuming no rotation) in the rotation-angle power spectrum $C_L^{\alpha
    \alpha}$ for CMB-S4 (with effective noise of $1.81\,\mu$K-arcmin, and a resolution of $1.4'$), along with the predicted 95\% CL bound on the amplitude $A$ of the scale-invariant spectrum.  The improvement on the current constraints is several orders of magnitude.

\subsection{Dark matter}
\label{sec:dm}
\def\vg#1{[{\color{red}{\bf VG:} {\it #1}}] }

CMB measurements have great power in testing dark matter (DM) models and constraining parts of the parameter space that are inaccessible to laboratory experiments.   
In particular, the CMB directly probes the physics of cosmological DM throughout cosmic history and does not rely on assumptions about the local DM phase-space distribution within the Milky Way. 
Given current null results of targeted searches for well-motivated candidate models, broad scans of all possibilities are warranted; CMB-S4 will enable such an approach to the DM problem. We highlight three broad classes of DM models that are of particular interest to searches with CMB-S4.\footnote{In this document, we omit  discussion of DM annihilation, since the detection sensitivity is mostly saturated by current and upcoming measurements \cite{2018arXiv180807445T}.}
In Sect.~\ref{subsec:dm_baryons}, we discuss light DM that interacts with baryons, in Sect.~\ref{subsec:dark_radiation}, we discuss models that involve DM interactions with other new light degrees of freedom, and in Sect.~\ref{subsec:dm_axions}, we focus on ultra-light axion-like DM particles. 

\subsubsection{Scattering with baryons}
\label{subsec:dm_baryons}

Traditional nuclear-recoil-based direct-detection experiments are only sensitive to WIMPs (weakly-interacting massive particles) with masses above about a GeV, and with such low interaction cross-sections\footnote{For reference, one of the current largest experiments of this class, Xenon1T, is sensitive to cross-sections roughly in the range $10^{-47}$--$10^{-31}\,$cm$^2$ \cite{2018PhRvD..97k5047E,2018PhRvD..97l3013K}.} that they can penetrate the heavy shielding of detector targets \cite{2013arXiv1310.8327C,2018PhRvD..97k5047E,2018PhRvD..97l3013K}. 
Reducing the amount of shielding can lift the ``ceiling'' on the sensitivity of direct searches toward higher cross-sections \cite{Hooper:2018bfw}, and new strategies are being explored to expand their sensitivity to low DM particle masses, below a GeV \cite{Battaglieri:2017aum}.
An entirely complementary way to probe sub-GeV DM is to search for evidence of its interactions in cosmological data.
Since lower DM particle masses translate to a higher number density of scattering centers, the CMB is particularly sensitive to light particles.
In addition, the CMB is sensitive to all cross-section magnitudes near and above the nuclear scale.

\begin{figure}[t]
\begin{center}
\includegraphics[width=0.5\textwidth]{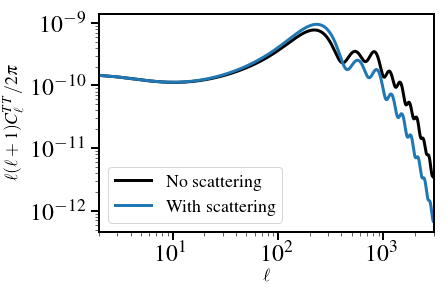}
\includegraphics[width=0.4\textwidth]{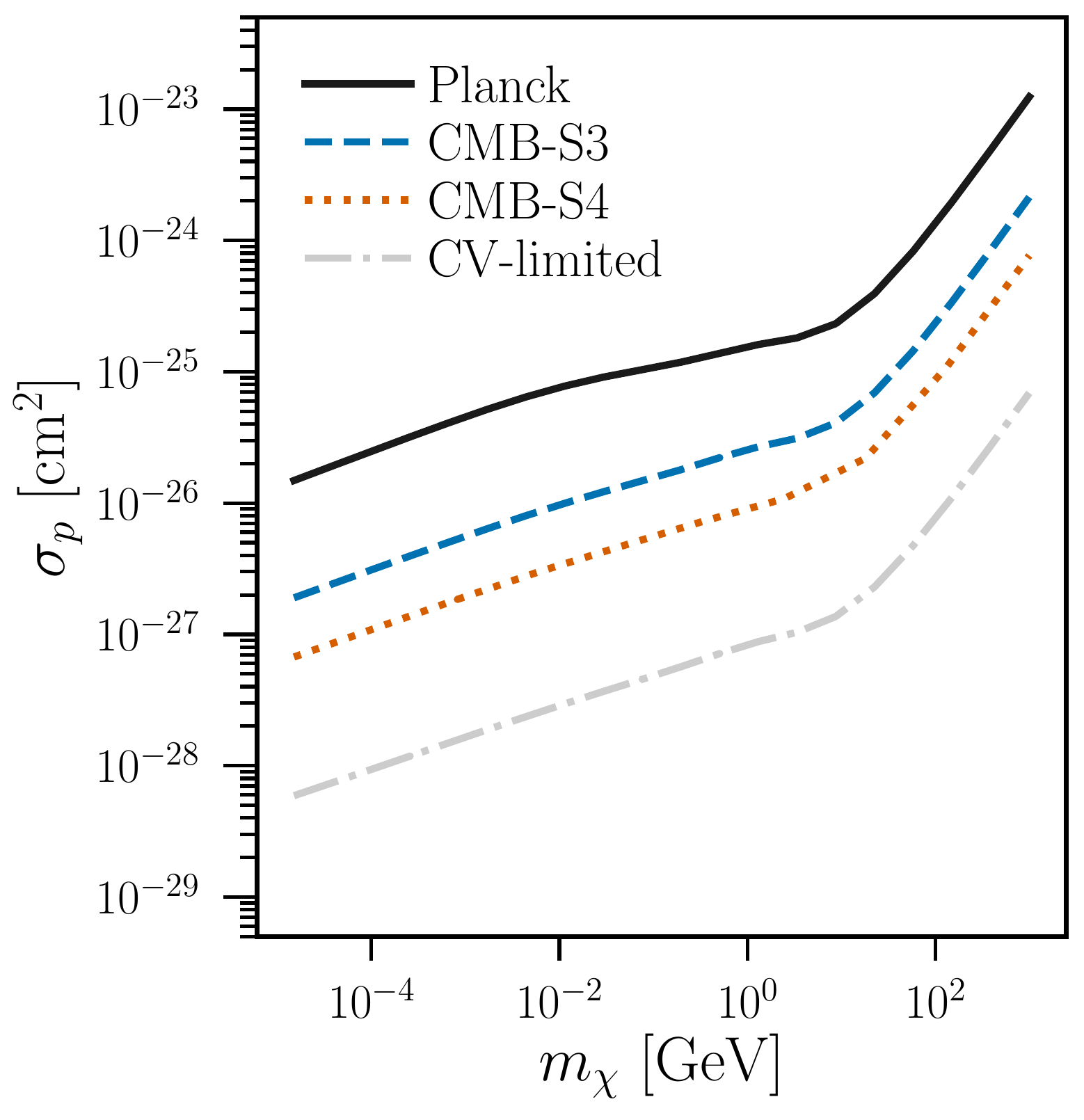}
\caption{\textit{Left:} Illustration of the effect of a velocity-independent spin-independent contact interaction between dark matter and baryons (with a cross-section 100 times higher than the current upper limit from {\planck}) on the CMB temperature power spectrum (blue), compared to the CDM case (black). \textit{Right:} Upper limits on the DM-proton interaction cross-section as a function of DM mass, for spin-independent velocity-independent scattering. In case of a null detection, areas above the curves are excluded at the 95$\%$ confidence level.
Shown are the current limits from {\planck} \cite{2018PhRvL.121h1301G} and forecasts for Stage-3 experiments (CMB-S3, such as AdvACT \cite{2016ApJS..227...21T}), CMB-S4, and a cosmic-variance-limited experiment (CV-limited); see also \cite{2018PhRvD..98l3524L}.}\label{fig:DM_baryons}
\end{center}
\end{figure}

In the scenario in which DM scatters with protons in the early Universe, a drag force between the two cosmological fluids damps the acoustic oscillations and suppresses power in density perturbations on small scales. 
As a result, the CMB temperature, polarization, and lensing power spectra are suppressed at high multipoles, with respect to those in a $\Lambda$CDM universe (see the left panel of Fig.~\ref{fig:DM_baryons} for illustration).
This effect has been used to search for evidence of DM-proton scattering for the case of heavy DM, using CMB and Lyman-$\alpha$ forest measurements \cite{2002astro.ph..2496C,2004PhRvD..70h3501S,Dvorkin:2013cea}. 

Recently, Ref.~\cite{2018PhRvL.121h1301G} presented the first cosmological search for DM particles with any mass down to a keV (orders of magnitude below the mass thresholds of direct-detection experiments), and was followed by a number of related studies \cite{2018PhRvD..98h3510B,2018PhRvD..97j3530X}.
In particular, Ref.~\cite{2018arXiv180108609B} used {\planck} data to derive the first CMB limits on the non-relativistic effective theory of DM-proton scattering---a framework developed by the direct-detection community to characterize all available phenomenologies for scattering through a heavy mediator \cite{Fan:2010gt,Fitzpatrick:2012ix,Anand:2013yka,Dent:2015zpa}.
Similarly, Refs.~\cite{2018PhRvD..98l3506B,2018PhRvD..98b3013S,2018PhRvD..97j3530X} have searched for interactions that can be parameterized by a power-law dependence of the interaction cross-section on the relative particle velocity, and reported improved limits on a wide range of models.
Furthermore, Ref.~\cite{2018PhRvD..98l3506B} developed an improved treatment of non-linear effects that arise in the calculation of post-recombination scattering signals.
This and related studies have enabled a robust investigation of a scenario where only a fraction of DM interacts---and tightly couples---with baryons (while the rest behaves just like the standard cold DM fluid), leading to the first robust cosmological limits on an interacting DM sub-component \cite{2018PhRvD..98l3506B, 2018arXiv180511616D}.
In addition, analyses of CMB data provided essential consistency-tests of recent claims that the anomalous 21-cm signal reported by the EDGES collaboration \cite{2018Natur.555...67B} could be explained with late-time DM-baryon scattering \cite{2018Natur.555...71B}; see, for example, Refs.~\cite{2018PhRvD..98j3529K,2018PhRvD..98b3013S}.

These state-of-the-art CMB tests of DM-baryon interaction physics, which resulted in a broad exploration of new parameter space, relied primarily on moderate-resolution temperature measurements from {\planck}.
Since the DM signal is increasingly prominent at higher multipoles (see Fig.~\ref{fig:DM_baryons}), high-resolution measurements with CMB-S4 will thus substantially improve sensitivity to DM-baryon interactions. 

In Fig.~\ref{fig:DM_baryons}, we present current and projected upper limits on the cross-section for scattering on protons, as a function of DM mass, for a spin-independent velocity-independent interaction (chosen as our fiducial model).
Areas above the curves are excluded at the 95$\%$ confidence level.
We compare current limits obtained from {\planck} (from Ref.~\cite{2018PhRvL.121h1301G}) with forecasts for Stage-3 experiments (with a white noise level of 10$\mu$K-arcmin and resolution of $1.4^\prime$), CMB-S4 (with specifications listed in Sect.~\ref{chap:referencedesign}), and a cosmic-variance-limited experiment (with vanishing noise levels). 
We note the large improvement factor that CMB-S4 delivers over the current limits, for the entire DM mass range considered.
 
Most of the constraining power in the case of CMB-S4 comes from the lensing anisotropy, while the temperature and polarization anisotropies contribute roughly equally to the projected constraint \cite{2018PhRvD..98l3524L}.
Furthermore, Ref.~\cite{Ade:2018sbj} has shown that an increase in sky coverage beyond about $10\%$ (keeping all other parameters fixed) only marginally improves the projected sensitivity.
Finally, Ref.~\cite{2018PhRvD..98l3524L} has also shown that DM-baryon scattering is easily distinguishable from most other new-physics effects sought by CMB experiments (the neutrino mass, new light degrees of freedom, and DM annihilation) once the lensing anisotropy is measured at the level of CMB-S4. 
Therefore, beyond its ability to cover vast open portions of DM parameter space, CMB-S4 also holds promise as a DM discovery tool.

Small-scale CMB anisotropy measuements enabled by CMB-S4 will probe DM-baryon interactions at the time when the Universe was much less than a thousand years old.
These measurements will be sensitive to particle masses outside the detection limits of most existing direct-detection experiments; furthermore, unlike all Earth-based experiments, CMB analyses are independent on the assumptions about the local phase-space distribution of DM.
Compared to the Simons Observatory forecasts \cite{Ade:2018sbj}, CMB-S4 will further improve the sensitivity to DM scattering by a factor of 4--5 (for a DM mass of 1\,GeV and a velocity-independent interaction)---sufficient to enable signal confirmation and detailed subsequent studies, in the case of a marginal detection with future Simons-Observatory data.
On the other hand, CMB-S4 will provide important consistency checks for small-scale-structure probes of DM microphysics, enabled by upcoming galaxy surveys such as LSST.
In particular, DM interactions can leave imprints on satellite galaxy populations \cite{2019arXiv190410000N} and other collapsed structures in the local Universe \cite{2019arXiv190201055D}, many of which will be measured in detail in the coming decade of observations.
However, tests of DM microphysics with small-scale-structure tracers face modeling and simulation challenges; CMB measurements that can capture the same physical effects on large scales and in the early Universe will be pivotal for robust inference of DM particle properties from observational data sets in the future.

\subsubsection{Interactions with dark radiation}
\label{subsec:dark_radiation}

The exquisite sensitivity of the CMB to the depth and size of the DM gravitational potentials near the surface of last scattering makes it a particularly good probe of any new physics affecting the clustering of DM on large scales at early times. This sensitivity to DM density fluctuations is extended to lower redshifts via the weak gravitational lensing that CMB photons experience as they propagate to us. Similar to how tight coupling with photons inhibits the growth of baryon fluctuations until the epoch of hydrogen recombination, DM interacting with light (or massless) dark radiation (DR) at early times experiences a suppressed growth of structure due to the dark-radiation pressure opposing gravitational infall. Models where such interactions arise are diverse in their particle content (see e.g., Refs.~\cite{Aarssen:2012fx,Cyr-Racine:2013fsa,Buen-Abad:2015ova}) and have been invoked to explain the apparent low amplitude of matter fluctuations measured by certain weak-lensing surveys \cite{Lesgourgues:2015wza,Chacko:2016kgg,Buen-Abad:2017gxg}. They can also naturally arise in the context of self-interacting DM, which has been proposed to address possible anomalies on subgalactic scales \cite{Bullock:2017xww}. 

The impact of DM-DR interaction on the CMB has been studied in detail in Refs.~\cite{Cyr-Racine:2013fsa,Krall:2017xcw} (see also Ref.~\cite{Cyr-Racine:2015ihg} for a detailed derivation of the relevant Boltzmann equations). In short, the presence of extra DR mimics the presence of extra neutrino species (see Sect.~\ref{sec:science_light_relics}) and affects the expansion history of the Universe, possibly modifying the epoch of matter-radiation equality, the CMB Silk damping tail, and the early integrated Sachs-Wolfe effect. However, unlike standard free-streaming neutrinos, the DR forms a tightly coupled fluid at early times, leading to distinct signatures on CMB fluctuations which include a phase and amplitude shift of the acoustic peaks (see Refs.~\cite{Bashinsky:2003tk,Follin:2015hya,Baumann:2015rya}). In addition, the DR pressure prohibits the growth of interacting DM fluctuations on length scales entering the causal horizon before the epoch of DM kinematic decoupling. This weakens the depth of gravitational-potential fluctuations on these scales, affecting the source term of CMB temperature fluctuations. Finally, the modified matter clustering in the Universe due to the interaction with DR affects CMB lensing. For interacting DM models that are still allowed by the current \planck\ data, this latter effect is where CMB-S4 can significantly improve the constraints on these non-minimal DM theories.

Given the large array of possible DM theories to constrain, it is useful to pick a simple benchmark DM-DR model to assess the sensitivity of CMB-S4 to the effects described in the previous paragraph. A particularly useful model is one in which the interaction between DM and DR is mediated by a particle with mass of order a few MeV. Such theories have been put forward in the context of self-interacting DM \cite{Tulin:2012wi,Tulin:2013teo,Kaplinghat:2015aga} in order to obtain a velocity-dependent self-interaction cross-section that can fit both the observed density profiles of dwarf galaxies and of large galaxy clusters. For these models, the momentum-transfer rate between the DR and DM in the early Universe can be written as $\dot{\kappa}_{\rm DR-DM} \propto \Omega_{\rm DM}h^2 a_4 (1+z)^4$, where $a_4$ is a parameter that controls the strength of the DM-DR interaction that depends on the exact Lagrangian used (see Ref.~\cite{Cyr-Racine:2015ihg}). Within this class of models, the most relevant parameters that can be constrained with CMB-S4 are the interaction strength $a_4$, the fraction of DM interacting with DR $f_{\rm int}$, and the amount of DR present in the Universe $\Omega_{\rm DR}h^2$. 

\begin{figure}[t]
\begin{center}
\includegraphics[width=0.48\textwidth]{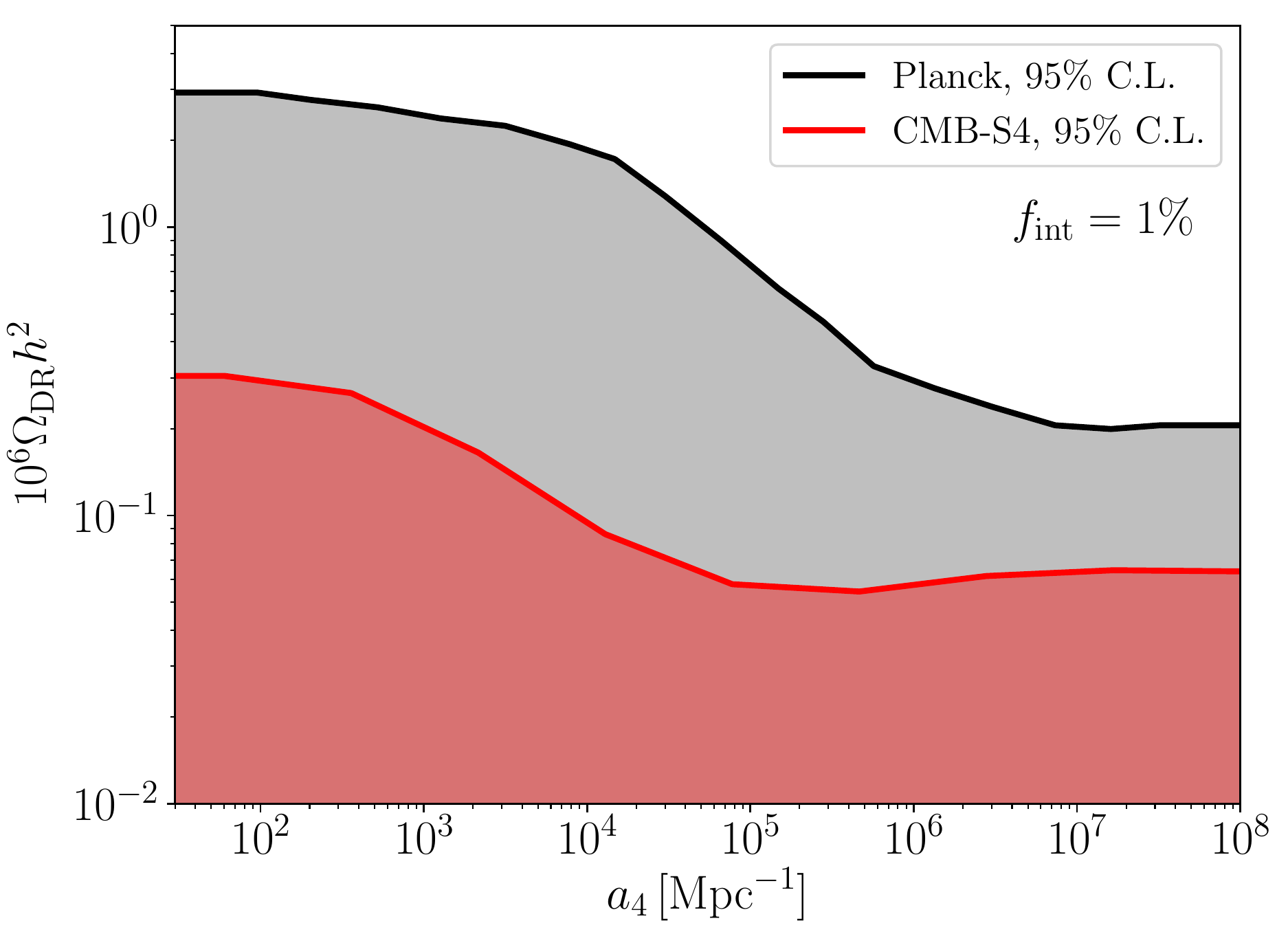}
\includegraphics[width=0.48\textwidth]{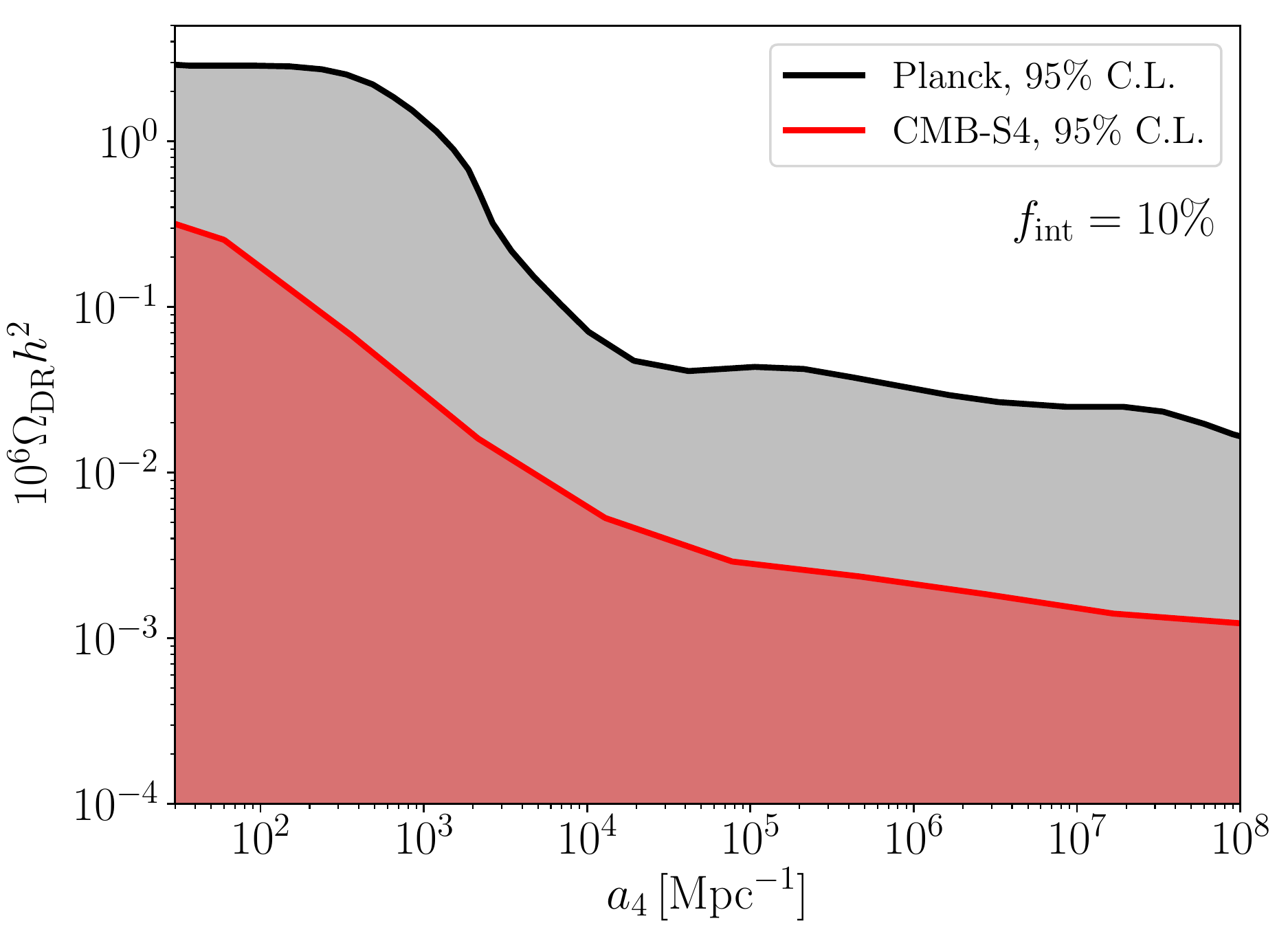}
\caption{Dark matter-dark radiation interaction sensitivity of \planck\ (adapted from Ref.~\cite{Cyr-Racine:2013fsa}) and an experiment like CMB-S4; for this plot, we assumed a configuration of CMB-S4 similar to that outlined in the CDT. The x-axis corresponds to the interaction strength, while the y-axis corresponds to the abundance of dark radiation. We show two different fractions of DM interacting with the DR, as indicated on each panel. The shaded regions are allowed at the 95$\%$ confidence-level. Note the different scales for the y-axes on each panel.}\label{fig:DR_vs_a4}
\end{center}
\end{figure}

In Fig.~\ref{fig:DR_vs_a4}, we show the projected sensitivity of a CMB-S4-like experiment to the presence of DM-DR interactions in the $\Omega_{\rm DR}h^2$--$a_4$ plane, for two values of $f_{\rm int}$, namely 1\% and 10\%. 
For low values of the interaction strength $a_4$, the DR decouples very early from DM and essentially behaves like an extra free-streaming relativistic species, and the constraint on $\Omega_{\rm DR}h^2$ thus becomes equivalent to that on $\Delta N_{\rm eff}$ discussed in Sect.~\ref{sec:science_light_relics}. As the DM-DR interaction strength is increased, $\Omega_{\rm DR}h^2$ becomes more severely constrained, to ensure that DM kinematic decoupling occurs before the modes probed by the CMB enter the causal horizon. In addition, the constraints become stronger as the fraction of interacting DM increases. 

Looking ahead, the main difficulty in constraining theories in which DM interacts with DR using CMB-S4 will be modeling of non-linearities in the CMB lensing power spectrum, which can be important for multipoles above $500$ (in the reconstructed lensing-potential power spectrum). Recent progress in this respect is already promising \cite{Vogelsberger:2015gpr}. The CMB constraints on DM-DR interaction could be further improved by combining CMB-S4 with probes of small-scale structure, such as the Lyman-$\alpha$ forest or the luminosity function of Milky Way satellite galaxies (see e.g., Ref.~\cite{Schneider:2016uqi}).

\subsubsection{Axion-like particles}
\label{subsec:dm_axions}
 The QCD axion and other axion-like particles (ALPs) are well-motivated DM candidates and can contribute to the DM density (see Ref.~\cite{Marsh:2015xka} for a recent review). One example is ultralight axions (ULAs), which are non-thermally created via vacuum realignment and have a distinctive phenomenology. ULAs are predominantly non-thermal, nonrelativistic, and do not contribute to $N_{\rm eff}$.\footnote{Note, however, that if these axions have couplings to ordinary matter, as described in Sect.~\ref{sec:science_light_relics}, then a second, relativistic population of axions is created.}

The ULAs we consider here are motivated by string theory and are associated with the geometry of the compact spatial dimensions. These axions can contribute either to the dark-matter or dark-energy budget of the Universe, depending on their particular mass, which sets the time at which the axions begin to coherently oscillate and redshift as matter.  We consider axions within a range of masses $10^{-33}\,\mathrm{eV}\leq m_{\rm a}\leq 10^{-20}\,\mathrm{eV}$, with negligible couplings to the Standard Model particles.  We can compare these assumptions to those used in Sect.~\ref{sec:science_light_relics}; the contribution of thermal axions to $\Neff$ applies to any mass $m_{\rm a}\lesssim 1\,\mathrm{eV}$, including the well known QCD axion, but depends in detail on the couplings to the Standard Model particles and on the reheat temperature.  Cosmological constraints on axion dark matter are thus complementary in the space of masses and couplings to the constraints on a thermal population of axions.

At the moment, the CMB (through measurements of CMB lensing, temperature, and polarization \cite{Hlozek:2017zzf}, using tools developed in Refs.~\cite{Hlozek:2014lca,Hlozek:2016lzm}), offers the best gravitational probe of the ULA density in the regime $10^{-33}\,{\rm eV}\leq m_{\rm a}\leq 10^{-25}\,{\rm eV}$. We choose a fiducial value of the axion energy density consistent with these constraints. In this window, ULAs cannot be all of the dark matter, but can be a significant component, with density comparable to baryons and neutrinos. Given the rich spectrum of particles in the Standard Model, there is no reason to expect a trivial dark sector, and it is important to explore the power of the CMB to constrain any particle species with a potentially detectable density.

CMB-S4 will push the upper edge of the CMB ULA window to $m_{\rm a}\sim 10^{-23}\,{\rm eV}$. At higher masses still, ULAs will alter pulsar timing signatures \cite{DeMartino:2017qsa,Porayko:2018sfa}, suppress the clustering of neutral hydrogen at high-$z$ (an observable probed by measurements of the Lyman-$\alpha$ forest flux power spectrum \cite{Irsic:2017yje,Kobayashi:2017jcf,Armengaud:2017nkf,Leong:2018opi,Nori:2018pka}), alter the mass spectrum of black holes through Penrose processes \cite{Arvanitaki:2009fg,Arvanitaki:2010sy,Marsh:2015xka,Stott:2018opm,Kitajima:2018zco} with implications for Laser Interferometer Gravitational-Wave Observatory (LIGO) event rates, and lead to observable gravitational-wave signatures in the {\it Laser Interferometer Space Antenna (LISA)\/} band \cite{Baumann:2018vus,Arvanitaki:2010sy,Stott:2018opm,Kitajima:2018zco} (see Ref.~\cite{Grin:2019mub} for a recent review of ULA gravitational signatures).

\subsubsection{Constraints on cold axion energy density}
\label{sec:ax_adiabat}
The degeneracies of ULAs with other cosmological parameters, such as $N_\mathrm{eff}$ or $m_\nu$, vary depending on the axion mass \cite{Hlozek:2016lzm}. Dark-energy like axions with masses around $10^{-33}\,{\rm eV}$ change the late-time expansion rate and therefore the sound horizon, changing the location of the acoustic peaks. There are thus degeneracies of the ULA density $\Omega_{\rm a}$ with the matter and curvature content. Heavier axions ($m_{\rm a} \gtrsim 10^{-26}\,{\rm eV}$) affect the expansion rate in the radiation era and reduce the angular scale of the diffusion distance, leading to a boost in the higher acoustic peaks, which has a degeneracy with $N_{\rm eff}$. 

In the matter power spectrum, and thus CMB lensing power, light axions suppress clustering, suggesting a degeneracy with effects of massive neutrinos that must be broken to make an unambiguous measurement of neutrino mass using the CMB. The above-mentioned effects in the expansion rate break this degeneracy for some axion masses. There remains a significant degeneracy between $\Omega_{\rm a}$ and the sum of the neutrino masses ($\Sigma_{i}m_{\nu_{i}}$) and massive neutrinos if $m_{\rm a} \lesssim 3\times 10^{-29}\,\mathrm{ eV}$. Effort should be made to break these (and other) degeneracies using other data sets to establish if CMB-S4 data will show the unambiguous signature of neutrino mass, or hint at some other new physics.
\begin{figure}[t]
\begin{center}
\begin{tabular}{cc}
\includegraphics[width=0.4\textwidth]{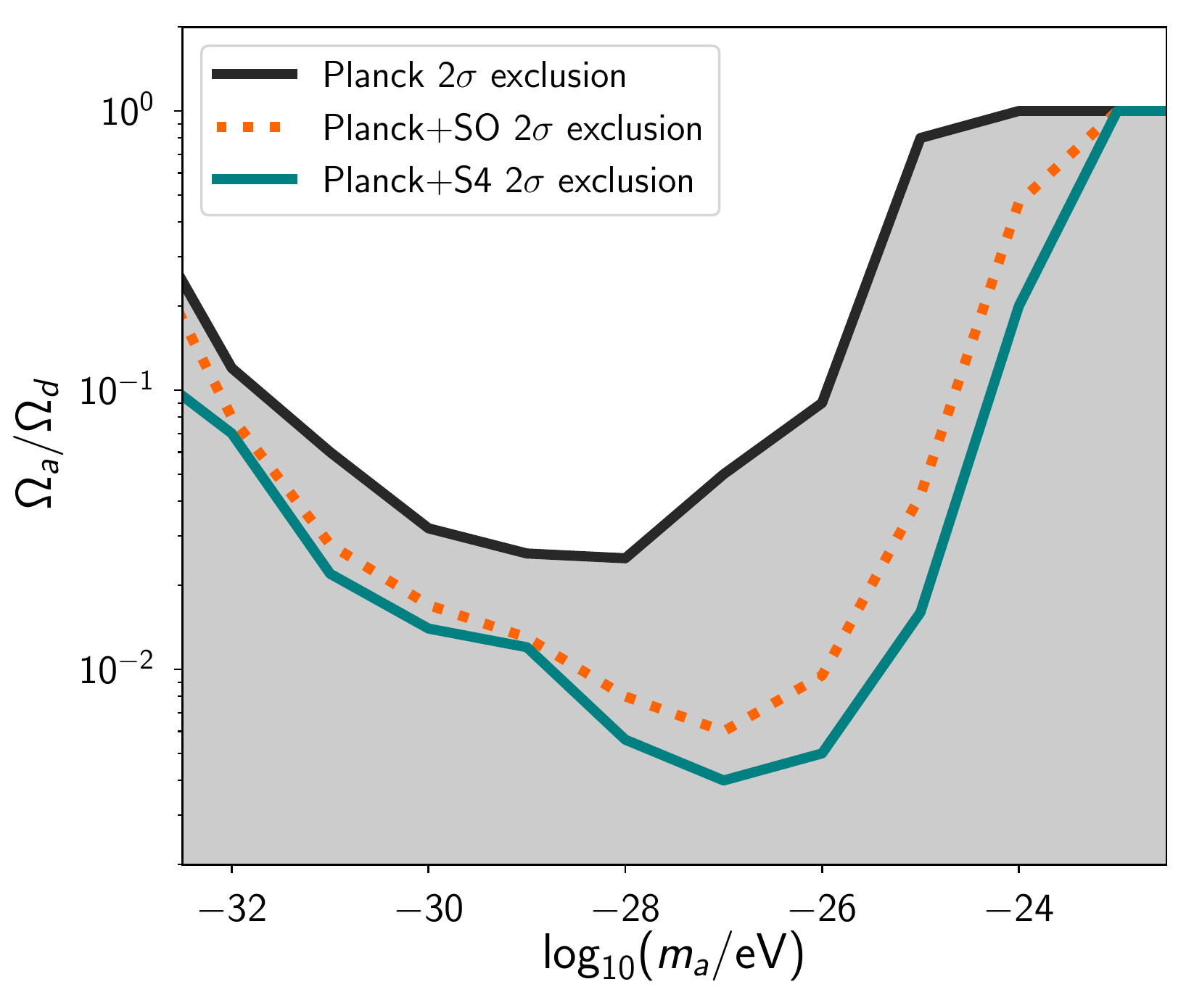} &
\includegraphics[width=0.4\textwidth]{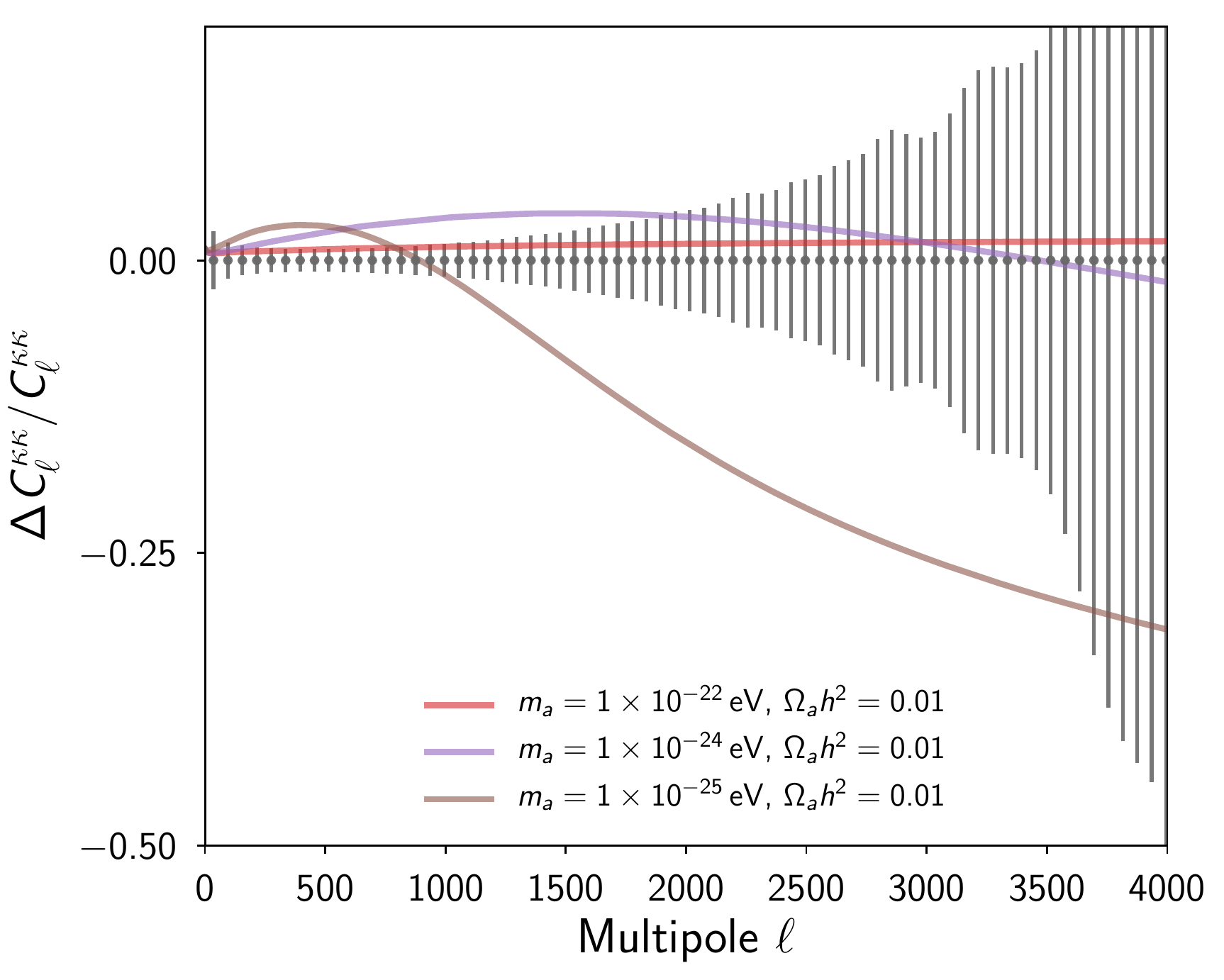} \\
\end{tabular}
\caption{Constraints on ultra-light axions (ULAs). Left: Fisher-forecasted $2\sigma$ exclusion regions for the ULA mass fraction $\Omega_{\rm a}/\Omega_{\rm d}$ for the {\it Planck\/} alone, {\it Planck\/} + Simons Observatory (SO) (as discussed in Ref.~\cite{2018arXiv180807445T}), and CMB-S4, where $\Omega_{\rm d}=\Omega_{\rm a}+\Omega_{\rm c}$.  Right: Residual in lensing-convergence power spectrum $C_{L}^{\kappa \kappa}$ for different models including ULA DM, compared with Fisher-forecasted errors for CMB-S4. \label{fig:DM_axions}}
\end{center}
\end{figure}

We show the forecasted sensitivity to the axion energy density from CMB-S4, including lensing in the left panel of Fig.~\ref{fig:DM_axions} (for fixed neutrino mass of $\Sigma m_\nu = 0.06\,{\rm eV}$). Adding information from the lensing reconstruction using CMB-S4 (or SO) will improve sensitivity to axion DM significantly. A percent-level measurement of the lensing deflection power at multipoles $\ell > 1000$ leads to an improvement in the error on the axion energy density of a factor of around 8 relative to the current \planck\ constraints, for an axion mass of $m_{\rm a}=10^{-26}\,{\rm eV}$.  This will allow us to test if DM has different particle components at the percent level. Furthermore, since $\Omega_{\rm a}\propto f_{\rm a}^2$, this improves the expected constraint on the axion decay constant from $10^{17}\,{\rm GeV}$ with \planck\ to $10^{16}\,{\rm GeV}$ with CMB-S4, testing the ``string axiverse" scenario~\cite{Arvanitaki:2009fg}. 

Using only \planck\  data, ULAs are degenerate with CDM at $m_{\rm a}=10^{-24}\,\mathrm{eV}$, and constraints are still weak at $m_{\rm a}=10^{-25}\,\mathrm{eV}$. CMB-S4 could make a $>5\sigma$ detection of departures from CDM for masses as large as $m_{\rm a}=10^{-25}\,\mathrm{eV}$, and could improve the lower bound on DM particle mass to $m_{\rm a}=10^{-23}\,\mathrm{eV}$ and fractions ${\cal O}$(10\%). Realizing this sensitivity will require better modeling of the non-linear clustering of axions \cite{Marsh:2016vgj}, building upon and applying the results of previous work \cite{Marsh:2013ywa,Schive:2014hza,Schive:2014dra,Schive:2015kza,Robles:2014ysa,Veltmaat:2016rxo,Du:2016aik,Zhang:2016uiy,Chen:2016unw,Schive:2017biq,LinaresCedeno:2018oso,Du:2018zrg,DeMartino:2018zkx,Lin:2018whl,Robles:2018fur,Veltmaat:2018dfz,Nori:2018hud}.

\subsubsection{Axion isocurvature}
\label{sec:axioniso}
The axion decay-constant, $f_{\rm a}$, specifies the scale at which the underlying $U(1)$ symmetry is broken. If $H_{\rm I}/2\pi<f_{\rm a}$, then this symmetry is broken during inflation, and the axion acquires {\it uncorrelated\/} isocurvature perturbations \cite{Axenides:1983hj,Fox:2004kb,Hertzberg:2008wr}.\footnote{We ignore the case where $H_{\rm I}/2\pi>f_{\rm a}$, since then no isocurvature fluctuations are excited. The limit $r_{0.05}\lesssim0.1$ \cite{Aghanim:2018eyx} implies that isocurvature is produced if $f_{\rm a}>1.6\times 10^{13}\,{\rm GeV}$. This accounts for the QCD axion in the ``anthropic'' window (roughly half of the allowed range of $f_{\rm a}$ on a logarithmic scale), axions with GUT-scale decay constants (such as string axions~\cite{Svrcek:2006yi,Arvanitaki:2009fg}) and axions with lower $f_{\rm a}$ in models of low-scale inflation.} The uncorrelated CDM isocurvature amplitude is bounded by \planck\ to be $A_{\rm a}/A_{\rm s}<0.038$ at 95\%~CL \cite{Ade:2015lrj}. While axion isocurvature probes the same inflationary energy scale probed by primordial $B$ modes, searches for isocurvature are independent of other constraints on the tensor-to-scalar ratio, offering a distinct test of inflationary physics, if a significant fraction of the DM is composed of axions.

The axion isocurvature amplitude is:
\begin{equation}
A_{\rm I} = \left(\frac{\Omega_{\rm a}}{\Omega_{\rm d}}\right)^2\frac{(H_{\rm I}/M_{\rm Pl})^2}{\pi^2(\phi_{\rm i}/M_{\rm Pl})^2} \, ,
\label{eqn:iso_amplitude}
\end{equation}where $\Omega_{\rm d}=\Omega_{\rm a}+\Omega_{\rm c}$ is the combined total primordial relic density parameter in ULAs and standard CDM.
The initial axion displacement, $\phi_{\rm i}$, fixes the axion relic abundance such that $\Omega_{\rm a}=\Omega_{\rm a} (\phi_{\rm i},m_{\rm a})$ \cite{Preskill:1982cy,Abbott:1982af,Dine:1982ah,Turner:1983he,Steinhardt:1983ia,Marsh:2010wq}. Thus, if the relic density and mass can be measured by independent means, such as direct detection, a measurement of the axion isocurvature amplitude can be used to measure the energy scale of inflation, $H_{\rm I}$.

We forecast the errors on axion isocurvature for the baseline CMB-S4 experiment with a $1\,\mu$K-arcmin noise level and a $1^{\prime}$ beam; the isocurvature limit will be improved by a factor of approximately $5$ compared to \planck, allowing for detection of axion-type isocurvature at 2$\sigma$ significance in the region $0.008<A_{\rm I}/A_{\rm s}<0.038$. If the QCD axion is all of the DM, axion direct-detection experiments (e.g., ADMX \cite{Asztalos:2009yp} already in operation, or CASPEr \cite{Budker:2013hfa}, in progress) and CMB-S4 could probe $H_{\rm I}$ in the range
\begin{eqnarray}
 2.5\times 10^6\lesssim &H_{\rm I}/\text {GeV}&\lesssim 10^{7} 
~~~\text{(S4+ADMX)}, \\
 10^{8}\lesssim &H_{\rm I}/\text {GeV}&\lesssim 4\times 10^{9}
~~~\text{(S4+CASPER)},
\end{eqnarray}
where we have used the standard formulae relating the QCD axion mass and relic abundance to the decay constant \cite{Fox:2004kb}. {Combining axion DM direct detection with CMB-S4 isocurvature measurements allows a unique probe of low-scale inflation, inaccessible to searches for tensor modes.}

We now consider isocurvature fluctuations in ULAs, which have a number of distinctive features the CMB \cite{Marsh:2013taa,Marsh:2014qoa,Hlozek:2016lzm,Hlozek:2017zzf,Hlozek:2014lca,Marsh:2013ywa}. We fix the fiducial ULA fraction to 1\%, such that $\Omega_{\rm a}$ and $m_{\rm a}$ can be separately measured using the CMB-S4 lensing power, and thus using Eq.~(\ref{eqn:iso_amplitude}), a measurement of $A_{\rm I}$ is a measurement of $H_{\rm I}$. In contrast to the QCD axion, there are masses ($m_{\rm a}\lesssim 10^{-26}\text{ eV}$) for which tensor modes impose a stronger constraint on $H_{\rm I}$ than isocurvature. There are also regions of overlap between possible tensor and isocurvature measurements, where CMB-S4 could be used to determine isocurvature and axion parameters, yielding an independent measurement of $H_I$:
\begin{equation} 2.5\times 10^{13}\lesssim H_{\rm I}/\text{GeV}\lesssim 10^{14}\,
\text{(ULAs, CMB-S4 alone)}\,\,.
\end{equation}
A broader range of scenarios is considered in Ref.~\cite{Abazajian:2016yjj}. Already with \textit{Planck} data, the combination of lensing reconstruction and isocurvature limits gives the posterior range for the tensor-to-scalar ratio of $r<0.01$ for $m_{\rm a}=10^{-24}\,{\rm eV}$ \cite{Hlozek:2017zzf}.

\newpage

\section{Mapping matter in the Cosmos}

\begin{shaded}
\emph{The third science theme relates to surveying the contents of the Universe revealed by the millimeter-wave sky.}

The deep and wide CMB-S4 field will provide a unique census of a large fraction of the sky
at centimeter to millimeter wavelengths.  Matter in the Universe shows up in this
census in multiple ways: the gravitational potential coherently shears the image of the CMB fluctuations,  some CMB photons are scattered by electrons along the way, and many objects emit their own radiation.  All of these help us to map out the matter in the cosmos.

The main value of CMB-S4 for non-CMB scientists  will come from
the deep and wide survey (although the ultra-deep survey will also be useful for these communities).
In the reference design, CMB-S4 will map $70\%$ of the sky in total
intensity and linear polarization, with a cadence of around 2 days, reaching a $5\sigma$ point-source 
noise well below 1\,mJy in the final coadded maps,  along with corresponding maps of gravitational lensing 
and Compton-$y$.

The CMB-S4 survey will complement and enhance the LSST optical survey of the same region
\cite{Ivezic:2008fe}, the DESI spectroscopic survey \cite{DESI:2016},
the {\it eROSITA\/} all-sky X-ray survey \cite{Merloni:2012}, and other
planned and yet-to-be-imagined surveys from both ground- and space-based
facilities.  By going dramatically deeper than previous CMB surveys and covering
a large fraction of the sky, the potential for new discoveries is high.
This section provides an overview of some of the products to be
derived from CMB-S4 that will be of significant utility to the broader
astronomical community.

With these data, CMB-S4 will be able to probe physical processes that govern and regulate galaxy formation in various classes of objects, and probe the circumgalactic and intracluster mediums.  Secondary CMB anisotropies are an excellent probe of reionization because scattering of photons by free electrons affects the observed temperature and polarization in a predictable fashion that is sensitive to its duration and morphology (or ``patchiness'').
CMB-S4 will leave a legacy that includes catalogs of extragalactic sources, maps of extragalactic integrated density and pressure, and multifrequency images of Galactic polarization.

These data sets  will be exploited by astronomers and astrophysicists over the decades to come.

\end{shaded}

\subsection{Extragalactic component maps}
\label{sec:componentmaps}

With the broad frequency coverage of CMB-S4, we can use internal-linear-combination techniques to produce maps of isolated sky signals. 
These products will include maps of the CMB temperature and polarization
anisotropies, the reconstructed CMB gravitational-lensing potential,
Compton-$y$ maps, the cosmic infrared background,  
and Galactic synchrotron and dust (in both intensity and polarization). 
Of particular utility to the broader astronomical community will be the
lensing map and the Compton-$y$ (or tSZ) map; we detail their properties below.  

\textbf{Gravitational Lensing Map:} One of the main goals of the CMB-S4 project is to extract cosmological information from a reconstructed map of the large-scale structures responsible for gravitationally lensing the CMB.  This will constitute a map of all matter between us and the CMB, including dark matter, with a broad redshift weighting that peaks near redshift $z = 2$.  The map will cover the entire footprint of the CMB-S4 survey and will represent the true mapping of matter, with map-level signal-to-noise ratio exceeding unity, from the largest scales in the survey down to scales of approximately 12 arcminutes (representing multipole $L \simeq 1000$).  Even on smaller scales, there will still be significant statistical information.

Since the contribution to CMB lensing is very broad as a function of  redshift, including appreciable weight at $z < 1$ where many optical and other surveys are sensitive, this map will have significant potential for cross-correlation with data at a range of wavelengths.  Thus far, CMB lensing maps have been studied in conjunction with a very diverse group of data sets, including (in order of decreasing wavelength): 
extragalactic radio catalogs \citep[][]{Smith:2007rg, Hirata:2008cb, Allison:2015fac};
maps of the Sunyaev-Zeldovich effect \citep[][]{Hill:2013dxa};
the cosmic infrared background \citep[][]{Holder:2013hqu,Ade:2013aro,Ade:2013hjl,vanEngelen:2014zlh};
far-infrared galaxies \citep[][]{Bianchini:2014dla};
mid-infrared galaxies \citep[][]{Bleem:2012gm,Ade:2013tyw};
mid-infrared quasars \citep[][]{Geach:2013zwa};
optical redshift surveys   \citep[][]{Pullen:2015vtb}, including filaments \citep[][]{He:2017owu} and voids \citep[][]{Cai:2016rdk};
optically-selected galaxy clusters \citep[][]{Ade:2013tyw};
optically-selected quasars  \citep[][]{Sherwin:2012mr,Ade:2013tyw};
optical weak lensing  \citep[][]{Hand:2013xua,Liu:2015xfa}; 
optical photometric catalogs \citep[][]{Baxter:2016ziy,Giannantonio:2015ahz};
the Lyman-$\alpha$ forest \citep[][]{Doux:2016xhg};
X-rays \citep[][]{Hurier:2017rgp};
and $\gamma$-rays  \citep[][]{Fornengo:2014cya}. 
Additionally, CMB lensing has been used to measure the masses of SZ-selected \citep[][]{Baxter:2014frs,Ade:2015fva} and optically-selected \citep[][]{Madhavacheril:2014slf}  galaxy clusters and groups.

The  lensing map from CMB-S4  will have much lower noise levels than those used in each of these earlier studies, and will additionally have very wide sky coverage. Because it
is primarily based on CMB polarization data rather than temperature data, it should also be
more robust against possible foreground contamination.
Future analyses can thus be performed at much higher signal-to-noise ratio, with fewer
concerns about contamination.  Furthermore, many new data sets in many wavelength ranges will be available on the timescale of CMB-S4 that will  both be deeper and  have wider sky coverage than many of the studies carried out so far, including: LSST \citep{Ivezic:2008fe} for both photometric galaxy catalogs and weak lensing; DESI \citep{Levi:2013gra} for spectroscopic samples; and sensitive maps of ionized gas with the SZ effect from CMB-S4 itself.  In addition to being used in delensing the CMB-S4 map of the primary CMB, the CMB-S4 lensing map will also provide the ability to delens the maps from future satellite CMB missions, which are unlikely to have the same internal sensitivity to lensing.  The map of CMB lensing from CMB-S4 will thus represent a lasting, legacy value to the subsets of the extragalactic astronomical community working in nearly every wavelength range.

\label{subsubsec:ymap}

\textbf{Compton-\emph{y} map:} Multi-frequency CMB temperature data enable the reconstruction of tSZ (Compton-$y$) maps by taking advantage of the known, unique tSZ spectral function.  The first such maps have been produced recently using data from {\it Planck\/} \cite{Aghanim:2014xxi,Hill:2013dxa, Aghanim:2015xxii, Khatri2016}.  Here, we forecast the reconstruction of a Compton-$y$ map using data from the CMB-S4 large-aperture telescopes (LATs) in combination with {\it Planck\/} (due to the large-scale atmospheric noise from the ground, {\it Planck\/} will remain useful at low $\ell$).  We use a harmonic-space internal-linear-combination (ILC) method to obtain post-component-separation noise curves for the tSZ signal.  The sky model includes essentially all important contributions to the microwave sky (Galactic and extragalactic), and is identical to that used in Ref.~\cite{2018arXiv180807445T}.  The CMB-S4 LAT noise properties are described elsewhere in this document.  We include {\it Planck\/} data from 30--353 GHz, assuming white noise and Gaussian beams.  The ILC approach includes the option of ``deprojecting'' particular contaminants with a specified SED, at the cost of increased statistical noise.  We specifically consider deprojecting the CMB, a fiducial cosmic infrared background (CIB) spectrum (modified blackbody with temperature of 19.6\,K and spectral index of 1.2), or both.

\begin{figure}[t]
\begin{center}
\includegraphics[width=0.8\textwidth]{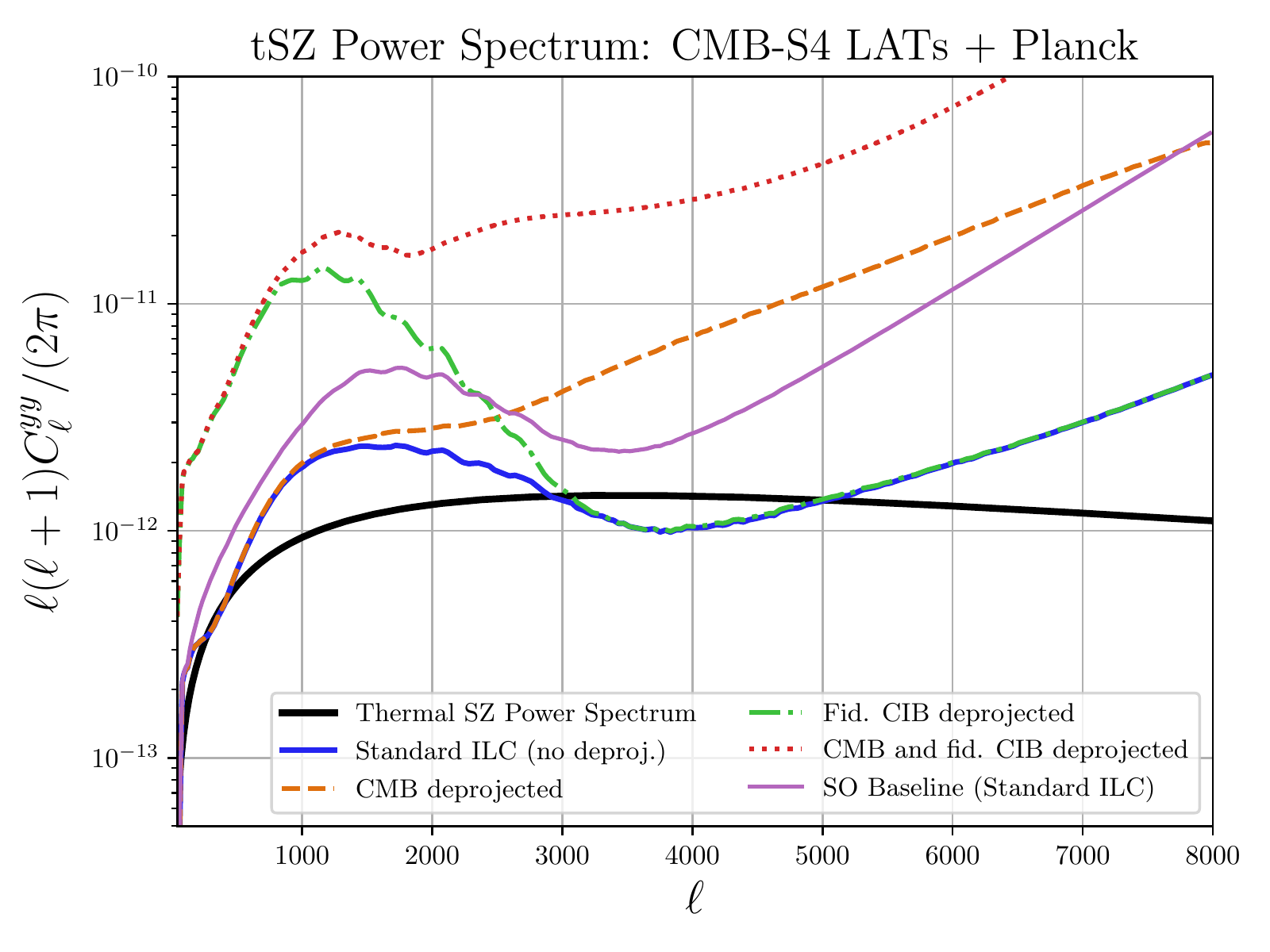}
\caption{Post-component-separation noise for the Compton-$y$ map reconstructed from the CMB-S4 LATs and {\it Planck}.  The solid black curve shows the tSZ power-spectrum signal.  The solid blue curve shows the ILC reconstruction noise, while the other curves show the noise levels for various foreground-deprojection options.  The thin, solid magenta curve shows the ILC reconstruction noise for the Simons Observatory baseline configuration. \label{fig:s4_yy}}
\end{center}
\end{figure}

The results of this analysis are shown in Fig.~\ref{fig:s4_yy}.  The figure shows the post-component-separation noise (per mode) of the Compton-$y$ map produced from CMB-S4 LAT and {\it Planck\/} data for various ILC-based foreground-cleaning options, in comparison to the tSZ power-spectrum signal.  The CMB-S4 noise is low enough that the Compton-$y$ field will be mapped on a mode-by-mode basis for multipoles $3000 \lesssim \ell \lesssim 5000$, even when explicitly deprojecting the CIB foreground.  In fact, even with the expected level of atmospheric noise, CMB-S4 still improves enough over {\it Planck\/} to allow mapping at $\ell \approx$ few hundred.  The figure also shows a similar noise curve for the baseline Simons Observatory (SO) configuration \cite{2018arXiv180807445T}, demonstrating the significant gains that CMB-S4 will achieve.  The total signal-to-noise ratio of the CMB-S4 tSZ power-spectrum measurement is 1570 (standard ILC), 570 (CMB deprojected), 1500 (CIB deprojected), or 130 (CMB and CIB deprojected); for this estimate, we neglect the trispectrum contribution to the covariance matrix, considering Gaussian errors only.  We also conservatively assume $f_{\rm sky} = 0.45$.  For the standard ILC, this is an improvement over the current {\it Planck\/} measurement by nearly two orders of magnitude.  Moreover, this estimate only captures the Gaussian contributions to the tSZ signal, which is a lower bound on the total information content in the $y$-map, given the significant non-Gaussianity of the tSZ field.  The legacy value of this map will be immense, including cross-correlations with optical, infrared, X-ray, and intensity-mapping data sets.

\subsection{Galaxy formation and evolution}
\label{subsec:GF}

The small fraction of CMB photons scattered during their cosmic journey create secondary anisotropies that encode the evolving spatial distribution and thermal energy of diffuse ionized gas throughout the Universe.  Because the CMB is a nearly uniform back light on small scales, angular features in the scattering distortions and anisotropies generated can be used to probe {\em all\/} the ionized gas within a given region, not only that gas dense enough to emit in the X-rays or which happens to contain the particular ionic species that generate observable absorption features along the particular lines of sight to background sources in instrument-specific redshift windows. With its unprecedented sensitivity and angular resolution, CMB-S4 will deliver transformative high signal-to-noise ratio maps of ionized gas over a wide range of scales and redshifts, from ionized bubbles during reionization to the intracluster medium of the most massive present-day halos. These maps, alone and in combination with large-scale structure probes at other wavelengths, will allow us to finally address a crucial longstanding question: {\em how does the energy injected by galaxies and supermassive black holes affect the surrounding baryonic reservoir of diffuse gas driving their evolution?}

In this section, we describe in detail the new physical insights and model constraints that CMB-S4 will provide by addressing this question, and present new quantitative forecasts for constraints on state-of-the-art numerical models of galaxy formation and reionization.  We emphasize that due to the rich nature of this science, the forecasts presented here represent only a lower bound on the impact of CMB-S4 in this area; the range of applications is vast, and will certainly expand further by the start of the CMB-S4 survey.

\subsubsection{Intracluster medium and circumgalactic medium}
Understanding the physical processes that govern and regulate galaxy formation over cosmic time is a central goal in astrophysics.  It has long been known that star formation is an inefficient process: less than 10\% of the cosmic abundance of baryons has been converted into stars over the past 13.8\,Gyr (e.g., \cite{Fukugita-Peebles2004,Gallazzi2008}).  Yet early hydrodynamic simulations of galaxy formation predicted much larger star-formation efficiencies; over the past two decades it has become clear that this discrepancy is likely resolved by {\em feedback\/} processes that prevent the over-cooling of gas in dark matter halos (e.g., \cite{Schawinski2007}).  However, the exact mechanism(s) and even the basic energetics of these feedback processes remain poorly constrained.  On the scale of galaxy groups and clusters, the most important source of feedback---i.e., energy and/or momentum injected into the intragroup/intracluster medium (ICM)---is expected to be that due to active galactic nuclei (AGN), supermassive black holes whose accretion disks drive powerful winds and jets of radiation into their surroundings.\footnote{At high redshift ($z \gtrsim 2$), other sources of feedback (e.g., from star formation) may be of similar importance in groups and proto-clusters.}  At lower mass scales (e.g., Milky-Way-sized halos), feedback due to supernovae and stellar winds may play the dominant role in injecting feedback energy into the circumgalactic medium (CGM).  This energetic feedback profoundly alters the thermodynamic structure of the ICM and CGM, heating the gas (thereby preventing star formation) and driving it to large halo-centric radii, sometimes entirely ejecting it from the halo.  Such processes are likely responsible for the ``missing baryons'' problem, i.e., the empirical fact that low-redshift galaxies (such as the Milky Way) have a baryon deficit relative to the cosmological abundance determined from the CMB and BBN \cite{Bregman2007}. 
Moreover, this re-distribution of the gas leads to non-negligible changes in the dark matter distribution on scales below $\approx 10$ Mpc, thus altering the matter power spectrum.  The amplitude of this effect is the largest source of theoretical systematic uncertainty in cosmological constraints derived from ongoing and upcoming weak-lensing surveys (e.g., \cite{Eifler2015}).  Thus, constraining the properties of AGN and supernova feedback is essential not just for understanding galaxy formation, but also for enabling next-generation constraints on dark energy and neutrino physics.  Beyond feedback, the role of non-thermal pressure support sourced by gas bulk motions and turbulence in stabilizing the ICM and intragroup medium must be precisely determined in order to calibrate biases in hydrostatic-equilibrium-based estimates of cluster masses (e.g., \cite{Nagai:2007mt, 2013ApJ...777..151L, 2016MNRAS.455.2936S}); however, the amplitude of this effect is currently better understood in hydrodynamical simulations than that of AGN feedback (e.g., \cite{Lau2009,Nelson2014,Emberson2018}), and thus we focus on the latter here.

The CMB offers multiple powerful tools with which to constrain feedback processes and non-thermal pressure support in galaxies, groups, and clusters.  The tSZ and kSZ 
effects, which are sourced by the Thomson-scattering of CMB photons off free electrons, directly probe the electron thermal pressure and momentum, respectively.  Given external knowledge of the peculiar velocity field (e.g., from galaxy redshift surveys), the kSZ signal is a measure of the electron density.  Both the tSZ and kSZ signals are redshift-independent, a unique property that is shared by few probes in observational astrophysics.  The tSZ signal allows the thermal pressure of ionized gas to be measured over a wider range of halo masses, redshifts, and halo-centric radii than any other probe.  The kSZ signal is essentially the only observational tool that can directly measure the ionized gas distribution with few assumptions over a wide range of halo masses, redshifts, and halo-centric radii.  Moreover, the small-scale properties of both signals are strong probes of feedback.  Qualitatively, AGN and supernova feedback tends to flatten the gas pressure and density profiles of halos, while decreasing the integrated pressure and density content within the virial radius of lower-mass objects (due to the ejection of gas from these shallower potential wells, in comparison to the deeper wells of larger halos).  However, the details of these predictions, including their mass, redshift, and radial dependence, depend strongly on the exact feedback model and implementation.  In most simulations, the relevant physics is ``sub-grid’’; thus, it is essential to directly measure these quantities in order to correctly calibrate the energetics of these processes.

Here, we demonstrate explicitly how CMB-S4 will achieve this goal.  We focus on extracting the tSZ and kSZ signals of halos of various masses and redshifts via cross-correlations of the deep and wide CMB-S4 LAT survey with galaxy catalogs from the DESI, Baryon Oscillation Spectroscopic Survey (BOSS), and Sloan Digital Sky Survey (SDSS) spectroscopic surveys (LSST and other photometric surveys will also provide useful catalogs for such measurements) and cluster samples from the CMB-S4 tSZ-selected catalog (the {\it eROSITA\/} X-ray-selected catalog will also be useful in this context).  We closely follow the methodology of Ref.~\cite{Battaglia2017} to forecast constraints on the stacked electron-density and pressure profiles of these halo samples.  We consider four example sets of halos, with numbers drawn from DESI \cite{DESI:2016}, BOSS/SDSS \cite{Dawson:2012va}, or the CMB-S4 tSZ cluster catalog:
\begin{itemize}
    \item low-redshift BOSS/SDSS luminous red galaxies (LRGs), $z=0.2$, $M_{200c} = 10^{13} \, {\rm M}_{\odot}$, $N = 2.5 \times 10^5$;
    \item high-redshift DESI LRGs, $z=1$, $M_{200c} = 10^{13} \, {\rm M}_{\odot}$, $N = 2.5 \times 10^5$;
    \item low-redshift CMB-S4 clusters, $z=0.2$, $M_{200c} = 10^{14} \, {\rm M}_{\odot}$, $N = 1.5 \times 10^3$;
    \item high-redshift CMB-S4 clusters, $z=1$, $M_{200c} = 10^{14} \, {\rm M}_{\odot}$, $N = 1.1 \times 10^3$.
\end{itemize}
In all cases, we choose narrow mass and redshift bins: $\Delta \log_{10}(M_{200c}) = 0.1$ and $\Delta z = 0.1$.  The LRG samples from these optical surveys cover the entire window $0.2 < z < 1$, so stacked measurements similar to those shown here can be performed over this entire redshift range.  Samples of other galaxy types (e.g., emission-line galaxies) will extend to even higher redshift, enabling similar analyses.  The CMB-S4 cluster counts assumed here are drawn from Fig.~\ref{fig:s4_galaxy_clusters}, and will enable galaxy cluster studies from $0.1 < z < 2$.  Halo masses inferred from CMB-S4 CMB lensing data (and LSST weak lensing at low-$z$) will be essential in this work, relating the observed baryonic profiles to the underlying matter-density field.

We apply the component-separated CMB temperature (for kSZ) and Compton-$y$ (for tSZ) noise curves described in the forecasting methodology appendix to a stacked aperture photometry estimator to obtain error bars on the projected gas density (kSZ) and projected thermal gas pressure (tSZ) profiles of the halos.  Note that we use the ``velocity-reconstruction'' kSZ estimator here, which uses spectroscopic redshift information to weight the CMB map in a manner that is highly robust to foreground contamination \cite{Ho:2009iw,Schaan:2015uaa}.  Our kSZ forecasts assume that the noise in the electron-density profiles is dominated by the noise (instrumental and foreground) from the CMB data rather than noise in the velocity reconstruction; an approximate estimate indicates $\approx 10$\% residual noise in the velocity reconstruction using DESI.  Future simulation work will be necessary to precisely calibrate this effect.

The electron-pressure and density-profile error bars calculated using the methodology described above are contrasted with theoretical predictions extracted from six cosmological hydrodynamics simulations (for more details on the simulations, see Appendix~\ref{chap:forecasting}): BAHAMAS (fiducial,``high-AGN,'' and ``low-AGN'' models) \cite{McCarthy2017}, simulations of Refs.~\cite{Battaglia2010,Battaglia2012}, EAGLE \cite{Schaye2015}, and IllustrisTNG-300 \cite{Weinberger2017,Barnes2018,Springel2018}.
We rescale all electron-density and pressure profiles to the same background cosmological parameter values ($\Omega_{\rm b} = 0.04898$, $\Omega_{\rm m} = 0.3111$, $h=0.6766$ \cite{Aghanim:2018eyx}), so that only the astrophysical model differences are reflected in the forecasts.  These simulations all differ in significant ways (even at the level of using grid- or particle-based codes), particularly in their implementations of sub-grid models for AGN and stellar feedback.  As a fiducial model, we simply take the mean of the predictions from all six simulations for each observable.  We also perform a smooth extrapolation of the simulation predictions to large radii (beyond $2 r_{200c}$, within which the one-halo term dominates), which is necessary for the line-of-sight projection calculations.  However, the extrapolation is not significant, since we assume that the virial shock leads to a steep decline in the profiles at $2.5 r_{200c}$.

\begin{figure}[!ht]
    \centering
    \textbf{Low-redshift BOSS/SDSS LRGs} \\[0.12 in]
    \includegraphics[width=0.48\textwidth]{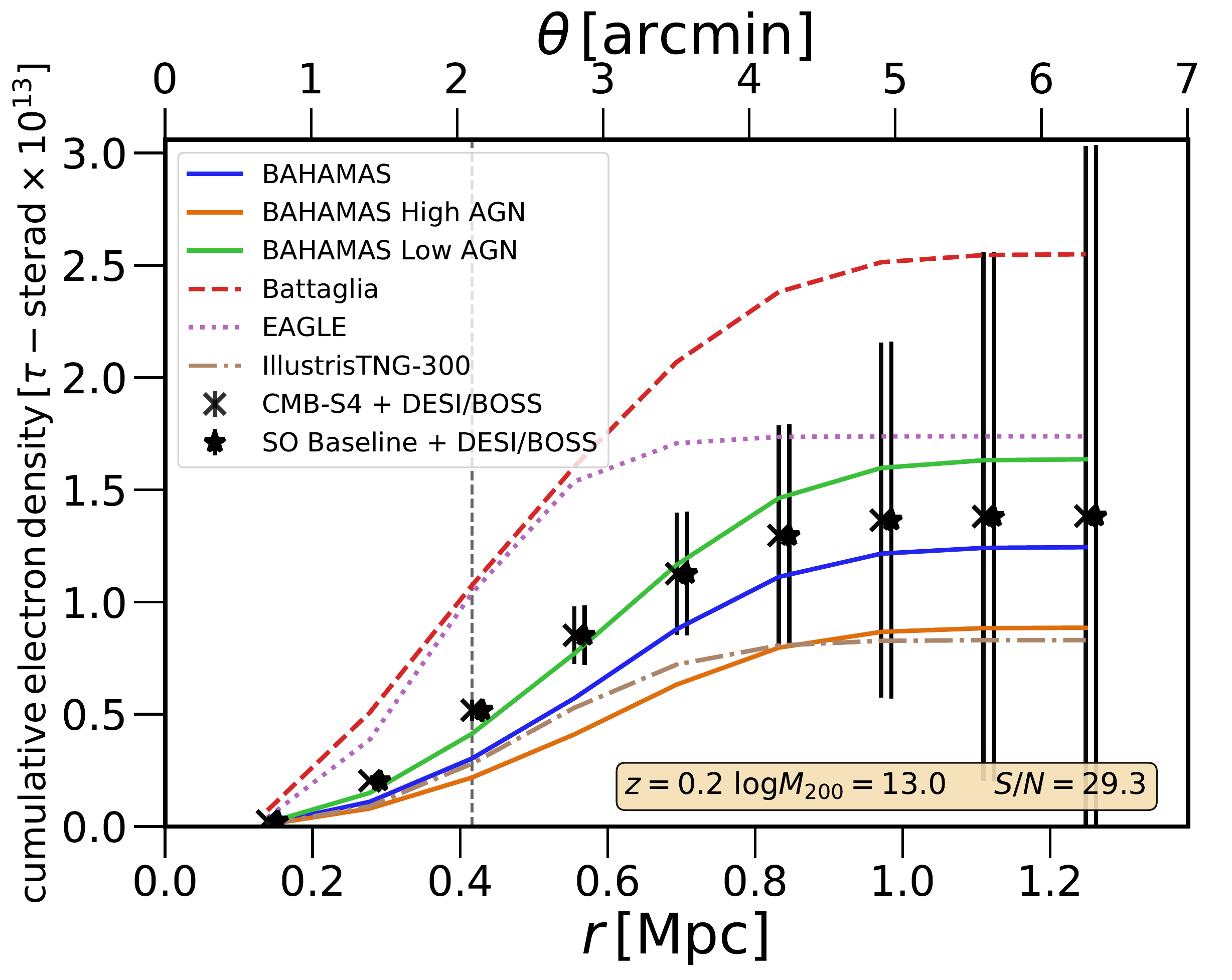}
    \includegraphics[width=0.48\textwidth]{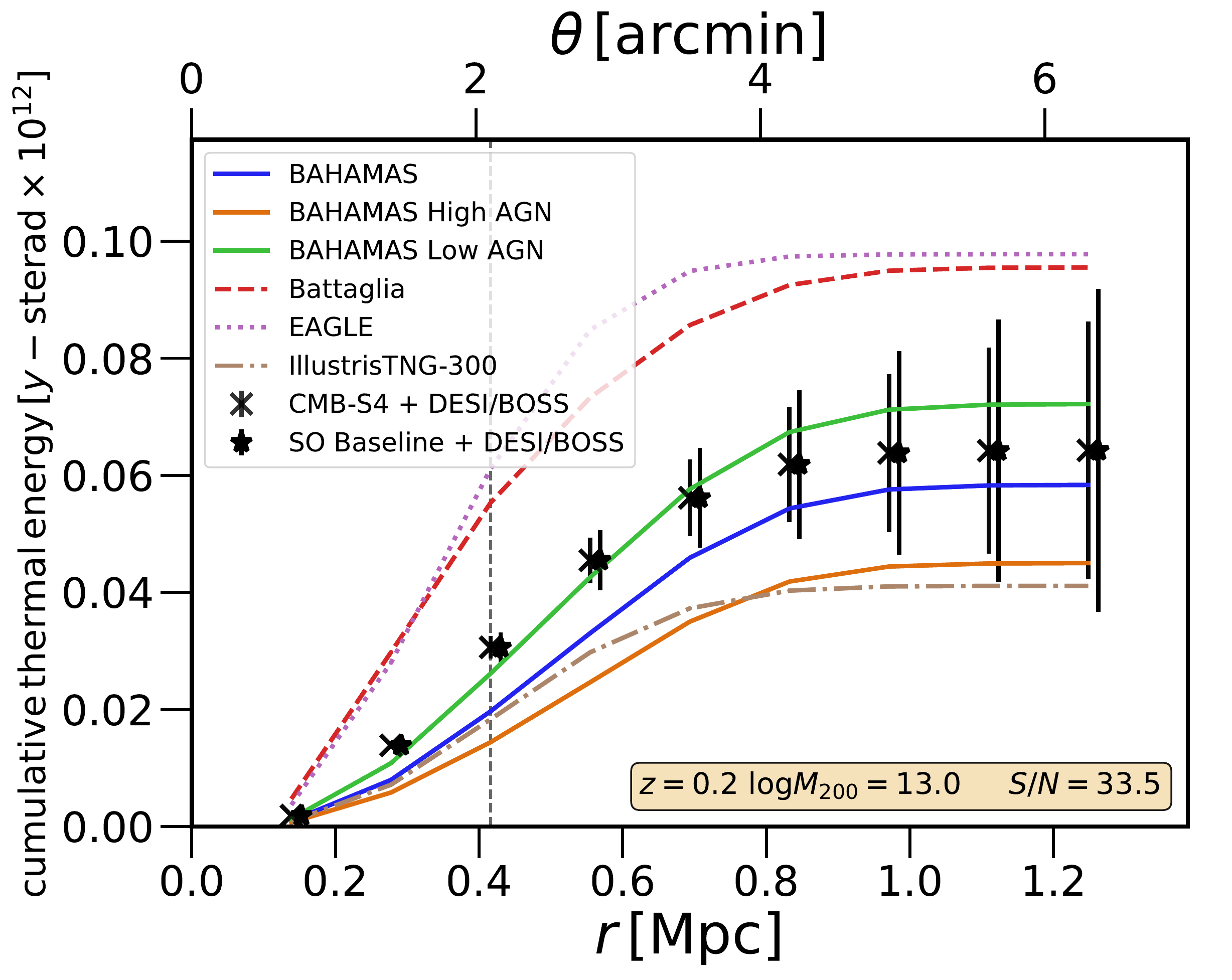} \\[0.2 in]
    \textbf{High-redshift DESI LRGs} \\[0.12 in]
    \includegraphics[width=0.48\textwidth]{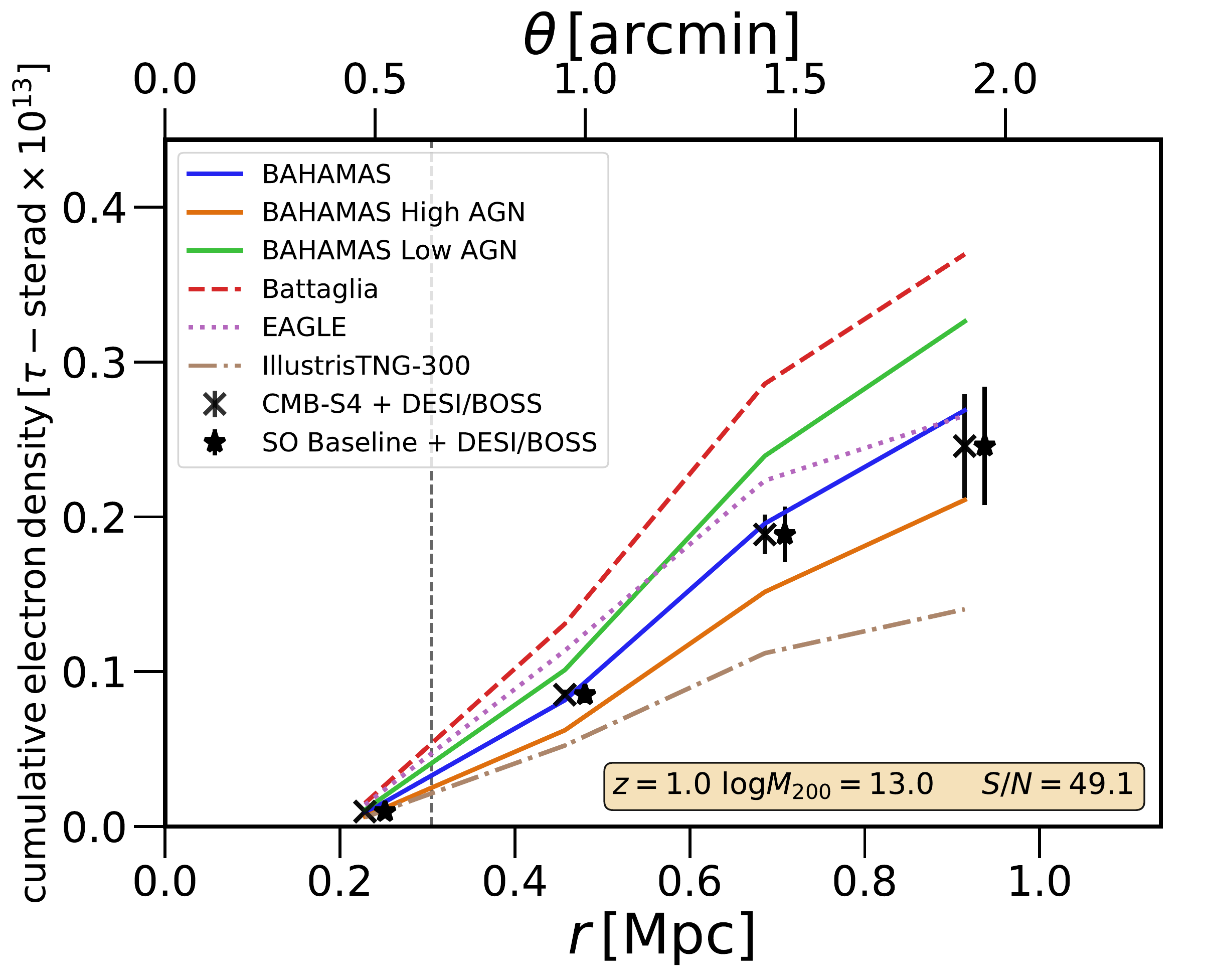}
    \includegraphics[width=0.48\textwidth]{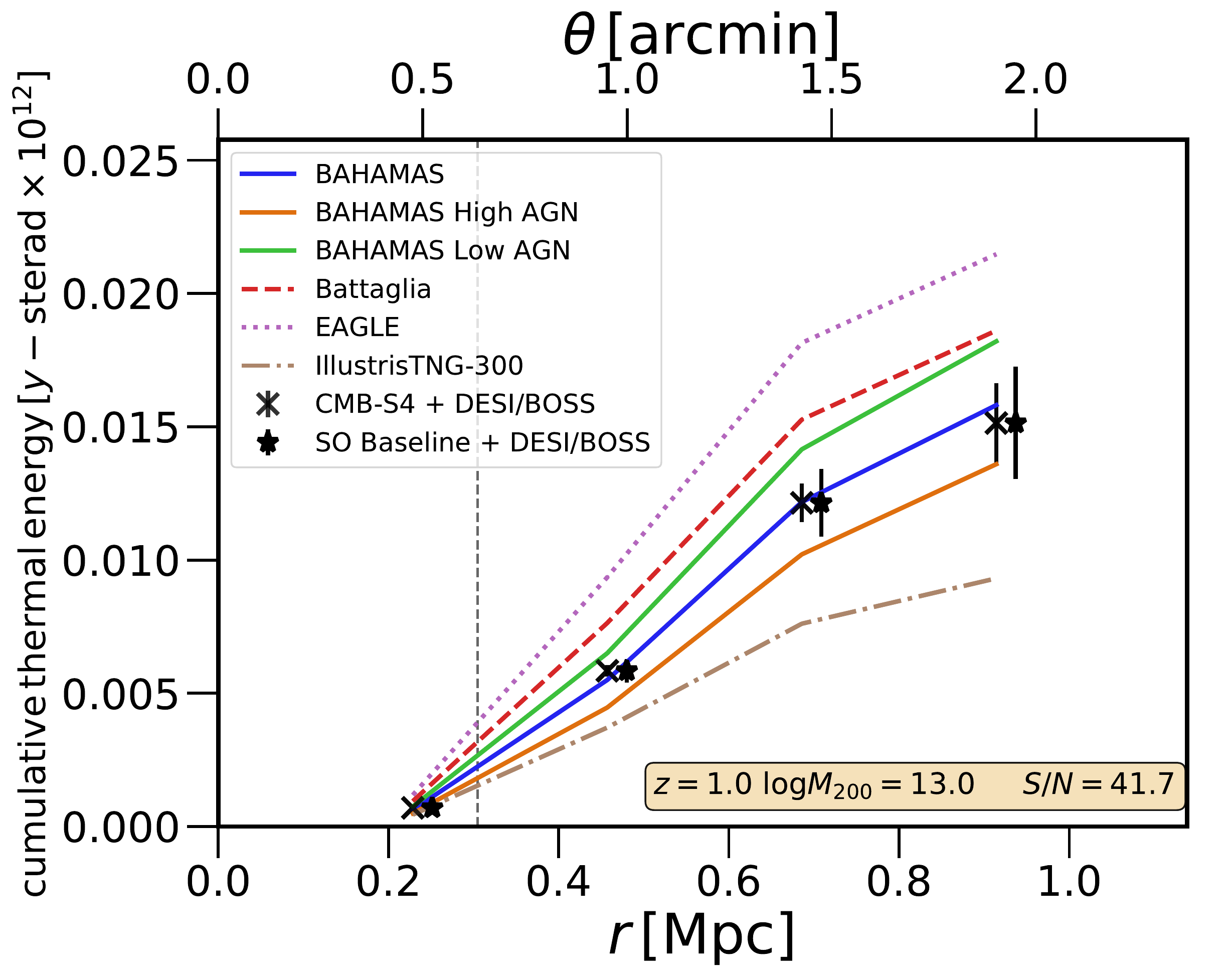}
    \caption{CMB-S4 constraints on the cumulative electron-density (left) and thermal-energy (right) profiles will distinguish between  feedback models.
    Top row: Stacking $N = 2.5 \times 10^5$ BOSS and SDSS LRG halos of average mass $M_{200c} = 10^{13} \, {\rm M}_{\odot}$ at $z=0.2$.
The left panel is extracted from the kSZ signal and the right panel from the tSZ signal.
The lines come from density and pressure profiles around such halos measured in six cosmological hydrodynamics simulations: BAHAMAS \cite{McCarthy2017} (fiducial blue, ``high-AGN'' orange, ``low-AGN'' green); Battaglia et al.~\cite{Battaglia2010,Battaglia2012} (red); EAGLE \cite{Schaye2015} (magenta); and IllustrisTNG-300 \cite{Weinberger2017,Barnes2018,Springel2018} (brown).
The data points average the predictions, and show error bars determined via stacked aperture photometry applied to component-separated maps from CMB-S4 LAT and \emph{Planck\/} data (or SO and \emph{Planck} data).  The dashed vertical lines denotes $r_{200c}$.  The insets give the CMB-S4 forecast signal-to-noise ratio.  The error bars are highly correlated due to the photometry method, but the models can nonetheless be distinguished at high significance.
Bottom row: The same, inferred by stacking $N = 2.5 \times 10^5$ DESI LRG halos of average mass $M_{200c} = 10^{13} \, {\rm M}_{\odot}$ at $z=1$.   We emphasize that ionized gas properties in the low-mass, high-redshift regime shown in the bottom row cannot be easily measured with any other astrophysical probe.
}
    \label{fig:s4_prof_m13_z02}
\end{figure}

\begin{figure}
    \centering
     \textbf{Low-redshift CMB-S4 clusters} \\[0.15 in]
    \includegraphics[width=0.48\textwidth]{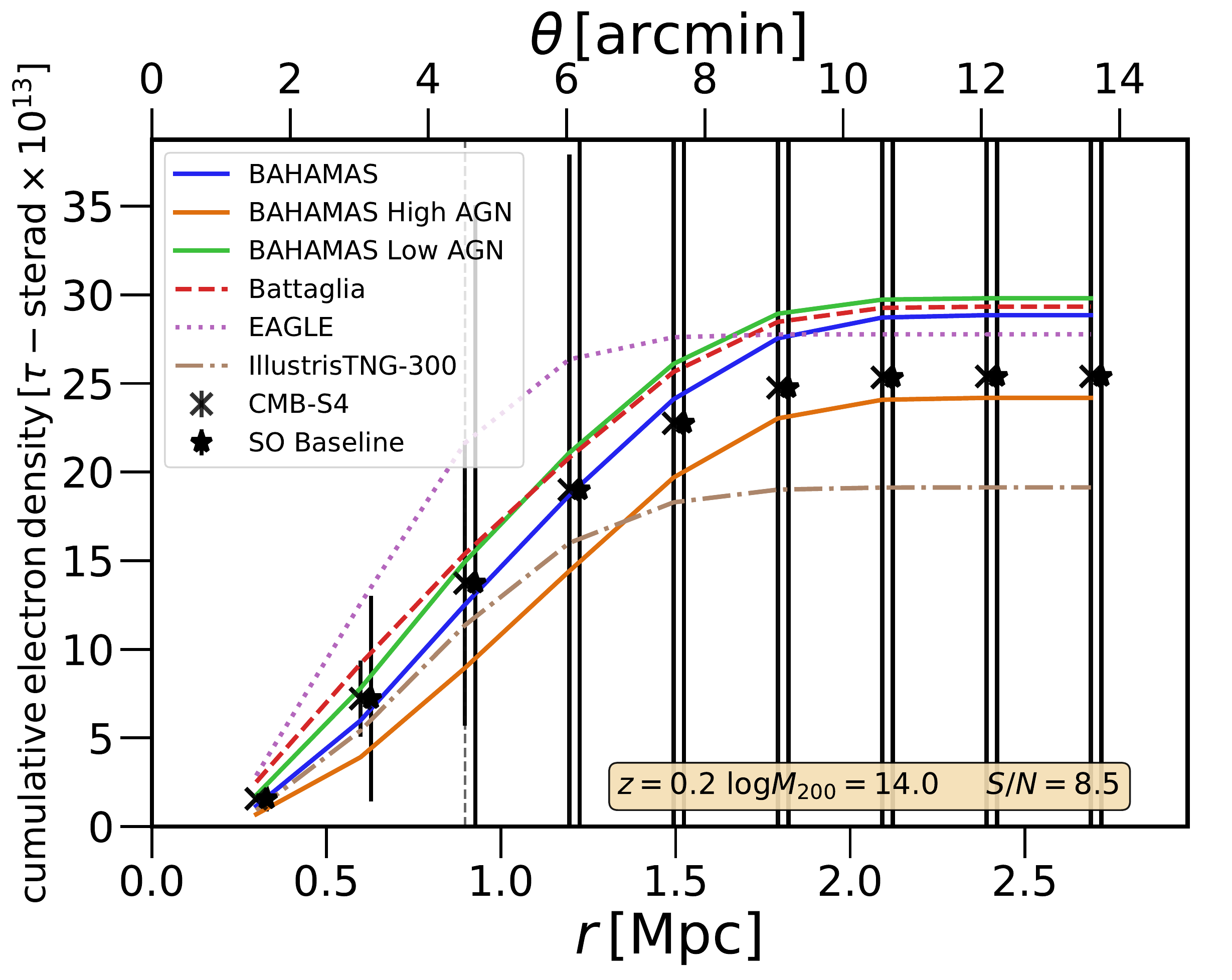}
    \includegraphics[width=0.48\textwidth]{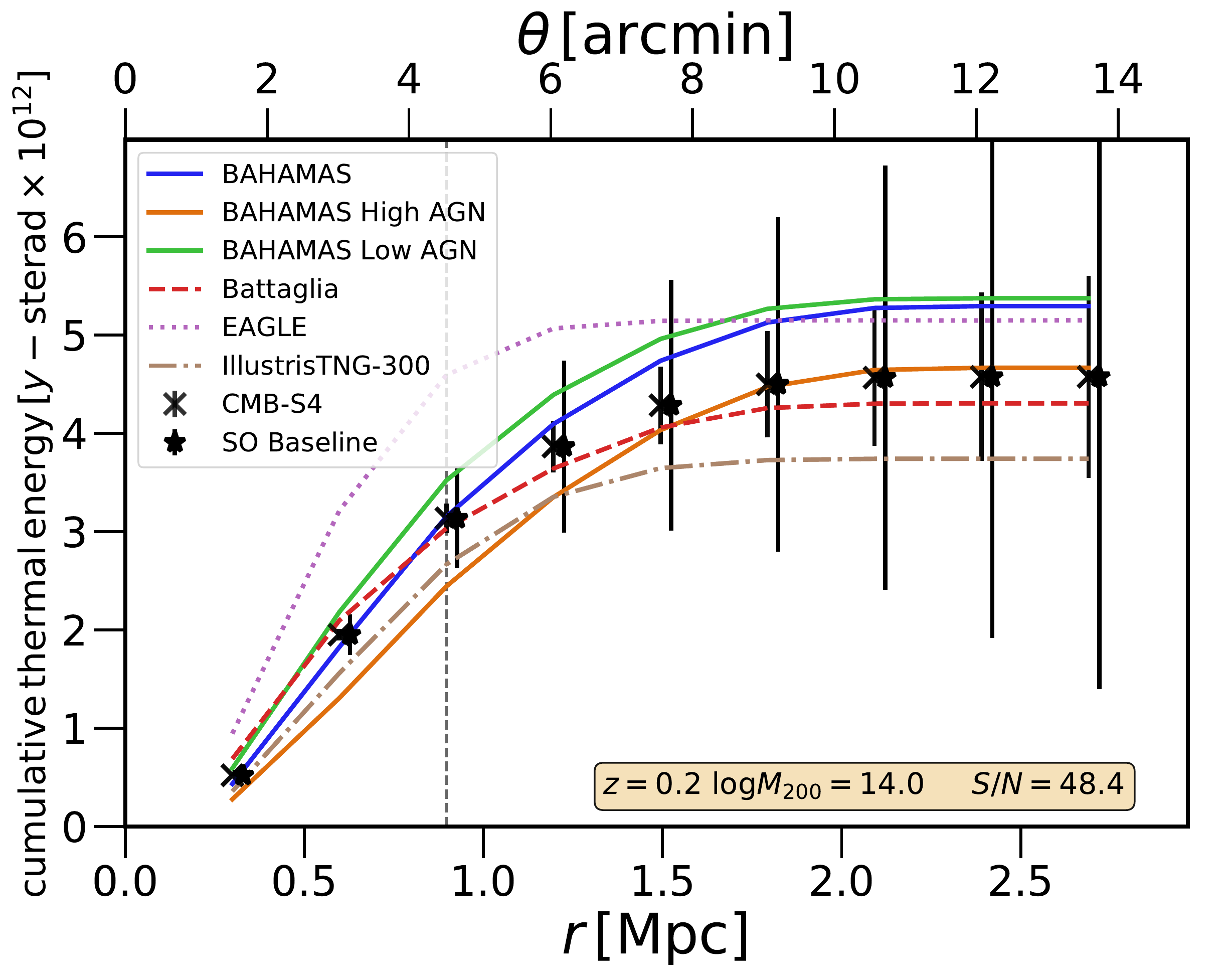} \\[0.3 in]
    \textbf{High-redshift CMB-S4 clusters} \\[0.15 in]
    \includegraphics[width=0.48\textwidth]{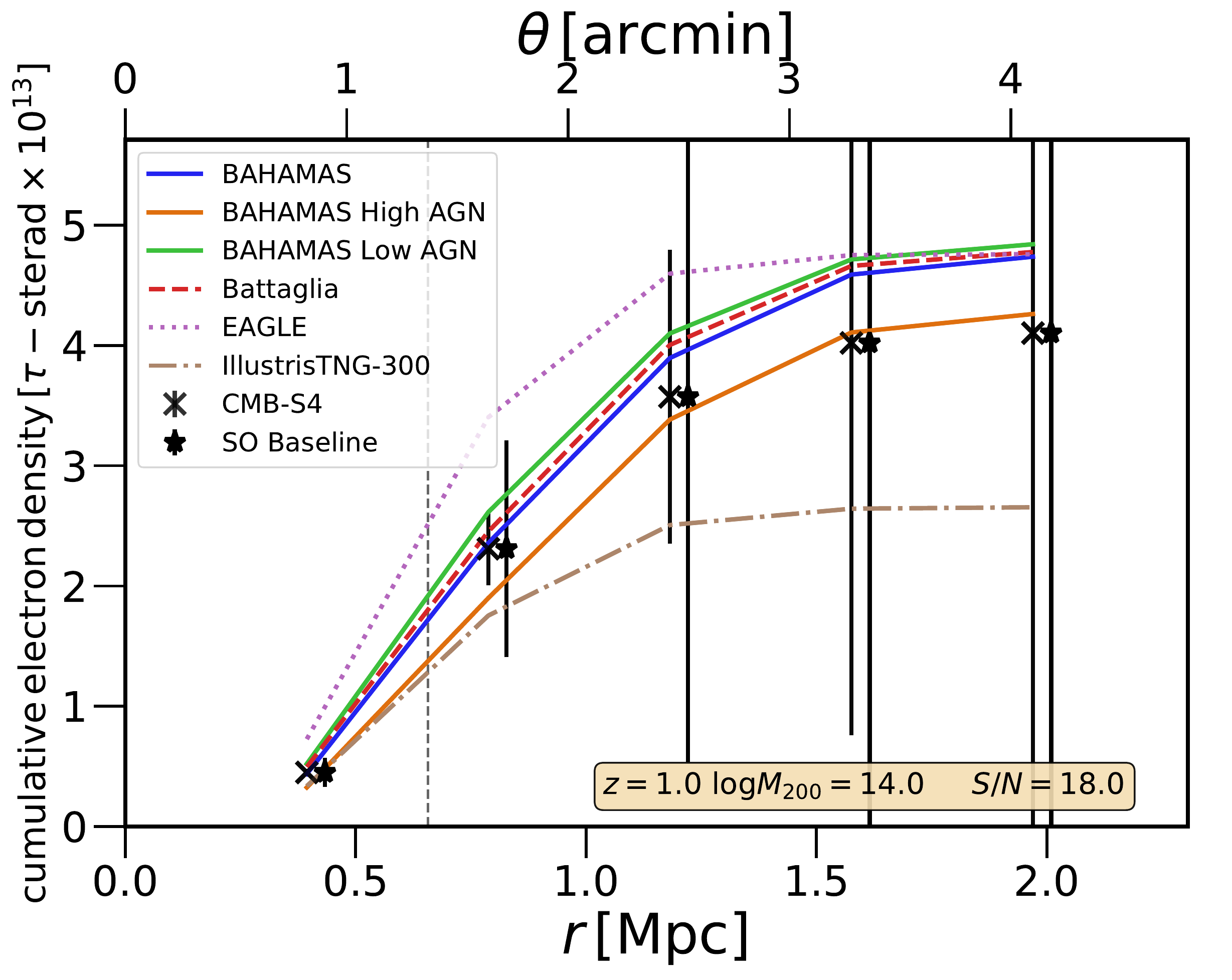}
    \includegraphics[width=0.48\textwidth]{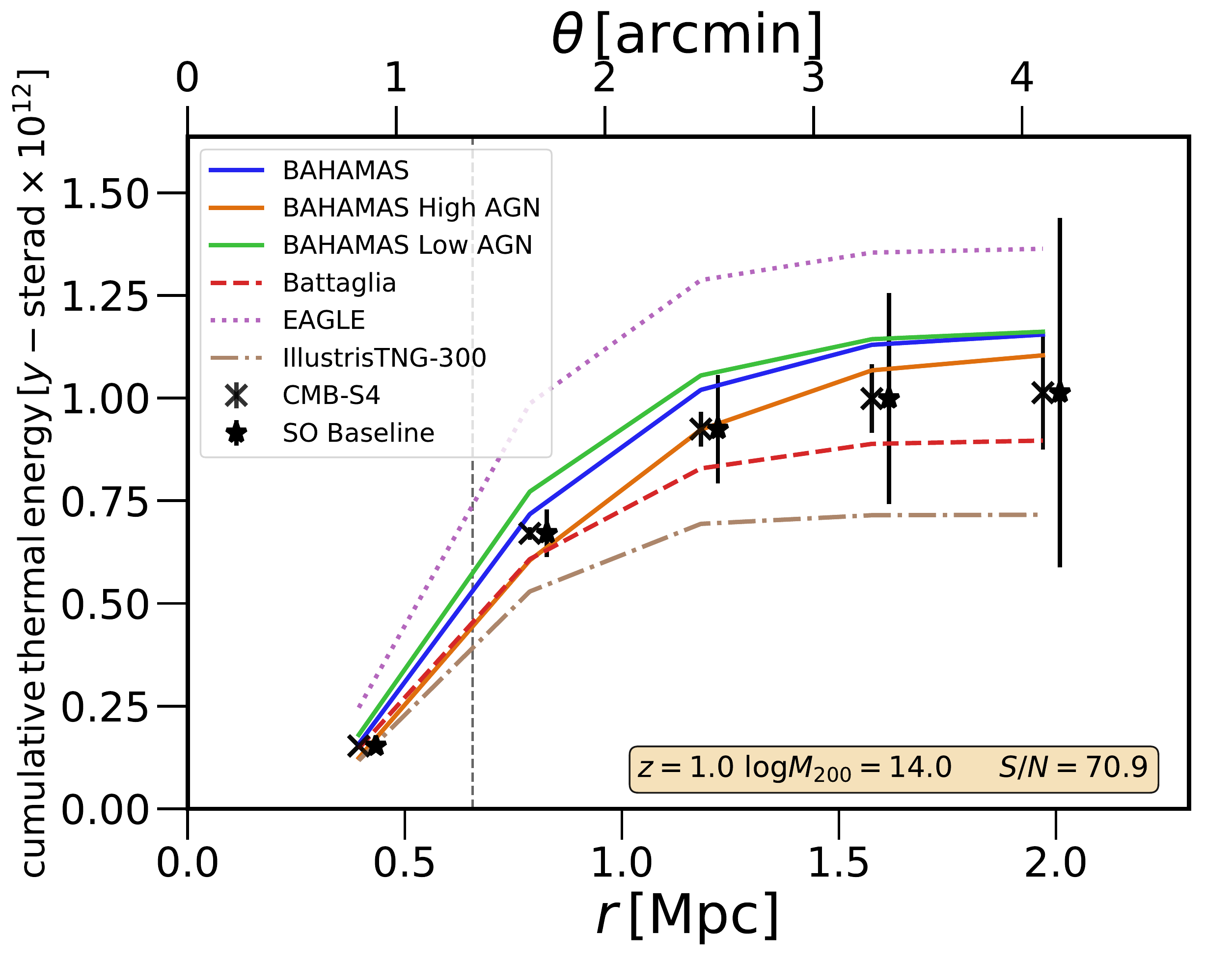}
    \caption{Top: CMB-S4 constraints on the cumulative electron-density (left) and cumulative thermal-energy (right) profiles inferred by stacking $N = 1.5 \times 10^3$ CMB-S4 clusters of average mass $M_{200c} = 10^{14} \, {\rm M}_{\odot}$ at $z=0.2$. Bottom: The same inferred by stacking $N = 1.1 \times 10^3$ CMB-S4 clusters of average mass $M_{200c} = 10^{14} \, {\rm M}_{\odot}$ at $z=1$.  The panels, curves, and data points with error bars are analogous to those shown in Fig.~\ref{fig:s4_prof_m13_z02}.  As in the previous figure, the signals are detected at high significance, and the galaxy-formation models can be distinguished.  X-ray constraints at small radii will be complementary to those shown here.}
    \label{fig:s4_prof_m14_z1}
\end{figure}

The results of this analysis are shown in Figs.~\ref{fig:s4_prof_m13_z02} and \ref{fig:s4_prof_m14_z1}.  Each figure corresponds to a different choice of halo mass and redshift, with the left panel showing the cumulative electron-density profile and the right panel showing the cumulative electron thermal-pressure profile (i.e., the cumulative thermal energy in electrons).  Here, ``cumulative'' means ``integrated within an aperture,'' where the angular size of the aperture in arcminutes is given on the top axis of each figure.  The profiles asymptote to constant values as the full electron content or thermal energy content of the halo is enclosed.  The electron-density profiles are shown in terms of $\tau$, the Thomson scattering optical depth, while the electron pressure profiles are shown in terms of Compton-$y$, the line-of-sight integral of the electron pressure.  The figures also show the forecast signal-to-noise ratio in each case, which range from $\approx 10$ to greater than $100$.  All of the profile forecasts and error bars are convolved to an effective beam of FWHM = 1.4 arcmin (the CMB-S4 LAT resolution at 145 GHz).  The figures also show error bars for the baseline configuration of the Simons Observatory (SO) \cite{2018arXiv180807445T}, including the effect of the smaller number of clusters for the $M_{200c} = 10^{14} \, {\rm M}_{\odot}$ halos (6--7 times fewer clusters for these cases).

It is evident from the figures that the galaxy-formation models considered here can be distinguished at high significance with CMB-S4.  Particularly important in this regard is the high angular resolution of the CMB-S4 LAT, which permits access to the inner regions of the halos at low-$z$ and is roughly matched to the virial scale at high-$z$.  No other astrophysical observables are sensitive to these thermodynamic properties of the gas over such large ranges in mass, redshift, and halo-centric radius.  X-ray observations are limited at large cluster-centric radii, at low halo masses, and at high redshifts.  However, measurements with {\it eROSITA\/} will be strongly complementary to those from CMB-S4; the combination of the two facilities will enable gas pressure and density profiles to be measured from the core to the outskirts of galaxy groups and clusters.  Absorption line studies (e.g., with the Cosmic Origins Spectrograph on the {\it Hubble Space Telescope\/}) are limited to comparatively small samples of halos at low mass and low redshifts; however, they are potentially complementary to CMB-S4 measurements on stacked samples of galaxies at lower masses than considered in Figs.~\ref{fig:s4_prof_m13_z02} and \ref{fig:s4_prof_m14_z1}, where galaxy-formation models yield even larger differences in their predictions than seen in the figures.

\begin{figure}[!ht]
    \centering
    \includegraphics[width=0.75\textwidth]{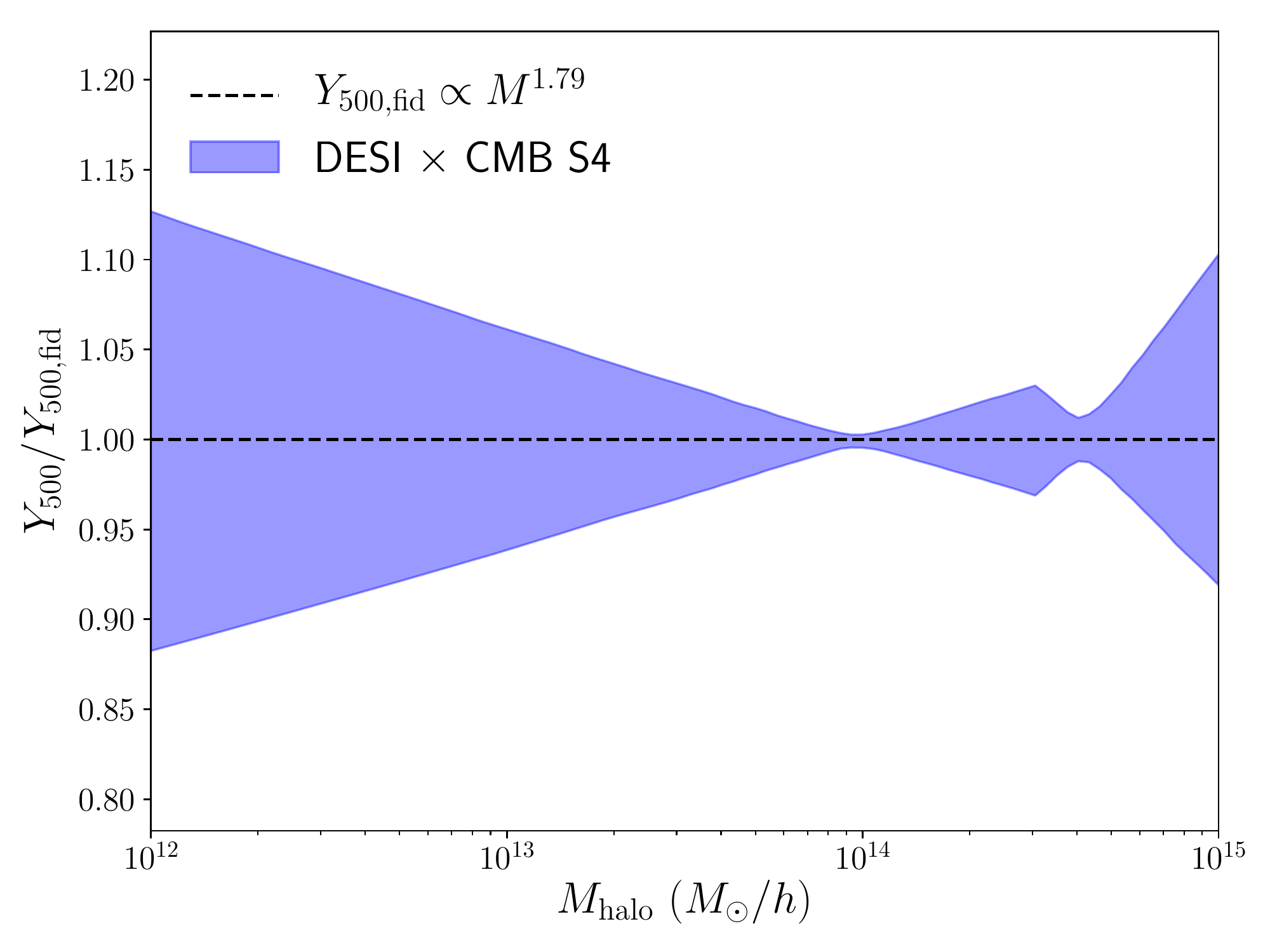}
    \caption{Fractional constraints on the integrated pressure versus halo mass relation from a joint analysis of CMB-S4 and DESI data.}
    \label{fig:s4_YM}
\end{figure}

The profile-based constraints shown here are only a small fraction of the total information content in the CMB-S4 dataset relevant to galaxy formation.  CMB-S4 will yield tight constraints on the integrated Compton-$y$ and $\tau$ signals as a function of halo mass and redshift (``scaling relations'').  The slopes of these relations are highly sensitive to feedback models \cite{Planck2013LBG,LeBrun2015,Hill2018SDSS}.  An optimal analysis will simultaneously incorporate constraints from the spatial distribution (profiles) and mass- and redshift-dependences of these signals.  As an illustration of the constraining power of CMB-S4 in this regard, Fig.~\ref{fig:s4_YM} shows forecasted constraints on the integrated Compton-$y$ signal as a function of halo mass, derived from the combination of CMB-S4 data with the DESI bright galaxy sample.  The plot shows the fractional constraint on the deviation of the $Y$-$M$ relation from a fiducial $M^{1.79}$ power-law model.  The constraints span the range of halo masses from Milky-Way-sized galaxies to massive clusters, with sub-percent constraints near a pivot halo mass of $10^{14}\,{\rm M}_{\odot}$.  These improve over current consraints by more than an order of magnitude; at Milky Way mass scales, no constraints currently exist (e.g., \cite{Hill2018SDSS}).

Further information can be extracted by cross-correlating the CMB-S4 kSZ and tSZ data with quasars, different galaxy types (e.g., red or blue), weak-lensing maps, or filament catalogs.  In the kSZ case, some of these cross-correlations will rely on a different estimator than that used here, in which the multifrequency CMB-S4 data are employed to extract the kSZ effect on small scales via component separation, filtering, and squaring of the map; forecasts indicate $S/N \approx$ few hundred for CMB-S4 in combination with LSST or {\it SPHEREx\/} using this method \cite{Hill:2016dta,Ferraro:2016ymw}.  In addition, the large-scale (two-halo) regime of the tSZ signal contains additional unique information about the evolution of the bias-weighted average thermal electron pressure of the Universe \cite{Vikram2017}.  Finally, CMB-S4 will make the first detection of the polarized SZ effect \cite{Hall2014,Deutsch2018}, a significant milestone in the eventual use of this signal as an astrophysical and cosmological probe.  Overall, the forecasts presented here are only a first step toward exploring the rich astrophysics of galaxy formation accessible with CMB-S4.

\subsubsection{Patchy reionization}

The patchiness of reionization leaves its imprint on the CMB through the kSZ effect, which refers to blackbody temperature fluctuations induced by a combination of coherent bulk flows on larges scales and variations in the electron density on small scales.  These ``patchy kSZ'' anisotropies have only recently been used to place constraints on the duration of reionization \cite{Reichardt2012,Zahn2012,George2015,Adam:2016hgk}. While incremental gains in precision are expected from experiments that are underway or begin built (e.g., \cite{Calabrese:2014gwa,2018arXiv180807445T}), CMB-S4 will provide definitive reionization constraints from the high-$\ell$ CMB, due to its much higher sensitivity. 
The patchy nature of reionization also generates polarization fluctuations (e.g., \cite{Dvorkin:2008tf,2009PhRvD..79j7302D}). Recent work shows this signal may be detectable by CMB-S4 through the $B$-mode power at $\ell \approx 50$--$500$ or by explicit reconstruction of the optical depth at the map level \cite{Roy2018}. Additionally, it is possible to correlate these patchy polarization anisotropies induced by patchy reionization with other tracers of large scale scale structure such as the cosmic infrared background and CMB lensing \cite{2018PhRvD..97l3523F,2018arXiv180801592F}.

Here we highlight the power of CMB-S4 through the patchy kSZ effect with forecasts based on simulations for which the predicted shape and amplitude of the kSZ power spectrum respond naturally to the variation of unknown physical parameters over their possible values. The simulation techniques are described in Refs.~\cite{AlvarezAbel2012} and \cite{2016ApJ...824..118A}. The spectra are simulated at discrete ``step sizes''  as two parameters of the underlying model are varied away from their fiducial values: the ionization efficiency (or number of atoms ionized per atom in halos above the minimum mass), $\zeta$; and the mean free path of ionizing photons, $\lambda_{\rm mfp}$. The reionization history for each of the simulations is used to determine the Thomson scattering optical depth, $\tau_{\rm es}$, and the duration of reionization, $\Delta{z}\equiv z_{75}-z_{25}$, the redshift interval over which the volume filling factor of ionized regions evolved from 25 to 75 per cent.  We then calculated derivatives of the power spectrum with respect to these parameters, $\tau_{\rm es}$ and $\Delta{z}$, by finite differencing.  We impose a Gaussian prior on the late-time ``homogeneous'' kSZ contribution. This term is already known at the roughly 10\% level, accounting for astrophysical and cosmological uncertainties \cite{Shaw2012,Park2018}, and will be known better than this in the CMB-S4 era (e.g., by using the measurements described in the previous subsection).  Additional uncertainty due to marginalization over non-kSZ foregrounds is incorporated with the ``Deproj-0'' (standard ILC) CMB-S4 LAT + \emph{Planck} noise curves for the reference CMB-S4 configuration.

\begin{figure}[t]
    \centering
    \includegraphics[width=0.75\textwidth]{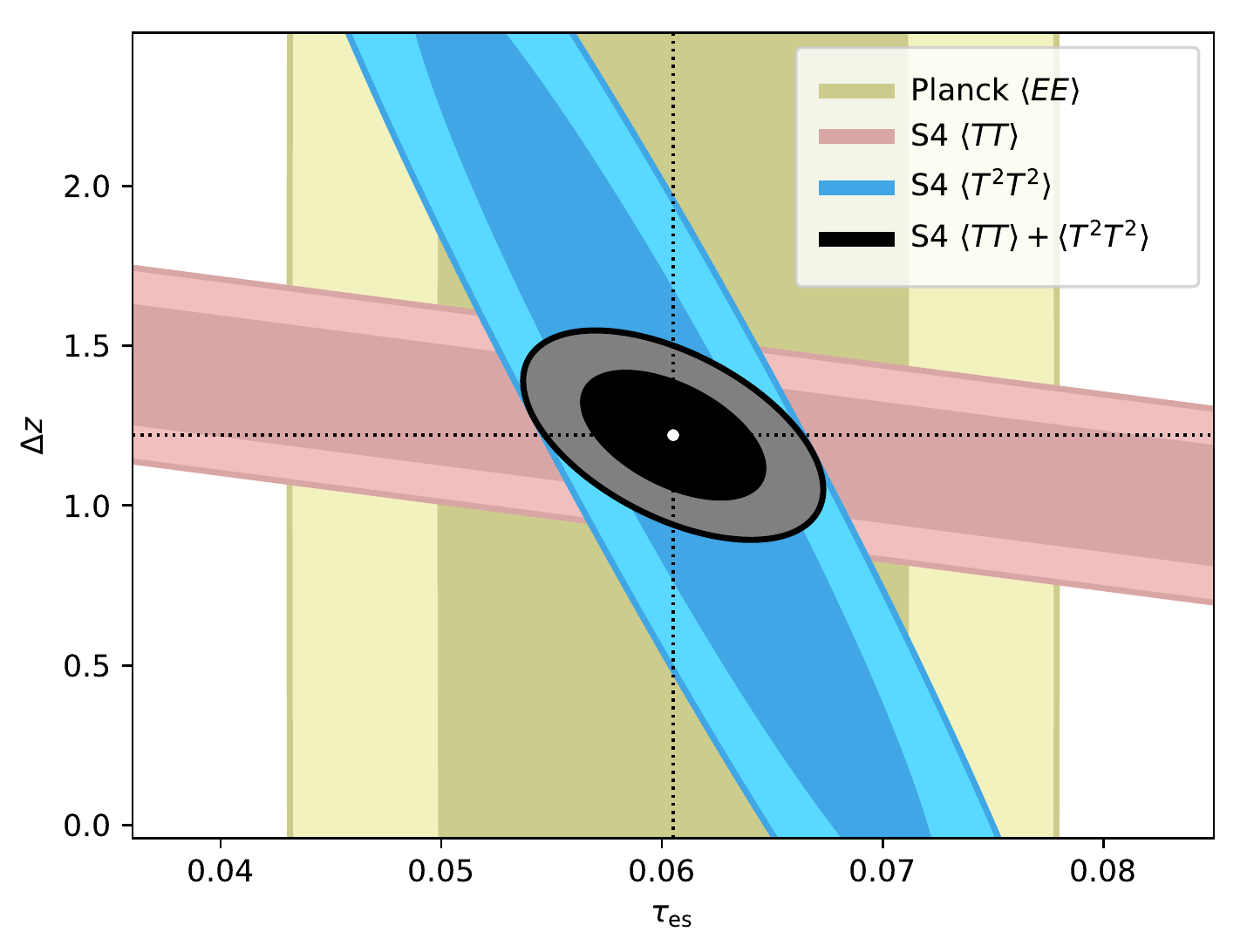}
    \caption{CMB-S4 constraints on the optical depth and duration of reionization in a joint analysis using the kSZ power spectrum and four-point function.}
    \label{fig:s4_ksz_eor}
\end{figure}

The results are shown in the right-hand panel of Fig.~\ref{fig:s4_ksz_eor}, where we have also included the kSZ four-point function, which provides additional information to separate early- and late-time kSZ contributions \cite{SmithFerraro2017}, by using the same excursion set simulated kSZ maps described above. The power spectrum is sensitive to the duration of reionization, but this is highly degenerate with the optical depth. However, the prior knowledge of the optical depth from \emph{Planck} large-scale $EE$ power spectrum measurements, which is included in our analysis via a prior $\sigma(\tau) = 7\times{10}^{-3}$, breaks this degeneracy.  Thus we find $\sigma(\Delta{z}) \approx 0.14$ from the kSZ power spectrum. The four-point function, when combined with the \emph{Planck} $EE$ constraint on the optical depth, performs worse on the duration than the power spectrum, as seen in the figure. However, when combining the power spectrum and four-point constraints, we find that the $\tau$ measurement from CMB-S4 alone {\em improves on the current \emph{Planck} $EE$-based $\tau$ constraint by a factor of 2--3}, with $\sigma(\tau)\approx 2,5\times{10}^{-3}$. Note that this estimate marginalizes over an unknown contribution from the low-redshift kSZ signal (albeit with a roughly 10\% prior) and also assumes we have no knowledge of other foregrounds. It is thus possible that tighter constraints will result from improved measurements of low-redshift kSZ signals, as well as improved measurements of high-$\ell$ foregrounds.

Finally, we note that alternative scenarios, such as those in which rare quasars or very high-redshift sources play a significant role, are described by models with different parameters than we have assumed here. Thus, our forecasts by definition are uncertain and depend on the assumed model and even the fiducial parameters within that model.  The high precision of the model parameter constraints we have forecasted in the standard UV-dominated scenario implies that we will be able to rule out or confirm, with high significance, other more exotic reionization scenarios such as those that include early X-ray binaries, population~III sources, or rare quasars. This will especially be the case when CMB-S4 is combined with external data sets such as 21-cm and Lyman-$\alpha$ emitter surveys, together with independent CMB-based constraints on the optical depth from the large-scale $EE$ power spectrum. Such discoveries are an inevitable byproduct of probing the high-precision, high-redshift frontier with CMB-S4.

\label{subsec:LC}

\subsection{Extragalactic Legacy Catalogs}

\subsubsection{Galaxy clusters}

\begin{figure}
\begin{center}
\includegraphics[width=0.5\textwidth]{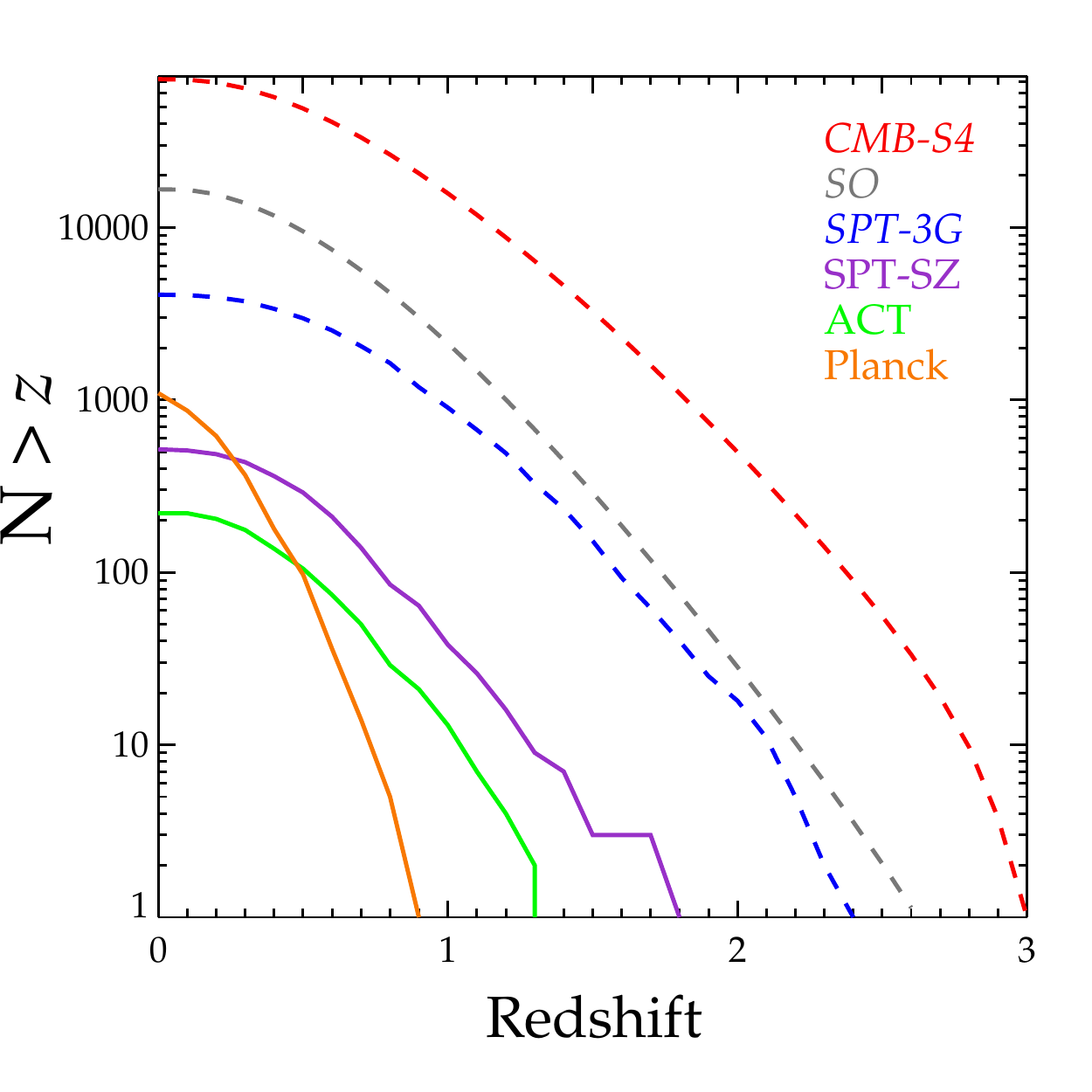}
\caption{Cumulative  number of clusters above a fixed redshift for published (solid \cite{Ade:2015xxvii,Bleem:2014iim,Hilton:2017}), and upcoming (dashed \cite{Benson:2014qhw,2018arXiv180807445T}) CMB cluster samples. CMB-S4 will discover an order of magnitude more of the highest-redshift ($z>1.5$) clusters than previous surveys. }
\label{fig:s4_galaxy_clusters}
\end{center}
\end{figure}

Clusters of galaxies, the largest gravitationally bound systems in the Universe, are powerful probes of both cosmology and astrophysics. 
The deep and wide millimeter-wave CMB-S4 sky survey will enable the identification of over 70,000 such systems at high significance ($>5 \sigma$) via measurements of the thermal Sunyaev-Zeldovich (tSZ) effect \cite{Sunyaev:1972eq}. 
Cluster detection and characterization via the tSZ effect is highly complementary to techniques at other wavelengths (e.g., \cite{Carlstrom:2002,Weinberg:2012es}) because tSZ observables provide both low-scatter mass proxies and a detection method that is independent of cluster distance (since it is a spectral distortion of the CMB).  As shown in Fig.~\ref{fig:s4_galaxy_clusters},
CMB-S4 will discover \textit{an order of magnitude more high redshift systems} ($z>1.5$) than ongoing and upcoming CMB experiments \cite{Niemack:2010wz,Benson:2014qhw,2018arXiv180807445T}. 
CMB-S4's deep multi-band observations will enable separation of the tSZ signal from contaminating radio and infrared emission from cluster members (Sect.~\ref{subsubsec:ymap}), which will be particularly important for low-mass, high-redshift clusters. 

This large and well-characterized cluster sample, in conjunction with data from LSST,
will significantly improve constraints on cosmological models (Sect.~\ref{subsec:DE}, \cite{Madhavacheril:2017}) and, when further combined with data from the next generation ESA X-ray mission \textit{Athena} (as well as other proposed missions, including NASA's \textit{Lynx} and \textit{ Origins Space Telescope}), will also offer an unprecedented opportunity to understand the assembly and evolution of the massive galaxies that reside in these systems, as well as the effects of astrophysical feedback on the intracluster medium (Sect.~\ref{subsec:GF}).

\subsubsection{Dusty star-forming galaxies }

The unique combination of resolution, depth, and area covered make CMB-S4 ideal for
constructing catalogs of extragalactic mm-wave sources.  
A remarkable and largely unanticipated result (but see, e.g.,
\cite{Negrello2007}) from the previous generation of CMB surveys was the discovery of large numbers of strongly lensed dusty star-forming galaxies (DSFGs, \cite{Vieira:2009ru}), the   
massive galaxies that make up the high-redshift ($z > 1$) component of the cosmic infrared background. 
Because these objects are selected by their dust emission, which is re-radiated UV emission from intense star formation, this population of objects is complementary to samples of galaxies found from their stellar light at optical and near-IR wavelengths (rest-frame UV at high redshift) with instruments such as \textit{Hubble} and {\it JWST}. 
The DSFG sample from the first generation South Pole Telescope (SPT) survey includes the most distant and massive 
halo known within the epoch of reionization ($z$$=$$6.9$; \cite{strandet17,marrone18}), and a galaxy cluster forming just 1.5\,Gyr after the big bang \cite{miller18}. 
Most of these sources are 
gravitationally lensed by intervening galaxies; this magnification
enables them to be studied in detail with ALMA \cite{weiss13,strandet16}, providing fast redshift determinations and increased effective spatial resolution.
These first massive galaxies mark the sites of the largest overdensities in the cosmic web \cite{marrone18} and trace the formation of the dusty, molecular interstellar medium from the metal-poor and chemically simple raw materials present in the first galaxies. 
With ALMA, the lensed systems are also useful as probes of the dark-matter distribution in the foreground lensing halos \cite[e.g.,][]{hezaveh16}.
In addition to ALMA, these sources will be excellent targets to followup with {\it JWST}, and the catalog will also have strong synergies with surveys such as those carried out with LSST, {\it WFIRST}, SKA, and {\it eROSITA}. 

\begin{figure}
\begin{center}
  \includegraphics[width=\textwidth]{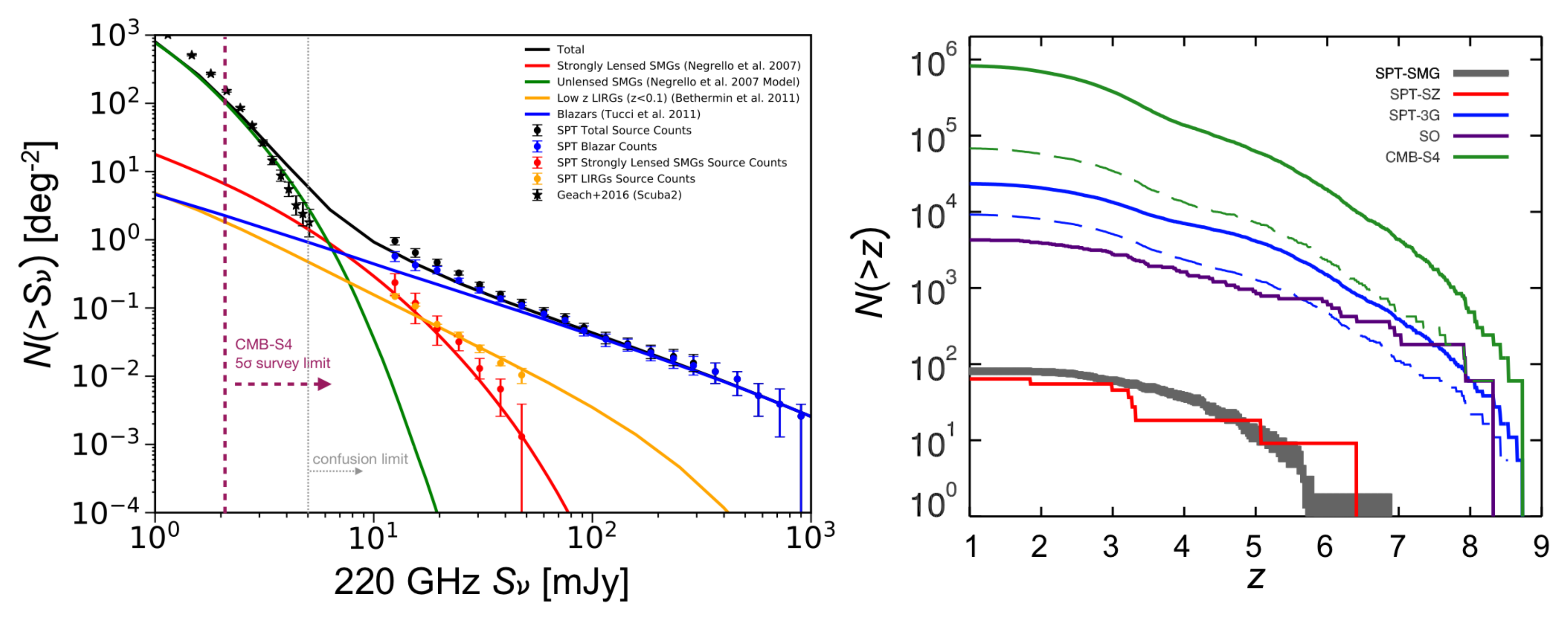}
  \end{center}
\caption{Left: Source density in the mm sky from SPT-SZ, SCUBA-2, and various
galaxy-evolution models. With the expected 220-GHz detection threshold shown by the 
short-dashed line, CMB-S4 will detect over 100,000 extragalactic sources, including more than 10,000 strongly-lensed galaxies and thousands of galaxies at $z\,{>}\,7$.  Right: Expected number of DSFGs in the CMB-S4 survey. The \textit{gray} line shows the measured distribution of DSFGs from the SPT-SZ survey \cite{strandet16}, while the \textit{red} line shown the predicted distribution from the phenomenological model of Ref.~\cite{bethermin2015}. Also shown are the predicted numbers DSFGs that will be detected by the SO (\textit{purple}), SPT-3G (\textit{blue}), and CMB-S4 (\textit{green}). The dashed lines show the number of sources detected above the confusion limit for a 6-m (\textit{green dashed line}) and 10-m (\textit{blue dashed line}) telescope.
\label{fig:ngts}}
 \end{figure}

The sample of DSFGs from the CMB-S4 survey is expected to contain
over 100,000 sources, of which around 10,000 should be strongly lensed (see Fig.~\ref{fig:ngts}). 
Based on source counts and redshift distributions measured with SPT-SZ data \cite{bethermin2015, strandet16}, we expect thousands of these sources to lie at $z\,{>}\,7$, more than 100 to be at $z\,{>}\,8$, and around 1,000 to be associated with high-redshift protoclusters (Fig.~\ref{fig:ngts}).  The beamsize of CMB-S4 is well-matched to the expected scales of protocluster regions, building a catalogue far superior to the prelimary study of peaks in the CIB made possible using {\it Planck}'s all-sky survey \cite{PlanckHighz}.  The CMB-S4
high-redshift sources will open up a new window onto massive galaxy formation in the epoch of reionization, provide an important complement to the LSST and {\it WFIRST\/} surveys, and be prime targets for followup with ALMA and {\it JWST}. 
With the unique resource of the CMB-S4 dusty source catalog, we can directly observe the onset of dust production in the Universe, identify the first massive protoclusters of galaxies that represent the predecessors of the SZ cluster sample, and provide a well-understood target list for the high-redshift ALMA user community to explore galaxy formation.

An intriguing possibility is that CMB-S4 might be the first wide survey that
can select both proto-clusters and genuine clusters of galaxies, using their
different spectral signatures.  Statistical samples of both kinds of source,
and those intermediate between them, have never been available before.

Finally, the deep, high-resolution, multi-band data from 
CMB-S4 will be a rich source of ``serendipitous'' science, examples of
which have already been demonstrated with the Atacama Cosmology Telescope (ACT), {\it Planck}, and SPT
(e.g., strongly lensed DSFGs and mm transients). 

\subsubsection{Radio sources}

The broad spectral coverage of CMB-S4 will also provide high signal-to-noise ratio intensity and polarization measurements of radio sources. 
Applying the model of Tucci et al.\ \cite{Tucci:2011} with a confusion limit of 1\,mJy at 90\,GHz, over  100{,}000 AGN are expected to be detected at $>5\sigma$ in the deep and wide survey field. 
An additional 5{,}000 sources will also be detected above $5\sigma$ in polarization \cite{Puglisi:2017}. 
This source sample, selected from the highly-uniform CMB-S4 data set covering 70\% of the sky,  will be an order of magnitude larger than that from upcoming CMB surveys (e.g., Ref.~\cite{2018arXiv180807445T}), and will be a valuable tool for statistical studies of AGN populations (\cite{Murphy:2010, Mocanu:2013, Marsden:2013, Aghanim:2016xxvi, Partridge:2017}). 
The broad spectral coverage will enable measurements of source SEDs and---for the brightest polarized sources---Faraday rotation measures.  
These data will also be highly complementary to higher-resolution observations at other wavelengths (e.g., VLA/VLASS, ASKAP/EMU, SKA, and {\it eROSITA\/}).
In the latter case, CMB-S4 data will provide an excellent opportunity both via individual and stacked source measurements, to characterize the radio/sub-millimeter properties of over 1 million AGN to be detected in the {\it eROSITA\/} X-ray survey \cite{Kolodzig:2013}.

While the majority of radio sources detected by CMB-S4 will be flat or falling-spectrum blazars, new and rare categories of sources may emerge from the large source sample.
As an example of such discovery space, Ref.~\cite{PlanckIntXLV}  recently reported a small intriguing sample of low-$z$ AGN with rising high-frequency spectra that may contain cold intrinsic dust; such AGN---which will be more easily identified with the significantly higher resolution and deeper CMB-S4 data---could provide a view of the transition stage between the star-formation and AGN phases of such galaxies.  
CMB-S4 will similarly provide crucial higher-frequency data for source populations detected at other wavelengths. 
One interesting case will be the characterization of narrow-line Seyfert~1 (NLS1) galaxies for which there is limited high-frequency data (see e.g., Ref.~\cite{Lahteenmaki:2017}). 
These galaxies are assumed to be young AGN; however, based on 37-GHz observations, their detection rate is much higher than assumed by many models
\cite{Lahteenmaki:2017,Lahteenmaki:2018}.
Observations also indicate that---contrary to expectations---relativistic jets
can be generated in such sources, regardless of their radio-loudness
classification based on low-frequency data. 
A new catalog of southern NLS1
sources has recently been released \cite{Chen:2018} for which
higher-frequency data are needed. 
CMB-S4 will be able to provide SED determinations, as well as information on the time variability of the sources (and many other well-detected galaxies, since CMB-S4 will be a powerful time-domain survey of the mm-wave sky). 
By combining observations of  the NLS1s with data on other types of young AGN (such as compact
steep-spectrum sources, GHz-peaked spectrum sources, high-frequency
peakers, and compact symmetric objects), CMB-S4 will enable studies of the launching of 
relativistic jets, as well as the  evolutionary paths that young AGN take on their way to becoming fully-evolved, powerful radio sources.

\subsection{The Milky Way Galaxy}

\textbf{Galactic polarization:}
One area in Galaxy-scale science for which the deep polarization data from CMB-S4 will be particularly impactful is in the characterization of the interstellar medium (ISM), the polarization of which is a particularly pernicious foreground in the search for primordial $B$ modes (Sect.~\ref{sec:gw}). 
The ISM is known to be highly turbulent, compressible, and magnetized and while 
phenomenological models provide some insight into how foreground turbulence affects the dust polarization signal, they can not yet convincingly explain a number of 
current observations, such as the ratio of $E$ modes to $B$ modes. Excitingly, synthetic polarization maps from MHD turbulence simulations coupled with CMB observations are now beginning to shed light on the properties of the turbulent ISM in our galaxy.

The $E$-mode and $B$-mode power depend on the detailed properties of the turbulent density fluctuations in the ISM and it is not possible to characterize these fluctuations without knowledge of the sonic/Alfv{\'e}nic Mach numbers in the ISM. To extract this information, CMB observations are coupled with other diffuse ISM tracers (e.g., 21\,cm or H\,$\alpha$) to obtain velocity statistics. 
Current results are limited by \textit{Planck}'s sensitivity and resolution, but with future CMB-S4 polarization data and velocity information we will finally be able to place tight constraints on the parameters surrounding cosmic-ray acceleration, star formation, and diffuse ISM structure formation.

Another outstanding mystery, in the area of the star formation, is {whether
the magnetic fields that thread molecular clouds are strong enough to
inhibit movement of gas across magnetic field lines, thereby reducing the
efficiency with which stars form}. By studying the linearly-polarized
radiation from dust grains, which tend to align with respect to their local
magnetic field \cite{andersson2015}, CMB-S4 will create highly-detailed (${<}\,1$-pc resolution) maps of the magnetic field morphology for nearly a thousand molecular
clouds.  Using these polarization maps we will
measure the disorder in the magnetic field direction
\citep{Davis1951, Chandrasekhar1953,
Hildebrand2009, Houde2009}, the degree of polarization \citep{Fissel2016},
and determine whether there is alignment between the orientation of the magnetic
field and cloud structure \citep{Planck:XXXV}. Applying the same analysis
techniques to both the clouds observed with CMB-S4 and ``synthetic observations''
of numerical simulations, CMB-S4 can constrain the magnetization levels within
molecular clouds.  We can then compare the magnetization to the star-formation
efficiency measured with {\em Herschel}, {\em Spitzer}, and
the {\em Wide-field Infrared Survey Explorer (WISE)}.  The large number of molecular clouds that can be mapped by CMB-S4 is
crucial, since dust polarization is only sensitive to the magnetic field
component parallel to the sky and so large numbers of molecular clouds are
needed to correct for this degeneracy \citep{King2018}. {The results from CMB-S4
will place strong constraints on the dynamical importance of magnetic
fields in different stages of the star-formation process.}

\textbf{Galactic intensity:} 
Deep, multi-frequency maps over about half of the sky will be a useful resource for
tracing the census of star-formation, including seldom-studied regions away
from the Galactic plane.  There is also the potential for adding information
on particluar classes of star (e.g., those with disks).  Additionally, an exciting
new possibility enabled by these CMB-S4 data is the search for variable
Galactic sources, which we discuss in the next section.

\newpage
\section{The time-variable millimeter-wave sky}

\begin{shaded}
\emph{The fourth and final science theme relates to time-variable sources.}
  
CMB-S4 will provide a unique platform to conduct a wide-field time-domain survey in the millimeter band, covering over half of the sky to few-mJy depths every two days.
In this waveband, the time-variable sky is largely unexplored, with the exception of a shallow survey by {\it Planck}, surveys of the Galactic Plane (such as with the JCMT), targeted measurements of a few individual sources, and a single survey by SPTpol \citep{Whitehorn2016}; this is largely the result of limited observing time and fields of view for mm-band instruments (e.g., ALMA), which tend to focus on high-resolution observations of known objects.
Despite this, a wide variety of sources are either known or believed to have particularly interesting time-variability in bands observed by CMB-S4. 
Expected sources include tidal disruption
events, nearby supernovae, X-ray binaries, and classical novae. Particularly
good candidates are $\gamma$-ray bursts and active galaxies, such as the time-variable
blazar that was identified as a possible source of high energy neutrinos. 
The combination of high sensitivity and wide area for CMB-S4 will open a new
window for time domain astronomy and multi-messenger astrophysics.
\end{shaded}

\label{sec:transient}

\subsection{Gamma-ray bursts}
Gamma-ray bursts (GRBs) are one of the primary time-domain science targets for CMB-S4. The spectrum of GRB afterglows is well-described by a self-absorbed synchrotron process, with a broad emission peak from approximately 100\,GHz to 1\,THz \citep{GRBbook}, with emission lasting on the order of one week. The existence of so-called orphan afterglows from bursts without detected prompt $\gamma$-ray emission---either because of the $\gamma$-ray instrument field of view, misalignment of the jet with Earth, or absorption of the primary $\gamma$-ray emission
---is a generic prediction of GRB models, but none have ever been detected, despite a number of possible candidates (e.g., Refs.~\cite{RadioOrphanGRBs,OpticalOrphanGRBs}).
The main obstacle to their detection has been that, at both short and long wavelengths, the sources are very dim and, at the frequencies where they are bright enough to be detectable (for example, in the millimeter band), either few or no blind surveys have been conducted.

CMB-S4's observing strategy and sensitivity are expected to change this picture dramatically, delivering a factor of 2000 improvement on the only previous time-domain millimeter survey \citep{Whitehorn2016}, which had a candidate detection, and gives an expected 1700 afterglow detections from a population model of on- and off-axis bursts (PSYCHE, \cite{Ghirlanda:2013}) over a 7-year CMB-S4 survey. Other theoretical predictions find
that at all times during the survey there should be an ongoing detectable GRB afterglow 
\citep{Metzger2015}.  
Detection of these objects would serve a number of scientific goals:
\begin{itemize}
\item{confirm measurements of the beaming angle of GRBs from jet breaks and thus the total energy budget for GRBs in the Universe;}
\item{improve modeling of off-axis emission, and connect with gravitational-wave sources;}
\item{constrain the existence of a large population of $\gamma$-dark GRB-like objects, which are potential sources of the TeV--PeV diffuse neutrino background observed by IceCube;}
\item{potentially detect afterglows from GRBs made by population-III stars at high-$z$, during and prior to reionization.}
\end{itemize}

Note that uncertainties in this last item are large, but nevertheless,
CMB-S4 will enter the range of required sensitivities \cite{pop3grbs}.
Detection of even one of these would provide a wealth of information about
the early Universe, while a non-detection would constrain models of the
first generation of star formation.

\subsection{Fast transients}

Fast radio bursts (FRBs) are a striking astrophysical phenomenon of unknown
origin. They have been seen serendipitously in radio-frequency observations
with Areceibo, the Canadian Hydrogen Intensity Mapping Experiment (CHIME), Green Bank, and Parkes radio telescopes (e.g., Refs.~\cite{Keane2015,CHIME:2018,Ravi2019}), with around
20 detected to date. Given this event rate and estimates of sky coverage, the
full-sky rate of FRBs is roughly 5000 per sky per day, if they are isotropically
distributed. They are remarkably bright: the known examples all have a minimum
flux density ranging from 0.3 to 2\,Jy at frequencies of a few GHz,
plus the original 
``Lorimer burst'' with flux of 30\,Jy \cite{Lorimer2007}. They also have a very short duration; 
many last for less than 1\,ms, with durations up to 5\,ms. Most have no useful
polarization information, although a few (such as FRB150807 \cite{Ravi2016})
have been observed to have linear polarization at various levels.

FRBs are consistent with having random sky locations (with the exception of two
repeating sources \cite{Spitler2016,CHIME2019}). They also have dispersion measures ranging from
375 to $1600\,{\rm pc}\,{\rm cm}^{-3}$. This leads to an arrival-time dependence on frequency proportional
to $\nu^{-2}$, which is very well measured. Electrons in the Milky Way give
a total dispersion measure of around $100\,{\rm pc}\,{\rm cm}^{-3}$,
supporting the idea that FRBs are
at cosmological distances of order 1\,Gpc. Their frequency spectrum is
unknown but limited evidence in the 1--2\,GHz range suggests 
consistency with a flat spectrum in $S_\nu$. 
If these intriguing sources do have a flat spectrum, some will be potentially
detectable in microwave-background experiments, with flux densities greater than a
few mJy at microwave frequencies. 
Current TES-based
detectors have a sampling time of order 2\,ms, meaning that FRBs would generally
be visible only in a single time sample if their microwave and radio durations 
are comparable. A point-source FRB signal would appear to be a single bright
microwave burst at a random position in the focal plane with the spatial profile
of the experiment's beam shape. If its spectrum is fairly flat, it would
also appear in all channels of a multichroic detector simultaneously, and in
both linear polarization channels. In the time stream, the burst itself will
appear as a spike, with a subsequent decay according to the detector 
time-constants; this decay behavior distinguishes a sky signal from electronic
glitches. 

The detection of FRBs at microwave wavelengths would establish their
frequency spectrum, while upper limits on their rate would significantly
constrain the flux distribution.  Either outcome would contribute substantially
to understanding these mysterious extragalactic events.

\subsection{Protostellar variability} 

Variability studies
offer unique insights into the mass-accretion history of protostars.
There is a fundamental disconnect between steady-state star-formation models
and observed protostars---we know that at least some protostars go through
explosive burst events during their evolution and that this may be important
for the overall mass assembly (e.g., Ref.~\cite{Armitage2015}).
Circumstantial evidence suggests
that most, if not all, protostars go through these stages; however,
given the rarity of these large bursts, to catch one requires monitoring
a huge number of protostars (thousands) over many years.  CMB-S4 provides
that opportunity.

A survey with the James Clerk Maxwell Telescope \cite{Johnstone2018}
has found that about 10\% of protostars show low-level accretion variability,
over timescales of a year or so \cite{Mairs2017}.  This strongly suggests
that we are probing physical conditions in
the inner, planet-forming, disk.  In at least one case quasi-periodicity is
seen over many years, suggesting a link to long-lived structure (i.e.,
companions or planets) within the disk.
Additionally protostars are expected to be very active and a handful of
extremely powerful flares have been observed in the radio
\cite{Forbrich2017}.  Magnetic fields
are believed to play an important role in funneling material from the disk
onto the star (the last step for accretion).  Detailed measurements of the
properties of magnetic reconnection events are required to make significant
progress in our understanding of this complex phenomenon.  Follow-up
spectroscopic studies of flares discovered with CMB-S4 could unveil the
astrochemsitry of the time-varying signatures of volatile species and
UV and X-ray dissociation, as well as changes in the protoplanetary disk
snow-line through evaporation of ices.

Studying protostellar variability in this waveband is a new research area.
The examples we give illustrate that there are important physical processes that
can be discerned through measurements of variability and transient phenomena
within the Milky Way.  At millimeter wavelengths, so far only one modest
survey (the JCMT Transient Survey) has monitored star-forming regions, and
has just begun to produce results.  To draw statistically significant
conclusions will require a much larger survey, just like CMB-S4 will provide.

\subsection{Accreting binary systems}

Within our Galaxy, accretion in binary systems can lead to strongly time-variable
emission.  Observations have shown that classical novae could be easily detectable
in CMB-S4, with fluxes observed to be hundreds of mJy at mm-wavelengths 
\citep{Ivison1993, Chomiuk2014},
while flaring events have been observed in X-ray binaries to be tens of mJy 
\citep{Paredes2000}.

\subsection{Solar System}

\label{sec:solar_system}
\begin{figure}[t]
\begin{center}
\includegraphics[width=0.8\textwidth]{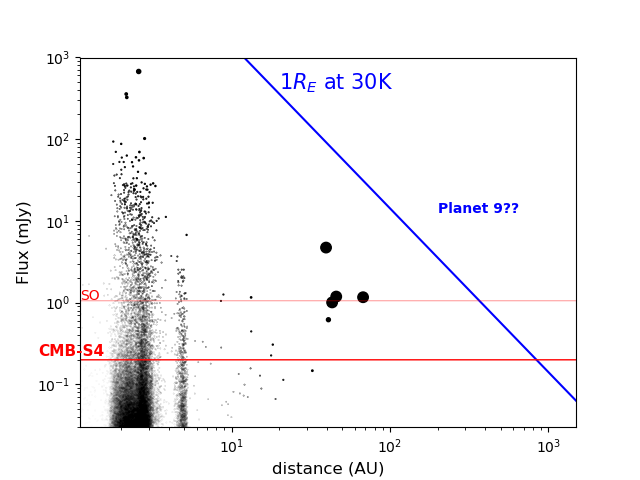}
\caption{Millimeter-wave Solar System.  Red circles show predicted asteroid flux densities at
150\,GHz, crosses show estimated flux densities of known dwarf planets at 90, 150, and 220\,GHz, and
diagonal lines show the flux density of any possible Earth-sized planet in the outer
Solar System in the CMB-S4 bands. Planet 9 is estimated to be several Earth masses at
a distance of several hundred AU.}
\label{fig:s4_solar_system}
\end{center}
\end{figure}

CMB-S4 will survey about 70\% of the sky approximately every two days. Moving objects,
such as asteroids, dwarf planets, or as yet undiscovered bodies such as Planet Nine,
can be detected in difference images using their long-wavelength thermal emission. 
The expected mm-wave Solar System can be seen in Fig.~\ref{fig:s4_solar_system}. For
mm-wave imaging of particular objects, an instrument like ALMA would be far more
sensitive; the particular benefit of CMB-S4 will be nearly daily measurements of
large numbers of bright objects over several years, along with wide sky coverage for
discovery of new or unexpected objects. 

Using thermal emission rather than reflected light, mm-wave measurements are less
sensitive to albedo effects, and the wavelengths are long enough that molecular
absorption and emission is negligible. Furthermore, the fall-off in flux as a function
of distance is less severe, with the main effect being the usual 1/$d^2$
dependence; optical
reflected light gets an additional factor of 1/$d^2$ from the distance to the Sun.
The temperature of Solar-System objects in equilibrium
with solar irradiance scales only as $d^{-1/2}$, while objects with substantial internal
heat or radioactive heating would decrease even more slowly. 

Asteroids show substantial time variability due to rotation and viewing-angle effects.
Using CMB-S4, more than 1000 known asteroids will be detected, with many of these having
high enough signal-to-noise ratio to chart the flux as a function of rotational phase. 
Comparisons with measurements in the optical band, and also the infrared (since many of these
objects have been measured as part of the {\it NEOWISE\/} project \cite{Masiero2018}), will yield new 
insights into the surface properties and geometry of these objects
(e.g., Ref.~\cite{Mueller2017}). 

Dwarf planets, such as Pluto, will be easily detected, with measurements possible out
to roughly 100\,AU. Extending further, CMB-S4 will be able to detect any possible
planets of Earth radius or larger at about 1000\,AU. The recent excitement about
a possible Planet Nine \cite{Batygin2016}
(nominally 5--15\,${\rm M}_\oplus$ at several hundred AU) 
has made it clear that there is substantial discovery space
for new planets in the outer Solar System. Assuming an Earth-like composition, 
1\,${\rm M}_\oplus$ of rocky material should have an equilbrium temperature of ${>}\,30\,$K just from
radiogenic heat. An atmosphere would further enhance the mm-wave flux.
The mm-wave-sky monitoring performed by CMB-S4 offers the potential to
search for a wide variety of moving objects.

\subsection{Multi-messenger astrophysics}

A wide-area sensitive mm-wave survey can play an important role in multi-messenger astrophysics,
in concert with both high-energy neutrino searches and gravitational wave astronomy. 

The IceCube neutrino source TXS 0506+056 appeared to be associated with a blazar 
\citep{IceCube2018} that is mm-bright.
With CMB-S4, this source would have had nearly daily flux measurements over many years, as well as
many other similar sources that could be used to characterize the statistics of variability.

The first binary neutron star merger, GW170817, was not visible at mm-wavelengths
\citep{alexander2017}, most likely due to the low density of the environment for that particular
event. It is expected that at least some future events should be in denser environments that will
enhance the mm-wave flux \citep{berger2014}. If there is no other detectable emission, CMB-S4 would 
provide arcminute localization as part of regular survey operations.
In addition, the mm-wave light curve can be compared with 
emission at other wavelengths to better understand gravitational-wave events.

We do not know what fraction of future GW sources will turn out to be
optically-obscured, but visible in the millimeter---so it seems
wise to have an instrument that is regularly scanning the sky in this
waveband.  Provided that the GW source is in the southern hemisphere,
then it {\it will\/} be in the CMB-S4 deep and wide survey field, and observed every few
days as a matter of normal survey operations.  {\it This same leveraging of the
CMB-S4 wide survey applies to all other future examples of transient source
for which astronomers would like rapid follow-up observations.}

\eject

\chapter{Science and Measurement Requirements} 
\label{chap:sciencemeasurement}

\section*{Introduction}
The science portfolio for CMB-S4 is rich and broad, spanning areas of
fundamental physics through astrophysics. Here we identify four of the
most compelling and unique areas of expected scientific output from
CMB-S4 to define the Level 1 (L1) Science Goals that will drive the
measurement requirements and instrument design. We then present a
high-level overview of the process by which we have determined these
measurement requirements, under cost constraints, to achieve our
L1 goals. The quantitative flowdown work is described in Appendix~A.
Here we summarize key results from that appendix and present our
measurement requirements. The measurement requirements driven by our four
L1 goals enable the great variety of additional science goals of the
CMB-S4 science program that we presented in Chapter 1. 

\section{Primary science goals}

 There are two primary quantitative science targets, related to primordial gravitational waves and light 
 relics. 

\noindent{\bf Primordial Gravitational-Wave (PGW) Science Goal: }

\begin{adjustwidth}{0.5in}{0.5in}
{\em If $r=0$, achieve a 95\% confidence upper limit of $r \le 0.001$.
  If $r \ge 0.003$, achieve a 5$\sigma$ detection.} 

{Motivation}: All inflation models that naturally explain the observed deviation from
scale invariance and that also have a characteristic scale equal
to or larger than the the Planck mass predict $r \ge 0.001$. A well-motivated sub-class
within this set of models predicts $r = 0.003$ to $0.004$. A characteristic
scale near the Planck mass arises in many models whether they
emerge from string theoretic considerations, effective
field theory, or a minimal-new-physics approach (Higgs inflation), precisely
because of the role gravity plays in the origin of the scale.
An upper limit at $r=0.001$ would point us toward
more complicated solutions that introduce a non-Planckian scale. 
The observed departure from scale invariance is a potentially
important clue that strongly motivates exploring gravitational wave amplitudes down to $r = 10^{-3}$.
\end{adjustwidth}

{\bf Light-Relics (LR) Science Goal: }

\begin{adjustwidth}{0.5in}{0.5in}
{\em Achieve $\Delta N_{\rm eff} < 0.06$ at 95\% confidence.}

{Motivation}: We have the opportunity with CMB-S4 to detect new light particles thermally produced in the
early Universe. The contribution to $N_{\rm eff}$ depends on both the
nature of the particle and the energy at which it was in equilibrium
with Standard Model particles. A natural target is to search for new particles back to before the QCD phase transition. With CMB-S4, any particle
that was in thermal equilibrium at the beginning of the QCD phase transition can be ruled 
out at 95\% confidence. While that sensitivity is not sufficient to detect a real scalar 
at an epoch earlier than the QCD phase transition, the sensitivity of CMB-S4 allows a further two order of magnitude improvement in energy sensitivity to either a Weyl fermion or vector particle.
\end{adjustwidth}

We also have two Legacy Survey science goals that we have used to define
measurement requirements.

{\bf Galaxy-Clusters (GC) Science Goal: }

\begin{adjustwidth}{0.5in}{0.5in}
{\em For galaxy cluster searches, achieve a lower mass limit that is
below $10^{14}\,{\rm M}_\odot$ at $z \geq 2$.}

{Motivation:} Galaxy clusters in the local Universe appear to have formed the bulk of their stars
at $z\approx2$--3. A catalog at these redshifts will provide new views on the astrophysics
of galaxy clusters. This sensitivity will allow
views of clusters similar to massive clusters that we see
at $z\approx 0.5$, but at an earlier stage in their development when
they were forming their stars.
\end{adjustwidth}

{\bf Gamme-Ray Burst (GRB) Science Goal: }

\begin{adjustwidth}{0.5in}{0.5in}
{\em Measure many gamma-ray burst afterglow light curves.}

{Motivation:} Gamma-ray burst afterglows contain a wealth of information about the central engine
and the surrounding medium. The peak wavelength of the emission evolves with time, passing through 
mm-wavelengths a few days after the burst. Measurements made at this time will provide key information
in particular about the density of the surrounding medium, uniformly for all bursts in the survey area.
\end{adjustwidth}

There are numerous additional science goals, presented in the previous
chapter. Those additional goals are all enabled by the CMB-S4 survey, but none 
are design drivers. 

\section{Flowdown to measurement requirements}

\begin{figure}[th]
\begin{center}
\includegraphics[width=5.75in]{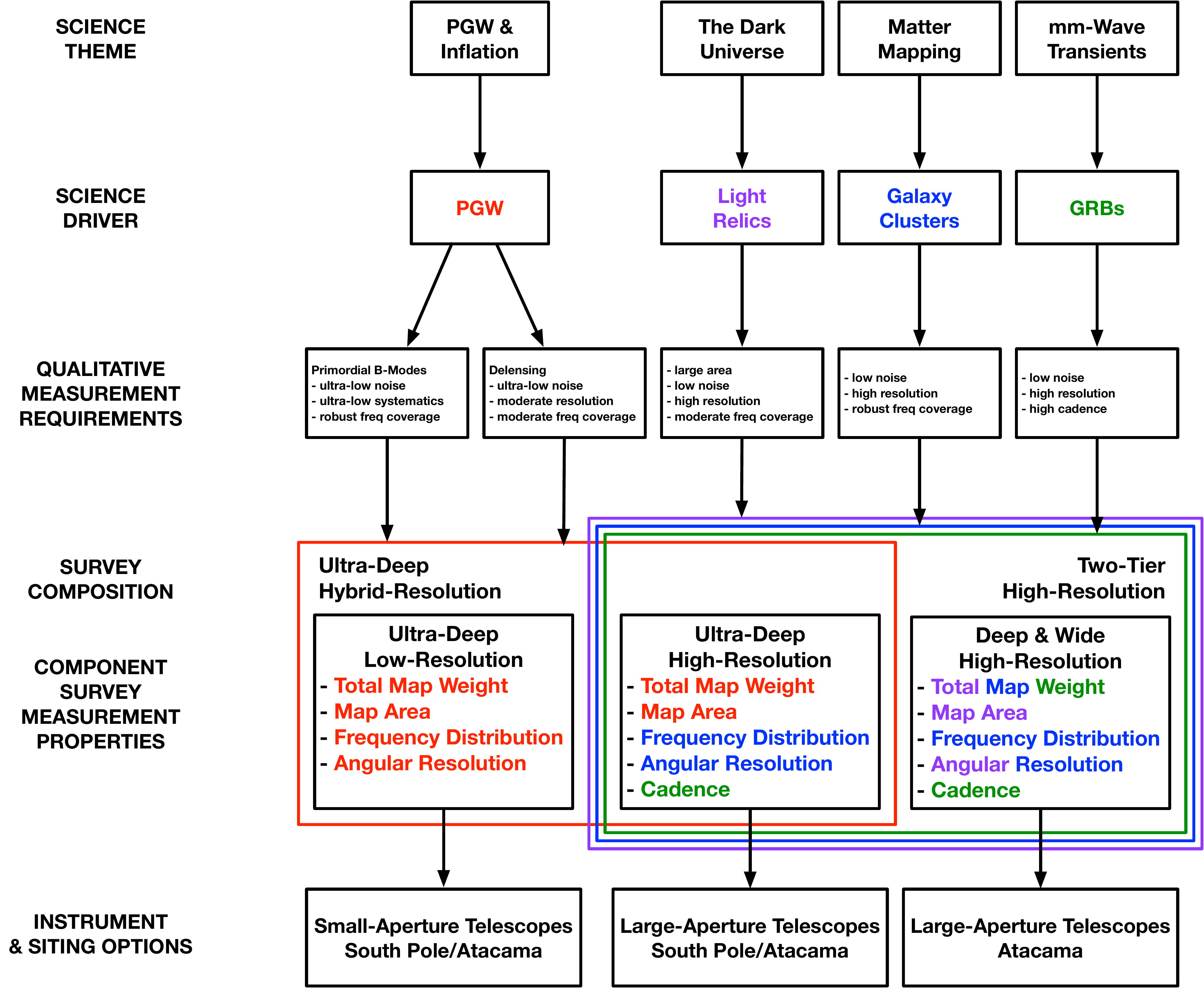}
\end{center}
\caption{Each science theme has one science topic contributing to the
  definition of measurement requirements and instrument
  requirements. The choice of a combination of low-res maps and
  high-res maps is driven entirely by PGW. The choice of a two-tiered
  survey is driven by the combined requirements of the PGW and Light
  Relics drivers. The connections of all the drivers to the
  measurement properties are indicated with color coding. Some
  measurement properties are written out in more than one color to
  indicate that they are influenced by more than one science driver.}
\label{fig:flowdown}
\end{figure}

Figure~\ref{fig:flowdown}  presents an overview of how key measurement
properties are derived via a flowdown process from our four key science drivers.
The work developing the quantitative
relationships summarized here is presented in greater detail in Appendix~A.

\subsection{Overview}

The Primordial Gravitational-Waves and Light-Relics science drivers lead
us to a two-tiered survey, with a small ultra-deep field contained
within a much larger deep field. For the Light-Relics target,
reduction of sample variance drives us toward observing as much sky as
possible. Information about $N_{\rm eff}$ primarily comes, at CMB-S4
noise levels, from the $TE$ spectrum. Sample variance in this power
spectrum is determined by sky coverage and the amplitudes of the $TT$
and $EE$ spectra---amplitudes that are sufficiently large as to make
sample variance the dominant statistical source of uncertainty on the
relevant angular scales. For the Primordial Gravitational Wave science, which depends on
measurement of the $BB$ spectrum, the relevant sample variance is 
much smaller and can be reduced further by delensing. Galactic
foregrounds are also much brighter relative to the signal of interest,
which motivates finding the cleanest regions of sky. While the optimal
strategy depends in detail on how well we can delens, and on how well
we can clean the foregrounds, current forecasting points to the need
to go very deep on a small ($f_{\rm sky} \approx 0.03$) patch of sky. 

Therefore the full CMB-S4 data set consists of an Ultra-Deep Field
with many mm-wave bands over a fairly small patch of sky, and a
Deep and Wide Field (shallower, though still unprecedentedly deep), covering a large
fraction of the entire sky ($f_{\rm sky} \approx 0.7$).

The Ultra-Deep field must be covered by both low-resolution and high-resolution 
instruments, 
driven by the unique demands of
isolating degree-scale CMB $B$ modes, and of estimating the contribution
to these $B$ modes from the gravitational lensing of $E$ modes. The
rationale for this split is given in more detail in subsequent chapters, as are instrumental
considerations that drive the bifurcation into small- and large-aperture telescopes (SATs and LATs). 

With the basic scheme in place of a survey with deep and ultra-deep fields, with
the ultra-deep field having both high- and low-resolution components, we can now list
the important measurement properties for each of these three
components of the CMB-S4 survey: map area; total statistical weight; distribution
of weight across frequency; angular resolution; and cadence.  In the subsections that follow, we
describe our
process of flowing down science requirements to requirements for these
measurement properties,  
for each of
the four science drivers in turn.

\subsection{Measurement properties required to achieve PGW Science Goal}

 Meeting our Primordial Gravitational-Wave goals will require significant advances, 
 not only in raw sensitivity but also in control of foregrounds and instrumental systematics. As discussed in
 detail in the CMB-S4 Science Book and in previous publications 
 \cite{Abazajian:2013vfg}, attaining the desired levels of sensitivity to the signatures
 of gravitational waves in the CMB necessitates first and foremost at least an 
 order-of-magnitude increase in the raw number of detectors on the sky compared to Stage-3 experiments. These works also noted
 that foreground mitigation will be crucial for CMB-S4, particularly in the pursuit of the gravitational-wave 
 signal. To minimize contamination from Galactic foregrounds, it is clear that multiple frequency
 channels are required. It is known from analysis of BICEP/Keck and \planck\ data \cite{Ade:2015tva}
 that, if unsubtracted and unmodeled, Galactic dust imparts a bias to the measurement of the 
 tensor-to-scalar ratio $r$ at 150\,GHz at a level two orders of magnitude above the target 
 $\sigma(r)$ for CMB-S4. Synchrotron emission is expected to impart a similar bias at lower frequencies.
 At least one frequency channel is required to remove each of these contaminants, and multiple
 channels will be needed for each component if the behavior of these foregrounds is not perfectly 
 uniform across the survey. Galactic foregrounds are also reduced by observing a patch or patches
 of sky with as low as possible column density of Galactic material. This consideration---and optimizations
 of raw sensitivity---drive a survey for the degree-scale gravitational-wave signal to small patches of sky 
 ($f_{\rm sky} \approx 0.03$).

 The other major foreground for gravitational-wave searches, one which cannot be mitigated with 
 frequency coverage, is the signal from gravitational lensing. Scalar density perturbations in the
 early Universe produce only even-parity (``$E$-mode'') patterns in the CMB polarization to first order, 
 while gravitational-wave perturbations also produce odd-parity (``$B$-mode'') patterns
 \cite{Kamionkowski:1996zd,Seljak:1996gy,Zaldarriaga:1997ch}. This fact is essential to our 
 ability to detect a gravitational-wave signal in the presence of the much larger signal from scalar
 density perturbations. Gravitational lensing of the $E$ modes, however,
 produces a secondary source of $B$ modes, at an amplitude significantly larger than the
 target CMB-S4 sensitivity to the gravitational-wave signal at degree scales. The lensing process can effectively 
 be inverted with sufficiently high-fidelity measurements of the
 $E$-mode signal, to estimate the gravitational
 lensing potential, and a map-level prediction of the lensing
 contamination can then be accounted
 for in gravitational-wave searches \cite{Knox:2002pe,Kesden:2002ku}. 
 The level of this so-called ``delensing'' required for a small-area survey and the CMB-S4 target
 $\sigma(r)$ requires deep, high-resolution data over the same patch of
 sky observed for the degree-scale $B$ modes.

 The above measurement challenges were taken into account in the optimization exercises described in
 the CMB-S4 Science Book and the CDT report. We have extended that
 work for this report,  and it is presented in a self-contained manner (rather than
 as an update) in this chapter with details in Appendix~A. 

If the true value of $r$ is $\gtrsim 0.01$
then it may be the case that we
can improve our characterization of the primordial $B$-mode signal by
covering more than 3\% of the sky. Thus we leave open the possibility
of broadening our Ultra-Deep Field in order to re-optimize if Stage
III data, or early data from CMB-S4, indicate a large value
of $r$. 

As indicated in Fig.~\ref{fig:flowdown} the PGW science goals 
are the primary drivers for many of the CMB-S4 measurement properties. We now
review the measurement properties of the Ultra-Deep surveys and how the
requirements on them are determined by the PGW science requirements. 

{\bf Map area}: Assuming $r=0$, our semi-analytic framework formally leads
to an optimal map area that is less than 1\% of the total sky. There
are several effects, not included in the analytic optimization
procedure that argue against targeting less than 1\% of the sky, as
discussed in Appendix~A.  These include the cut-sky impact on $E$/$B$ decomposition
and risks in becoming too heavily dependent on
delensing. Folding these considerations into our semi-analytic
results, we have chosen 3\% as a nominal target field size for many of
our optimization studies. 
\commentout{
Guided by this analytic result, we used an observing
simulator to produce coverage maps, from Chile and from Pole, that are as compact as possible,
taking into account the finite size of the field of view and the need
to avoid observing near the Sun and the Moon, and, from each
site. The Pole allows for a greater concentration of the observing
weight in a small region. Our deepest simulated observations from the
Pole cover about 3\% of the sky, and we use 3\% as a nominal sky
fraction for many of our optimization studies. }

Broader sky coverage has advantages if $r$ turns out to be
sufficiently large. Thus we considered both a ``Pole Deep'' survey of
about 3\% of the sky, and a ``Pole Wide'' survey of about 7\% of the
sky. To cover more sky may require observations from Chile. 
Based on the studies in Appendix~A, we may deploy
some SATs to Chile if $r \gtrsim 0.01$. In this happy case, we will
have the opportunity to make high-significance detections from two
different sites using different observing strategies. Our
reference design takes advantage of our access to these two different
sites by building in deployment flexibility. The baseline plan is for
all SATs to go to Pole, but if we find indications that $r$ is
sufficiently large, then we can also send some SATs to Chile.

{\bf Degree-scale total map weight and frequency distribution}:
Our semi-analytic framework can be used to
calculate $\sigma(r)$ from observations of a given amount of sky, as a
function of total detector effort (or, equivalently, total map
weight), optimally distributing the detectors across the low-resolution
frequency bands and a high-resolution band for delensing. 
The result of this calculation for $f_{\rm sky} = 0.03$ is that achievement
of $\sigma(r) = 5 \times 10^{-4}$
requires $1.8\times 10^6$ 150-GHz-equivalent detector
years. Roughly seventy percent of this (1.2 million) is allocated to the
degree-scale observations (with the remainder allocated for delensing)
with a particular distribution across frequency as described in Appendix~A. 

In reality one is not able to choose the number of detectors
in each frequency band in the continuously variable manner
assumed in the semi-analytic calculations.
For the reference design a realistic mapping of detectors into
dichroic optics tubes has been carried out while seeking to maintain
the band distribution as
determined in the optimization calculations. This departure from the
optimal distribution results in an increased requirement for the total
number of 150-GHz-equivalent detector years for the degree-scale survey.
The result is the reference design described in
Sect.~\ref{sec:refdessum}, with expected measurement performance as 
given in Table~\ref{tab:satreq}, with a total effort of $1.5\times
10^6$ 150-GHz-equivalent detector years delivering a total map weight, 
from all frequency channels
combined, of 38 $\times 10^6 \mu$K$^{-2}$. 

\begin{table}[t!]
\begin{center}
\begin{tabular}{| l | c c c c c c c c c |}
\hline
Frequency & 20 & 30 & 40 & 85 & 95 & 145 & 155 & 220 & 270 \\
\hline
Angular resolution (arcmin) & 11.0 & 72.8 & 72.8 & 25.5 & 25.5 & 22.7 & 22.7 & 13.0 & 13.0 \\
Total survey weight / $10^6$ ($\mu$K$^{-2}$) & 0.12& 0.69 & 0.43 & 11.0 &
                                                                   14.1&
                                                                         5.7
                                   & 4.8 & 0.71 & 0.24\\
\hline
$Q/U$ rms ($\mu$K-arcmin)  & 8.4 & 3.5 & 4.5 & 0.88 & 0.78 & 1.2 & 1.3 & 3.5 & 6.0 \\
\hline
\end{tabular}
\caption{Ultra-Deep degree-scale map noise and angular resolution
  measurement requirements. Survey weight is $2A/N^2$ where $A$ is the
  effective sky area (in particular, $f_{\rm sky}^{\rm noise}$ as
  defined in Appendix~A) and $N$ is the $Q/U$ map noise
level. This quantity is linear in detector count and run time. Note
that, as we will see in the next chapter, the 20-GHz channel is on an
LAT rather than an SAT.
}
 \label{tab:satreq}
\end{center}
\end{table}

{\bf Degree-scale angular resolution}: 
The telescope aperture determines the angular resolution as a function
of frequency, with lower frequencies having a lower angular resolution.
The aperture size is set to allow 30\,GHz and higher
frequencies sensitivity to the primordial 
$B$-mode recombination bump at   $\ell \approx 100$. 
Lower resolution than achieved by the reference design aperture at 30 and 40\,GHz would result in reduced
performance. As detailed in Appendix~A, a 20-GHz channel on the SAT
had insufficient sensitivity at $\ell \gtrsim 100$ due to insufficient
resolution. Therefore the 20-GHz channel was moved to the delensing
LAT. 

We find the map noise levels at the given frequencies and angular
resolutions in Table~\ref{tab:satreq}
are sufficient for achieving our science requirements.

{\bf Hi-res ultra-deep total map weight and angular resolution}:
The delensing total map weight is informed by our semi-analytic
optimization. For observing 3\% of the sky, one can see from
Fig.~\ref{fig:sigr} that achieving our target $\sigma(r)$ requires delensing at a level
such that the residual lensing $B$ modes have an rms amplitude about 30\% of the 
uncleaned amplitude, corresponding to a residual lensing $B$-mode power 
that is about 10\% of the uncleaned level.
From Fig.~\ref{fig:lens_res} one can see that 
this level of delensing requires a map noise of about 0.7 $\mu$K-arcmin in the delensing survey.

The above is all for an idealized case of a delensing observation
with a $1^\prime$ beam and no foreground contamination. In practice we need
greater total weight to clean or constrain foregrounds, as we discuss below.

In Fig.~\ref{fig:lens_res} we show forecasts for the amount of residual lensing $B$-mode
power as a function of delensing map noise and angular resolution,
again for coverage of 3\% of the sky. We see that a requirement to
clean out approximately 90\% of the lensing-induced
$B$-mode power is not a strong driver for angular resolution, and it
is a strong driver for low map noise. Angular resolution for the 
delensing LAT reference design is driven primarily by our galaxy
cluster science goal.

{\bf Hi-res ultra-deep frequency distribution}:
We have not performed a rigorous optimization of the allocation of detectors
across frequency for the delensing LAT focal plane, as we have done
for the SATs. For estimation of the lensing potential, and then of the
lensing $B$ modes, we expect foreground contamination to be
a much lower fraction of the signal of interest than is the case for
the degree-scale observations. 

Our reference design has 95- and 145-GHz channels as the main CMB
channels, and then a pair of lower frequencies to guard against
synchrotron and free-free, and a pair of higher frequencies to guard
against dust. Forecasts for lensing performance in the presence of
foregrounds are made with the ILC method described in Appendix~A. 

We see the reference design allocation of detectors across frequency 
as a highly conservative one regarding control of foreground
contamination. Allocation of detectors across
frequency for the delensing LAT is primarily driven by our galaxy
cluster science goal.

\begin{table}[t!]
\begin{center}
\begin{tabular}{| l | c c c c c c c |}
\hline
Frequency (GHz) & 20 & 30 & 40 & 95 & 145 & 225 & 270 \\
\hline
Angular resolution (arcmin) & 11.0 & 7.3 & 5.5 & 2.3 & 1.5 & 1.0 & 0.8 \\
White noise level ($\mu$K-arcmin) & 8.4 & 5.0 & 4.5 & 0.68 & 0.96 & 5.7 & 9.8 \\
\hline
\end{tabular}
\caption{Measurement properties of a dedicated delensing survey that
  meets our measurement requirements.}
\label{tab:delensspec}
\end{center}
\end{table}

\subsection{Measurement properties required to achieve LR Science Goal}

To reach the light relics goal, a highly precise measurement of the CMB power spectrum is required, in particular for the cross-power between temperature and polarization ($TE$).  The precision of this measurement is primarily set by the uncertainty in the maps and the number of independent samples (i.e., the area of sky for the survey). 

The uncertainty in the polarization maps is expected to be dominated by instrument noise rather than by
foreground uncertainties. This is because the dominant small-scale foregrounds 
for CMB-S4 (thermal SZ and dusty star-forming galaxies) are largely unpolarized; 
radio sources can be polarized, but at the sensitivity of CMB-S4 the bright sources can
be individually identified and removed (or masked), with the remaining number of undetected
sources expected to be too small to contribute any contamination. 

{\bf Map area}: The measurement property to which the standard
deviation on $N_{\rm eff}$ is most sensitive is map
area. Semi-analytic estimates of $\sigma(N_{\rm eff})$ as a function
of sky coverage, at fixed total map statistical weight, show rapid
improvement as $f_{\rm sky}$ increases, as shown in
Appendix~A in Fig.~\ref{fig:Neff_fsky}. 

Thus motivated, we explored observing strategies with successively
smaller minimum observing elevations, which allow progressively
greater sky coverage. There is a frequency-dependent noise penalty 
for observing at lower elevations, through higher air mass, due
to the greater background loading of the detectors. Our observing
strategy simulations took these noise penalties into account, allowing
for the construction of multi-frequency maps of noise variance. These in turn were
reduced to estimates of errors on CMB power spectrum measurements, and
from there to estimates of $\sigma(N_{\rm eff})$. 

Exploration of observing strategies led to the discovery that avoiding
the galaxy was more harmful than helpful, due to the impact of the
time it takes to halt a scan and reverse its direction. The result for
these galaxy-avoiding maps was they were effectively anti-apodized,
with coverage building up near the edges. We have thus opted to scan
right through the galaxy. In addition to opening up opportunity for a
wide range of Galactic science, this scan strategy also allows us to
make a choice of Galactic cut, for the $N_{\rm eff}$ science, after we
have gathered the data.

The result of our analysis is Fig.~\ref{fig:Neff_WAFTT} showing $\sigma(N_{\rm eff})$
for three different choices of minimum elevation angle and as a
function of the Galactic cut. The corresponding amount of sky used for
light relic science is shown in Table~\ref{tab:fsky_WAFTT}. For a $30^\circ$
minimum elevation angle we can achieve our goal with a 13 to 14\%
galaxy cut. This means we cut out from our map the regions with
Galactic contamination that are at the same level as the worst 13 to
14\% of the whole sky. We believe that such a cut is sufficient for
protecting us from significant foreground-modeling-induced biases in
estimates of $N_{\rm eff}$. We plan to investigate further with
map-based simulations.  We have not considered even lower elevation angles due to the rapidly
increasing air mass penalties and concerns about sidelobe
pickup from the ground.

For the above reasons, our reference design survey assumes the sky coverage that comes from a $30^\circ$
minimum elevation angle. The coverage is inhomogeneous and thus
leads to different measures of the amount of sky covered, as described
in Appendix~A. For what we call $f_{\rm sky}^{\rm noise}$ we have a
total coverage of 71\%, or 29,000\,${\rm deg}^2$. After the 13 to 14\% Galactic cut this leaves
about 62\% for determination of $N_{\rm eff}$. Other science goals,
with greater sensitivity to Galactic foreground contamination, are likely to require larger Galactic masks. 

\begin{table}[t!]
\begin{center}
\begin{tabular}{| l | c c c c c c |}
\hline
Frequency (GHz) & 30 & 40 & 95 & 145 & 220 & 270 \\
\hline
Angular resolution (arcmin) & 7.4 & 5.1 & 2.2 & 1.4 & 1.0 & 0.9 \\
Total survey weight ($TT$)/$10^6$ [$\mu$K$^2$]  & 0.22 & 0.68 & 26.3 &
                                                                     26.3
                                     & 2.2 & 0.38\\ 
White noise level for $TT$ ($\mu$K-arcmin) &21.8 &	12.4&	2.0&	2.0&	6.9&	16.7 \\
  White noise level $E$/$B$ ($\mu$K-arcmin) & 30.8&	17.6&	2.9&	2.8&	9.8&	23.6\\

\hline
\end{tabular}
\caption{Deep and Wide Field map noise and angular resolution
  requirements.}
\label{tab:lasreq}
\end{center}
\end{table}

{\bf Frequency distribution, total map weight, and angular resolution}: The light relics science goal does not require a heavy amount of
foreground cleaning. The major driver of frequency distribution is
galaxy clusters. Given the distribution of detectors across frequency,
and the exercise described above to determine the map area, we
conclude we need a total map weight of 56 $\times 10^6 \mu$K$^2$. Some
of this statistical weight is spent on regions of the sky where
Galactic emission is sufficiently bright that it is useless for light
relics science, as discussed above. 

Figure~\ref{fig:Neff_BeamNoiseFsky} studies the impact on changes to
noise level and beam size on $\sigma(N_{\rm eff})$. Of course, lower
noise is better, as is decreased beam size. Given that the reference
design hits our $\sigma(N_{\rm eff})$ target, rather than surpasses
it, we can not tolerate single changes such as an increase in noise or
an increase in beam size. However, a decrease in beam size would allow 
for an increase in noise, or vice versa. We have not yet attempted a
joint optimization of cost across all of these degrees of freedom,
constrained to fixed $\sigma(N_{\rm eff})$.

\subsection{Measurement properties required to achieve GC Science Goal}

Galaxy clusters in the local Universe appear to have formed the bulk of their stars at $z\approx2$--3.
Building a nearly mass-selected galaxy cluster catalog that extends to this period is a high
priority for CMB-S4.

Measuring high-redshift galaxy clusters requires high sensitivity at several frequencies over 
a wide area. The clusters are detected using the distinctive frequency
dependence of the thermal SZ effect. Removal of the cosmic microwave
background and cosmic infrared background fluctuations is done 
using a combination of spatial and spectral filtering, requiring at least
three frequencies at roughly comparable depth for measuring the separate 
components. 

The CMB-S4 galaxy cluster program builds on Stage-3 experiments:
SO will survey a comparable sky area with lower sensitivity, while SPT-3G is surveying about 20 times less
area but with projected sensitivity that is similar to CMB-S4. 

Assembling a galaxy cluster catalog at $z\approx2$--3 requires enough sky area
to have a sufficiently large cosmological volume, 
and enough sensitivity to detect clusters with low enough masses that the total cluster number density is sufficiently high. 
Galaxy clusters are rare, with the number density of the most massive clusters at any redshift
exponentially suppressed, so the sensitivity is particularly important. 

The sensitivity to galaxy clusters is a combination of raw detector noise, beam size, and frequency 
coverage. Everything else being held equal, lower noise straightforwardly improves sensitivity. 
At fixed detector noise, sensitivity will increase as the beam size is reduced as long as the
cluster is not well-resolved; clusters at $z\approx2$--3 will have their virialized regions subtending
roughly $2^\prime$. Multiple frequencies improve sensitivity by allowing the removal of the cosmic microwave
background primary fluctuations and the cosmic infrared background fluctuations, both of which are
sufficiently bright that in any single frequency map they will swamp the signal from $z\approx2$--3 clusters.

\subsection{Measurement properties required to achieve GRB Science Goal}

CMB-S4 will be a sensitive mm-wave survey of the time-variable sky. While there are many expected
sources, a particularly interesting target will be gamma-ray burst (GRB) afterglows. Many GRBs have
been observed to be mm-luminous, and there is the possibility of detecting off-axis GRBs where the
prompt gamma-ray emission is either too faint or relativistically beamed away from our line of sight.

The measurement requirements
are that CMB-S4 have good point-source sensitivity with a cadence such that locations are re-visited 
several times per week.

Lower noise flux levels are helpful, allowing a larger volume to be probed. 
The minimum flux sensitivity is set by
the range of observed fluxes for pointed follow-up of GRBs, which typically 
range from 1--10\,mJy at frequencies of 90--230\,GHz \citep{deUgarte:2012}. 

The time evolution of GRBs is such that there may be substantial evolution within the first day for some
sources \citep{Laskar:2018}, but the typical light curve evolves over a timescale of roughly 1 week. 
This sets a requirement on cadence that a wide sky area is surveyed multiple times per week.
More sky area simply increases the probability of catching a GRB within the field. 

Theoretical estimates \citep{Metzger2015}
are that there is at least one observable mm-wave GRB afterglow on the sky at any given moment at a flux density
around 1\,mJy at 150\,GHz. 
The counts are expected to be in the Euclidean regime, where the number of sources
above a flux limit $N(>S)$ scales as $S^{-3/2}$. At a sensitivity higher than 10\,mJy the rate would be such that
there would be at best a handful of events detectable per year.

\chapter{Instrument Overview}  
\label{chap:instrumentoverview}

\section{Requirements and design drivers}

The measurements outlined in the previous chapter
require about an order of magnitude improvement in measurement accuracy compared with Stage-3 experiments. Such a large step in performance brings significant challenges, the most important of which is sensitivity. Detectors for ground-based CMB experiments are background-limited, so CMB-S4 must have an enormous number of detectors; roughly 500k detectors will be needed on telescopes with angular resolution ranging from tens of arcminutes for measuring $r$, to around $1.5 \textrm{ arcmin}$ for measuring $N_{\textrm{eff}}$.

High raw sensitivity is only useful if systematic errors can be controlled. Beyond astrophysical foregrounds, which will be managed with broad frequency coverage, the major sources of systematic errors are atmospheric brightness fluctuations, pickup of unwanted signals from the ground, and polarization errors due to gain and beam shape differences between the two detectors in a polarimeter. Systematic errors are the principal driver for observing from sites with a stable atmosphere, and for using small telescopes, which are easier to shield, for measuring large angular scales. Only small telescopes have demonstrated control of systematic errors at a level that is needed to constrain $r$, so the reference design uses small telescopes for the $r$ survey.

Many systematic errors can be measured and corrected, so their impact is primarily through unmodeled residuals. Simulations for the CDT report demonstrated that residual, noise-like, additive, systematic errors, both correlated and uncorrelated across different frequencies, with various angular power spectra, should be $\lesssim10\%$ of the noise to ensure $\lesssim10^{-4}$ bias in $r$. Residual systematic errors at this percentage level are consistent with what has been achieved in Stage-3 experiments \cite{Array:2015xqh}. A similar percentage error is expected at CMB-S4 sensitivity levels, because the various techniques that are used to model or filter the low-order effects of systematic errors are limited by the noise. In this case, CMB-S4 will not require vastly smaller absolute systematic errors than Stage-3 experiments, as long as the errors are stable. Much of the detailed design effort for CMB-S4 will focus on demonstrating that systematic errors can be controlled at an acceptable level, but our working assumption for the reference design is that Stage-3 technology is sufficient; we just need more of it.

\section{Instrument configuration}

CMB-S4 is a single experiment with a mix of large and small telescopes at the South Pole and in the Chilean Andes. Both sites offer excellent observing conditions at millimeter wavelengths, and both have well-established infrastructure that has supported many CMB experiments \cite{Ogburn2012, Carlstrom2011, Fowler:2007dn, Barron2014}. Chile is higher, so the dry component of the atmosphere is more transparent, but the South Pole has consistently low precipitable water vapor and a more stable atmosphere, with substantially lower $1/f$ noise due to blobs of water vapor drifting by (e.g., see Ref.~\cite{kuo2017}).  

A measurement of $N_{\textrm{eff}}$ requires arcminute resolution over about 70\% of the sky, but only about $10\%$ of the sky is accessible from the South Pole, while about $80\%$ is accessible from Chile, so the two large telescopes for the $N_{\textrm{eff}}$ survey must be in Chile. There is no strong driver for opening up a third site to increase the survey area, but we would likely take advantage of such a site if it were developed outside the scope of CMB-S4.

A measurement of primordial gravitational waves requires a survey using small telescopes for the primordial signal and one large telescope for delensing. The optimal siting of these telescopes will depend on the value of $r$, the nature of polarized galactic foreground emission, and the actual performance of the instruments in the field. For very low values of $r$, the smallest, deepest, sky patch is required, and the ability to observe a single patch at all times from te South Pole favors siting all the SATs there. For larger values of $r$ we can take advantage both of the larger sky area available from the Atacama to observe multiple patches, and of the differences between the two sites to help control systematics, so a split distribution would be preferable. For the reference design we baseline siting all the SATs at the South Pole to ensue that we meet our L1 science requirements for any value of $r$, but we cost and schedule siting the SATs both at the South Pole only and equally split between the South Pole and Atacama to ensure that a wide range of options remain viable. The delensing LAT is sited at the South Pole in either case. We know that additional information on all of the issues informing this decision will become available during the CMB-S4 construction project from both existing experiments and from the planned South Pole and Simons Observatories (with the latter pair being particularly informative about comparable instrument performance), and we will continue to update the siting plan based on these new data. Finally we include the option within the baseline plan of moving SATs from the South Pole to the Atacama during operations should results from early operations indicate that doing so would be beneficial---for example, if an early detection were made of a moderate value of $r$ and the priority became to control foreground and/or site-specific systematics and to reduce sample variance.

The small telescopes are based on designs that have been used for many years by the BICEP/Keck experiments, but with dilution refrigerators, for continuous operation, and three telescopes per cryostat to reduce the number of telescope mounts. The large telescopes are essentially copies of what is being implemented for Simons Observatory, but with more sub-field cameras to completely fill the telescope field of view with detectors. All of the telescopes have horn-coupled transition-edge-sensor detectors, with time-domain-multiplexed readout, because these technologies are scalable and have a strong heritage in Stage-3 experiments. Small residual systematic errors require well-controlled beams, so horns are an obvious choice. Windows, cold optics, and thermal filters are copies of Stage-3 designs. 

The raw data rate for CMB-S4 will be approximately $20\textrm{ TB/day}$, about the same as LSST, so transmission, storage, and analysis will be challenging, but manageable. Real-time monitoring, flagging, and calibration of time-domain data will all have to be more automated than in existing experiments, and reduction to well-characterized maps will have to be efficient, so that systematic errors can be identified quickly and corrected.

The flow down of measurement requirements to instrument requirements is still at an early stage, especially for requirements associated with instrument systematic errors, e.g., ground pickup and polarization errors, where it is the residual error, after modeling and filtering, that matters. Quantitative estimates of residual systematics require fairly detailed simulations; in some cases, errors must be applied in the detector timestreams and propagated all the way through to science analysis. Detailed simulations are needed to develop requirements for sidelobe level, which drives the design of baffles and shields for the telescopes, and for instrumental and cross polarization errors, which mainly affect the design of the horn to detector coupling scheme. We do not have to freeze all the instrument requirements before detailed design begins, because we already have some understanding of how to allocate systematic errors to subsystems from Stage-3 experiments, but developing simulation tools that accurately capture the effects of instrument errors is a high priority for CMB-S4.
\section{Reference design summary}
\label{sec:refdessum}

The reference design for CMB-S4 is a point design that establishes the feasibility of the experiment and provides a basis for preliminary costing and project planning; it is not the final design. Many details will change, based on the results of technology development and design activities, but the overall scope and general characteristics of the experiment will not change.

The reference design has 18 0.5-m-class refractors (small-aperture telescopes, or SATs) for measuring large angular scales in the $r$ survey.
The basic characteristics of the SATs are shown in Table~\ref{tab:SATproperties}.  It also has 
 three 6-m crossed-Dragone telescopes (large-aperture telescopes, or LATs) for measuring arcminute scales for the $N_{\textrm{eff}}$ survey, and for delensing the $r$ survey.  The basic characteristics of the LATs are shown in Table~\ref{tab:LATproperties}.

 The full system includes over 500k horn-fed TES detectors with time-division multiplexing (TDM) readout, 
 with band centers ranging from 20 to 270~GHz. 
 To keep the number of telescopes reasonable, most of the camera pixels are dichroic; 
 each pixel has two frequency bands and two polarizations, for a total of four detectors per pixel. 
 SATs deployed in Chile have fast polarization modulators to reduce the effects of atmospheric brightness fluctuations, but the SATs at the South Pole do not.

\begin{table}[htbp!]
\begin{center}
\begin{tabular}{ |l||c|c||c|c||c|c||c|c|| } 
\hline
 Property &\multicolumn{2}{c||}{LF} &  \multicolumn{2}{c||}{CF High} & \multicolumn{2}{c||}{CF Low} & \multicolumn{2}{c||}{HF}\\ \hline
 Center frequency (GHz) & 30 & 40 & 85 & 145 & 95 & 155 & 220 & 270 \\ 
 Primary lens diameter (cm) &55 & 55 & 55 & 55 & 55 & 55 & 44 & 44  \\
 FWHM (arcmin) &72.8 & 72.8 & 25.5 & 25.5 & 22.7 & 22.7 & 13 & 13  \\ 
 Fractional bandwidth & 0.3 & 0.3 & 0.24 & 0.22 & 0.24 & 0.22 & 0.22 & 0.22  \\ 
 NET ($\mu$K$\sqrt{\rm s}$) per detector & 177 & 224 & 270 & 238 & 309 & 331 & 747 & 1281 \\
 $N_{\rm det}$ per optics tube & 288 & 288  & 3524 & 3524 & 3524 & 3524 & 8438 & 8438  \\ \hline
 $N_{\rm tubes}$  & \multicolumn{2}{c||}{2} & \multicolumn{2}{c||}{6} & \multicolumn{2}{c||}{6}  & \multicolumn{2}{c||}{4}   \\  \hline
 $N_{\rm wafers}$  & \multicolumn{2}{c||}{24}  & \multicolumn{2}{c||}{72} & \multicolumn{2}{c||}{72}  & \multicolumn{2}{c||}{36}   \\  \hline
  $N_{\rm wafers}$ total  & \multicolumn{8}{c||}{ 204} \\ \hline
 $N_{\rm detectors}$  & 576 & 576  & 21144 & 21144 & 21144 & 21144 & 33752 & 33752  \\ \hline
 $N_{\rm detectors}$ total  & \multicolumn{8}{c||}{ 153232} \\ \hline
 Data rate (18 optics tubes) & \multicolumn{8}{c||}{ 1.7 TB/day} \\ \hline
\end{tabular}
\end{center}
\caption{Small-aperture telescope (SAT) receiver properties. \label{tab:SATproperties}}
\end{table}

\begin{table}[htbp!]
\begin{center}
\begin{tabular}{ |l||c||c|c||c|c||c|c|| } 
\hline
 Property & ULF &  \multicolumn{2}{c||}{LF} & \multicolumn{2}{c||}{MF} & \multicolumn{2}{c||}{HF}\\ \hline
 Center frequency (GHz) & 20 & 27 & 39 & 93 & 145 & 225 & 278 \\ 
 FWHM (arcmin) &10.0 & 7.4 & 5.1 & 2.2 & 1.4 & 1.0 & 0.9  \\ 
 Fractional bandwidth & 0.25 & 0.22 & 0.46 & 0.38 & 0.28 & 0.27 & 0.16  \\ 
 NET ($\mu$K$\sqrt{\rm s}$) per detector & 438 & 383 & 250 & 302 & 356 & 737 & 1840 \\
 $N_{\rm detectors}$ per tube & 160 & 320  & 320 & 3460 & 3460 & 3744 & 3744  \\ \hline
 $N_{\rm wafers}$ per tube & 4 & \multicolumn{2}{c||}{4} & \multicolumn{2}{c||}{4}  & \multicolumn{2}{c||}{4}   \\  \hline
 \hline
 \multicolumn{8}{l}{\  } \\
 \multicolumn{8}{l}{Chile (Wide Field Survey -- 2 LATs } \\ \hline
 $N_{\rm tubes}$  per LAT & 0 & \multicolumn{2}{c||}{2} & \multicolumn{2}{c||}{12}  & \multicolumn{2}{c||}{5}   \\  \hline
 Data rate (2 LATs) & \multicolumn{7}{c||}{ 10.8 TB/day} \\ \hline
 \hline
\multicolumn{8}{l}{\  } \\
 \multicolumn{8}{l}{South Pole (Delensing Survey -- 1 LAT)  } \\ \hline
 $N_{\rm tubes}$   & 1& \multicolumn{2}{c||}{2} & \multicolumn{2}{c||}{12}  & \multicolumn{2}{c||}{4}   \\  \hline
 Data rate (1 LAT) & \multicolumn{7}{c||}{ 5.0 TB/day} \\ \hline
 \multicolumn{8}{l}{\  } \\
 \multicolumn{8}{l}{Total (3 LATs)  } \\ \hline
 $N_{\rm detectors}$  & 160 & 1920  & 1920 & 124560 & 124560 & 52416 & 52416  \\ \hline
 $N_{\rm detectors}$ total & \multicolumn{7}{c||}{ 357952 } \\ \hline
 $N_{\rm wafers}$  & 4 & \multicolumn{2}{c||}{24} & \multicolumn{2}{c||}{144}  & \multicolumn{2}{c||}{56}   \\  \hline
 $N_{\rm wafers}$ total & \multicolumn{7}{c||}{ 228 } \\ \hline
\end{tabular}
\end{center}
\caption{Large-aperture telescope (LAT) receiver properties. \label{tab:LATproperties}}
\end{table}

All the raw data from the Chile site (approximately $1\textrm{ Gbps}$ or 4\,PB/yr) will be returned 
to the US promptly for analysis and permanent storage. Only a small subset of the South Pole data can be returned via satellite, 
with the full data set being stored on disk and shipped to the US every austral summer. 
This difference in data link bandwidth means that while Chile needs only modest local computing resources, 
the South Pole site may have to support reduction to maps.

\section{Manufacturability and reliability}

The CMB-S4 reference design is based on a subset of Stage-3 technologies that worked well, were relatively straightforward to manufacture, and appear to be scalable, so we are confident that the reference design can be built, and that it will meet the performance requirements. What has to change for CMB-S4 is the level of quality control, in particular a better understanding of the requirements for subsystems, thorough testing, and testing on time. CMB-S4 cameras contain an enormous number of parts which must be tested cold, and at varying levels of integration, so a large part of the project planning is focused on testing.

The scale of CMB-S4 will require a more organized approach to manufacturing, most obviously for the detectors and readout electronics. Anti-reflection-coated lenses, cryostat windows, and thermal filters, will also be needed in quantities that will be a challenge for suppliers of Stage-3 experiment parts. All of these items are included in pre-project technical development or in the early stages of on-project development.
\section{Design approach}

CMB-S4 is a large experiment that involves complex technologies ranging from large precision structures to micro fabrication of superconducting detectors. Matching experts with critical design tasks will require a mix of in-house designers, in universities and national laboratories, and industry partners. Stage-3 experiments followed a similar approach, so we have substantial experience in choosing what to do in-house and what to contract out. The large telescopes, and mounts for the small telescopes, are probably best designed by industry, e.g., through design/build contracts. Cameras, data acquisition systems, and analysis tools involve more specialized technologies and will probably be designed in-house, though many parts will be manufactured under industry contracts.

In-house design work will be done by a mix of university researchers and national laboratory staff. National laboratories will focus on cryostats, detectors and readout, and data management, drawing on expertise in universities through subcontracts. DOE project funds will go to a lead national laboratory, and then to other national laboratories and universities via subcontracts. NSF-funded universities will focus on telescopes, sites, simulations, and analysis. NSF project funds will go to a lead institution, and then to universities via subcontracts.

Pre-project technical development for CMB-S4 is focused on reducing the highest risks (see Sect.~\ref{sec:risk}). Our top priority is demonstrating detector noise performance appropriate for CMB-S4, at full wafer scale, with two different readout technologies. We also have some development work on technology options and simulations that will be needed to support the detailed design of CMB-S4. Development activities must be complete before final design starts, which occurs at different times for different subsystems in the reference design's technically limited schedule. Technical development is currently funded through LDRDs and DOE awards to national laboratories, and we have submitted an NSF proposal for project development funds.

\section{Options}  

Viable options are an important part of any risk mitigation plan. The highest risks for CMB-S4 are in detectors and readout, so our options include two readout technologies and an alternative approach for detector fabrication. In some cases, options can reduce the overall cost of the project, though a more likely outcome is improved performance for a given cost. Collaborations tend to grow, and this often brings new options because new partners typically want to contribute their own designs and technologies. Exploring new possibilities helps to advance technology in the field, which expands the broader impact of the project. While options are generally a good thing, they can derail the overall effort; a clear, timely, selection process is critical to avoid wasting resources on bad ideas.

Options are a central component of the pre-project development for CMB-S4. We are actively working on: re-use of existing telescopes and site infrastructure; \fmux{} versus \umux{} readout; kinetic-inductance detectors; lenslet coupling; planar antennas; and commercial detector fabrication. Details are given in Appendix~\ref{app:options}. In a technically limited schedule, the time for development is short, so it is critically important to meet goals for making down-select decisions. A simple, open process is needed to develop selection criteria, review the work that has been done, and recommend a path forward. The process must also include an a priori plan for dealing with the very likely event of no clear preference. The best solution is sometimes to implement multiple options, but this should be decided well in advance to minimize disruption.


\eject
 
\chapter{Reference Design}
\label{chap:referencedesign}

The reference design for CMB-S4 is a point design that establishes the feasibility of the experiment and provides a basis for preliminary costing and project planning.  This design and the technologies used for its components are in advanced states of readiness.  However, given the ongoing development of the present generation of CMB experiments and the heritage of technical innovation from the US CMB community we expect to capitalize on technology advances for the final experiment.  

In this section we describe the technical details of the reference design.  We organize this according to the major subsystems which are: (1) Sites; (2) Large-aperture telescopes; (3) Small-aperture telescopes; (4) Detectors and readout; (5) Data acquisition and control; (6) Data management; and (7) Integration and commissioning.   Each subsection provides a discussion of the design drivers, a description of the design, and a presentation of implementation details.

\section{Sites}
\label{sec:sites}

The CMB-S4 scientific requirements necessitate a site with access to wide area sky coverage and a site that allows 
for the small area ultra-deep survey.  Chile and the South Pole are the only mature sites that can meet these two needs.   
In this section we describe the flow down from the science requirements through 
the instrument design to the site infrastructure requirements needed to support the CMB-S4 experiment. 

\subsection{Site locations: design drivers and overview}

\begin{figure}
\centering
\includegraphics[width=\textwidth]{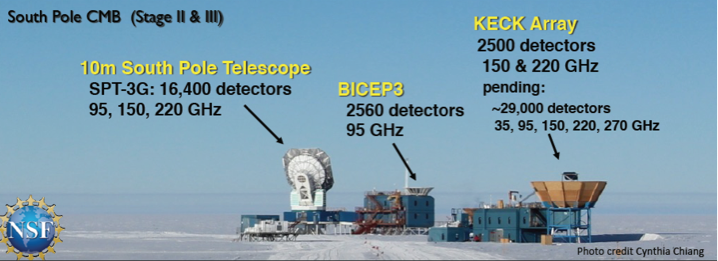}
\caption{\label{fig:spnow}
South Pole mm-wave instrumentation site as it currently exists. 
CMB-S4 would expand on the existing infrastructure at this site. 
}
\end{figure}

\begin{figure}
\centering
\includegraphics[width=\textwidth]{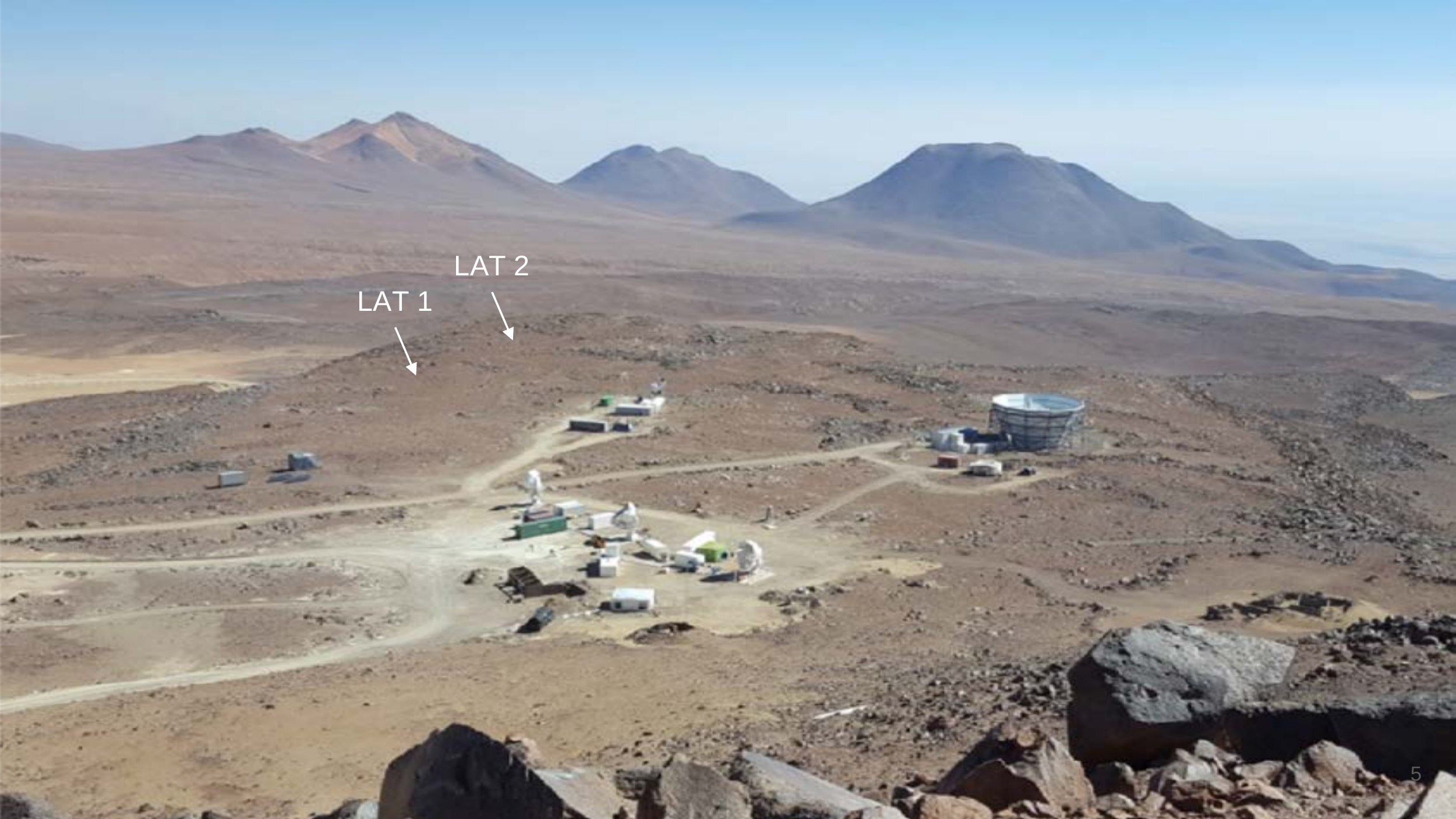}
\caption{\label{fig:chiles4}
CMB observatories in Chile are all at this Cerro Toco site.  The white arrows indicate possible LAT 
locations that would not conflict with the preliminary Simons Observatory (SO) instrumentation layout.
}
\end{figure}

State-of-the-art CMB observations require the highest and driest sites on Earth to achieve high atmospheric transmittance and low atmospheric emission within the spectral bands required.  
The two best sites developed for CMB are South Pole in Antarctica and Cerro Toco in the Chilean Andes (their current states being shown in Figs.~\ref{fig:spnow} and \ref{fig:chiles4}). 
We baseline using the Cerro Toco site in Chile, but will conduct an
early trade study to confirm this choice, as compared with using other nearby Chilean sites. 
The atmospheric transmittance at the relevent frequencies has been studied extensively (e.g., Refs.~\cite{Lane1998,ChileSiteStudy}). 
Both of these sites have been in use for several decades and have hosted many CMB telescopes (see, for example, \cite{Ogburn2012, Carlstrom2011, Fowler2007, Arnold2014,Harrington2016}). 
Known information about the atmospheric conditions at these sites was used in the sensitivity forecasts described in Appendix A. 

The Chilean Atacama site, at a latitude of 23$^\circ$ South, has access to more than 80\% of the sky,
enabling observations of the 70\% of the sky required by many of the science goals discussed above. 
The South Pole site, on the other hand, is better suited to deep integration on a small portion of the sky. 
As described in Chapters 2 and 3, this makes the South Pole site particularly well-suited to making observations on the ultra-deep field.  
The reference design presented here achieves the CMB-S4 science goals with the configuration of telescopes shown in Table~\ref{tab:siteconfig}. 
If $r$ is measured between now and the Preliminary Baseline review (DOE CD-1/3a, currently planned for Q3 2021), 
we will re-evaluate this configuration in the context of the measured value of $r$. 
Up until that point, both sites will be designed to be capable of hosting all the SATs. 

\begin{table}
\begin{center}
\begin{tabular}{|l|c|c|c|}
\hline 
Site & Number of LATs & Number of SAT cryostats & Number of individual SATs \\
\hline 
Chile & \nlatchile & \nsatcryochile & \nsattubeschile \\
South Pole & \nlatsp & \nsatcryosp & \nsattubessp \\
\hline
\end{tabular}
\caption{
\label{tab:siteconfig}
Number of LATs and SATs at each site.
}
\end{center}
\end{table}

The long history of CMB observations from these sites has led to significant infrastructure buildup at each, 
and a solid understanding of the additional infrastructure required to enable CMB-S4.  
Large-aperture and small-aperture telescopes of similar size to those required for CMB-S4 have already been deployed to each site;  the primary new features of CMB-S4 which drive project and site requirements are the number of detectors per telescope, and the number of small-aperture telescopes that will be deployed.  
These demand a larger-scale integration effort, higher power requirements, and higher data bandwidths than previously fielded experiments at each site. 
Options to re-use infrastructure are discussed in Sect.~3.6. 
We assume here that we will not re-use any existing CMB instrumentation infrastructure at the site, and cost the site development appropriately. 
The infrastructure required at each site is described below. 
Two critical inputs from the instrument to the infrastructure required are the overall electrical power that will be used and data rate that will be generated. 
Data rate calculations are shown in Chapter 3. 
A summary of the electrical power requirements is in Table \ref{tab:sitepowerbudget}. 
Implementation of the power generation necessary to supply this power will be an important long-lead time item to consider in site construction. 

\begin{table}
\begin{center}
\begin{tabular}{|l|c|c|}
\hline 
& \multicolumn{2}{c|}{Average power} \\ \hline 
Equipment &  Chile & SP \\ 
& Total [kW] & Total [kW] \\ 
\hline 
LATs (including receiver and readout but not DAQ) & 210 & 105 \\ 
SATs (including receiver and readout but not DAQ) & 0 & 264 \\ 
Cooling system & 30 & 14 \\ 
DAQ, compute, and office & 44 & 34 \\ 
Data management & 10 & 50 \\ 
Site power & 30 & 20 \\ 
\hline
Total & 304 & 487 \\ 
\hline
\end{tabular}
\caption{
\label{tab:sitepowerbudget}
Power budget for each site.
Note that requirements will be re-evaluated in the context of any redistribution of SATs.
}
\end{center}
\end{table}

It will be critical to begin planning for long lead-time site infrastructure, including foundations, buildings and power, at both sites to ensure they are ready in time for telescope deliveries.
	
\subsection{Chilean site}

A possible layout of the Chilean site is shown in Fig.~\ref{fig:chiles4}. 
This layout uses the area that has been studied by existing CMB instruments. 
The maximum horizon blockage seen within this area (presented by Cerro Toco) is 15$^\circ$. 
Rock in this area is generally appropriate for construction of the necessary foundations, 
and the instruments will not present significant horizon blockage to existing instruments. 

The critical schedule path for the site work in Chile is either the telescope foundation fabrication 
or the political negotiations necessary to establish a legal presence in Chile and grant that 
legal entity the right to build on the site. 
If either these tasks or the inputs necessary to complete these tasks are delayed, the readiness of the Chilean site will be delayed. 
Note, however, that site readiness is not currently on the critical path of the project. 

The following subsections are organized according to the work breakdown structure (WBS) for site work in Chile. 
While most of this WBS organization is similar to that for the South Pole (Sect.~\ref{sec:sitesp}), there are important differences. 
The low-elevation facility necessary in Chile is functionally equivalent to the South Pole station where people live and eat while deployed. 
The safety requirements at both sites are critical, with Chile having the added issue of high elevation, and the project having to supply its own safety infrastructure. 
 
\subsubsection{Management, personnel, and safety}
\label{sec:chilemng}

Management, Personnel and Safety includes some of the most important parts of the site work package. 
These are critical to support the successful integration and commissioning of the instruments, and will lay the groundwork for the operations phase of the project. 

\paragraph{Political and legal issues:}
The major tasks that need to be done to have formal approval of the project's siting in Chile are as follows: 
(1) determine who will represent CMB-S4 legally in Chile; 
(2) establish that legal presence, and ensure they can represent CMB-S4; 
(3) determine desired location of site; 
(4) determine if our agreement will be with Parque Astron{\'o}mico or CONICyt; 
(5) negotiate agreement with correct government agency; 
(6) determine outreach plan; and
(7) sign off on agreement for CMB-S4 in Chile. 
Items (1)--(5) could each take many months. 
Item~(3) can be in parallel with others, but otherwise they are somewhat required to be in series. 
The CMB-S4 team aims to mitigate the schedule risk by aiming to retire the first four steps in this negotiation as part of the interim project office (IPO) work between CD-0 and CD-1. 

\paragraph{Outreach to the Chilean community:}
Public outreach within Chile by CMB-S4 is critical to satisfy the requirements of the Chilean government and to mitigate the risk of Chilean community members objecting to our use of their resources. 
We will develop an outreach program that is visible and has the backing of the Chilean government.

\paragraph{Personnel:}
Personnel here can be separated into US management and engineering personnel, 
who are needed early in the project to shepherd the political and design process, 
and Chilean on-site management, engineering, and technical personnel who 
will implement plans on-site. It is key that the relevent civil and mechanical engineering 
personnel be identified on-schedule 
so that design schedule risk can be retired as soon as possible after receiving the relevant input from: (1) the vendors supplying the LAT and SAT mount; and (2) the geotechnical study of the area once the site layout is established. This engineering effort 
will be coordinated with Chilean contractors who have important expertise. 

\paragraph {Safety:}
The high-elevation safety requirements set by Chilean law 
will be complied with; this requires a safety officer knowledgeable about those requirements, and the 
presence on-site of a safety officer during major construction, integration, and commissioning activities. 
There is significant local expertise (because of previous astronomy and mining projects) to fill this position. 
A safety policy will be developed consistent with Chilean law and constraints coming from 
the relevant US institutions, and that safety policy will be reviewed by an external panel and the safety officer.  

\subsubsection{Site infrastructure and logistics}
\label{sec:chileinfr}

This work package contains a large number of separate tasks. 
None of them are schedule-critical provided site permission 
(described in the section on political and legal issues above) is obtained in a timely manner. 
The major tasks that need to be done to have formal approval of the project's siting in Chile include plans for: 
(1) accessing the site with people and equipment; (2) a site layout; (3) power generation; (4) electrical distribution; (5) data/communication to and within the site; (6) non-telescope foundations, buildings and interconnects; (7) cooling system; (8) site monitoring for data quality; (9) tools and supplies; (10) telescope infrastructure; and (11) the low-elevation facility.

Item (4) is discussed in more detail in Chapter 3. 
We note here simply that there is a path to full-bandwidth communication from the Chilean site to North America and beyond with ms-level latency. 
Power generation in the baseline design is achieved through diesel generators, though we note that there are risks associated with this that could be mitigated by implementation of solar and wind power. 
Some other important points are discussed in more detail below. 

\paragraph{Low-elevation facility:}
Personnel working at the Chilean site will commute daily to the site from the town of San Pedro de Atacama, or somewhere in the vicinity. 
Because of the tourist economy in San Pedro, there are a number of places capable of housing and providing food for a deployment team. 
Requirements for the low-elevation facility should be developed early so that we can reach an agreement with one of these establishments and retire cost risk associated with this activity. 

\paragraph{Telescope infrastructure and non-telescope foundations, buildings, and interconnects:}
The telescope infrastructure itself, for the purpose of this work package, is simply the mechanical interface to the telescope. 
This is the foundation that provides the required stability to meet the long-term tilting requirement of the telescope, and to prevent the telescope from falling over in wind, snow load, or earthquake conditions. 
The foundation design will be significantly different at the two sites, since at the South Pole ice pads are used, while in Chile the foundation will be cement anchored to rock.
For this reason, foundation design is part of the site work package. 
The metal anchoring ring (foundation insert) that inserts in the foundation and provides the attachment points for the telescope is part of the LAT and SAT work packages for their respective foundations. 
The site team is responsible for providing a summary of the possible environmental conditions to the telescope vendors. 
The telescope vendors will use that input to calculate possible loads on the foundation created by the telescope in all environmental conditions. 
The LAT and SAT vendors will then be expected to provide an interface specification with the design of that metal foundation insert and the loads that they have calculated. 
The site team is then responsible for contracting the design of a foundation that provides the necessary anchoring. 

The other buildings (such as labs for receiver assembly, office to house personnel, and structures to house utilities), the foundations associated with them, and the interconnects between them, all need to be laid out and specified based on the instrument requirements. 
Establishing these instrument requirements in time for the necessary procurement 
will be an important task for each instrument team and will require close interfacing between the instrument teams and the site design team.

\subsection{South Pole}
\label{sec:sitesp}

At the South Pole, site infrastructure for previous experiments has been designed and installed collaboratively with NSF and their primary 
contractor for Antarctic logistics, currently ASC.  We expect to follow the same path for CMB-S4's site infrastructure requirements, as detailed below.
Outdoor construction is done in the austral summer (November--February), while interior work can be done in the austral winter (February--November).  
Cargo arrives at the Pole in the austral summer, either by C-130 or via a month-long land traverse from McMurdo.  Cargo gets to McMurdo either via air (C-17 or C-130) throughout the season, or by ship late in the austral summer.  The slower methods (ship and traverse) are preferable from a cost standpoint, but the air routes are quite capable;  SPT, for example, was shipped entirely by air transport to Pole.  We currently baseline a mixture of air 
transport and traverse for CMB-S4. 

\subsubsection{Laboratory building}
CMB-S4 will require a central space (used for the LAT and SATs) with a variety of lab facilities, such as clean room space for assembling focal planes,
a small shop for mechanical work, an electronics space for assembling, debugging and repairing electronics, and a room to house a data analysis cluster
and data storage.  The scale of this building is expected to be similar to the Dark Sector Lab (DSL) that currently houses BICEP and is attached to SPT, or the Martin Pomerantz Observatory (MAPO) which is attached to the Keck Array.  

\subsubsection{Telescope buildings}
A longstanding practice for winterover telescopes at the South Pole is to have all drive components and receivers accessible from a heated space.
BICEP is built into the top of DSL.  SPT is built on a separate tower, with a heated equipment room beneath it from which the receiver can be accessed when the telescope is docked.  The Keck Array is on a tower with a room below the telescope, similar to SPT, but connected to a different building (MAPO).

All the new CMB-S4 telescopes will be placed on towers, with heated equipment rooms below them, attached to the new laboratory building mentioned above by enclosed, heated walkways.  This is critical to enabling ongoing summer and winter maintenance and operations.  The towers will be designed
to meet specifications spelled out by the telescope contractor, but will not be part of the telescope test build in the 
United States.
They will be installed 
on site prior to the telescopes' arrival.  
We have budgeted for these towers and equipment rooms by scaling the cost of the SPT versions of these to today's dollars.  

\subsubsection{South Pole Station support}
NSF's Admundsen-Scott South Pole Station supports all the science projects located there.  The elevated station can house approximately 155 people;  overflow buildings have been used in busy construction seasons, boosting the available population above 200.  Winter population has recently been between 40 and 50, but has been much higher in the past during periods of major construction.  

NSF provides housing, meals, cargo services, and
data transmission for all projects.  Large projects such as CMB-S4 are charged a per-person rate for flights, housing and meals, and a rate
per pound for cargo depending on the transport type.  We have used the NSF-provided rates for these in our budget.
Data transmission is currently provided as a service at no cost to the projects, but is limited to existing channels.

As shown in Table~\ref{tab:sitepowerbudget}, the CMB-S4 average power requirement is nearly 500\,kW. This will likely require new power generation equipment at the South Pole.  We are working with NSF to better understand the currently available power limit, 
and to ensure CMB-S4's future power requirements can be met.

\section{Large telescopes}
\label{sec:largetelescope}

\subsection{Requirements and design drivers}

The 
wide field survey
drives the key measurement requirements for the large telescopes (see Chapter~\ref{chap:instrumentoverview}):
\begin{enumerate}
\item angular resolution $<1.5\,\textrm{arcmin}$ at 150\,GHz;
\item noise level of a few $\mu\textrm{K-arcmin}$ at 95 and 150\,GHz over approximately 70\% of the sky;
\item measurements at frequencies up to at least 270\,GHz to control foregrounds and enable astrophysics science goals.
\end{enumerate}

The angular resolution and frequency range, combined with typical characteristics of the atmosphere at a good millimeter-wavelength site, and a typical CMB experiment scan strategy, lead to the telescope design requirements in Table~\ref{tab:LargeTelescopeDesignRequirements}. The delivered half wavefront error (HWFE), or equivalent surface error, ensures diffraction-limited performance at $\lambda=1\,\textrm{mm}$: the scan pointing knowledge ensures that mixing of $E$ to $B$ modes due to pointing errors does not limit measurements of $r$ \cite{Hu:2002vu}; the scan following error is small enough to ensure uniform coverage of the target field; the scan speed is fast enough to freeze atmospheric brightness fluctuations (typically at a height of 1\,km, and drifting along at $10\,\textrm{m\,s}^{-1}$, so the speed on the sky is $0.6\,\textrm{deg s}^{-1}$); the scan turn around time allows efficient observations (scans are typically a few tens of seconds, so any wasted time must be just a few seconds); and the emissivity is small enough to ensure that loading on the detectors is dominated by the CMB and atmosphere, not the telescope.

Achieving the low noise requires $\approx250,000$ detectors, but the large telescopes in Stage-3 experiments only support $\approx10,000$ detectors. Large telescopes are expensive, so a design with many more detectors per telescope is critical for controlling the overall cost. A wide-field telescope design can support $\approx100,000$ detectors; just three of these telescopes will be needed 
for both the 
deep and wide field
survey and for the ultra-deep survey.

Low noise is only useful if systematic errors are also small. Systematic errors are dominated by unwanted pickup through the telescope sidelobes. Ground pickup is the main effect, but pickup from the galaxy, Sun, and Moon are also a concern. Minimizing the telescope sidelobes is therefore a key design driver. Sidelobe calculations are notoriously difficult and inaccurate, so the design has to rely heavily on practical experience from Stage-3 experiments. Small sidelobes require: (i) an unblocked (i.e., offset) optical configuration; (ii) low illumination of the mirror edges (i.e., small spillover); (iii) small scattering due to gaps between mirror panels; (iv) large clearances to avoid clipping of the beam by telescope structures; and (v) absorbing baffles and reflective shields to terminate unwanted sidelobes on stable surfaces, or on the cold sky. The large telescopes in Stage-3 experiments all have reflective, comoving shields that wrap around the lower part of the telescope; CMB-S4 telescopes will have more extensive shielding around the entire telescope.

Spillover onto warm surfaces increases loading on the detectors, which in turn increases the noise. Spillover is always a challenge, but the requirements for CMB-S4 are similar to those for Stage-3 experiments, so the same general approach can be used: a cold stop to limit the size of the beam on the mirrors; and oversized mirrors to direct any spillover onto the cold sky.

Polarization errors are not a strong design driver for the large-telescope mirror configuration, 
because those polarization errors are generally much smaller than 
those associated with the planar antennas or horns that feed the detectors.

If CMB-S4 is to achieve its science goals in a reasonable time, the observing efficiency must be high, so the telescopes must be reliable and easy to maintain. In practice, this means locating the drive mechanisms and electronics in protected areas, with easy access, providing lifting equipment to support the removal and installation of heavy motors, gearboxes, and brakes, and planning for on-site replacement of the axis bearings.

\begin{table}
\centering
\begin{tabular}{|l|l|l|}
\hline 
Parameter & Value & Notes\\
\hline
Aperture diameter$^{a}$ & 5.5\,m & 1.5\, arcmin beamwidth at 150\,GHz\\
HWFE & $<37\,\mu\textrm{m rms}$ & 80\% Strehl ratio at $\lambda=1\,\textrm{mm}$\\
Scan pointing knowledge & $<5\,\textrm{arcsec rms}$ & $<1/10\textrm{th}$ beamwidth at $\lambda=1\,\textrm{mm}$\\
Scan following error & $<10\,\textrm{arcsec rms}$ & 1/4 beamwidth at $\lambda=1\,\textrm{mm}$\\
Scan speed AZ(EL) & $3(1.5)\,\textrm{deg s}^{-1}$ & $\approx1\,\textrm{deg s}^{-1}$ on the sky to freeze atmosphere\\
Scan turn around time & 2\,s & requires $6\,\textrm{deg s}^{-2}$ acceleration\\
Emissivity & $<0.01$ & cf.~a few \% atmospheric loss at mm wavelengths\\
\hline 
\end{tabular}
\caption{Large telescope design requirements. \label{tab:LargeTelescopeDesignRequirements}
Note: $(a)$
the physical aperture should be a little larger, e.g., 6\,m, in order to reduce spillover and scattering.}
\end{table}

\subsection{Reference design summary}

The large telescopes are a 6-m crossed-Dragone design on a conventional elevation over azimuth mount, with the instrument boresight aligned with the elevation axis. This alignment facilitates instrument rotation and enables partial telescope boresight rotation, because the elevation axis can rotate a full $180^\circ$, from horizon to horizon, enabling two distinct telescope orientations at every azimuth position. The design copies the SO-LAT and CCAT-prime, and was adopted because this is the only high-throughput design that will be demonstrated on the timescale of CMB-S4. 

The reference design has two new large telescopes in Chile to support the 
deep and wide survey.
An option is to re-use the SO-LAT and CCAT-prime, in which case no new large telescopes would be built for Chile. 
A third large telescope is needed at the South Pole to support the ultra-deep survey. 
The South Pole large telescope has the same optical configuration as the SO-LAT, 
but with some modifications to the mount to accommodate logistics constraints and operation in the cold. 

Since the telescopes are copies of the SO-LAT and CCAT-prime, no prototyping is needed, and procurement contracts can be awarded early, so telescope testing can be completed before cameras are delivered.

\begin{figure}
\begin{center}
\includegraphics[width=6in]{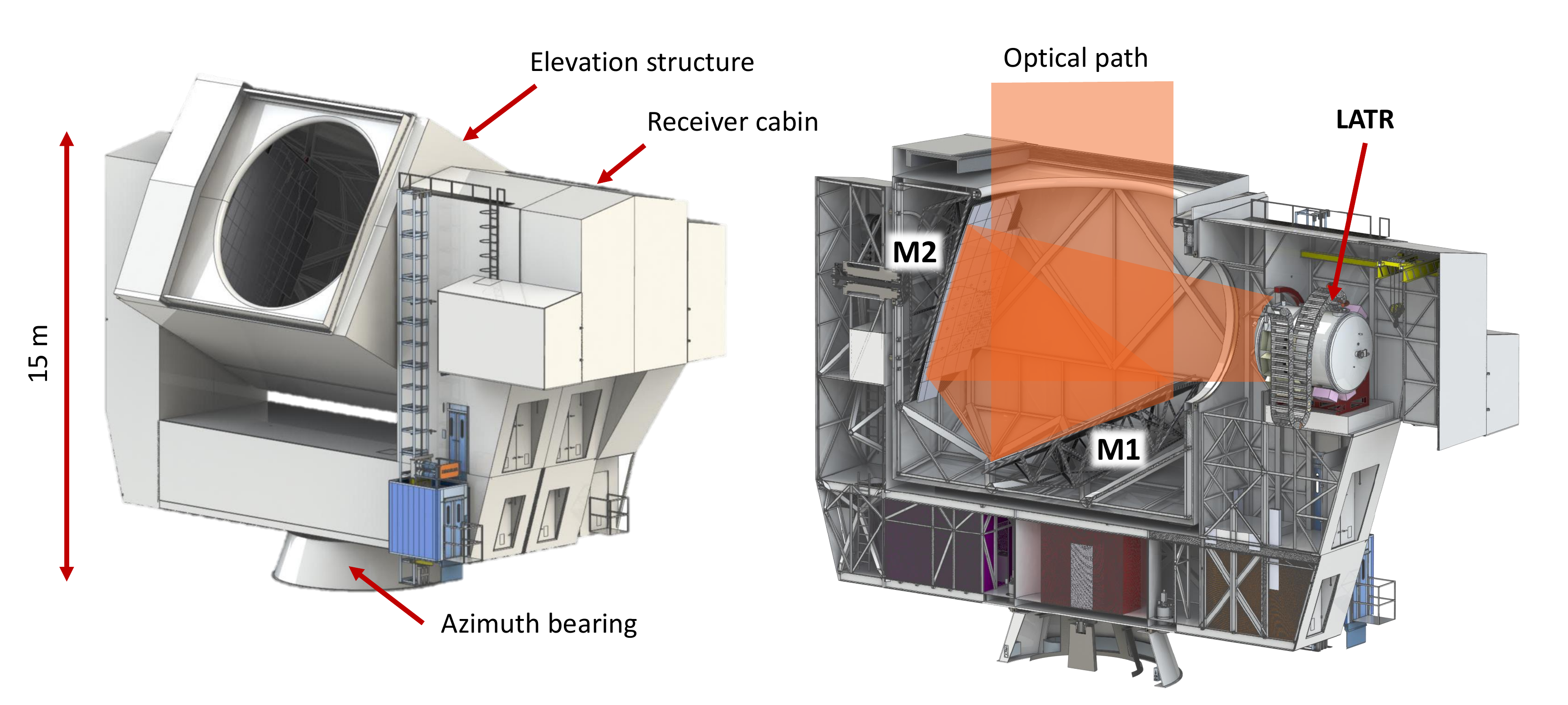}
\end{center}
\caption{
The CMB-S4 Large Aperture Telescope reference design is based on the design of the SO-LAT and CCAT-prime telescopes \cite{parshley:2018a,Galitzki:2018wvp}.  The mirrors (M1 and M2) are completely enclosed in the co-moving elevation structure (left), which will improve sidelobe mitigation compared to existing telescopes. The Large Aperture Telescope receiver (LATR, right) is aligned with the telescope elevation axis.  This enables two new and valuable features for suppressing systematics compared to existing large telescopes: (1) the LATR can either track the elevation angle of the telescope or to be rotated independently of the telescope angle; and (2) partial boresight rotation of the telescope can be achieved because the telescope boresight can be rotated from one horizon to the other, between 0 and 180$^\circ$.
}
\label{fig:LAT}
\end{figure}

\subsection{Optics}

The high-throughput off-axis crossed-Dragone (CD) telescope design was first proposed by Corrado Dragone \cite{Dragone:1978a}. This configuration has previously been used for CMB measurements by the QUIET \cite{Bischoff:2013} and ABS \cite{Kusaka:2018} projects. Larger-aperture CD designs were recently shown to be capable of achieving many times larger throughput than existing 5--10\,m microwave telescopes, particularly when combined with refractive optics tubes \cite{Niemack:2015iae}. In 2017 a 6\,m CD design that includes coma corrections \cite{Dragone:1983a} was adopted for CCAT-prime and the Simons Observatory (Fig.~\ref{fig:LAT}) \cite{parshley:2018b, Galitzki:2018wvp, parshley:2018a}, and both telescopes are scheduled to begin observations in Chile in 2021.

\begin{figure}
\begin{center}
\includegraphics[height=3in]{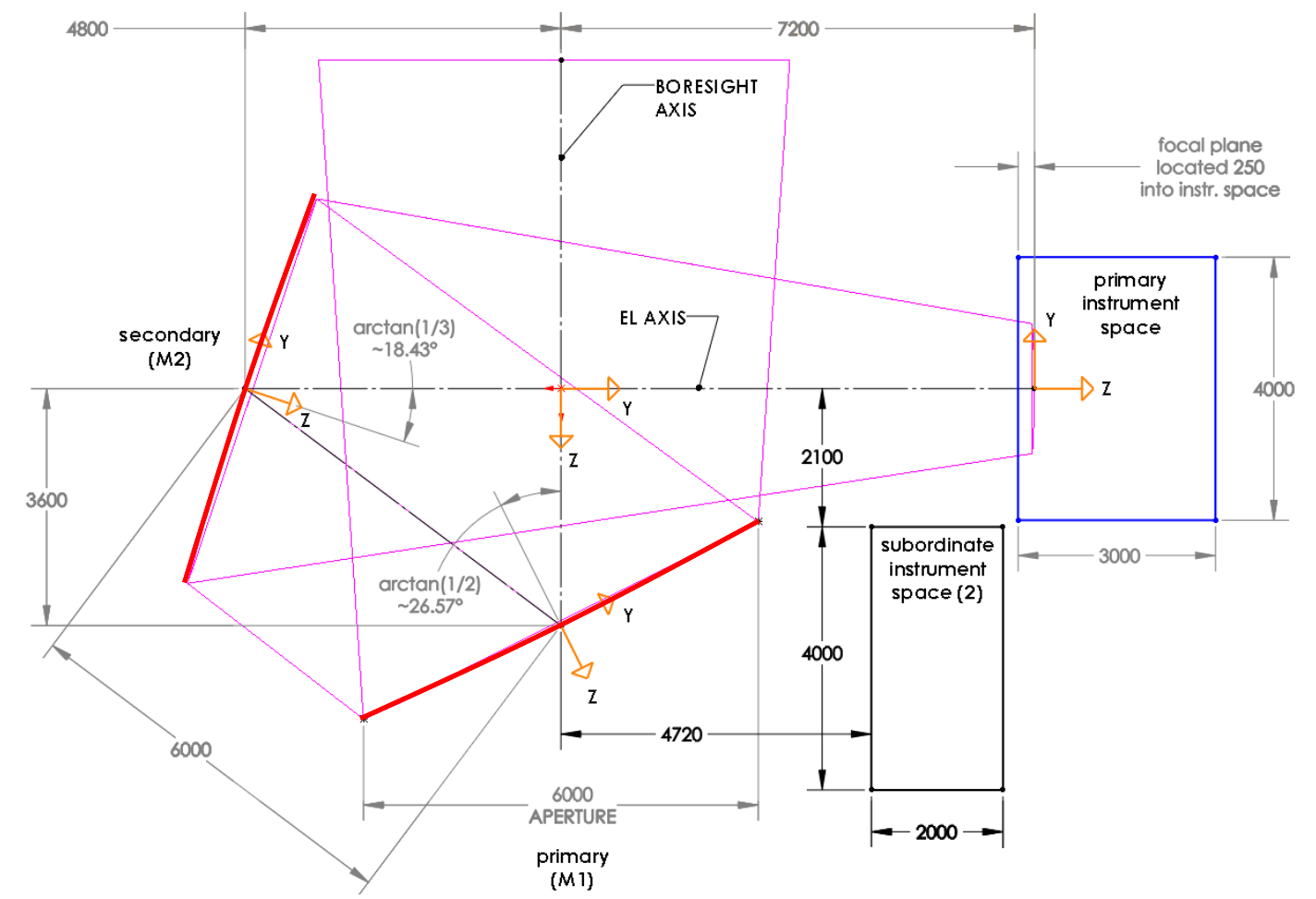}
\includegraphics[width=2.in]{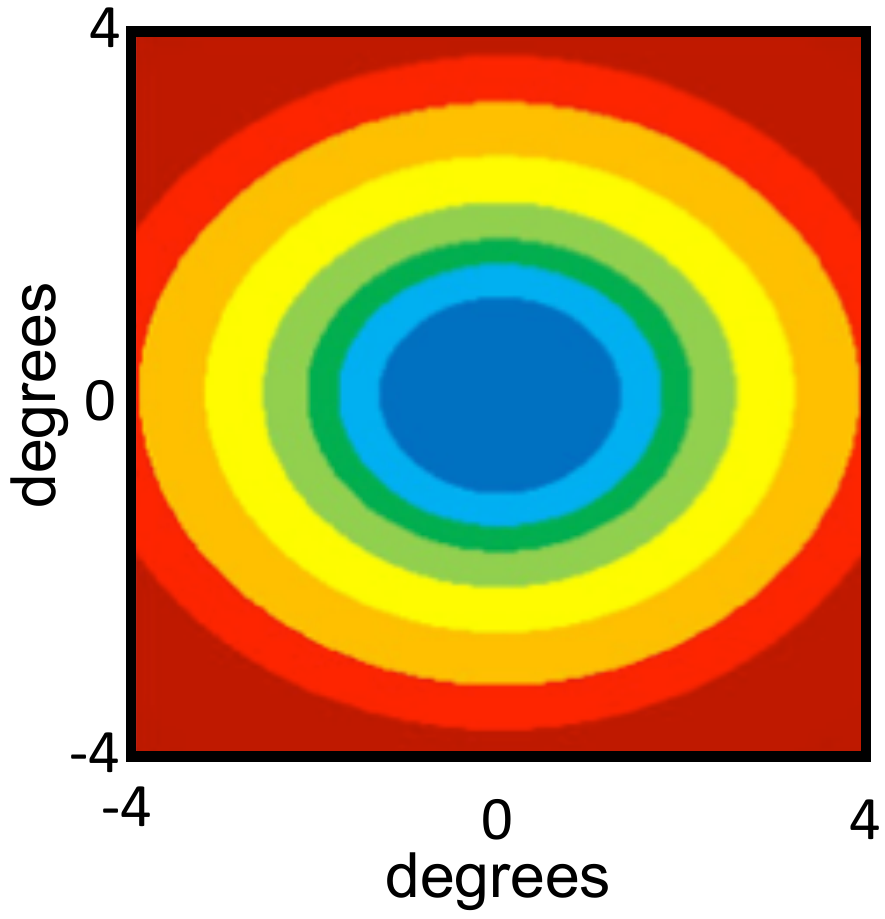}
\end{center}
\caption{
Mirror optics definition (left) for the CCAT-prime and SO-LAT \cite{parshley:2018b}.  The pink rays indicate the geometric keepout zones for a 7.8$^\circ$ diameter FOV illuminating the full 6-m aperture primary mirror. The primary and secondary instrument spaces are also shown. The mirrors alone provide diffraction-limited image quality across much of the usable FOV as shown by the contours (right). The contour colors (blue through red) show the regions where the Strehl ratio is $> 80$\% at 870, 490, 345, 230, 150, 100, and 75\,GHz \cite{parshley:2018a}. Refractive optics can improve the optical performance beyond what is shown here within individual optics tubes. 
}
\label{fig:LAToptics}
\end{figure}

The CD layout offers several advantages compared to other telescope designs. For CMB-S4 the most important advantages are the large diffraction-limited field of view (FOV) combined with the compactness of the two mirror configuration (Figs.~\ref{fig:LAT} and \ref{fig:LAToptics}). While modestly larger diffraction-limited FOVs have been achieved with other telescope configurations, including three mirror anastigmat designs, the alternative designs were not nearly as compact and use of a third mirror will degrade the sensitivity of the detectors due to the extra mirror emission.

The polarization properties of the CD configuration have been shown to be superior to Gregorian and Cassegrain designs, 
which is in part due to the smaller angles of incidence on the reflecting surfaces \cite{Tran:2008}. 
Systematics associated with the 
telescope mirrors are much smaller than systematics caused by the 
lenses, filters, and feedhorns inside the cameras \cite{Gallardo:2018rix}.

\begin{figure}
\begin{center}
\includegraphics[width=\textwidth]{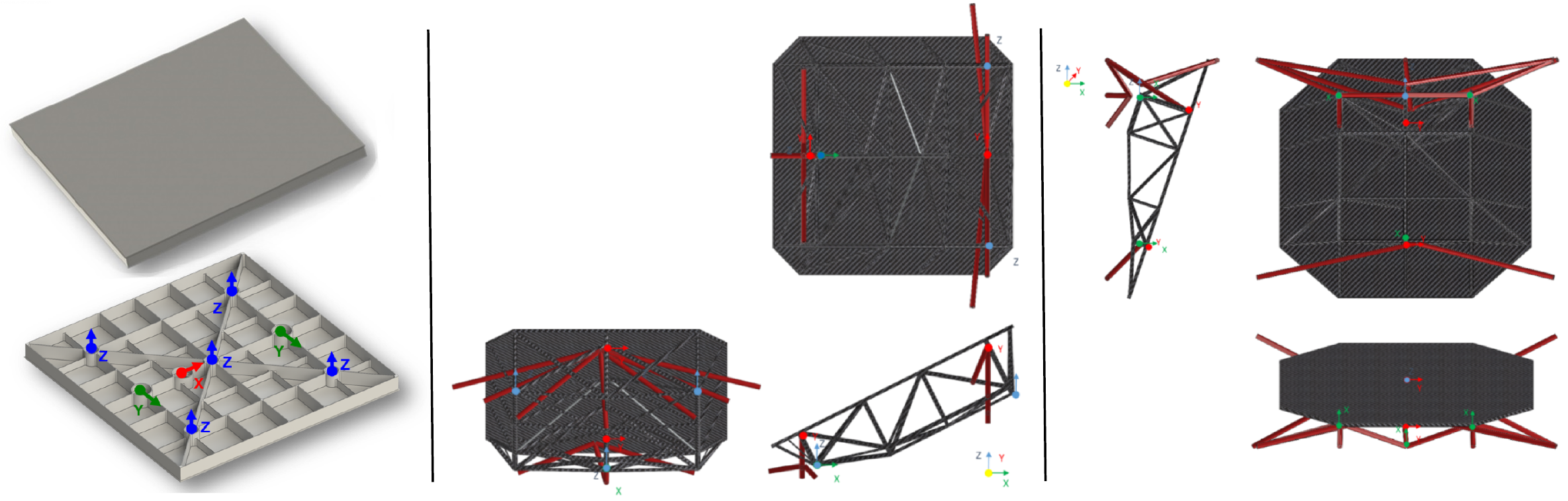}
\end{center}
\caption{
{\it Left:} Mirror panel views, reflecting surface (top), backside (bottom).  The panel is 700\,mm on a side and significantly light-weighted from a solid block of aluminum.  Eight adjusters locate the panel.  The five $z$-axis adjusters allow for some compensation of low order distortions. {\it Middle:} Three views of a preliminary carbon fiber bus structure for the primary mirror panels. The red elements are tuned CTE beams that minimize the half wavefront error and pointing error due to imperfect mirror alignment from thermal effects.
{\it Right:} Similar views of the carbon fiber bus structure for the secondary mirror panels.
}
\label{fig:mirrors}
\end{figure}

The mirrors are each approximately 6\,m in diameter and will be segmented into many panels (Fig.~\ref{fig:mirrors}) \cite{parshley:2018a}. All mirrors on CMB telescopes near this diameter have been segmented thus far. The sidelobe pickup associated with the gaps between the mirrors is sufficiently small that it will not contaminate the high multipole $E$-mode polarization,  lensing $B$ modes, or  temperature signals that will be the foci of measurements with these telescopes. The gaps between the mirrors for the CCAT-prime/SO-LAT design are required to be $<$0.5\% per mirror at the nominal observing temperatures. This requirement both minimizes loss in the gaps and minimizes the amplitude of the sidelobes due to the panel gaps. The mirrors are supported by carbon fiber bus structures as shown in Fig.~\ref{fig:mirrors} to minmize variations in HWFE and pointing error due to thermal variations. Table \ref{tab:HwfeBudget} shows the HWFE budget.

The reference design provides a CD FOV that is $>$2\,m diameter. An FOV this large can make coupling to arrays of detectors challenging. A natural approach to take advantage of such as large FOV is to split it into multiple independent optical paths or optics tubes \cite{Niemack:2015iae,Dicker:2018ytl}. These optics tubes are designed to be modular and easily replaceable, which facilitates deployment of different frequencies in each optics tube. Each tube has an independent image of the primary, or Lyot stop, to control illumination of the telescope mirrors. The stop is cooled to 1\,K to minimize excess background loading on the detectors. The reference Lyot stop design would geometrically illuminate 5.5\,m of the 6.0-m aperture primary mirror. Thus, the mirror will provide a 0.25-m baffle in radius to control diffracted spillover, and designs for low precision baffles to extended the mirrors are being studied now \cite{Gallardo:2018rix}.

\begin{figure}
\begin{center}
\includegraphics[height=1.85in]{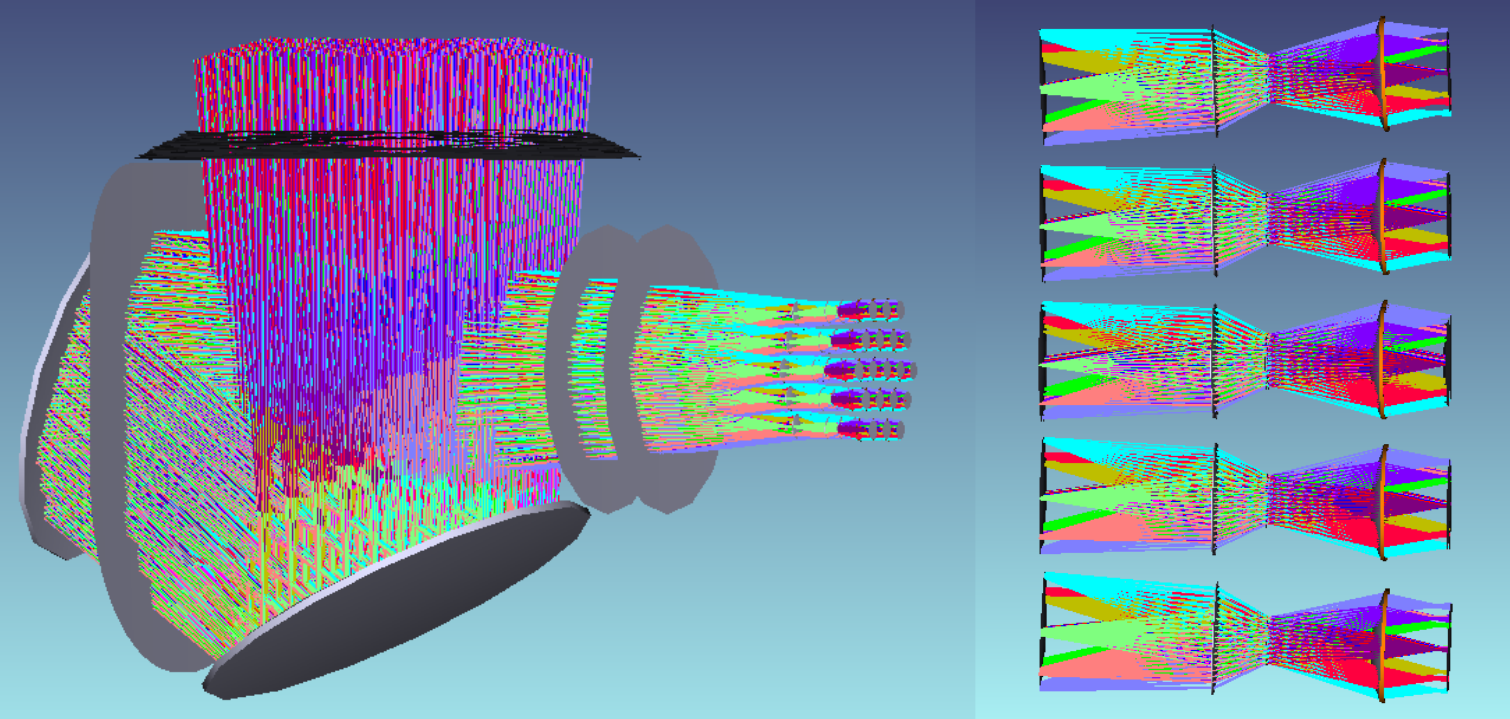}
\includegraphics[height=1.85in]{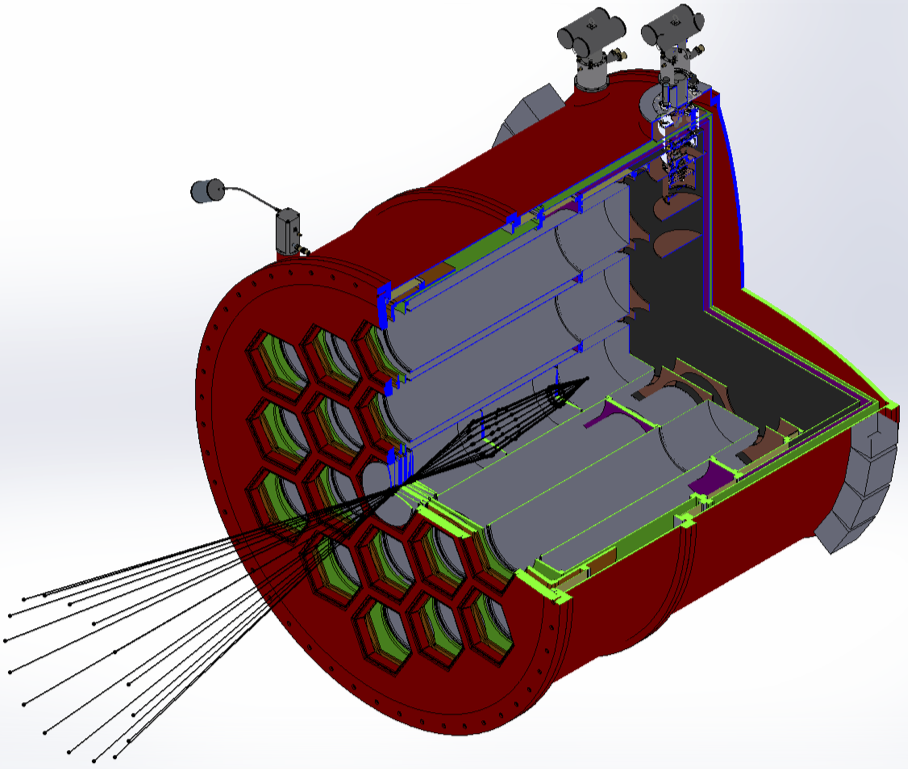}
\end{center}
\caption{
{\it Left:} Raytrace of the CD telescope design with 19 optics tubes that are each designed to illuminate 5.5\,m of the 6-m aperture telescope. {\it Middle:} Side view showing five of the 19 optics tubes. 
{\it Right:} Preliminary design for a 19-optics-tube cryostat developed by the Simons Observatory collaboration.
}
\label{fig:LAT19tubes}
\end{figure}

Several different sizes of optics tubes have been studied for the SO-LAT/CCAT-prime telescope design, ranging from 0.2-m to 2-m diameter optics. There are a variety of practical constraints on the detector array size (detectors are typically fabricated on 150-mm wafers), availability of optics materials (e.g., the maximum single crystal silicon diameter is 450\,mm), vacuum windows, filters, and mechanical structure between tubes that influence the optimal optics tube size. A few of the optics tube options that were studied and compared are discussed in \cite{Dicker:2018ytl,Hill:2018rva}. These and related cost versus sensitivity analyses led to the selection of optics tubes with 450-mm spacing for SO-LAT/CCAT-prime. A preliminary design for a 19-optics-tube camera for these telescopes is shown in Fig. \ref{fig:LAT19tubes}.

The detector arrays within each optics tube will be feedhorn-coupled transition-edge sensor detectors. Feedhorns are one of the most mature technologies for controlling polarization systematics. They have circular beams and small polarization errors. By coupling the feedhorns through a cryogenic Lyot stop, they can be packed much more closely together, which improves the overall instrument sensitivity, but can significantly increase detector count. Roughly half of the beam from each feedhorn falls on the Lyot stop (see Table \ref{tab:OpticalLoss}), but the sensitivity degradation is small because the stop is very cold.

\begin{table}
\centering
\begin{tabular}{|l|c|}
\hline 
Subsystem & HWFE ($\mu\textrm{m rms}$)\\
\hline 
Primary & 18\\
Secondary & 15\\
Telescope alignment & 21\\
Camera filters & 10\\
Camera lenses & 10\\
Camera alignment & 10\\
Total & 36\\
\hline 
\end{tabular}
\caption{HWFE budget. \label{tab:HwfeBudget}
Telescope errors include gravity, 15-K ambient temperature change, 1.4-K temperature gradients (in direct sunlight), $\textrm{9\,m\,s}^{-1}$ wind (3rd quartile in Chile), and 0 scan acceleration.}
\end{table}

\begin{table}
\centering
\begin{tabular}{|l|l|l|l|}
\hline 
Subsystem & Loss & Temperature & Notes\\
 & 30/40/95/150/220/270\,GHz & (K) & \\
\hline 
Telescope mirrors & & & \\
\hspace{3mm}Spillover & 0.02 & 300 & 2 mirrors\\
\hspace{3mm}Ohmic loss & 0/0/0.004/0.01/0.02/0.03 & 300 & 2 mirrors\\
\hline
Camera window & 0.001/0.002/0.005/0.01/0.015/0.02 & 300 & 1/2 inch UHMWPE\\
\hline
Filters & & & \\
\hspace{3mm}Alumina & 0.0002/0.0003/0.0006/0.001/0.0015/0.002 & 80 & 2\,mm thick\\
\hspace{3mm}IR blockers & 0.004 & 300, 80, 40, 4 & 4 blockers\\
\hspace{3mm}Low-pass & 0.03 & 4, 1, 0.1 & 3 filters\\
\hline
Lenses & 0.006/0.009/0.02/0.03/0.05/0.06 & 4, 1, 1 & 3 lenses\\
\hline
Feedhorn & 0.72/0.51/0.75/0.57/0.35/0.20 & 1 & Spillover on cold stop\\
\hline
Total & 0.77/0.57/0.82/0.67/0.48/0.36 & &\\
\hline 
\end{tabular}
\caption{Optical loss budget. \label{tab:OpticalLoss}
See Table~\ref{tab:filter_temp} for filter details.}
\end{table}

\subsection{Mount}

The mount is a fork-style elevation over azimuth design, sized to accommodate the crossed-Dragone optical layout. The mount has three main parts: a fixed support cone at the base; the yoke, which rotates in azimuth and contains all the science equipment; and the elevation housing that supports the mirrors.  The design allows for elevation angles past zenith, which is useful for measuring systematic errors, and can even be driven to $-90^\circ$ to facilitate maintenance. The chief ray between the secondary and the focal plane is along the elevation axis, so the camera can be attached to the yoke structure. Gravity acts mostly in-plane for the secondary mirror, minimizing gravitational deformation. The overall size of the structure is approximately $23\,\textrm{m long}\times8\,\textrm{m wide}\times16\,\textrm{m tall}$, with the elevation axis $\approx11\,\textrm{m}$ above ground for the Chile version. The total weight is 210\,T (see Table~\ref{tab:TelescopeMass}).

The elevation housing contains the primary and secondary mirrors.  The beam from the sky enters through a shutter that can be closed during bad weather, reflects off of the primary, over to the secondary, and then exits at a right angle to the entrance beam through the 4-m ID elevation bearing. The focal plane is $\approx3\,\textrm{m}$ beyond the bearing.

The mount is a welded steel structure, split up into transportable sections. The mirrors have machined aluminum panels, $\approx0.7\,\textrm{m}$ on a side, mounted on a CFRP back up structure.  Each panel has three adjusters to set the in-plane position, and five adjusters to set the height of the reflecting surface. The five height adjusters allow some correction of low-order machining errors. The back up structure is kinematically attached to the elevation housing via a 6-point connection with three vertical and three horizontal constraints.

The outside of the structure is covered with aluminum-clad insulated panels to reduce pointing errors and HWFE due to thermal gradients, which are driven mainly by solar heating. In order to provide a reasonable operating environment for equipment and personnel, the base of the yoke is temperature controlled to $15\pm5\,\textrm{C}$ using heaters and waste heat from the science equipment with outside air for cooling. The elevation housing and yoke arms, including the camera space, are actively vented to keep the structures at outside temperature; any electronics attached directly to the camera are also at outside temperature.

Telescope servo cabinets, instrument electronics, and helium compressors are all located inside the base of the yoke, so if the camera does not have a rotator to track parallactic angle, camera connections do not have to pass through a cable wrap. The camera space has a 6-T gantry crane for loading/unloading, positioning, and assembly/disassembly work while the camera is installed.

The scanning requirements of the telescope drive the choice of bearings and drives. 
The drive has two motor/gearbox units for the elevation axis, and 4 motor/gearbox units for the azimuth axis. 
The azimuth drives are on the base of the yoke, around the support cone, 
and the elevation drives are in the yoke arm opposite the camera. 
Preliminary modal analyses of the structure suggest a first mode with a locked azimuth rotor at 2.6\,Hz and 
a locked rotor elevation mode at 3.0\,Hz \cite{parshley:2018b}; 
both of these frequencies have increased as the design has matured.

The pointing performance of the telescope (see Table~\ref{tab:PointingErrorBudget}) is limited mainly by deformation due to wind forces and insolation. The effects of slowly varying thermal deformation are partly corrected by continuously measuring the locations of the many point sources in each survey patch.

The telescope design meets survival requirements for Chile (see Table~\ref{tab:SurvivalRequirements}). Survival requirements for the South Pole are less stringent, except for the lower ambient temperature. Design modifications for the South Pole are discussed in Sect.~\ref{sec:ModificationsForSouthPole}.

\begin{table}
\centering
\begin{tabular}{|l|c|}
\hline 
Component & Mass (T)\\
\hline 
Mirrors$^{a}$ & 5\\
Camera & 5\\
EL structure & 50\\
Yoke & 135\\
AZ cone & 15\\
Total & 210\\
\hline 
\end{tabular}
\caption{Telescope mass. \label{tab:TelescopeMass}
Note: $(a)$
total for primary and secondary, including adjusters and back up structure.}
\end{table}

\begin{table}
\centering
|\begin{tabular}{|l|c|}
\hline 
Contribution & Pointing error (arcsec rms)\\
\hline 
1.4-K temperature gradients$^{a}$ & 2.5\\
$9\,\textrm{m\,s}^{-1}$ wind$^{b}$, 15-K ambient temperature change & 0.5\\
Random errors$^{c}$ & 0.8\\
Radio pointing$^{d}$ & 1\\
Total & 2.9\\
\hline 
\end{tabular}
\caption{Scan pointing knowledge error budget. \label{tab:PointingErrorBudget}
Notes: $(a)$ In direct sunlight;
$(b)$ 3rd quartile in Chile;
$(c)$ bearing wobble, tiltmeter error, encoder coupling;
$(d)$ the pointing offset is measured every $\approx1/10\,\textrm{hr}$ using radio observations of bright point sources.}
\end{table}

\begin{table}
\centering
\begin{tabular}{|l|c|c|}
\hline 
Load case & Chile & South Pole\\
\hline 
Wind speed (ms$^{-1}$) & 69 & 35\\
Seismic acceleration ($g$) & 1/3 & 0\\
Ambient temperature (C) & $-30$ & $-90$\\
Ice (cm) & 1 & 1\\
Snow$^{a}$ (cm) & 120 & 0\\
\hline 
\end{tabular}
\caption{Survival requirements. \label{tab:SurvivalRequirements}
Note: 
$(a)$ $100\,\textrm{kg\,m}^{-3}$.}
\end{table}

\subsection{Baffles and shields}

\begin{figure}
\includegraphics[width=0.5\textwidth]{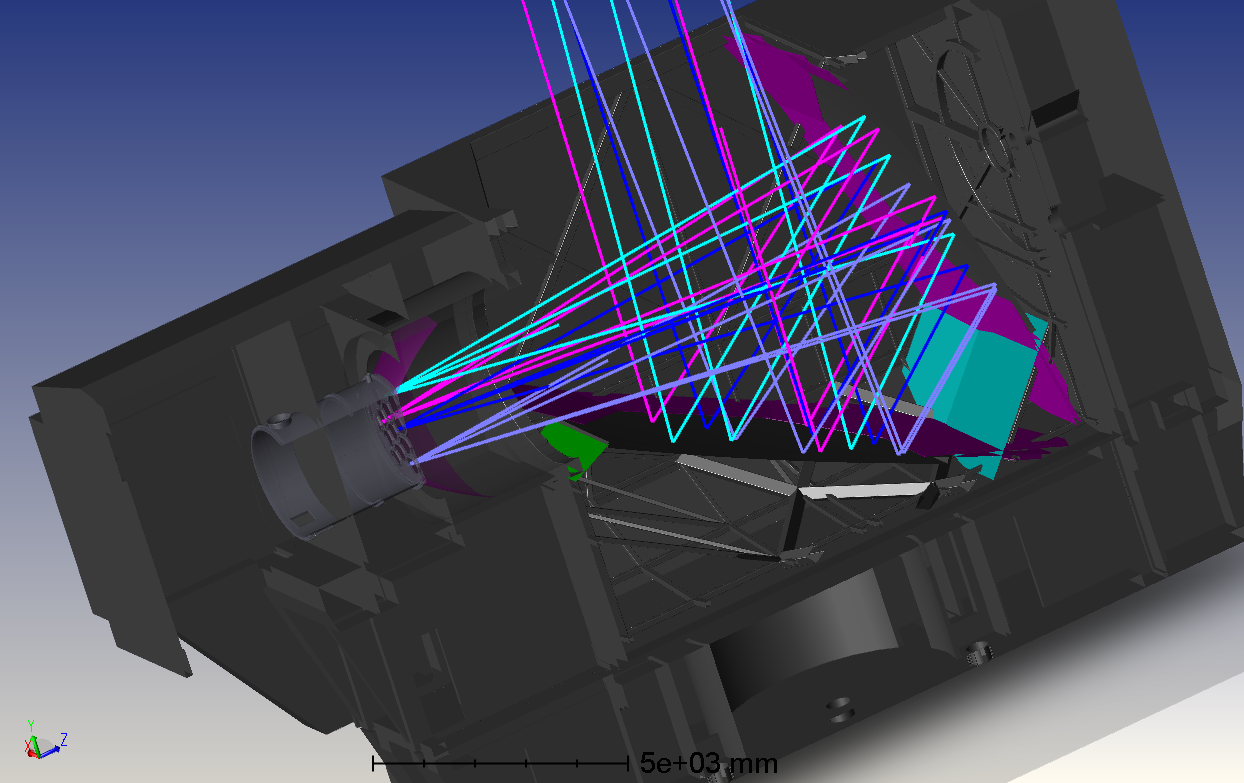}
\includegraphics[width=0.46\textwidth]{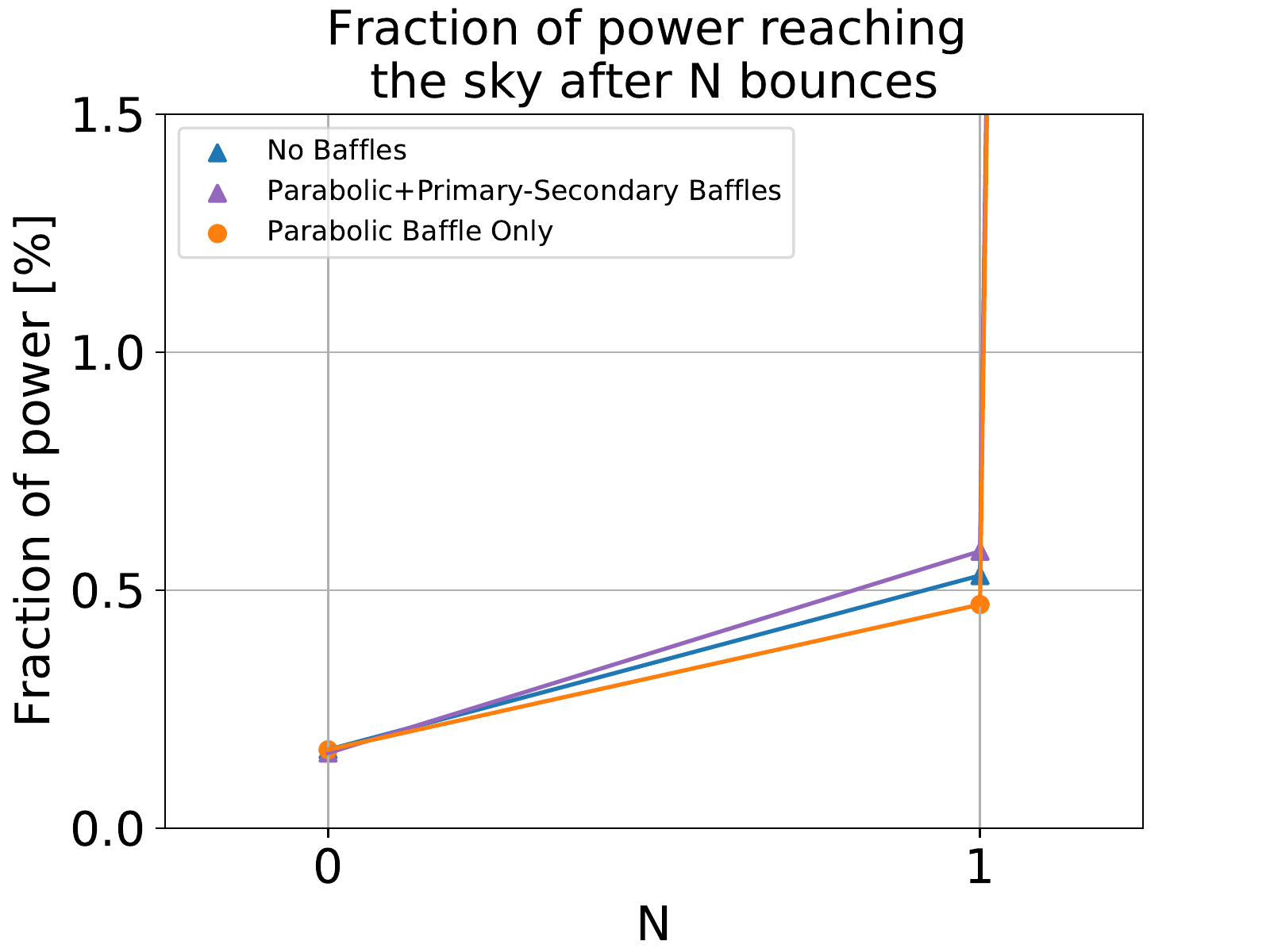}
\includegraphics[width=0.5\textwidth]{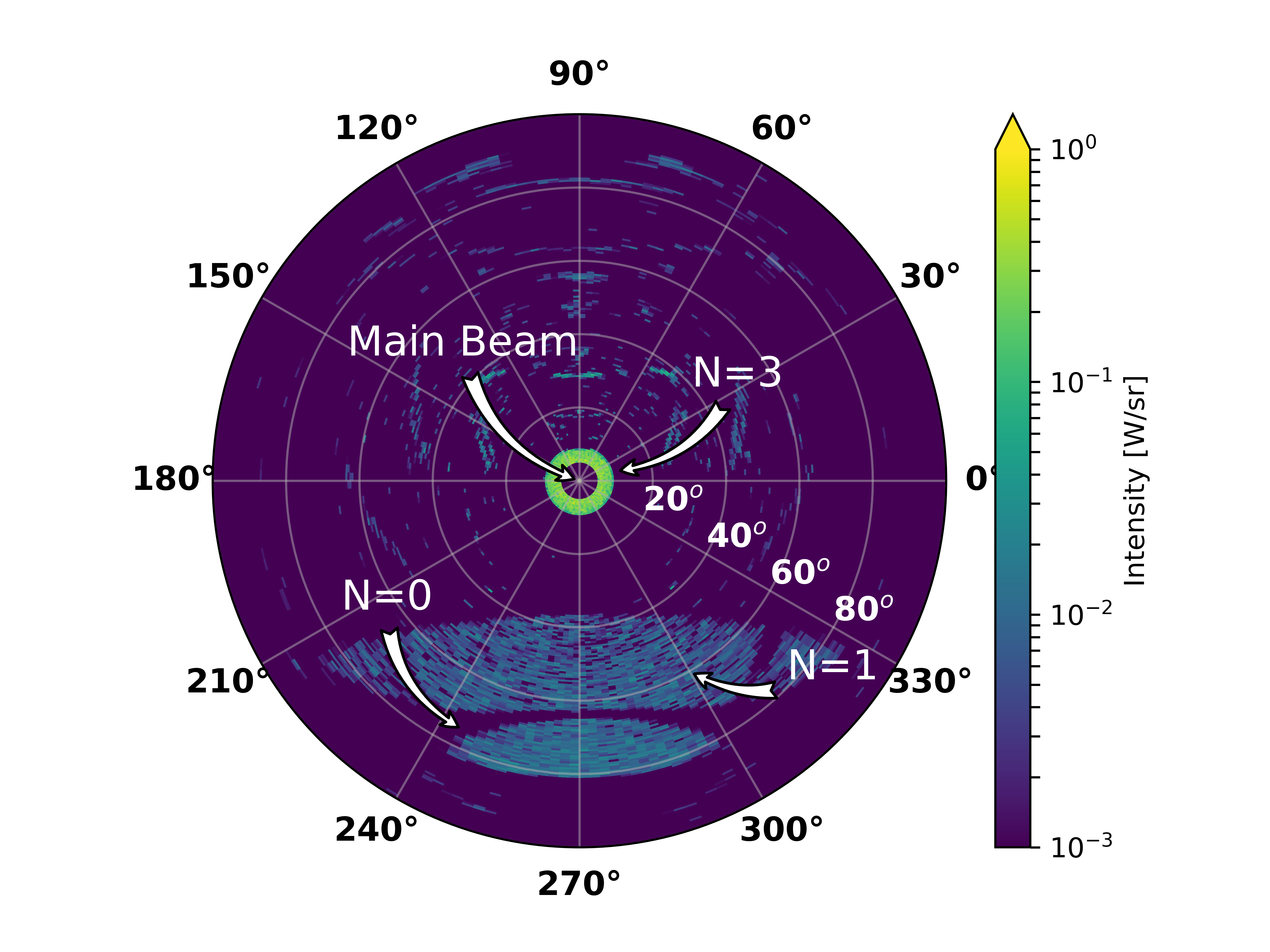}
\includegraphics[width=0.46\textwidth]{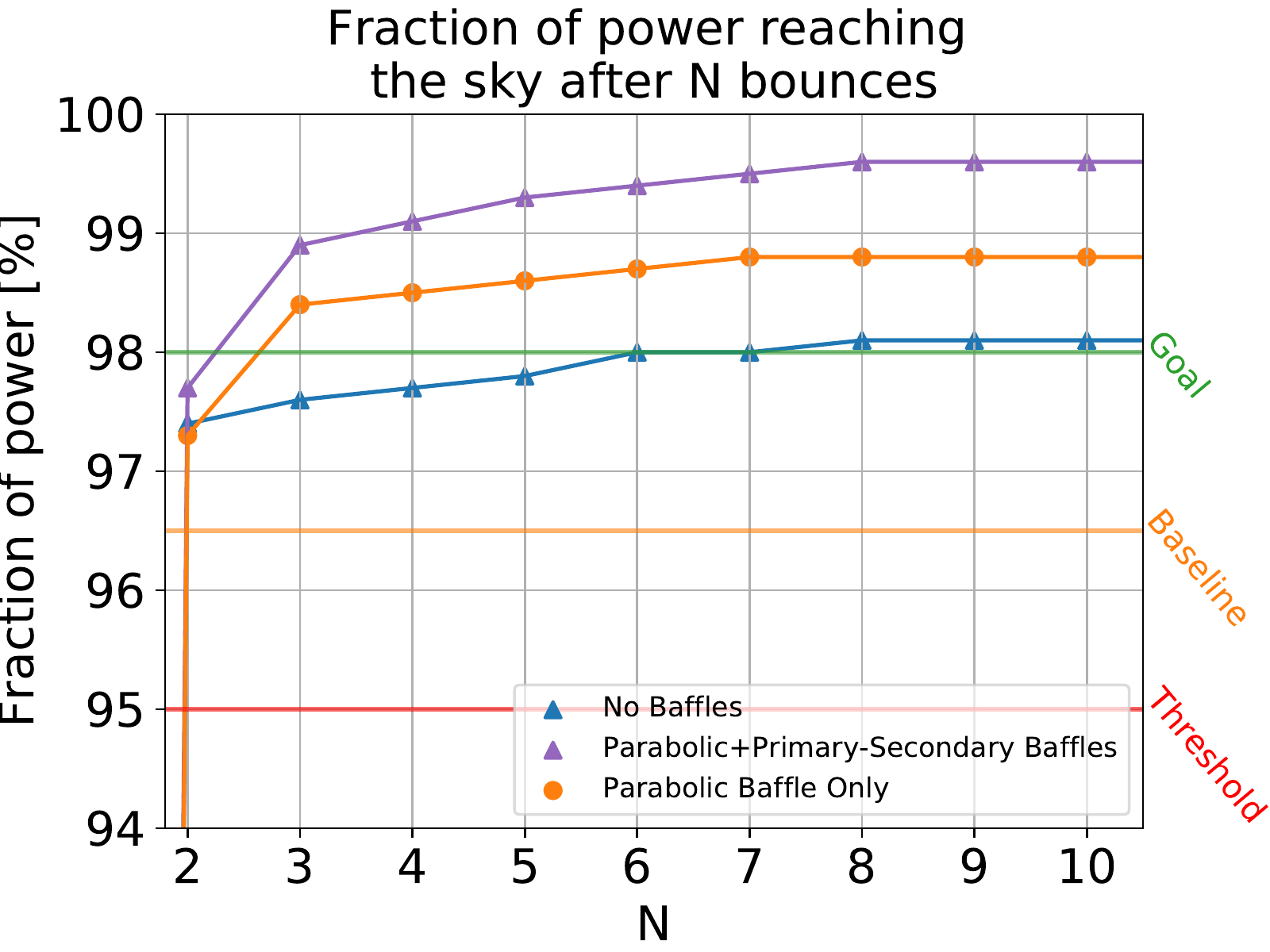}
\caption{Preliminary baffle and shielding studies for the Large Aperture Telescope using non-sequential ray tracing \cite{Gallardo:2018rix}.
The upper left shows a baffle configuration for the Large Aperture Telescope that includes a parabolic baffle near the receiver plus large secondary and primary guard ring baffles. Current analyses suggest that the parabolic baffle (pink, left) is important, but the large guard ring baffles (green, teal, and also pink, right) will not be needed. The position of 4 fields being launched from the focal plane of the telescope is also shown. 
We note that the large co-moving ground shield from the elevation structure will provide more baffling than in any previous large CMB telescope. 
Upper right (lower right) shows the fractional power that reaches the sky from the center field of the center camera versus number of bounces for a perfectly reflective telescope solid model for N in the interval 0--1  (2--10) bounces. Lower left shows the power density at the sky for N=3 bounces for a configuration with a parabolic baffle alone. Power density at the sky is normalized to have an injected power equal to $1\,\rm W$. In this configuration, one sidelobe is expected between 60 and 80 degrees, corresponding to the direct line of sight from the receiver camera to the sky. The parabolic baffle forms a ring around the main beam at N=3, which contains most of the difference in power between the parabolic baffle and the no-baffle model. Additional analyses that build on these will inform the desired finish (generally absorbing vs. reflective) and design for each of the baffles.}
\label{fig:baffles}
\end{figure}

Errors in the shape of the beam convert temperature fluctuations on the sky to polarization, so control of systematic errors at a level appropriate for CMB-S4 requires accurate knowledge of the beam. Main-beam errors are a tractable problem for the large telescopes because the beam shape can be measured directly, using a bright source, e.g., a planet, or by filtering the CMB observations with templates that capture the effects of beam errors. In addition, CMB fluctuations on arcminute scales are small, so systematics due to beam errors are also small.

Unwanted pickup through far sidelobes is more complicated, because far sidelobes are difficult to measure or model, especially in polarization. Far sidelobes are generated by scattering from mirror panel gaps, clipping of the beam as it passes by telescope structures, and unwanted lines of sight, e.g., directly from the camera to the sky. The far sidelobes cause unwanted pickup from the ground, Sun, Moon, and galaxy. Recent work using a combination of measurements and ray-tracing has been successfully used to model far sidelobes for ACT \cite{Gallardo:2018zfa}, and the same techniques are being applied to CCAT-prime and the SO-LAT \cite{Gallardo:2018rix}. 

Scattering from panel gaps is the dominant source of far sidelobes in telescopes with segmented mirrors. The sidelobes are sharp, arc-like features, that typically account for 1\% of the total response \cite{Rojas:2016mmq}. If the sidelobes fall on the ground, they cause enormous pickup, so large CMB telescopes always have comoving ground shields. The shields are usually reflective to direct sidelobes to the cold sky; an absorptive shield would significantly suppress far sidelobes, giving a cleaner beam, but at the expense of increased loading on the detectors, which would result in lower sensitivity. The mirrors in the CCAT-prime and SO-LAT designs are completely enclosed in a metal box, which naturally acts as a comoving shield. In this respect, the design has better shielding than existing Stage-3 large telescopes.

Low spillover is critical for reducing the loading on the detectors. The part of the beam that does spill beyond the mirror edges is usually managed with reflecting baffles that direct the spillover to the sky, rather than to warm absorbers, essentially trading detector loading for sidelobes. Fig.~\ref{fig:baffles} shows a spillover simulation in which rays are transmitted from the camera windows. With no baffles, the spillover on warm surfaces is $\approx2$\%, but adding baffles around the camera, and behind the mirrors, reduces the warm spillover to $<1$\%.

An issue that has not yet been fully resolved is whether or not the large telescopes need a fixed ground shield surrounding the entire telescope, as was done for ACT \cite{Fowler:2007dn}. Simulations of ground pickup during pre-project technical development will inform this decision; our expectation is that fixed ground shields will not be needed, but they are included in the cost estimate for the reference design. The ground shields are simple static structures, so the cost impact is small.

\subsection{Control and monitoring}
\label{sec:rd:latcam:monitor}

The large telescopes have a stand-alone control system that follows a path commanded by the observatory control system, with offsets based on a generic set of scan patterns. The telescope control system works in (AZ, EL) coordinates; conversion from celestial coordinates is handled at higher level in the observatory control system using an open-source astrometry package. The telescope control system commands the motor control loops to move within safe limits, and monitors the actual positions of the telescope axes and the status of the drive systems.  

Encoder positions, with accurate timestamps, and a summary of the drive status and faults, are reported at 200\,Hz, and are available to the observatory control system within 1\,s, which is fast enough to support decisions that are needed to control observations, e.g., waiting for the telescope to acquire a specific position before starting a scan. Encoder positions, and information about the status of the drive, are recorded continuously, even when observations are not running.

The telescope control system provides tools for debugging drive problems, and for acceptance testing, e.g., plots of axis position and following error vs. time, plots of following error spectral density, and measurements of the drive transfer functions.

\subsection{Assembly and test}

The large telescopes will be assembled on-site by a contractor. The contractor could be the telescope designer, a construction company, or some combination of these. 

The telescope mirrors need to be assembled and aligned before they are installed in the telescope structure. Assembly and alignment could be done at the factory, as for SO-LAT and CCAT-prime, but if we decide to fly mirrors to the South Pole, they will have to be broken down to fit in a C-130 aircraft, in which case some assembly and alignment will be needed on site. Even if mirrors are aligned at the factory, the contractor will measure the surface profile immediately before installation using photogrammetry or a laser tracker. Photogrammetry measurements of the ALMA antennas achieved $30\,\mu\textrm{m/6 m}$ rms error \cite{Mangum:2006dg}; a laser tracker is a factor of a few better, with $20\,\mu\textrm{m}$ peak error on a 6-m part. Mirror surface measurements will be made at the typical observing temperature in Chile, or at ambient austral summer temperature at the South Pole. Following installation in the telescope, the mirror positions will be aligned, again using photogrammetry or a laser tracker.

After assembly, the drive system functions will be tested, and the following error will be measured at the axis encoders to demonstrate that the drive meets the requirements for fast scanning. The pointing stability will be demonstrated using observations of bright stars with a small optical telescope, which is fast, easy, tests many telescope and observatory systems, and quickly reveals problems that are difficult to see with slower radio measurements of the pointing. Drive system functional testing and following error testing are tasks for the contractor, but pointing measurements will require support from CMB-S4.

Final alignment of the telescope optics will be based on out-of-focus holography measurements using a planet as the source. With a signal to noise ratio of 100 at $\lambda=3\,\textrm{mm}$, out-of-focus holography can achieve a few $\times10\,\mu\textrm{m}$ rms for the lowest-order modes of the wavefront error \cite{Nikolic:2006me}. Holography measurements will be done by CMB-S4 using either a commissioning camera or the full science camera.
\subsection{Safety}

Telescope safety is managed by a safety-rated programmable logic controller that monitors axis limit switches and speed sensors, e-stop switches, lockouts, and access interlocks. Lights on the outside of the telescope indicate when it is safe to approach. Enclosed spaces have fire and carbon-monoxide sensors, and there are cameras for remote monitoring. The telescope design follows fire and electrical codes (NFPA 70 and similar IEC standards) and structural safety standards (DIN 18800).
\subsection{Modifications for the South Pole}
\label{sec:ModificationsForSouthPole}

The SO-LAT and CCAT-prime were designed for operation in Chile, but constraints for the South Pole are different, so some design changes are needed.
\begin{enumerate}
\item Transportation. The telescope structure and mirrors may have to be broken up into 
small pieces that will fit in a C-130 aircraft. 
Adding bolted connections to the structure generally increases the mass and may also result in lower stiffness. 
It may be possible to send larger pieces by traverse, but this will involve a 1-year delay because the 
traverse arrives at the South Pole fairly late in the austral summer. 
The reference design plan assumes that the large telescope parts will be transported by air.

\item Foundation. Telescopes at the South Pole must be mounted on towers to give a reasonable lifetime before snow starts to bury the structure. Existing telescopes are on $5\textrm{--}8\textrm { m}$ towers which sit on wood rafts. Addition of a tower reduces the stiffness of the telescope structure, resulting in worse pointing and scanning performance. The CMB-S4 large telescopes have roughly the same mass and envelope as the existing 10-m SPT, so the reference design assumes a similar tower and raft.

\item Operation in the cold. Temperatures at the South Pole can be as low as $-80\,\textrm{C}$ (cf.\ $-20\,\textrm{C}$ for Chile), so bearings, drive motors, electronics, and wiring that are appropriate for Chile may not work at the South Pole. These elements can be heated to address the lower temperatures after enclosing the bearings and isolating the instrument space from ambient temperature using a large conical baffle between the instrument front plate and the bearing at the elevation structure. Heating in general results in thermal deformations, which can affect pointing and will be addressed through a combination of modeling and measurements during commissioning.

\item De-icing. At the South Pole, telescope mirrors generally have to be heated about 1\,K above ambient to prevent icing during the long polar night. The temperature rise is small, so thermal deformation is not an issue, but provision must be made to apply heat uniformly over the back of the mirror. For the machined aluminum panels on the 10-m SPT, we typically apply $50\,\textrm{Wm}^{-2}$ to keep the surface clean.

\item Power. Electrical power for the SO-LAT/CCAT-prime is Chilean standard 400V, 3-phase, 50Hz, but the South Pole is US standard 480V, 3-phase, 60Hz. Some changes will be requred in the electrical distribution and in equipment with large motors, e.g., helium compressors.
\end{enumerate}

\section{Cameras for large telescopes}
\label{sec:cryostatslargetelescopes}

\subsection{Requirements and design drivers}

The key measurement requirement for the large telescope cameras is the map noise, because this determines the number of detectors needed at each observing frequency (see Chapter~\ref{chap:instrumentoverview}). The field of view of the telescope described in Sect.~\ref{sec:largetelescope} is large enough to accommodate the required detector counts if two-color pixels are used, e.g., 220/270, 95/150, and 30/40\,GHz. In practice, some field of view is lost to dead space between the camera optics tubes and between detector wafers in each optics tube, so it becomes challenging to accommodate enough pixels. Achieving the required mapping speed (i.e., enough detectors with sufficiently low noise), without adding an additional, expensive, large telescope, is the principal design driver for the cameras.

An obvious approach for maximizing the number of detectors in the cameras is to pack the optics tubes and detector wafers as close as possible. A single hexagonal detector wafer per optics tube is attractive in this respect, but it requires a large number of small tubes. For the reference design, we have adopted a scheme with three hexagonal wafers and three smaller rhomboid-shaped pieces per tube, but this choice will be revisited during detailed design.

The mapping speed can be increased by making the pixels smaller, so more pixels can be squeezed into the telescope field of view, but the noise in adjacent pixels becomes correlated; there is essentially no improvement in mapping speed for pixel spacings smaller than $\approx1F\lambda$ \cite{Hill:2018rva}, so the reference design has $1F\lambda$ pixel spacing.

Achieving sufficiently low detector NET values requires careful control of losses in the camera optics. Small losses require thin lenses (which generally means smaller diameter lenses) made of low-loss material, e.g., Si, with good anti-reflection coatings, e.g., moderate-bandwidth machined structures.

Stage-3 experiments use either 250-mK sorption fridges or 100-mK dilution fridges. Sorption fridges must be cycled, which reduces observing efficiency, but dilution fridges operate continuously. Observing efficiency drives mapping speed, which is critical for CMB-S4, so the reference design uses dilution fridges. The lower base temperature for a dilution fridge reduces phonon noise in the detectors, but the improvement in overall noise is small if the detectors are background limited. Fridges dominate the site power requirements, so the number of fridges impacts the cost of operations. 

Polarization errors are usually dominated by the detectors (e.g., differential gain errors between $Q$ and $U$ detectors, ortho-mode transducer alignment errors, and leakage at the horn to detector wafer interface). Small, stable polarization errors can be corrected by filtering the CMB measurements, so absolute polarization errors in CMB-S4 can be similar to those in Stage-3 experiments.

Cool down time is a serious concern, because the cameras for CMB-S4 large telescopes are much bigger than Stage-3 cameras. A cooling time longer than a couple of weeks makes installation and commissioning difficult, and it can severely impact the overall efficiency of the experiment, because cameras do have to be warmed up occasionally for maintenance. Handling is also a concern for large cameras. Minimizing the mass and size of the cameras makes for easier operations and a simpler telescope interface.


\subsection{General configuration}

\begin{figure}
\begin{center}
\includegraphics[width=5in]{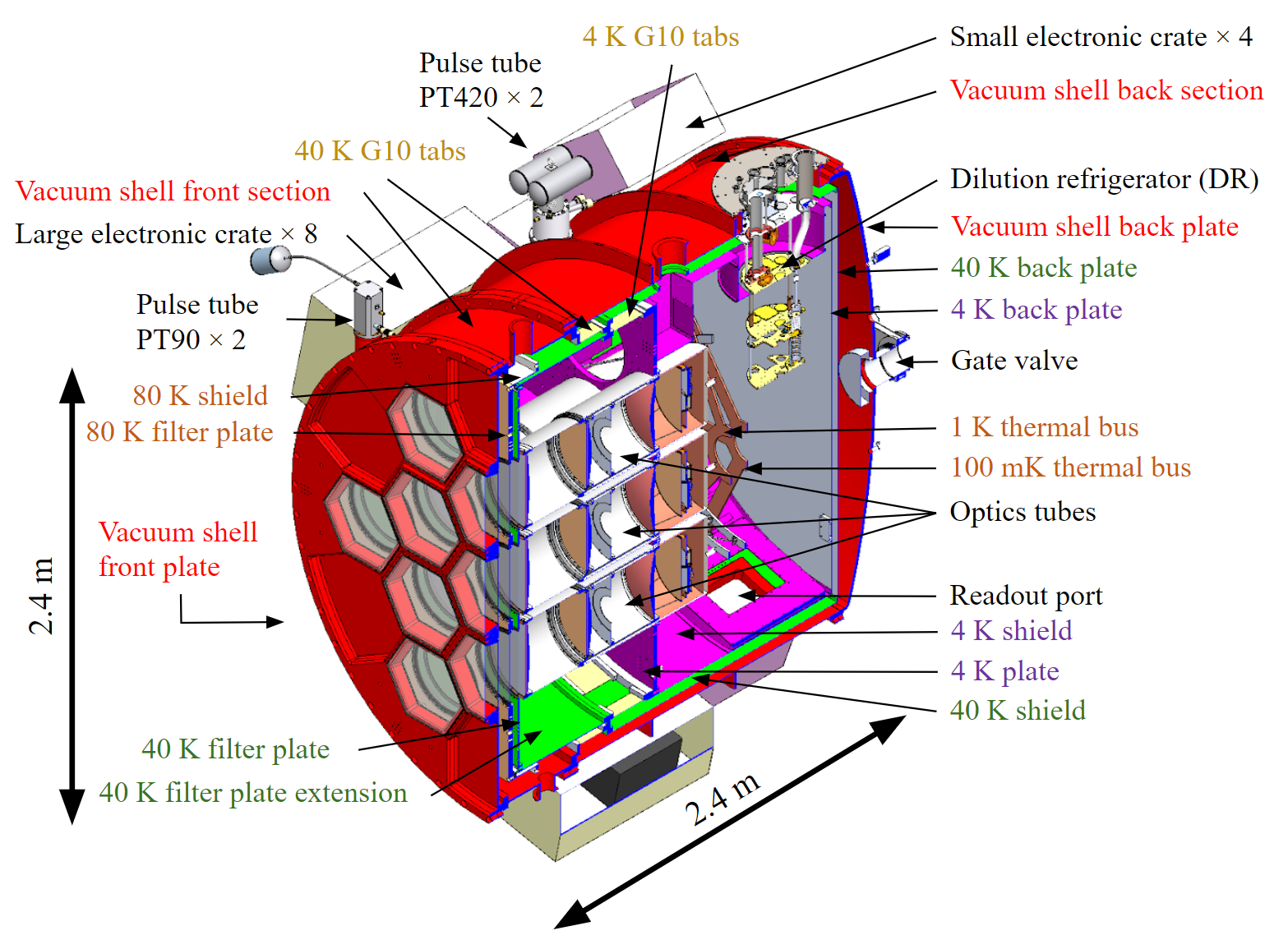}
\end{center}
\caption{
SO 13-tube camera design.
}
\label{fig:LATR}
\end{figure}

The camera has a 2.6-m diameter cryostat containing 19 close-packed optics tubes with 
the detector distribution shown in 
Table~\ref{tab:LATproperties}.
The design is based on the SO 13-tube camera shown in Fig.~\ref{fig:LATR}, but with a slightly larger cryostat and a full hexagonal close-packed array of optics tubes. Most of the details presented in this section are from the SO 13-tube camera design.

The design is modular, in that all the optics tubes have similar mechanical, thermal, and electrical interfaces. Replacement of an optics tube is relatively straightforward, so the assignment of bands to telescopes can be adjusted to deal with unexpected problems, e.g., complicated spectral behavior in foregrounds. The modular design also allows several teams to work independently on parts for one camera, so fabrication and testing can proceed quickly.

The 19-tube version of the camera will require some structural changes to the front plate, to handle the larger force on the plate, and a custom forging may be needed, because of the size, so the cost will be higher. Internal components, e.g., radiation shields, will have the same general configuration as in Fig.~\ref{fig:LATR}, but some parts will require design changes to accommodate 19 optics tubes. The cooling capacity may also have to be increased by adding additional pulse tube refrigerators, more heat straps, and possibly changing to a higher capacity dilution refrigerator. The modifications will not be trivial, and they will result in higher mass and cost.

\begin{figure}
\begin{center}
\includegraphics[width=5in]{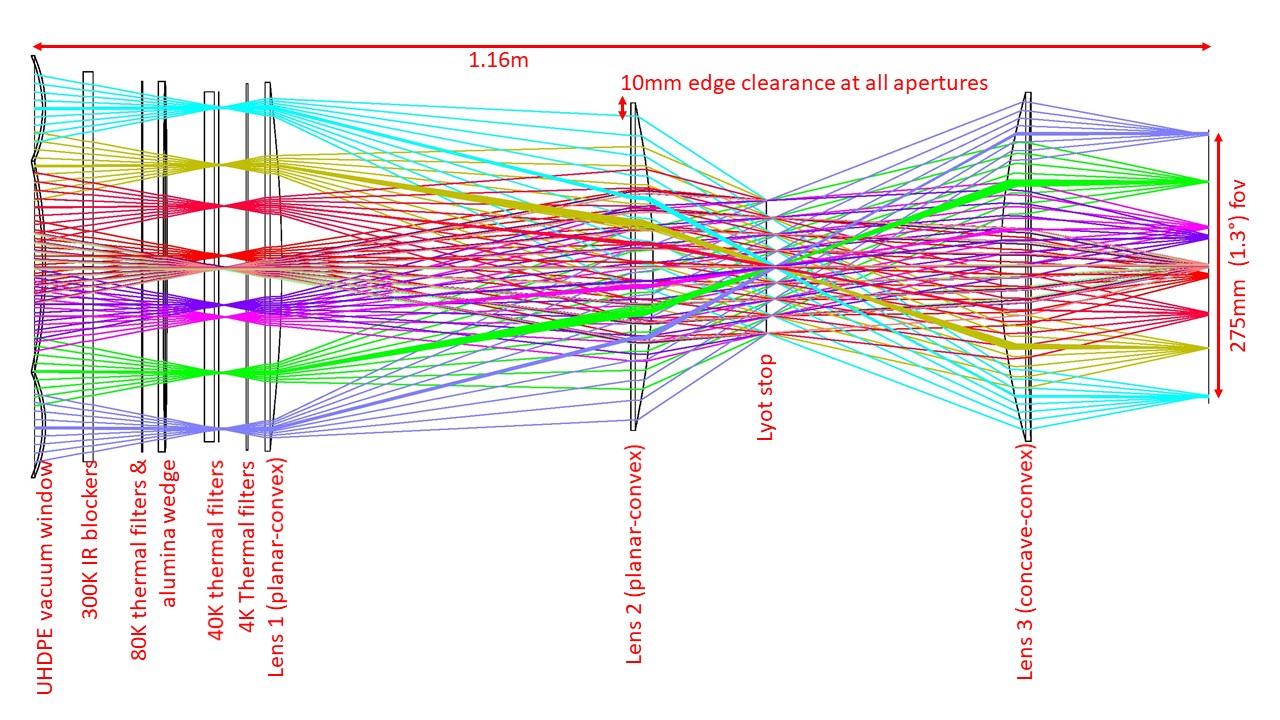}
\includegraphics[width=5in]{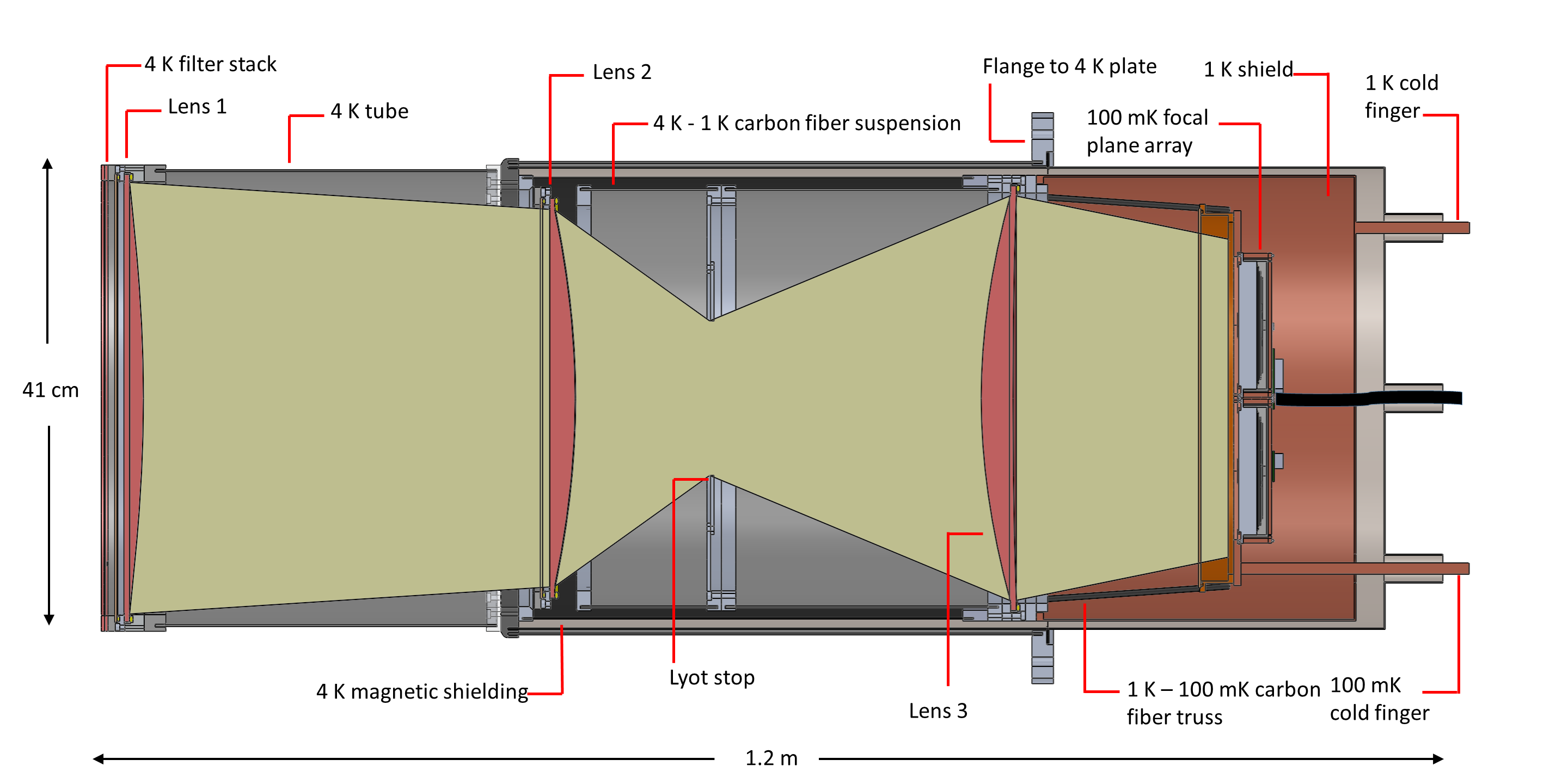}
\end{center}
\caption{
Preliminary optics tube designs for the LAT receiver \cite{Dicker:2018ytl}.  The top shows a ray trace and includes labels for all the optical elements.  The bottom shows much of the internal structure of the mechanical design that will be used to support, thermally isolate, and cool the optics and detector arrays.
}
\label{fig:optics_tube}
\end{figure}

\subsection{Cold optics}

The cold optics are comprised of silicon lenses, IR-blocking and band-defining filters, and spline-profiled feedhorns.   

Silicon lenses were selected due to the combination of a high index of refraction ($n = 3.4$) as required by the optical design, 
low loss tangent ($\tan \delta \sim 10^{-5}$), 
and ability to mitigate reflections using the proven approach of machined metamaterial antireflection coatings.  
These AR coatings have demonstrated control of reflections to $3 \times 10^{-3}$, 
minimizing systematics effects and nearly eliminating sensitivity losses due to reflectance in silicon lenses.  
The LATR requires lenses with diameters up to 40\,cm, while silicon is available up to 46\,cm diameter.   
Fabrication of lenses at the production rate required for CMB-S4 is on track to be demonstrated in early 2019 for Simons Observatory.

The IR-blocking filters are composed of metal mesh shaders which reflect IR radiation used in conjunction with alumina filters which absorbs the light.  The alumina filters  will be AR coated using a laminate of plastic layers as was demonstrated on SPT-3G.  The metal mesh filters are fabricated using standard lithographic techniques on dielectric substrates.   Three successive low-pass edge filters will be used to block radiation outside the passband of each dichroic detector array. This eliminates susceptibility to blue leaks in the detectors.   The low pass edge filters are fabricated by combining a number of patterned mesh filters into a laminated stack.  

Finally, radiation arriving at the focal plane is coupled to the detectors through a monolithic array of spline-profiled feedhorns fabricated by machining aluminum. These feeds offer tight control over beam systematic effects, ability to optimize the coupling efficiency, and ability to produce these arrays at a relatively low cost per horn.  This approach has been used SPTPol, MUSTANG-2, and is being scaled up for Simons Observatory.  Production at the required rate for CMB-S4 is a low risk proposition.

\subsection{Cryogenics}

\subsubsection{Refrigerators}

\begin{table}
\label{tab:fridges}
\centering
\begin{tabular}{|l|l|l|rrrrr|}
\hline
Fridge & Type & Quantity & \multicolumn{5}{c|}{Cooling capacity per fridge}\\
\hline
 & & & 80\,K & 40\,K & 4\,K & 1\,K & 100\,mK\\
\hline
PT90 & Pulse tube & 2 & 90\,W & & & & \\
PT420 & Pulse tube & 2 & & 55\,W & 2\,W & & \\
LD400 & Dilution & 1 & & & & 25\,mW & 400\,$\mu$W\\
\hline
\end{tabular}
\caption{Refrigerators.}
\end{table}

The detectors are cooled to $\approx100\,\textrm{mK}$, with intermediate stages at 1, 4, 40, and 80\,K for radiation shields, filters, and wiring thermal intercepts \cite{Zhu2018}. The 80-K stage has two single-stage Cryomech PT90 pulse tubes, and the 40-K and 4-K stages have two PT420 pulse tubes (see Table~\ref{tab:fridges}). Two additional pulse tubes can be installed in the camera if a faster cooling time is needed. The 100-mK stage is cooled by a BluFors LD400 dilution refrigerator with an intermediate stage at 1\,K. The dilution fridge has its own PT420, which provides additional cooling power for the camera 40-K stage. Flexible copper braids are used to connect the refrigerators to the camera cold stages. All the refrigerators must be kept within $\approx60^\circ$ of vertical to maintain cooling capacity.

\subsubsection{Heat pipes and switches}

Roughly 1500\,kg needs to be cooled by the PT420 4-K stages, which have limited cooling power, especially at temperatures $\gtrsim100\,\textrm{K}$. During cooldown, when the camera's 40-K stage is near its base temperature, the 4-K stage will still be at $\approx200\,\textrm{K}$, so we would like to transfer heat efficiently between the two stages at high temperature and switch off the transfer at temperatures $\lesssim$ 40\,K. Possible solutions include a heat pipe filled with high pressure nitrogen gas \cite{Shukla2015} and pre-cooling using liquid nitrogen. Pre-cooling is used in many laboratories, but is awkward at a telescope site, because it requires large quantities of liquid nitrogen, so we will use heat pipes charged with 1000\,psi of nitrogen at room temperature.

Heat switches can be used to significantly reduce the cooling time by temporarily creating strong thermal links between the 1-K and 4-K stages, and between the 100-mK and 1-K stages. The heat switches must have small conductance in the open state and high conductance in the closed state. Gas-gap and mechanical heat switches are viable options; both are available commercially, and samples are being tested.

Gas-gap heat switches have a small helium adsorption pump that can actively fill or evacuate a small cavity between two copper rods inside a thin-walled stainless steel tube. The open conductance is reduced to the minimal conduction through the small cross-sectional area of the stainless steel tube. The closed conductance is several orders of magnitude larger, since the small gap between the cold and warm copper rods is filled with helium gas, which conducts heat across the gap efficiently.

Mechanical heat switches have an actuator to make or break a physical connection. The open conductance can be essentially zero, and the closed conductance can be very high as long as enough pressure can be developed at the interface. Thermal boundary resistance is significant issue at lower temperatures \cite{Coppi:2018rrp}, so extensive testing of several mechanical switches is currently underway.

\subsubsection{Optics tubes}

\label{sec:opt}

The optics tubes contain all optical and detector components between 4\,K and 100\,mK. Each tube is self contained, so it can be installed as a single unit (see Fig.~\ref{fig:optics_tube}). An optics tube has three silicon lenses \cite{Dicker:2018ytl}. The first lens is at 4\,K, and is supported from the cryostat 4-K plate by an $\approx1$-m long aluminum tube which has a thin wall to keep the mass low. The tube is fabricated from 1100-H14 or 6063-O aluminum, as a compromise between strength and thermal conductivity, and it is lined with Vacuumschmelze A-4K magnetic shielding material. The second lens, Lyot stop, and third lens are all at 1\,K, and are supported from the 4-K tube by a thermally-isolating carbon fiber tube. The inside surfaces of all the tubes that surround the beam are blackened or have annular baffles. The detector assembly is supported by a carbon fiber truss attached to the 1-K structure, with a second layer of magnetic shielding around the detectors. Thermometer wires, cold finger connections to the camera cold stages, and detector readout cables run through the back of the optics tube.

\subsubsection{Heat loads}

\begin{figure}
\centering
\includegraphics[width = 0.6\linewidth]{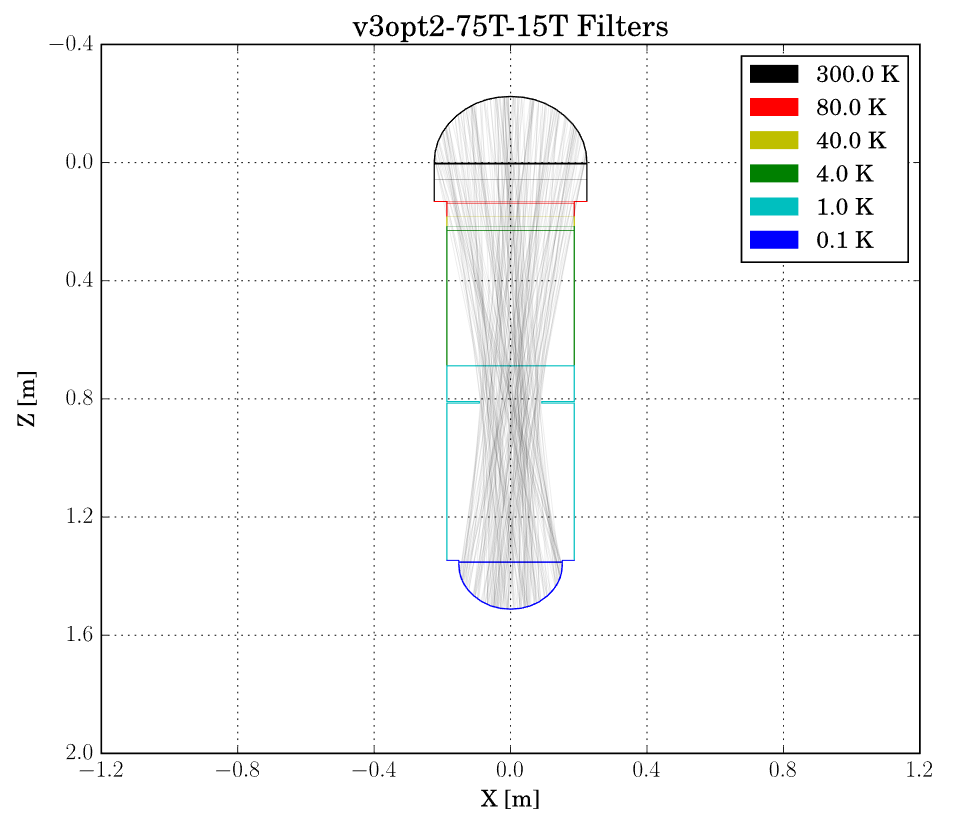}
\caption{Thermal model of an optics tube showing all the filter elements and tube walls at each temperature stage. The light lines do not represent the actual ray trace; they merely highlight the location of the Lyot stop.}
\label{fig:ray}
\end{figure}

The total power emitted and absorbed at each temperature stage has been estimated using a radiative transfer simulation based on numerical ray optics \cite{Zhu2018}. The simulation captures the optics tube geometry (see Fig.~\ref{fig:ray}) and spectral properties of the tube walls and filters. Calculated heat loads for an optics tube are shown in Table~\ref{tab:load} and filter temperatures are shown in Table~\ref{tab:filter_temp}. Heat loads for the complete camera are given in Table~\ref{tab:latloading} \cite{OrlowskiScherer2018,Coppi:2018rrp}. 

\begin{table}
\label{tab:load}
\centering
\begin{tabular}{|l|lllll|}
\hline
Temperature stage & 80\,K    & 40\,K     & 4\,K      & 1\,K      & 100\,mK\\
\hline
Optical load & 3.90\,W & 2.0\,mW & 27.6\,mW & 29\,$\mu$W & 312\,nW \\
\hline
\end{tabular}
\caption{Heat loads for a single optics tube.}
\end{table}

\begin{table}
\label{tab:filter_temp}
\centering
\begin{tabular}{|l|llllllll|}
\hline
Stage & 300\,K & 80\,K & 80\,K & 40\,K & 4\,K & 4\,K & 1\,K & 100\,mK\\
\hline
Filter type$^{a}$ & IRB & IRB & AF & IRB & IRB & LPF & LPF& LPF\\
Cutoff$^{b}$ ($\mu$m) & 27 & 40 & & 300 & 660 & 800 &1470 & 1610\\
Filter center temperature (K) & 297 & 252 & 81 & 74 & 62 & 12 & 1.1 & 0.105\\ \hline
\end{tabular}
\caption{Optics tube filter temperatures.
Notes: 
$(a)$ IRB=metal-mesh infrared blocking filter, AF=alumina filter, and LPF=low-pass filter;
$(b)$ LPF cutoffs are for 90/150-GHz optics tubes.}
\end{table}

\begin{table}
\label{tab:latloading}
\centering
\begin{tabular}{|cccccc|c|}
\hline
Stage & Support & Radiative & Optical & Readout & Total & Cooling capacity\\
\hline
80\,K & 1.3\,W  & 7.2\,W & 50.7\,W & N/A    & 59.2\,W & 180\,W \\
40\,K & 6.0\,W & 4.6\,W & 13.2,W & 22.3\,W   & 46.1\,W & 110\,W\\
4\,K & 0.84\,W & 0.01\,W & 0.36\,W & 0.73\,W & 1.94\,W & 2.0\,W\\
 1\,K & 5.01\,mW & 0.01\,mW & 6.46\,mW & 6.33\,mW & 19.82\,mW & 25.0\,mW\\
100\,mK & 68.6\,$\mu$W & 0.1\,$\mu$W & 0.5\,$\mu$W & 45.7\,$\mu$W & 115.0\,$\mu$W & 400\,$\mu$W\\
\hline
\end{tabular}
\caption{Heat loads for the complete camera.}
\end{table}


\subsection{Mechanics}

\subsubsection{Vacuum shell}

\begin{figure}
\begin{center}
\includegraphics[width = .75\linewidth]{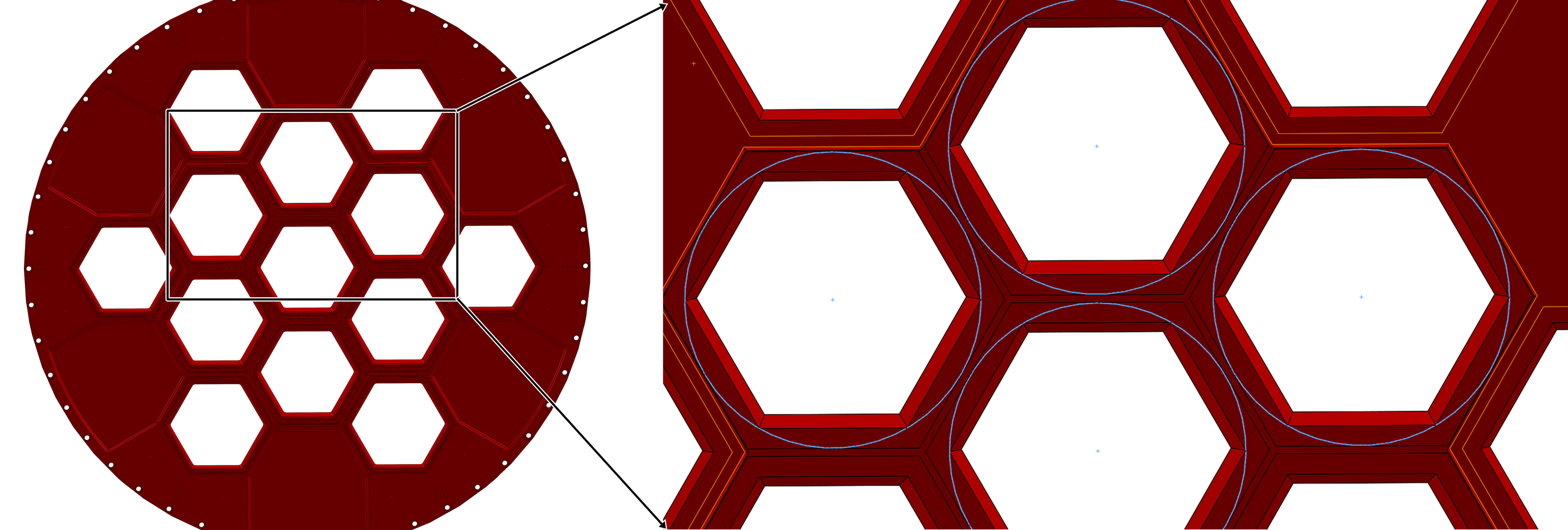}
\end{center}
\caption[example] 
{Front plate detail showing the material saved by using hexagonal windows. The gray circles are the size of the hole of equivalent minimum beam clearance at the outside surface of the front plate. Making the windows inscribed hexagons instead of the circles which circumscribe them adds a significant amount of material at the weakest point in the fplate.}
\label{fig:hexwindow}
\end{figure}

\begin{figure}
\begin{center}
\includegraphics[width = .75\linewidth]{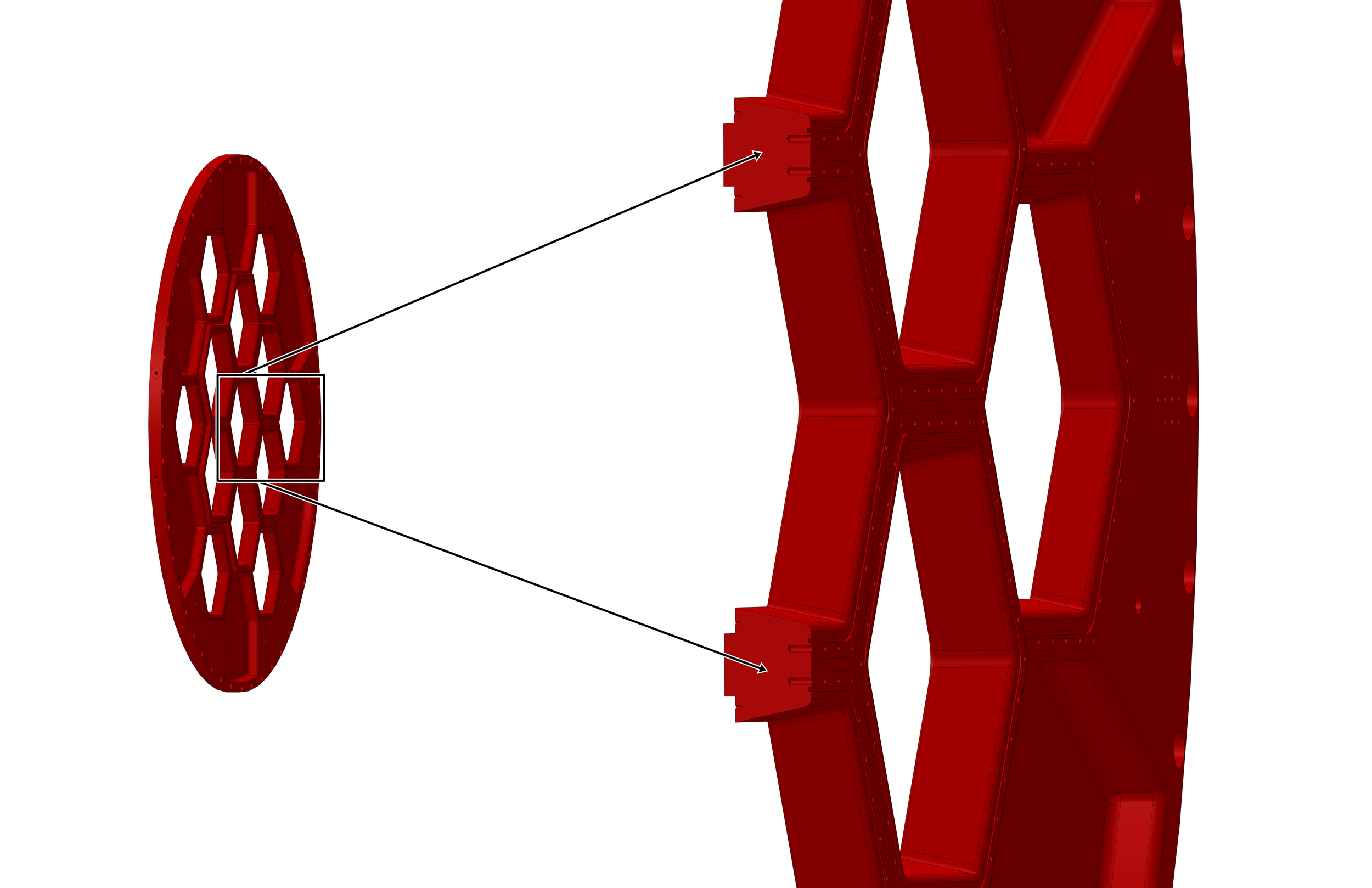}
\end{center}
\caption{Cross-section of the front plate showing tapering of the window holes to match the beam divergence. Light is entering the cryostat from the right, and the detector arrays are on the left.}
\label{fig:Taper}
\end{figure}

The camera vacuum shell presents a significant challenge, because the front plate is large and flat, it has many large holes for the windows, and deflections must be $\lesssim2\,\textrm{cm}$ to avoid a touch to the 80-K stage behind. Stresses are highest at the center of the plate, so this region is $\approx1\,\textrm{cm}$ thicker. The holes for the windows follow the hexagonal shape of the beam, which allows more material to be left in the front plate (see Fig.~\ref{fig:hexwindow}). The window holes also have tapered walls that match the beam divergence, again to allow more material between windows (see Fig.~\ref{fig:Taper}) \cite{OrlowskiScherer2018}. The plate is machined from a single billet of 6061-T6 aluminum, with aggressive lightweighting on the outside face to keep the overall mass of the camera reasonable (see Fig.~\ref{fig:LATR}). The front plate is 7\,cm thick, has a total mass of 350\,kg, and deforms 1\,cm. The complete vacuum shell conforms to the ASME VIII-2 pressure vessel code.

\subsubsection{Windows}

The cryostat windows are made of 3-mm thick ultra high molecular weight polyethylene, anti-reflection coated with porous teflon sheets that are hot-pressed onto the window surfaces. The windows are clamped to the vacuum shell front plate with hexagonal metal rings.

\subsubsection{Stage supports}

The 40-K and 4-K stages (labelled shields and plates in Fig.~\ref{fig:LATR}) are nested cylindrical cans, supported by rings of G10-CR tabs that provide high axial and lateral stiffness, with low thermal conductivity. The tabs accommodate differential thermal contraction between stages by flexing inwards \cite{OrlowskiScherer2018}. The 1-K and 100-mK stages are OFHC copper bus bars supported by carbon fiber trusses \cite{Zhu2018}. The complete assembly of cold stages and optics tubes has a total mass of $\approx2\,\textrm{T}$, which is much larger than in existing CMB cameras, but the large size of the stages results in a stiff support. The lowest natural frequency of the cold assembly, with 13 optics tubes installed, is $\approx19\,$Hz. Stiffening the 4-K plate has the potential to significantly increase this frequency \cite{OrlowskiScherer2018}. A lower natural frequency makes the camera more sensitive to vibrations, which typically couple to the detectors and readout electronics through magnetic fields, so the optics tubes have two layers of magnetic shielding.

\begin{table}
\centering
\begin{tabular}{|l|r|l|}
\hline 
Component & Mass (kg) & Notes\\
\hline 
Vacuum shell & 1310 &\\
Refrigerators & $\approx200$ &\\
80-K stage & 248 &\\
40-K stage & 451 &\\
4-K stage & 225 &\\
Optics tube & 77 & times 13 (19) optics tubes\\
Total & 3435 (3897) & with 13 (19) optics tubes\\
\hline 
\end{tabular}
\caption{Camera mass. \label{tab:CameraMass}}
\end{table}

\subsubsection{Assembly procedure}

Camera assembly requires a 4-m high bay, so the optics tubes can be inserted vertically. The assembly procedure begins with bolting together the two halves of the vacuum shell and mounting the shell in its cart. The 40-K radiation shield is assembled, wrapped in multi-layer insulation, and inserted vertically into the shell. The 4-K plate is lowered into place, the PT-420 coolers are installed, and then the 4-K radiation shield is installed. The 40-K and 80-K filter plates are bolted in place, the PT-90 coolers are installed, the optics tubes are inserted, and then the vacuum shell front and back plates are attached.

\subsubsection{Vacuum system}

The camera has a turbo pump attached to an ISO-150 gate valve on the back of the vacuum shell. The gate valve closes automatically in the event of a power loss. There are KF-50 ports on the front and back, for gauges and additional pumps, and pressure relief valves to protect the windows in case the vacuum shell becomes pressurized.


\subsection{Control and monitoring}

The cameras have a stand-alone control system that manages the vacuum and fridge systems, heaters, and temperature monitors. The control system is essentially a wrapper for various commercial units; it provides local control and monitoring for testing purposes, and it accepts high-level commands from the observatory control system for automatic and remote operations. Temperatures, vacuum sensors, and the status of the various fridge and compressor controllers are sampled every few seconds and reported continuously, even when the camera is not observing.


\subsection{Prototypes, assembly, and testing}

The cameras are complicated systems, and while we can rely heavily on the work done by SO, the CMB-S4 and SO cameras are not identical, 
so prototypes will be needed. The reference design includes the following two levels of prototyping
\begin{enumerate}
\item A single optics tube, with detectors, in a small test cryostat that can be cycled quickly. The goal here is to check fit, function, and performance of everything in an optics tube. Tests will include operating temperatures and heat loads (using thermometers on key components), passbands (using a Fourier transform spectrometer), beam shapes (using a chopped thermal source on an xy stage), polarization errors (using a chopped thermal source with a rotating wire grid), and noise. Optical tests will require a cold neutral density filter, e.g., at the cold stop, or bolometers with a high-$T_{\rm c}$ TES that can handle room temperature loading.
\item A prototype of the final camera design, including a full-size vacuum shell with fridges and radiation shields, at least one optics tube, and heaters to simulate heat loads from missing wiring and windows. In this case, testing will focus on temperatures and heat loads. 
\end{enumerate}

The single-tube prototype cryostat will become a commissioning camera that will be used for checking telescope mechanical and software interfaces, and for measuring optical loading and telescope beam shapes. For this purpose, the single-tube prototype cryostat will be mounted in an adapter that simulates the mechanical interfaces of the full-size camera. Since the tests that will be done with the commissioning camera address telescope design issues, it will be sufficient to perform the tests on just one of the large telescopes.

The cooldown time for a complete camera is several weeks, so the Stage-3 practice of using the camera to test subsystems is impractical; the camera must be assembled with a working cryostat and fully tested optics tubes and detectors. Optics tubes will be tested in a suite of four small test cryostats, like the one used for testing the prototype optics tube. Each test cryostat will be supported by a small team that will run tests, analyze the results, and fix problems. Testing will likely be done at several different locations, but will be part of a single quality assurance process with common procedures, software, and documentation.

Each camera vacuum shell will initially be assembled with port cover plates, and leak checked. Radiation shields, fridges, thermometers and heaters, monitor and readout wiring, and windows will be installed, and the camera will be cooled to check cooling rates and base temperatures. After fixing any cryogenics issues, any and problems with thermometry and wiring, the optics tubes will be installed (with cold neutral density filters if the bolometers do not have a 2nd TES with high $T_{\rm c}$) and the complete camera will be cooled. The tests that were done on each optics tube will be repeated, but for a subset of the detectors, with the goal of finding problems that appear only in the complete camera (e.g., excess noise when all the electronics are operating). Testing at this level will be part of the same quality assurance process used for the individual optics tubes. After testing, the camera will be warmed up and the parts will be prepared for shipping. Our intent is to ship the cameras essentially complete, to minimize re-assembly work on site.

The main constraint for assembling and testing the camera is the long cooldown time. A full thermal cycle will take a few weeks, so it is impractical to include many cycles in the plan.

\section{Small telescopes and cameras
\prelim{ ({\it A. Kusaka, J. Kovac and C-L. Kuo})}}
\label{sec:smalltelescopes}

\subsection{Design drivers}

The ultra-deep survey, which targets the signature of primordial gravitational
waves, drives the key measurement requirements for the small telescopes.

The measurement challenges for this survey, outlined in Sect.~2.3, differ from
those of the deep and wide
survey in ways that set unique design drivers
for the small telescopes.  To achieve the science goal of measuring $r$ to
the required precision ($\sigma(r) = 0.0005$ for a non-detection) ultimately
demands measuring primordial $B$-mode polarization patterns with uncertainties
of $< 10$\,nK at angular scales of a few degrees on the sky.  To do this requires
an instrumental design that enables sufficient control of statistical noise fluctuations,
galactic foregrounds, and instrumental systematics, all of which are made
harder at degree scales by $1/f$ contributions to instrumental and atmospheric
noise fluctuations and by the red spectrum of many foreground and systematic
effects.

Small aperture telescopes offer intrinsic advantages for meeting the 
unique requirements of measuring $r$, which include:
\begin{enumerate}
\item \textbf{efficiency} to test, integrate, and deploy large arrays of detectors, due
to their small size and large optical throughput;
\item \textbf{stability} of fully cryogenic telecopes, including their beams and internal systematics;
\item \textbf{calibrators} can be aperture-filling or easily placed in the far-field;
\item \textbf{modulators} can be aperture-filling and fully cryogenic;
\item \textbf{Mounts} can be compact and allow full boresight rotation;
\item \textbf{sidelobes} can be suppressed with unobstructed on-axis optics and superior
baffling and shielding.
\end{enumerate}
Additionally, small-aperture telescopes from the BICEP/Keck series at the South Pole
have produced all the leading $r$ constraints so far from ground-based $B$-mode measurements 
through Stages 1, 2, and 3.  Small aperture telescopes, for example the ABS experiment, 
have also had the greatest success among telescopes in Chile 
in achieving sensitivity on degree scales, by using half-wave plate modulators
to mitigate $1/f$ noise from the less stable atmosphere there. 
The intrinsic advantages of the small-aperture approach, combined with the
experience that over the past decade, only small-aperture telescopes have demonstrated
sufficient performance in degree-scale $B$-mode measurements to allow direct scaling
that can meet the measurement requirements of CMB-S4, clearly dictate the use of
small-aperture telescopes for the reference design.

\begin{figure}
\begin{center}
\includegraphics[width=5in]{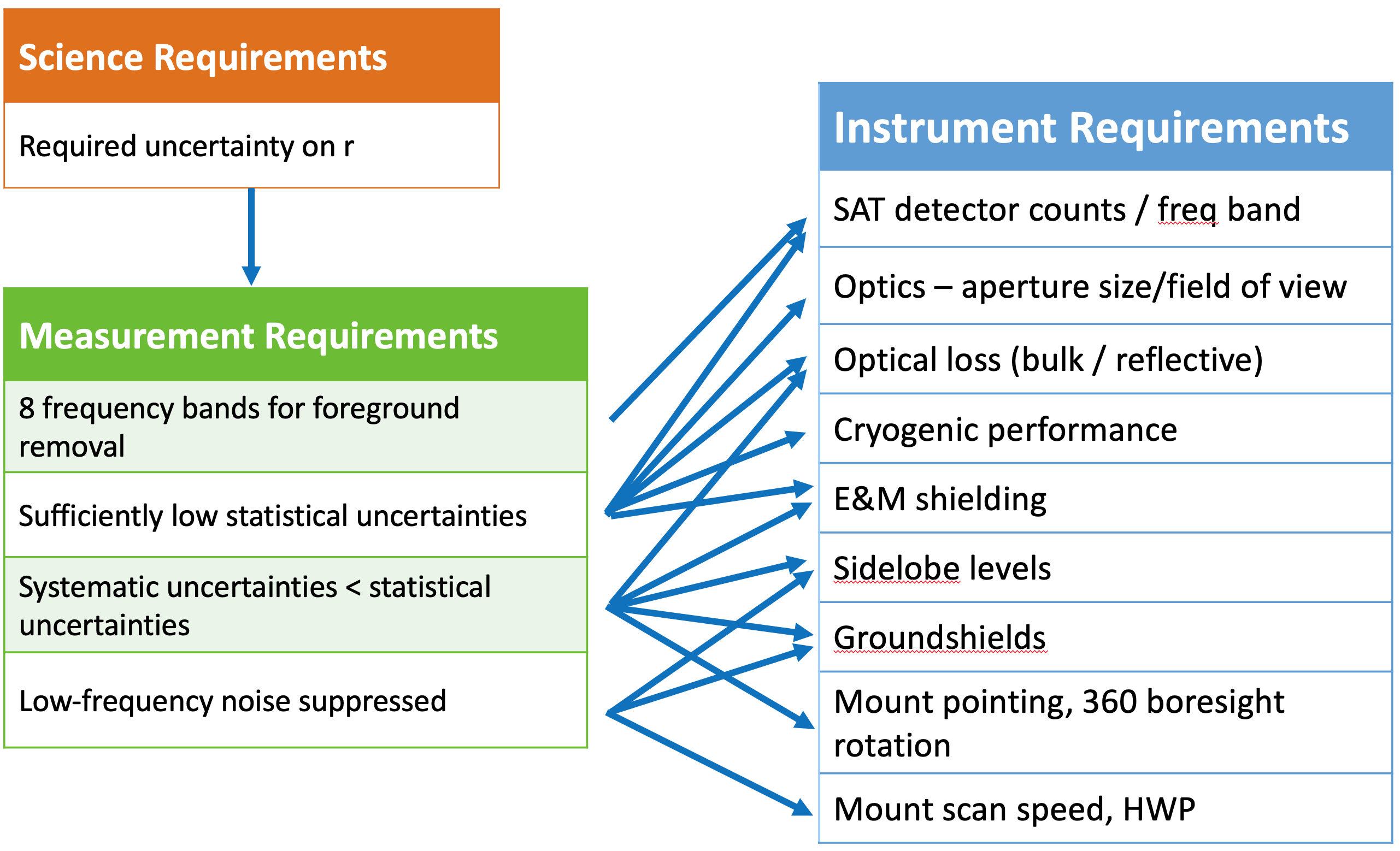}
\end{center}
\caption{
The small-telescopes instrument requirements, detailed in this subsection, are motivated
by the unique measurement challenges set by measuring $r$ to the required uncertainty 
using ultra-deep $B$-mode measurements at degree scales.
}
\label{fig:SAT_flowdown}
\end{figure}

The detailed design choices for the small-aperture telescopes are motivated by
the requirements flowdown illustrated in Fig.~\ref{fig:SAT_flowdown}.
In considering changes for the reference design compared to the small-aperture
telescopes that have achieved previous deep $r$ measurements, we have incorporated
new technologies---for example dichroic detectors, dilution refrigerators, and 
(if small-aperture telescopes are deployed to Chile) cryogenic half-wave
plate modulators---where there is a consensus that
they promise improved performance while adding little technical risk.  In making design choices
we have distinguished between \emph{engineering issues}, those that can be
fully developed and demonstrated in the lab to retire risk, and \emph{science issues},
those whose impact on successfully meeting the measurement and science requirements
must be judged with comparison to direct experience of making deep $B$-mode maps.
For example, cryostat design is primarily an engineering issue because 
we are confident our design choices can be fully validated in the lab.
Examples of science issues include beam and sidelobe optical performance,
polarization modulation approach, ground pickup and shielding, and other systematic
effects, and for design choices that impact these issues 
we have endeavored to stay close to and to build upon proven experience.

\subsection{Reference design summary}
\label{sec:rd:sat:summary}

Table~\ref{tbl:SAT_tubes_detectors} summarizes the optical design features and distribution of detectors
between optics tubes. The lower frequency tubes use alumina lenses, while the highest frequency tubes
use silicon lenses for their known low loss.  
We rely on proven experience from South Pole and Chile to guide our choice
of polarization modulation technique for the small-aperture telescopes.
For South Pole, the polarization signal is measured by differencing pairs
of orthogonal detectors, modulated by scanning the telescope in azimuth.
Any telescopes deployed to Chile at 85\,GHz and above
will be outfitted with
continuously-rotating half-wave plates (HWPs), described in Sect.~\ref{sec:sat:hwp},
to suppress atmospheric $1/f$ noise and instrumental systematics.  The atmosphere is 
sufficiently stable in the lowest frequency bands that a HWP is not needed.

\begin{table}[t]
  \begin{center}
    \begin{tabular}{|ccccccc|}
\hline
	    Bands & Lenses & \parbox{5em}{\centering Field\\of view} & \parbox{5em}{\centering Min.\ edge\\taper} & \parbox{5em}{\centering Modulation \\ (Pole/Chile)} & Detectors / tube & Tubes \\
      \hline
      30 / 40 & $2\times$ 55\,cm Al & $29^\circ$ & $-9.3\,$dB & scan & 576 & 2 \\
      85 / 145 & $2\times$ 55\,cm Al & $29^\circ$ & $-6.2\,$dB & scan / HWP & 7048 & 6 \\
      95 / 155 & $2\times$ 55\,cm Al & $29^\circ$ & $-8.4\,$dB & scan / HWP & 7048 & 6 \\
      220 / 270 & $3\times$ 44\,cm Si & $35^\circ$ & $-13.4\,$dB & scan / HWP & 16876 & 4 \\
\hline
      \multicolumn{5}{|r}{total:} & \multicolumn{2}{l|}{153,232 detectors, 18 tubes} \\
      \hline
    \end{tabular}
  \end{center}
  \caption{\label{tbl:SAT_tubes_detectors} Summary of small-aperture telescopes for the reference design.}
\end{table}

\subsection{Optics}
\label{sec:sat:optics}

\begin{figure} [t]
\begin{center}
\includegraphics[width=4in]{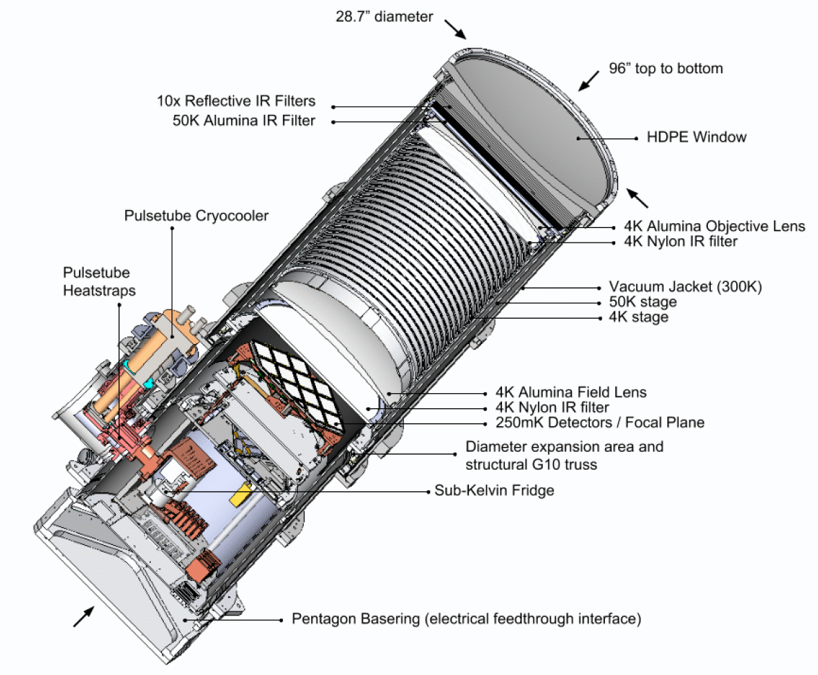}
\end{center}
\caption{
The CMB-S4 small-telescope optics are simple cryogenic refractors of approximately half-meter
aperture, a design based on heritage from the cryogenic refractors of the BICEP series 
of telescopes which have proven their performance in 
deep $r$ measurements through Stages 1, 2, and 3.
Shown here is the BICEP3 telescope, which has been observing since 2015, 
as an example of an existing instrument that illustrates the essential optics
design elements (including lenses, cold baffles, filters, and vacuum window) 
that all the CMB-S4 small telescopes will contain.
}
\label{fig:SAT_opticsB3}
\end{figure}

\begin{figure} [t]
\begin{center}
\includegraphics[height=3in]{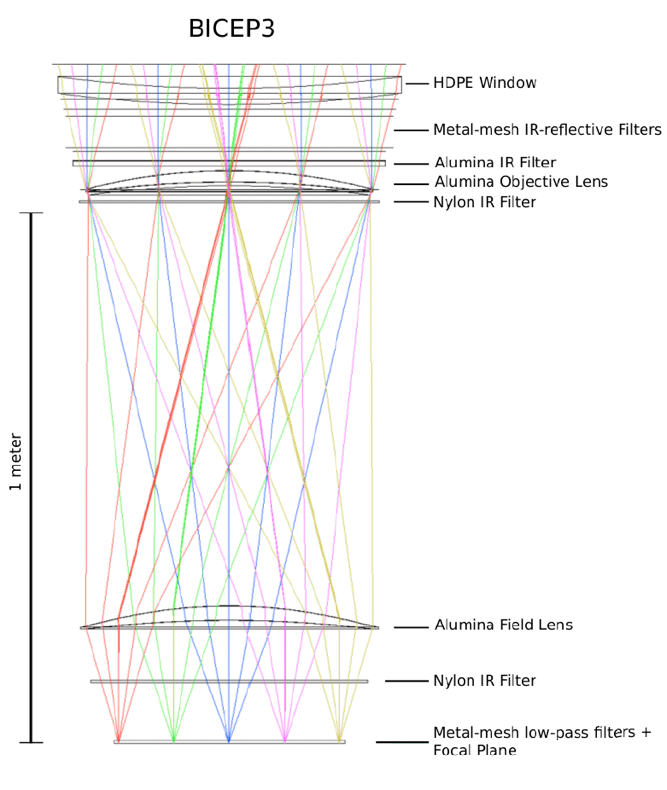}
\includegraphics[height=2.9in]{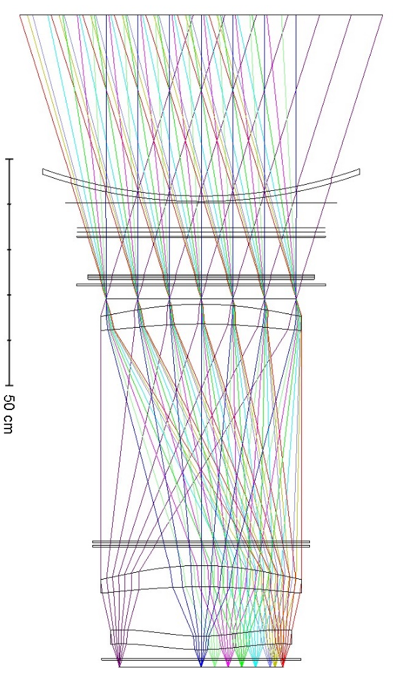}
\end{center}
\caption{
The detailed optical designs for the small telescopes are 
based on the realized 2-lens (BICEP, left panel), 
and 3-lens (SO, right panel) refractors that offer large optical throughput 
(etendue), symmetric main beams, and excellent polarization properties. Refractors are significantly 
more compact compared to crossed-Dragone telescopes with the same primary aperture, and offer
advantages of symmetry and sidelobe control. 
}
\label{fig:SAT_raytrace}
\end{figure}

The measurement requirements described above can be economically met with simple on-axis refractors. 
The 2-lens refractive optics of the BICEP series of experiments---BICEP1, BICEP2, Keck Array, BICEP3, and now 
BICEP Array---changed little over the past 14 years between the generations.  The only notable upgrade was 
the increase in aperture/throughput that drives the change of the lens material from high-density polyethylene (HDPE) to alumina from BICEP3 onwards. 
The choice of on-axis refractive optics was re-affirmed by the SO collaboration who, after extensive comparisons
between refractors with the leading alternative, crossed-Dragone reflectors, has also selected refractors for
their compactness, lower cost, and easier sidelobe control. 

The lenses must be anti-reflection (AR) coated to prevent reflections at percent or better levels. The SAT reference
design baselines dichroic detectors which require similarly broadband cryogenic AR coatings on the optical elements. 
For silicon lenses, subwavelength features can be cut into the lens surface with a custom three axis silicon dicing
saw. These features constitute a metamaterial that behaves as a simple dielectric coating with excellent optical and 
mechanical properties. For the alumina lenses, this can be done by gluing or laminating appropriate dielectric layers 
onto the lens surfaces. In particular, laser-treated epoxy coating has been applied successfully in BICEP3 (95\,GHz) and
POLARBEAR2 (95/150\,GHz). 

The SAT reference design will have 55-cm alumina optics at low frequencies and 44-cm silicon optics at high frequencies. 
This distribution was driven by the resolution requirements at low frequency, and uncertainties in loss and anti-reflection 
coating of alumina at high frequencies.
These designs and aperture sizes are
based on optics that have been fully demonstrated in BICEP3 and/or realized for the SO design. 

An aperture size larger than $\approx 20 (\lambda/2\,{\rm mm}) {\rm cm}$ is required to resolve the degree-scale $B$ modes. 
Beside the resolution requirements, a larger aperture offers larger optical throughput which proportionally reduces 
the cost of cryogenics and telescopes given the same number of detectors. On the other hand, the size of the vacuum 
window that can be reliably built and the availability of the lens material present practical limitations. 
Low-loss alumina lenses are available up to 80\,cm in diameter.  The Stage-3 experiment BICEP Array will adopt 55-cm 
alumina optics at all frequencies.  For the CMB-S4 SAT reference design, silicon lenses are chosen for high frequency bands 
for two main reasons.  First of all, high purity-silicon has superior transmittance compared to alumina.  Additionally, 
broadband AR coating of silicon has been demonstrated and verified up to 240\,GHz.  

In addition to the lenses, each of these optics tubes requires a vacuum window and several thermal filters. 
The vacuum window consists of HDPE anti-reflection coated with layers of expanded polytetrafluoroethylene (PTFE) or HDPE.  They can now
be made with $>60$\,cm diameter and $<3$\,cm of thickness, which contribute only a small fraction of in-band 
optical loading. For a 55-cm aperture, more than 100W of thermal radiation is entering the cryostat through each of 
the vacuum windows. This must be significantly reduced by thermal filtering. In Stage-3 experiments such as 
BICEP3, SPT-3G, and POLARBEAR2, adequate thermal filtering is achieved by a combination of scattering and absorptive 
filters. The scattering filters consist of stacks ($\approx$ 10 layers) of thinly sliced foam sheets, sometimes known 
as the RT-MLI (radio-transparent multi-layer insulation). The remaining infrared loading can be absorbed by IR-opaque 
alumina filters, whose high thermal conductivity also makes them very effective heat sinks.  In BICEP3, a 
foam stack reduces the incoming thermal loading by 90\,W and an AR-coated 
alumina filter absorbs and dumps 12\,W of the remaining power 
into the first stage (40\,K) of the pulse tube cooler. 
In the BICEP series of experiments, additional nylon filters 
were used on the 4-K stage to reduce loading onto the sub-Kelvin stage, and low-pass edge filters
were used directly above the focal plane to reduce out-of-band direct illumination of bolometer islands.
Given the much larger cooling power of dilution refrigerators, and the bolometer island shielding offered by
a feedhorn architecture, both of these final layers of filtering can likely be demonstrated in lab testing
to be unneccessary, but to keep the reference design maximally conservative it retains these filters for now.
Overall, the vacuum window and thermal filtering required by 
CMB-S4 SAT do not go beyond the level already demonstrated in BICEP3 and therefore present very low risks. 

Another key decision made for the SAT reference design is related to the pixel density. As is well known, 
when the spillover illumination of the feeds was terminated at sufficiently low temperature ($<2\,$K), 
formal sensitivity optimization tends to favor feeds with diameters smaller than or comparable to the size of a 
diffractive spot ($f\lambda$), given a fixed optical throughput. However, the corresponding larger field strength 
at the edge of the aperture and the integrated spillover power terminated inside the optics tubes tend to 
generate near- and far-sidelobes with non-trivial polarization asymmetries
through diffraction and/or small-grazing angle reflections/scattering. 
This corresponds to more demanding requirements for the cold aperture stop and the absorptive baffles, 
and for the beam measurement and control of temperature to polarization leakage in analysis.  
The BICEP series of experiements have used conservative edge tapers that are in the $-8$ to $-23\,$dB range.
For the SAT reference design, we have relaxed this slightly to allow edge tapers as high as $-5.7\,$dB
(see Table~\ref{tbl:SAT_tubes_detectors}),
which are still relatively conservative.
There are ongoing efforts to make the cold absorbers more mechanically robust
under cryogenic conditions, and better matched in polarization response. 
However, with this conservative choice of moderate packing density for the 
feeds, the SAT cold stop, baffling, and systematics contol and analysis strategies 
likely would not require anything beyond that has been demonstrated in Stage-3 experiments.

\subsection{Half-wave plates}
\label{sec:sat:hwp}

In the event that SATs are deployed to Chile, those at 85\,GHz and above will be equipped with 
continuously-rotating cryogenic half-wave plates (HWP) 
to reduce the impact of the higher level of atmospheric noise fluctuations at that site.
The HWP can also be used to eliminate the effects of 
instrumental polarization by optical elements between the HWP and the detectors.
The HWP is located near the cold aperture, on the sky side of 
all the lenses, therefore eliminating the optical systematics due to the lenses
and detectors. 
Reduction of instrumental polarization reduces temperature-to-polarization leakage,
and the effects of polarized sidelobes generated by effects on the detector side of the HWP.

\begin{figure}
\begin{center}
  \includegraphics[width=0.45\textwidth]{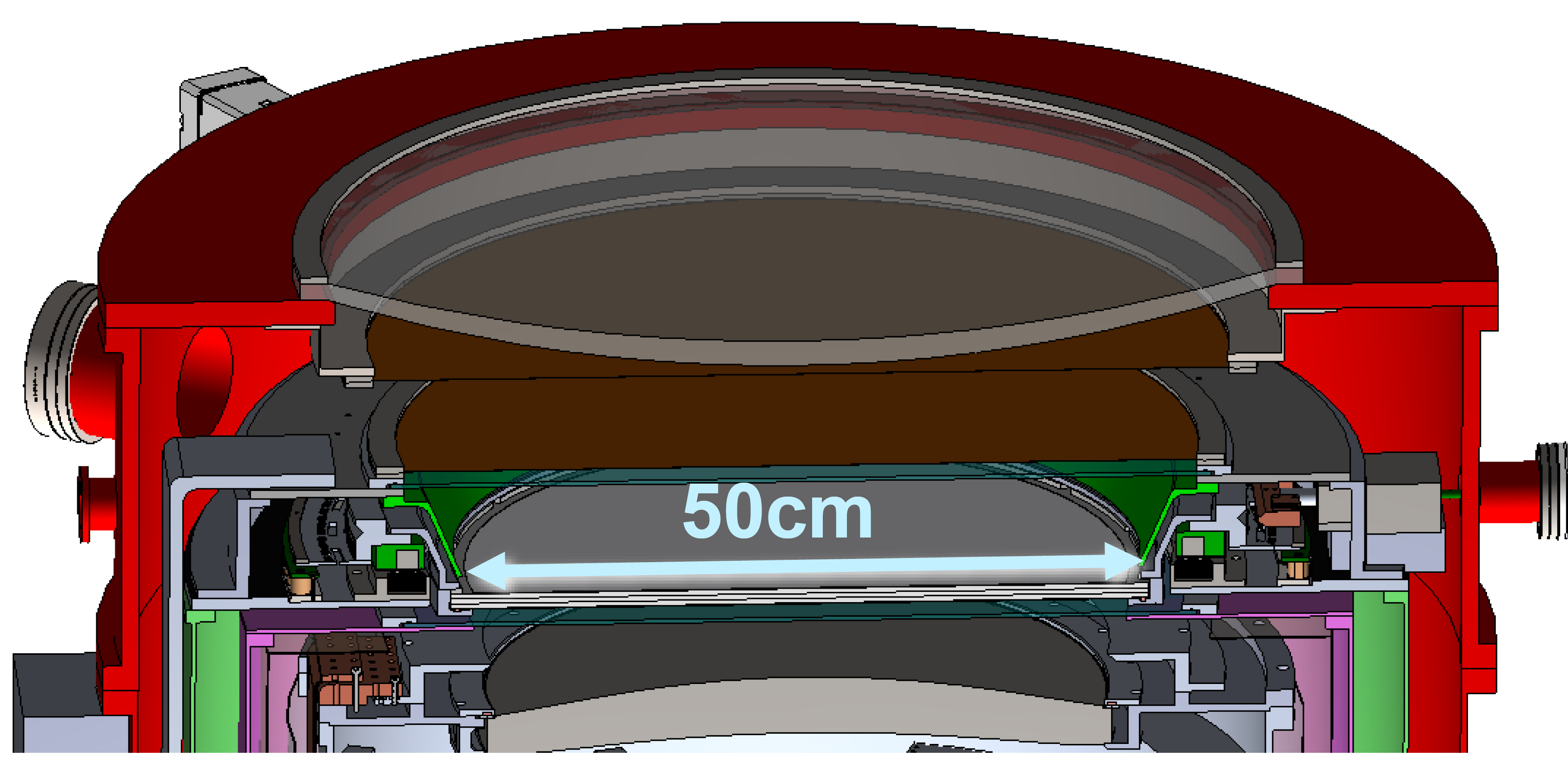}
  \hspace{5mm}
\includegraphics[width=0.45\textwidth]{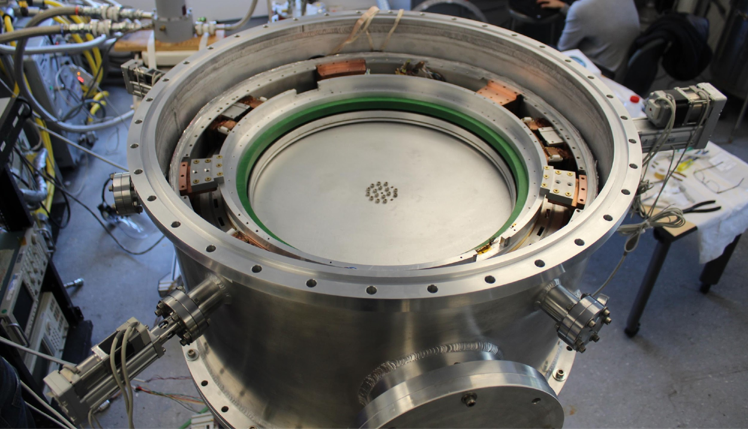}
\end{center}
\caption{ %
  {\it Left}: drawing of the SO cryogenic half-wave plate section.  The HWP system consists of the sapphire HWP and rotation
  mechanism adopting superconducting mag-lev bearing.  The optical clear aperture is 50\,cm in diameter.
  {\it Right}: a photo of the Simons Array cryogenic half-wave plate system,
  from which the SO system is derived.
}
\label{fig:sat_chwp}
\end{figure}

Our baseline HWP design is similar to that of the SO small-aperture telescopes,
which is derived from the HWP implementation in the Simons Array receiver
(Fig.~\ref{fig:sat_chwp}).
The optical element consists of three-layer stack of A-cut birefringent sapphire,
with an anti-reflection coating similar to that of Alumina lenses.
The three-layer stack allows a wide-enough modulation bandwidth
for the dichroic detectors.
The HWP is located on the 45-K stage.
Since the sapphire loss-tangent rapidly decreases as a function of temperature,
the thermal emission from the HWP is negligible below 100\,K.
The rotation mechanism consists of superconducting mag-lev bearing
and electromagnetic drive, achieving non-contact rotation.
This allows smooth rotation and long-lifetime operation
in the vacuum and cryogenic environment.

\subsection{Cryostat}

\begin{figure}[t]
\begin{center}
\includegraphics[height=2.0in]{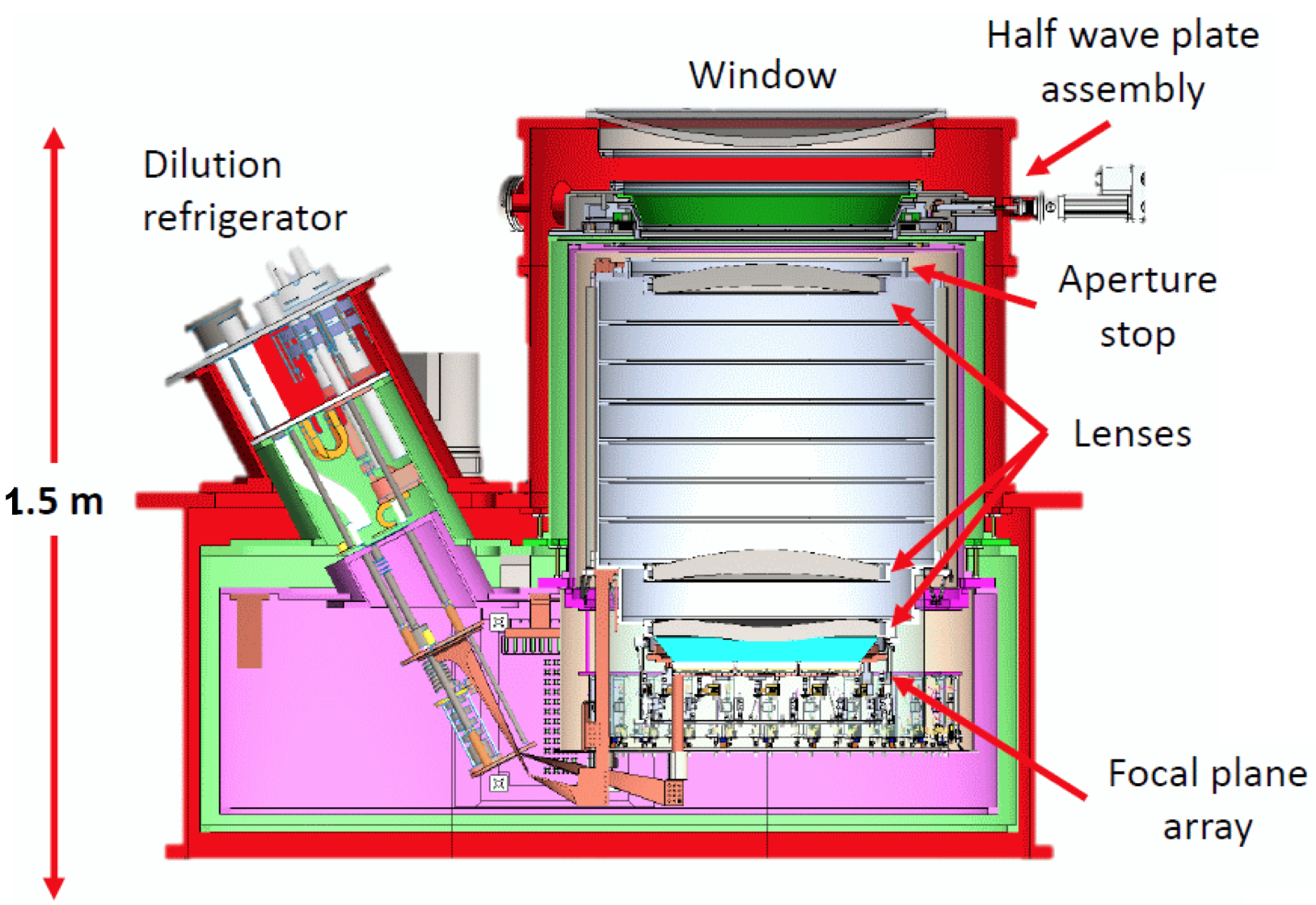}
\includegraphics[height=3.0in]{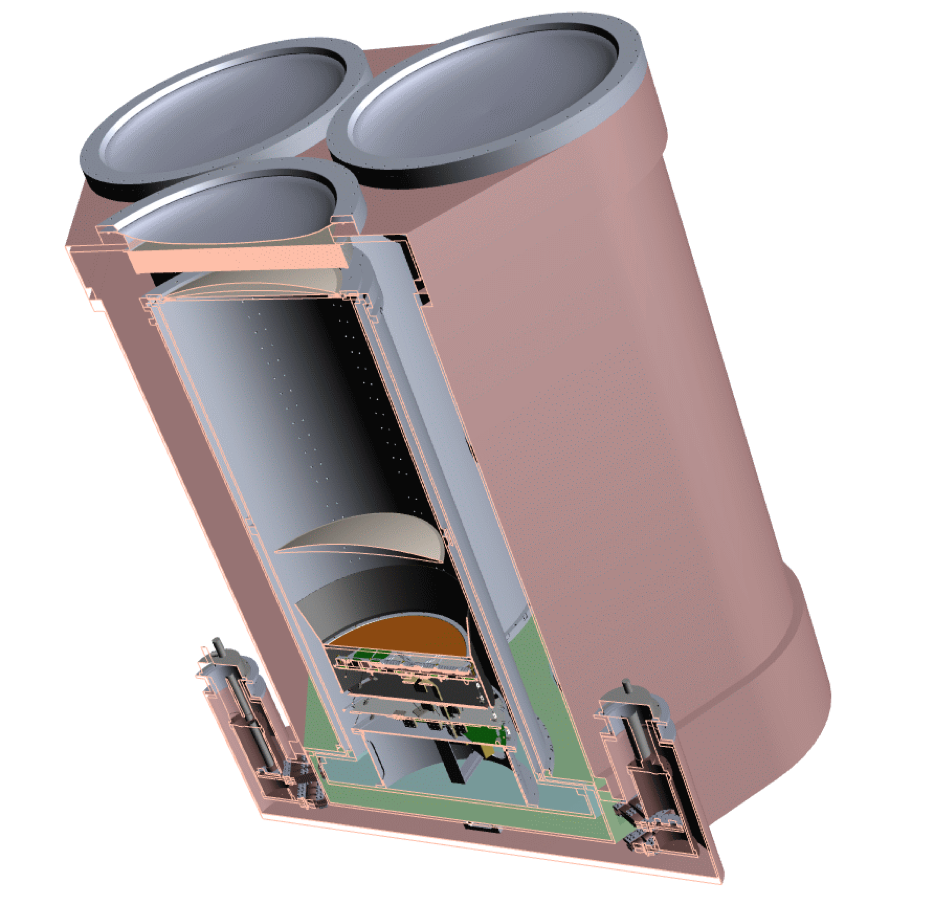}
\end{center}
\caption{
On the left, the SO cryostat, illustrating a design that uses a dilution refrigerator to cool focal
plane and 1-K optics.  On the right, the reference design ``3-tube'' cryostat.
It accommodates three cryogenic optics tubes each fitting within 
a 70-cm diameter $\times$ 150-cm long envelope, cooled to 4\,K by the three pulsetubes located
around the lower perimeter of the cryostat and to 1\,K and (for the focal planes)
	100\,mK by the diliution refrigerator.  The optics tube shown is the low-frequency, alumina lens 
version, derived from the BICEP Array design.  The high-frequency, silicon lens version 
will fit in the same volume, with bolt-on extensions between the cryostat and
each vacuum window to accommodate the HWPs if deployed to Chile.
}
\label{fig:SAT_cryostat}
\end{figure}

\subsubsection{Cryostat design drivers}

The cryostat consists of lenses and an aperture stop that are cooled to $\approx$1\,K, and a focal plane maintained at $\approx$100\,mK by means of a dilution refrigerator. Maintaining these temperatures rather than higher ones offers sensitivity gains, and fabrication margin for the superconducting detectors. 

We considered two alternative designs for the cryostat: (1) a cryostat with a single optics tube and focal plane assembly; and (2) a cryostat that contains three optics tubes and focal plane assemblies. We refer to these as the ``single-tube" and ``3-tube" systems. 
In both cases, three tubes share a common telescope mount.
Our choice was based on the following design drivers. 

\begin{description}
\item[Electrical Power:] The system of pulse-tube and dilution refrigerator coolers is the 
largest power consumer during operation, and supply of power is among the most expensive 
logistics requirements at either site.  
In particular, limitation in power at the South Pole may be a driving constraint.
The power consumption per unit of sensitivity is among the primary figures of merit.
The 3-tube cryostat allows more efficient use of power, and 
thus has a better power consumption figure of merit.
\item[Modularity:] The modularity of the system is a key element during 
the integration and testing.  While a single-tube system allows more modular 
integration and deployment, the total number of units to deploy is smaller with the 
3-tube design, and they still offer a large degree of modularity compared with the LAT cameras.
\item[Risk:] Commissioning the cryogenic systems is one of the largest schedule risks.  
When optimizing the system for power consumption, 
the 3-tube system has larger margin in cooling capacity, and thus has 
less risk in cryogenics.  On the other hand, all the pulse-tube-cooled small-aperture systems 
deployed in the past (BICEP series, ABS, and CLASS) are single-tube cryostats, and this 
heritage translates to reduced risk.  
\item[Baffling:] A single-tube system must have larger separation between the three cryostats'
vacuum windows (on a single telescope mount), 
allowing each tube to have a dedicated forebaffle centered on that tube. 
These centered, symmetric forebaffles are advantageous for the single-tube system.  
A 3-tube system would have a common and proportionally larger forebaffle,
leading to higher cost and a construction challenge.  
On the other hand,
since the separation among the vacuum windows in a 3-tube system is smaller, the ground screen 
would also be smaller. 
\item[Cost:] A 3-tube cryostat will be operated with a single dilution refrigerator
offering cost-savings relative to the single-tube option, which requires three times more refrigerators. 
\end{description}
Based on these design drivers, and specifically because of lower power consumption and cryogenic risk the reference design uses the 3-tube system. 
However, we maintain the 
single-tube system as an option.  Among the near-term development items 
is the investigation of the cryogenics significantly contributing to this \emph{engineering}
decision in the future.

Among the various multi-tube cryostat options, the {\it three}-tube was judged optimal 
because 
the 1-K stage and 4-K stage cooling power of the BlueFors SD400 and Cryomech PT420
are well-matched to support three optics tubes.  Increasing 
the number of close-packed optics tubes per cryostat beyond three likely requires 
additional cryocooling, and will also lead to a
larger common forebaffle.
For these reasons we limited
the number of tubes per cryostat
to three.

\subsubsection{General configuration}

The general configuration of a SAT receiver will consist 
of a 2-meter cryostat with three optical tubes. 
The main shell will have three windows and a common 40\,K 
radiative shield. 
Inside this shield, we will place three receivers that will have 
separated 4\,K shields and optics. 
The size of the single optical tube will be around 70\,cm, 
comparable with the present Stage-3 small-aperture instruments. 
This design will allow different teams to work in parallel on 
optics tubes that fill a single cryostat design, which would be identical for
the South Pole and Chile small telescopes.
A key factor of the optical tube design will be the 
modularity of the mechanical, optical and readout parts. 
That will allow a fast and straightforward replacement of parts 
and also the capability to reconfigure the instrument observation 
bands to enhance multi-frequency coverage and map depth. 
Standardization of the parts that involve cabling, mechanical supports, 
and detector module interfaces has demonstrated advantages of cost
savings and efficiency and is standard practice within in Stage 2 and Stage 3
experiments (e.g., Keck Array, BICEP Array, SO) that have deployed or
are planning to deploy small-aperture telescopes at multiple frequencies.

\subsubsection{Cryogenics}
The detectors are cooled to $\approx$100\,mK, with intermediate stages
at 1, 4, and 45\,K for radiation shields,
optical filters, and wiring thermal intercepts. 
The 45-K and 4-K stages have one PT420 pulse tube
supplemented (to add margin) 
by one single-stage PT60 pulse tube (see Table~\ref{tab:sat_cooler}).
The 100-mK stage is cooled by a BlueFors
SD400 dilution refrigerator with an intermediate stage at 1\,K.
The dilution fridge has its own PT410, dedicated for the
dilution fridge system.
Flexible copper braids are used to connect
the refrigerators to the cryostat cold stages.
All the refrigerators must be kept within $\approx45^\circ$ of vertical to
maintain cooling capacity.

\begin{table}
  \label{tab:sat_cooler}
  \centering
  \begin{tabular}{|l|l|l|r|rrrr|}
    \hline
    Fridge & Type & Quantity & Power & \multicolumn{4}{c|}{Cooling capacity per fridge}\\
    & & & & 45\,K & 4\,K & 1\,K & 100\,mK\\
    \hline
    PT60 & Pulse tube & 1 & 3.3\,kW & 20\,W & & & \\
    PT420 & Pulse tube & 1 & 12.5\,kW & 55\,W & 2\,W & & \\
    PT410 & Pulse tube & 1 & 8.4\,kW & 40\,W & 1\,W & & \\
    SD400 & Dilution & 1 & 2.0\,kW & & & 25\,mW & 400\,$\mu$W\\
    \hline
  \end{tabular}
  \caption{Refrigerators.  The PT410 is only used for the dilution refregerator system. }
\end{table}

The reference design cryogenic architecture for a SAT receiver includes 
three optical tubes inside a single 40\,K shield volume,
cooled by the PT420 and PT60 pulse tube first stages. 
The PT420 pulse tube second stage will cool the 
three tubes down to 4\,K.  The PT410 will provide 
the base temperature to run the dilution refrigerator for cooling 
the insert sub-kelvin stages and the focal plane to an operating 
temperature of 1\,K and 100\,mK, respectively. 
The required cooling power for the SAT can be extrapolated 
considering similar configuration experiment design and the BICEP Array receiver. 
We are expecting to have roughly 50\,W at 40\,K and 1.5\,W at 4\,K that 
can easily be managed by the combination of Cryomech PT420 and PT60. 
The dilution refrigerator will have stages at 1\,K and 100\,mK respectively. 
Again comparing with similar instrument designs we are 
expecting around 0.5\,mW at 1\,K and 45\,$\mu$W at 100\,mK, 
well within the capabilities of a BlueFors SD400 Dilution Refrigerator. 

In order to mitigate input thermal loads, the 40-K and 4-K radiative 
shields have to be wrapped with Multi-Layer Insulation aluminized mylar. 
Moreover, the mechanical supports between the stages have stringent 
requirements in terms of their thermal conductivity. 
The current generation of cryogenic instruments for CMB uses 
G10 fiberglass supports (sometimes supplemented by titanium tension members)
above 30\,K and the composite carbon fiber ones below 30\,K. 
These solutions will be implemented in the SAT design. 
The choice of materials for the shields and the thermal straps is crucial. 
Based on past experience, we will largely use high-conductivity 1100 Aluminum for the 
radiative shields and 101 Oxygen-free Copper for cold plates and thermal straps. 
For the design of the heat straps, depending on the temperature stage, 
we will use copper braids or foil shims. Their design maximizes the 
thermal conductivity allowing the differential thermal contraction 
due to the cooling process to take place without damaging the cryogenic system. 

Superconducting devices such as TES bolometers and
KIDs are sensitive to the presence of external magnetic fields. 
Also, the readout architecture for TES bolometers is based on the use of 
Superconducting Quantum Interference Devices (SQUIDs), 
which are very sensitive magnetometers. 
These millimeter-wave sensors and the cold readout 
electronics are therefore quite sensitive to electromagnetic pickup induced by 
even very weak magnetic fields. 
The Earth's magnetic field (modulated by the telescope scan), as 
well as fields produced by the instrumentation surrounding the experiment, 
are potential sources.
This pickup can result in the production 
of artifacts in the CMB temperature and polarization maps that can be very hard 
to disentangle and remove. A common strategy for mitigating these effects is 
to implement magnetic shields in the cryostat mechanical design, a technique 
that is currently adopted in all the Stage-3 CMB experiments such as 
BICEP Array, SPT-3G, and Advanced ACT. 
The shielding strategy varies from experiment to experiment but it is 
based on the interleaving of high permeability and superconductive material layers. 
The baseline design for the SAT magnetic shield is assumed to copy that of BICEP Array, 
which uses a high-permeability (Amuneal A4K) cylinder around the 40-K tube combined with a Niobium superconductive 
cup at 300\,mK (or at 1\,K for S4), to reduce magnetic fields by a factor of about 200 at the focal plane. 
Additional shielding layers are usually added locally in the detectors/SQUID areas. 
Looking again at the BICEP Array architecture, 
its FPU detector module design provides an additional factor of 500 
shielding with the use of an A4K planar sheet and a Niobium enclosure; a similar
shielding factor at the detector modules is assumed to be achievable for the S4 reference design. 
The use of finite-element model simulation software (e.g., COMSOL Multiphysics) 
can help optimize the shielding configuration.

\subsubsection{Mechanics}

Survival requirements on the vacuum jacket and vacuum window include:
safe storage to temperatures of $-90\,$C and maintenance of vacuum to temperatures of $-40\,$C.
In past experiments the latter has been straightforward to achieve with silicone elastomeric o-rings.
In operation the cryostats will be contained in an environmental enclosure
which surrounds the telescope mount, with heated airflow directed around
the vacuum window to eliminate frost/snow accumulation, so exposure of
cryostat components to colder ambient temperatures while in operation is not
expected.

Within the cryostat, pointing rigidity of the optical tubes is maintained
by arranging for kinematic constraint of the thermal standoffs, with rigid triangulated
G10/carbon fiber truss elements supporting bottom end of the 50-K, 4-K, and 1-K stages
and Ti-Al-4V tensile members providing radial constraint at the top (open) end of each
tube \cite{Crumrine2018}.
Lenses are supported from optics tubes using tangential-blade standoffs 
which provide thermal conductance while maintaining concentricity and
relieving radial strain from differential thermal contraction.
Based on pointing performance of Keck Array and BICEP3, we expect pointing effects from
gravitational deflections of the optical elements within the cryostat 
to be easily limited to $<15$\,arcsec with such a mechanical support system, and to be
repeatable.  The cryostats themselves are supported in the telescope mount
with rigid triangulated space frame members which connect hard points near
the top and bottom vacuum flanges to mounting points on the inner diameter of the
boresight rotation stage that surround the cryostat's center of gravity.  

The total cryostat mass is estimated to be 1800\,kg (approximately three times that
of a BICEP Array cryostat).  SAT cryostat assembly (both in North America
and in the field) will take place on dedicated wheeled lab stands designed to distribute their
mass over a $2\times2$\,m footprint, minimizing requirements for specially-reinforced lab floors.
Assembly of the cryostats requires a well-controlled 2-ton hoist with a hook height of $\ge12.5\,$feet.

\subsubsection{Control and monitoring}

The small-aperture telescope cryostats share the same control and monitoring
interface as the LAT cameras (described briefly in Sect.~\ref{sec:rd:latcam:monitor}).
We monitor temperature sensors, vacuum gauge and valve, and the status of the refrigerator and compressor.
Control interfaces are used for heaters, the vacuum valve, refrigerator, and compressor.

Control and monitoring for each SAT integrates into the overall observatory
control software in the form of one or more subsystems, as outline in
Sect.~\ref{sec:acq:hk}.
The requirements for timing synchronization are modest and the monitoring
data rate is insignificant in comparison to the bolometer data.

\subsection{Mount}

\begin{figure}
\begin{center}
\includegraphics[height=2.4in]{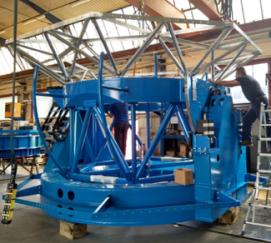}
\includegraphics[height=2.7in]{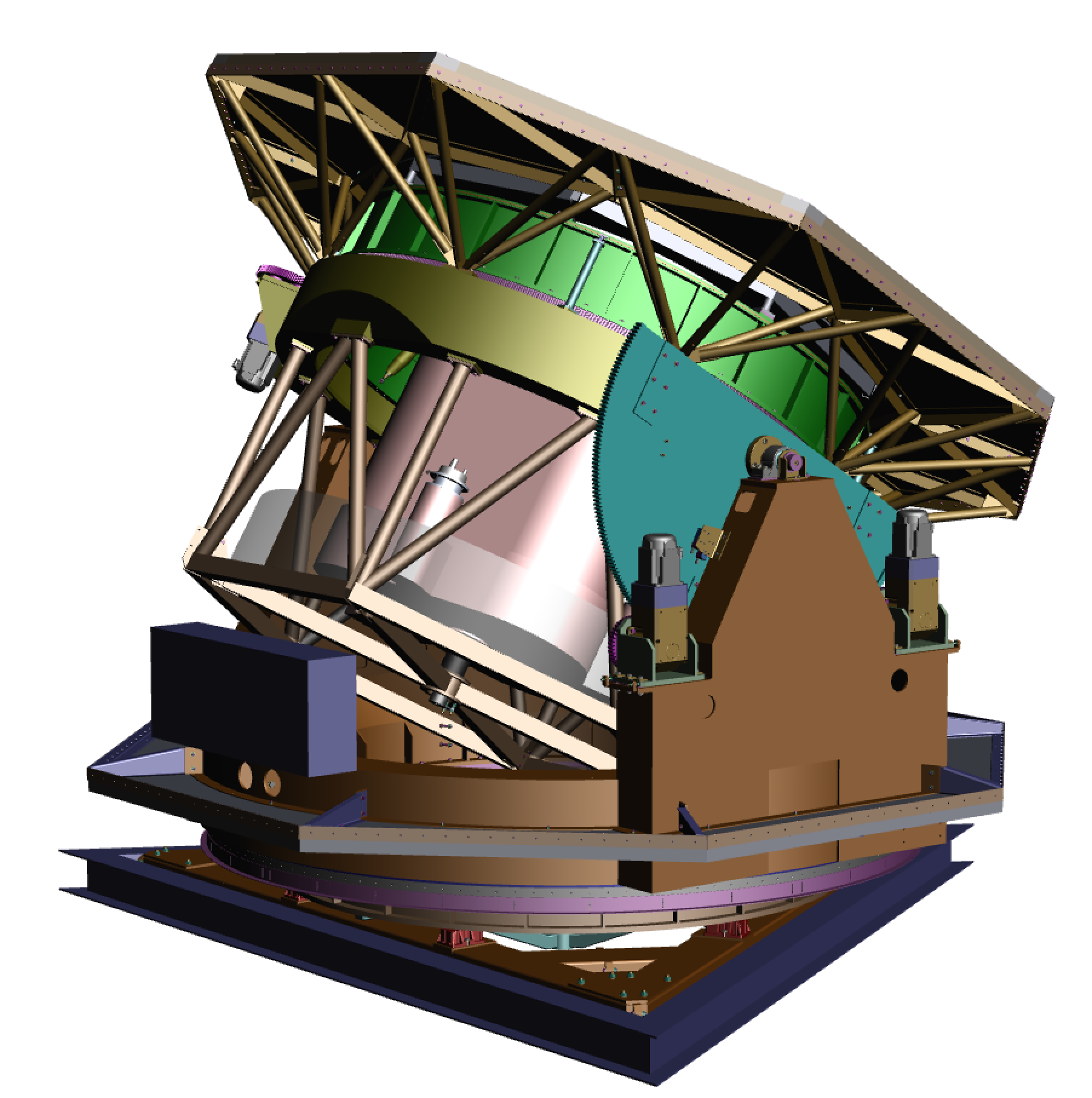}
\end{center}
\caption{
The CMB-S4 small-aperture telescope reference design telescope mount is based
on the existing BICEP Array mount design (left).  Although originally designed
for four individual small-aperture telescope cryostats (each with a single
pulsetube), its size is well-suited to mount the single 3-tube cryostat with
three pulsetubes and dilution refrigerator (right).  It allows three axis motion
with infinite rotation (using rotary feedthroughs) on the azimuth and boresight
axes, and a design that allows for full environmental enclosure with space for
interior access to all critical components.
}
\label{fig:SAT_mount}
\end{figure}

The CMB-S4 reference design mount for the small-aperture telescopes is a 3-axis
fast-scanning mount, based on the existing design of the (Stage 3)
BICEP Array mount (see Fig.~\ref{fig:SAT_mount} and Ref.~\cite{Crumrine2018}).

The CMB-S4 mount must be capable of carrying the SAT
cryostats and their associated electronics and cryogenic equipment and
of scanning them on the sky rapidly enough to make differential 
degree-scale measurements in the presence of $1/f$ instrumental noise and
changing atmosphere and ground signals.
It must achieve an accuracy of pointing knowledge while scanning that ensures sufficient
control of systematic uncertainties, both in mixing of $E$ to $B$ modes 
due to pointing reconstruction errors \cite{Hu:2002vu} and in
cross-referencing maps across SAT frequencies and with the LAT data
for accurate galactic foreground and lensing separation.
It must provide 3-axis rotation, including full rotation around the boresight
axis to allow both for suppression of instrumental systematics in polarization
measurement and for consistency checks offered by a full set of null tests \cite{Ade:2015fpw}.
It must provide an environmental enclosure to protect the cryostats,
electonics, and other elements from extremes of temperature
or snow exposure and to offer a stable operating environment with
reliable interior access to all critical components.  The design requirements
are summarized in Table~\ref{tab:SmallTelescopeDesignRequirements}

\begin{table}
\centering
\begin{tabular}{|l|l|l|}
\hline 
Parameter & Value & Notes\\
\hline
Mass of instrument & up to 4500\,kg & includes cryostat, DR system, electronics, forebaffle\\
Motion & 3 axis & full boresight rotation of instrument and forebaffle\\
Scan pointing knowledge & $<15\,\textrm{arcsec rms}$ & $<1/20\textrm{th}$ beamwidth at $\lambda=1\,\textrm{mm}$\\
Scan speed AZ/EL/TH & $5/1/1\,\textrm{deg s}^{-1}$ & $\approx3\,\textrm{deg s}^{-1}$ on the sky for fast diff. measurements\\
Scan accel. AZ/EL/TH & $3/1/1\,\textrm{deg s}^{-2}$ & turnaround efficiency\\
Range AZ/EL/TH & $\infty/45\dots110/\infty$ & continuous AZ desirable\\
Shipping envelope & standard double pallet & deployment via C-130 / standard vehicles\\
Mount mass & $< 25$\,tons & includes instrument, comoving forebaffle and scoop\\
Survival: wind & $70\,{\rm m}\,{\rm s}^{-1}$ & Chile dominates\\
Survival: seismic & $0.3\,$g & Chile dominates\\
Survival: temperature & $-90\,$C & Pole dominates\\
\hline 
\end{tabular}
\caption{Small-telescope mount design requirements. \label{tab:SmallTelescopeDesignRequirements}}
\end{table}

The reference design mount illustrated in Fig.~\ref{fig:SAT_mount} 
meets the design requirements using the existing BICEP Array mount design.
Although this mount was originally designed for four 
individual small-aperture telescope cryostats, each with a single
pulsetube, its size is well-suited to mount the single 3-tube cryostat with
three pulsetubes and dilution refrigerator.  This mount 
includes two separate rotary unions which allow continuous rotation 
about the azimuth axis and the array boresight axis without the need for cable wraps. 
These rotary unions each contain 10 helium channels. 
Eight of these connect the pulse tubes and their compressors, 
while two channels serve as pressure guards. 
An additional nitrogen channel provides a pressurized environment 
on front end of the cryostats which prevents water absorption 
into the window material. Slip rings at the ends of the unions 
additionally provide data and power connections to electronics 
across separately rotating stages of the mount. 
These rotary unions allow the helium compressors required to operate 
the pulse tube coolers to sit below the mount structure in the 
stationary equipment room. 
Helium lines route upwards into the lower (ground fixed) 
half of the first rotary union and then out through the upper 
half which rotates in azimuth along with the receivers. 
The hoses from the upper half are then routed through a short 
cable chain that provides flexure when rotating in elevation. 
The second rotary union is then similarly connected between 
the elevation and boresight stages.
Modifications to the current mount design will be required
mount the dilution refrigerator gas handling system 
inside the envelope of the instrument package which
co-rotates on the boresight stage, 
and engineering work will be required to ensure this
system can operate while tilting up to $45^\circ$ from vertical.

The mount provides a flexible environmental seal that fully 
encloses the components seen in Fig.~\ref{fig:SAT_mount}, while
exposing only the vacuum window, forebaffle, and co-rotating shields
(see Fig.~\ref{fig:SAT_shields}).  Access to the telescope for service
is from an equipment room below which houses helium compressors
and control computers, similar to existing BICEP and Keck Array facilities
currently in use at the South Pole.  The access passage through the azimuth bearing
is large in diameter, and is designed to accommodate installation
of cryostats and other components as well as personnel access,
all within the interior environmental space of the observatory
complex.  Alternatively, cryostats may be lifted into the mount from above
using a crane.

\subsubsection{Control and monitoring}

The large and small telescopes share a common control and
monitoring interface (see Sect.~\ref{sec:largetelescope}).
Drive amplifiers, brushless servo motors, reducers and drives, 
encoders, emergency stop and safety interlock systems 
are specific to each telescope design.  Based on the systems
developed for the Keck Array and BICEP Array mounts, 
we expect these drives to consume 10\,kW peak power.

\subsection{Ground shields and exterior baffles}

\begin{figure}
\begin{center}
\includegraphics[width=6in]{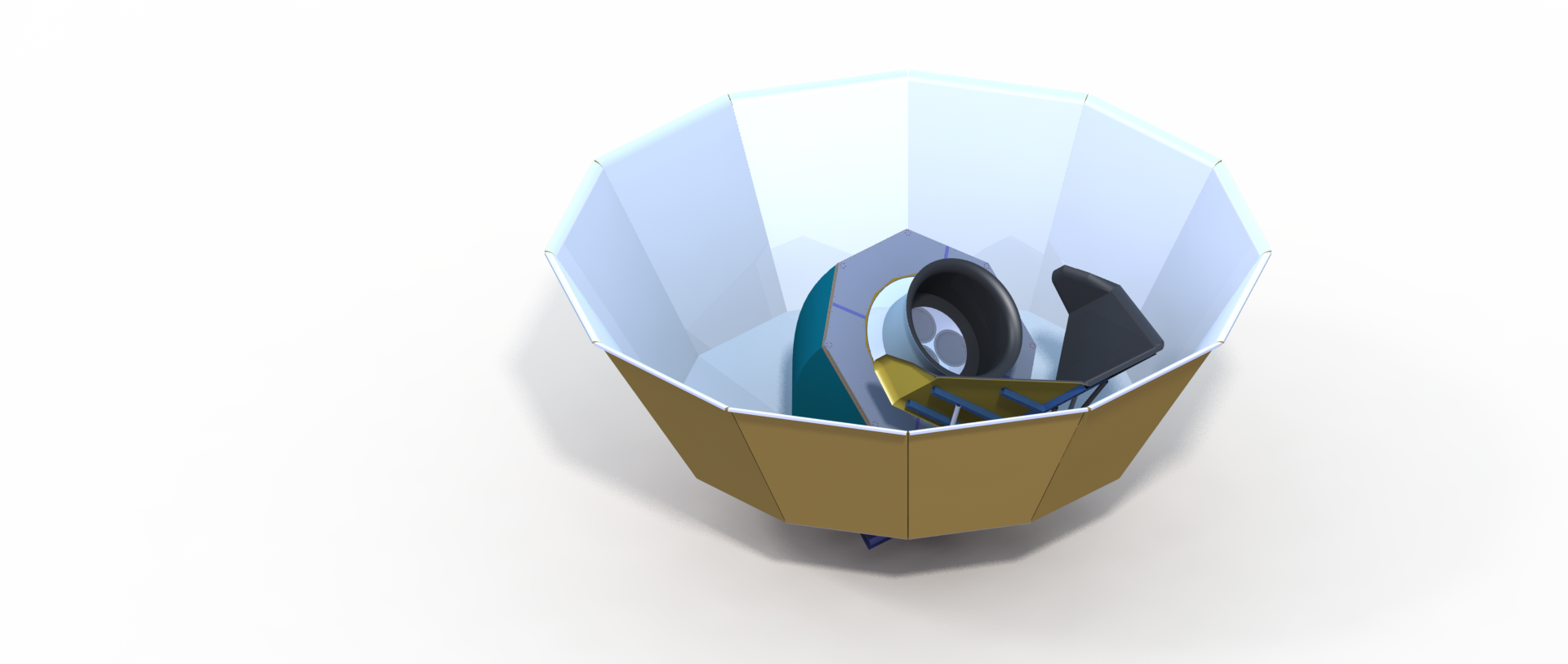}
\end{center}
\caption{
Exterior view of the small-telescope mount and shielding.  The 3-tube cryostat is surrounded 
by a co-moving absorptive forebaffle.  The mount sits within a large 
reflective ground shield.  An additional co-rotating ``scoop'' is used 
to keep the shield dimensions reasonable.  
A flexible environmental seal surrounds the mount structure.
}
\label{fig:SAT_shields}
\end{figure}

Extended beam response can couple to the warm ground or the
Galaxy/Sun/Moon during CMB scans, potentially adding structure
at the degree scales relevant for the r measurement.  A key
advantage of the small-aperture approach is that the entire telescope
can be enclosed in multiple levels of shielding, significantly
reducing the effect of sidelobes.  Indeed, all of the leading $r$
constraints to date have used similar shielding strategies to absorb
sidelobes and/or reflect them to the cold sky.  

The first level of baffling is a co-moving absorptive cylinder which
extends around the telescope aperture.  Sidelobes---usually at
$\gtrsim \SI{10}{\degree}$ scales---that terminate on this surface see
a constant-temperature load, and therefore do not contribute to
degree-scale structure in the map (although the extra loading slightly
decreases sensitivity).  Like in the BICEP experiments, we will also
use an additional fixed, reflective ground shield that surrounds the
telescope mount.  While no rays from the aperture couple directly to
the ground shield, those that diffract over the lip of the co-moving
forebaffle do; the ground shield ensures that rays must diffract twice
before terminating on the warm ground.  A ray that couples to the
ground shield can add azimuth-fixed structure to the map, which must
then be filtered from the timestreams.  Finally, to keep the ground
shield dimensions reasonable, we use a tertiary shield that co-rotates
in azimuth with the mount (but is fixed in elevation), which partially
surrounds the telescope to couple to sidelobes at the lowest
elevations.  Figure~\ref{fig:SAT_shields} illustrates the three levels
of shielding around the 3-tube cryostat.

The dominant sources of sidelobes in small-aperture telescopes will be
different from those in the large-aperture case.  Instead of mirror
scattering and panel gaps, the most important effects for the SATs
will be: diffraction at the cold stop and primary aperture; scattering from
the window, filters, and other optical elements near the aperture; and
non-sequential reflections between optical elements.  We expect to
model the beams using Zemax and GRASP simulations to demonstrate that
we can meet the stringent systematics requirements.  These models will
then be confronted with \textit{in situ} measurements, both of the sidelobes
themselves and of the amplitude and temporal stability 
of the ground pickup signals on the relevant angular scales,
taken with and without various levels of shielding.

\subsection{Assembly and test plan}

Because integrating a single Small Aperture Telescope optics tube, focal plane (FP) 
and cryostat is expected to take on the order of 6--9 months and because there are 
six SAT 3-tube Systems totaling 18 tubes, the plan is to 
create three independent integration and test (I\&T) facilities.  
In addition to documenting and reviewing assembly/test plans for the 
focal planes and optics/cryostat assemblies, we will be defining 
the necessary infrastructure requirements for these I\&T sites.  

After the sites are chosen, SAT specific modifications will have to be completed.  
It is expected that clean rooms will be required for focal-plane assembly.  
In addition, vacuum and cryogenic related systems will be required for cold testing.  
Finally, high bay space with adequate crane coverage and a hook height of at least 12.5\,feet
is required for handling the large cryostats and assembled optics tubes.  
Each of these facilities will be occupied for a period of approximately 
2 years including time for testing and crating parts and assemblies.

The first of these three SAT I\&T sites will be utilized to initiate the assembly/test process.  
The first focal plane (FP) integration and test process will be used to capture necessary 
improvements and refine the FP I\&T procedures prior to passing them on to the other two sites.  
The optics/cryostat integration procedures will be tested and refined during a careful 
first integration of the optics and cryostat.  Again, test procedures will be exercised 
and refined during pumping, cool-down and trial operation of the first cryostat.  
All these refined procedures and lessons learned will be utilized in integrating the remaining SAT systems.

Should throughput for the FPs be sufficient at one or two facilities, it would be 
possible to reduce the number of facilities for this portion of the I\&T scope.  
However, to match the fabrication schedule for the remaining parts, we believe three sites are necessary.  
This plan allows all SAT systems to be integrated, tested and crated within somewhat less than 2 years.  
It also allows for contingency time in the event of problems arising during 
testing of the FPs or cryostat/optics assemblies.

To summarize the integration process, the FP hardware and detectors are assembled in a 
clean room and cold tested to ensure a high percentage of detectors are working.  
The tested FP assemblies are then delivered to the optics tube assembly area where 
they are temporarily installed on the back of the tube.  Three-tube structures are 
then installed in the integrated cryostats (with pulse tubes and vacuum equipment).  
The integrated cryostat is then pumped and cooled to operating temperature which is 
expected to take approximately 3 weeks time. A quick functionality test (to be defined) 
is performed to ensure proper operation. If problems arise, the system must be allowed 
time to come back to room temp so that the cryostat can be opened and the necessary repairs made.  
This could extend the test time by 2--3 months.

Once all systems are verified to be operating correctly, the cryostat will be opened 
to allow the FP to be removed and proper shipping supports installed before resealing 
and installing in the shipping crate.  Once at Site (Pole or Chile), time will be 
required to remove the cryostat assemblies from their crates, open and remove 
shipping blocks, install cryogenic supports, re-install the FPs and button 
everything back up for a final cool-down and functionality test. 
This is expected to take up to 2 months time at site.

\section{Detectors and readout \prelim{({\it K. Irwin, C. Chang, and A. Lee})}}
\label{sec:detectorsreadout}

\subsection{General description}\label{subsec:generaldescription}

The detectors and readout for CMB-S4 include the systems that couple RF from telescope at the focal plane, define the frequency bands, detect the RF power, and readout the resulting signals to room temperature.

CMB-S4 will use feedhorn-coupled transition-edge sensor (TES) bolometers as the focal plane detector technology. The pixels will employ a multichroic architecture with good performance over a 2.3:1 optical bandwidth. This dichroic scheme allows a single pixel to simultaneously observe in two distinct optical passbands, efficiently utilizing focal plane real estate without substantially burdening readout or optical coating technologies.

The detectors will couple to the telescopes through monolithic arrays of machined spline-profiled feedhorns. Feedhorns are a well developed technology providing high-efficiency telescope coupling with circular beams and low side-lobes. The spline-profiled shape provides the required 2.3:1 optical bandwidth with minimal crosspolar leakage. The waveguide output of the horn couples to a broadband planar orthomode transducer (OMT), which separates the orthogonal polarization components of the incoming radiation onto two pairs of transmission lines. The superconducting transmission lines feed diplexer filter banks that channelize the signal into the two optical passbands. For CMB-S4, the bandwidth of each optical passband may be as low as 15--20\%. Though potentially smaller than currently fielded systems, this bandwidth can still be achieved with demonstrated technologies. The diplexer output feeds a $180^\circ$ hybrid tee, which eliminates coupling to higher order modes in the waveguide providing single-moded performance over the full pixel optical bandwidth. The difference output of the hybrid tee is transmitted to a TES bolometer where the optical signal is thermalized and measured. Summing the outputs of a  bolometer pair within each pixel measures the total intensity of the incoming radiation, and differencing
the two measures the Stokes $Q$ parameter. The full linear polarization of a given location on the sky is measured by combining measurements of multiple polarimeters rotated relative to each other or by a continuously rotating half-wave plate.

\begin{figure}[h!]
\centering
\includegraphics[width=0.30\textwidth]{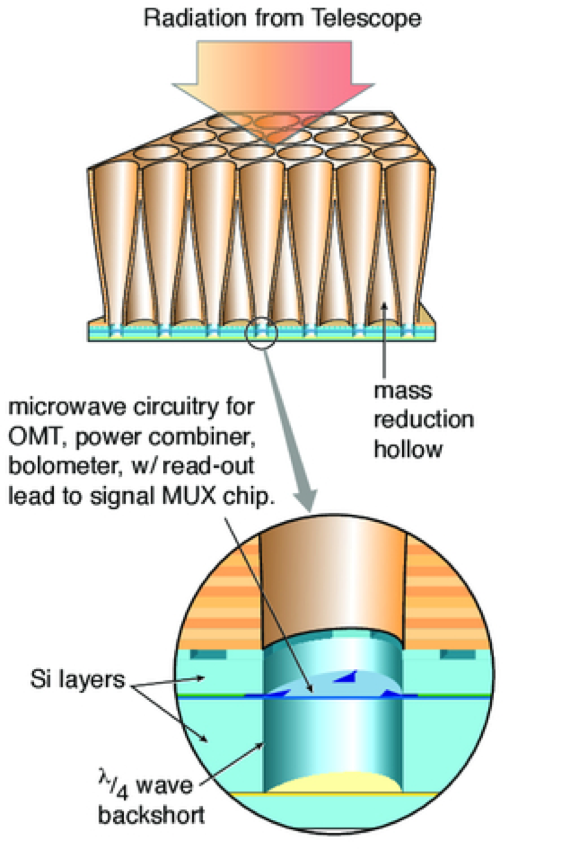}
\caption{Schematic diagram of a horn-coupled focal plane \cite{2016SPIE.9914E..0DR}.}
\label{fig:horn_OMT_schematic}
\end{figure}

The optical signal is measured using a transition-edge sensor (TES) bolometer, a highly sensitive thermometer consisting of a thin superconducting film weakly heat-sunk to a bath temperature much lower than the superconductor $T_{\rm c}$. 
TES dynamics are well understood, providing good descriptions of the noise and response for real devices. For CMB-S4, the TES detectors are designed so that the sensitivity is dominated by the (irreducible) statistical fluctuations of the absorbed photons. For ground-based experiments, this noise is typically $O(10)~{\rm aW}/\sqrt{{\rm Hz}}$, though values vary depending on platform/site, observation frequency/bandwidth, and the instrumental throughput/efficiency. Apart from photon noise, TES bolometers also measure fluctuations in the thermal carriers of the TES weak thermal link. The CMB-S4 TES bolometers will be designed and fabricated with parameters tailored for individual passbands at each site. These parameters will be chosen to provide the detectors with sufficient dynamic range to accommodate variations in weather and to provide sufficient thermal isolation so that the thermal fluctuation noise is much less than the photon noise. This optimization process is well developed from Stage-2 and Stage-3 experiments. Together with sufficiently low noise readout electronics, the TES detectors for CMB-S4 will operate with nearly ``background limited'' sensitivities. 
Despite the nearly background limited TES detector performance, over 500,000 detectors are 
required to achieve sufficient sensitivity to realize the CMB-S4 science targets. 
The CMB-S4 detector count is 5--10$\times$ larger than focal planes currently being developed, which will make it the most sensitive ground-based CMB experiment in the world.

The TES bolometers for CMB-S4 will have two superconducting transition sensors in series, each with different superconducting transition temperatures. 
This design allows the TES bolometers to have two modes. 
One mode uses a sensor with higher superconducting transition, allowing the  
TES bolometers to operate under a high optical load from calibration equipment. 
The second mode uses a sensor with a lower superconducting transition temperature. 
In this mode, the dynamic range of the detectors are tuned to achieve background limited sensitivity for astronomical observation. 
In this observation mode, a sensor used for the calibration source does not interfere with observation 
as that sensor is superconducting with zero resistance. This technique has been implemented successfully across multiple deployed CMB experiments. Aluminum is used for sensors with higher superconducting transition temperature (approximately 1.2\,K) and manganese doped aluminum will be used for observation sensors (0.16\,K). 

The TESs will be read out using superconducting quantum interference devices (SQUIDs). 
SQUIDs have a large noise margin over the detector noise, 
a critical feature for operating large arrays of detectors at sub-kelvin temperatures. 
In this Reference Design, we use time-division multiplexing (TDM), where a group of detectors is 
arranged into a two-dimensional logical array.  Each column of detectors shares a 
dedicated readout amplifier chain, and only one row of the array is routed to the 
amplifiers at any given time. The various rows are addressed cyclically in rapid succession to 
record the entire array. In TDM, the first-stage SQUID is wired in parallel with a 
Josephson junction switch, and the series voltage sum of all such units in the column is amplified 
by a series SQUID array (SSA) for transmission to the warm electronics.  
During multiplexing all but one of the switches are closed to short out the inactive SQUIDs, 
so that only a single first-stage SQUID feeds the SSA at any given time. The bandwidth of each 
pixel is limited to below the Nyquist frequency of the sampling, so that the signal in each 
pixel can be faithfully demultiplexed.

The focal plane multiplexing readout will be composed of three components: a DC wafer, multiplexing chips, and a critical line (CL) wafer.
The DC wafer will have inductors and resistors for biasing the detectors, but also function as a wire-routing wafer to connect the detector signal lines through the inductors/resistors to the SQUIDs and to connect the addressing lines.
The multiplexing chips will be mounted to the DC wafer and connected by superconducting aluminum wires.
The CL wafer will have pockets to get the bias lines out from the DC wafer and the addressing lines from the mux chips. It will also serve as a platform for the interconnects to get the signals from the focal plane to the warmer temperature stages.

\begin{figure}[!ht]
  \centering
  \includegraphics[width=0.9\textwidth]{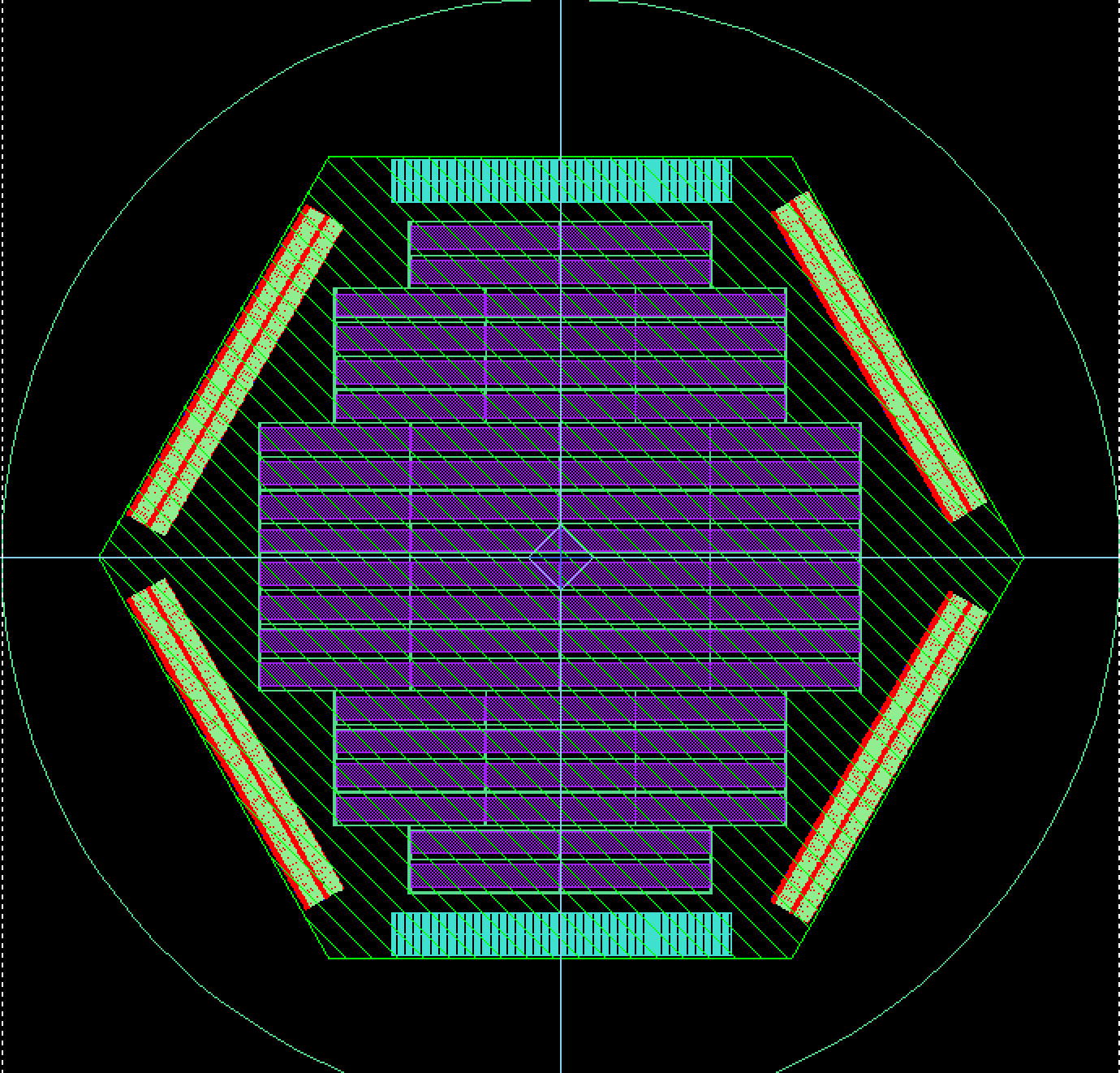}
  \caption{\label{fig:DCwafer}The figure represents a DC wafer with 32 multiplexing chips for allowing 2048-detector readout capability. The addressing lines will be routed out from top or bottom of this wafer and the bias lines will be routed out from the left/right vertices.}
\end{figure}

The CMB-S4 detector arrays will be assembled into closely packed universal focal-plane modules (UFMs).  The UFMs will provide a universal cryomechanical interfaces so they can be used in both the large and small-aperture CMB-S4 telescopes, but there will be different versions of the UFM for different detector wiring at different frequencies, etc. Each UFM will carry a 150-mm detector array, a feedhorn array and associated optical coupling parts, and the cryogenic components for multiplexing the 1000+ detectors, in a 20--40\,mm-thick hexagonal package, 
depending on the frequency.  Each UFM will mount to a focal plane base (FPB) plate via a flange which provides an optical reference surface (i.e., the flange is at a well-defined distance below the phase centers of the feedhorn array).  Flanges from neighboring UFMs interlock (reference SPT-3G and PB2) to allow closer packing between arrays, given edge-to-edge distance of 125\,mm.     

The number of detectors on one wafer varies depending on the frequency, and on the horn size
which is different for the LAT and SAT.  Thus a variety of 
different design layouts are needed for the cryogenic multiplexing components. 
Even more designs are needed for detector wafers, as the thermal conductances 
will also be a function of the size of the telescope, as well as the site. 

Tables that list the band center frequencies, the number of wafers per optics tube, 
the number of TESes per wafer, the number of SATs and LATs, the number of wafers, 
and the total number of detectors are shown in 
Tables~\ref{tab:SATproperties} and \ref{tab:LATproperties}.
Note that the focal planes as designed include partial-wafer sub-arrays, some shaped as half-hexagonal arrays, 
others as rhomb-shape sub-arrays.  The TESs on three rhombs add to the same number as a hexagonal wafer, 
and, similarly, the TESs on two half-hexagons add to the same number as a hexagonal wafer.  
The wafer counts in the tables referenced above add the partial wafer detector counts into an 
equivalent full wafer count.

\subsection{Detectors}\label{subsec:detectors}

Detectors for CMB-S4 will be fabricated on silicon wafers using well tested micro-fabrication techniques built on the foundation of processes that were established for Stage-2 and Stage-3 experiments. These fabrication processes will be optimized so that $>90$\% of the detectors on an array are operational (i.e. can be electrically connected, optically coupled, and biased).

\begin{figure}[h]
    \centering
    \includegraphics[height = 1.5in]{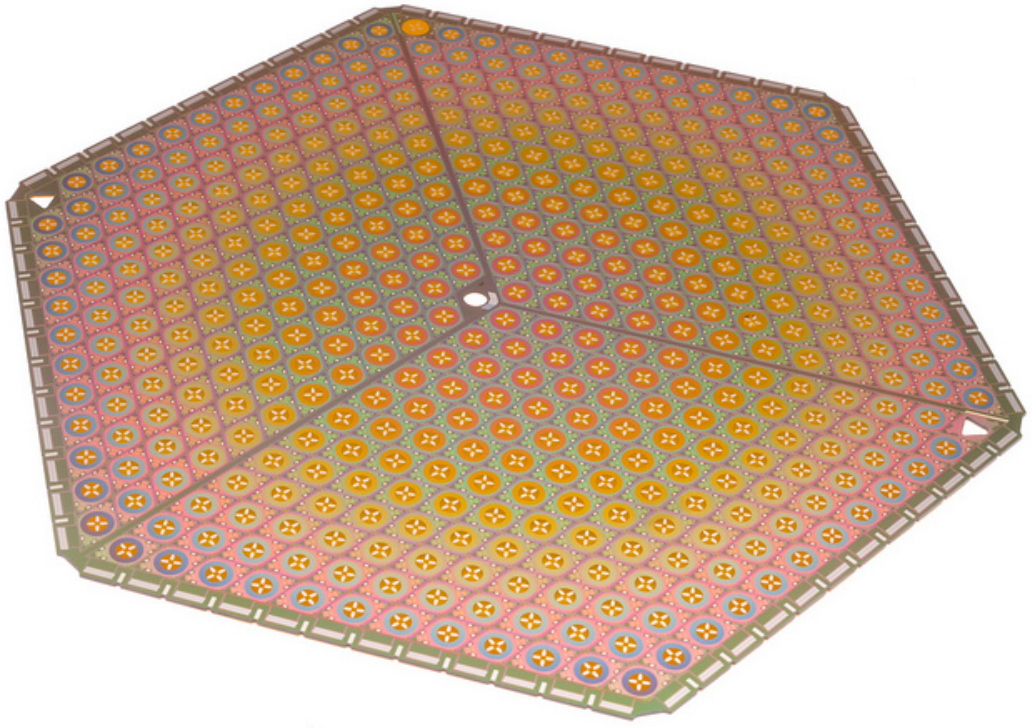}
    \includegraphics[height = 1.5in]{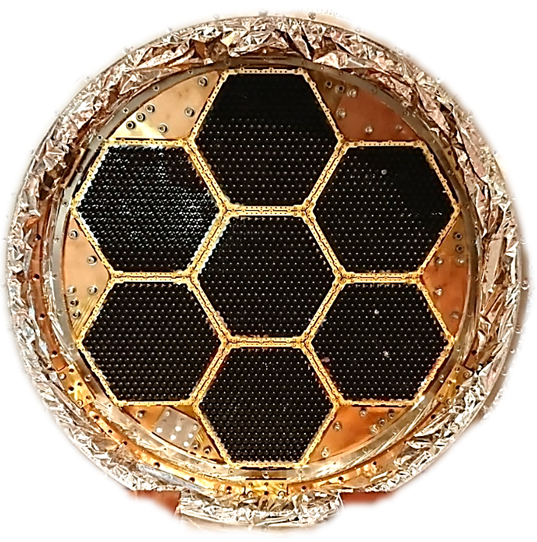}
    \caption{(Left) Photograph of an Advanced ACTPol detector array fabricated by NIST on a 150-mm diameter silicon wafer.
    (Right) Photograph of a POLARBEAR-2 focal plane. The diameter of the focal plane is approximately 400\,mm. The focal planes for the small aperture telescopes will be constructed from from combinations of hexagonal, half-hexagon, and rhomb sub-arrays that vary with the frequency band.   A hexagonal focal plane for the large aperture telescopes will be constructed from three close-packed hexagonal wafers with three smaller rhombuses at the edges. The three smaller rhombuses will be fabricated on a single wafer.}
    \label{fig:NISTArray}
\end{figure}

\paragraph{Detector array layout:}
The CMB-S4 detector arrays will be fabricated on 150-mm diameter wafers. 
A combination of hexagonal, half-hexagon, and rhomboidal tiles that comprise $1/3$ of a hexagonal wafer will be cut from the 150-mm diameter wafers such that they can be assembled in a close-packed form to make a large focal plane as shown in Fig.~\ref{fig:NISTArray}.

\begin{figure}[h]
    \centering
    \includegraphics[height = 1.5in]{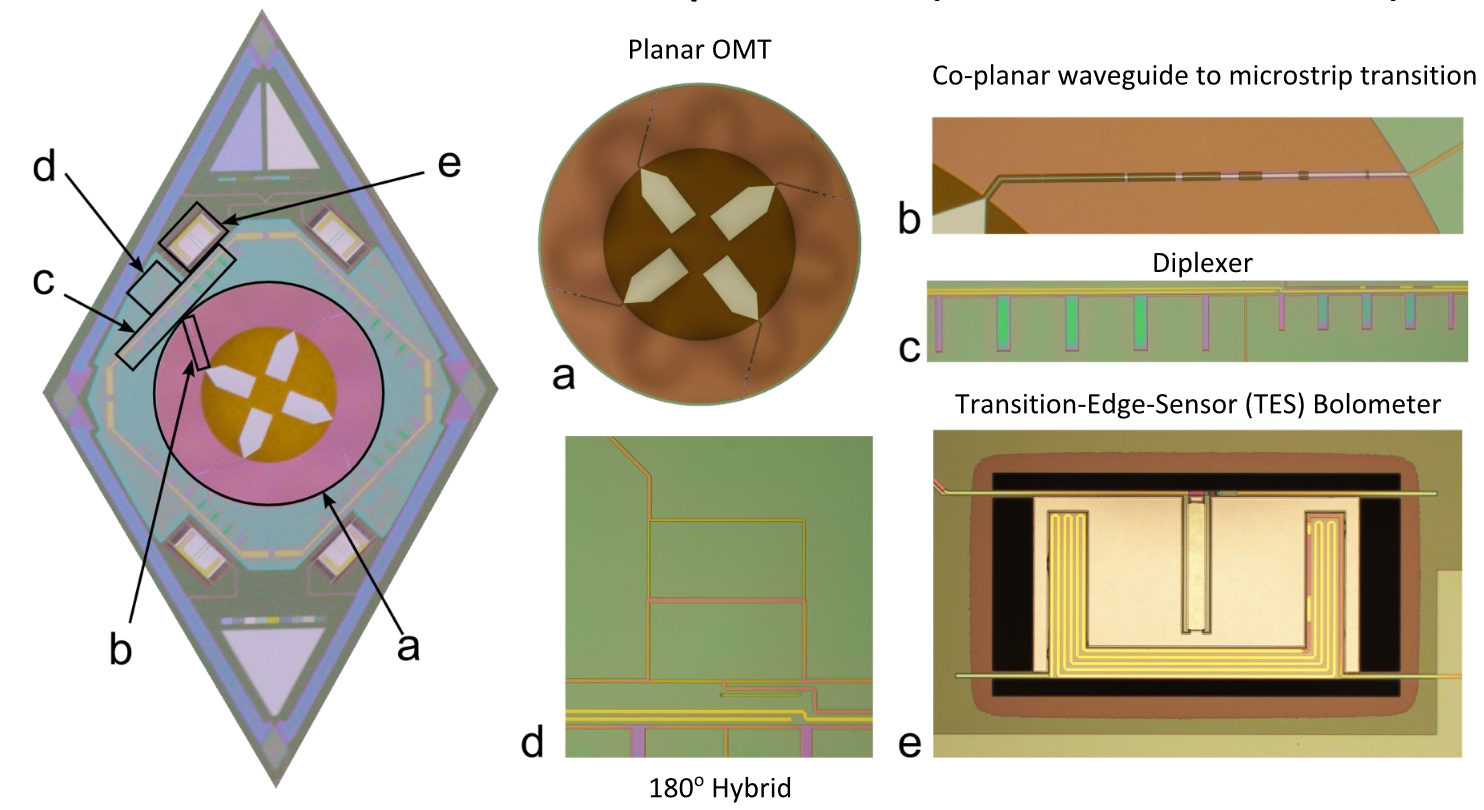}
    \caption{Optical microscope image of an Advanced ACTPol detector pixel highlighting several of the key components. Magnified images of the major pixel components include: (a) the planar orthomode transducer; (b) the coplanar waveguide to microstrip transmission line; (c) the band-defining in-line stub filters; (d) the 180 degree hybrid tee; and (e) one of the AlMn TESs.}
    \label{fig:AdvACTPixel}
\end{figure}

Pixels are arranged on a hexagonal tile to maximize packing density and yield. 
The pixel count per hexagonal tile will range from approximately fifty to five hundred, depending on the observation frequency of the detector array.
Higher frequency bands will have more pixels per array.
Each pixel consists of an OMT 
to optically couple to the pixel, radio-frequency (RF) circuits to control the RF signal (band pass filters, mode-selector and cross-under), TES bolometers to measure the incident signal, and microwave transmission lines that connect these elements as shown in Fig.~\ref{fig:AdvACTPixel}.
Each pixel on the wafer will be sensitive to two frequency bands and two orthogonal linear polarizations. 
Thus, each pixel will contain four TES bolometers. 
Each wafer will have approximately 10\% additional TES bolometers that are not connected to optical coupling element to monitor detector sensitivity to the environment. 
In total, there will be approximately a few hundred to several thousand TES bolometers on a hexagonal tile. 
The electrical connection to the detector wafer is made with rows of wire bond pads placed at the perimeter of the hexagonal detector tile.

\paragraph{Fabrication plan:}
\begin{figure}[h]
    \centering
    \includegraphics[width = 5in]{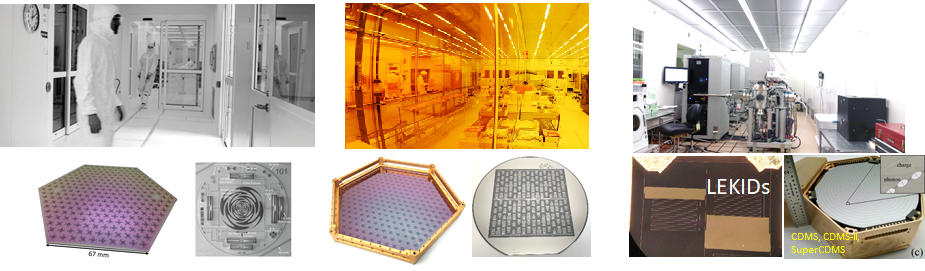}
    \caption{Photographs of nanofabrication facilities and devices fabricated at: 
    (Left) ANL
    (Center) LBNL/ U.C. Berkeley with 
    (Right) SLAC}
    \label{fig:FabFabrication}
\end{figure}

The detector fabrication processes will be built on the foundation of detector fabrication processes established during Stage-3 experiments. The baseline plan utilizes teams of two fabrication engineers carrying batches of 150-mm detector wafers through the fabrication process with multiple, standardized quality-control measurements.

The detectors are fabricated using several cycles of material deposition, patterning, and etching to build up a complete device.
Superconducting materials (niobium, aluminum and manganese-doped aluminum) and normal metals (gold and palladium) will be sputter- or e-beam evaporation deposited. Dielectric films will be deposited with plasma enhanced chemical vapor deposition.
Features will be defined using stepped lithography for frontside patterning and contact-alignment lithography for backside patterning. Chemical wet etch, plasma etching and ion-milling will be used to define lithographed features.  
A deep reactive silicon etch will be used to release membranes for the OMT and the TES bolometer thermal isolation. 

The detectors will be produced in nanofabrication facilities with class 100 clean rooms based at several national laboratories. These facilities have an extensive heritage in superconducting transition-edge sensor fabrication, having fabricated multiple Stage-3 CMB instruments, SuperCDMS, X-ray spectrometer instruments, etc. Photos of a few of these facilities and examples of fabricated detectors are shown in Fig.~\ref{fig:FabFabrication}.

\paragraph{Fabrication quality control:}
One critical challenge for detector fabrication is to develop robust fabrication procedures that guard against process deviation. To address this, CMB fabrication facilities have been tracking processes throughout fabrication and ensure that the fabrication tools receive regular maintenance. All fabrication and metrology steps will be tracked in a computer-based logging system. 
The status of the fabrication tools and clean room environment, including chamber base pressure, plate temperature, clean room temperature, and humidity will be recorded each time a tool is used. 
The data will be continually compared to historical norms to identify deviations from process stability. 
This monitoring process has been implemented successfully in detector fabrication for Stage-3 experiments.
While the tooling may differ between fabrication facilities, all detector arrays will be required to meet the same performance specifications, ensuring uniformity across all detector arrays.

Each detector wafer and detector pixel design includes test structures that can be used to monitor the fabrication processes and material characteristics. Sacrificial test wafers will also be carried through some process steps to monitor process steps.
These methods have been proven to be useful in producing consistent detector arrays for Stage-3 fabrication. 
Fabrication engineers will use inspection tools including profilometers, reflectometers, ellipsometers, sheet resistance, and film stress monitors on these test structures to monitor fabrication processes attributes such as film uniformity in deposition, etch rate, surface roughness, dielectric constant of the dielectric film, sheet resistance, and film stress. Additionally, materials will be tested at cryogenic temperatures. Variable temperature four-wire resistance measurements will be used to measure the transition temperature of superconducting materials. Low-frequency ($\approx2$--8\,GHz) diagnostic microwave resonators will be used to evaluate the loss in the dielectrics and superconducting materials.

\paragraph{Detector characterization:}

Detector characterization is an essential part of the fabrication process. Characterizing test structures, single pixels, and full arrays provides critical feedback to the fabrication process and ensures that the performance and uniformity targets are met. While finished detector arrays are tested at room temperature for connectivity, much of the detector characterization requires cryogenic temperatures to verify detector performance at operating temperatures. Here we discuss tests during the fabrication process, single pixel testing before array fabrication, and how characterized data are used to monitor detector fabrication performance. Further details on full the detector assembly and its testing are discussed in  
Sect.~\ref{section:assembly_test}.

First, single pixels will be tested in a ``dark'' environment where there is no optical signal on the detectors. Dark tests provide the fabrication team with information about detectors such as: superconducting transition temperature of the TESs, superconducting transition shape of TESs, resistance of the TESs, saturation power of the TES bolometers, time constants of the detectors, and noise performance. These single pixel and full array testing results will be continually fed back to fabrication teams, who will make adjustments to processes to ensure that detectors achieve the specified performance. For example, if the time constant of the detectors is too fast, the thickness of the heat capacitor material (palladium) can be increased to add more heat capacity to the TES bolometers, slowing the time constant.

Single pixels will then be tested in an ``optical'' environment, where a blackbody signal illuminates the detectors. 
Optical tests will provide the fabrication team with information about the RF circuit such as optical efficiency, antenna pattern, polarization purity, and bandpass filter performance. 
As with dark testing, the fabrication team will make adjustments to the detector design and/or fabrication method as needed to ensure that the detectors are within specification. 
For example, if the frequency range of the bandpass filter is slightly high, it is possible to make a minor correction by changing the thickness of the dielectric material without changing the detector layout.

Single pixel cryogenic detector characterization is carried out using sub-Kelvin cryostats configured for specific measurements (e.g. a dark cryostat with SQUID readout for measuring the transition temperature, resistance, and saturation power of released TES.) 
Each of the candidate fabrication facilities has a suite of cryostats to allow for detailed characterization of their fabrication processes and the resulting detectors.

\subsection{Detector RF coupling}\label{subsec:detectorrfcoupling}

The CMB-S4 reference design uses smooth-wall, wide-band, spline-profiled feedhorns for the detector RF coupling. The feedhorn array sits above the detector wafer, and couples the light onto the orthogonal fins of the detector OMT (Fig.~\ref{fig:horn_OMT_schematic}). Feedhorns are a mature technology that has been demonstrated on many CMB experiments with published CMB polarization science results across a wide range of frequencies, including ACTPol, AdvACT, SPTPol, ABS, and CLASS. 

\begin{figure}[h!]
\centering
\includegraphics[width=0.89\textwidth]{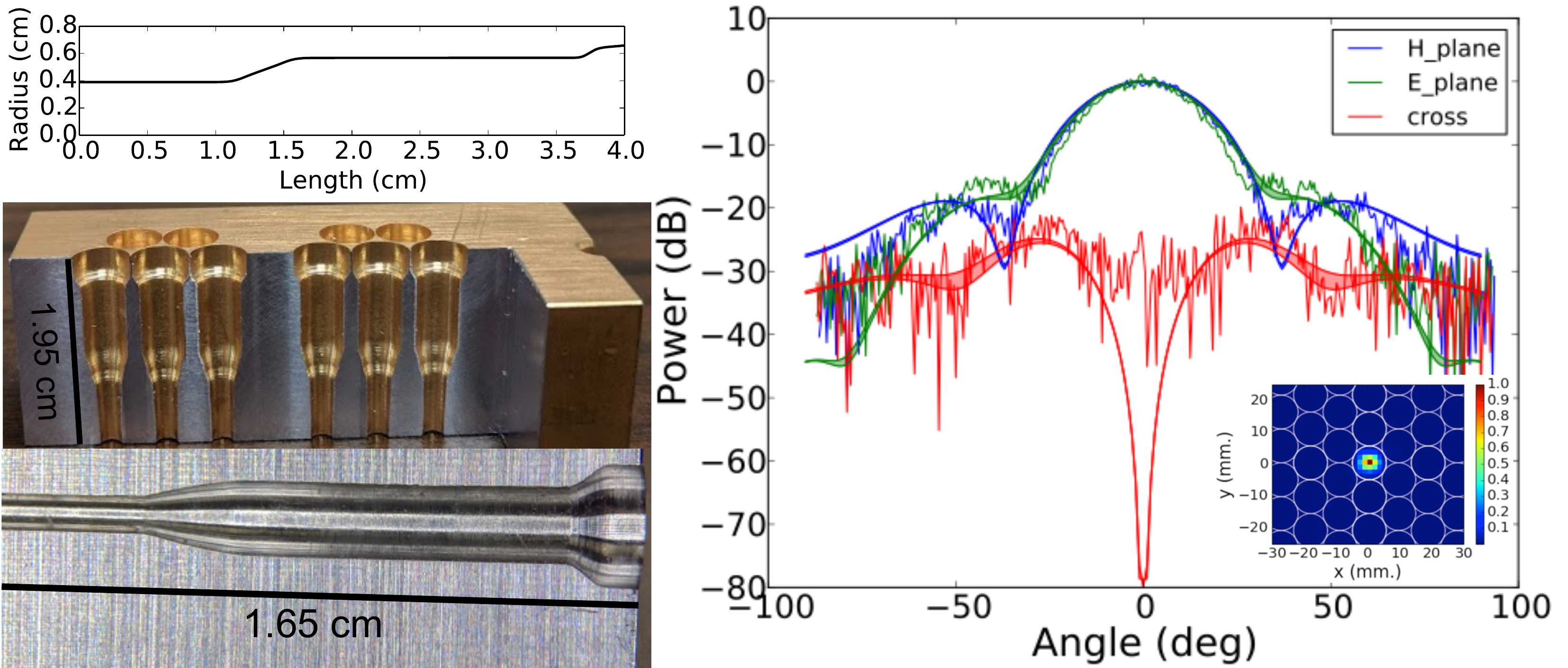}
\caption{Spline-profiled feedhorn profiles across the planned CMB-S4 frequency bands are shown on the left. The top profile is the AdvACT 27/39-GHz design, which was direct-machined into SiAl alloy. The middle profile is a Au-coated Al feedhorn with the same design as the 90/150-GHz AdvACT feedhorn. The bottom profile is an Al SO 220/280-GHz design. The figure on the right is a measurement of the beams of the AdvACT 150/230-GHz feedhorn at 150\,GHz with its simulations. The inset shows a 2D beam measurement of the Al 90/150-GHz design at 125\,GHz. The feedhorn beams can be measured at room temperature for fast turnaround and match their simulations well.}
\label{fig:feedhorn}
\end{figure}

Spline-profiled feedhorns have good performance across wide bandwidths, making them ideal for multichroic pixels. AdvACT has fielded spline-profiled feedhorns on multichroic arrays at 27/39\,GHz, 90/150\,GHz, and 150/230\,GHz, which spans almost the entire planned frequency range for CMB-S4. Figure~\ref{fig:feedhorn} shows several spline-profiled feedhorn designs across the CMB-S4 frequency bands and examples of beam maps showing their performance. Spline-profiled feedhorns maintain high coupling efficiency with small apertures, enabling denser pixel packing and thus increased sensitivity. Because their profile designs can be tuned between beam coupling efficiency and beam systematic control, spline-profiled horns allow for greater design control and balanced performance \cite{Simon_SPIE_2016,Yassin_2007,Zeng_2010,Granet_2004}. 

Spline-profiled feedhorns can also be made monotonically increasing in diameter so that they can be direct-machined into metal, significantly reducing fabrication time and cost. We plan to use custom drill and reamer sets to direct-machine the feedhorn profiles into Al~6061. Al~6061 is cost-effective, low-density, and has high machinability \cite{Simon_SPIE_2018}. A CMB-S4-like 90/150-GHz feedhorn array with 507 feedhorns would only take about 20 hours to machine. To efficiently machine a large number of arrays, the feedhorn profiles for a batch of several horn arrays are drilled into a single block of Al, and then individual arrays are cut from the block. The feedhorn array production can be readily performed with the tools available in most machine shops, so it can be spread across multiple machine shops. Because Al is superconducting at the operational temperature of the array, it requires the addition of a normal metal for heat sinking, so we will gold-plate the arrays after fabrication. This fabrication process is simple and as such has an expected yield of 100\%.

We will use non-contact metrology on cross-sectioned feedhorns to verify that the machined profiles match the designed profiles and beam maps to check that the feedhorn performance matches simulations \cite{Simon_SPIE_2018}. One large advantage of feedhorns is that they can be tested at ambient temperature prior to integration with the detector array. This enables rapid turnaround on array testing and reduces risk by only integrating fully vetted horn arrays with the detector arrays. Before production begins, the drill and reamer set is vetted with non-contact metrology measurements, and the performance of a single feedhorn is measured with beam maps. After the arrays are produced, we will measure several horns at several frequencies on each array to verify performance. Tool wear on the drill and reamer set can be assessed within batches with the beam maps and between batches with non-contact metrology.

The horn arrays are aligned with the detector stack by a pin and slot to account for the differential thermal contraction between the Al feedhorn array and the Si detector stack. The detector stack is comprised of a photonic choke to reduce leakage at the interface between the horn array and the detector stack \cite{Wollack_2010}, a waveguide interface plate, the detector wafer, and a $\sim$quarter-wave backshort. Behind the backshort is a Cu heat clamp plate for heat sinking that bolts directly to the metal feedhorn array. The feed array also has additional mounting features machined directly into its design for mounting onto the focal plane. This array package design is based on designs used for ACTPol, AdvACT, and SPTPol.

Wide-band, spline-profiled feedhorns are a mature technology. Smooth-walled Al feedhorns have been fielded on BLASTPol \cite{Dober_2014}, and Simons Observatory (SO) will use Al feedhorns for the 220/280-GHz arrays and half of the 90/150-GHz arrays. The main challenges associated with Al feedhorn arrays are achieving the necessary machining tolerances in the feedhorn profiles and the alignment of the feedhorn array with the detector stack, especially at the highest frequencies (220/280\,GHz). Simulations using the scatter measured in the a 220/280-GHz feedhorn profile from non-contact metrology measurements show that the current level of variation in the profile is negligible to the feed's performance \cite{Simon_SPIE_2018}. Additional simulations show that the SO 220/280-GHz array must be aligned with the detector stack to $\approx10\,\mu$m. Differential thermal contraction can cause greater misalignment and add additional mechanical stress to the array, but these risks can be retired with cryogenic testing in the design phase. Two prototype SO feedhorn arrays are currently in production and will be tested with full detector arrays by the end of 2018. In 2019, SO will produce, test, and integrate approximately 30 Al feedhorn arrays, further demonstrating the necessary production capacity and performance of this technology for use in CMB-S4.

\subsection{Cold readout electronics}\label{subsec:coldreadoutelectronics}

\paragraph{Cold Readout Introduction}
\label{sec:ColdReadoutIntro}

The readout of the individual detectors will be multiplexed at the lowest temperature stage using time-division SQUID multiplexing (TDM). In TDM a group of detectors is arranged into a 2D logical array, in which each column of detectors shares a dedicated readout amplifier chain. The various rows are addressed cyclically in rapid succession to record the entire array. When a row of the array is actuated, a single TES bolometer in each column is read out by its associated column amplifier. The signal from each bolometer is low-pass filtered before multiplexing to a frequency below the Nyquist frequency of the sample rate, so that the signal can be reconstructed with acceptable degradation from aliasing. Time-division SQUID multiplexers have extensive on-sky heritage in the CMB community, including instruments such as the BICEP/Keck cameras and the MBAC and ACTpol cameras on ACT.

\begin{figure}[htb]
   \centering
   \includegraphics[width=3in]{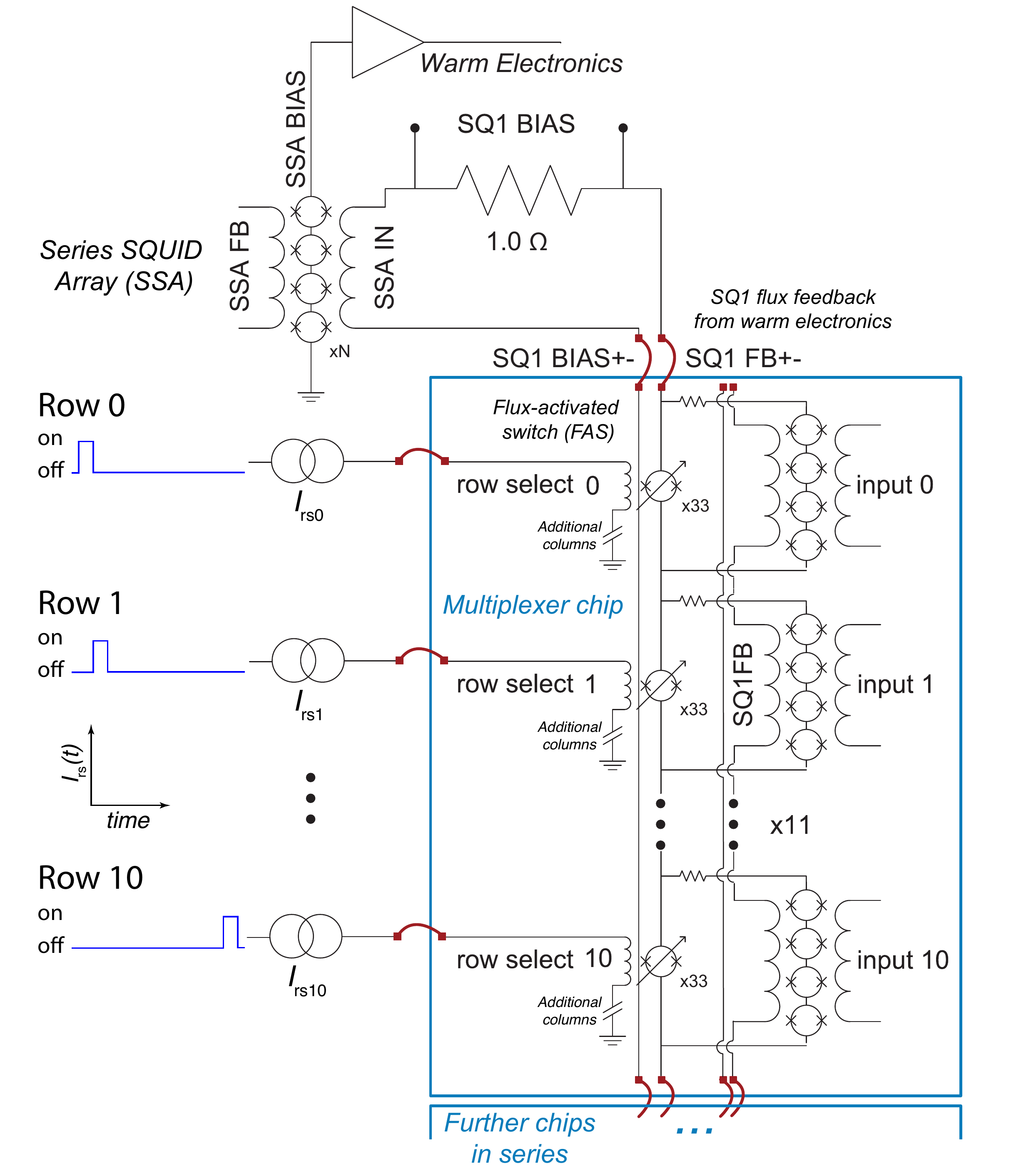}
   \caption{Schematic illustration of a single column of a time division SQUID multiplexer. 
   Each TES is coupled inductively to a first stage SQUID (SQ1).  
   All SQ1s in a column are wired in series to the input of a SQUID array amplifier (SAA), but at any given time all but one row of SQ1s is bypassed by a flux-activated switch (FAS). 
   The various row-select lines are biased in sequence with low-duty-cycle square waves, as shown at left.
   }
   \label{fig:tdm_schem}
\end{figure}

Figure~\ref{fig:tdm_schem} shows the schematic of one column of a time division SQUID multiplexer.  
The multiplexer chips outlined in blue in this figure are located at 100\,mK, whereas SQUID Array Amplifiers (SAAs) are heatsunk to $\approx$4\,K.  
A detector bias/filter chip (not shown), which sits between the multiplexer and the detector, contains shunt resistors that provide the TES voltage bias and inductors to limited the bandwidth to below the Nyquist frequency of the sampling.
The multiplexer footprint in the focal plane is 1.8\,mm$^2$/channel. 
The size of the bias/filter chip is driven by pixel bandwidth and TES sensor impedance; in practice, however, the silicon area required is roughly the same per pixel as the SQUID multiplexer chip. Superconducting wire bonds and superconducting circuit board traces pass signals between the detector wafers, filter chips, and multiplexing chips; links to the rest of the amplifier chain may be made with non-superconducting pin-socket interconnects.

\begin{figure}[htb]
   \centering
   \includegraphics[width=3in,trim={0 0 10.5cm 0},clip]{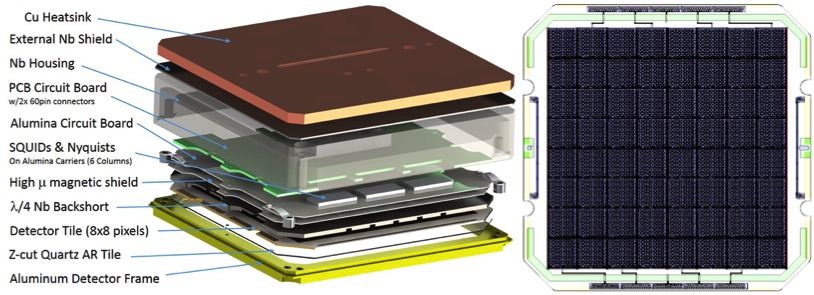}
   \caption{Behind-wafer TDM packaging implemented for BICEP3.  
   Figure reproduced from Ref.~\cite{hui2016bicep3}.
   }
   \label{fig:tdm_package}
\end{figure}

\paragraph{Packaging}
The small per-element footprint of TDM enables the readout to lay in-plane behind the detector wafers.  
This arrangement enables the close-packing of detectors to efficiently use the available focal plane.  
The approach has been used in BICEP3 \cite{hui2016bicep3} (see Fig.~\ref{fig:tdm_package}), and the same concept 
is implemented in the CMB-S4 reference design.  
Given the readout footprint, 150-mm wafers with up to 2,000 channels per wafer may be read out with one layer of multiplexing components, and this channel count is adequate for all the array types in the CMB-S4 reference design.

\paragraph{Fabrication plan}
Many tens of thousands of TDM SQUID channels have been fabricated and successfully deployed in astronomical instruments.  
The multiplexer design and fabrication are mature. TDM fabrication will be done at class 100 clean room facilities at the National Laboratories. The baseline fabrication plan utilizes teams of two fabrication engineers carrying batches of 150-mm TDM wafers through the fabrication process, with multiple standardized quality-control measurements.

Superconducting metals (niobium, aluminum, and niobium trilayer tunnel junctions) and normal metals (gold) will be sputter deposited. Dielectric films will be deposited with plasma enhanced chemical vapor deposition. 
Features will be defined using stepper lithography. Both chemical wet etch and reactive ion etching will be used to define lithographed features. 

\paragraph{Fabrication Quality Control}
All fabrication and metrology steps will be tracked in a computer-based logging system. 
Status of the fabrication tools and clean-room environment,  including chamber pressure, platen temperature and clean-room temperature and humidity will be recorded each time a tool is used. 
Data will be continually compared to historical norms to identify deviations from process stability. 

Each TDM wafer and 32-channel TDM chip includes test structures that allow monitoring of fabrication processes and material characteristics. Sacrificial test wafers will also be carried through some fabrication steps for process monitoring. Fabrication engineers will use inspection tools including profilometers, reflectometers, ellipsometers, resistance sheet and film stress monitors on these test structures to monitor fabrication processes such as etch rate, surface roughness, dielectric constant of dielectric film, sheet resistance and film stress.

One critical challenge for detector fabrication is to develop robust fabrication procedures that guard against process deviation.
Fabrication tools will require regular maintenance by highly skilled engineers.

\paragraph{Cryogenic Multiplexer Screening}

After fabrication, a 150-mm SQUID multiplexer wafer is diced into individual time-division multiplexer chips. The reference design implements a multiplexing factor of 
64, with each readout column incorporated on two 32-channel TDM chips. 
Bias/filter chips will match this channel count and approximate form factor.

Every 32-channel TDM chip will be screened in a fast-turnaround 4-K test facility. Only chips with 100\% working channels will be accepted. Conservatively assuming a 67\% yield of 32-channel chips with all working channels, each successful 150-mm wafer will provide 4,096 channels of TDM readout (128 $\times$ 32-channel chips). Bias/filter chips contain much simpler passive components, and are thus expected to be fabricated with higher yield and throughput. Occasional batch-screening of these chips in a similar fast-turnaround cryogenic system is adequate to guard against drifts in component values.

Based on previous experience, a production rate of 10 science-grade wafers/year/FTE may be assumed. 

\subsection{Warm readout electronics}\label{subsec:warmreadoutelectronics}

\paragraph{Time-division SQUID multiplexing} 
The time-division SQUID multiplexing (TDM) readout system includes room temperature electronics to control and bias the multiplexer, provide a bias to the bolometers, acquire and filter data from each bolometer in the array, and exchange commands and bolometer data with the observatory data acquisition (DAQ) system. 
 
The warm electronics provide low-noise current and flux biases that configure the multi-stage SQUID amplifier into a working state. The SQUID series array (SSA), the last stage of the cryogenic amplifier shown in Fig.~\ref{fig:tdm_schem}, is provided with a current bias to activate the SSA and a flux bias to place it in its linear regime. A second current bias is provided to activate each column's first-stage SQUIDs and flux-activated switches. The warm electronics also manage the TDM switching among the various rows in sequence. A fast-switching address card asserts a single row-select bias line at a time, which drives a single flux-activated switch (FAS) in each column normal. This serves to connect that row's first-stage SQUIDs to the readout chain while the remaining rows are shorted out by the other rows' inactive switches. The row select timing is chosen so that all rows are visited at a cadence of $\approx 20$\,kHz.  
 
The warm electronics also operate a separate closed-loop servo to linearize and record data from each bolometer in the array.  As each row is activated by its FAS, the warm electronics apply an appropriate nulling feedback signal (SQ1FB) to each column to keep the first-stage SQUID in that column in a linear regime. The SSA output voltage forms an error signal input to  a PID loop which calculates the appropriate feedback signal for the next visit.  The feedback and error signals are buffered, filtered, and reported to the observatory DAQ, typically at a few hundred Hertz. Mature software exists to tune and optimize the performance of this multi-stage multiplexer system.
  
In addition to providing the SQUID currents described above, low-noise bias cards in the warm electronics provide detector biases to the TES sensors in the array. These are absent in Fig.~\ref{fig:tdm_schem}, but would be connected at the right in the diagram. Scripts running in the warm electronics optimize the bias levels in the SQUID-based amplifier system and tune the bias levels of the TES array in a matter of seconds.

TDM systems have been used to control and readout large TES arrays in sub-millimeter astronomy (SCUBA-2, Zeus2, and SOFIA), CMB experiments (ACT, BICEP/Keck, CLASS, SPIDER, \dots) and also for ``homeland security.'' These systems operate at sea level and high altitude sites, in the stratosphere and at the South Pole controlling bolometers in instruments as large as 10,000 bolometer elements.

\paragraph{Low-risk path to modernization}
  The present generation of TDM warm electronics is already capable of handling the bolometer arrays proposed for S4 without a redesign.  However, the system is based on field-programmable gate arrays (FPGAs) which are near the end of their market cycle.  These FPGAs should be replaced by modern units, a straightforward design effort.  The optical fiber drivers in use for communication between the warm electronics and the computer directly connected to it are now obsolete and should also be replaced.  This provides an opportunity to to move to a more modern communication format.  The design effort for this change is also modest.
  
  The present electronics have been used to control arrays with as many as 64 rows, sufficient for the  2000-channel  and smaller  arrays in the S4 reference design.  Moving to  systems larger that $64\times 64$ (4096) elements would require faster settling times in the cryogenic SQUID circuits than those presently in use. This would induce an undesirable performance coupling between the warm and cold electronics, and so is not incorporated into the reference design.
 
 The present electronics are fabricated on ordinary multilayer circuit boards by commercial vendors and housed in an industry standard 6U crate, with minor mechanical modifications for direct electrical coupling to the cryostat.  This system will be simple to manufacture on the necessary scale.

\begin{figure}[!ht]
{\includegraphics[width=15cm]{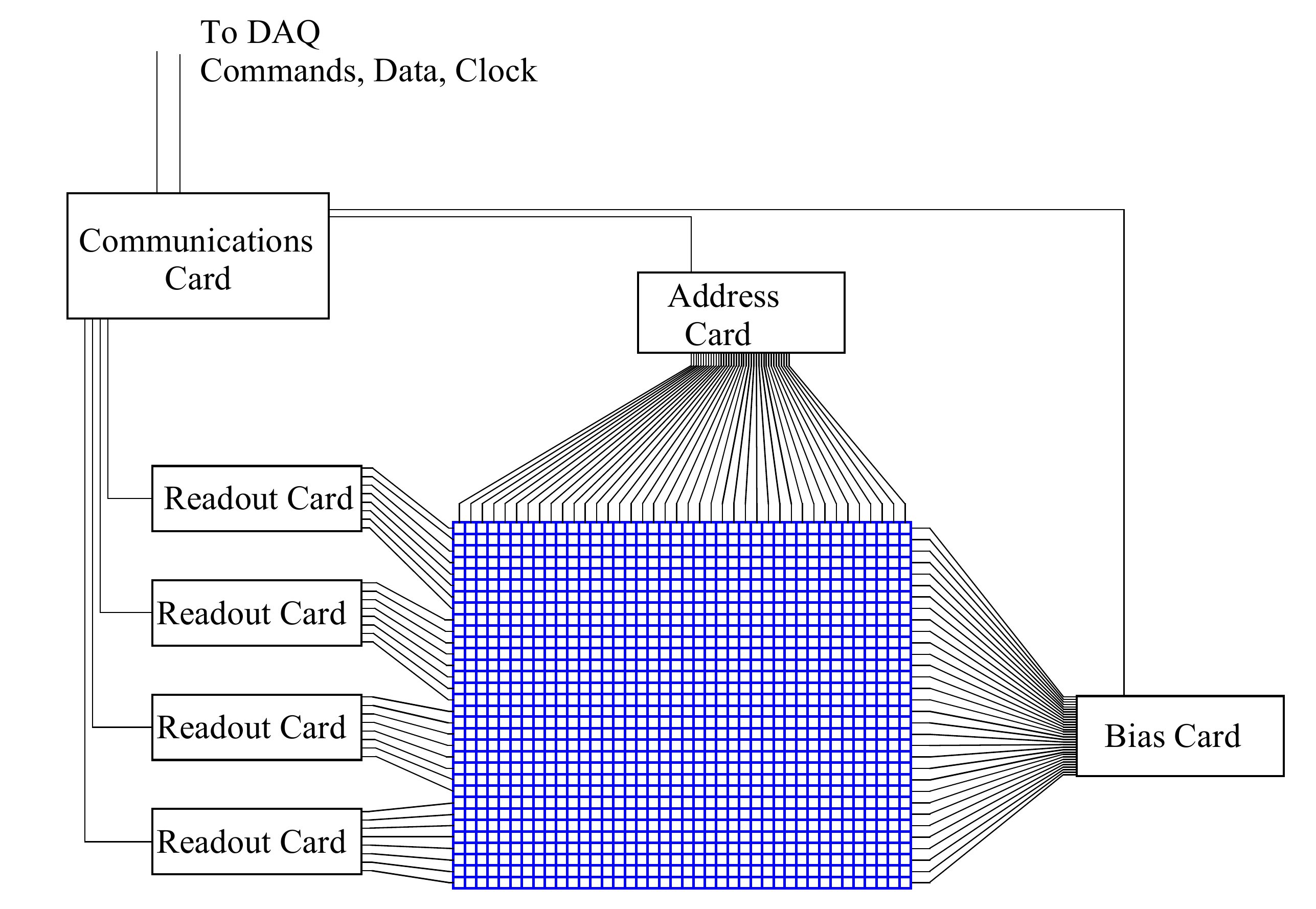}}
{\caption{Block diagram of one  bolometer array controlled and read out by one box of warm TDM electronics. Typically all boxes attached to a given cryostat are synchronized at the 50-MHz ADC clock or at the data strobe rate, which is often $\approx 1$\,kHz.  No technical innovation is needed to use the present electronics to control a $64\times 64$ element array.  Command, data and clock connections to the Observtory DAQ are via optical fiber.  The system requires three twisted pairs of wires per column to provide SSA bias and feedback and a SQ1 feedback signal.  Twenty pairs are required to operate a 64-way multiplexer in the \textit{column direction}, providing the Row Select signals.  The number of wires required for bolometer bias depends on anticipated element uniformity.  The resulting cryogenic harnesses of a few hundred wires are constructed of low thermal conductivity wire woven as twisted pairs.
The system is comprised of several Readout and Bias cards, one Address card and one communication card in a rack.  The Communication card is attached to a PC interposed between it and the Observatory DAQ. 
}\label{fig:block}}

\end{figure}

\subsection{Modularity, packaging/test plan}\label{subsec:modularityassemblytest}
\label{section:assembly_test}

\begin{figure}
\centering
\includegraphics[width=4in]{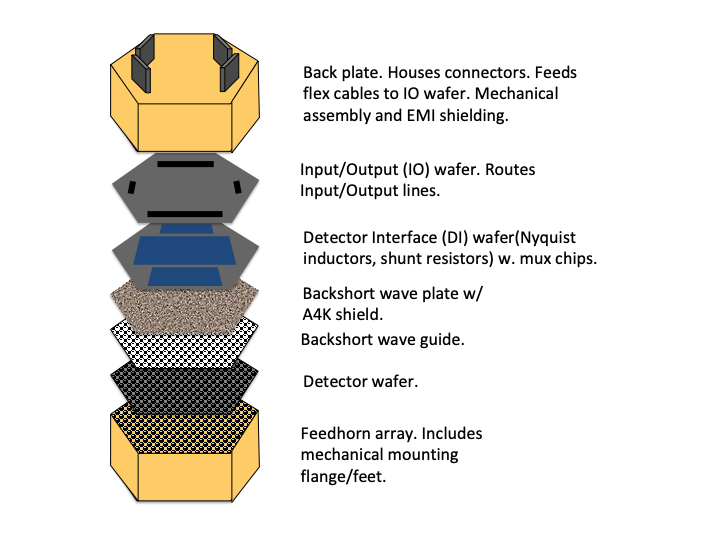}
\caption{Schematic of the CMB-S4 detector module stack consisting of a feedhorn array, detector array, backshort wafers, and the CRS. The CRS consists of the DI wafer with mux chips and IO wafer.}
\label{fig:cartoonstack}
\end{figure}

Each CMB-S4 detector module is assembled in a vertical stack outlined in Fig.~\ref{fig:cartoonstack}. The stack consists of the following three parts.
\begin{itemize}
\item Detector stack: the detector array, backshort wave guide and backshort end plate, with the latter including cryogenic magnetic shielding.
\item Cold readout stack (CRS): the DI wafer, multiplexer chips, and IO wafer. This part houses all of the sub-kelvin multiplexer components. It is assembled first and screened at 4\,K using a bank of test resistors. We require a CRS to have $>95\%$ operational yield.
\item Mechanical housing: feedhorn array and back plate, with the latter including micro-D connectors to connect wiring to the 4-K stage. This part provides the EMI shielding and mechanical structure for the module. It is mounted directly to the Au-plated Cu plates of the focal plane stage in the optics tubes.
\end{itemize}
The semi-hexagonal and rhombus shaped modules will have an identical vertical stack with appropriate overall form factor.

Modules are packaged as follows.
\begin{itemize}
\item The CRS is assembled from cryogenic components that have passed screening. The chips and wafers are glued together using cryogenic epoxy and the entire assembly is wirebonded using Al wirebond. As discussed above, the CRS will be tested at 4\,K with a bank of calibrated resistors to verify high operational yield.
\item The detector stack and CRS are assembled on top of an appropriate feedhorn array. Custom copper beryllium springs \cite{Ward16} will keep the stacks on the feedhorn and provide thermal contact to the temperature bath. Au wirebonds will be used to provide additional heat sinking of stack components.
\item The detector array is wirebonded to the CRS using Al wirebond. The IO wafer is then wirebonded, with Al wirebond, to flexible circuit cables \cite{Pappas16}, which are glued by rubber cement to the IO wafer.
\item The entire module is enclosed by the backplate with the flexible circuit lines passing through appropriate slots. The flex lines are glued to the backplate and Al wirebonded to the PCB housing the micro-D connectors.
\end{itemize}
Packaging will take place at a number of facilities with required assembly tools including: custom mounting jigs; automatic wirebonders; and technical staff associated with maintaining and operating these tools. Facilities equipped with these resources exist at the national labs and some universities. Module packaging is expected to take one day per module. All detector and readout components will be tracked using a suitable database that collects 
all the information related to testing and device history. Each device will be issued its own tracking number and traveler document.

All assembled modules will be tested twice, once ``dark’’ and once ``optical.’’ Dark tests include measurements of IV characteristics versus stage temperature, operational yield, noise (including low frequency), and optical efficiency (via a cold thermal load). Optical tests will measure the detector optical bandpass for approximately 30\% of the array. This approach exploits the fact that the polarized beam is determined by the feedhorn array and alignment of the OMT to the feedhorn assembly, which can be completely measured using room temperature measurements during feedhorn evaluation and module assembly. As a result, the only required cryogenic ``optical’’ measurement is spectroscopy to evaluate the bandpass. 

The full suite of tests will take approximately one month and requires 24 test cryostats, each operating three modules at a time. These cryostats will be primarily located at national labs, with a few potentially at universities. The test cryostats, including readout, will be procured, verified and cross calibrated prior to the start of production and all testing results will be stored in the database and reviewed by the project fabrication team.

\section{Data acquisition and experiment control \prelim{ ({\it L. Newburgh and N. Whitehorn})} } 
\label{sec:daq}

The reference data acquisition and experiment control subsystem is closely based on that used in existing Stage-3 CMB experiments, with a clear technical path to handling the higher data rates from CMB-S4.
This subsystem encompasses the acquisition of high-rate (7\,Gbit/s across both sites, 3--4\,Gbit/s per site) sample data from the detector arrays, the acquisition of low-rate data principally from thermal, position, and pointing sensors attached the telescope (broadly described as ``housekeeping'' hereafter), initial storage of data to on-site disk, the provision of timing and frequency reference signals, telescope pointing and cryogenic control, and real-time monitoring systems.

In the following, each CMB-S4 site is considered a completely independent entity; the logistical differences between a polar and a Chilean site both impose somewhat different requirements and make any unified control complicated with little return.
The systems in both cases are either similar or identical, but are not planned to be connected.

The total expected data rate to disk from CMB-S4, summed across both sites, is approximately 8\,Gbit/s, dominated by bolometer data.

\begin{figure}
\begin{center}
\includegraphics[width=5in]{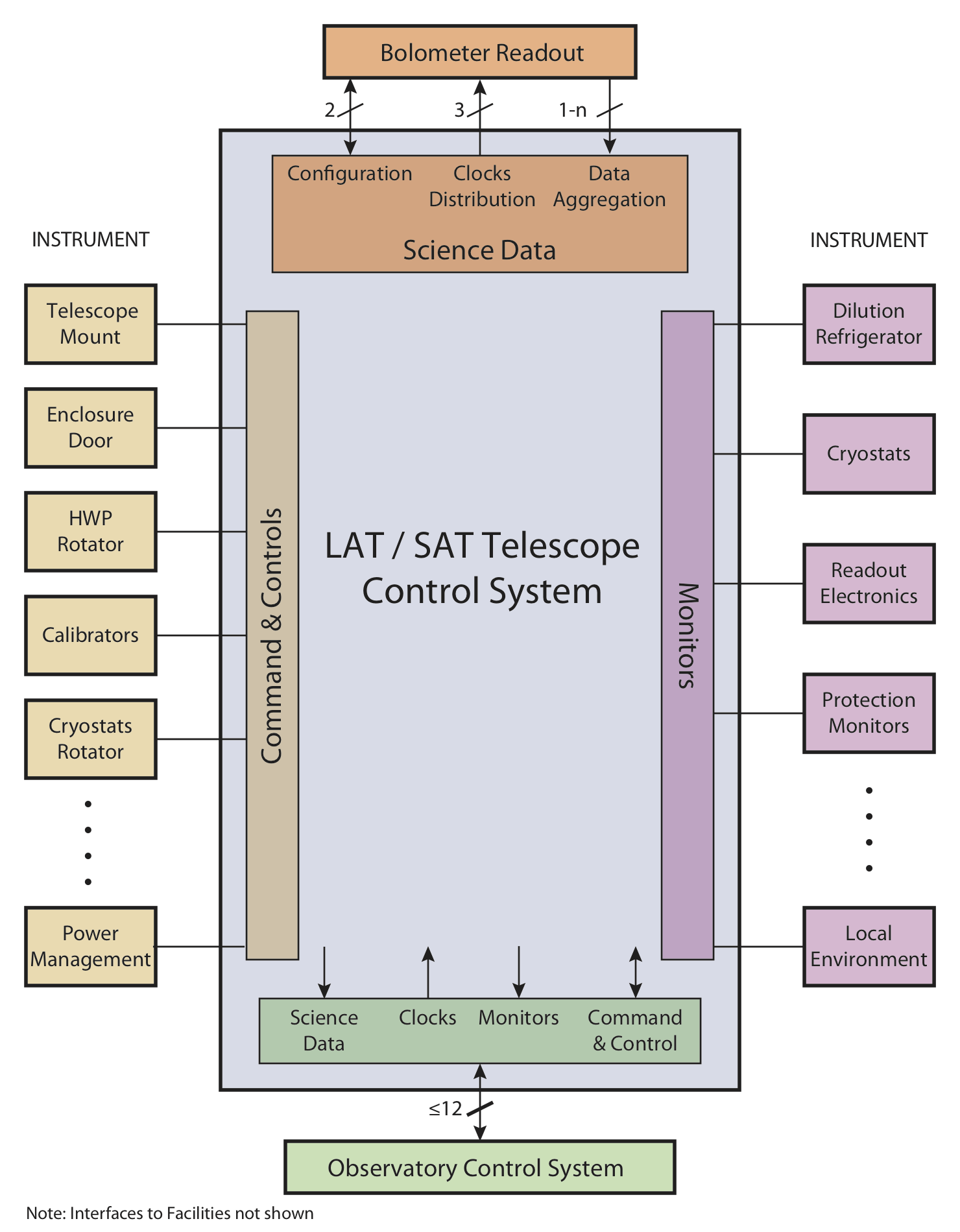}
\end{center}
\caption{Telescope platform detail of the observatory control system, showing high-level control of a telescope platform, warm readout, and housekeeping systems.}
\label{fig:DAQTelSys}
\end{figure}

\subsection{Data acquisition}

\subsubsection{Bolometer data acquisition}

Following the designs in existing projects, the warm readout electronics 
(Sect.~\ref{subsec:warmreadoutelectronics})
will be connected to Ethernet through several intermediate control computers (L1 aggregators) implementing the interface to the warm electronics (either Ethernet or a custom format implemented through a PCI-E card, depending on the modernization discussed in the previous section). These convert the data to a generic format and collate the responses from multiple boards, then forward them to another system (L2 aggregator) by Ethernet, which in turn assembles final data frames from an entire focal plane. As we do not record phase information, there is no clear benefit to taking data jointly from multiple telescopes and we plan to treat each telescope as an independent entity for the purposes of data acquisition.

The intermediate format used by the L1 and L2 aggregators is baselined as the common format currently used by SPT-3G, POLARBEAR2, and Simons Observatory, but may be adapted to CMB-S4 pursuant to the needs of the instrument. This format will, to the extent possible and to the extent to which the L2 aggregator and later data acquisition steps depend on it, be independent of the employed readout technology, limiting the impact of changes in readout design (Appendix~\ref{app:options}) to hardware and software changes on the L1 aggregator points.

The L2 aggregator forms the only serial chokepoint in the data flow and needs to be able to aggregate at minimum one focal plane's worth of data in one place. This architecture has been demonstrated on SPT-3G with 16,000 channels. The highest-burdened serial stage of this data collation uses CPU time linearly with channel count and is a factor of three from the limit of a 2017-era CPU, allowing scaling to single receivers of 50,000 detectors, comparable to that planned for CMB-S4 
(Table~\ref{tab:LATproperties}), 
without any further engineering effort.

Similarly, we are currently baselining the data acquisition code written for SPT-3G and POLARBEAR2 (and planned for use with Simons Observatory), which appears capable of handling the high data rates expected for CMB-S4 in limited testing. Extensive software research and development work for bolometer data acquisition is not expected.

After the L2 aggregator, data is forwarded to a first-level disk cache for later non-real-time unification with housekeeping data, compression, and archival. 
This follows the successful strategy employed by all Stage-3 instruments, which 
have demonstrated this to be a highly parallel process.  It can thus
be straightforwardly scaled to CMB-S4 data rates with current commercially available 
processing capabilities.

\subsubsection{Aggregation of bolometer and housekeeping data}
\label{sec:acq:hk}

In addition to bolometer data, we will record data from a variety of
housekeeping systems, which include telescope mount encoders and tilt meters,
cryostat thermometry, a weather station, and other monitors of the observatory
status. A computer running acquisition for subsystems will be responsible for packing
its own data into timestamped containers that are sent via Ethernet to the
first-level disk cache. The container data structure supports time-ordered
data with different sample rates; most of the housekeeping systems will be
sampled much slower ($\approx1$\,Hz) than bolometer data. All subsystems will be synchronized via IRIG-B002, as discussed in Sect.~\ref{sec:daq:sync}, so timestamps
can be used to identify and aggregate simultaneous data. It is not necessary for this process to occur in real time (though it will be nearly real time, in
practice), so it is mostly insensitive to lags in delivery from particular
subsystems. The bolometers dominate the overall data rate, so the additional
housekeeping subsystems will not be a substantial burden on the network or
the first-level disk cache.

After the first-level disk cache, we plan a third data aggregation step (L3) that happens asynchronously using an on-site computing cluster.
At L3, the auxiliary data streams are merged into the same files as the high-rate bolometer data so that all the data at a particular time are colocated for correlation analyses during processing.
A second copy of the auxiliary data will be stored to a database for retrieval by tasks that do not also require the bolometer data and can benefit from a more leightweight interface, such as site monitoring.

\subsubsection{Live monitoring of bolometer data, auxiliary data, and alarms}

The data monitoring subsystem is a server/client system that
aggregates data from each of the subsystems that communicates with the
OCS, and presents it over a web interface for monitoring of data
quality and site status.  The server provides a common API for each
connected client to supply and receive monitoring information from
other subsystems, and to issue alarms.  Each telescope site maintains
one monitoring server, and clients can connect from anywhere to
request data, often from multiple sites.  This system is separate from
the main data management system described in Sect.~\ref{sec:datamanagement},
as it largely handles data provided in near-real-time, although it
should also be able to retrieve archived data should a client request
it.

For example, one such client would be a housekeeping consumer that
monitors the state of each cryogenic system.  This client provides a
web interface to view housekeeping data retrieved from the server, and
issues alarms if temperatures are out of scope.  An alarm may be
issued as an email to the operations team, a text message or a
telephone call to the site, or a simple warning on the web interface.

A separate monitor/viewer is maintained for bolometer data at each
telescope site, either streaming in near-real-time, or as snapshots of
specific statistics (e.g., average current noise in a particular
frequency range, depth into TES transition).  These data can be used
during calibration runs, and for monitoring data quality during normal
observing periods.  For example, one might issue an alarm if too many
channels are outside of the normal operating range on the TES
transition.  Subsets of the data aggregated by this monitor are also
sent to the central monitoring server.

\subsection{Observatory control}

\subsubsection{Main observatory control system}

\begin{figure}
\begin{center}
\includegraphics[width=5in]{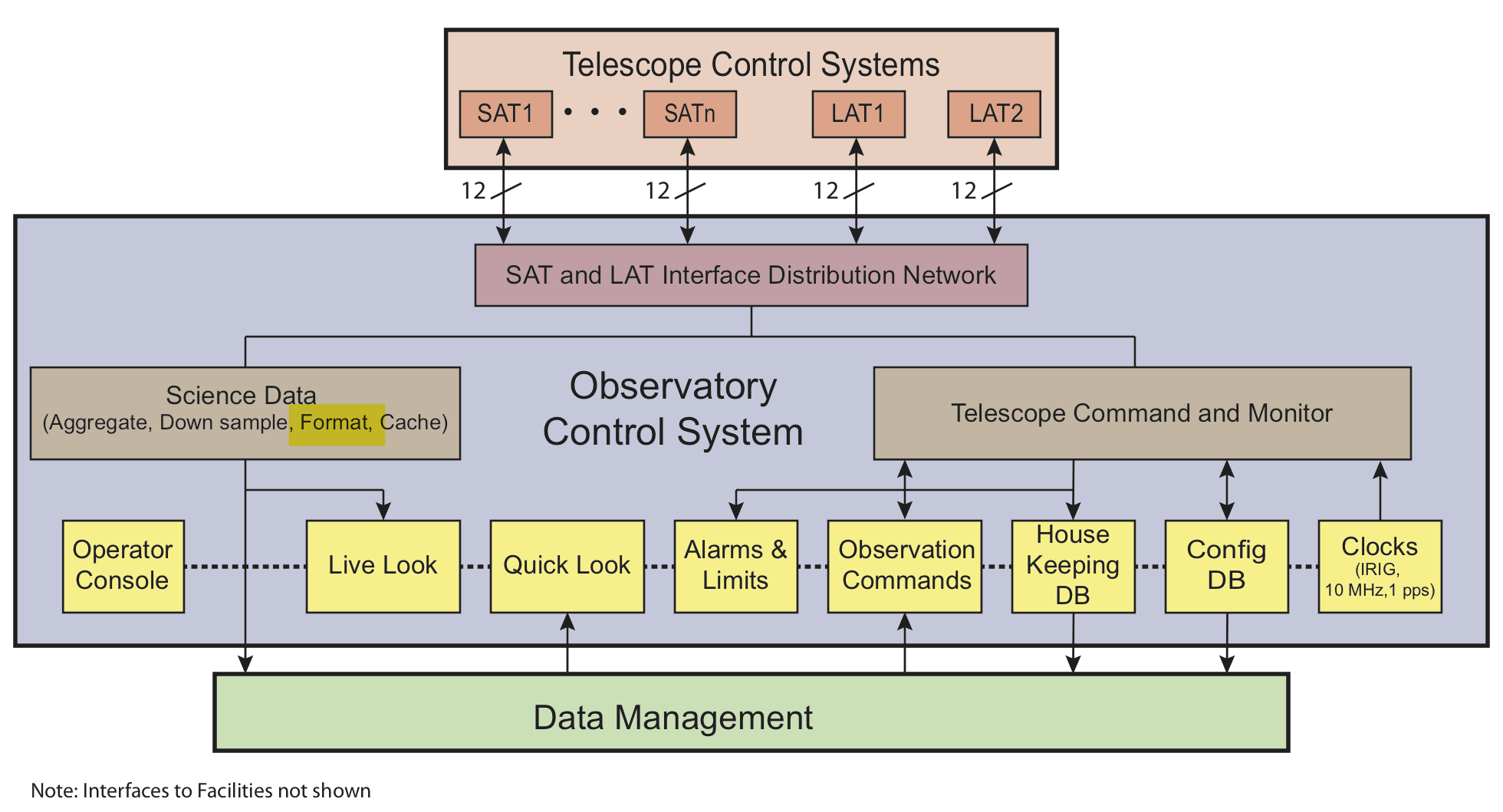}
\end{center}
\caption{Overview of the observatory control system, including the alarms, data acquisition control, and monitoring, including the interface to all telescope platforms.}
\label{fig:DAQObsCntl}
\end{figure}

The Observatory Control System (OCS) consists of software to
facilitate both automated and user-interactive coordination of
operations involving multiple telescope platforms, including all on-board subsystems (warm readout control, telescope encoder, half-wave-plate, etc).  The full set of
subsystems are likely to include a combination of commercial devices
alongside specialized hardware designed for CMB-S4 or adapted from
existing CMB projects.

While individual telescope subsystems within the observatory may produce large
volumes of data, or may need to operate in hard real-time to meet
performance requirements, there are no interactions between subsystems
that require exchange or analysis of large amounts of data in order to
maintain instrument performance.  Furthermore, the synchronization
requirements, when coordinating operations of various systems, are
easily achieved assuming NTP-synchronized clocks and typical Ethernet
latencies.  As a result, we are free to adopt a simple OCS
architecture wherein all devices accept commands and exchange
low-speed data on the observatory network using a simple interface
language defined by the OCS.  All devices will produce status
information, such as the success or failure of requested operations,
in an OCS-defined format.  Control software will operate by
translating high-level requests (such as a request to observe a
certain sky field for a certain amount of time) into a sequence of
lower-level tasks, and then initiating and monitoring the execution of
those tasks on the relevant subsystems.  Any on-going device operations
that are unrelated to specific observing requests will be initiated
and monitored through the same system, but by different, persistent
high-level control daemons within OCS.

In practice, the implementation of such an OCS involves creating, for
each device, an associated piece of server software that can accept
requests formatted according to the OCS protocol.  For non-commercial,
ad-hoc devices, such a server can be written as the primary means of
control for the device.  In the case of commercial devices, the
associated server might consist of a simple translation of OCS
requests into requests on the device's native high-level interface.
Suitable libraries and example code will be provided by OCS
maintainers for use in the development of subsystem servers.  The OCS
command and data exchange protocol will be implemented on top of
remote procedure-call and publish-subscribe systems supplied by open
source, community-maintained software libraries.

\subsubsection{Telescope control}

The telescope control system is comprised of an isolated computing
environment that communicates directly with the telescope on-board
computer (provided by the telescope manufacturer).  This system
receives pointing data and diagnostics from the telescope at a fixed
rate, and schedules observations by providing a sequence of scan
profiles to control the telescope position and velocity.  Commanding
to the telescope is asynchronous with data acquisition to avoid
reliance on strict real-time processing, while allowing the on-board
control loop to run at a fast enough rate (typically 100\,Hz) for
smooth telescope operation.  The control system interfaces with the
OCS to receive high-level scheduling commands, send status messages,
and process alarms.

\subsection{Hardware---timing and synchronization}
\label{sec:daq:sync}
Synchronization for each observatory will be accomplished with IRIG-B002 timestamps distributed to the bolometer readout, telescope encoders, and other housekeeping subsystems. A timing card specified for use by most Stage-3 experiments will provide absolute GPS timestamps and can free-run without a GPS lock with minimal drift (5 parts in $10^{11}$) over 24 hour time scales. It is configurable with outputs of IRIG-B timestamps, a 10-MHz oscillator, and a pulse-per-second (PPS). With an Allen variance on the 10-MHz oscillator of $<10^{-11}$\,s, these achieve the stringent requirements necessary for frequency-based readout as well as the $1-\mu$s absolute timing accuracy sufficient for all readout and subsystems baselined in the reference design. There are additional options we are likely to explore, in particular PTP networking protocol across each site. In particular, we are not baselining a synchronization pulse or word to be distributed across the site, however individual subsystems (e.g., synchronization between readout crates on a platform) may be required and specified by the subsystem, and that synchronization signal could be timestamped and recorded as well.

\begin{figure}[htbp!]
\begin{center}
\includegraphics[width=5in]{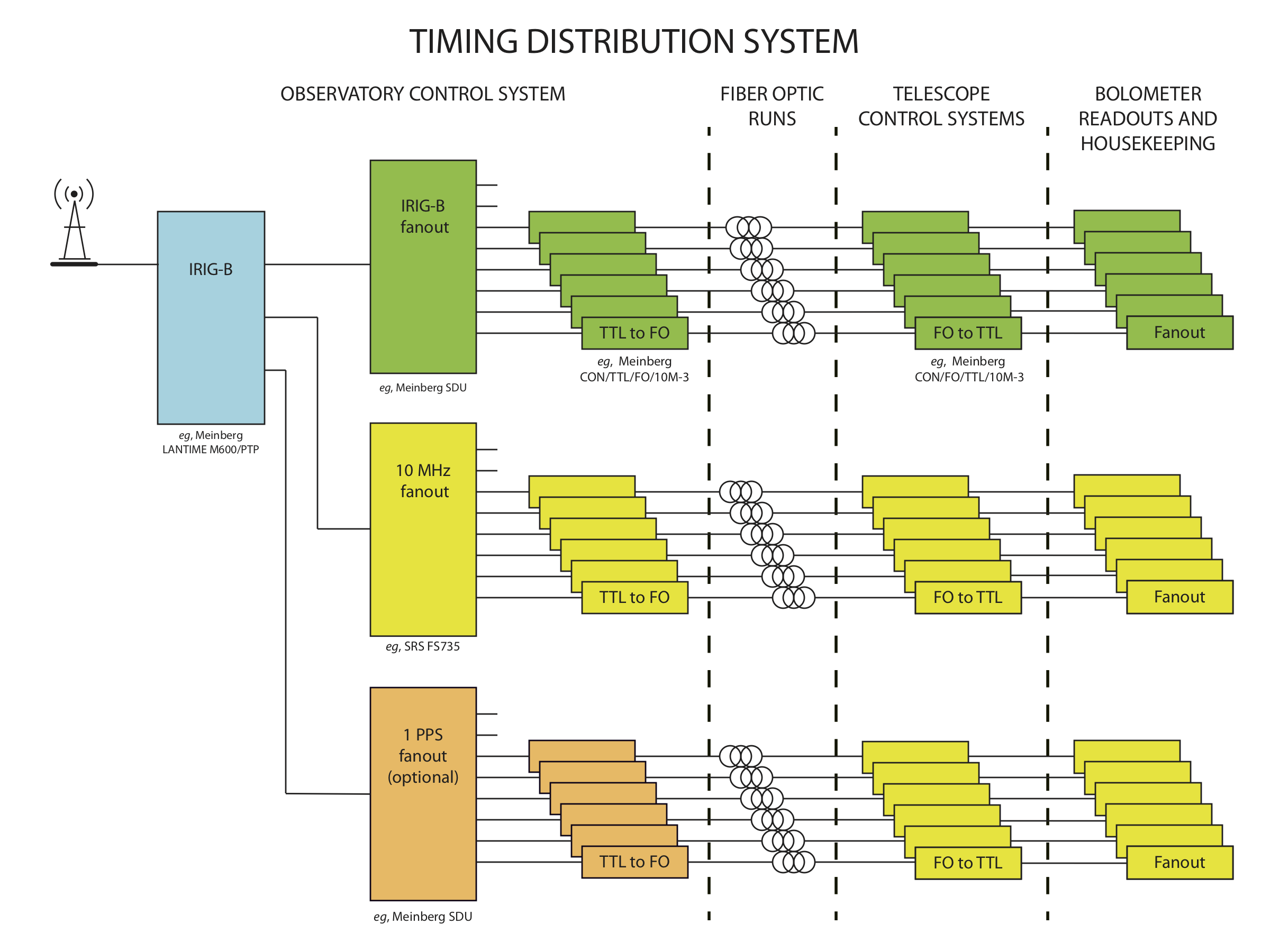}
\end{center}
\caption{Distribution of timing signals (IRIG timestamps, 10-MHz clock line, optional pulse-per-second) for a given observatory.}
\label{fig:DAQtiming}
\end{figure}

\subsection{Scaling and technical readiness from Stage-3 experiments}

In large part, data acquisition technologies developed for Stage-3 experiments can be applied to CMB-S4 without fundamental changes in approach (and in some cases can be reused as-is), limiting technical risk in this sector to a low level. The only major uncertainty is the communications protocol between TDM electronics and L1 DAQ computing, which needs to be developed. As this communication does not approach any fundamental limits (bandwidth over fiber optic cables, hard latency requirements, etc.) this task is, like many of the others described here, a low-risk technical engineering task. The demonstrated performance of the technologies outlined here as part of Stage-3 experiments thus provides a high technical readiness level for CMB-S4.


\section{Data management \prelim{ ({\it T. Crawford and M. Hasselfield})} } 
\label{sec:datamanagement}

The data-management subsystem deals with the transport, storage, and reduction of the instrument data, the generation and reduction of synthetic data sets, and the internal and external distribution of the derived data products. The remit of the data-management subsystem extends from the raw data recorded on disk at the two sites (where the remit of data acquisition is defined to end) to the production and distribution of well-characterized sky maps in the various CMB-S4 observing bands (where the various science analyses are defined to start). The subsystem includes dedicated on-site hardware (storage and analysis servers), computational resources in the U.S. (cycles, storage, and network bandwidth), software (including modules, pipelines and frameworks), data distribution (internal and external), and subsystem management. 

The biggest project cost will be the personnel required to create, maintain, and execute the data analysis software stack. A smaller effort will be devoted to the systems for storing the data on site, transmitting it to the U.S., and distributing it to the collaboration and community. The smallest component will be subsystem management, including technical and project leadership. In addition to the managers, engineeers, and scientists included as part of the project cost, it is assumed that additional scientists in the CMB community---at both national labs and universities---will provide significant uncosted resources in both construction and operations.

The computational resources required for the data-management subsystem are dominated by the unprecedented scale of the storage and cycles required for data analysis in the U.S. However, following the example of the {\it Planck\/} satellite mission (the closest analog to CMB-S4 in data management), our assumption is that these resources will be acquired through agreements with major national computing facilities, and they are therefore assumed not to be part of the project cost \footnote{The one exception to this for {\it Planck\/} was the need to augment the standard NERSC disk quota; however under NERSC's storage plan for the next decade \cite{lockwood2017} this would not be necessary for CMB-S4.}. A much smaller on-project hardware component will provide the dedicated data storage, processing and transport resources required at each site.

\subsection{Design drivers}

The primary driver for the design of the data-management subsystem is the unprecedented volume of the CMB-S4 data set (Table~\ref{tbl:dv}).

\begin{table}[htbp!]
\begin{center}
\begin{tabular}{|c|c|c|c|c|c|c|}
\hline
Telescopes & Detectors & Sampling rate & Data rate & Transmission & Storage & Samples \\
 & & (GHz) & (Gbps) & (TB/day) & (PB) & ($\times10^{15}$)\\
\hline
18 $\times$ SAT  & 153360 & 100 & 0.46 & 1.69 & 4.31 & 3.4 \\
1$\times$LAT (delensing surv.) & 114432 & 400 & 1.36 & 5.04 & 12.87 & 10.1 \\
2$\times$LAT (Legacy surv.) & 243520 & 400 & 2.90 & 10.72 & 27.38 & 21.5 \\
\hline
Total & 511312 & \dots & 4.72 & 17.44 & 44.56 & 35.0 \\
\hline
\end{tabular}
\caption{The reference design data rates and volumes for 7 years of observing by the 18 small and 3 large aperture telescopes (SATs and LATs respectively), assuming that the data can be compressed to 35\% of their raw volume for transmission and storage, and only uncompressed on the fly for analysis.}
\end{center}
\label{tbl:dv}
\end{table}

This volume---comparable to that of the triggered LHC or LSST---is orders of magnitude greater than current CMB experiments. The US data center must have sufficient archival storage to hold the full data volume, enough spinning disk to hold as much of the data as is required at any one time for analysis, and enough cycles to support the full data analysis program, all available for the duration of both the construction and operations phases. This translates to needing exa- to zetta-scale computing resources for the next 15 years, which can only realistically be provided by national computing resources such as the DOE high performance computing (HPC) and the NSF high throughput computing (HTC) facilities.

The data-management software stack must be robust, validated and verified, and able to run efficiently at appropriate scale on whatever computing architecture(s) are available to us at any epoch. Both the scale of the data to be processed, and the computational architecture on which that processing will occur, will evolve significantly over the lifetime of the project, and this evolution must be accounted for in both provisioning the computational resources and planning the software lifecycle.

Additional constraints are imposed by the bandwidth available from the remote observing sites and the need to deliver data products at the cadences required for monitoring data quality and supporting transient studies (taken to be daily here, but subject to further input from the scientific community). Both sites must have sufficient reliable local resources to store data pending transfer to the primary data center in the U.S. (including failsafe storage), and to perform any data reductions required prior to transfer. 

\subsection{Reference design}

\begin{figure}[htbp]
\includegraphics[width=0.95\textwidth]{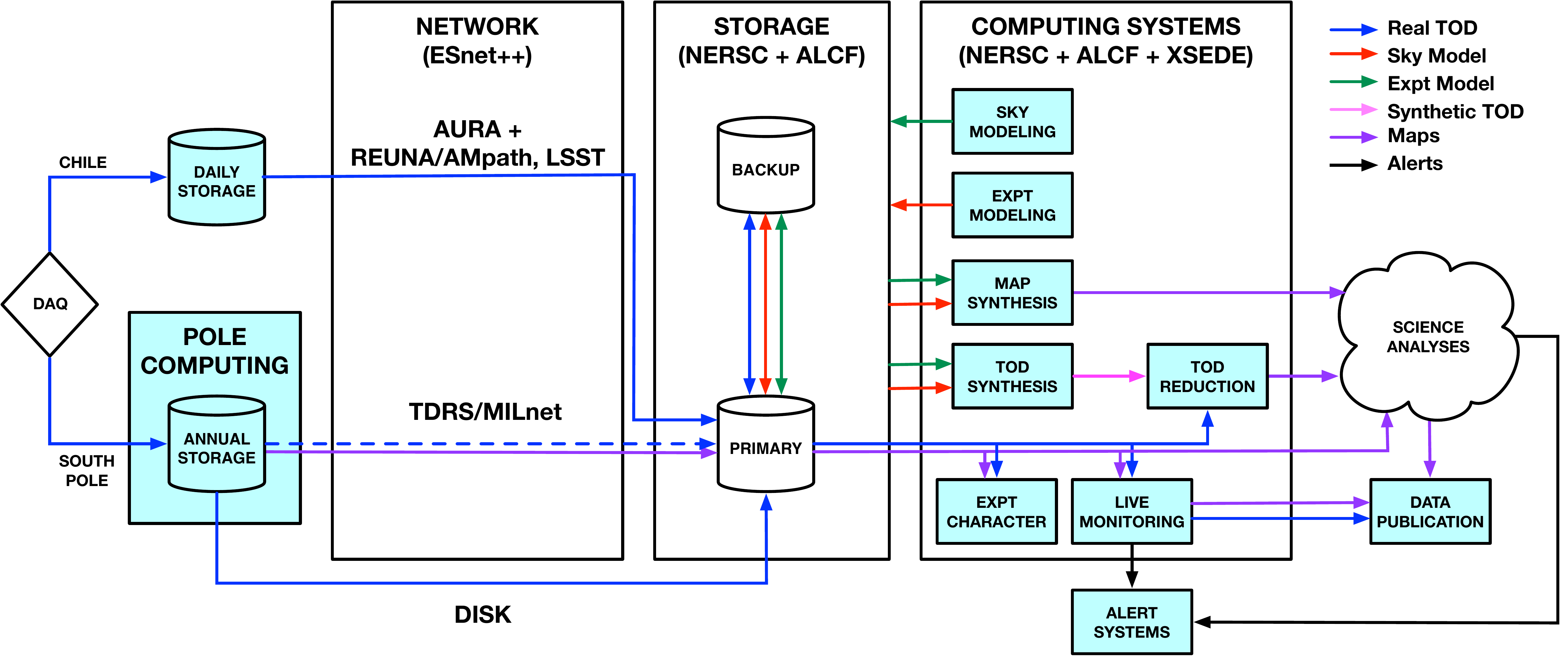}
\caption{Schematic view of the data-management subsystem, spanning the range from Data Acquisition to Science Analyses, with on-project cost areas highlighted in cyan. Note that the named networking, storage and compute resources are indicative and anticipated, not yet confirmed.}
\label{fig:dmschema}
\end{figure}

Figure~\ref{fig:dmschema} shows a schematic view of the reference design data-management subsystem, including on-site resources, data-transport systems, archival storage, computing systems, the overall software stack, and data distribution within and outside of the project. 

\subsubsection{On-site resources}

The reference design assumes sufficient network bandwidth from Chile to return the data to the primary U.S. data center essentially in real time, following the models of both the Simons Observatory and LSST. As such, we only need to provide a sufficient local storage to hold a few days data in the event of a temporary network outage. However this is not assumed to be the case from the South Pole, and instead we follow the model of the South Pole Telescope and BICEP/Keck experiments, with a small subset of the raw data, together with some reduced data products, transferred over the network to the US each day, and the entire data set stored on site and shipped once a year. This means that the South Pole site must be provided with not only sufficient disk to store and transport a year's data, but also with sufficient computing resources to perform all time-critical data reductions---primarily to generate the data products that will be sent over the network each day, including live monitoring for data quality and transient studies. This piece of the data-management system requires close coordination with the site infrastructure systems to ensure that sufficient space and power are available for the on-site computing resources, especially at the South Pole. Taking conservative estimates, the reference design includes 100\,TB of storage in Chile, and 5\,PB of storage and 10\,Tflop/s of computing at the South Pole.

\subsubsection{Data transport}

\begin{figure}[htbp!]
\centering
\includegraphics[width=0.6\textwidth]{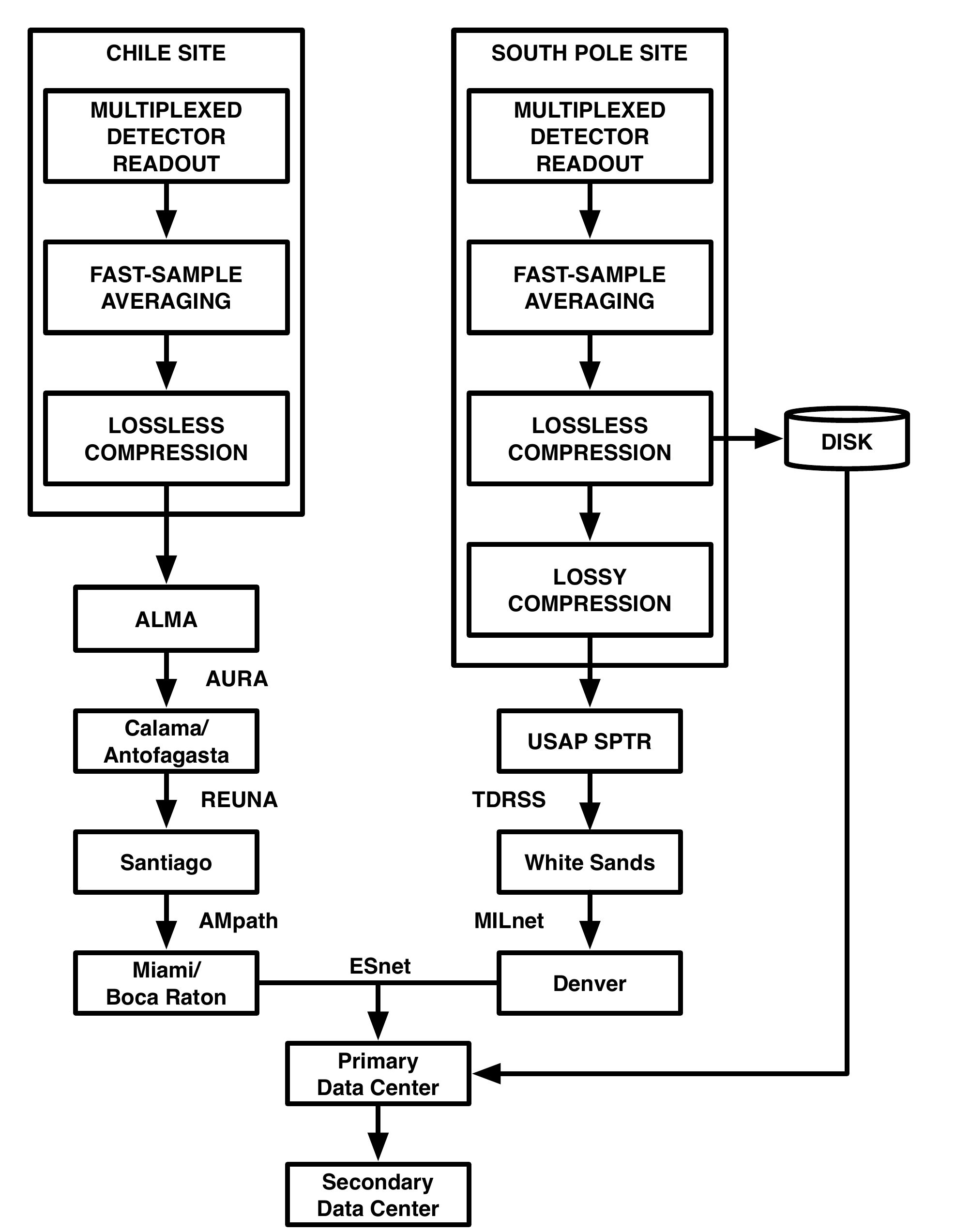}
\caption{Schematic view of data-transport paths from acquisition to the U.S. data centers for the Chile and South Pole sites.}
\label{fig:dmtransport}
\end{figure}

Raw data files will be registered on-site as they are gathered and this information will accompany the files to the primary U.S. data center. For the Chilean site, a fiber-optic network connection will be made to the nearby ALMA Array Operations Site as part of site development. From there, several possible paths are available to provide the necessary high-bandwidth connection to ESnet and hence to the data centers, including the current REUNA/AMpath networks and the possibility of sharing dedicated LSST fiber. From the South Pole, daily transfers of reduced data products will use the TDRS satellite link and MILnet. These transfer capabilities are controlled by the limited number of satellites in orbits visible from the South Pole and the limited fraction of the day during which they are above the horizon. As noted above, current network data-transfer capabilities (about 100\,GB/day) are not sufficient for the full CMB-S4 data rate. For the reference design we will therefore plan to store each year's data on site and transfer them to the U.S. at the start of each Austral summer, however we will also continue to work with NSF and DOE to investigate possibilities for implementing a connection to the US data center with sufficient bandwidth to stream the data. From the primary U.S. data center the data will be transferred over ESnet to the secondary center to provide a secure backup. The various reference design paths are illustrated in Fig.~\ref{fig:dmtransport}.

\subsubsection{Data centers}

The U.S. data centers must provide archival storage for the entire CMB-S4 data volume (including all derived data products and Monte Carlo simulation sets) at two distinct sites, sufficient spinning disk to hold the data volume required for the largest single analysis, and sufficient cycles for all of the data processing, including both the many re-processings of the real data and the generation of Monte Carlo simulation sets. Given the CMB-S4 data volume and project lifetime, these resources can only realistically be provided by national computing facilities. Such facilities have the additional advantages of being cost-free to the project, and of having long-term development plans that will see their resources steadily increase over the lifetime of the project---with new single-site DOE HPC systems being fielded approximately every 4 years, and more continuous growth of the distributed NSF HTC systems. However, given the variety of architectural responses to energy-constrained computing, and the increasing effort required to port to and optimize for them, we should also to limit the number of architectures we will use by adopting centers whose development paths are as similar as possible. The reference design therefore assembles the necessary resources from the DOE's National Energy Research Scientific Computing Center (NERSC) and Argonne Leadership Class Facility (ALCF), and the NSF's Extreme Science and Engineering Discovery Environment (XSEDE). Local computing resources may also be used by individual members of the collaboration for smaller-scale analyses, but their deployment is outside of the remit of the data-management subsystem except insofar as the data are made available from the data centers.

Given its ability to provide all the required types of resource, its co-location with ESnet, its 20 year history of supporting the global CMB data analysis community, and its open access policies, the reference design adopts NERSC as the primary U.S data center. Since so many existing CMB experiments already use NERSC resources, this will also allow us to leverage community-wide data and software. The secondary archive is then located at ALCF, with a very high bandwidth connection with NERSC over ESnet, and additional cycles are provided by both ALCF (with the most similar systems to NERSC) and XSEDE. Having cycles at multiple facilities mitigates against the risks posed by down-time or queue congestion at any one facility; having both HPC and HTC cycles allows us to map our processing to the most appropriate resource for a particular pipeline (specifically depending on whether the pipeline requires the inter-process communication supported by HPC but not HTC). Having cycles at a limited access facility devoted to ``heroic'' computations (ALCF) also provides additional resource security for our most intensive computations. ALCF also hosts many of the large-scale structure simulations being used to generate mock catalogs for experiments like LSST, {\it WFIRST}, and {\it Euclid}, which we hope to leverage in our sky modeling to support joint cross-correlation studies.

Since these are national facilities that allocate their finite resources to a large number of projects, we must mitigate the risk of insufficient allocations for CMB-S4. Following the example of the {\it Planck\/} satellite mission, we will seek a formal memorandum of understanding with DOE to guarantee long-term and sufficient NERSC access. With this in hand, we will then seek complementary agreements with DOE and NSF for long-term access to ALCF and XSEDE.

As noted above, both the maximum data volume and the computing architecture will evolve over the lifetime of the mission, so we must be confident that we will have sufficient resources at each epoch. One key milestone is to demonstrate that we can process the full CMB-S4 data volume before operations commence; in order to reach this goal we plan to double the data volume being synthesized and reduced every two years. Table~\ref{tbl:cr} maps the resources needed for the 2020, 2022, 2024, and 2026 (i.e., construction project) data volume challenges to those known or anticipated on the various generations of NERSC flagship system that will be used, illustrating that the requirements can be met at NERSC (even while anticipating resources elsewhere too). Note that each of these runs require a significant fraction of a top 10 HPC system, illustrating the need for resources of this type. Since this will involve three generations of HPC system at NERSC, we will interleave the data volume challenges with architecture challenges, porting to and optimizing for a new architecture in each of the intervening years. With this schedule, the architecture challenges would be in 2021 (Perlmutter), 2023 (Aurora at ALCF), and 2025 (NERSC-10).

\begin{table}[htbp!]
\begin{center}
\begin{tabular}{|c|c|c|c|c|c|}
\hline
Year & 2020 & 2022 & 2024 & 2026 \\
\hline
NERSC system & Cori (KNL) & Perlmutter & Perlmutter & NERSC-10 \\
Cycles (Pflop/s) & 30 & 150 & 150 & 1000 \\
Memory (PB) & 1 &  & 4 & 8 \\
Disk (PB) & 30 & 150 & 150 & 300 \\
\hline
Data fraction & 1/8 & 1/4 & 1/2 & 1 \\
Samples (x$10^{15}$) & 0.9 & 1.8 & 3.5 & 7.0 \\
Total cycles (Eflop) & 2.7 & 5.5 & 11 & 22 \\
System fraction & 3\% & 1\% & 2\% & 1\% \\
Peak memory (PB) & 0.8 & 1.6 & 3.2 & 6.4 \\ 
System fraction & 80\% & 40\% & 80\% & 80\% \\
\hline
\end{tabular}
\caption{Estimated computational resources available at NERSC and required by CMB-S4 to scale the data-management software to the full data volume prior to first light, assuming characteristic observing efficiency, operations per sample (including both synthesis and reduction), computational efficiency, and memory high-water mark.}
\label{tbl:cr}
\end{center}
\end{table}

Finally we note that this combination of network, storage and cycle resources, spread across multiple facilities, and connecting a major scientific instrument to a distributed collaboration, is conceptually what is sometimes called a ``data superfacility." We will therefore work with the superfacility experts at NERSC and elsewhere to see how we can leverage their work to make this a seamless computational ecosystem.

\subsubsection{Software}

The development and deployment of the CMB-S4 software stack comprises the bulk of the data-management subsystem. As with many aspects of CMB-S4, the primary challenge here is that of scaling existing technologies to unprecedented levels rather than developing entirely new ones, and hence more about efficient, scalable, implementations than algorithms per se. In addition, given the long-term nature of the project and the expected evolution of (and possible revolution in) computer architectures over its lifetime, porting and optimizing the software stack to new state-of-the-art systems will be an equally critical task. Note that even if existing system had the longevity to support CMB-S4 throughout its lifetime, they do not have the capacity or capability we will ultimately need, so that progression through multiple generations of system must be accepted as an inevitable feature of any data-management plan.

Another way in which CMB-S4 data management will necessarily differ from most (if not all) previous CMB experiments will be in the quality of our software engineering. Rather than the traditional patchwork of single-author/single-user tools developed on a best-effort basis and strung together into a loose sequence of analysis steps with untracked intermediate data products passed between the individuals running them, we will need a software development team producing a coherent suite of stable, tested, documented, modules that can be assembled into efficient, seamless, pipelines whose full provenance can be recorded at any run. While this is approach is largely unfamiliar to the CMB community, it is widely used elsewhere and we will take advantage of the lessons learned and adopt established best practices.

The software can be divided into three broad categories---data synthesis, data reduction, and the overall infrastructure to support these.

\paragraph{Data synthesis:}
Synthetic data play key roles in CMB studies, including in the design and development of experiments, the validation and verification of their reduction and analysis pipelines, and the quantification of uncertainties in and removal of biases from their science results using Monte Carlo methods. Data synthesis includes modeling both the experiment and the sky it will observe, along with generating synthetic data sets both in the map- and the time-domains.

While the development of detailed models of individual sky components is outside of the realm of the data-management subsystem, it does include the assembly of such models into total skies to be as inputs for synthetic data sets. The total sky includes the temperature and polarization components of the scalar, tensor and non-Gaussian CMB; the extragalactic foregrounds (and their associated lensing of the CMB); and the galactic foregrounds. Realistic sky models will be particularly important for optimizing the frequency coverage of the various telescopes, and for the validation and verification of key science analysis algorithms and implementations including foreground cleaning and delensing. We will seek to support joint analyses of CMB-S4 with large-scale structure observations by adopting common extragalactic skies with experiments such as LSST, {\it WFIRST}, and {\it Euclid}.

Experiment modeling provides a parametric model of the entire experiment, comprising the instrument optics and electronics, and the observation environment and survey strategy. The parameterization should be rich enough to support the most detailed (and therefore computationally expensive) time-domain syntheses, while also providing standard reduced representations sufficient for approximate (but computationally cheaper) map-domain and Fisher forecasting approaches. 

The data synthesis step applies the experiment model to the sky model to generate the data that would be gathered by the instrument making its observation of that sky. This includes signals from the sky, the atmosphere, and the ground, possibly modulated by a half-wave plate, convolved with each detector's beam and bandpass, converted to a voltage by the detector, and fed through the readout system, and incorporating all of the detector-detector correlations, for example from multichroic pixels or highly multiplexed readout. The complexity of the processing and scale of the data necessarily introduce a trade-off between computational cost and realism, reflected in the prioritization of both map- and time-domain syntheses (with time-domain syntheses also requiring reduction to maps before analysis). The development and deployment of these two paths will be coupled to take advantage of the strengths of each approach and the ability to cross-check them against one another. Specifically we will ensure that each approach is able to synthesize data from the same combinations of experiment and sky model.

\paragraph{Data reduction:}
The data reduction pipelines form the core of the data-management subsystem. Comprising experiment characterization, live monitoring, and map-making, they must produce all of the data products required to monitor the performance of the instruments and to support all of the various science analyses, and be able to do so with the required cadence. Note that both the characterization and monitoring pipelines must run near-autonomously and, for the South Pole data, they must be able to be executed on site.

Experiment characterization covers both the measurement of instrument properties and the reconstruction of its observations. This category is particularly important during the deployment phase, and includes detector liveness checks; preliminary noise, calibration, and systematics measurements; and telescope pointing and beam checks. The software for these tasks must be flexible and easy to adapt on the fly in response to unexpected features the data. Most of the development and implementation of this software is expected to be conducted by junior scientists at universities.

The live monitoring pipeline includes all time-critical processing, including both near-real-time checks on instrument performance and data quality and the generation of the daily data products needed to support transient searches, together with the alert systems associated with each of these. Performance/quality alerts will be passed directly to the instrument teams. The exact cadence and content of the transient searches will be determined in consultation with the wider community; for the reference design we baseline the production of daily maps of both the wide-area and the ultra-deep (delensing) LAT surveys, with the former generated at NERSC and the latter at the Pole and included in the daily satellite data-transfers. These will then be passed to a dedicated collaboration working group to identify transients and source alerts through standard astronomical alerts systems.

The map-making pipeline includes the pre-processing needed to make the raw data match the data model assumed by any particular map-making algorithm, the map-making itself, and the characterization of the resulting maps. Pre-processing includes such steps as finding and flagging spikes or jumps in the data, narrow-band filtering of spectral-line-like contamination from, e.g., detector sensitivity to the cryocooler or drive system, and calibration to convert the raw data voltages to temperatures. This part of the pipeline also needs to be flexible enough to support data exploration and the mitigation of unanticipated systematic effects. The map-making itself can use filter-and-bin, destriping, or maximum likelihood algorithms, each of which has a different data model and therefore requires somewhat different pre-processing. The characterization of the map can range from a simple weight map to a full covariance matrix to a suite of corresponding Monte Carlo realizations. This pipeline is also often iterative, with the map produced being used to refine the pre-processing steps. We anticipate producing maps at each observing frequency, for each survey (SAT, LAT delensing, LAT wide-area), for a range of data cuts in both time and across the focal plane(s); each map-making will necessarily include data from multiple telescopes, and possibly from multiple sites, though neither of these is anticipated to pose qualitatively different challenges to the single telescope, single site, analyses we are used to. The diversity of instrument and survey types in CMB-S4 (low- and high-resolution telescopes, small- and large-area surveys) dictates flexibility in map-making algorithms and implementations; the reference design therefore includes all types of map-making and allows for separate optimization for HPC and HTC architectures, with the goal of conducting direct comparison of the various approaches' products and performance.

\paragraph{Infrastructure:}
The overall data-management software stack will consist of the individual software modules described above, assembled into task-specific pipelines, which are then executed within architecture-specific frameworks, together with the databases needed to support them. 

Pipelines will be built to execute standard sequences of data synthesis and reduction operations, and the critical features will be their overall efficiency and their adaptability. In order to avoid the prohibitive IO overhead of writing out a synthetic data set and reading it back in for reduction, when synthesizing and reducing mock data such pipelines will instead run on-the-fly. Since such data will not (in general) be saved to disk, such pipelines will also incorporate the ability to synthesize a data set once and reduce it multiple times (for example employing different algorithms and/or implementations and/or data cuts) within a single job execution, even when the reduction is destructive.

The frameworks will enable the pipelines to run within any HPC or HTC environment available to us (subject to basic sufficiency requirements), with the goal of being able to take advantage of all of the resources that we have access to in order to minimize any bottlenecks from overall resource limits or from congestion on shared systems. The baseline will be built on the existing TOAST HPC and SPT-3G HTC frameworks, originally developed for the {\it Planck\/} and SPT experiments respectively, and the goal will be to be able to use the same pipelines on both HPC and HTC systems wherever possible. Note that this will only apply to pipelines that require no inter-process communication, for example filter-and-bin map-making. Enabling this hybridization is an open research and development project; for the reference design we assume that at very least any module that operates on a small subset of the data (as is often the case in pre-processing and some types of map-making) will be able to be run in either framework, and that we will have the tools needed to translate in situ between the frame-based (SPT-3G) and distributed (TOAST) data models.

Databases will be employed to track data provenance, support data selection for specific reductions, and enable data distribution to remote sites. They will also be used to record the iterations of the instrument model, which will continuously be refined with the analysis of laboratory data, commissioning test data, and ultimately the real data themselves, and to deliver these to the data synthesis codes. Finally they will be used to record the input parameters, intermediate and final data products, and the code module, pipeline and framework versions used for each analysis to support traceability and reproducibility.

\subsubsection{Data distribution and publication}

All CMB-S4 collaboration members will have access to the raw data and to intermediate data products up to and including the well-characterized single-frequency maps. The main mode for the distribution of data to the collaboration will be via the primary data center, where all collaboration members will also be eligible to have user accounts and have access to significant compute resources. Given its volume, typically we expect users to bring their code to the time-domain data, and only to transfer map-domain data to their local computing resources (though these can obviously be processed at NERSC too). Although they lie outside of the remit of the data-management subsystem, we anticipate that additional derived data products (e.g., component-separated maps, maps of the lensing potential, power spectra and likelihood codes) will be similarly distributed.

Public releases of CMB-S4 data products will occur at regular intervals through well-established CMB data archives. Maps of all flavors (single-frequency, foreground-cleaned CMB, lensing potential, etc), catalogs of galaxies and galaxy clusters, the CMB power spectra, and cosmological parameter likelihood codes will be archived at NASA's Legacy Archive for Microwave Background Data (LAMBDA), including both the real data and subsets of the supporting suites of Monte Carlo map realizations. Time-domain data and the full Monte Carlo suites, together with the code used to process them, will be made available through public directories at NERSC, where users will also have access to the computational resources needed to make practical use of such large data sets. We will also make the data-management software stack (complete with full documentation) public wherever possible both through a public repository on a version control system such as github and as a standard installation at NERSC.

\subsubsection{Management}

The management of the CMB-S4 Data Management team will include both the formal WBS elements of software control, resource management, and subsystem interfaces, and the informal responsibilities of team building and the synchronization of activities and products with the project and collaboration schedules.

The CMB-S4 DM software stack will be fully version controlled using standard repository management tools such as git. All code delivery will include validation and verification, unit tests, and documentation, and the full stack will be subjected to nightly builds for continuous integration. We will also adopt a limited set of programming languages (with formal criteria for expanding that set), and develop---or adopt---a common software style guide. We will ensure that there are regular training opportunities in the software standards as part of the process for bringing in new team members, or as refreshers for existing members, as well as in coding for new architectures and tools such as debuggers and profilers.

The entire data-management plan rests on the allocation of sufficient computational resources to meet our needs across a range of facilities; the acquisition, allocation, and management of these resources will be a key activity. As noted above, we will pursue formal agreements with DOE and NSF to safeguard our access to the necessary resources across the various facilities, and for each facility and resource we will develop an annual burn-down plan centered on the major production runs scheduled for that year to ensure that these are not limited by insufficient resources. Where possible, the resources needed for the core activities will be allocated to a project production account (which a limited number of core developers will have access to) rather than to one or more individual user accounts in order both to ring-fence the resources and avoid being dependent on single users for production runs.

The data-management subsystem has critical interfaces both within the project and with the overall collaboration. Within the project, these interfaces are with data acquisition (corresponding to the hand-off of data from the telescope to the data centers), and with site infrastructure (ensuring that sufficient space and power are available for the on-site computing resources, and coordinating the transfer of disks to/from the Pole). These interfaces will be mediated within the overall WBS structure through the L2 leads and the project office. There are also equally critical interfaces with the collaboration's technical working groups, to develop a detailed, accurate, experiment model, and analysis working groups, to ensure that the overall suite of data products are necessary and sufficient. During the construction project, it will also be critical for the data-management subsystem to deliver simulated data sufficient for the working groups---both technical and analysis---to support instrument and observation design, and to validate and verify the software stacks for the various science analyses. These interfaces will be mediated through the Technical and Science Councils, as well as through a dedicated and cross-cutting simulations working group that will coordinate data science challenges alongside the data volume and architecture challenges.

The overall data-management team will be built on the convergence of the separate Stage 3 teams, leveraging both pre-project research and development funds and off-project cross-cutting collaborations. As the project matures these will coalesce around the various Level 3 activities (broadly synthesis, reduction, and infrastructure), while maintaining their interconnectedness through cross-cutting activities such as the annual production runs. We will make use of a variety of communication channels (from wikis to email to slack channel and telecons), as well as face-to-face and virtual hackathons and training events to develop the team, and, where appropriate, adopt an agile software development approach.

The project will face numerous review gates, and the data-management subsystem schedule must not only be synchronized to its own reviews, but also to the range of data needs and requirements imposed by those of all of the other subsystems. Here we will make the annual scaling and architecture challenges do double-duty as also producing state-of-the-art data sets for the project and collaboration as a whole, as well as responding to more limited but higher cadence requirements through the data or software needed for smaller and/or simpler simulations, including directly in the map domain.

\section{Integration and commissioning}
\label{sec:integrationcommissioning}

Integration and commissioning are the final activities before the completion of the project and the start of science observations. 
We will integrate and commission each large telescope (LAT) and small telescope (SAT) separately, 
allowing for a phased start of science observations.
Integration and commissioning is done separately, by largely separate teams, at each site. 
To the extent that I\&C requires support personnel from within the collaboration, 
careful coordination will be required to ensure that support personnel are available. 

\subsection{Scope}

Integration and commissioning (\ic), as described here, refers only to on-site \ic\ activities.
We assume here that the following have occurred off-site prior to the start of I\&C on-site.
\begin{itemize}
\item The LAT has been installed on-site and validated with the vendor. 
\item The LAT receiver (LATR) has been integrated and tested off-site, and minimally disassembled for shipping. This disassembly is assumed to include removal of the focal planes and placement of shipping braces so that the optics tubes can be shipped inside the receivers.  
\item The SAT receiver (SATR) and telescope mount have been already tested together off-site, then minimally disassembled for shipping. This disassembly is assumed to include separation of the SATR from its mount and removal of the focal plane arrays from the SATR.
\end{itemize}

\textbf{Integration} of an SAT involves installing the telescope mount, installing the three FPUs into the SATR, and then installing the SATR on the mount.  Integration of an LAT begins with the installation and use of a commissioning receiver to test the LAT;  this is followed by the assembly of the LATR (similar to the SATR), and installation of the LATR on the LAT.

\textbf{Commissioning} begins after the end of integration. 
This includes the execution of validation tests to demonstrate that the receiver and telescopes are functional, and the execution of initial calibration measurements. 
Commissioning ends when functionality is demonstrated and science observations can begin. 
Note that functionality being demonstrated does \textbf{not} mean that all calibrations are complete;  ongoing calibrations are required throughout the observing period.

\subsection{LAT integration}

LAT Integration includes two distinct categories of activities: (1) using a commissioning receiver to validate the integrated performance of the LAT; and (2) integrating the LATR, cooling it down in the LAT receiver cabin, and connecting it to all utility, control and data acquisition connections. 

\subsubsection{Commissioning camera}

Prior to integration, the LAT telescope performance 
will be demonstrated in cooperation with the vendor following the requirements in the LAT contract. 
This should include demonstration of pointing using a star camera. 
Then, integration using a commissioning receiver begins. 
The commissioning receiver will arrive with its focal plane removed. 
It will be re-assembled and installed and cooled down on the LAT.
This will provide a platform for testing mm-wave performance of the LAT and the interface between the LAT and all of the cryogenic and detector readout systems before installing the LATR. 
A list of tests to be done with the commissioning receiver will be developed by the LAT and LATR teams prior to the start of construction of the commissioning receiver to ensure it has the needed functionality. 

\subsubsection{Installing the LATR}

Integration of the LATR is then an intense period that requires 12--17 people on-site for about 80 days. 
Upon arrival, the LATR will first be re-assembled, then installed in the LAT receiver cabin to be cooled down.  
The detailed list of work to be done during that time is shown in Table~\ref{tab:ltcint}. 
At the South Pole, it is critical that this integration work be completed during the Austral summer.
Note that since the cooldown takes 40 days, if a problem is found that requires warming up and cooling back down again, it will delay the completion of integration into the winter season, to be completed by winter crew. 
In Chile, the constraints on total duration are less critical. 
However, a failed cooldown would still be a significant problem in terms of schedule delay.

\begin{table}
\begin{center}
\begin{tabular}{|l|l|l|l|l|}
\hline
Duration & Receiver tasks & Detector tasks & Other tasks & Number\\
days & & & & of people \\
\hline
5 & Receiver unpacking & Setup clean-room & Prep LAT space & 12 \\
3 & Remove shipping braces & Check detector modules & Continue & 12 \\
17 &  Install FPUs, close receiver & Install detectors in FPUs & Prep LAT, DAQ, DM & 17 \\
2 &Install LATR receiver cabin & Setup, test det readout & DAQ, DM & 13 \\
2 & Pump out receiver & Continue & Continue & 13 \\
40 & Cooldown, cryo validation & Continue & Continue & 17 \\
14 & Receiver basic validation & Detector basic validation & Commissioning prep& 13 \\
\hline
78 & Total & & & \\
\hline
\end{tabular}
\caption{\label{tab:ltcint}
Task, durations, and staffing levels during LAT integration. 
}
\end{center}
\end{table}

\subsection{SAT integration}

Prior to SAT integration, the entire SAT system will have been tested in the United States.
On-site, the tasks to be completed for integration, and their approximate resource requirements, are shown in Table~\ref{tab:satint}. 

\begin{table}
\begin{center}
\begin{tabular}{|l|l|l|l|l|}
\hline
Duration & Receiver tasks & Detector tasks & Other tasks & Number\\
days & & & & of people \\
\hline
3 & Receiver unpacking & Setup clean-room & Setup SAT mount & 10 \\
3 & Remove shipping braces & Check detector modules & Continue & 10 \\
9 &  Install FPUs, close receiver & Install detector in FPUs & Prep LAT, DAQ, DM & 12 \\
2 & Pump out receiver & Setup, test det readout & Continue & 10 \\
2 & Install SATR on mount & Setup, test det readout & DAQ, DM & 10 \\
20 & Cooldown, cryo validation & Continue & Continue & 10 \\
14 & Receiver basic validation & Detector basic validation & Commissioning prep& 10 \\
\hline
53 & Total & & & \\
\hline
\end{tabular}
\caption{\label{tab:satint}
Task, durations, and staffing levels during SAT integration. 
This is for a single SAT integration. 
For multiple SAT integrations, resource estimates should be approximately doubled, but calendar days do not need to be, since things can be done in parallel.
}
\end{center}
\end{table}

\subsection{Commissioning}

The initial validation of individual components (e.g., detectors) will be completed before components are shipped to the sites.  This section only covers the validation of the full integrated instrument. There are six main phases of validation during the commissioning stage prior to the beginning of scientific observations, which are listed below.

\paragraph{Integrated telescope and instrument validation:} During the first phase, we will confirm that the DAQ, data archiving, and communication systems integrated with the receiver are operational and test the basic functionality of the instrument.

\paragraph{Confirm that the detectors and readout are operational:} 
Next, we will perform current versus voltage (IV) curves to measure the saturation power 
and verify that it is in the nominal range. We will measure the time constants 
to verify that they are in the nominal range, using
bias steps to get the electrical time constants and 
a chopped thermal calibration source to measure optical time constants.
Using elevation nods, beam filling thermal loads (e.g., with a sparse wire grid), or galactic source observations
we will measure the relative 
responsivity of the detectors 
and confirm that 
it is within the nominal range. Finally, using rotating wire grid we 
will measure the relative polarization angles of the detectors to better than
2$^{\circ}$ uncertainty. If SATs are deployed Chile, another set 
of polarization angle measurements with the HWPs rotating will be performed to verify 
that the polarization angle is stable to 1$^{\circ}$.

\paragraph{Validate initial data timestreams:} Next we will take initial timestream data to verify that there are not excess glitches, SQUID jumps, or other issues. Less than $<10$\% of the timestream should require masking for glitches, etc.

\paragraph{Measure the noise performance of the system:}
We will take noise spectra curves to characterize the noise performance of the timestreams 
to verify that they are in the nominal range. 
Noise measurements will be performed both 
while the telescopes are stationary and performing various scan patterns. If any excess pickup is detected, we will power cycle various components 
to determine the source of the excess noise and mitigate it. 
Additionally, if SATs with polarization modulators are deployed to Chile, 
we will repeat the noise tests with their polarization modulators on and 
off to verify that there is less than an order of magnitude of 
extra noise in the noise curve measurements.

\paragraph{Measure the pointing and beams:} Prior to integration with the receivers, the LAT mirrors will be mapped out with photogrammetry to align the mirror panels to 25$\,\mu$m half wavefront error (HWFE). Next, the LAT platform pointing will be validated with a star camera to the pointing specifications laid out in the vendor contract. After full integration is complete, planet scans will be used to veify the LAT pointing to 20\,arcsec and beams to $\approx-10\,$dB. The SAT pointing will be measured to 4\,arcmin with a combination of star cameras and planet scans. The SAT beams will be verified to $\approx-10\,$dB with planet scans. Additional near-field measurements of the LATR and SATs with artifical sources could be used to verify that these measurements match those that will be taken in North America prior to deployment.

\paragraph{Survey validation:}
At the end of commissioning, we will aim to validate instrument performance through a measurement of the instrument noise-equivalent temperature (NET) within a factor of a few via CMB field observations.

\subsection{Calibration}
During science observations, we will perform calibration measurements to provide the calibrations necessary for scientific analyses. There will be two sets of calibration hardware, one for each site. We note that the calibration requirements are dependent on the instrument configurations and require in-depth studies to fully define. The requirements presented here are based on existing instrument requirement studies extrapolated to the scientific goals and estimated noise performance of CMB-S4. In this section, we present a baseline plan for calibration that is designed to meet the most stringent requirements. The final requirements will be determined by an instrument flowdown analysis before the CMB-S4 project baseline design.

\paragraph{Bandpass (same for LAT and SAT):}
The bandpass requirements are set by the requirements on foreground removal. 
The bandpasses of the detectors must be known to $<0.5$\% uncertainty \cite{Ward:2018fjf}.
Fourier-transform spectrometers (FTSs) will be used to characterize and validate the 
designed bandpasses of the detectors in situ in the field. 
Using different optical setups to couple the FTS to the telescope optics will enable us to 
use the same FTS design for measurements of the SATs and LATs. 
To achieve the low uncertainty required on the bandpass measurements, 
the transfer functions of the FTSs must be well understood. 
We can measure the FTS transfer functions by coupling two FTSs together and 
measuring their responses. To track the atmospheric-generated variations in the 
bandpasses, we will use a combination of thermometers, and barometers, 
and telescope elevation nods calibrated to existing radiometers.

\paragraph{Beams:}
The polarized and unpolarized beams of the instruments can be decomposed into three types: the main beam, near sidelobes, and far sidelobes. Each of these requires different methods of calibration or measurement that can include both celestial sources and artificial sources. Observations of planets or the Moon can be  
performed whenever the relevant sources are sufficiently above the horizon;
measurements with artificial sources require dedicated setups, typically done once a year if needed.
Near-field and mid-field beam measurements (with artificial sources) can be used to validate the 
beam performance and to inform the beam modeling, while far-field beam measurements feed directly 
into the analysis. The far-fields of the LATs (10s of km) are much further than 
those of the SATs (100s of m). 
Tower-mounted sources can be used in the SAT near/mid-field and 
the LAT near field, while drone-mounted sources can reach the 
SAT far-field and the LAT mid-field.

\textit{Polarized beams:} Polarized sources mounted on a drone can be used for polarized 
beam measurements, e.g., \cite{Nati2017}. The drone can be tethered or coupled with a weather balloon for further 
stability. These measurements could be further supplemented with polarized beam measurements 
from tower-mounted polarized sources. These sources and the tower/drone mounting can 
be also be used for polarization angle calibration. 

\textit{LAT main beams and near sidelobes:}
The LAT main beams and near sidelobes will be calibrated with planets. Uranus and Neptune will 
be used for the main beam down to $\approx-30\,$dB, and current large aperture experiments have used 
Venus, Jupiter, Saturn, and Mars to reach the $\approx-50\,$dB requirement with radial binning. 
While we expect that celestial sources will be sufficient for the LAT calibration 
requirements, tower-mounted thermal sources could be used for deeper measurements of the LAT near-field beams.

\textit{SAT main beams and near sidelobes:}
A combination of planets and artificial sources will be used to calibrate the SAT main 
beams and near sidelobes. Jupiter and Venus can be used to map main beam response 
down to $\approx-30\,$dB. Deeper measurements to the $\approx-50\,$dB requirement will be 
achieved using tower-mounted thermal sources coupled to a far-field flat mirror as was done with BICEP. 
At the South Pole, the planets are low in elevation, so can only be observed if the fixed ground 
shields are designed to be lowered.

\textit{Far sidelobes:} Far sidelobes will be characterized 
using the Sun and Moon in Chile. At the South Pole, tower-mounted sources 
will be used for response above the elevations of the Sun and Moon. 
Strong broadband noise sources offer higher signal-to-noise than thermal sources 
for these measurements. Many far-sidelobe effects can be mitigated with proper 
Sun and Moon avoidance in the scan strategy.

\paragraph{Polarization angle:}
To determine the polarization angle of the instruments, we will use a combination of self-calibration, observations of polarized astronomical sources like Tau~A and Cen~A, and artificial sources. Any polarization angle mis-calibration could give a bias on $r$. For CMB-S4, this means that the polarization angle requirement is $\lesssim0.1^{\circ}$ \cite{TechBookarXiv170602464A}.

\textit{Self-calibration:} In standard models of cosmology, the EB and TB power spectra should be zero. Any mis-calibration in the polarization angle causes non-zero EB and TB signals. Self-calibration assumes null EB and TB spectra, and uses a minimization of these to recover the polarization angle. However, foregrounds can contaminate EB and TB. The extent to which CMB-S4 will be able to use self-calibration is currently under study.

\textit{Astronomical sources:} Tau~A has been measured to $\approx0.33^{\circ}$ and is available to both the SAT and LAT instruments in Chile. The LATs will also be able to observe Cen~A, which has been measured to $\approx1^{\circ}$, from both Chile and the South Pole.

\textit{Artificial sources:}
Given the uncertainties in astronomical sources and the uncertainty in the level of self-calibration we can achieve with foregrounds, we require the artificial sources used for polarization angle calibration to be able to reach the requirement of $\lesssim0.1^{\circ}$. Polarized sources mounted on a tower or a drone can currently achieve $\lesssim0.5^{\circ}$ uncertainties in polarization angle calibration and are limited by the local gravity reference to $\approx20\,$arcsec. However, further refinements and integrating these technologies with a star camera could open the possibility to achieve the required $\approx0.01^{\circ}$ uncertainty, e.g., \cite{Nati2017}. The ideal polarization calibrator would measure the polarization angle in the far-field. A drone would enable polarization angle calibration in the SATs far-field and the LAT mid-field. This means that the SAT polarization angle can be more precisely calibrated with artificial sources than the LAT, but additional precision in the LAT polarization angle can be gained by cross-calibrating the LATs and SATs.

Additional relative (SATs and LATs) and absolute (SAT) polarization angle calibration to supplement these measurements could be achieved with a sparse or full wire grids. 
The balloon-borne Spider experiment used this method to 
achieve $\approx0.2^{\circ}$ precision on absolute polarization angles in pre-flight calibrations.

\paragraph{Pointing:}
We will use observations of Jupiter, Venus, and the Moon in combination with a star camera to determine the SAT pointing model to $<50\,$arcsec. At the South Pole, where astronomical sources are only available at low elevations, the full SAT pointing model will be reconstructed from the star camera calibrations. Because they have higher resolution, the pointing model for the LATs will be derived using the point sources detected during nominal observations to $<15\,$arcsec. The per-detector pointing of both the SATs and the LATs can be further refined by cross-correlating with the CMB.

\paragraph{Time constant:}
The time constant requirements are set by the beam and pointing requirements. 
The effects on the beam and pointing decrease for smaller time constants, 
so faster detectors have less-stringent requirements. 
For TES detectors, optical time constants are a function of optical loading, 
which requires that they be measured regularly.

The optical time constants can be calibrated by analyzing beam-smearing from planet observations.
We will also consider using a chopped thermal calibrator coupled to a small
fraction of the telescope's throughput (as done by SPT), to measure and track
changes in optical time constants.  Electrical bias steps can also be used to monitor changes
in time constant.  

\paragraph{Gain:}
The season-long, focal-plane average detector response can be calibrated by comparison of 
temperature anisotropy or polarization anisotropy maps with those made by Planck.  
The detector-to-detector relative gain 
response and the temporal variation of the gain must be calibrated at the percent level. 
These can be monitored via regular galactic source and
planet calibrations.
A variety of other methods can be used to determine relative gain response, 
including sparse wire grid measurements and elevation nods (modulating the atmospheric signal). 
We will also track the temporal variations in gain with bias steps. 
We will also consider using chopped thermal calibrator source signals as an additional option for 
monitoring temporal gain variations.

\eject

\chapter{Science Analyses}
\label{chap:scienceanalysis}

\begin{figure}
\begin{center}
\includegraphics[width=5in]{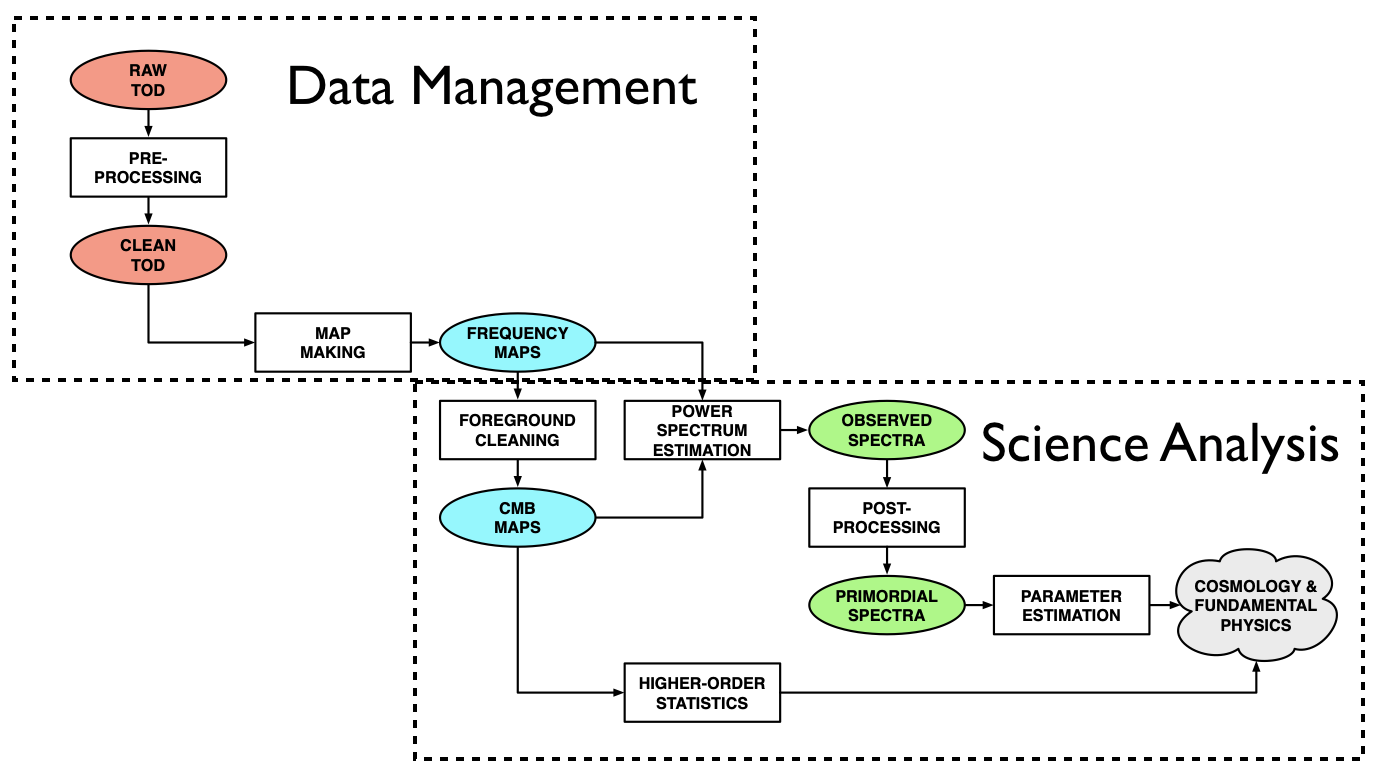}
\end{center}
\caption{Schematic view of the CMB-S4 data analysis pipeline (from figure~88 of the
CMB-S4 Science Book), with boxes illustrating which elements are grouped under
Data Management (Sect.~\ref{sec:datamanagement}) and which elements are grouped under
Science Analyses (this chapter).}
\label{fig:pipeline}
\end{figure}

For the purposes of this document, the analysis of CMB-S4 data is divided into two 
stages: (1) the synthesis of raw detector, pointing, and housekeeping data into 
single-frequency maps of the sky; and (2) further processing of those maps into data products
such as CMB lensing maps, temperature, polarization, and lensing power spectra,
and catalogs of galaxy clusters and emissive sources, and using these downstream
data products to derive scientific results. In this document, the first stage of analysis is
part of the Reference Design under the Data Management subsystem
(Sect.~\ref{sec:datamanagement}), while the second
stage of analysis is referred to as ``Science Analysis'' and is treated in this chapter
(see Fig.~\ref{fig:pipeline} for a graphical illustration of this division).

The vast array of CMB-S4 science goals discussed in Chapter~\ref{chap:science}
will correspondingly necessitate many different types of post-map analyses, and a full
accounting of all these types of analyses and the different options for carrying them out
would fill a 200-page document by itself. Here we limit ourselves to short discussions
of the analyses involved in some of the key CMB-S4 science goals, including the pursuit
of a background of primordial gravitational waves, the search for light relics in the early
Universe, measuring the mass of the neutrinos, constraining the nature of dark energy,
learning about galaxy formation and evolution, and producing legacy catalogs of galaxy
clusters and emissive sources. In the following sections, we describe the major analysis
steps involved in post-map CMB data analysis and how they relate to these primary
science goals, and we discuss major outstanding algorithm choices for certain of the
analysis steps. For much more detail on all of these steps, and on the synthesis of raw
data to maps, see Chapter 8 of the CMB-S4 Science Book \cite{Abazajian:2016yjj}.

\section{Reconstruction of the CMB lensing potential}
\label{sec:analysislensing}

Crucial to nearly all CMB-S4 science goals will be a maximally accurate and precise 
estimate of the CMB lensing potential $\phi$. The power spectrum of the lensing 
potential will be the primary CMB-S4 observable used to constrain the nature of 
dark energy (Sect.~\ref{sec:de}) and the sum of the neutrino masses (Sect.~\ref{sec:mnu}).
A clean and precise estimate of the lensing potential is also crucial to reaching 
CMB-S4 goals on the tensor-to-scalar ratio $r$ describing the amplitude of a background
of primordial gravitational waves (PGW). Delensing, or using the measured (lensed) CMB temperature and
polarization fields and an estimate of $\phi$ to reconstruct the unlensed fields will
be a major part of the PGW analysis pipeline, but it is also
a promising avenue to improving constraints from the primary CMB temperature and 
$E$-mode polarization power spectra, such as the number of light relic species in the early
Universe and the shape of the primordial density perturbation spectrum.

The current state of the art in lensing reconstruction from CMB data is the quadratic estimator 
method \cite{Hu:2001kj}, \cite{Okamoto:2003zw}.
These quadratic estimators are the first step in an iterative estimation of the true likelihood, and in
the weak-lensing limit they are nearly optimal. To harness the full power of CMB-S4 lensing information,
however, it will likely be necessary to develop lensing estimators that more closely 
approximate the maximum likelihood solution. Development of maximum-likelihood algorithms
or their equivalent is underway \cite{Millea:2017fyd,Caldeira:2018ojb}, but they have yet to
be demonstrated on large CMB data sets.

Another outstanding question, particularly for delensing, is what information to include in 
the $\phi$ estimate. The baseline assumption in the CMB-S4 Science Book and in many other
publications is that the $\phi$ reconstruction for CMB-S4 will come entirely from the combination
of CMB $E$-mode and $B$-mode polarization, from the so-called $EB$ estimator. This is partially because
at CMB-S4 noise levels the $EB$ estimator has the lowest noise, but it is also because any
estimators involving temperature are potentially contaminated by foregrounds, particularly 
at high multipoles. But there is still significant information to be gained from the temperature field, 
even at CMB-S4 noise levels (cf.\ figure~43 in Ref.~\cite{Abazajian:2016yjj}), and recent work has 
hinted at ways to mitigate foreground contamination in temperature-based lensing estimators
\cite{Madhavacheril:2018bxi}. As shown in Ref.~\cite{Manzotti:2017oby}, information from tracers of $\phi$
other than the CMB can also improve delensing results, even at CMB-S4 noise levels. One 
analysis challenge will be to build the optimal combination of all these $\phi$ estimators that 
is also robust to foreground contamination.

\section{Power-spectrum estimation}

\subsection{Power-spectrum estimation methods}

Early measurements of CMB temperature anisotropy, with comparatively few map pixels or angular modes
measured, often used maximum-likelihood methods to produce maps of the sky (e.g., Ref.~\cite{Wright:1996dk}) and
either a direct evaluation of the full likelihood or a quadratic approximation to that likelihood (e.g., Ref.~\cite{Bond:1998zw}) to go from 
maps to angular power spectra. With the advent of the \wmap\ and \planck\ space 
missions, which would map the entire sky at sub-degree resolution, it became apparent that computing
resources could not compete with the $\mathcal{O}({\cal N}^{\,3})$ scaling of the full-likelihood approach 
(e.g., Ref.~\cite{Borrill:1998tn}). The solution for power spectrum analysis
that has been adopted by most current CMB experiments is a
Monte-Carlo-based approach advocated in Ref.~\cite{Hivon:2001jp}. In this approach, a biased estimate of
the angular power spectrum of the data is obtained by simply binning and averaging the square 
of the spherical harmonic transform of the sky map. That estimate (known as the 
``pseudo-$C_\ell$ spectrum'') is related to the unbiased 
estimate that would be obtained in a maximum-likelihood procedure through the combined effect
of noise bias, sky windowing, and any filtering applied to the data before or after mapmaking
(including the effects of instrument beam and pixelization). These effects are estimated by ``observing''
and analyzing simulated data and constructing a matrix describing their net influence on simulated data. 
This matrix is inverted, and the inverse matrix is applied to the pseudo-$C_\ell$s to produce the 
final data product. Some version of this Monte-Carlo treatment is likely to be 
adopted for CMB-S4. 

\subsection{Contamination}

At CMB-S4 noise levels, as important as raw sensitivity will be control of contamination to 
power spectrum measurements. This is especially true in the PGW-targeted analysis of degree-scale
$B$-mode polarization, in which the signal of interest is known to be significantly smaller than 
the two main sources of contamination, Galactic foregrounds and lensed $E$-mode power
(e.g., Ref.~\cite{Ade:2018gkx}). Additional sources of contamination to $B$-mode searches are intrinsic
to observing a finite patch of sky and to pseudo-$C_\ell$ methods, including ambiguous (not clearly
$E$ or $B$) modes around the edges of the sky patch and spurious $B$ modes introduced in analysis.
Mitigation strategies have been developed for all of these contaminants, but it remains to be seen
exactly which methods will be most appropriate for CMB-S4 data. 

To separate the CMB signal from the contaminating foregrounds, 
data from multiple bands will be combined, either 
in a cross-spectrum analysis or by making linear 
combinations of maps in different bands to produce a ``pure-CMB'' map for power spectrum estimation.
In either case, an underlying model of foreground behavior is assumed---even if that model is simply
an assumption regarding the level to which the spectral behavior of foregrounds varies over the sky.
There are two challenges related to uncertainties in foreground modeling: one statistical and one
systematic. The statistical issue is simply how to propagate the statistical uncertainty on the foreground 
model to uncertainties on cosmological parameters. In explicitly parameterized foreground models, 
this happens automatically through the covariance resulting from the fit. For non-parametric models,
this covariance can be assessed through Monte-Carlo methods, but making many independent 
realizations of large-scale Galactic foregrounds is problematic because of the strongly non-Gaussian
behavior of these foregreounds.

The primary question for 
delensing and dealing with spurious or ambiguous $B$ modes is whether to subtract a 
statistical estimate of their contribution to the $B$-mode power spectrum or to attempt 
to construct a phase-ful estimate of the realization of contamination in the actual data.
The latter strategy is in general leads to smaller residuals (because there is no sample
variance in the estimate of contamination) but is more computationally intensive. For
delensing, it has been determined that for a small-area survey which measures a 
comparatively small number of sky modes to high precision, by-realization cleaning
is mandatory, and all the delensing techniques described in the previous chapter are 
in this category. Spurious/ambiguous $B$ modes can be dealt with either by estimating 
the statistical bias to the final $B$-mode spectrum or by constructing
a matrix representing the effect of any analysis steps on the true sky \cite{Ade:2014xna}. The latter
approach involves constructing an ${\cal N}_\mathrm{pixel}$-by-${\cal N}_\mathrm{pixel}$ matrix, equal in size to the 
full pixel-pixel covariance, and will not be feasible for high-resolution CMB-S4 data but could be 
used in analyzing lower-resolution data.

Perhaps the most important aspect of any contamination mitigation is the estimate of the 
residual contamination after cleaning. If this estimate is inaccurate, it will lead directly to biases
in the final determination of the amplitude of the cleaned power spectrum. In the case of PGW
searches, this leads directly to a bias in the tensor-to-scalar ratio $r$. To avoid this bias, 
all of the contamination-mitigation
methods discussed above will need to effectively marginalize over uncertainties in our knowledge
of the contamination sources and mechanisms.

\section{Component maps and cross-correlation with other surveys}

It has become increasingly clear, even since the publication of v1 of the CMB-S4 Science Book,
that a wealth of exciting science will come from the combination of CMB-S4 data with survey data
from other instruments at other wavelengths. These contributions will come both in the area of 
constraints on cosmological parameters and in increased knowledge of processes on smaller 
scales and in our local Universe. As shown in Chapter~\ref{chap:science}, cross-correlations
with optical data in particular can lead to improved constraints on the growth of structure, the 
sum of the neutrino masses, the nature of dark energy, and the primordial power spectrum, as 
well as informing our understanding of galaxy formation and evolution.
Fully exploiting this area of science will require building analysis pipelines jointly with experts in
data from other surveys. 

Among the challenges and questions in this area of analysis will be
whether to perform the analyses in real or harmonic space, how to estimate the covariance 
of the cross-correlations, and, as in CMB power spectrum analyses, how to deal with contaminating
signals and estimate the residual contamination after cleaning. This last point will be key to 
obtaining precise and accurate determinations of cosmological parameters and constraints on 
galaxy feedback models through correlations of optical observables with maps of CMB lensing
and the tSZ and kSZ effects. In particular, contamination from the tSZ is currently the limiting
factor in correlation analyses of optical data and CMB lensing \cite{Abbott:2018ydy,Baxter:2018kap}, so the foreground
mitigation schemes discussed in Sect.~\ref{sec:analysislensing} are also critical to cross-correlation.
In general, foreground treatment will be a key aspect of the production of component maps
such as the Compton-$y$ map and the CMB blackbody map (see Sect.~\ref{sec:componentmaps}). As with
power spectrum estimation, it will be crucial not only to minimize the contamination in the 
component maps but also to have accurate estimates of the residual contamination and to 
propagate these estimates through any cross-correlation analysis involving component maps.

\section{Parameter estimation}

The final step in the analysis of a CMB data set is the estimation of cosmological parameters from
the various post-map statistics discussed above.
This involves estimating the likelihood of the data
given a model parameterized by the standard six $\Lambda$CDM parameters, possible extensions
of the cosmological model, and any nuisance parameters involving the instrument, foregrounds, and
other sources of systematic uncertainty. The current industry standard for this part of the analysis are
Markov-chain Monte-Carlo (MCMC) methods, in particular the implementation in CosmoMC
\cite{Lewis:2002ah}, and it is expected that CMB-S4 will use similar methods.

\section{Creating source and cluster catalogs}

An additional post-map product of interest for CMB-S4 is the location and properties of compact
sources, in particular clusters of galaxies identified through the thermal SZ effect. The standard 
practice for extracting SZ clusters from multifrequency millimeter-wave maps is through the application
of a Fourier-domain spatial-spectral filter \cite{Melin:2006qq}.
The computational effort involved in this step is small compared to the estimation of power spectra 
and higher-order correlations, and the algorithms are well-developed and fully implemented for 
multi-frequency data sets (e.g., Ref.~\cite{Ade:2013skr,Bleem:2014iim})---however, the cluster density
could be high enough in CMB-S4 data that approaches more sophisticated than the simple matched
filter (e.g., Ref.~\cite{Pierpaoli:2004bp}) could be required to maximize cluster yield.

\section{Transient and time-domain analyses}

One class of analyses that is not represented in Fig.~\ref{fig:pipeline} and does not 
strictly come under the heading of post-single-frequency-map analysis is the area of 
time-domain and transient astronomy. As discussed in Sects.~\ref{sec:solar_system} and \ref{sec:transient}, CMB-S4
will be a rich data set for detecting mm-wave GRB afterglows, monitoring the light
curves of AGN, and discovering new planets. These science goals will require analyses of
the time-ordered detector data that are very different from the standard pipeline that will
create single-frequency maps of full-depth data. In particular for GRB afterglows, science
yield will be maximized by a quasi-real-time alert system linked to the transient alert 
mechanisms in the wider community. This will require on-site computing and analysis
software that runs autonomously.

\section{Sky simulations}

Simulations will be a key aspect of the entire CMB-S4 analysis pipeline, both in the stages
handled by the data management subsystem and in post-map analyses. As discussed
in Sect.~\ref{sec:datamanagement}, we assume here that the software for synthesizing
mock skies into simulated time-ordered detector data and maps is part of the data 
management subsystem, but that the creation of the simulated skies themselves will occur 
outside of the data management sub-system. 
Sky simulations that will be useful for CMB-S4 fall into two main categories: (1) simulations 
of the polarized mm-wave emission from our own galaxy; and (2) simulations of the (mostly) 
unpolarized emission from extragalactic sources. These present different challenges and
will likely come from independent sources.

Comprehensive, accurate, and flexible simulations of Galactic emission will be critical
for reaching the CMB-S4 goals in the search for primordial gravitational waves.
Key challenges in the generation of simulations of Galactic emission include:
\begin{itemize}
\item the level of coherence of diffuse emission across observing frequencies, as any decoherence will limit the efficacy of cleaning a foreground from one observing band using the measurement in a different band;
\item the existence or not of a simple parametric emission law for each component emission, such as power laws (for synchrotron) or modified blackbody emission (for dust components);
\item the absolute level of foreground emission (in particular for those components that do not scale simply as a function of frequency, such as the superposition of many individual sources with a specific emission law each);
\item whether or not emissions for which the level of polarization is unknown or unclear 
must be modeled and treated for \cmbexp\ or can be safely neglected;
\item The level at which foregrounds can be treated as Gaussian random fields, which is an assumption of certain foreground cleaning approaches.
\item the reliability of models based on observations at angular resolution lower than that of CMB-S4, integrated in broad frequency bands, and with a sensitivity limit at least an order of magnitude worse than what will be achieved with CMB-S4.
\end{itemize}

Simulations of extragalactic emission will be important for estimating contamination to 
small-scale power spectra and lensing maps, and they will play a crucial role in validating
analysis methods and verifying results in the area of cross-correlation with surveys at other
wavelengths. This requires simulated maps at many wavelengths with the same underlying
initial conditions and large-scale structure fluctuations, and with the proper correlations between
observables such as CMB lensing potential, tSZ and kSZ, galaxy positions, and galaxy 
weak-lensing shear. A key challenge in the area of extragalactic simulations will be the 
competing needs of high accuracy at small angular scales---which can only be achieved 
with full hydrodynamical simulations---and the ability to create quickly many realizations with different
cosmologies and galaxy-formation parameters.

\section{Implementation}

As described in Sect.~\ref{sec:datamanagement}, the volume of raw, time-ordered data in CMB-S4 will
necessitate both the provisioning of unprecedented computing resources and innovation in 
CMB data processing algorithms. Once the raw data have been reduced to single-frequency
maps, however, the data volume for CMB-S4 will no longer be significantly large than that 
of \planck. Implementation of the standard methods in the field for post-map processing is 
not expected to overly tax the computing resources of the field. If minimum-variance methods
are used for certain post-map analysis steps---for instance the simultaneous estimation of 
the CMB lensing potential, the unlensed fields, and cosmological parameters as described
in \cite{Millea:2017fyd}---or if full pixel-pixel matrix methods are used for spurious/ambiguous 
$B$-mode cleaning, the requirement for computing resources could be higher.

\section{Validation and verification}

The importance of the key CMB-S4 science goals and the challenge of not only achieving
the necessary sensitivity but also demonstrating that the final signal is not significantly 
affected by contamination or bias, compels us to set up a framework in which we can validate
any software used for the key science analyses and verify that the algorithms and implementation
thereof produce unbiased results with the expected precision on simulated data. This process
is already happening in the PGW forecasting working group through a series of data challenges,
and we expect to extend the data challenge framework to key results from the large-area, 
large-aperture surveys in the near future.

\eject

\chapter{Project Overview  \prelim{ ({\it J. Yeck})} }  
\label{chap:project}

\section{Introduction}

The CMB-S4 Collaboration and a pre-Project Development Group of experienced project professionals jointly contributed to the development of a Work Breakdown Structure (WBS), Organization, Cost Book, Resource Loaded Schedule, and Risk Registry.  The reference design and project baseline prepared for this document is the basis for subsequent design and project development work to be completed by the Interim Project Office and the Collaboration during 2019--2020.  A permanent Integrated Project Office will be established in 2020 to manage the construction phase which starts in 2021.  

The CMB-S4 project total estimated cost is currently \$591.6M (fully loaded and escalated to the year of expenditure) including a 35\% contingency budget.  In-kind contributions delivered by Private and International partners are expected and would reduce the total cost to NSF and DOE.  Critical R\&D is in progress, funded by DOE. There is a pending proposal to NSF for Design and Development support under the NSF MSRI-R1 program. An eight-year construction project is anticipated (2021--2029) with a transition to operations starting with completion of commissioning of the 1st telescope in 2026.  Long lead procurements for the construction project will start in 2021.

\section{Scope, work breakdown structure, and cost}

The CMB-S4 Work Breakdown Structure (WBS) includes twelve major categories as shown in Table 6.1.  The  distribution of WBS elements by funding agency will be discussed with the Joint Coordination Group established by the funding agencies.  As these plans mature there will be a general understanding of the proposed responsibilities of each funding agency and partners providing in-kind contributions.  The scope distribution will leverage the capabilities of universities, national laboratories, and industry.

The cost estimate shown in Table~\ref{tbl:wbs_cost} includes all Materials \& Services (M\&S) and Salaries, Wages and Fringe Benefit (SWF) costs for the project.  The M\&S costs and labor resources are estimated at the lowest (task) level in the Project Schedule.   The costs in the schedule are given in FY19 dollars. Appropriate overhead and escalation are done external to Primavera, within the Cobra Project cost management and reporting software that will eventually be used to compute earned value.  It is foreseen that all Project tracking and reporting will be done using Cobra and Primavera software for the duration of the Project.  The cost estimate is the full cost, i.e., does not take credit for contributions from collaborating institutions supported by private and international partners, e.g., Large Aperture Telescopes currently under construction in Chile as part of the Simons Observatory, and Large and Small Aperture Telescopes proposed by international collaborators.  The value of in-kind contributions could reduce the total cost of the CMB-S4 project by 20--25\%.

Scientific labor resources not in project management roles, i.e., scientific and technical development work, are provided to the Project through research program support outside of the Project. This is the traditional funding model used for DOE and NSF joint projects including LSST, ATLAS, and CMS.  The approach equalizes the treatment of the collaboration's experimental physicists with respect to their cost to Projects, regardless of their funding source. 

The cost contingency estimate was constructed using input from subject matter experts with previous experience in previous CMB experiments and similar NSF MREFC projects and DOE MIE projects.  The current estimate of contingency budget or reserve is 35\% of the base cost estimate.  As the design, cost estimates, and schedules mature the contingency as a percentage of the base cost estimate is expected to decrease to 30\% or less.  The target range for the start of the CMB-S4 construction project is 25--30\%.

The contingency estimate was compared to similar large research infrastructure projects sponsored by the DOE and NSF and is reasonably consistent with other projects at this early stage of project development.

\begin{table}[htp]
\begin{center}
\begin{tabular}{|l|r|}
\hline
\multicolumn{1}{|c|}{WBS Level~2 Element} & \multicolumn{1}{|c|}{Total Cost (\$M)} \\
\hline
\multicolumn{2}{|c|}{Total Estimated Cost (TEC)} \\
\hline
1.01 -- Project Management & 19.6 \hspace*{0.35in} \\
1.03 -- Detectors & 39.5 \hspace*{0.35in} \\
1.04 -- Readout & 59.9 \hspace*{0.35in} \\
1.05 -- Module Assembly \& Testing & 31.8 \hspace*{0.35in} \\
1.06 -- Large Aperture Telescopes & 86.5 \hspace*{0.35in} \\
1.07 -- Small Aperture Telescopes & 52.3 \hspace*{0.35in} \\
1.08 -- Observation Control \& Data Acquisition & 13.9 \hspace*{0.35in} \\
1.09 -- Data Management & 26.9 \hspace*{0.35in} \\
1.10 -- Chile Infrastructure & 38.1 \hspace*{0.35in} \\
1.11 -- South Pole Infrastructure & 37.0 \hspace*{0.35in} \\
1.12 -- Integration \& Commissioning & 7.7 \hspace*{0.35in} \\
\hline
Direct TEC & 413.2 \hspace*{0.35in} \\
TEC Contingency (35\%) & 144.6 \hspace*{0.35in} \\
Total TEC & 557.9 \hspace*{0.35in} \\
\hline
\hline
\multicolumn{2}{|c|}{Other Project Cost (OPC)} \\
\hline
1.01 -- Project Management & 7.0 \hspace*{0.35in} \\
1.02 -- Research \& Development & 24.2 \hspace*{0.35in} \\
\hline
Direct OPC & 31.2 \hspace*{0.35in} \\
OPC Contingency (35\%) -- excludes R\&D & 2.5 \hspace*{0.35in} \\
Total OPC & 33.7 \hspace*{0.35in} \\
\hline
\hline
\multicolumn{2}{|c|}{Total Project Cost (TPC)} \\
\hline
TEC + OPC with contingency & 591.6 \hspace*{0.35in} \\
\hline
\end{tabular}
\caption{CMB-S4 WBS structure and cost.}
\label{tbl:wbs_cost}
\end{center}
\end{table}

The resource-loaded schedule determines the annual funding profile.  The schedule is technically driven starting in 2022, i.e., the technically efficient schedule unconstrained by funding in any year.  The resulting funding profile is provided in the following table.

\begin{table}[htp]
\begin{center}
\begin{tabular}{|c|r|r|r|r|r|r|r|r|r|r|c|}
\hline
         & FY19 & FY20 & FY21 & FY22 & FY23 & FY24 & FY25 & FY26 & FY27 & FY28 & Total \\
\hline
OPC & 4.3 & 11.4 & 17.9 & 0.0 & 0.0 & 0.0 & 0.0 & 0.0 & 0.0 & 0.0 & 33.6 \\
TEC & 0.0 & 0.0 & 7.1 & 80.8 & 127.7 & 184.0 & 93.5 & 50.6 & 13.5 & 0.5 & 557.9 \\
TPC & 4.3 & 11.4 & 25.0 & 80.8 & 127.7 & 184.0 & 93.5 & 50.6 & 13.5 & 0.5 & 557.9 \\
Operations & \dots & \dots & \dots & \dots & \dots & \dots & \dots & 1.0 & 5.0 & 10.0 & 16.0 \\
\hline
\end{tabular}
\caption{CMB-S4 funding profile.}
\label{tbl:profile}
\end{center}
\end{table}

\subsubsection{WBS dictionary}

The WBS Dictionary defines scope of each CMB-S4 WBS element as described in Table~\ref{tbl:wbs_dict}.

\begin{table}[htp]
\begin{center}
\begin{tabular}{|p{2.25in}|p{4in}|}
\hline 
WBS Element & Description \\
\hline
1.01 -- Project Management & Labor, travel, and materials necessary to plan, track, organize, manage, maintain communications, conduct reviews, and perform necessary safety, risk, and QA tasks during all phases of the project. Overall project Systems Engineering is a subsection of this wbs element. However, subsystem-related management and support activities for planning, estimating, tracking, and reporting as well as their specific EH\&S and QA tasks are included in each of the subsystems. \\
\hline
1.02 -- Research \& Development & Labor, travel and materials necessary to support development of a Conceptual Design (pre CD-1 for DOE).  Activities include design of detector wafers and the readout systems, data management, optical design, and aspects of the cryostat design for both large and small telescopes. \\
\hline
1.03 -- Detectors & Labor, materials, and equipment associated with the design, fabrication and testing of the detector wafers. \\
\hline
1.04 -- Readout & Labor, materials, and equipment associated with the design, fabrication and testing of the detector readout system. \\
\hline
1.05 -- Module Assembly \& Testing & Labor, materials, and equipment associated with the design, parts fabrication, assembly and testing of the detector modules. \\
\hline
1.06 -- Large Aperture Telescopes & Labor, materials, and equipment associated with the design, prototyping, materials selection, construction and certification for the Large Aperture Telescope (LAT) System.  \\
\hline
1.07 -- Small Aperture Telescopes & Labor, materials, and equipment associated with the design, prototyping, materials selection, construction and certification for the Small Aperture Telescope (SAT) System. \\
\hline
1.08 -- Observation Control \& \;\;\;\;\;\;\;\;\;\;\;\; \mbox{\;\;\;\;\;\;\;\;\; Data Acquisition} & Labor, materials, and equipment associated with the design, construction, certification, and delivery of the control systems for the observatories and data acquisition. \\
\hline
1.09 -- Data Management & Labor, materials, and equipment associated with the design, construction, certification, and delivery of the data management system. \\
\hline
1.10 -- Chile Infrastructure & Labor, travel, and materials necessary to plan, track, manage, maintain communications, conduct reviews, and perform necessary safety monitoring on site including management of all shipping of CMB-S4 components to Chile and oversite of construction activities on site. \\
\hline
1.11 -- South Pole Infrastructure & Labor, travel, and materials necessary to plan, track, manage, maintain communications, conduct reviews, and perform necessary safety monitoring on site including management of all shipping of CMB-S4 components to the South Pole and oversite of construction activities on site. \\
\hline
1.12 -- Integration \& Commissioning & On Site Integration and Commissioning of the CMB-S4 telescopes and infrastructure in Chile and the South Pole. \\
\hline
\end{tabular}
\caption{WBS dictionary.}
\label{tbl:wbs_dict}
\end{center}
\end{table}

\subsubsection{Scope management plan}

The scope of the CMB-S4 project is defined to meet the scientific and technical requirements.  The reference design and emerging conceptual design meets these requirements as demonstrated by the flow down of science requirements to the proposed project scope.  The delivery of the project scope will be time phased and elements of the project scope will be added and deleted as necessary and in accordance with the configuration management plans.  Changes to the project scope will be subject to formal change control procedures and approved at the appropriate level of authority.  For example, minor changes can be approved by the Level~2 or Level~3 managers, major changes by the Project Director and/or Project Manager, and finally changes affecting the overall scientific performance or key performance parameters will need to be approved by the funding agencies and partners.

Scope contingency options will be identified and managed including the timing of decision points, both for scope reductions and scope restoration or scope additions.  A scope management plan will be developed describing this process.

\subsection{Organization}

CMB-S4 is both a scientific collaboration and a nascent DOE/NSF project. While these are certainly tightly-coupled, they do have different roles and responsibilities; the overall organization of CMB-S4 therefore decouples into the organization of the collaboration and the project.

\subsubsection{Collaboration organization}

During the 2013 Snowmass particle physics project planning exercise, the US CMB community came together and recognized:
\begin{enumerate}
\item that realizing the extraordinary scientific potential of the CMB would require an increase in the sensitivity of our instruments corresponding to moving from tens to hundreds of thouands of detectors;
 \item that the cost of an experiment of this scale would limit us to a single instance, in contrast to the long history of having multiple completing experiments at any epoch;
 \item that the challenges of fielding such an experiment would primarily be associated with scaling existing technologies (in hardware and data management) to unprecedented levels; and
 \item that meeting these scaling challenges would require adding the unique capabilities of DOE laboratories to the long-standing NSF program.
\end{enumerate}
The community therefore proposed CMB-S4 to the 2014 Particle Physics Project Prioritization Process (P5) as single, community-wide, experiment, jointly supported by DOE and NSF. After P5 recommended CMB-S4 under all budget scenarios, in 2015 the CMB community started holding biannual workshops---open to CMB scientists from around the world---to develop and refine the concept. At the request of DOE and NSF, in late 2016 the Astronomy and Astrophysics Advisory Committee (AAAC) convened a Concept Definintion Taskforce (CDT) to conduct a CMB-S4 concept study, and the resulting report was unanimously accepted in late 2017. 

\begin{wrapfigure}{r}{0.55\textwidth}
\centering
\includegraphics[width=0.5\textwidth]{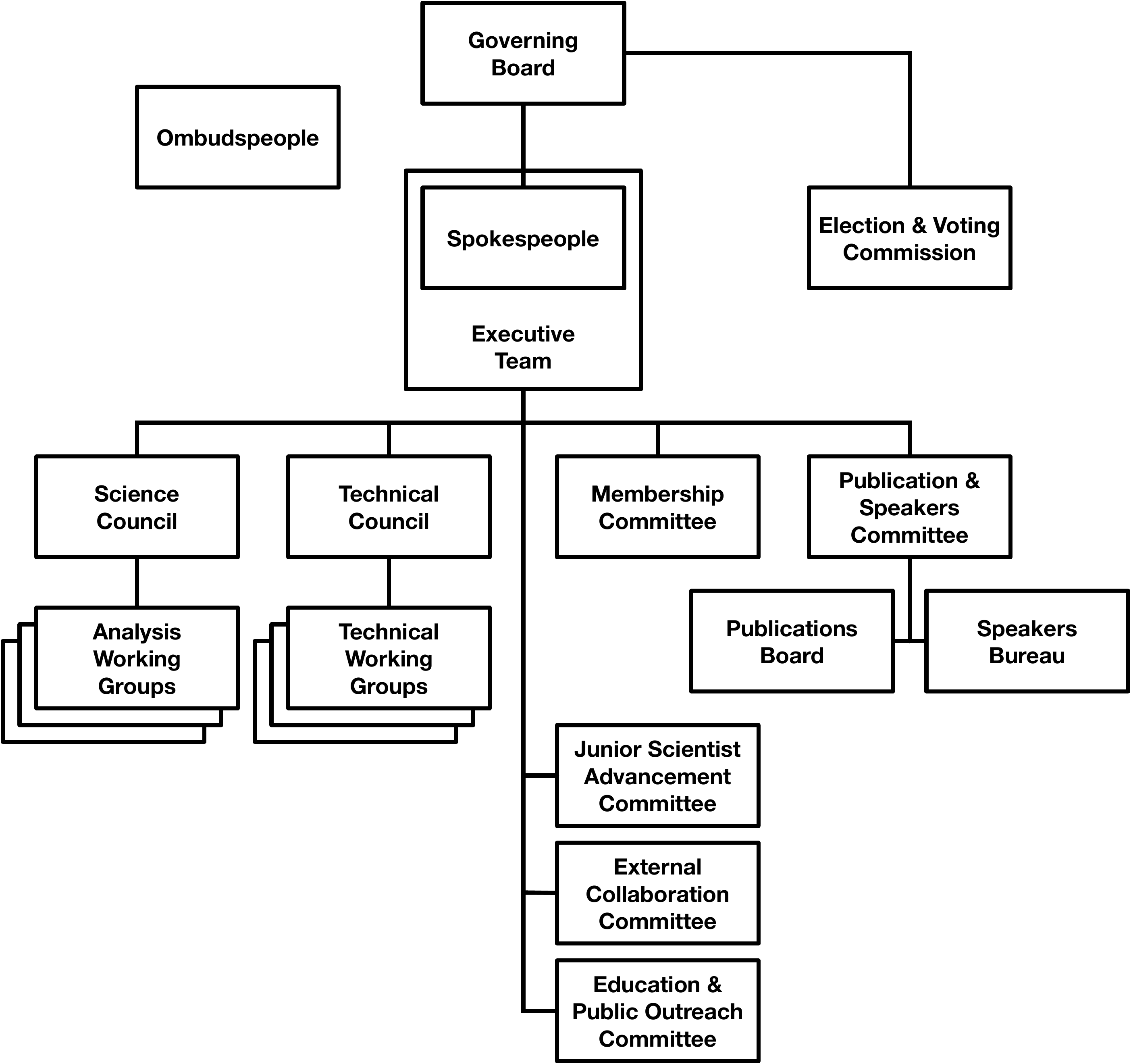}
\vspace*{0.1in}
\caption{Organizational chart of the CMB-S4 collaboration.}
\label{fig:collab_org}
\end{wrapfigure}

One recommendation of the CDT report was that the community should organize itself into a formal collaboration, and an Interim Collaboration Coordination Committee was elected to coordinated this process. The resulting draft bylaws were refined at the Spring 2018 community workshop, and overwhelmingly ratified on March 19th 2018, bringing the CMB-S4 collaboration into being, and the first elections for the various officers of the collaboration were completed by the end of April 2018.

Figure~\ref{fig:collab_org} shows the organizational structure of the CMB-S4 collaboration. The Governing Board sets policy and provides oversight to an Executive Team led by two equal co-Spokespersons which is responsible for the day-to-day management of the collaboration. A number of Councils, Committees, and Working Groups then carry out the necessary work to enable the overall scientific goals of the collaboration. As of summer 2019 the collaboration has 198 members, 71 of whom hold positions within the organizational structure. These members represent 11 countries on 4 continents, and 76 institutions comprising 16 national laboratories and 60 universities. It should be noted that collaboration members from both national laboratories and universities are engaged in the entire scope of work, and that US institutional affiliation does not automatically map to specifically DOE or NSF scope.

\subsubsection{Project organization}

The CMB-S4 project organization including lines of authority, communication, oversight and advisory committees will be consistent with the expectations for large research infrastructure projects sponsored by the NSF and DOE.  This includes a core project office reporting to institutions directly accountable to NSF and DOE for the successful delivery of the CMB-S4 project.  The lead institutions will establish oversight committees and will work with the project office to ensure the necessary advisory committees are effective.

The general approach is to develop and deliver the CMB-S4 project with an Integrated Project Office and project organization established by the institutions leading the NSF CMB-S4 MREFC project and the DOE CMB-S4 Major Item of Equipment (MIE) project.  The lead NSF institution will be the NSF partner for the MREFC Cooperative Agreement, and the lead DOE institution will be a DOE M\&O contractor (National Laboratory) identified by DOE as the lead integrator for the MIE project.

The organization and management approach will adhere to the following principles:

\begin{enumerate}
\item one experiment undertaken by a single collaboration and run as one project;
\item joint NSF and DOE oversight and management, currently a Joint Coordination Group (JCG), with a lead agency to be defined prior to approval of the integrated project baseline;
\item lead institutions accountable to NSF and DOE for the MREFC and MIE projects, respectively;
\item lead institutions establish a project governance agreement that engages all major institutional partners, e.g., an Integrated Project Steering Committee (IPSC) comprised of the major institutions delivering project scope including universities, DOE labs, and private and international partners;
\item a single Integrated Project Office established by the lead institutions and their oversight council, e.g., the IPSC, with clear reporting lines to NSF and DOE;
\item clarity in the NSF and DOE scope of work to ensure direct lines of accountability to each agency and clearly defined management interfaces;
\item integrated Project Office prepared agreements for contributions by private and international partners; and,
\item common management systems that meet the needs of both funding agencies, e.g., project cost and schedule tools, systems engineering processes, etc.
\end{enumerate}

\begin{figure}[htbp]
\begin{center}
\includegraphics[width=0.98\textwidth]{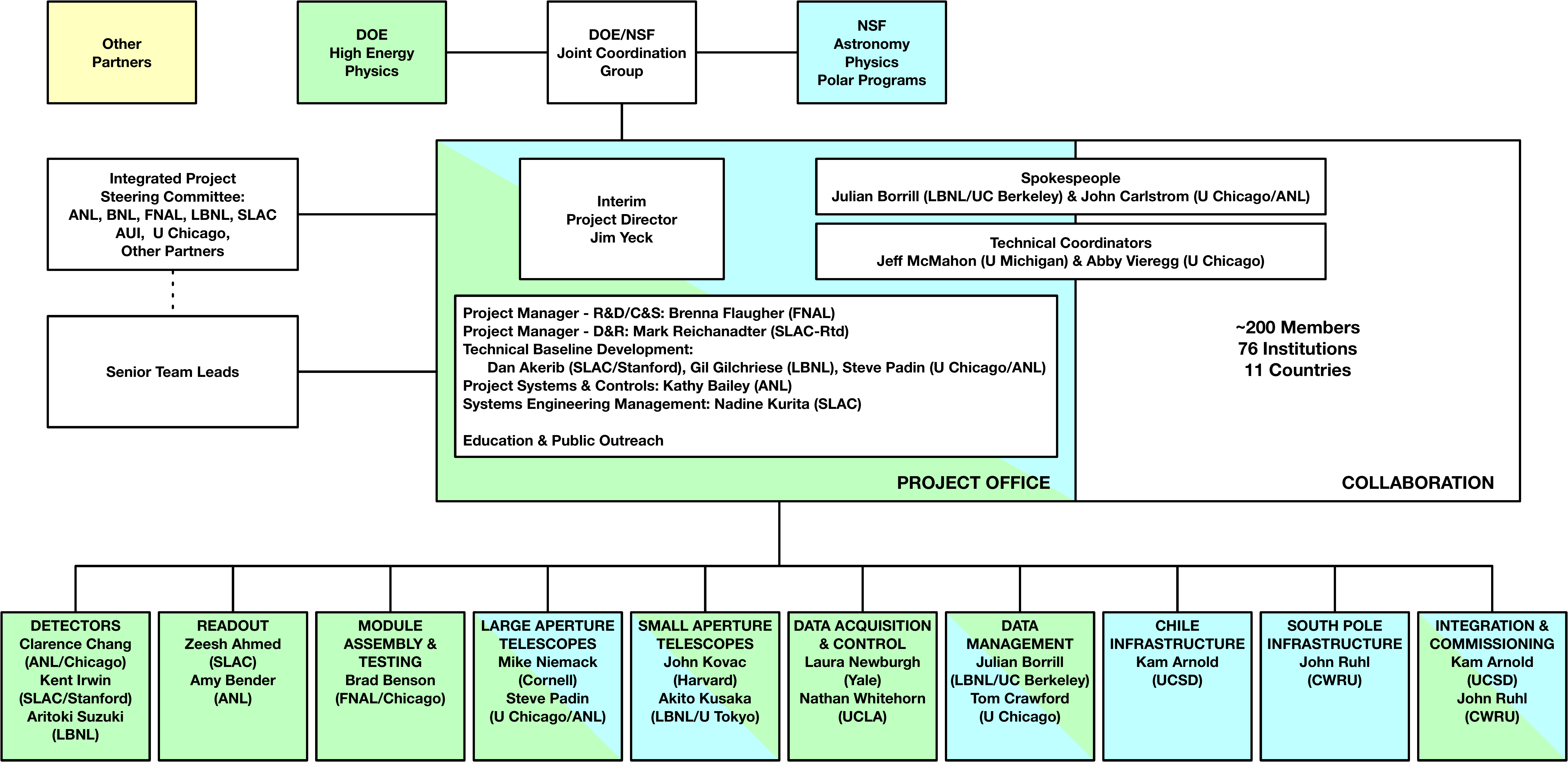}
\caption{Organizational chart of the interim project office.  The figure includes a notional distribution of project scope by funding agency (NSF = blue, DOE = green, Other = yellow).  We are actively pursuing partners who could make significant scope contributions in areas aligned with their expertise.}
\label{fig:prj_org}
\end{center}
\vspace {-10pt}
\end{figure}

Prior to the establishment of the lead institutions, an Interim Project Office (see Fig.~\ref{fig:prj_org}) will coordinate the development of management plans that adhere to the principles described above and prepare for the project execution phase.  The Interim Project Office will report monthly progress on project development efforts and R\&D activities to the Joint Coordination Group established by NSF and DOE.  The Interim Project Office is led by the Interim Project Director.  The transition to the construction phase will include the appointment of a permanent management structure.

The Interim Project Office, with the support of the Collaboration, will continue to mature the experiment design and project execution plans during 2019--2020.  The Interim Project Office will transition into a permanent Integrated Project Office in 2020 to prepare for the construction phase which starts in 2021.  A key feature of the organization is the role of collaboration members in the Project Office, primarily as leaders of the Level~2 systems.  The Level~2 managers are supported by engineering and project management professionals.  The NSF/DOE scope distribution will promote the engagement and participation of universities and national laboratories.  Graduate students, postdocs, professional technicians and engineers are expected to be involved in all aspects of the project.

\subsubsection{Private and international partners}

The CMB-S4 Project is a collaborative project, with the scientific Collaboration members serving in technical leadership roles in the project.  This is similar to many successful projects including IceCube, ATLAS, and CMS.  The lines of accountability for project delivery are clearly defined within the project organization.  The project office is responsible for forming partnerships with key stakeholder institutions including DOE National Laboratories, universities, and potential collaborating observatories/projects such as the Simons Observatory, South Pole Observatory, and the CCAT-prime project.  Partnerships are also expected to include foreign institutions participating in the CMB-S4 Science Collaboration and contributing to the CMB-S4 Project.

\subsubsection{In-kind contributions}

The CMB-S4 project is expected to include significant contributions from collaborating institutions supported by funding agencies other than NSF and DOE.  These ``in-kind'' contributions will be defined as deliverables to the project.  The collaborating institutions agree to deliver items, e.g., instrumentation and effort, required for the success of the CMB-S4 project.  The actual cost of each item is the responsibility of the collaborating institution providing the In-kind deliverable.  The current best estimate of the value of in-kind contributions is 20--25\% or the total cost of the project.  This includes both existing infrastructure, telescopes currently under construction, and telescopes and instrumentation proposed by international collaborators.

The CMB-S4 Project Director (PD) is responsible for ensuring the successful delivery of all in-kind contributions required for the CMB-S4 project.  The PD is supported in this role by the NSF, DOE, international funding agencies, private foundations, the collaboration and their elected spokesperson, the host institutions, and the Project Office and Level~2 and Level~3 managers. The management approach engages all of these parties in the process of defining and ensuring the delivery of in-kind contributions from partners. 

In-kind contributions will be defined in Memoranda of Understanding (MoU) and/or Statements of Work (SoW) executed between the Project Office and the contributing institutions.  The MoUs or SoWs define the in-kind contributions for each collaborating institution including the schedule milestones for the institution's deliverables.  The Project Director is responsible for sign off on the in-kind deliverables. 

Milestones for the in-kind deliverables will be defined in the CMB-S4 project schedule.  Progress against these milestones is evaluated during monthly schedule reviews by the Project Office and Level~2 managers.  Large deliverables are the terminal milestone for a sequence of lower level milestones.  This procedure includes determining the completion forecast for each milestone and taking corrective action when needed.
 
The project management approach is the same for the entire project: in-kind scope delivered by partners or scope supported by NSF or DOE.  The only difference is in the tracking of actual costs, in-kind actual costs are tracked by the collaborating institution providing the in-kind deliverable and are not tracked by the Project Office.

\section{Cost, schedule, and risk}

The project has developed a task based detailed resource loaded schedule which was reviewed by an external panel of experts in December 2018.  The estimate follows the guidance in the NSF Large Facilities Manual, NSF 17-066 and the Project Management for the Acquisition of Capital Assets, DOE 413.3b.

For the task-based estimate, tasks were defined at the lowest level elements of the WBS. The entire ensemble of tasks represents all the required resources, activities, and components of the entire project. Each of the tasks are scheduled and estimated by the teams using accepted techniques. The estimates are documented with a Basis of Estimate (BOE) developed by the subsystems leads and stored in a set of google documents which can be ingested into the PrimaveraTM scheduling program.  To facilitate proper integration into the CMB-S4 project control system, standard PrimaveraTM layouts are used to enter the information into the database. The schedule has 1110 activities, 1928 relationships, 5 Level~1, 20 Level~2 and 299 Level~3 Milestones for the CMB-S4 project.  

\subsection{Cost}

Each detailed task-based cost estimate corresponds to a task in the project schedule. For that specific task, resources and their quantities are assigned from a standardized list of resources. The list includes multiple resource classes in each of the categories: labor, materials/non-labor, or travel. A task estimate consists of the number of hours of each labor resource class, the base-year dollar cost of each materials/non-labor resource class, the number of trips for each travel resource class, and the basis for each estimate. 

\subsection{Schedule}

\begin{figure}[htbp]
\begin{center}
\includegraphics[width=1.0\textwidth]{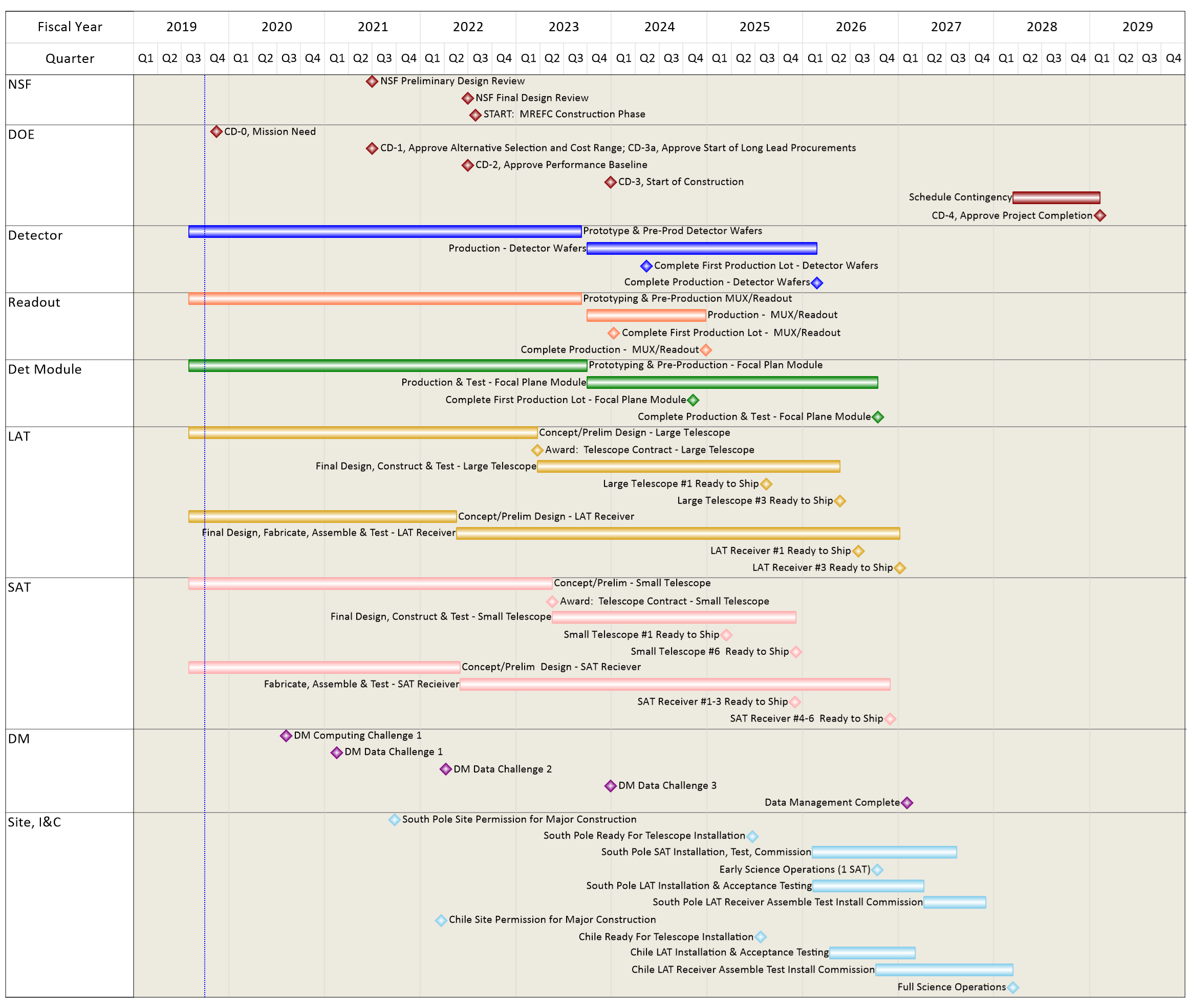}
\caption{CMB-S4 schedule and milestone summary.}
\label{fig:schedule}
\end{center}
\end{figure}

Table~\ref{tbl:schedule} shows the proposed the NSF Level~1 Milestones along with the corresponding DOE Critical Decision Milestones and Fig.~\ref{fig:schedule} shows a summary of the schedule and high level milestones.  The Interim Project Office, jointly supported by NSF and DOE, will further define the schedule for NSF and DOE reviews and approvals with two guiding objectives, a technically driven schedule and coordinated agency reviews and approvals.  This approach is necessary for a single integrated project and the clear delineation of scope and responsibilities for each funding agency and partner.  The critical path is driven by fabrication of detector wafers and delivery of assembled and tested detector modules to the large aperture telescopes.

The schedule development strategy is to define a schedule that is consistent with the funding potentially available during FY2019-FY2021, and subsequently  technically driven.  The project is working towards an early completion milestone that contains limited schedule float and a year of schedule float following this early project complete milestone is included in the overall project complete milestone (CD-4).  The Interim Project Office will continue to optimize the schedule and include explicit float for activities that are not on the critical path.  The best opportunity to improve on the schedule is to reduce the time required to deliver the full quantity of the Detectors and Readout (D\&R) components.  This is a major focus of the R\&D program supported by the DOE.  The Interim Project Office formed a D\&R Task Force in early 2019 to evaluate existing fabrication and testing capabilities and to provide recommendations on production plans  A formal review of the resulting detector fabrication plan will be completed in mid-2019.

\begin{table}[htp]
\begin{center}
\begin{tabular}{|l|c|}
\hline
NSF Level~1 Milestone (DOE Critical Decision) & Schedule (FY) \\
\hline
Pre-Conceptual Design (CD-0, Mission Need) & Q3 2019 \\
Preliminary Baseline (CD-1/3a, Cost Range/Long-Lead Procurement) & Q3 2021 \\
Preliminary Design Review (CD-2, Performance Baseline) & Q2 2022 \\
Final Deign Review (CD-3, Start of Construction) & Q4 2023 \\
Completion of 1st Telescope (CD-4a, Initial Operations) & Q2 2026 \\
Project Completion(CD-4, Operations) & Q1 2029 \\
\hline
\end{tabular}
\caption{Funding agency milestones.}
\label{tbl:schedule}
\end{center}
\end{table}

\subsection{Reporting and contingency management}

The Project will provide reports on a regular basis to NSF and DOE management.  The objective of the reporting is to compile essential technical, cost, schedule and performance data into reports to aid in the monitoring and management of the Project.
 
All cost account managers (CAMs) will submit monthly written narrative reports to the Project Office detailing specific progress on the pertinent subsystems.  These reports summarize the activities of the previous month, describe the activities planned for the upcoming month, and include comments and concerns.  In addition, performance reporting, including cost and schedule variance reporting, are submitted by the CAMs.  These are collected and summarized in a corresponding monthly report prepared by the CMB-S4 management team.   This report outlines progress, problems, and budget and schedule status, including comparisons of projected status versus actual status.  

Requests for cost and/or schedule contingency usage will be included in the monthly reports and will generate a change request which must be approved by (depending on the amount requested) the project change control board, project management, laboratory and agency program officers prior to allocation of funds. 

\section{Risks and opportunities}
\label{sec:risk}

\subsection{Risk and opportunity management plan}

The CMB-S4 Risk and Opportunity Management Plan, describes the continuous risk and opportunity management (RM) process implemented by the project.  RM is a disciplined approach to managing project risks throughout the life cycle of the project.  This plan is consistent with DOE O413.3B, ``Project Management for the Acquisition of Capital Assets,'' and the NSF 17-066, ``NSF Large Facilities Manual.''  The plan establishes the methods of assessing CMB-S4 project risk and opportunities for all subsystems as well as the system as a whole. Project risk and opportunity are managed throughout the life of the project, from development through construction and commissioning phases.

The primary goal is to manage the risks and opportunities associated with the development and construction of CMB-S4 and focus on understanding, reducing, or eliminating identified risks. Project risks and opportunities are centrally managed, but are the result of project-wide integrated and quantitative assessment which supports management decision-making. The statistical analysis of the residual risk after the planned mitigations informs the project contingency analysis for both cost and schedule.

Current and comprehensive risk updates provide management with additional information in preparing for and reacting to contingent events and adverse outcomes to planned events. The process also provides a uniform language for tracking risk elements and communicating that information. The Risk Registry documents the risk assessment, mitigation strategy, and the residual risk after mitigation.  It also includes information about all identified risks within the project.  The registry has incorporated lessons learned in several recent  projects. Risk Review Board meetings with the project leads will be held on a regular basis to review critical project risks, updates to the registry and status on mitigations. The Risk Register is maintained by the Systems Engineering and the risk management execution is owned by the project management.

The current risk registry for CMB-S4 is being used to define the R\&D programs to mitigate risks and to develop the baseline plan for the overall project. A series of risk management meetings were held with each WBS Level~2 system leads and their designated technical experts.  The primary purpose of these meetings were to discuss ``why'' risk management is an essential tool for all levels of project management, as it helps the team communicate and work together to reduce the negative impacts and increase positive impacts.  The other goal was the ``how'' of risk management which included identifying risks, developing informative risk statement, accessing the current probability and impacts of the risks and their possible mitigations.  Risk  review board meetings will be held regularly to discuss the top risks of the program and update status and changes to the risk registry, as well as action tracking.

\subsection{Risk/opportunity register}

The CMB-S4 risk registry has 140  risks identified and assessed.  There are four (4) risks that are currently assessed at critical and 38 risks at high.  
Table~\ref{fig:RiskStats}  provides the current assessment summary for the identified risks and Table~\ref{fig:Risks} summarizes the top risks of this program.  
The project is working on mitigations to ensure that these risks are lowered to reasonable levels that are consistent with our overall project timeline and performance.  

\begin{table}[htbp]
\begin{center}
\includegraphics[width=0.9\textwidth]{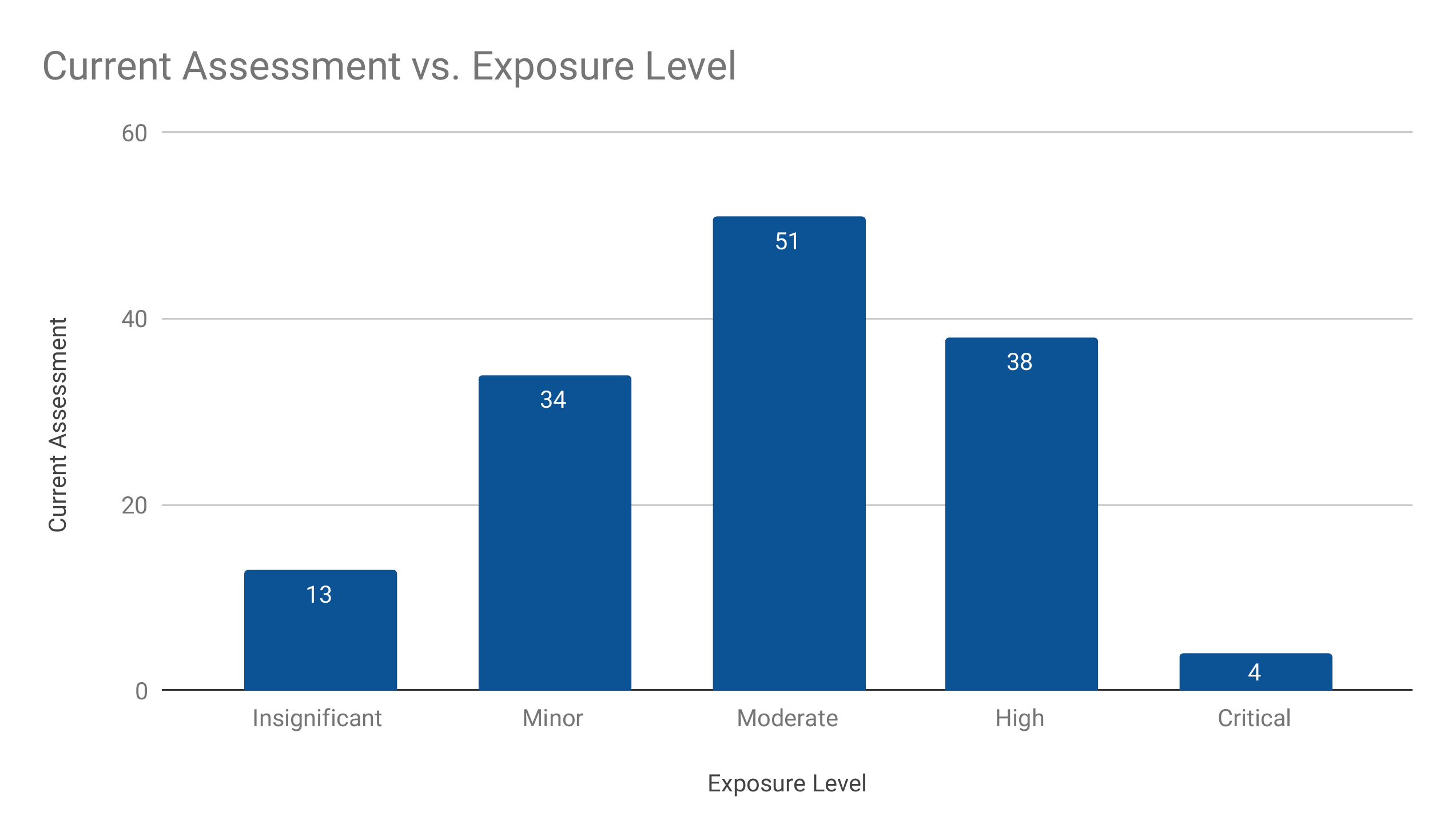}
\caption{CMB-S4 risk registry current assessment exposure level.  This table shows the current assessment exposure level of the risks that have been identified.}
\label{fig:RiskStats}
\end{center}
\end{table}

\begin{table}[htbp]
\begin{center}
\includegraphics[width=1\textwidth]{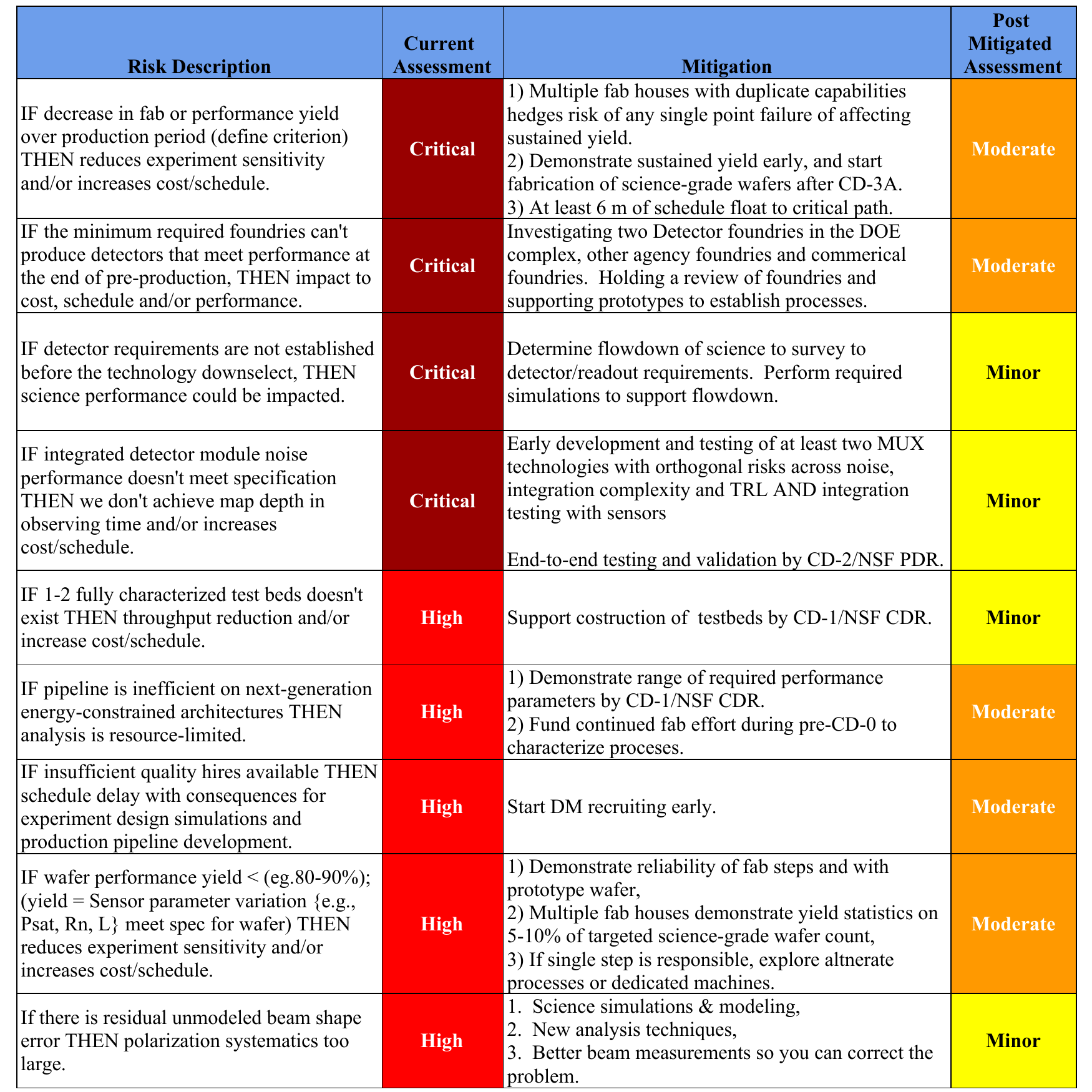}
\caption{CMB-S4 Risk Registry.  This table is a summary of the major risks that have been analyzed by the collaboration.  A description of the risk and its impact are presented in the first two columns.  The `current assessment' is derived from the product of a probability factor and an impact factor.  The planned mitigation action is listed in the third column, followed by a projected post-mitigated risk assessment in the fourth column.}
\label{fig:Risks}
\end{center}
\end{table}

In summary, CMB-S4 will require a large increase in the number of detectors and associated readout components and camera optical elements as compared to Stage-3 experiments.  Improving reliability and throughput is important. Technology development is being targeted to mitigate cost, schedule, and performance risk, as well as to exploit opportunities to reduce cost. 

\section{Value management---alternative analysis and selection}
\label{sec:VM}
The work proposed here will be evaluated through a well-established value engineering process. The first of three phases in this process is the Functional Analysis to define the technical scope, including current design decisions in the context of the full project, as well as known requirements (performance, operational, environment, etc.). Each of the functions will be classified as basic or secondary through their connection to the flow down requirements. The second phase is the Alternative Analysis, which establishes which among the space of solutions could best meet project objectives relative to a reference design. For each of these options weighted criteria are defined (i.e. requirements, cost, schedule, production, operations, risks, complexity, flexibility, maintainability, safety, development cost/schedule) that will be used to evaluate the approaches. These considerations inform the weaknesses/risks that will be studied in a prioritized way through the above scenarios, subject to cost and schedule constraints. Through this phase, implications on cost, schedule, risks, and opportunities are evaluated and updated. Finally, the third phase is the Implementation in which the results of the Alternative Analysis with regard to scoring against the weighted criteria will be presented to the Project Office and/or a designated review team to arrive at a decision to either select or reject a design, or to do further work. Through this process, we will manage risk, maximize opportunity, and promote discussion and support for the final decisions.

\section{Operations plan}

\subsection{Transition to operations}

Completion of commissioning of the first telescope marks the start of CMB-S4 operations and completion of commissioning of the last telescope is the completion of the construction project.  The transition from commissioning to operations will occur over a 2--3 year period and will involve a significant change in staffing at the sites, data centers, and partner institutions. A successful operations phase will require transitioning key personnel from Integration \& Commissioning to operations and hiring new staff with skills that are appropriate for operations. A transition plan will be developed to cover this period.

\subsection{Operations}

The basic operations model for CMB-S4 will be observations with multiple telescopes and cameras distributed across two sites, with observing priorities and specifications optimized for the CMB-S4 science goals, and data from all instruments shared throughout the entire CMB-S4 collaboration. Scientists working at laboratories and universities will coordinate the observations, monitor the data, design and implement the data pipeline, and carry out science analyses. Instruments at both sites will collect data nearly continuously. These data will include calibrations as well as CMB observations. Nearly all observations will be automated, so that local operators on the sites will not be needed during routine observations. 
The two CMB-S4 sites though remote, are both sufficiently well-established that fielding the CMB-S4 instrumentation does not represent a large risk, and costs and schedule estimates can be predicted from past experience. Site-specific planning and management are required because there are many site-specific issues such as power generation, safety monitoring, data storage and transfer.  The Chile site is at very high altitude (5200\,m, barometric pressure about half sea-level) and must deal with snow removal, but is accessible year round. Staff and visitors are housed in San Pedro de Atacama (2400\,m, population 2500) and make the one hour drive up to the site as needed. The South Pole gets very cold ($-80$\,C), is not accessible mid February through late October, and suffers shipping restrictions on the size and weight of parts, but the small winterover staff is housed within 1~km from the telescopes, and is always available.  At both sites, major maintenance is generally in summer when the weather is worst for observing. 

The operations cost is based on a preliminary bottom-up estimate that includes management, site staff, utilities, instrument maintenance, data transmission, data products, pipeline upgrades, collaboration management, and science analysis. The effort is roughly 30 FTE/year for data products and pipeline upgrades, 15 FTE/year for science analysis, and 25 FTE/year for management and site support. The annual operations cost is \$32M in 2019 dollars, excluding 20 FTE/year of scientist effort supported by DOE research funds. If the DOE-supported scientist effort is included, annual operations are 10\% of the construction cost, which is typical for an observatory.

\section{R\&D and pre-conceptual design}

The CMB-S4 Collaboration is preparing for CD-1/PDR on the timescale of April 2021.   R\&D and pre-Conceptual design work is aimed at reducing risk and firming up cost and schedule estimates.  

Funding to support these efforts has been provided by DOE and funding has been requested from NSF through a MSRI-R1 proposal.  

R\&D is focused on Detectors and Readout while the pre-Conceptual Design studies are aimed at 5 distinct areas: Cold Optics, Ground Pickup Sidelobes and Beam systematics, LAT Cryostat Design, SAT Cryocooler test and Data Management. 
 
The interim project office includes an R\&D and pre-Conceptual Design manager that works with the collaboration Technical Coordinators (see Fig.~\ref{fig:prj_org}).  Progress on R\&D and pre-Conceptual Design work is reported to the manager and to the interim project office on a monthly basis.  

A short summary of the R\&D and pre-Conceptual Design efforts is given below.
\begin{description}
\item[Detectors and Readout:] The highest technical risk to the CMB-S4 project was identified by the CDT and the December 2018 review as detector fabrication and testing capacity.  ANL and LBNL/HYPRES have already demonstrated the capability to fabricate sinuous antenna detectors that are compatible with fmux readout.  The near term R\&D will focus on bringing these fabrication facilities up to speed to fabricate another possible combination, OMT horn-coupled detectors that are compatible with umux (and dfMUX) readout, optimized to operate at the planned 100mK focal plane temperature.   The long term goal is to have the performance of a CMB-S4 wafer and detector module fully characterized and understood by October 2020.
\item[Cold Optics:] CMB-S4 requires more cold optical elements that any previous CMB experiment.  These include lenses, filters, half-wave plates (HWP), and baffling materials. The lenses, half-wave plates, and absorbing filters require high quality anti-reflection (AR) coatings. The goal of this effort is to demonstrate implementations of these technologies to enable performance forecasting and technology selection by CD-1. 
\item[Ground Pickup \& Sidelobes (SATs and LATs):] Systematic errors due to ground pickup, and pickup from the Sun, Moon, and galaxy are a critical issue for CMB-S4 and have been identified as one of the most significant project risks. Pickup adds noise and biases in measurements of $r$ and $N_{\rm eff}$, and must be carefully controlled if CMB-S4 is to meet its ambitious science goals. This effort is focused on modeling pickup for the CMB-S4 large and small telescopes, using a mix of physical optics and ray tracing tools, with inputs and techniques verified by sidelobe measurements from Stage-3 telescopes and laboratory measurements. 
\item[LAT Cryostat Design:] The key design driver for the large telescope cameras is accommodating enough detectors to achieve the required mapping speed. Existing large camera designs, e.g., the Simons Observatory LAT cryostat, are too small.  This effort is aimed at developing a new concept for CMB-S4, including consideration of the optimum pixel spacing, optics tube diameter and spacing, and detector wafer configuration and will develop a preliminary solid model of the new camera concept that will provide a basis for detailed design.
\item[SAT Cryocooler Test:] The cooling capacity of 4-K and 40-K stages are the driver for the cryostat design of CMB-S4 SAT. The cryocooling system is the dominant contributor to the power budget at the site.  This activity covers a quick test of the cooling capacity for currently available dilution refrigerators and pulse tube systems and will enable an early start to the SAT cryostat design effort.  
\item[Data Management:] CMB-S4 faces the twin challenges of controlling systematic effects to unprecedented precision and the extraordinary volume of the data to be processed. This effort addresses activities associated with these challenges that are on the critical path to CD-1 either because they are required to inform the baseline design (systematics simulations) or because they have critical milestones on that timescale (data processing at scale and on new computing architectures).
\end{description}

\appendix
\appendixpage
\addappheadtotoc  
\chapter{Science Forecasting}
\label{chap:forecasting}

Here we present the forecasts that form the basis of our flowdown from science requirements to measurement requirements. We begin in Sect.~\ref{sec:small-area} with the forecasts for constraints on the tensor-to-scalar ratio, $r$. In Sect.~\ref{sec:light-relic} we present the forecasts for constraints on the light relics parameter $N_{\rm eff}$. Forecasts related to our two design-driving legacy survey goals are presented in Sect.~\ref{sec:ClustersFlowdown} and~\ref{sec:GRBflowdown}.

\vskip 12pt
\section{Ultra-deep field targeting the degree-scale signature of gravitational waves}
\label{sec:small-area}

Using simulations to optimize the design of a CMB experiment inevitably involves a trade-off between the degree of detail that the simulations are able to capture and the computational (and human) cost of generating and analyzing them. This trade-off includes the choice of domain in which the simulation is generated, ranging from the most detailed but most expensive time domain, through the map domain, to the most simplified, but most flexible, spectral domain.  Inclusion of additional detail can help to validate general results, to explore their sensitivity to assumptions about foreground models, sky coverage, and instrumental noise and systematics, and in more mature stages of design, can inform specific instrument and survey strategy choices.

To ensure realism, our general forecast/simulation approach has been an
iterative one. We rely on a closed forecasting loop to tie the semi-analytic
tools, which allow for fast optimizations, with map-based studies, which can
include multiple layers of additional complexity. Our measurement requirements
and the baseline experiment configurations which can achieve them are
established as a result of multiple passes through this loop. The main steps
describing this process are as follows.
\begin{enumerate}
       \item Develop a (semi-) analytic spectral forecast that makes use of noise performance that is informed by scaling from actual analyses of real experiments from time-streams to power spectra.
       \item Use this forecasting tool to optimize the allocation of detector effort across frequencies, determining certain baseline ``checkpoints" in survey
definition space.
       \item Validate these checkpoint configurations with standardized, version-numbered map-based data challenges.  If independent analyses show recovery of science parameters from these challenge maps that does not match analytic forecasts (either in terms of variance or bias), we revise the forecasts accordingly.
       \item Iterate between steps 1 and 3, injecting increasing realism in the form of: (a) sky model complexity informed by the latest data and modeling efforts; (b) survey coverage based on proven observing strategies; and (c) systematics whose form, parameterization, and likely amplitude is likewise guided by real-world experience.
\end{enumerate}

For the CMB-S4 Science Book \cite{Abazajian:2016yjj} an $r$-forecasting machinery was assembled based on
scaling the bandpower covariance matrices and noise spectra of published BICEP/Keck analyses.
Given a defined set of bandpasses, and assumptions about foreground power spectra,
this semi-analytical approach is capable of optimizing allocation of detectors across the sky (sky coverage), frequency channel, and
and delensing vs. degree-scale surveys, for the lowest $\sigma(r)$ at fixed effort,
where ``effort''= total number of 150-GHz-equivalent detector-years of observation.
(It makes sense to define ``effort'' in these units since it is equivalent to focal plane area, which is in turn the strongest driver of overall project cost.)

For the CDT report we extended this work to map-domain simulations in order to be able to capture additional complexities that cannot be represented in the spectral domain, while remaining computationally tractable. These complexities include:
\begin{itemize}
  \item non-Gaussianity and statistical anisotropy of the Galactic foregrounds;
  \item instrumental systematic effects;
  \item inconsistency between the data and the assumptions (either explicit or implicit) of any given analysis method;
  \item foreground contamination in the delensing map.
\end{itemize}
We also used these simulations to validate the spectral domain forecasts for configurations where the approaches are directly comparable.

For this Decadal Survey Report, we have continued the practice of using the semi-analytic approach to explore a wide number of options, while
using sets of map-based simulations at specific checkpoints to calibrate the semi-analytic approaches and confirm their validity.
Here we review the methods used to explore parameter space for the ultra-deep field, including map level noise simulations, sky models, and observation strategy. We also describe our approach to modeling instrumental systematics, the delensing map, and the analysis methods. We present results of our calculations that guide our flowdown to the measurement requirements.

A strength of CMB-S4 is the access to two different observing sites with complementary capabilities. The sites differ significantly in the range of sky area on which coverage can be concentrated, with the Chilean site allowing for greater sky coverage, and the Pole site allowing for more concentrated sky coverage. We present here forecasts for $\sigma(r)$ as a function of the distribution of SATs across site, forecasts that inform our choice of a baseline configuration, and motivate the flexible approach to deployment that is part of our reference design.

In Sect.~\ref{sec:semianaflowdownallocation} we present our flowdown to sky coverage, over-all sensitivity level for both the SATs and the delensing LAT, and for the allocation of detectors across SAT frequencies using our semi-analytic optimization framework.
In the next two subsections we build on these results with improved realism in two different ways.
In Sect.~\ref{sec:cdtrepsims} we present map-based simulations as a cross check on the semi-analytic results, and as an examination of robustness of the forecasted performance given a variety of foreground models as well as systematic errors of instrumental origin.
To provide the calculations we need as a basis for our optimization of SAT and delensing LAT siting, we return to our semi-analytic framwork. We present these calculations in Sect.~\ref{sec:DSRupdates} for survey coverage maps that include constraints imposed by each site and the finite extent of the instrument field of view.

\vskip 9pt

\subsection{Flowdown to total number of detector years and allocation of detectors across frequency and delensing effort: semi-analytic calculations}
\label{sec:semianaflowdownallocation}

To obtain the optimal allocation of detectors across frequencies, we use the
aforementioned performance-based forecasting framework. This semi-analytic tool
is grounded in published BICEP/Keck achieved performances, in the form of
bandpower covariance matrices and noise angular power spectra from end-to-end
analyses of multiple on-sky receiver years at
\{95, 150, 220\}\,GHz \cite{Ade:2018gkx}.
For projections we assume that we can scale down the noise
based on increased detector count and integration time and that we can apply
beam-size and NET rescaling to estimate the achieved performance at other
frequencies. This ``achieved performance'' approach automatically builds in all real world
inefficiencies, including (but not limited to) imperfect detector yield,
non-uniform detector performance, read-out noise, observing inefficiency,
losses due to timestream filtering, beam smoothing, and non-uniform sky
coverage. A detailed presentation of the framework and optimization process is
given in Ref.~\cite{rforecast_paper:2019}.

\begin{figure*}[htbp]
\begin{center}
\includegraphics[height=3.2in]{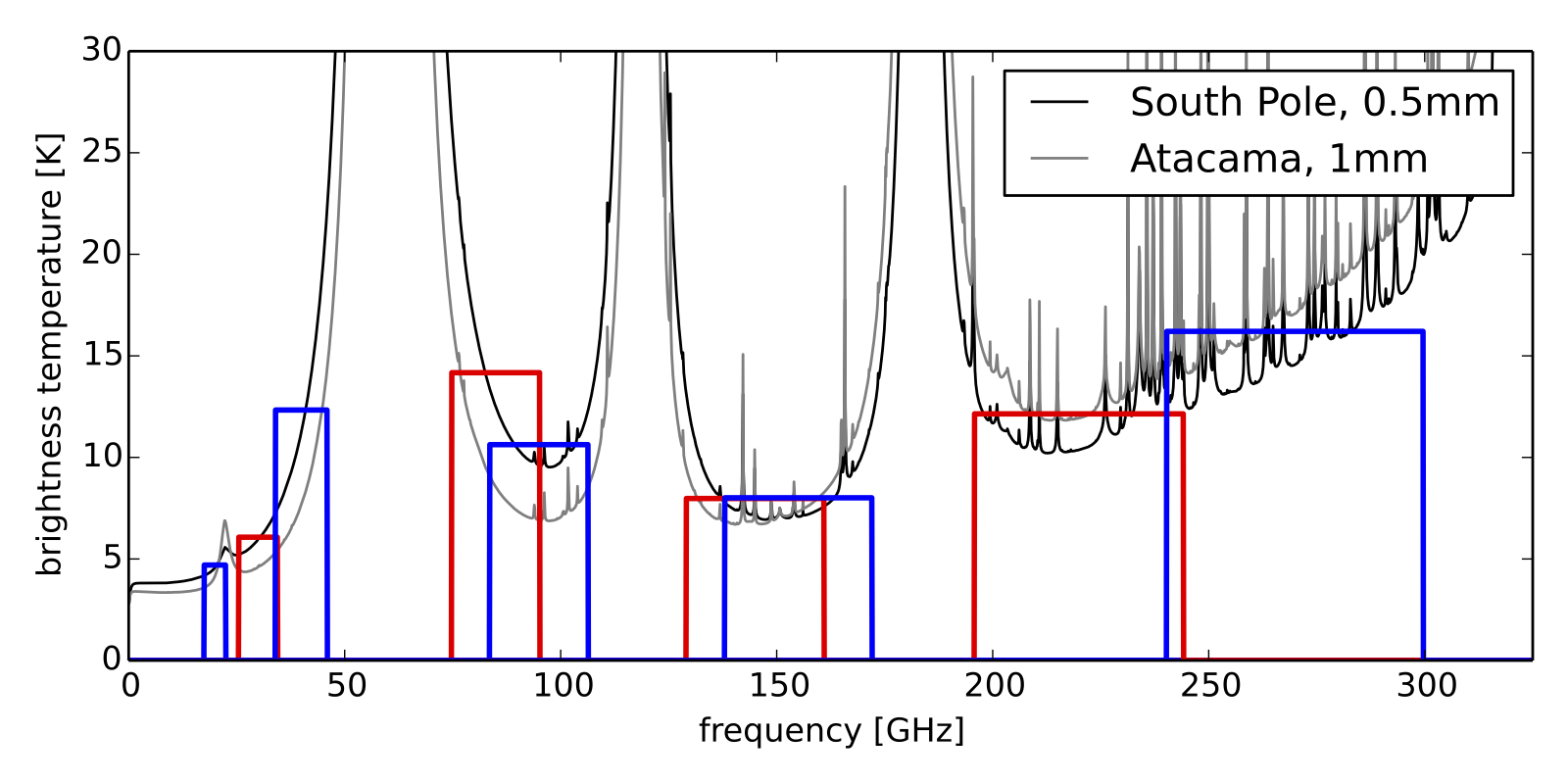}
\caption{Calculated atmospheric brightness spectra (at zenith) for the South Pole at 0.5\,mm PWV and Atacama at 1.0\,mm PWV (both are near median values). Atmospheric spectra are generated using Ref.~\cite{paine_scott_2018_1193646}. The tophat bands are plotted on top of these spectra, with the height of each rectangle equal to the band-averaged brightness temperature using the South Pole spectrum.}
\label{fig:bpass}
\end{center}
\end{figure*}

To span the four available atmospheric windows (Fig.~\ref{fig:bpass}) and
have enough channels to mitigate against complex foregrounds, we assume eight
channels at \{30, 40, 85, 95, 145, 155, 220, 270\}\,GHz, which are placed on
small aperture telescopes. In addition, we also include a 20-GHz channel on a
large aperture telescope. This latter inclusion is the result of insight
gained from an early iteration through the forecasting loop, which demonstrated
that for certain foreground models sizeable biases were present due to
synchrotron residuals. The CDT strawperson design was updated accordingly to
mitigate against such biases, with the 20-GHz channel being placed on a large
aperture telescope due to resolution constraints.

The procedure used to come up with the split in each window was to separate the overlapping bands as far as possible while still keeping the calculated
per-detector NET within 10--15\% of the NET for a detector that spans the full
window. 
The ideal per-detector  NETs were calculated with {\tt NETlib.py}
\footnote{\url{cmb-s4.org/wiki/index.php/New\_NET\_Calculator\_and\_Validation}} at
Pole and Chile, using the 10-year MERRA2 median atmospheric profiles. We use
the mean over the two sites, which are \{214, 177, 224, 270, 238, 309, 331, 747, 1281\} $\mu {\rm K}_{\rm CMB}\sqrt{{\rm s}}$ for our nine channels.
These NETs are calculated for a 100\,mK thermal bath, as opposed to 250\,mK for
the Science Book, and are therefore lower. Note that this is the only departure from  achieved performance.
We want to emphasize that these NET
numbers are only used to determine the appropriate scalings between different
channels to allow current achieved performance numbers to be applied to proposed CMB-S4 instrument
configurations, and not to calculate ab-initio sensitivities. The scaling procedure is presented in detail in Ref.~\cite{rforecast_paper:2019}.

\begin{figure}[htbp]
\begin{center}
\includegraphics[height=2.7in]{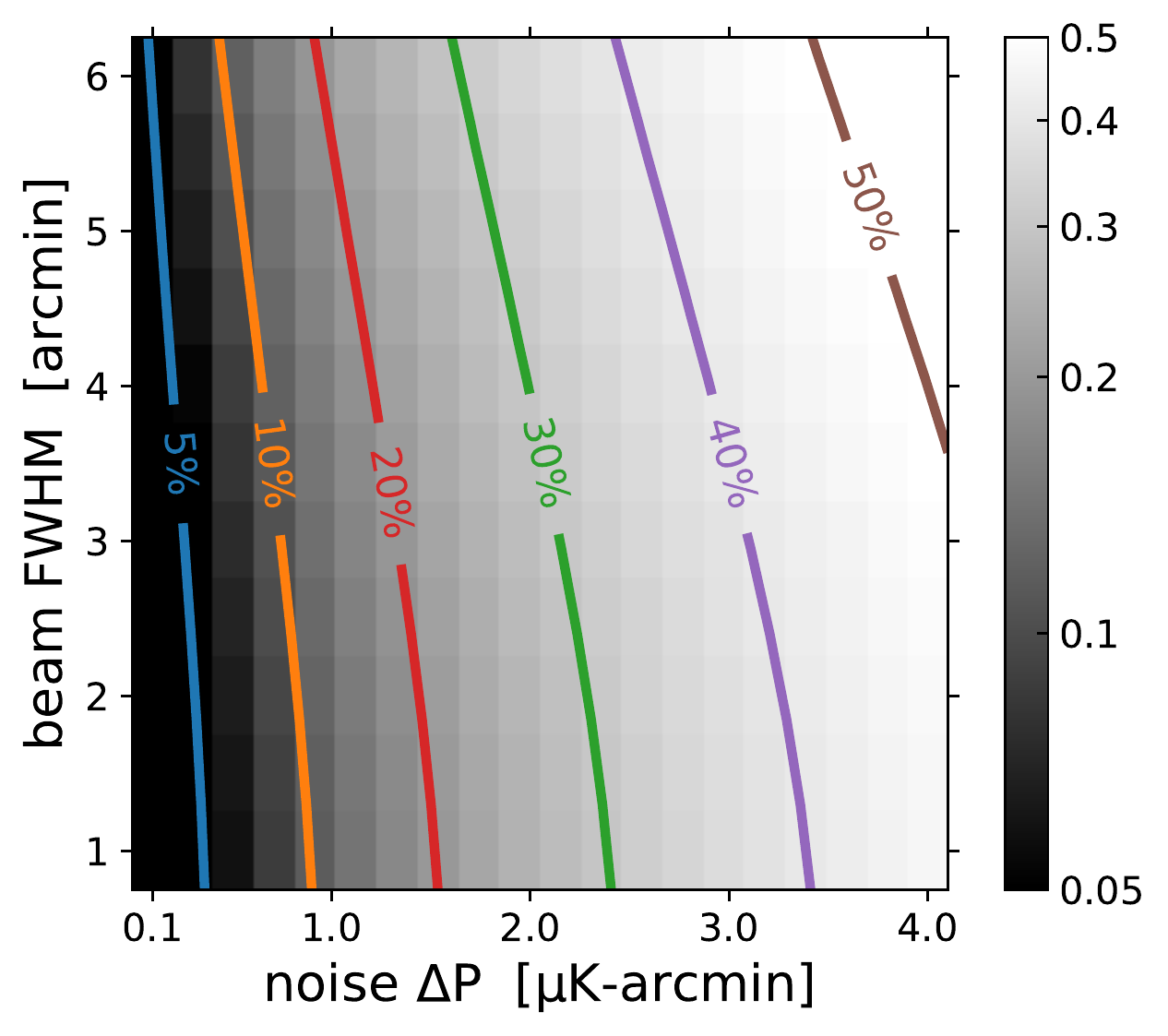}
\caption{Forecasted lensing $A_{\rm L}$ residual (grey scale plus colored contours as labeled) using the $EB$-only iterative delensing \cite{Smith:2010gu}, as a function of the beam full width half maximum and noise level in $Q$ and $U$.}
\label{fig:lens_res}
\end{center}
\end{figure}

The optimization process specifically includes the need to delens and assigns
a fraction of the detectors for that purpose. We assume a separate
high-resolution instrument dedicated to measuring the intermediate- and small-scale
information necessary to construct a template of lensing B modes, so that their effect can be removed.
In this initial semi-analytic optimization process, the delensing
instrument is assumed to have 1-arcminute resolution and detector weight at a
single frequency with mapping speed equivalent to that of the 145-GHz channel.
The translation between detector effort and map noise in the delensing
instrument is based on the method used for the low-resolution instrument,
but without certain non-idealities specific to low-resolution instruments and
low-$\ell$ analysis (such as mode removal and non-uniform coverage)
\cite{rforecast_paper:2019}. Following the formalism in
Ref.~\cite{Smith:2010gu}, we convert the map noise in the delensing map to a
delensing efficiency, or equivalently a fractional residual in lensed B-mode
power, as is shown in Fig.~\ref{fig:lens_res}.

\begin{figure}[htbp]
\begin{center}
\includegraphics[height=3.5in]{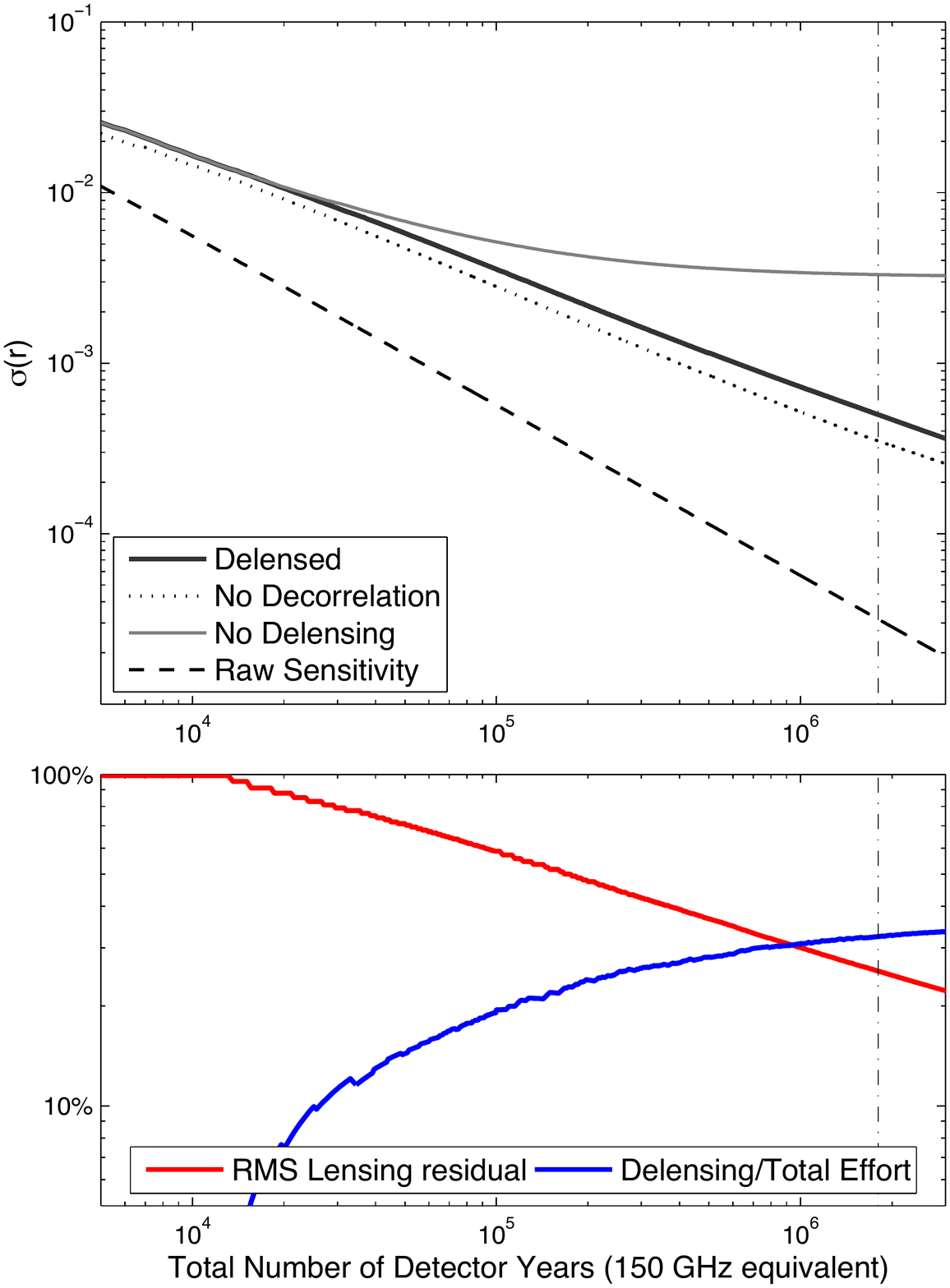}
\includegraphics[height=3.5in]{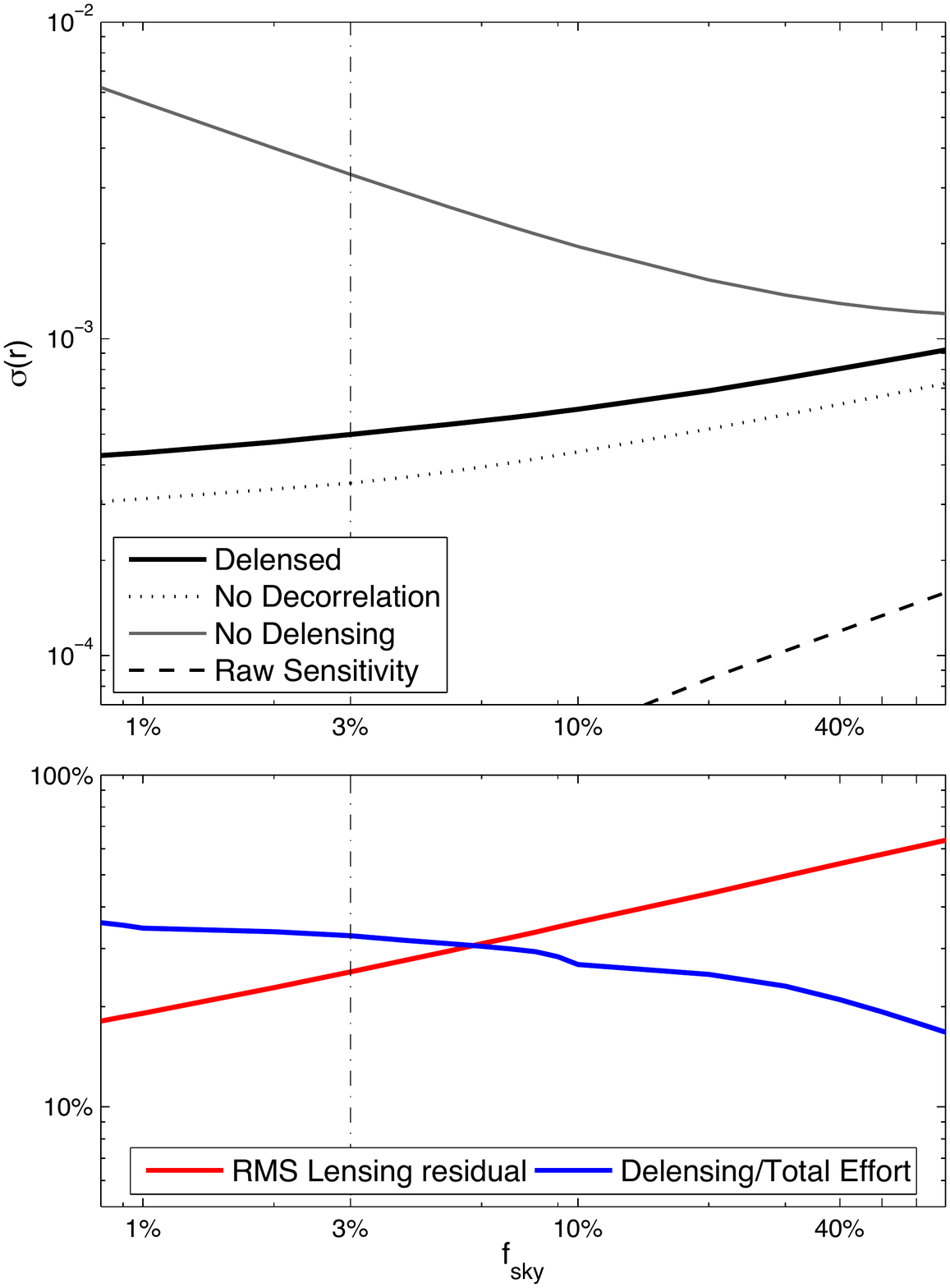}
\caption{Top panels: forecasted uncertainty on $r$ as a function of effort (left) and sky fraction $f_{\rm sky}$ (right). The left panel is for 3\% sky fraction, whereas the right panel is for $1.8\times 10^6$ detector years of effort, as represented by the vertical dashed lines. We included in solid black the case of full delensing, while allowing for decorrelation of the foregrounds, in solid grey the case without delensing, in dotted grey the case where no decorrelation is allowed in the model, and in dashed black the raw sensitivity in the absence of foreground and lensing.
Bottom panels: For the delensed case, we show the fraction of effort spent on removing the lensing sample variance and the resulting rms lensing residual.}
\label{fig:sigr}
\end{center}
\end{figure}

The trade-off between raw sensitivity, ability to remove foregrounds, and
ability to delens results is a complicated optimization problem with respect
to sky coverage. Figure~\ref{fig:sigr} (right), shows the $r$ sensitivity forecast for
CMB-S4 as a function of the observed sky fraction for the case that we only
have an upper limit ($r=0$). We note that for an initial detection the
optimization process requires a deep survey that targets as small an area as
possible. This conclusion of course depends on the forecasting assumptions; to
that end we would like to draw attention to several key factors. First, holding
the desired constraint on $r$ fixed, the level to which we rely on delensing
to decrease sample variance increases appropriately at smaller sky fractions.
For example, as shown in Fig.~\ref{fig:sigr} (right), achieving the forecasted sensitivity on $r$ for a
field targeting 1\% of the sky will require an $> 80$\% reduction in the map rms
level of the CMB lensing B modes. While from a sensitivity standpoint it is
possible to achieve these levels, the extent to which systematic effects and
small-scale foregrounds will need to be constrained may become too stringent.
Second, the current optimization assumes identical foreground behaviour across
the sky (equivalent to that in the BICEP2/Keck region), while in reality the
average amplitude, and possibly the complexity, of foregrounds increase as
larger sky areas are targeted. This effect would steepen the optimization curve at
high sky fractions and increase our preference for small sky, but it is
important to be cautious until we know more about foregrounds at these
sensitivity levels. Third, a practical consideration for the robustness of the
final $r$ result is its reproducibility across the sky. It is therefore useful
to observe multiple roughly 1\% patches from which we can derive separate cosmological constraints. Finally, the technical aspect of $E$/$B$ separation heavily disfavors
patches smaller than about 1\% of the sky due to cut-sky effects. Balancing the
forecasting results with these concerns, we have chosen $\approx3$\% as
the default sky fraction for CMB-S4 $r$ constraints (assuming a true value of
$r = 0$).

\begin{figure}[htbp]
\begin{center}
\includegraphics[height=2.7in]{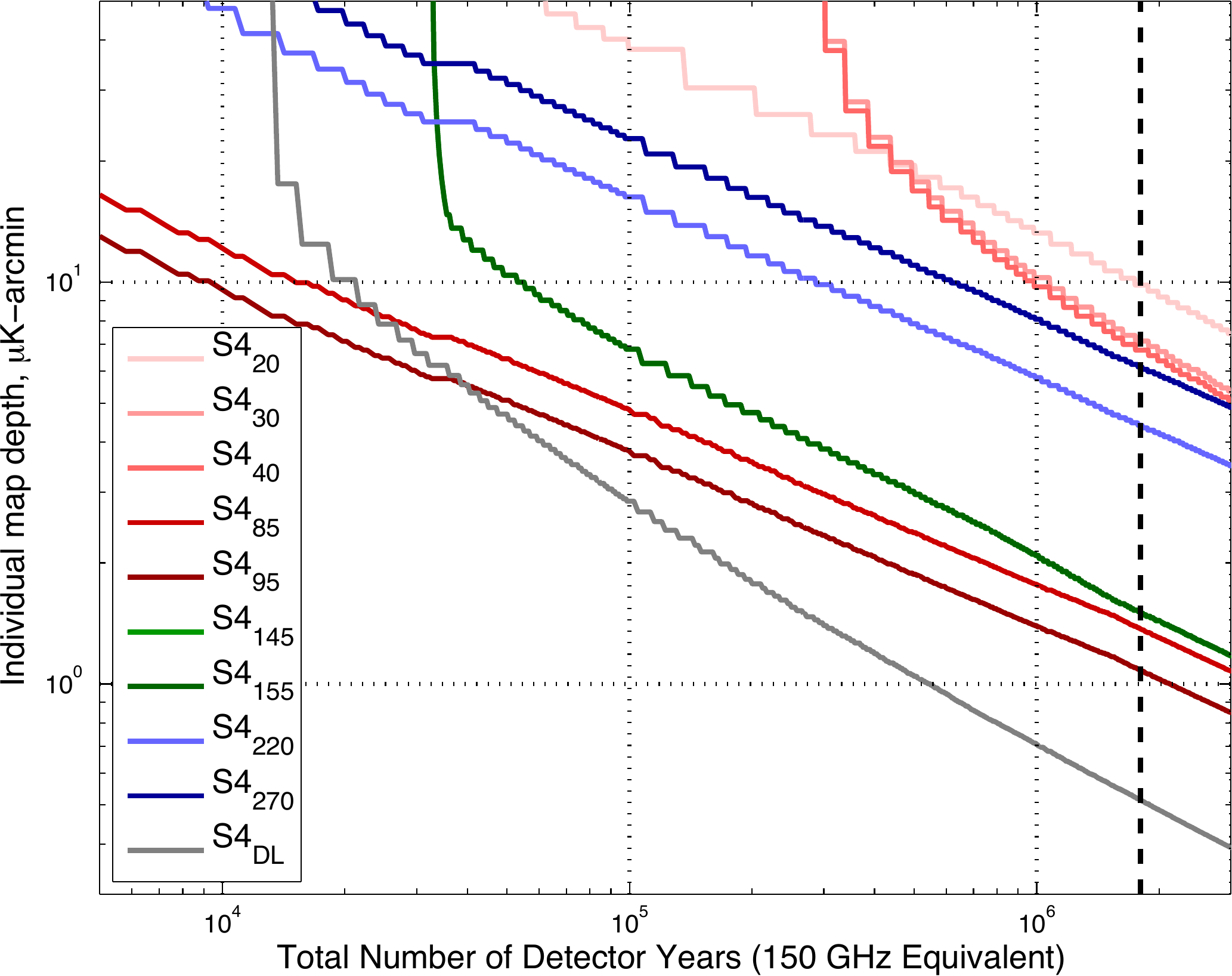}
\caption{Optimized map-depth in each of
the small-aperture channels as well as in the delensing channel, for an
$f_{\rm sky}=3$\%.}
\label{fig:mapdepth}
\end{center}
\end{figure}

For this choice of sky fraction, we find that we need $\approx 1.8\times10^6$
150-GHz-equivalent detector-years (or $\approx 1.2\times10^6$ under more optimistic foreground assumptions) to reach our science requirements, as can be
seen in Fig.~\ref{fig:sigr}. Roughly 30\% of this effort is dedicated towards
the delensing portion of the observations, yielding a 30\% rms lensing residual, with
the rest of the effort dedicated towards degree scale component separation.
The specific optimal allocation across frequencies and delensing effort and corresponding map depths are shown in Fig.~\ref{fig:mapdepth}.
This configuration provides a starting point which leads to the reference design.

\subsection{Cross-checks with map-based simulations}
\label{sec:cdtrepsims}

In this section, we present the next step in our iterative forecasting,
where we generate and re-analyze map based simulations.
As described earlier in this Appendix, this allows us
to cross-check the results of the semi-analytic calculations,
and also to probe the impacts of foregrounds and instrumental systematic effects.

\subsubsection{Map noise realizations}
\label{sec:noisim}

To produce map-level simulations it is necessary to translate the
BICEP/Keck noise bandpowers into a prescription for map noise.
We do this by fitting the $N_\ell$s to a $\hbox{white}+\ell^{\,\gamma}$ model
accounting for beam smoothing, etc.
For the small-aperture BICEP/Keck data, we find $\ell_\mathrm{knee} = $
50--60 with $\gamma$ of $-2$ to $-3$.
To translate to map noise levels, we must pick a specific
sky hit pattern. As we have argued in Sect.~\ref{sec:semianaflowdownallocation}, we
want a patch with 3\% area, we use here an idealized circular hit pattern
which has this effective area (shown in Fig.~\ref{fig:relhits}).
We then generate Gaussian noise realizations at each band and
divide by the square-root
of the assumed coverage pattern such that the noise ``blows up around the edge''
as it does in real maps.

\vskip 9pt
\subsubsection{Foreground models}
\label{sec:skymod}

To make simulated sky maps we add realizations of lensed CMB both without and with an $r$ component to models of the Galactic foregrounds. We used the following seven foreground models.

\begin{enumerate}
\setcounter{enumi}{-1}

\item Simple Gaussian realizations of synchrotron and dust with power-law angular power spectra at amplitudes set to match the observations in the BICEP/Keck field, and simple uniform SEDs (power law for synchrotron, modified blackbody for dust).

\item The PySM\footnote{\url{https://github.com/bthorne93/PySM\_public}} model {\tt a1d1f1s1}, where the letters refer to anomalous microwave emission, dust, free-free and synchrotron respectively, and the numbers are the base models described in Ref.~\cite{Thorne:2016ifb}.

\item The PySM model {\tt a2d4f1s3}, where the models have been updated to variants that are also described in Ref.~\cite{Thorne:2016ifb}.  Note that these include 2\% polarized AME, a curvature of the synchrotron SED, and a two-temperature model for dust.

\item The PySM model {\tt a2d7f1s3}, where the dust model has been updated to a sophisticated physical model of dust grains as described in Ref.~\cite{Hensley2015}.
This model is interesting in that it does not necessarily conform to the modified blackbody SED.

\item The dust in model~3 is replaced by a model of polarized dust emission that incorporates H{\sc i} column density maps as tracers of the dust intensity structures, and a phenomenological description of the Galactic magnetic field as described in Ref.~\cite{Ghosh:2017}.  The model is expanded beyond that described in the paper to
produce a modest amount of decorrelation of the dust emission pattern as a function of frequency motivated by the analysis of {\it Planck\/} data in Ref.~\cite{Aghanim:2017}.

\item A toy model where the dust decorrelation suggested in figure~3 of Ref.~\cite{Aghanim:2017} is taken at face value ($\mathcal{R}^{217\times353}_{80}=0.85$).
While such a model is not ruled out by current data it appears to be very hard to produce such strong decorrelation in physics-based models.  We also note that Ref.~\cite{Sheehy:2017} have re-analyzed the same {\it Planck\/} data and, while they find that the high level of decorrelation in this model is consistent with the data, their best fit to that same data has no decorrelation.

\item A model based on MHD simulations \citep{Kritsuk:2017} of the Galactic magnetic field, which naturally produces non-Gaussian correlated dust and synchrotron emission.

\end{enumerate}

Models~1 to 4 use the actual large-scale modes of the real sky as measured above the noise in the {\it Planck\/} data.  This means that these models are intrinsically ``single-realization,'' and this must be borne in mind when interpreting the results. Models~4 and 6 are not based on {\it Planck}, but still contain a fixed signal realization. Models~0 and 5 have different seeds for each signal map and include the (Gaussian) sample variance. The PySM models fill in the small-scale structure with power-law Gaussian extrapolations, while models~4 and 6 naturally produce non-Gaussian small-scale structure.  However, all of these models are consistent with current data, and
we should be careful not to necessarily associate nominal sophistication with greater probability to more closely reflect reality.

\vskip 9pt
\subsubsection{Delensing}

We have started to generate high-resolution simulated maps on which we can run explicit lensing reconstruction and then include that information in the analysis.
However, that process is not yet converged, and so for the present we approximate delensing by scaling down the $\Lambda$CDM lensing signal. As described in \ref{sec:semianaflowdownallocation}, the delensing efficiency at power spectrum level for a coverage of 3\% is predicted to reach $\approx90$\%, so in the map-based results presented here, we use an effective $A_{\rm L}$ of 0.1.

\vskip 9pt
\subsubsection{Instrumental systematics}
\label{sec:syst}

Control of instrumental systematics is a critical design consideration.  However, predicting and modeling these effects realistically is a difficult task that is dependent on actual instrument and survey design details, and in any case their impact on an actual result comes not through the modeled effects but through unmodeled residuals.  So far we have simulated various generic classes of additive systematic by injecting additional
noise-like components into the maps, and then re-analyzing them without knowledge of what was put in.  We have experimented with components that are both correlated and uncorrelated across frequency bands, and which have white, $1/\ell$, and $\hbox{white} + 1/\ell$ spectra, at varying levels compared to single-frequency map noise or, for correlated cases, combined map noise.  Examples of mechanisms that might produce map residuals within this class, after modeling them and either correcting or filtering their leading-order effects, include bandpass mismatches, beam and pointing variations, calibration variations, cross-talk effects, half-wave-plate leakages, ground pickup, and readout irregularities.

\vskip 9pt

\subsubsection{Analysis methods}

To make simulated maps the noise realizations described in Sect.~\ref{sec:noisim} are added to the sky models described in Sect.~\ref{sec:skymod}, and potentially also the systematics realizations described in Sect.~\ref{sec:syst}.
For each realization one then has a stack of multi-frequency $I/Q/U$ maps containing non-uniform noise, foregrounds and signal, and the challenge is to re-analyze them to recover the parameter of interest (in this case $r$).  This can be done by different teams using different methods, and could be done in a blind manner, although we have not done this yet.

So far we have experimented with two methods.  The first is a map-based ILC cleaning method \citep[e.g.,][]{Eriksen:2004jg}, which seeks the linear combination of maps that minimizes the remaining CMB signal, followed by a marginalization over residual foregrounds.  This method has the advantage that it does not need to know the bandpasses of the frequency channels.

The second method is an evolution of the parametric multi-component fit to the ensemble of auto- and cross-spectra as used for the BICEP/Keck analysis to date \citep{Ade:2015tva,Ade:2018gkx}.  This method fits the observed bandpowers to a model composed of the lensing expectation plus dust and synchrotron contributions and a possible $r$ component.  Dust and synchrotron each have an amplitude ($A_{\rm d}$ and $A_{\rm s}$), a spatial spectral parameter ($\alpha_{\rm d}$ and $\alpha_{\rm s}$),  and a frequency spectral parameter ($\beta_{\rm d}$ and $\beta_{\rm s}$).  We also allow dust/synchrotron correlation ($\epsilon$), and decorrelation of the dust patterns over frequency ($\Delta_{\rm d}$).

Both of these analysis methods are only close to optimal when the foreground behavior is close to uniform across the observing field.  For analysis of larger fields, algorithms that fit, for example, the frequency spectral indices individually in (large) pixels, will be required.

\vskip 9pt
\subsubsection{Results}

Table~\ref{tab:modres} summarizes the results of re-analysis of simulations of
$1.2\times10^6$ 150-GHz-equivalent detector-years and residual lensing power $A_{\rm L}=0.1$
as described in Sect.~\ref{sec:semianaflowdownallocation}.
We see that for $r=0$ the simple Gaussian foreground model~0 gives $\sigma(r)\approx5\times10^{-4}$, as expected from the semi-analytic calculations.
As we progress to the more complex foreground models, $\sigma(r)$ is generally in the range 5--8$\times10^{-4}$.
The bias remains below $1\sigma$ in all cases.
(These simulations are sets of 500 realizations, so the statistical uncertainty on the bias is $\approx 0.04 \sigma$.)
The strong decorrelation model 5 does significantly increase $\sigma(r)$, and,
while the parametric method is able to resist bias in this case, by construction information is lost.
In fact if one believed in such a scenario, re-optimization to concentrate the sensitivity at closer-in frequencies would be called for.

\newbox\tablebox
\newdimen\tablewidth
\def\leaderfil{\leaders\hbox to 5pt{\hss.\hss}\hfil}
\def\endCMBSfourtable{\tablewidth=\wd\tablebox
  $$\hss\copy\tablebox\hss$$
  \vskip-\lastskip\vskip -2pt}
\def\tablenote#1 #2\par{\begingroup \parindent=0.8em
  \abovedisplayshortskip=0pt\belowdisplayshortskip=0pt
  \noindent
  $$\hss\vbox{\hsize\tablewidth \hangindent=\parindent \hangafter=1 \noindent
    \hbox to \parindent{$^#1$\hss}\strut#2\strut\par}\hss$$
  \endgroup}
\def\doubleline{\vskip 3pt\hrule \vskip 1.5pt \hrule \vskip 5pt}

\begin{table}[htbp]
  \begingroup
  \newdimen\tblskip \tblskip=5pt
  \nointerlineskip
  \vskip 1mm
  \setbox\tablebox=\vbox{
    \newdimen\digitwidth
  \setbox0=\hbox{\rm 0}
  \digitwidth=\wd0
  \catcode`*=\active
  \def*{\kern\digitwidth}
  \newdimen\signwidth
  \setbox0=\hbox{+}
  \signwidth=\wd0
  \catcode`!=\active
  \def!{\kern\signwidth}
  \newdimen\decimalwidth
  \setbox0=\hbox{$3$}
  \digitwidth=\wd0
  \catcode`|=\active
  \def|{\kern\digitwidth}
  \halign{\hbox to 1.0in{#\leaderfil}\tabskip 1em&
    \hfil#\hfil&
    \hfil#\hfil&
    \hfil#\hfil&
    \hfil#\hfil&
    \hfil#\hfil\tabskip=0pt\cr
    \noalign{\doubleline}
    \noalign{\vskip 3pt}
    \omit&&\multispan2\hfil ILC\hfil&\multispan2\hfil Parametric\hfil\cr
    \noalign{\vskip -2pt}
    \omit&&\multispan2\hrulefill&\multispan2\hrulefill\cr
    \noalign{\vskip 2pt}
    \omit\hfil$r$ value\hfil&Sky model&$\sigma(r)\times10^{4}$&$r$ bias $\times10^{4}$&$\sigma(r)\times10^{4}$&$r$ bias $\times10^{4}$\cr
\noalign{\vskip 3pt\hrule\vskip 5pt}
0&       0&4.4 &$-0.2$ &5.7&!0.3\cr
\omit& 1&4.6 &!0.8 &6.4&!5.2\cr
\omit& 2&4.7 &!0.7 &6.5&!1.9\cr
\omit& 3&4.6 &!1.2 &6.7&!0.7\cr
\omit& 4&6.5 &!4.8 &8.3&$-7.7$\cr
\omit& 5\rlap{$^{\rm a}$}&18*&!17 &15&!0.2\cr
\omit& 6&4.8 &$-1.8$ &6.5&!1.8\cr
\noalign{\vskip 4pt}
\hline
\noalign{\vskip 4pt}
0.003& 0&6.6 &$-0.7$ &8.1&!0.4\cr
\omit&  1&6.9 &!0.9 &8.5&!5.4\cr
\omit&  2&6.5 &$-0.1$ &7.9&!1.9\cr
\omit&  3&7.0 &!1.4 &8.7&!0.9\cr
\omit&  4&11 &!7.1 &11&$-6.2$\cr
\omit&  5\rlap{$^{\rm a}$}&23 &!17 &!17* &!0.4\cr
\omit&  6&7.5 &$-0.2$ &8.6&!2.5\cr
\noalign{\vskip 3pt\hrule\vskip 3pt}}}
\endCMBSfourtable
\tablenote {{a}} An extreme decorrelation model---see Sect.~\ref{sec:skymod}. The parametric analysis includes a decorrelation parameter. No attempt is made in the ILC analysis to model decorrelation.\par
\endgroup
\vglue 14pt
  \caption{Results of two analysis methods applied to map-based simulations using our suite of sky models.
    All simulations assume an instrument configuration including a (high-resolution) 20-GHz channel, a survey of 3\% of the sky with $1.2\times10^6$ 150-GHz-equivalent detector-years, and $A_{\rm L} = 0.1$, as described in Sect.~\ref{sec:semianaflowdownallocation}.}
  \label{tab:modres}
\end{table}

Table~\ref{tab:sysres} summarizes the results of re-analysis of simulations including additive systematic effects, in different combinations of  uncorrelated and correlated contamination with varying spectra, added on top of foreground model 3.  The levels of systematic contamination for these simulations were chosen  to predict biases on $r$ of $\approx 1 \times 10^{-4}$ in semi-analytic forecasts.  We can see that the different combinations explored increase biases on $r$ by amounts that typically vary from 0.5--1.5$ \times 10^{-4}$ for the  two different analyses, over the different cases.  We find that to restrict bias on $r$ to this level, the sum of additive  contamination effects  needs to be controlled to 3--7\%  of the single-frequency survey noise, or (in the case of correlated systematics) 6--11\% of the total combined noise levels.  Such percentages are consistent with the upper limits currently achieved for residual additive systematic contamination compared to survey noise
by small-aperture experiements \citep[e.g.,][]{Array:2015xqh}.  Assuming CMB-S4 will include a sustained effort to continue to control, understand, and model systematic effects down to levels limited by survey noise, these percentages provide reasonable benchmark requirements.

\begin{table}[htbp]
\begingroup
\newdimen\tblskip \tblskip=5pt
\vskip 9pt
\nointerlineskip
\vskip 1mm
\footnotesize
\setbox\tablebox=\vbox{
 \newdimen\digitwidth
 \setbox0=\hbox{\rm 0}
 \digitwidth=\wd0
 \catcode`*=\active
 \def*{\kern\digitwidth}
 \newdimen\signwidth
 \setbox0=\hbox{+}
 \signwidth=\wd0
 \catcode`!=\active
 \def!{\kern\signwidth}
\halign{\hbox to 1.6in{#\leaderfil}\tabskip 1em&
    \hfil#\hfil\tabskip 0.5em&
    \hfil#\hfil\tabskip 2em&
    \hfil#\hfil\tabskip 0.5em&
    \hfil#\hfil\tabskip 1em&
    \hfil#\hfil\tabskip 0.5em&
    \hfil#\hfil\tabskip 0.5em&
    \hfil#\hfil\tabskip 0.5em&
   \hfil#\hfil\tabskip=0pt\cr
\noalign{\doubleline}
\noalign{\vskip 3pt}
\omit&\multispan2\hfil Uncorrelated\hfil&\multispan2\hfil Correlated\hfil&\multispan2\hfil ILC\hfil&\multispan2\hfil Parametric\hfil\cr
\noalign{\vskip -2pt}
\omit&\multispan2\hrulefill&\multispan2\hrulefill&\multispan2\hrulefill&\multispan2\hrulefill\cr
\noalign{\vskip 2pt}
\omit \hfil Systematic\hfil&A [\%]&B [\%]&A [\%]&B [\%]&$\sigma(r)\times10^{4}$&$r$ bias $\times10^{4}$&$\sigma(r)\times10^{4}$&$r$ bias $\times10^{4}$\cr
\noalign{\vskip 4pt\hrule\vskip 4pt}
None& 0& 0& 0& 0& 5.3& 0.0*& 7.2& 0.0*\cr
Uncorrelated white& 3.3& 0& 0& 0& 6.0& 0.84& 8.0& 0.63\cr
Uncorrelated $1/\ell$& 0& 6.8& 0& 0& 5.0& 0.99& 7.0& 0.85\cr
Correlated white& 0& 0& 5.8& 0& 6.3& 1.2*& 7.3& 1.4*\cr
Correlated $1/\ell$& 0& 0& 0& 11& 5.2& 1.0*& 6.7& 0.97\cr
Uncorrelated white + $1/\ell$& 1.6& 3.5& 0& 0& 5.6& 0.89& 7.5& 0.76\cr
Correlated white + $1/\ell$& 0& 0& 2.9& 5.3& 5.5& 0.98& 6.9& 1.0*\cr
Both, white + $1/\ell$& 0.8& 1.7& 1.5& 2.6& 5.6& 1.1*& 7.9& 0.98\cr
\noalign{\vskip 3pt\hrule\vskip 3pt}}}
\endCMBSfourtable
\endgroup
\vskip 5pt
\caption{Results from re-analysis of map-based simulations containing systematics.
  We report sky model 3 and $r=0$, with additive systematic effects in varying combinations, the amplitudes of which are specified as percentages of survey noise, for the white (A) and $1/\ell$ (B) components.
  (These results are for a slightly different configuration with $1.0\times10^6$ 150-GHz-equivalent detector-years and a low-resolution 20-GHz channel.)}
\label{tab:sysres}
\end{table}

Results of simulating systematic errors in the determination of bandpasses vary by analysis method.  The construction of the ILC method makes it largely insensitive to such uncertainties.  The parametric analysis, which includes specific models of the frequency spectra of different foregrounds, shows biases on $r$ at the $1 \times 10^{-4}$ level for uncorrelated random deviations in bandcenter determination of 0.8\%, or for correlated deviations of 2\%, which we adopt as reasonable benchmark requirements to accommodate a variety of both blind and astrophysical foreground modeling approaches.

\subsection{Flowdown to distribution of SATs and delensing LAT across sites}
\label{sec:DSRupdates}

The semi-analytic optimization over frequency bands described in
Sect.~\ref{sec:semianaflowdownallocation} made simplistic
assumptions regarding the effect of varying the observed sky area.
In this section we describe updates to the framework which attempt to take
into account the impact of realistic observing strategies, as well as a
slightly more conservative approach to our delensing forecasts.

\subsubsection{Calculation of noise levels}

In reality one is not able to choose the number of detectors
in each frequency band in the continuously variable manner
shown in Fig.~\ref{fig:mapdepth} above.
For the reference design a realistic mapping of detectors into
dichroic optics tubes has been carried out while seeking to maintain
the band distribution as
determined in the optimization calculations---this
results in the configuration described in Sect.~\ref{sec:refdessum}.
We then scale the BICEP/Keck noise bandpower covariance matrix
in the same way as described in Sect.~\ref{sec:semianaflowdownallocation}
according to the number of detector-years and ratio of NETs.
A further re-scaling is then applied to account for sky coverages,
as explained in the next section.

\subsubsection{Sky coverage effects}\label{sec:skycov}

\begin{figure}[th!]
\begin{center}
\includegraphics[width=3in]{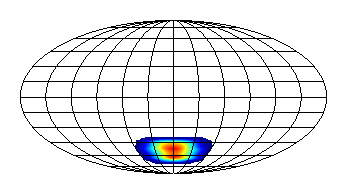} \\
\includegraphics[width=3in]{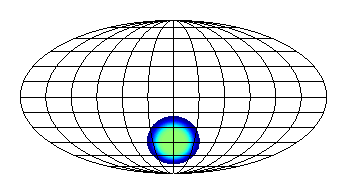}
\includegraphics[width=3in]{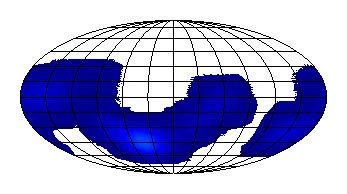}
\includegraphics[width=3in]{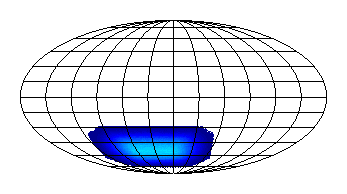}
\includegraphics[width=3in]{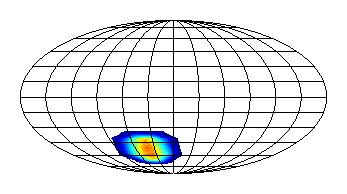}
\end{center}
\caption{Detector-second hit patterns on the sky for small aperture telescope
surveys.
top: the actual BICEP3 2017 hit pattern,
middle left: idealized circular pattern as used in Sect.~\ref{sec:cdtrepsims},
middle right: simulated ``Chile full'' pattern,
bottom left: simulated ``Pole wide'' pattern,
and bottom right: simulated ``Pole deep'' pattern.
Each pattern is normalized to the same sum and
the color scales are equal.
(The ``Chile deep'' and ``Chile shallow'' regions referred to in
the text are sub regions of the ``Chile full'' pattern.)}
\label{fig:relhits}
\end{figure}

The semi-analytic calculations of Sect.~\ref{sec:semianaflowdownallocation}
assumed a simplified re-scaling for sky area, while the map based simulations
of Sect.~\ref{sec:cdtrepsims} assumed an idealized circular sky patch which is
not actually achievable with a practical instrument from a site at any latitude.
Figure~\ref{fig:relhits} compares our prior assumptions to more realistic hit
patterns.
From Pole it is possible to concentrate the coverage onto
a compact region of sky,
but from Chile one has to observe different regions as the Earth turns, resulting
in a more extended coverage area.
The large instantaneous field of view of the SAT telescopes means
that there is minimum field size which can be achieved, and also that
there is always a strong ``edge taper'' in the coverage pattern.

We have performed a calculation which attempts to optimize simulated SAT
observations from Chile to produce the densest possible coverage
on a $\approx3$\% patch of low foreground sky resulting in the pattern
shown in the figure as ``Chile full.''
We segment this into its deepest part,
which we call ``Chile deep,'' and the remainder, which we call ``Chile
shallow.''

From Pole one can scan the same patch 24/7 with the size of
the observed patch basically controlled by the length of the scan
throw in Right Ascension.
A minimal length scan results in the pattern shown in the figure
as ``Pole deep.''
Lengthening the scan while remaining in low foreground sky results
in the pattern ``Pole wide.''
In the results below ``Pole deep'' and ``Pole wide'' are therefore
``either-or'' options.

Because the noise increases in regions with less observing time the
effective sky area for noise is larger than the effective sky area
for signal---and both of these also depend on the weighting applied
when analyzing the maps.
The patterns shown in Fig.~\ref{fig:relhits} have the effective sky fractions
reported in Table~\ref{tab:skyfrac}, assuming inverse noise variance weighting.

\begin{table}
 \centering
\begin{tabular}{|c|ccccc|}
\hline
   & \text{Pole deep} &  \text{Pole wide}&  \text{Chile deep}& \text{Chile shallow}  & \text{Chile full} \\ 
\hline
 $ f_{\rm sky}^{\rm noise} $ & 2.9 & 6.5 & 3.4& 20 & 18 \\ 
 $ f_{\rm sky}^{\rm signal} $& 1.9  & 4.3 & 2.4 & 10& 5.9\\ 
$ f_{\rm sky}^{\rm non-zero} $ & 5.0 & 12 & 5.0 & 47 & 52 \\ 
\hline
\end{tabular}
\vskip 5pt
  \caption{Effective sky fractions for signal and noise in \% for the
    observation patterns shown in Fig.~\ref{fig:relhits}, and the case of
    inverse noise variance weighting (i.e., equations~\ref{eq:scalsky}
    with $w_i = h_i$). We also report the fraction of sky with non zero
    coverage, $ f_{\rm sky}^{\rm non-zero}$. }
 \label{tab:skyfrac}
\end{table}

We can take account of the above effects by re-scaling the
BICEP/Keck bandpower covariance matrices (BPCMs) in a more sophisticated manner.
First, we need to scale the noise due to distributing the effort on a patch of sky larger than the original BK one.
The noise is scaled by the effective noise factor
\begin{equation}
 f_{\rm eff}^{\rm noise} =   \frac{\Omega_{\rm pix}}{4\pi} \frac{\sum_i  w_i^2 h_i^{-1} \sum_i  h_i}{\sum_i w_i^2},
\end{equation}
where $\Omega_{\rm pix}$ is the solid angle of a single pixel, $w_i$ are the weights for pixel $i$, and $h_i$ are the hit counts.

Second, since we observe a different number of modes, we need to scale the signal, noise and signal-cross-noise contributions of the BPCM by the factors

\begin{equation}
\label{eq:scalsky}
 f_{\rm sky}^{\rm noise} =   \frac{\Omega_{\rm pix}}{4\pi} \frac{(\sum_i  w_i^2 h_i^{-1})^2}{\sum_i w_i^4 h_i^{-2}},
\quad
 f_{\rm sky}^{\rm signal} =   \frac{\Omega_{\rm pix}}{4\pi} \frac{(\sum_i w_i^2)^2}{\sum_i w_i^{4}},
\quad
 f_{\rm sky}^{\rm cross} =  \frac{\Omega_{\rm pix}}{4\pi} \frac{\sum_i w_i^2 \sum_i  w_i^2 h_i^{-1}}{\sum_i w_i^4 h_i^{-2}}.
\end{equation}

We also need to take out the effect of these factors from the original BK BPCM.
In the BK analysis, the weights are the inverse noise variance, i.e., $w_i = h_i$.
In the case of CMB-S4, we will never be noise dominated, either due to an actual primordial signal, or due to the lensing and foreground residuals, so here we use the inverse variance of the total signal and noise to determine the optimal weighting.

\subsubsection{Delensing forecasts }
\label{sec:delens_new}

The CMB-S4 science goals can only be achieved if the majority of the lensing $B$ modes can be removed.
The optimization in Sect.~\ref{sec:semianaflowdownallocation} assumed
a single frequency channel assigned to the higher resolution delensing
observations.
The strength of polarized foregrounds at the relevant angular scales
is currently poorly constrained by data.
The reference design therefore includes some additional coverage at higher
and lower frequencies.

To forecast the delensing performance, we proceed in two steps. For a given LAT configuration and sky coverage, we derive the noise levels for an internal linear combination (ILC) that minimizes the variance of components with a frequency dependence that differs from that of a blackbody \cite{Tegmark:1995pn}. In this step, we assume that polarized foreground emission is dominated by Galactic synchrotron and thermal dust emission.
Using the ILC noise power spectrum, we then forecast the performance expected for iterative $EB$ delensing \cite{Smith:2010gu}, as shown in Fig.~\ref{fig:lens_res} above.

For the reference design for the Chile LATs described in Sect.~\ref{sec:refdessum} and a wide area survey covering 70\% of the sky, the two-step procedure predicts that 73\% of the lensing power can be removed in the ``Chile shallow'' region after seven years of observation.
Similarly, for the single LAT at the South Pole dedicated to delensing of the approximately 3\% ``Chile deep'' and ``Pole deep'' regions, we expect to be able to remove close to 90\% of the lensing power after 7 years of observation.

The numbers given above assume inverse noise variance weighting. For the reference design inverse noise variance weighting is typically suboptimal, and in all the forecasts presented below we employ weights that account for both signal (e.g., for $r=0$ lensing residual after foreground removal) and noise. For the same survey, this leads to slightly higher noise and lensing residuals, but
reduced $\sigma(r)$ due to reduced sample variance.

\subsubsection{Results}

The covariance matrices calculated as described above are used to produce the
results given in this section where the number, siting and coverage patterns
of the SATs are varied.
In all cases a delensing LAT at South Pole is assumed concentrating
its coverage on a small patch of sky, while delensing over larger
sky coverage is assumed to be available from the Chilean LATs.

As mentioned earlier, we split the Chilean coverage shown as ``Chile full'' in Fig.~\ref{fig:relhits} into a deep patch  ``Chile deep,'' where it overlaps with the ``Pole deep''
region, and call the remainder ``Chile shallow.''
We then make separate predictions for each region (with their very different
delensing levels) and add the $\sigma(r)$ results in simple inverse quadrature,
thereby making the small approximation of independence of the measured modes.
When we combine with Pole observations, we mimic a joint analysis by taking the sum of the ``Pole deep'' and ``Chile deep'' coverage maps, and computing the corresponding weights and lensing residuals.

Since some parts of the ``Chile shallow'' coverage lie closer to the galactic plane, we boost the foreground level by a factor of about 3 with respect to the deep patch.
This is based on analysis of maps with data driven foreground spatial variations.
We also apply Galactic cuts on top of the coverage maps, based on {\it Planck\/} polarization data.
We focus here on two cases, one where we mask out areas that are not part of the 58\% cleanest part of the full sky, and one where we use the 28\% cleanest.

\begin{figure}[htbp]
\begin{center}
\includegraphics[height=3.5in]{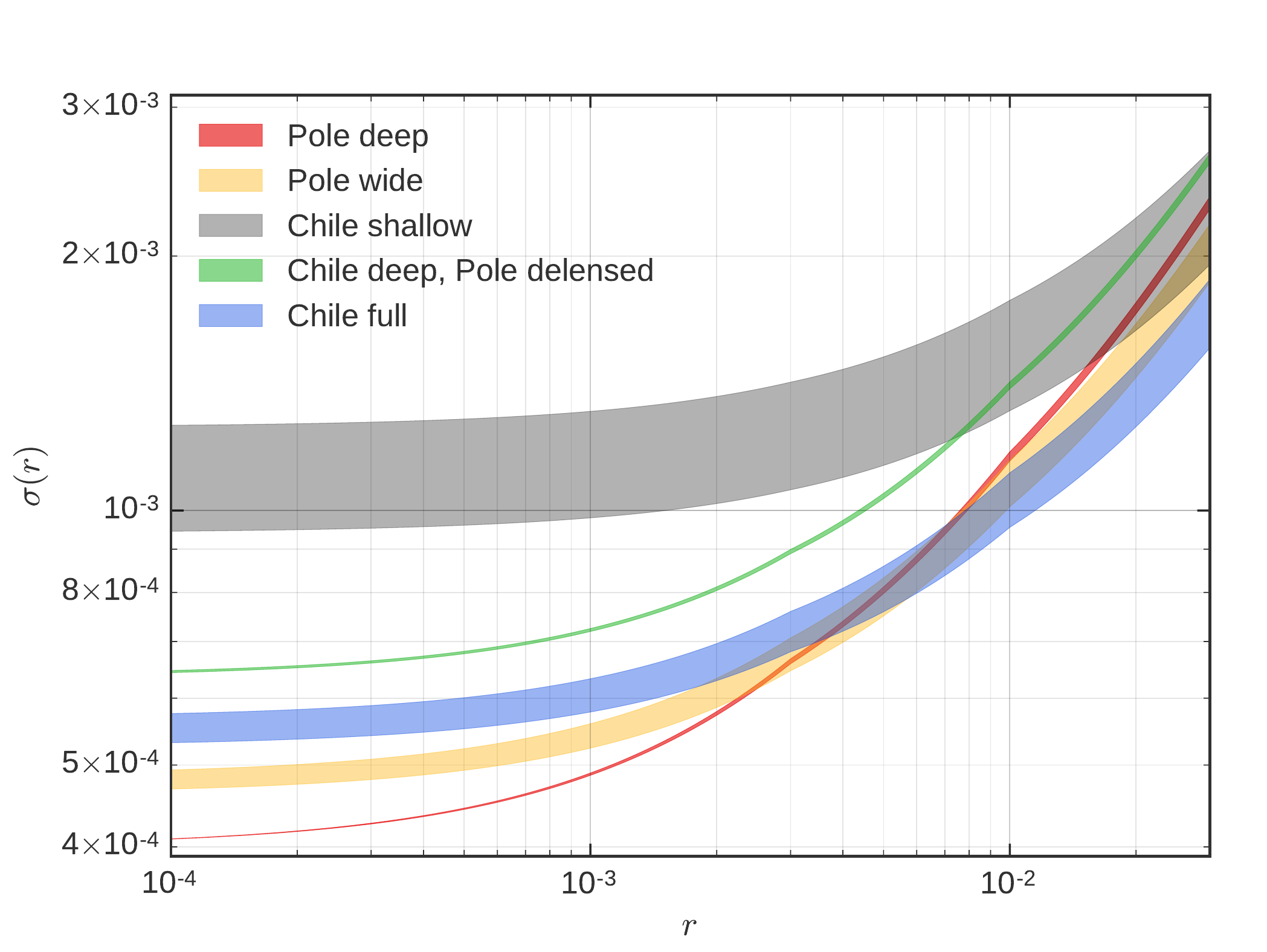}
\caption{Uncertainty on $r$ as a function of the value of $r$, for the different hit patterns shown in Fig.~\ref{fig:relhits}.
These results correspond to seven years of observation, assuming no
decorrelation parameters and observing efficiency in Chile equal to that at Pole.
Each band spans the range from a galactic cut which retains the cleanest 58\%
of the sky to one which retains the cleanest 28\% (based on {\it Planck\/} polarization data).
Note that we split the ``Chile full'' pattern into the deepest part (in green) that overlaps with the ``Pole deep'' map, thus delensed by the Pole delensing survey, and a shallow part that is delensed by the Chilean LATs (in gray). The combined ``Chile full'' (in blue) is the quadratic sum of these two.
}
\label{fig:sigrvsr_5curves}
\end{center}
\end{figure}

\begin{figure}[htbp]
\begin{center}
\includegraphics[height=4.7in]{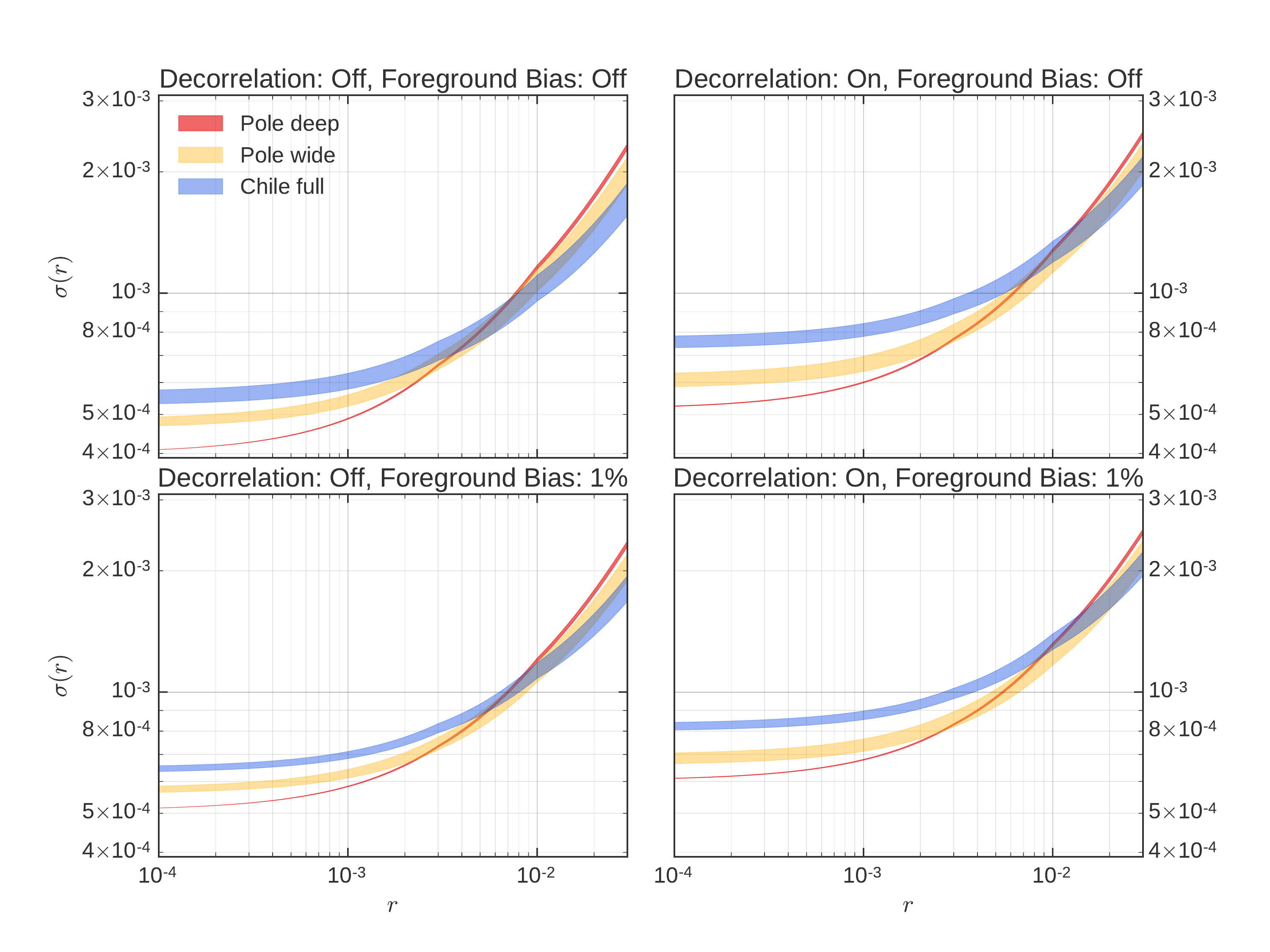}
\caption{Uncertainty on $r$ as a function of the value of $r$, for 18 tubes, for ``Pole deep,'' ``Pole wide'' and ``Chile full.'' The upper left is a subset of Fig.~\ref{fig:sigrvsr_5curves}. In the right panels, we turn on dust decorrelation ($\Delta_{\rm d}=0.97$, where $\Delta$ is defined in equation (F4) of the BK15 paper \cite{Ade:2018gkx}).
(Here we use a quadratic $\ell$-dependence, $g(\ell)=(\ell/80)^2$, as we did in the CDT report and the science book.)
In the lower panels, we add a foreground bias in quadrature to $\sigma(r)$. Its value is 1\% of the equivalent $r$ of the current foreground minimum of the BK15 data at $\ell=80$.
As in the previous figure, these results correspond to seven years of observation
and observing efficiency in Chile equal to that at Pole, and
each band shows different galactic cuts, based on {\it Planck\/} polarized foregrounds: the upper edge keeps the cleanest 28\% of the full sky, whereas the lower edge keeps the 58\% cleanest.
}
\label{fig:sigrvsr_4panels}
\end{center}
\end{figure}

\begin{figure}[htbp]
\begin{center}
\includegraphics[height=3.5in]{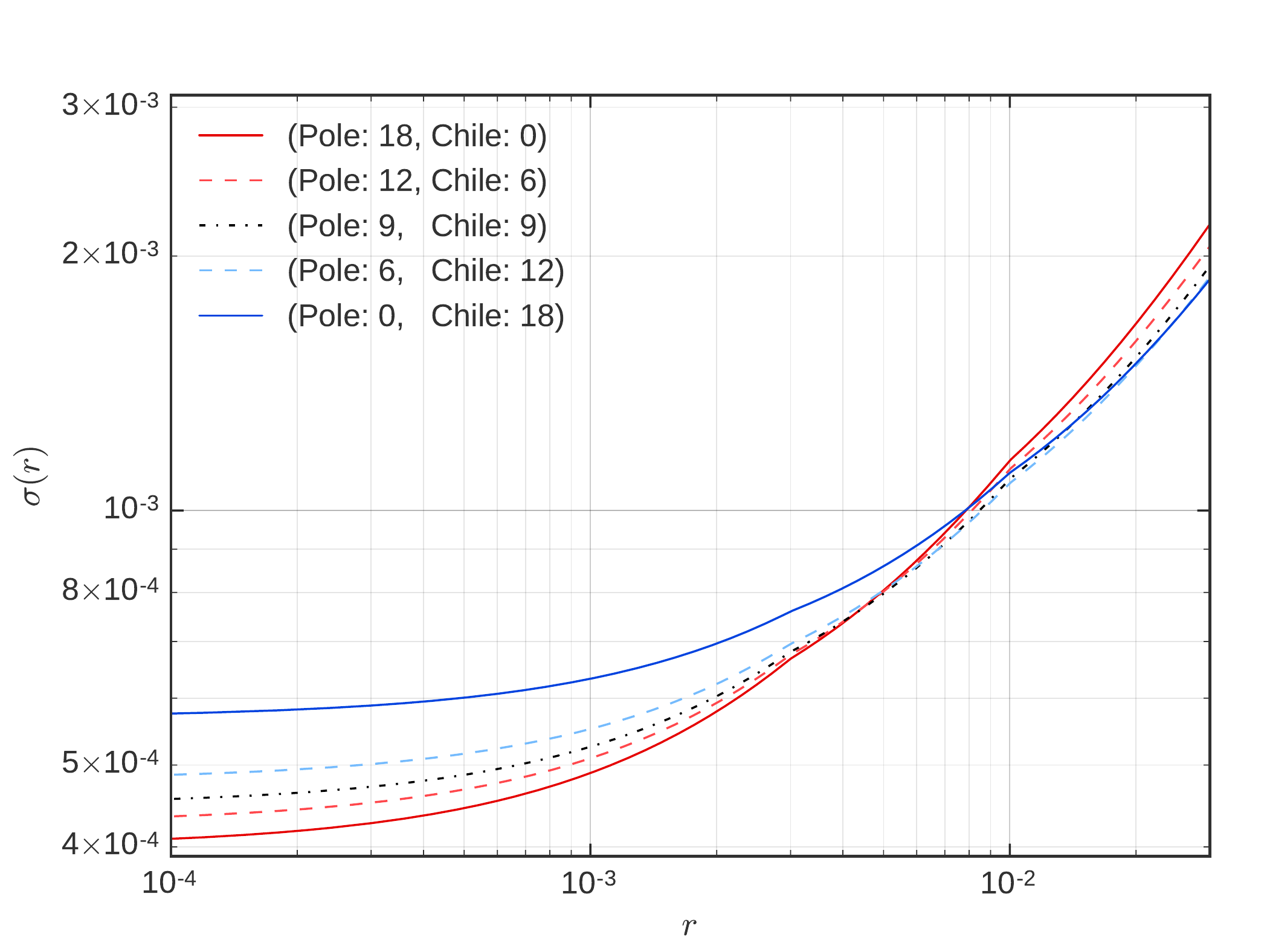}
\caption{Uncertainty on $r$ as a function of the value of $r$, for 18 optics tubes, and various splits between Pole and Chile siting.
As in the previous figures, these results correspond to seven years of observation
and observing efficiency in Chile equal to that at Pole.
For clarity, we only show the forecast using the 28\% cleanest polarized sky. Note that in the case where all the tubes are at Pole, we use the ``Pole wide'' pattern for $r\geq0.01$ since it yields better constraints.
}
\label{fig:sigrvsr_18tubes}
\end{center}
\end{figure}

In Figs.~\ref{fig:sigrvsr_5curves} and \ref{fig:sigrvsr_4panels}, we show the dependence of $\sigma(r)$ on $r$ for the different coverage masks.
We calculated these constraints for $r=0$, 0.003, 0.01 and 0.03, and show the linear interpolation between these points, for the different galactic cuts.
In figure \ref{fig:sigrvsr_18tubes}, we show how different sitings of a total of 18 tubes change our constraints on $r$.
Note that for $r\lsim0.003$ ``Pole deep'' is always better than ``Pole wide''
so we use the former in the results below when considering smaller values
of $r$.

In Tables~\ref{tab:sigr_7yr_nodecorr_0}--\ref{tab:sigr_7yr_nodecorr_0_58}
we present a set of $\sigma(r)$ results for seven years
of observations and $r$=0, while varying the number of optics tubes
in Pole and Chile over a wide range.
We show results for five different variants:
(i) with no decorrelation parameters and observing efficiency in Chile equal to that at Pole for the 28\% cleanest polarized sky;
(ii) assuming observing efficiency in Chile is half that at Pole;
(iii) assuming 1\% unmodeled foreground residual uncertainty;
(iv) assuming foreground decorrelation parameters are required in the re-analysis;
and (v) keeping the 58\% cleanest sky.
The results are moderately degraded in each of the first three variants.

While the forecast variant with 50\% observing efficiency from Chile has no impact on the sensitivity of the reference design, which locates all the small aperture
telescopes at the South Pole, it reinforces the preference for Pole-centric configurations.
Despite long histories of CMB observations at both sites, it is still quite difficult to make a clean comparison of their observing efficiencies.
Results from the BICEP/Keck program are responsible for leading constraints on $r$ for the last decade,
but it is not possible to disentangle the role of the observing site from other factors that contribute to the success of that program,
such as detector performance, instrument design, observing strategy, and operations management.
However, note also that even if we assume equal observing efficiency at both sites (Table~\ref{tab:sigr_7yr_nodecorr_0}), there is still a difference
between $\sigma(r)$ obtained for equivalent numbers of optics tubes in Chile vs Pole, due to sky fraction and delensing.

Note that the results where the total number of optics tubes is significantly
different to eighteen are subject to a caveat.
The delensing effort is assumed to be held fixed in these calculations---one LAT
at South Pole and two in Chile.
In principle as the total effort it varied away from the reference
design one should re-optimize the fraction of delensing effort as per Fig.~\ref{fig:sigr}.

In Tables~\ref{tab:95CL_7yr_nodecorr_0}--\ref{tab:95CL_7yr_nodecorr_0_58}  and Tables~\ref{tab:sigdeg_7yr_nodecorr_0.003}--\ref{tab:sigdeg_7yr_nodecorr_0.003_58} we
convert the results into 95\% confidence limit  and detection significance
under the scenario that $r=0$ and $r=0.003$ respectively, and show the same four variant cases. For these statistics, we assume the bandpowers have the model expectations values and then calculate the posterior probability distribution for $r$, marginalized
over all foreground parameters. We use the posterior obtained in the map based analysis for different masks to compute a functional form that depends on the noise residuals and the value of $r$. We scale these posteriors based on the noise sky fraction factors from \ref{eq:scalsky}. This distribution is significantly non-Gaussian, leading to detection
significance that is higher than we would naively calculate from $\sigma(r)$.
Combined detection significance for Deep and Shallow surveys is calculated by simply multiplying the posteriors.
Note that we do not quote these results in the main body of the report as we have not yet taken into account the
impact of expected fluctuations in the measured bandpowers. We plan to take these into account with simulations in future work.

\begin{table}
 \centering
\begin{tabular}{c|cccccc}
\hline\hline
   \text{Chile}$\backslash$\text{Pole} & 0& 6& 9& 12         & 18         & 30\\
\hline
   0 && \cellcolor[rgb]{1.000000,0.690888,0.000000} 6.3& \cellcolor[rgb]{0.915033,0.957683,0.000000} 5.0& \cellcolor[rgb]{0.653595,0.827477,0.000000} 4.5& \cellcolor[rgb]{0.444444,0.723312,0.000000} \textbf{4.0} & \cellcolor[rgb]{0.235294,0.619146,0.000000} 3.5 \\
   6 & \cellcolor[rgb]{0.994054,0.000000,0.000000} 12 & \cellcolor[rgb]{1.000000,0.949250,0.000000} 5.5& \cellcolor[rgb]{0.705882,0.853518,0.000000} 4.7& \cellcolor[rgb]{0.549020,0.775394,0.000000} \textbf{4.3} & \cellcolor[rgb]{0.392157,0.697270,0.000000} 3.9& \cellcolor[rgb]{0.183007,0.593105,0.000000} 3.5 \\
   9 & \cellcolor[rgb]{1.000000,0.392464,0.000000} 8.7& \cellcolor[rgb]{0.915033,0.957683,0.000000} 5.1& \cellcolor[rgb]{0.653595,0.827477,0.000000} \textbf{4.5} & \cellcolor[rgb]{0.496732,0.749353,0.000000} 4.1& \cellcolor[rgb]{0.339869,0.671229,0.000000} 3.8& \cellcolor[rgb]{0.183007,0.593105,0.000000} 3.4 \\
  12 & \cellcolor[rgb]{1.000000,0.568397,0.000000} 7.1& \cellcolor[rgb]{0.758170,0.879559,0.000000} \textbf{4.8} & \cellcolor[rgb]{0.549020,0.775394,0.000000} 4.3& \cellcolor[rgb]{0.444444,0.723312,0.000000} 4.0& \cellcolor[rgb]{0.287582,0.645188,0.000000} 3.7& \cellcolor[rgb]{0.130719,0.567064,0.000000} 3.4 \\
  18 & \cellcolor[rgb]{1.000000,0.875433,0.000000} \textbf{5.7} & \cellcolor[rgb]{0.601307,0.801435,0.000000} 4.4& \cellcolor[rgb]{0.444444,0.723312,0.000000} 4.1& \cellcolor[rgb]{0.339869,0.671229,0.000000} 3.8& \cellcolor[rgb]{0.235294,0.619146,0.000000} 3.6& \cellcolor[rgb]{0.130719,0.567064,0.000000} 3.3 \\
  30 & \cellcolor[rgb]{0.601307,0.801435,0.000000} 4.4& \cellcolor[rgb]{0.339869,0.671229,0.000000} 3.9& \cellcolor[rgb]{0.287582,0.645188,0.000000} 3.7& \cellcolor[rgb]{0.235294,0.619146,0.000000} 3.6& \cellcolor[rgb]{0.183007,0.593105,0.000000} 3.4& \cellcolor[rgb]{0.078431,0.541023,0.000000} 3.2 \\
\hline
\end{tabular}
 \caption{Combined $10^4 \times \sigma(r)$, assuming r=0 after 7 years of observation, keeping only the 28\% cleanest part of the sky, assuming no decorrelation and observing efficiency in Chile same as at Pole.}
 \label{tab:sigr_7yr_nodecorr_0}
 \centering
\begin{tabular}{c|cccccc}
\hline\hline
   \text{Chile}$\backslash$\text{Pole} & 0& 6& 9& 12         & 18         & 30\\
\hline
   0 && \cellcolor[rgb]{1.000000,0.690888,0.000000} 6.3& \cellcolor[rgb]{0.915033,0.957683,0.000000} 5.0& \cellcolor[rgb]{0.653595,0.827477,0.000000} 4.5& \cellcolor[rgb]{0.444444,0.723312,0.000000} \textbf{4.0} & \cellcolor[rgb]{0.235294,0.619146,0.000000} 3.5 \\
   6 & \cellcolor[rgb]{0.756196,0.000000,0.000000} 22 & \cellcolor[rgb]{1.000000,0.801615,0.000000} 5.9& \cellcolor[rgb]{0.810458,0.905600,0.000000} 4.9& \cellcolor[rgb]{0.601307,0.801435,0.000000} \textbf{4.4} & \cellcolor[rgb]{0.392157,0.697270,0.000000} 3.9& \cellcolor[rgb]{0.183007,0.593105,0.000000} 3.5 \\
   9 & \cellcolor[rgb]{0.931616,0.000000,0.000000} 15 & \cellcolor[rgb]{1.000000,0.875433,0.000000} 5.6& \cellcolor[rgb]{0.758170,0.879559,0.000000} \textbf{4.8} & \cellcolor[rgb]{0.549020,0.775394,0.000000} 4.3& \cellcolor[rgb]{0.392157,0.697270,0.000000} 3.9& \cellcolor[rgb]{0.183007,0.593105,0.000000} 3.5 \\
  12 & \cellcolor[rgb]{1.000000,0.054133,0.000000} 12 & \cellcolor[rgb]{1.000000,0.949250,0.000000} \textbf{5.4} & \cellcolor[rgb]{0.705882,0.853518,0.000000} 4.6& \cellcolor[rgb]{0.549020,0.775394,0.000000} 4.2& \cellcolor[rgb]{0.339869,0.671229,0.000000} 3.9& \cellcolor[rgb]{0.183007,0.593105,0.000000} 3.5 \\
  18 & \cellcolor[rgb]{1.000000,0.378931,0.000000} \textbf{8.8} & \cellcolor[rgb]{0.915033,0.957683,0.000000} 5.1& \cellcolor[rgb]{0.653595,0.827477,0.000000} 4.5& \cellcolor[rgb]{0.496732,0.749353,0.000000} 4.1& \cellcolor[rgb]{0.339869,0.671229,0.000000} 3.8& \cellcolor[rgb]{0.183007,0.593105,0.000000} 3.4 \\
  30 & \cellcolor[rgb]{1.000000,0.690888,0.000000} 6.2& \cellcolor[rgb]{0.653595,0.827477,0.000000} 4.6& \cellcolor[rgb]{0.496732,0.749353,0.000000} 4.2& \cellcolor[rgb]{0.392157,0.697270,0.000000} 3.9& \cellcolor[rgb]{0.287582,0.645188,0.000000} 3.7& \cellcolor[rgb]{0.130719,0.567064,0.000000} 3.3 \\
\hline
\end{tabular}
 \caption{Same as Table~\ref{tab:sigr_7yr_nodecorr_0}, but assuming 50\% Chilean efficiency.}
 \label{tab:sigr_7yr_nodecorr_0_half}
 \centering
\begin{tabular}{c|cccccc}
\hline\hline
   \text{Chile}$\backslash$\text{Pole} & 0& 6& 9& 12         & 18         & 30\\
\hline
   0 && \cellcolor[rgb]{1.000000,0.568397,0.000000} 7.0& \cellcolor[rgb]{1.000000,0.801615,0.000000} 5.9& \cellcolor[rgb]{1.000000,0.949250,0.000000} 5.5& \cellcolor[rgb]{0.915033,0.957683,0.000000} \textbf{5.1} & \cellcolor[rgb]{0.758170,0.879559,0.000000} 4.7 \\
   6 & \cellcolor[rgb]{0.985134,0.000000,0.000000} 13 & \cellcolor[rgb]{1.000000,0.690888,0.000000} 6.3& \cellcolor[rgb]{1.000000,0.912341,0.000000} 5.6& \cellcolor[rgb]{0.967320,0.983724,0.000000} \textbf{5.2} & \cellcolor[rgb]{0.862745,0.931642,0.000000} 4.9& \cellcolor[rgb]{0.705882,0.853518,0.000000} 4.6 \\
   9 & \cellcolor[rgb]{1.000000,0.324798,0.000000} 9.2& \cellcolor[rgb]{1.000000,0.801615,0.000000} 5.9& \cellcolor[rgb]{1.000000,0.949250,0.000000} \textbf{5.4} & \cellcolor[rgb]{0.915033,0.957683,0.000000} 5.1& \cellcolor[rgb]{0.810458,0.905600,0.000000} 4.9& \cellcolor[rgb]{0.705882,0.853518,0.000000} 4.6 \\
  12 & \cellcolor[rgb]{1.000000,0.487197,0.000000} 7.8& \cellcolor[rgb]{1.000000,0.875433,0.000000} \textbf{5.7} & \cellcolor[rgb]{0.967320,0.983724,0.000000} 5.2& \cellcolor[rgb]{0.862745,0.931642,0.000000} 5.0& \cellcolor[rgb]{0.758170,0.879559,0.000000} 4.8& \cellcolor[rgb]{0.653595,0.827477,0.000000} 4.5 \\
  18 & \cellcolor[rgb]{1.000000,0.636063,0.000000} \textbf{6.5} & \cellcolor[rgb]{1.000000,0.986159,0.000000} 5.3& \cellcolor[rgb]{0.915033,0.957683,0.000000} 5.0& \cellcolor[rgb]{0.810458,0.905600,0.000000} 4.9& \cellcolor[rgb]{0.705882,0.853518,0.000000} 4.7& \cellcolor[rgb]{0.653595,0.827477,0.000000} 4.5 \\
  30 & \cellcolor[rgb]{1.000000,0.949250,0.000000} 5.4& \cellcolor[rgb]{0.810458,0.905600,0.000000} 4.9& \cellcolor[rgb]{0.758170,0.879559,0.000000} 4.8& \cellcolor[rgb]{0.705882,0.853518,0.000000} 4.7& \cellcolor[rgb]{0.653595,0.827477,0.000000} 4.5& \cellcolor[rgb]{0.601307,0.801435,0.000000} 4.4 \\
\hline
\end{tabular}
 \caption{Same as Table~\ref{tab:sigr_7yr_nodecorr_0}, but assuming 1\% unmodeled foreground residual uncertainty.}
 \label{tab:sigr_7yr_nodecorr_0_fg}
 \centering
\begin{tabular}{c|cccccc}
\hline\hline
   \text{Chile}$\backslash$\text{Pole} & 0& 6& 9& 12         & 18         & 30\\
\hline
   0 && \cellcolor[rgb]{1.000000,0.419531,0.000000} 8.4& \cellcolor[rgb]{1.000000,0.608997,0.000000} 6.7& \cellcolor[rgb]{1.000000,0.801615,0.000000} 6.0& \cellcolor[rgb]{0.967320,0.983724,0.000000} \textbf{5.2} & \cellcolor[rgb]{0.601307,0.801435,0.000000} 4.4 \\
   6 & \cellcolor[rgb]{0.895937,0.000000,0.000000} 16 & \cellcolor[rgb]{1.000000,0.554864,0.000000} 7.3& \cellcolor[rgb]{1.000000,0.727797,0.000000} 6.2& \cellcolor[rgb]{1.000000,0.912341,0.000000} \textbf{5.6} & \cellcolor[rgb]{0.862745,0.931642,0.000000} 5.0& \cellcolor[rgb]{0.549020,0.775394,0.000000} 4.3 \\
   9 & \cellcolor[rgb]{1.000000,0.027067,0.000000} 12 & \cellcolor[rgb]{1.000000,0.608997,0.000000} 6.8& \cellcolor[rgb]{1.000000,0.801615,0.000000} \textbf{5.9} & \cellcolor[rgb]{1.000000,0.949250,0.000000} 5.4& \cellcolor[rgb]{0.810458,0.905600,0.000000} 4.9& \cellcolor[rgb]{0.549020,0.775394,0.000000} 4.3 \\
  12 & \cellcolor[rgb]{1.000000,0.270665,0.000000} 9.7& \cellcolor[rgb]{1.000000,0.653979,0.000000} \textbf{6.4} & \cellcolor[rgb]{1.000000,0.875433,0.000000} 5.7& \cellcolor[rgb]{1.000000,0.986159,0.000000} 5.3& \cellcolor[rgb]{0.758170,0.879559,0.000000} 4.8& \cellcolor[rgb]{0.549020,0.775394,0.000000} 4.2 \\
  18 & \cellcolor[rgb]{1.000000,0.487197,0.000000} \textbf{7.8} & \cellcolor[rgb]{1.000000,0.838524,0.000000} 5.8& \cellcolor[rgb]{1.000000,0.986159,0.000000} 5.3& \cellcolor[rgb]{0.862745,0.931642,0.000000} 5.0& \cellcolor[rgb]{0.705882,0.853518,0.000000} 4.6& \cellcolor[rgb]{0.496732,0.749353,0.000000} 4.1 \\
  30 & \cellcolor[rgb]{1.000000,0.764706,0.000000} 6.0& \cellcolor[rgb]{0.915033,0.957683,0.000000} 5.1& \cellcolor[rgb]{0.810458,0.905600,0.000000} 4.8& \cellcolor[rgb]{0.705882,0.853518,0.000000} 4.6& \cellcolor[rgb]{0.601307,0.801435,0.000000} 4.3& \cellcolor[rgb]{0.392157,0.697270,0.000000} 4.0 \\
\hline
\end{tabular}
 \caption{Same as Table~\ref{tab:sigr_7yr_nodecorr_0}, but assuming additional foreground decorrelation parameters.}
 \label{tab:sigr_7yr_decorr_0}
 \centering
\begin{tabular}{c|cccccc}
\hline\hline
   \text{Chile}$\backslash$\text{Pole} & 0& 6& 9& 12         & 18         & 30\\
\hline
   0 && \cellcolor[rgb]{1.000000,0.690888,0.000000} 6.3& \cellcolor[rgb]{0.862745,0.931642,0.000000} 5.0& \cellcolor[rgb]{0.653595,0.827477,0.000000} 4.5& \cellcolor[rgb]{0.444444,0.723312,0.000000} \textbf{4.0} & \cellcolor[rgb]{0.235294,0.619146,0.000000} 3.5 \\
   6 & \cellcolor[rgb]{1.000000,0.081200,0.000000} 11 & \cellcolor[rgb]{1.000000,0.949250,0.000000} 5.4& \cellcolor[rgb]{0.705882,0.853518,0.000000} 4.6& \cellcolor[rgb]{0.549020,0.775394,0.000000} \textbf{4.2} & \cellcolor[rgb]{0.339869,0.671229,0.000000} 3.8& \cellcolor[rgb]{0.183007,0.593105,0.000000} 3.5 \\
   9 & \cellcolor[rgb]{1.000000,0.460131,0.000000} 8.0& \cellcolor[rgb]{0.862745,0.931642,0.000000} 5.0& \cellcolor[rgb]{0.601307,0.801435,0.000000} \textbf{4.4} & \cellcolor[rgb]{0.444444,0.723312,0.000000} 4.1& \cellcolor[rgb]{0.287582,0.645188,0.000000} 3.7& \cellcolor[rgb]{0.130719,0.567064,0.000000} 3.4 \\
  12 & \cellcolor[rgb]{1.000000,0.622530,0.000000} 6.6& \cellcolor[rgb]{0.705882,0.853518,0.000000} \textbf{4.6} & \cellcolor[rgb]{0.496732,0.749353,0.000000} 4.2& \cellcolor[rgb]{0.392157,0.697270,0.000000} 3.9& \cellcolor[rgb]{0.287582,0.645188,0.000000} 3.6& \cellcolor[rgb]{0.130719,0.567064,0.000000} 3.3 \\
  18 & \cellcolor[rgb]{0.967320,0.983724,0.000000} \textbf{5.3} & \cellcolor[rgb]{0.496732,0.749353,0.000000} 4.2& \cellcolor[rgb]{0.392157,0.697270,0.000000} 3.9& \cellcolor[rgb]{0.287582,0.645188,0.000000} 3.7& \cellcolor[rgb]{0.183007,0.593105,0.000000} 3.5& \cellcolor[rgb]{0.078431,0.541023,0.000000} 3.2 \\
  30 & \cellcolor[rgb]{0.444444,0.723312,0.000000} 4.1& \cellcolor[rgb]{0.287582,0.645188,0.000000} 3.6& \cellcolor[rgb]{0.183007,0.593105,0.000000} 3.5& \cellcolor[rgb]{0.130719,0.567064,0.000000} 3.4& \cellcolor[rgb]{0.078431,0.541023,0.000000} 3.2& \cellcolor[rgb]{0.026144,0.514981,0.000000} 3.1 \\
\hline
\end{tabular}
\caption{Same as Table~\ref{tab:sigr_7yr_nodecorr_0}, but assuming we keep the 58\% cleanest part of the full sky}
 \label{tab:sigr_7yr_nodecorr_0_58}
\end{table}

\begin{table}
 \centering
\begin{tabular}{c|cccccc}
\hline\hline
   \text{Chile}$\backslash$\text{Pole} & 0         & 6         & 9& 12         & 18         & 30\\
\hline
   0 & & \cellcolor[rgb]{1.000000,0.595463,0.000000} 14& \cellcolor[rgb]{1.000000,0.949250,0.000000} 11 & \cellcolor[rgb]{0.810458,0.905600,0.000000} 9.7& \cellcolor[rgb]{0.549020,0.775394,0.000000} \textbf{8.6} & \cellcolor[rgb]{0.339869,0.671229,0.000000} 7.6 \\
   6 & \cellcolor[rgb]{0.979187,0.000000,0.000000} 26& \cellcolor[rgb]{1.000000,0.801615,0.000000} 12& \cellcolor[rgb]{0.862745,0.931642,0.000000} 10 & \cellcolor[rgb]{0.653595,0.827477,0.000000} \textbf{9.1} & \cellcolor[rgb]{0.496732,0.749353,0.000000} 8.3& \cellcolor[rgb]{0.287582,0.645188,0.000000} 7.4 \\
   9 & \cellcolor[rgb]{1.000000,0.324798,0.000000} 18& \cellcolor[rgb]{1.000000,0.949250,0.000000} 11& \cellcolor[rgb]{0.758170,0.879559,0.000000} \textbf{9.6} & \cellcolor[rgb]{0.601307,0.801435,0.000000} 8.8& \cellcolor[rgb]{0.444444,0.723312,0.000000} 8.1& \cellcolor[rgb]{0.287582,0.645188,0.000000} 7.3 \\
  12 & \cellcolor[rgb]{1.000000,0.527797,0.000000} 15& \cellcolor[rgb]{0.915033,0.957683,0.000000} \textbf{10} & \cellcolor[rgb]{0.705882,0.853518,0.000000} 9.2& \cellcolor[rgb]{0.549020,0.775394,0.000000} 8.6& \cellcolor[rgb]{0.392157,0.697270,0.000000} 7.9& \cellcolor[rgb]{0.235294,0.619146,0.000000} 7.2 \\
  18 & \cellcolor[rgb]{1.000000,0.764706,0.000000} \textbf{12} & \cellcolor[rgb]{0.705882,0.853518,0.000000} 9.3         & \cellcolor[rgb]{0.549020,0.775394,0.000000} 8.6& \cellcolor[rgb]{0.444444,0.723312,0.000000} 8.2& \cellcolor[rgb]{0.339869,0.671229,0.000000} 7.7& \cellcolor[rgb]{0.235294,0.619146,0.000000} 7.1 \\
  30 & \cellcolor[rgb]{0.705882,0.853518,0.000000} 9.3         & \cellcolor[rgb]{0.444444,0.723312,0.000000} 8.1         & \cellcolor[rgb]{0.392157,0.697270,0.000000} 7.8& \cellcolor[rgb]{0.339869,0.671229,0.000000} 7.5& \cellcolor[rgb]{0.235294,0.619146,0.000000} 7.2& \cellcolor[rgb]{0.130719,0.567064,0.000000} 6.8 \\
\hline
\end{tabular}
 \caption{Combined $10^4 \times 95$\% C.L., for r=0 after 7 years of observation, keeping only the 28\% cleanest part of the sky, assuming no decorrelation and observing efficiency in Chile same as at Pole.}
 \label{tab:95CL_7yr_nodecorr_0}
 \centering
\begin{tabular}{c|cccccc}
\hline\hline
   \text{Chile}$\backslash$\text{Pole} & 0         & 6         & 9         & 12         & 18         & 30\\
\hline
   0 & & \cellcolor[rgb]{1.000000,0.595463,0.000000} 14& \cellcolor[rgb]{1.000000,0.949250,0.000000} 11& \cellcolor[rgb]{0.810458,0.905600,0.000000} 9.7& \cellcolor[rgb]{0.549020,0.775394,0.000000} \textbf{8.6} & \cellcolor[rgb]{0.339869,0.671229,0.000000} 7.6 \\
   6 & \cellcolor[rgb]{0.726464,0.000000,0.000000} 46& \cellcolor[rgb]{1.000000,0.653979,0.000000} 13& \cellcolor[rgb]{0.967320,0.983724,0.000000} 10& \cellcolor[rgb]{0.758170,0.879559,0.000000} \textbf{9.4} & \cellcolor[rgb]{0.549020,0.775394,0.000000} 8.5& \cellcolor[rgb]{0.339869,0.671229,0.000000} 7.5 \\
   9 & \cellcolor[rgb]{0.910804,0.000000,0.000000} 31& \cellcolor[rgb]{1.000000,0.764706,0.000000} 12& \cellcolor[rgb]{0.915033,0.957683,0.000000} \textbf{10} & \cellcolor[rgb]{0.705882,0.853518,0.000000} 9.2& \cellcolor[rgb]{0.496732,0.749353,0.000000} 8.4& \cellcolor[rgb]{0.287582,0.645188,0.000000} 7.5 \\
  12 & \cellcolor[rgb]{0.997027,0.000000,0.000000} 24& \cellcolor[rgb]{1.000000,0.838524,0.000000} \textbf{12} & \cellcolor[rgb]{0.862745,0.931642,0.000000} 9.9         & \cellcolor[rgb]{0.653595,0.827477,0.000000} 9.1& \cellcolor[rgb]{0.496732,0.749353,0.000000} 8.3& \cellcolor[rgb]{0.287582,0.645188,0.000000} 7.4 \\
  18 & \cellcolor[rgb]{1.000000,0.324798,0.000000} \textbf{19} & \cellcolor[rgb]{1.000000,0.949250,0.000000} 11& \cellcolor[rgb]{0.758170,0.879559,0.000000} 9.5         & \cellcolor[rgb]{0.601307,0.801435,0.000000} 8.8& \cellcolor[rgb]{0.444444,0.723312,0.000000} 8.1& \cellcolor[rgb]{0.287582,0.645188,0.000000} 7.3 \\
  30 & \cellcolor[rgb]{1.000000,0.622530,0.000000} 13& \cellcolor[rgb]{0.810458,0.905600,0.000000} 9.6         & \cellcolor[rgb]{0.601307,0.801435,0.000000} 8.8         & \cellcolor[rgb]{0.496732,0.749353,0.000000} 8.3& \cellcolor[rgb]{0.392157,0.697270,0.000000} 7.8& \cellcolor[rgb]{0.235294,0.619146,0.000000} 7.1 \\
\hline
\end{tabular}
 \caption{Same as Table~\ref{tab:95CL_7yr_nodecorr_0}, but assuming 50\% Chilean efficiency.}
 \label{tab:95CL_7yr_nodecorr_0_half}
 \centering
\begin{tabular}{c|cccccc}
\hline\hline
   \text{Chile}$\backslash$\text{Pole} & 0         & 6         & 9         & 12        & 18        & 30\\
\hline
   0 & & \cellcolor[rgb]{1.000000,0.500730,0.000000} 15& \cellcolor[rgb]{1.000000,0.653979,0.000000} 13& \cellcolor[rgb]{1.000000,0.801615,0.000000} 12& \cellcolor[rgb]{1.000000,0.949250,0.000000} \textbf{11} & \cellcolor[rgb]{0.915033,0.957683,0.000000} 10  \\
   6 & \cellcolor[rgb]{0.967295,0.000000,0.000000} 27& \cellcolor[rgb]{1.000000,0.608997,0.000000} 13& \cellcolor[rgb]{1.000000,0.801615,0.000000} 12& \cellcolor[rgb]{1.000000,0.912341,0.000000} \textbf{11} & \cellcolor[rgb]{1.000000,0.986159,0.000000} 11& \cellcolor[rgb]{0.862745,0.931642,0.000000} 9.9 \\
   9 & \cellcolor[rgb]{1.000000,0.257132,0.000000} 20& \cellcolor[rgb]{1.000000,0.690888,0.000000} 13& \cellcolor[rgb]{1.000000,0.838524,0.000000} \textbf{11} & \cellcolor[rgb]{1.000000,0.949250,0.000000} 11& \cellcolor[rgb]{0.967320,0.983724,0.000000} 10& \cellcolor[rgb]{0.810458,0.905600,0.000000} 9.8 \\
  12 & \cellcolor[rgb]{1.000000,0.446597,0.000000} 16& \cellcolor[rgb]{1.000000,0.764706,0.000000} \textbf{12} & \cellcolor[rgb]{1.000000,0.912341,0.000000} 11& \cellcolor[rgb]{1.000000,0.986159,0.000000} 11& \cellcolor[rgb]{0.915033,0.957683,0.000000} 10& \cellcolor[rgb]{0.810458,0.905600,0.000000} 9.7 \\
  18 & \cellcolor[rgb]{1.000000,0.595463,0.000000} \textbf{14} & \cellcolor[rgb]{1.000000,0.875433,0.000000} 11& \cellcolor[rgb]{1.000000,0.986159,0.000000} 11& \cellcolor[rgb]{0.967320,0.983724,0.000000} 10& \cellcolor[rgb]{0.862745,0.931642,0.000000} 9.9         & \cellcolor[rgb]{0.758170,0.879559,0.000000} 9.5 \\
  30 & \cellcolor[rgb]{1.000000,0.875433,0.000000} 11& \cellcolor[rgb]{0.967320,0.983724,0.000000} 10& \cellcolor[rgb]{0.862745,0.931642,0.000000} 10& \cellcolor[rgb]{0.862745,0.931642,0.000000} 9.9         & \cellcolor[rgb]{0.758170,0.879559,0.000000} 9.6         & \cellcolor[rgb]{0.705882,0.853518,0.000000} 9.3 \\
\hline
\end{tabular}
 \caption{Same as Table~\ref{tab:95CL_7yr_nodecorr_0}, but assuming 1\% unmodeled foreground residual uncertainty.}
 \label{tab:95CL_7yr_nodecorr_0_fg}
 \centering
\begin{tabular}{c|cccccc}
\hline\hline
   \text{Chile}$\backslash$\text{Pole} & 0         & 6         & 9         & 12        & 18        & 30\\
\hline
   0 & & \cellcolor[rgb]{1.000000,0.338331,0.000000} 18& \cellcolor[rgb]{1.000000,0.541330,0.000000} 15& \cellcolor[rgb]{1.000000,0.653979,0.000000} 13& \cellcolor[rgb]{1.000000,0.912341,0.000000} \textbf{11} & \cellcolor[rgb]{0.758170,0.879559,0.000000} 9.5 \\
   6 & \cellcolor[rgb]{0.872152,0.000000,0.000000} 34& \cellcolor[rgb]{1.000000,0.487197,0.000000} 16& \cellcolor[rgb]{1.000000,0.622530,0.000000} 13& \cellcolor[rgb]{1.000000,0.764706,0.000000} \textbf{12} & \cellcolor[rgb]{1.000000,0.986159,0.000000} 11& \cellcolor[rgb]{0.705882,0.853518,0.000000} 9.3 \\
   9 & \cellcolor[rgb]{0.991080,0.000000,0.000000} 25& \cellcolor[rgb]{1.000000,0.554864,0.000000} 14& \cellcolor[rgb]{1.000000,0.653979,0.000000} \textbf{13} & \cellcolor[rgb]{1.000000,0.838524,0.000000} 12& \cellcolor[rgb]{0.967320,0.983724,0.000000} 10& \cellcolor[rgb]{0.705882,0.853518,0.000000} 9.2 \\
  12 & \cellcolor[rgb]{1.000000,0.202999,0.000000} 21& \cellcolor[rgb]{1.000000,0.608997,0.000000} \textbf{14} & \cellcolor[rgb]{1.000000,0.764706,0.000000} 12& \cellcolor[rgb]{1.000000,0.875433,0.000000} 11& \cellcolor[rgb]{0.915033,0.957683,0.000000} 10& \cellcolor[rgb]{0.653595,0.827477,0.000000} 9.1 \\
  18 & \cellcolor[rgb]{1.000000,0.446597,0.000000} \textbf{16} & \cellcolor[rgb]{1.000000,0.727797,0.000000} 12& \cellcolor[rgb]{1.000000,0.875433,0.000000} 11& \cellcolor[rgb]{1.000000,0.986159,0.000000} 11& \cellcolor[rgb]{0.810458,0.905600,0.000000} 9.8         & \cellcolor[rgb]{0.601307,0.801435,0.000000} 8.8 \\
  30 & \cellcolor[rgb]{1.000000,0.653979,0.000000} 13& \cellcolor[rgb]{1.000000,0.949250,0.000000} 11& \cellcolor[rgb]{0.915033,0.957683,0.000000} 10& \cellcolor[rgb]{0.810458,0.905600,0.000000} 9.8         & \cellcolor[rgb]{0.705882,0.853518,0.000000} 9.2         & \cellcolor[rgb]{0.549020,0.775394,0.000000} 8.5 \\
\hline
\end{tabular}
 \caption{Same as Table~\ref{tab:95CL_7yr_nodecorr_0}, but assuming additional foreground decorrelation paramaters.}
 \label{tab:95CL_7yr_decorr_0}
 \centering
\begin{tabular}{c|cccccc}
\hline\hline
   \text{Chile}$\backslash$\text{Pole} & 0         & 6& 9& 12         & 18         & 30\\
\hline
   0 & & \cellcolor[rgb]{1.000000,0.595463,0.000000} 14 & \cellcolor[rgb]{1.000000,0.949250,0.000000} 11 & \cellcolor[rgb]{0.810458,0.905600,0.000000} 9.7& \cellcolor[rgb]{0.549020,0.775394,0.000000} \textbf{8.6} & \cellcolor[rgb]{0.339869,0.671229,0.000000} 7.6 \\
   6 & \cellcolor[rgb]{1.000000,0.013533,0.000000} 24& \cellcolor[rgb]{1.000000,0.838524,0.000000} 12 & \cellcolor[rgb]{0.862745,0.931642,0.000000} 9.9& \cellcolor[rgb]{0.653595,0.827477,0.000000} \textbf{9.0} & \cellcolor[rgb]{0.496732,0.749353,0.000000} 8.2& \cellcolor[rgb]{0.287582,0.645188,0.000000} 7.4 \\
   9 & \cellcolor[rgb]{1.000000,0.419531,0.000000} 17& \cellcolor[rgb]{0.967320,0.983724,0.000000} 11 & \cellcolor[rgb]{0.705882,0.853518,0.000000} \textbf{9.3} & \cellcolor[rgb]{0.549020,0.775394,0.000000} 8.6& \cellcolor[rgb]{0.444444,0.723312,0.000000} 8.0& \cellcolor[rgb]{0.235294,0.619146,0.000000} 7.2 \\
  12 & \cellcolor[rgb]{1.000000,0.595463,0.000000} 14& \cellcolor[rgb]{0.810458,0.905600,0.000000} \textbf{9.8} & \cellcolor[rgb]{0.601307,0.801435,0.000000} 8.9& \cellcolor[rgb]{0.496732,0.749353,0.000000} 8.3& \cellcolor[rgb]{0.392157,0.697270,0.000000} 7.7& \cellcolor[rgb]{0.235294,0.619146,0.000000} 7.1 \\
  18 & \cellcolor[rgb]{1.000000,0.949250,0.000000} \textbf{11} & \cellcolor[rgb]{0.601307,0.801435,0.000000} 8.8& \cellcolor[rgb]{0.496732,0.749353,0.000000} 8.2& \cellcolor[rgb]{0.392157,0.697270,0.000000} 7.8& \cellcolor[rgb]{0.287582,0.645188,0.000000} 7.4& \cellcolor[rgb]{0.183007,0.593105,0.000000} 6.9 \\
  30 & \cellcolor[rgb]{0.549020,0.775394,0.000000} 8.5         & \cellcolor[rgb]{0.339869,0.671229,0.000000} 7.6& \cellcolor[rgb]{0.287582,0.645188,0.000000} 7.3& \cellcolor[rgb]{0.235294,0.619146,0.000000} 7.1& \cellcolor[rgb]{0.183007,0.593105,0.000000} 6.8& \cellcolor[rgb]{0.078431,0.541023,0.000000} 6.5 \\
\hline
\end{tabular}
 \caption{Same as Table~\ref{tab:95CL_7yr_nodecorr_0}, but assuming we keep the 58\% cleanest part of the full sky}
 \label{tab:95CL_7yr_nodecorr_0_58}
\end{table}

\begin{table}
 \centering
\begin{tabular}{c|cccccc}
\hline\hline
   \text{Chile}$\backslash$\text{Pole} & 0& 6& 9& 12         & 18         & 30\\
\hline
   0 && \cellcolor[rgb]{1.000000,0.726644,0.000000} 3.7& \cellcolor[rgb]{1.000000,0.989619,0.000000} 4.5& \cellcolor[rgb]{0.839216,0.919923,0.000000} 4.9& \cellcolor[rgb]{0.650980,0.826175,0.000000} \textbf{5.4} & \cellcolor[rgb]{0.415686,0.708989,0.000000} 6.0 \\
   6 & \cellcolor[rgb]{0.884045,0.000000,0.000000} 2.2& \cellcolor[rgb]{1.000000,0.920415,0.000000} 4.3& \cellcolor[rgb]{0.870588,0.935548,0.000000} 4.8& \cellcolor[rgb]{0.729412,0.865236,0.000000} \textbf{5.2} & \cellcolor[rgb]{0.556863,0.779300,0.000000} 5.6& \cellcolor[rgb]{0.352941,0.677739,0.000000} 6.1 \\
   9 & \cellcolor[rgb]{1.000000,0.025375,0.000000} 3.0& \cellcolor[rgb]{0.949020,0.974610,0.000000} 4.6& \cellcolor[rgb]{0.760784,0.880861,0.000000} \textbf{5.1} & \cellcolor[rgb]{0.635294,0.818362,0.000000} 5.4& \cellcolor[rgb]{0.494118,0.748051,0.000000} 5.8& \cellcolor[rgb]{0.290196,0.646490,0.000000} 6.3 \\
  12 & \cellcolor[rgb]{1.000000,0.685121,0.000000} 3.6& \cellcolor[rgb]{0.839216,0.919923,0.000000} \textbf{4.9} & \cellcolor[rgb]{0.666667,0.833987,0.000000} 5.3& \cellcolor[rgb]{0.556863,0.779300,0.000000} 5.6& \cellcolor[rgb]{0.431373,0.716801,0.000000} 5.9& \cellcolor[rgb]{0.243137,0.623053,0.000000} 6.4 \\
  18 & \cellcolor[rgb]{1.000000,0.948097,0.000000} \textbf{4.4} & \cellcolor[rgb]{0.666667,0.833987,0.000000} 5.3& \cellcolor[rgb]{0.541176,0.771488,0.000000} 5.7& \cellcolor[rgb]{0.447059,0.724614,0.000000} 5.9& \cellcolor[rgb]{0.337255,0.669927,0.000000} 6.2& \cellcolor[rgb]{0.164706,0.583991,0.000000} 6.6 \\
  30 & \cellcolor[rgb]{0.635294,0.818362,0.000000} 5.4& \cellcolor[rgb]{0.400000,0.701176,0.000000} 6.0& \cellcolor[rgb]{0.305882,0.654302,0.000000} 6.2& \cellcolor[rgb]{0.243137,0.623053,0.000000} 6.4& \cellcolor[rgb]{0.149020,0.576178,0.000000} 6.6& \cellcolor[rgb]{0.023529,0.513679,0.000000} 6.9 \\
\hline
\end{tabular}
 \caption{Combined detection significance for r=0.003 after 7 years of observation, keeping only the 28\% cleanest part of the sky, assuming no decorrelation and observing efficiency in Chile same as at Polee.}
 \label{tab:sigdeg_7yr_nodecorr_0.003}
 \centering
\begin{tabular}{c|cccccc}
\hline\hline
   \text{Chile}$\backslash$\text{Pole} & 0& 6& 9& 12         & 18         & 30\\
\hline
   0 && \cellcolor[rgb]{1.000000,0.726644,0.000000} 3.7& \cellcolor[rgb]{1.000000,0.989619,0.000000} 4.5& \cellcolor[rgb]{0.839216,0.919923,0.000000} 4.9& \cellcolor[rgb]{0.650980,0.826175,0.000000} \textbf{5.4} & \cellcolor[rgb]{0.415686,0.708989,0.000000} 6.0 \\
   6 & \cellcolor[rgb]{0.741330,0.000000,0.000000} 1.3& \cellcolor[rgb]{1.000000,0.823529,0.000000} 4.0& \cellcolor[rgb]{0.949020,0.974610,0.000000} 4.6& \cellcolor[rgb]{0.792157,0.896486,0.000000} \textbf{5.0} & \cellcolor[rgb]{0.619608,0.810550,0.000000} 5.5& \cellcolor[rgb]{0.384314,0.693364,0.000000} 6.0 \\
   9 & \cellcolor[rgb]{0.830527,0.000000,0.000000} 1.9& \cellcolor[rgb]{1.000000,0.878893,0.000000} 4.2& \cellcolor[rgb]{0.901961,0.951173,0.000000} \textbf{4.8} & \cellcolor[rgb]{0.745098,0.873049,0.000000} 5.1& \cellcolor[rgb]{0.588235,0.794925,0.000000} 5.5& \cellcolor[rgb]{0.368627,0.685552,0.000000} 6.1 \\
  12 & \cellcolor[rgb]{0.901884,0.000000,0.000000} 2.4& \cellcolor[rgb]{1.000000,0.948097,0.000000} \textbf{4.3} & \cellcolor[rgb]{0.854902,0.927735,0.000000} 4.9& \cellcolor[rgb]{0.713725,0.857424,0.000000} 5.2& \cellcolor[rgb]{0.556863,0.779300,0.000000} 5.6& \cellcolor[rgb]{0.337255,0.669927,0.000000} 6.1 \\
  18 & \cellcolor[rgb]{1.000000,0.025375,0.000000} \textbf{3.0} & \cellcolor[rgb]{0.949020,0.974610,0.000000} 4.6& \cellcolor[rgb]{0.760784,0.880861,0.000000} 5.1& \cellcolor[rgb]{0.635294,0.818362,0.000000} 5.4& \cellcolor[rgb]{0.494118,0.748051,0.000000} 5.8& \cellcolor[rgb]{0.290196,0.646490,0.000000} 6.3 \\
  30 & \cellcolor[rgb]{1.000000,0.837370,0.000000} 4.0& \cellcolor[rgb]{0.729412,0.865236,0.000000} 5.2& \cellcolor[rgb]{0.588235,0.794925,0.000000} 5.5& \cellcolor[rgb]{0.494118,0.748051,0.000000} 5.8& \cellcolor[rgb]{0.368627,0.685552,0.000000} 6.1& \cellcolor[rgb]{0.196078,0.599616,0.000000} 6.5 \\
\hline
\end{tabular}
 \caption{Same as Table~\ref{tab:sigdeg_7yr_nodecorr_0.003}, but assuming 50\% Chilean efficiency.}
 \label{tab:sigdeg_7yr_nodecorr_0.003_half}
 \centering
\begin{tabular}{c|cccccc}
\hline\hline
   \text{Chile}$\backslash$\text{Pole} & 0& 6& 9& 12         & 18         & 30\\
\hline
   0 && \cellcolor[rgb]{1.000000,0.634371,0.000000} 3.5& \cellcolor[rgb]{1.000000,0.851211,0.000000} 4.1& \cellcolor[rgb]{1.000000,0.975779,0.000000} 4.4& \cellcolor[rgb]{0.901961,0.951173,0.000000} \textbf{4.8} & \cellcolor[rgb]{0.729412,0.865236,0.000000} 5.2 \\
   6 & \cellcolor[rgb]{0.872152,0.000000,0.000000} 2.2& \cellcolor[rgb]{1.000000,0.809689,0.000000} 4.0& \cellcolor[rgb]{1.000000,0.961938,0.000000} 4.4& \cellcolor[rgb]{0.933333,0.966797,0.000000} \textbf{4.7} & \cellcolor[rgb]{0.823529,0.912111,0.000000} 5.0& \cellcolor[rgb]{0.666667,0.833987,0.000000} 5.3 \\
   9 & \cellcolor[rgb]{0.979187,0.000000,0.000000} 2.9& \cellcolor[rgb]{1.000000,0.906574,0.000000} 4.2& \cellcolor[rgb]{0.964706,0.982422,0.000000} \textbf{4.6} & \cellcolor[rgb]{0.870588,0.935548,0.000000} 4.8& \cellcolor[rgb]{0.760784,0.880861,0.000000} 5.1& \cellcolor[rgb]{0.635294,0.818362,0.000000} 5.4 \\
  12 & \cellcolor[rgb]{1.000000,0.482122,0.000000} 3.4& \cellcolor[rgb]{1.000000,0.989619,0.000000} \textbf{4.5} & \cellcolor[rgb]{0.901961,0.951173,0.000000} 4.8& \cellcolor[rgb]{0.823529,0.912111,0.000000} 5.0& \cellcolor[rgb]{0.729412,0.865236,0.000000} 5.2& \cellcolor[rgb]{0.588235,0.794925,0.000000} 5.5 \\
  18 & \cellcolor[rgb]{1.000000,0.809689,0.000000} \textbf{4.0} & \cellcolor[rgb]{0.901961,0.951173,0.000000} 4.7& \cellcolor[rgb]{0.807843,0.904298,0.000000} 5.0& \cellcolor[rgb]{0.745098,0.873049,0.000000} 5.1& \cellcolor[rgb]{0.666667,0.833987,0.000000} 5.4& \cellcolor[rgb]{0.541176,0.771488,0.000000} 5.6 \\
  30 & \cellcolor[rgb]{0.917647,0.958985,0.000000} 4.7& \cellcolor[rgb]{0.729412,0.865236,0.000000} 5.2& \cellcolor[rgb]{0.666667,0.833987,0.000000} 5.3& \cellcolor[rgb]{0.619608,0.810550,0.000000} 5.4& \cellcolor[rgb]{0.556863,0.779300,0.000000} 5.6& \cellcolor[rgb]{0.478431,0.740238,0.000000} 5.8 \\
\hline
\end{tabular}
 \caption{Same as Table~\ref{tab:sigdeg_7yr_nodecorr_0.003}, but assuming 1\% unmodeled foreground residual uncertainty.}
 \label{tab:sigdeg_7yr_nodecorr_0.003_fg}
 \centering
\begin{tabular}{c|cccccc}
\hline\hline
   \text{Chile}$\backslash$\text{Pole} & 0& 6& 9& 12         & 18         & 30\\
\hline
   0 && \cellcolor[rgb]{1.000000,0.025375,0.000000} 3.0& \cellcolor[rgb]{1.000000,0.685121,0.000000} 3.6& \cellcolor[rgb]{1.000000,0.823529,0.000000} 4.0& \cellcolor[rgb]{0.996078,0.998047,0.000000} \textbf{4.5} & \cellcolor[rgb]{0.760784,0.880861,0.000000} 5.1 \\
   6 & \cellcolor[rgb]{0.806741,0.000000,0.000000} 1.7& \cellcolor[rgb]{1.000000,0.583622,0.000000} 3.5& \cellcolor[rgb]{1.000000,0.795848,0.000000} 3.9& \cellcolor[rgb]{1.000000,0.920415,0.000000} \textbf{4.3} & \cellcolor[rgb]{0.933333,0.966797,0.000000} 4.7& \cellcolor[rgb]{0.713725,0.857424,0.000000} 5.2 \\
   9 & \cellcolor[rgb]{0.895937,0.000000,0.000000} 2.3& \cellcolor[rgb]{1.000000,0.726644,0.000000} 3.7& \cellcolor[rgb]{1.000000,0.865052,0.000000} \textbf{4.1} & \cellcolor[rgb]{1.000000,0.975779,0.000000} 4.4& \cellcolor[rgb]{0.870588,0.935548,0.000000} 4.8& \cellcolor[rgb]{0.666667,0.833987,0.000000} 5.3 \\
  12 & \cellcolor[rgb]{0.961348,0.000000,0.000000} 2.8& \cellcolor[rgb]{1.000000,0.795848,0.000000} \textbf{3.9} & \cellcolor[rgb]{1.000000,0.920415,0.000000} 4.3& \cellcolor[rgb]{0.980392,0.990235,0.000000} 4.6& \cellcolor[rgb]{0.823529,0.912111,0.000000} 4.9& \cellcolor[rgb]{0.635294,0.818362,0.000000} 5.4 \\
  18 & \cellcolor[rgb]{1.000000,0.482122,0.000000} \textbf{3.4} & \cellcolor[rgb]{1.000000,0.920415,0.000000} 4.3& \cellcolor[rgb]{0.964706,0.982422,0.000000} 4.6& \cellcolor[rgb]{0.870588,0.935548,0.000000} 4.8& \cellcolor[rgb]{0.745098,0.873049,0.000000} 5.1& \cellcolor[rgb]{0.572549,0.787113,0.000000} 5.6 \\
  30 & \cellcolor[rgb]{1.000000,0.892734,0.000000} 4.2& \cellcolor[rgb]{0.870588,0.935548,0.000000} 4.8& \cellcolor[rgb]{0.776471,0.888674,0.000000} 5.0& \cellcolor[rgb]{0.713725,0.857424,0.000000} 5.2& \cellcolor[rgb]{0.603922,0.802737,0.000000} 5.5& \cellcolor[rgb]{0.462745,0.732426,0.000000} 5.8 \\
\hline
\end{tabular}
 \caption{Same as Table~\ref{tab:sigdeg_7yr_nodecorr_0.003}, but assuming additional foreground decorrelation paramaters.}
 \label{tab:sigdeg_7yr_decorr_0.003}
 \centering
\begin{tabular}{c|cccccc}
\hline\hline
   \text{Chile}$\backslash$\text{Pole} & 0& 6& 9& 12         & 18         & 30\\
\hline
   0 && \cellcolor[rgb]{1.000000,0.740484,0.000000} 3.8& \cellcolor[rgb]{0.996078,0.998047,0.000000} 4.5& \cellcolor[rgb]{0.807843,0.904298,0.000000} 5.0& \cellcolor[rgb]{0.666667,0.833987,0.000000} \textbf{5.3} & \cellcolor[rgb]{0.321569,0.662115,0.000000} 6.2 \\
   6 & \cellcolor[rgb]{0.913777,0.000000,0.000000} 2.4& \cellcolor[rgb]{1.000000,0.975779,0.000000} 4.4& \cellcolor[rgb]{0.807843,0.904298,0.000000} 5.0& \cellcolor[rgb]{0.666667,0.833987,0.000000} \textbf{5.4} & \cellcolor[rgb]{0.494118,0.748051,0.000000} 5.8& \cellcolor[rgb]{0.274510,0.638677,0.000000} 6.3 \\
   9 & \cellcolor[rgb]{1.000000,0.431373,0.000000} 3.3& \cellcolor[rgb]{0.854902,0.927735,0.000000} 4.9& \cellcolor[rgb]{0.666667,0.833987,0.000000} \textbf{5.3} & \cellcolor[rgb]{0.541176,0.771488,0.000000} 5.6& \cellcolor[rgb]{0.400000,0.701176,0.000000} 6.0& \cellcolor[rgb]{0.196078,0.599616,0.000000} 6.5 \\
  12 & \cellcolor[rgb]{1.000000,0.823529,0.000000} 4.0& \cellcolor[rgb]{0.713725,0.857424,0.000000} \textbf{5.2} & \cellcolor[rgb]{0.556863,0.779300,0.000000} 5.6& \cellcolor[rgb]{0.447059,0.724614,0.000000} 5.9& \cellcolor[rgb]{0.305882,0.654302,0.000000} 6.2& \cellcolor[rgb]{0.117647,0.560554,0.000000} 6.7 \\
  18 & \cellcolor[rgb]{0.854902,0.927735,0.000000} \textbf{4.9} & \cellcolor[rgb]{0.494118,0.748051,0.000000} 5.8& \cellcolor[rgb]{0.368627,0.685552,0.000000} 6.1& \cellcolor[rgb]{0.274510,0.638677,0.000000} 6.3& \cellcolor[rgb]{0.164706,0.583991,0.000000} 6.6& \cellcolor[rgb]{0.000000,0.501243,0.000000} 7.0 \\
  30 & \cellcolor[rgb]{0.368627,0.685552,0.000000} 6.1& \cellcolor[rgb]{0.149020,0.576178,0.000000} 6.6& \cellcolor[rgb]{0.054902,0.529304,0.000000} 6.8& \cellcolor[rgb]{0.000000,0.501243,0.000000} 7.0& \cellcolor[rgb]{0.000000,0.494066,0.000000} 7.2& \cellcolor[rgb]{0.000000,0.481148,0.000000} 7.5 \\
\hline
\end{tabular}
 \caption{Same as Table~\ref{tab:sigdeg_7yr_nodecorr_0.003}, but assuming we keep the 58\% cleanest part of the full sky}
 \label{tab:sigdeg_7yr_nodecorr_0.003_58}
\end{table}

\subsection{Conclusion}

Here is a high-level summary of the key points regarding the survey targeting the signature of gravitational waves.

\begin{itemize}
\item The semi-analytic calculations described in
  Sect.~\ref{sec:semianaflowdownallocation} indicates that for a 3\% sky
  fraction $1.8\times10^6$ 150\,GHz-equivalent detector-years of observation
  are required to reach the science requirements, with 30\% of this
  assigned to the delensing observations.
  This calculation provides an optimal distribution of these detectors
  across frequency bands to achieve the goal in the presence of foregrounds.
\item The map based simulations described in Sect.~\ref{sec:cdtrepsims}
  confirm the $\sigma(r)$ results from the semi-analytic calculations.
  These simulations also indicate that bias in the recovered $r$ value
  is within $1\sigma$ for a suite of different foreground models.
  However, we note that foregrounds remain a serious issue which must be
  periodically revisited as the project progresses.
\item Additional map based simulations indicate that systematic bias
  on $r$ can be controlled to $<1\sigma$ provided fractional contamination
  levels similar to those already achieved by small aperture telescopes
  can be maintained.
\item Mapping the requirements from the semi-analytic calculations onto realizable
  instruments results in the reference design described in
  Sect.~\ref{sec:refdessum} for a seven-year survey period.
\item A larger fraction of the sky can be observed from Chile, but,
  due to the rotation of the Earth, one
  can concentrate the available sensitivity more deeply from the South Pole
  (see Fig.~\ref{fig:relhits}).
  In Sect.~\ref{sec:DSRupdates} we have extended the semi-analytic
  calculations to account for realistic observation patterns
  and probed the dependence of $\sigma(r)$ on $r$, finding that
  Pole is always favored in the limit of small $r$, with the cross over point
  depending on the specific assumptions (see Figs.~\ref{fig:sigrvsr_4panels}
  and \ref{fig:sigrvsr_18tubes}).
\end{itemize}

\newpage

\section{Deep and wide field for measuring light relic density}
\label{sec:light-relic}

The impact of a change to the light relic density $\Neff$ is most prominent in the damping tail region of the CMB power spectra.
The sensitivity of a CMB survey to $\Neff$ is therefore driven by its ability to map small scale fluctuations.
Measuring the relevant modes requires a low-noise, relatively high-resolution, and wide-coverage survey of temperature and polarization.
The impact of foregrounds limits the constraining power of the temperature auto spectrum, and so improvements in constraints on $\Neff$ tend to be driven by measurements of the $TE$ spectrum.
Exponential suppression of the primary CMB temperature and polarization on small angular scales favors observations of as much sky as possible, for fixed total observing time.

The CMB-S4 Science Book \cite{Abazajian:2016yjj} explored constraints on $\Neff$ over a wide range of experimental configurations.
Many of the qualitative lessons learned from that study remain unchanged.
However, for the present effort, we have significantly updated and improved the treatment of atmospheric effects, extragalactic foregrounds, and component separation.

\begin{figure}[t]
\begin{center}
\includegraphics[width=4in]{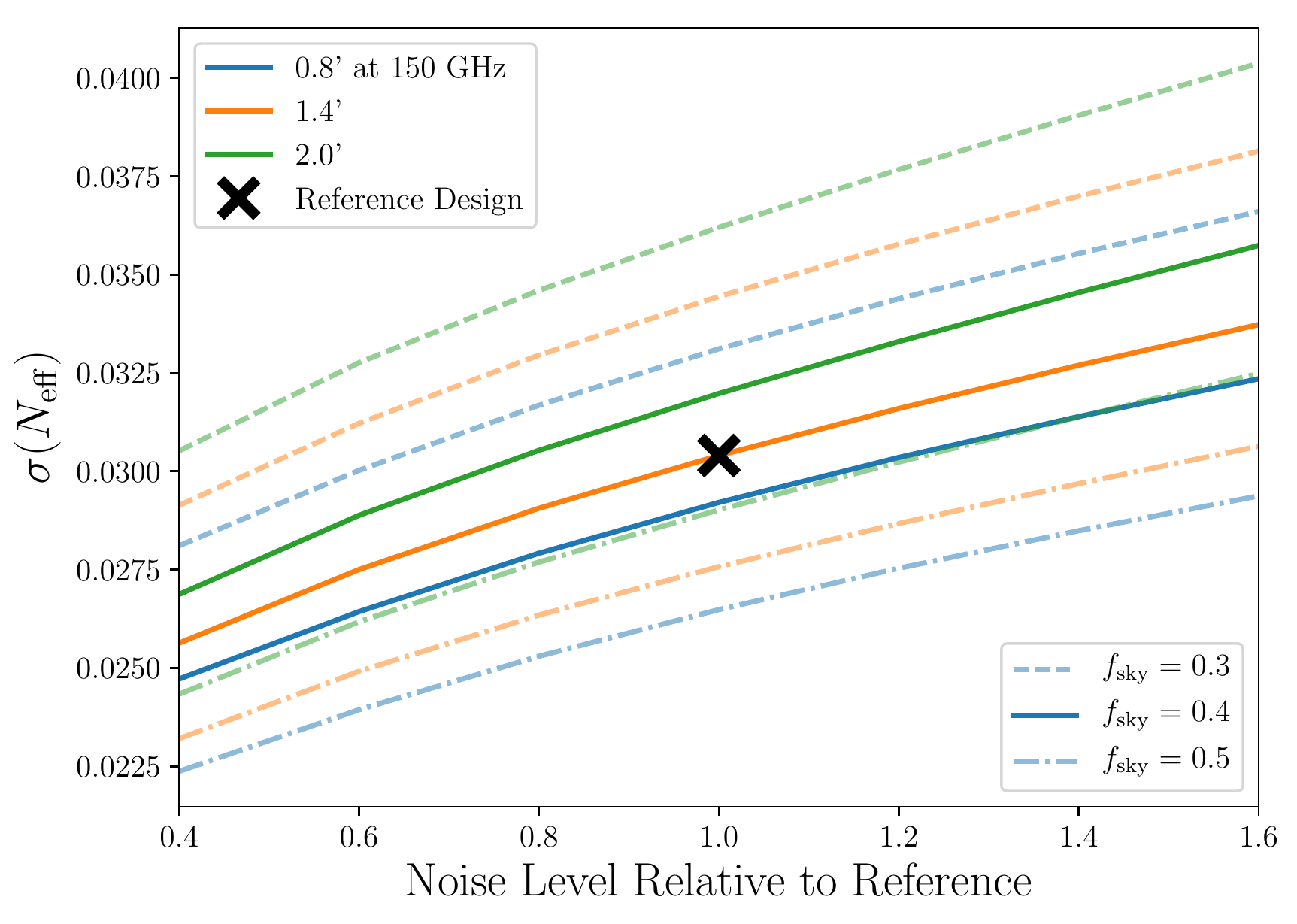}
\end{center}
\caption{Impact of changes to the noise level, beam size, and sky fraction on forecasted 1$\sigma$ constraints on $N_\mathrm{eff}$ with $Y_\mathrm{p}$ fixed by BBN consistency.  Changes to $f_\mathrm{sky}$ are taken here at fixed map depth.  The forecasts shown in this figure have less detailed modeling of atmospheric effects and foreground cleaning than those shown elsewhere.  The results should therefore be taken as a guide to how various experimental design choices impact the constraining power for light relics, but the specific values of the constraints should be taken to be accurate only at the level of about $10\%$.}
\label{fig:Neff_BeamNoiseFsky}
\end{figure}

In Fig.~\ref{fig:Neff_BeamNoiseFsky}, we show how the forecasted constraints on $\Neff$ are affected by changes to the noise level, beam size, and sky fraction.
As can be seen from the figure, reduction of the noise level, either through increased detector count or extended observing time, leads to improved constraints on $\Neff$, though significant improvements require fairly large improvements to the map depth of the survey.
This is due to the rapid drop in CMB power in the damping tail on small angular scales and to the impact of residual foregrounds, which act as additional noise.
Compared to the reference design, increasing the size of the telescope dish leads to a modest improvement in constraints on light relics, while reducing the dish size produces a slightly sharper drop in the constraining power.
Observing a larger fraction of the sky allows access to more independent small-scale modes, and results in tighter constraints on $\Neff$.
The dependence on sky fraction in Fig.~\ref{fig:Neff_BeamNoiseFsky} is shown at fixed map depth, but we will explore in much more detail below how the survey design affects light relics constraints.

\begin{figure}[t]
\begin{center}
\includegraphics[width=4in]{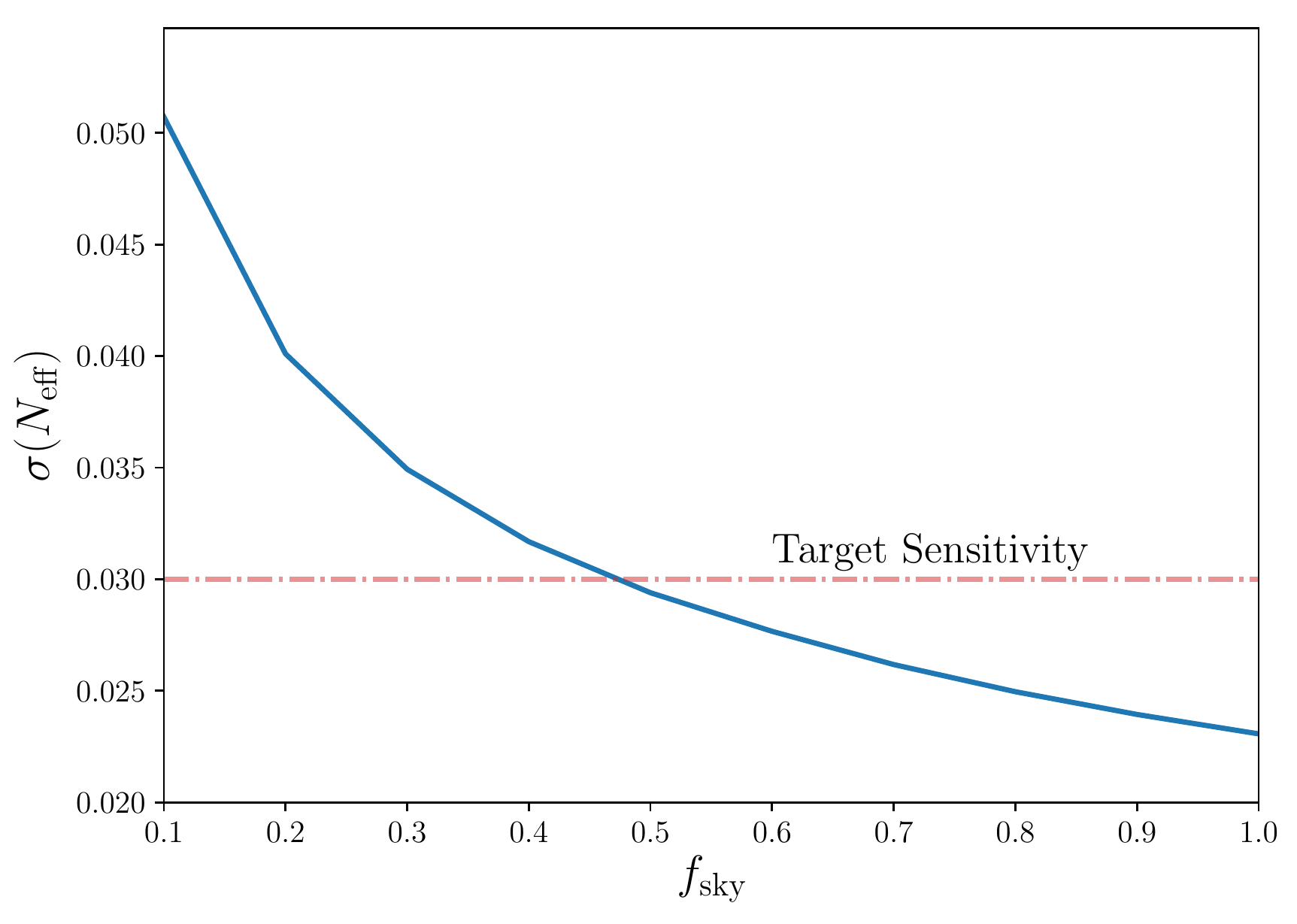}
\end{center}
\caption{Impact of changes to the sky fraction at fixed effort on forecasted 1$\sigma$ constraints on $N_\mathrm{eff}$ with $Y_\mathrm{p}$ fixed by BBN consistency.  The forecasts shown in this figure have less detailed modeling of atmospheric effects and foreground cleaning than those shown elsewhere and should be taken to be accurate only at the level of about $10\%$.}
\label{fig:Neff_fsky}
\end{figure}

Given the results shown in Fig.~\ref{fig:Neff_BeamNoiseFsky}, increasing the sky fraction seems to be the most promising route to reduce the constraints on $\Neff$.
At fixed observing time, surveying a larger sky fraction leads to an increased noise, though constraints on light relics still favor as wide a survey as possible (due to the rapid fall off of the damping tail).
The preference for larger sky fraction is demonstrated in Fig.~\ref{fig:Neff_fsky}, which for a simple modeling of the effects of atmosphere and foregrounds shows how the error on $\Neff$ is affected by changes to the observed sky fraction at fixed total effort.
Observing a larger sky fraction from the ground requires telescopes to point to lower elevations, and thus to observe through a greater column of atmosphere for portions of the survey.
We have therefore performed a more detailed study of the fraction of sky that can be realistically observed from Chile, taking into account the additional atmospheric loading when observing at lower elevations.

\subsection{Opportunistic scheduler for sky coverage}

In order to determine the map area and depth achievable from Chile, we built a set of observing schedules using an opportunistic scheduler for the observations.
First we tiled the sky in celestial coordinates with $10^\circ \times 20^\circ$ (RA$\times$Dec) tiles that overlap by half a tile in each direction.
Then we ran the scheduler with three choices of minimum observing elevation: 30, 40 and $50^\circ$.
We required a $30^\circ$ avoidance region around the Sun and the Moon.
The three schedules were run for each of two scan strategies designed with and without an additional ``elevation penalty'' which tells the scheduler to favor high elevation observations over low elevation observations. We refer to the strategy without the elevation penalty as the ``nominal schedule.''

We ran the scheduler on the full sky and subsequently masked pixels inside the galaxy from the resulting hit maps.
This strategy was chosen to avoid undesirable boundary effects, since adjusting the tile priority based on galaxy overlap made the final hit distribution around the masked area very uneven. The even-coverage approach led to slightly lower overall observing time but higher effective sky fraction.

Hits were binned into separate maps based on boresight elevation, to facilitate subsequent processing with elevation-dependent noise models.

\subsection{Atmospheric modeling and depth maps}\label{sec:atm_depth}

The elevation-binned hit maps were converted to full survey depth maps
in each frequency band as follows.  For each elevation bin, an
elevation-dependent noise model was used to convert observing time per
unit area to map depth in units of [$\mu$K-arcmin]$^{-2}$.  The
per-elevation depth maps were then summed to produce a single depth
map, in units of [$\mu$K-arcmin]$^{-2}$, for each frequency and each
choice of minimum elevation.  Intra-tube correlations were captured by
binning into cross-frequency depth maps (e.g., 90$\times$150\,GHz).
This computation is done for the white noise level and at $\ell =
1000,2000,3000$ to capture the impact of $1/f$ noise.

The noise model in $T$ includes white noise and atmospheric $1/f$
components.  The parameters describing the $1/f$ noise are based on
low-$\ell$ $TT$ spectra measured with ACTPol at 90, 145, and 225\,GHz.
These power spectra are measured from maps (and thus are expressed in
[$\mu$K-arcmin]$^2$), but are referenced to ACTPol array NETs to
produce effective atmosphere equivalent noise powers, in $\mu$K$^2\,{\rm s}$,
that can be combined with the CMB-S4 white noise NETs in each band.

In general, the rescaling of the $1/f$ noise spectrum to a different
instrument is complicated because of spatial correlations in the
atmospheric contamination.  However, our angular scales of greatest
interest are smaller than the ACTPol array diameter ($0.8^\circ$) and
in that regime it is reasonable to expect the atmospheric noise power
to average down according to the ratio of ACTPol and CMB-S4 focal
plane areas.  To extrapolate the $1/f$ noise power to other CMB-S4
bands, we use assume that variation of water vapor content is the
primary driver of small scale atmospheric noise and use the
AM\footnote{\url{https://www.cfa.harvard.edu/\~spaine/am/}} atmosphere
modeling code v9.2 (median parameters for Atacama, September-November)
to obtain a relative calibration between bands.  As the ACTPol noise
at 90\,GHz is 40\% higher than predicted by this modeling, we inflate
all lower frequency bands by 40\% as well.

To add elevation dependence, we assume that the $1/f$ power scales as
the square of the airmass. The scaling of the white noise level with
elevation is computed separately, using the detector noise model and
typical atmospheric loading provided by the AM model.

The polarization $1/f$ noise is not dominated by spatially correlated
atmosphere.  We assume in this case that $1/f$ noise scales with the
detector white noise, with a knee fixed at $\ell = 700$.  This is the
same assumption made in Simons Observatory LAT forecasting
\citep{2018arXiv180807445T}.

\begin{table}[]
\centering
\scriptsize
\begin{tabular}{|l||c|c|c|c||c|c|c|c||c|c|c|c|}
\hline
\textbf{Minimum Elevation}        & \multicolumn{4}{c||}{\textbf{50 degrees}} & \multicolumn{4}{c||}{\textbf{40 degrees}} & \multicolumn{4}{c|}{\textbf{30 degrees}} \\ \hline
\textbf{Galactic Cut (\%)}        & \textbf{30}  & \textbf{20} & \textbf{10} & \textbf{0} & \textbf{30}  & \textbf{20} & \textbf{10}  & \textbf{0} & \textbf{30}  & \textbf{20} & \textbf{10} & \textbf{0} \\ \hline 
          \multicolumn{1}{c}{ } & \multicolumn{12}{c}{\textbf{Nominal Schedule}}  \\ \hline
\textbf{Sky Fraction}                      & 0.39         & 0.45        & 0.52        & 0.57        & 0.48         & 0.55        & 0.63        & 0.68        & 0.54         & 0.62        & 0.69        & 0.76        \\ \hline
\textbf{Effective Sky Fraction for Noise}  & 0.35         & 0.41        & 0.47        & 0.52        & 0.44         & 0.51        & 0.58        & 0.64        & 0.50         & 0.57        & 0.65        & 0.71        \\ \hline
\textbf{Effective Sky Fraction for Signal} & 0.32         & 0.37        & 0.43        & 0.47        & 0.41         & 0.47        & 0.54        & 0.59        & 0.47         & 0.53        & 0.61        & 0.66        \\ \hline
          \multicolumn{1}{c}{ } & \multicolumn{12}{c}{\textbf{With Elevation Penalty}}  \\ \hline
\textbf{Sky Fraction}                      & 0.39         & 0.45        & 0.52        & 0.57        & 0.48         & 0.55        & 0.63        & 0.68        & 0.54         & 0.62        & 0.69        & 0.76        \\ \hline
\textbf{Effective Sky Fraction for Noise}  & 0.34         & 0.40        & 0.47        & 0.51        & 0.43         & 0.49        & 0.56        & 0.61        & 0.46         & 0.53        & 0.61        & 0.66        \\ \hline
\textbf{Effective Sky Fraction for Signal} & 0.31         & 0.36        & 0.42        & 0.46        & 0.38         & 0.44        & 0.50        & 0.55        & 0.40         & 0.46        & 0.53        & 0.58        \\ \hline
\end{tabular}
\caption{Sky fractions for various choices of minimum observing elevation and galactic cut for the two scan strategies discussed in the text.
The meaning of each type of sky fraction is the same as discussed in Sect.~\ref{sec:skycov}.}
\label{tab:fsky_WAFTT}
\end{table}

\subsection{Foregrounds and point-source removal}\label{sec:fg_ptsrc}

For the deep and wide field forecast, the sky model consists of a set of auto- and cross- power spectra for the set of instrument bands (similar to Ref.~\citep{Dunkley2013}).  It is built from power spectrum templates and source count models. In addition to the CMB, the foreground signals include tSZ, kSZ, Galactic cirrus, radio point sources, and dusty point sources.  We scale these with a single frequency dependence per component to translate the power spectra between the frequency bands of CMB-S4, which we treat as delta functions at the central frequencies.

The tSZ and kSZ power spectral templates come from the hydrodynamic simulations of \cite{Battaglia2012}.  The tSZ frequency dependence uses the non-relativistic formula, and the kSZ effect has the same frequency dependence at the CMB.
The Galactic cirrus template comes from the treatment of \citep{Dunkley2013}.  It is a power law $\propto \ell^{-0.7}$, using the amplitude they found for the clean portions of the sky.  The frequency dependence is $\propto \nu^{3.8}$ in flux density units.

Radio point-source counts come from \cite[]{Tucci:2011} at 148\,GHz (their model C2Ex).  We scale them to other bands as a power law with index $-0.5$ in flux density units.
Dusty point-source counts come from \cite{2012ApJ...757L..23B} at 217\,GHz, but we adjusted them on the faint end to match the SPT 220-GHz source counts \citep{Mocanu:2013}.  The dusty sources are scaled from 217\,GHz with a graybody spectrum with $\beta^{\rm CIB} = 2.1$ and $T^{\rm CIB} = 9.7$\,K.  The power spectral templates for the point sources are flat with an amplitude that depends on the integrated, squared flux density, weighted by the source counts.  We model the correlation between dusty sources and the tSZ using a halo model as in Ref.~\cite{2014A&A...571A..30P,Aghanim:2018eyx}.

We estimate the residual point-source power after masking while accounting for the instrument noise and beam.  In each band, and based on the total power spectrum of the sky model and the noise model,  we compute the variance after applying a filter optimized for point-source detection.  The variance is a function of the prospective flux cut, which sets the power of the point-source component.  We compare the filtered variance to the amplitude of a filtered point source to determine the signal-to-noise ratio of sources (as a function of the flux cut).  We choose the flux cut so that it self-consistently excludes sources with signal-to-noise ratio greater than 5.

For each source population, we identify the best band for excluding sources by scaling the per-band flux cuts back to the reference frequency and comparing them.  For radio sources, 90\,GHz finds the intrinsically faintest sources due to a combination of the band's noise, beam, and SED.  For dusty sources, 270\,GHz is the best.  Assuming these deepest source cuts determine the source mask, we finally compute the overall level of unmasked residual source power from each population and add it to the total sky model.

\subsection{Fisher forecasts}
For the forecasts, we begin with the multi-frequency model described in subsection~\ref{sec:fg_ptsrc} and noise spectra based on the depth maps~\ref{sec:atm_depth}. A harmonic space ILC algorithm \cite{Tegmark:1995pn} is used to derive foreground reduced temperature and polarization spectra and noise spectra. The component separated noise curves and extragalactic foreground residuals computed from the procedure above were then used in a Fisher forecast to determine the expected constraints on $\Neff$.
The CMB-S4 noise was combined with a model for the noise from the {\it Planck\/} satellite, in an inverse variance sum.
Doing so reduces the noise on the large scale temperature modes which would be contaminated by the atmosphere for CMB-S4 alone.
Contamination from extragalactic residuals was then added, which acts much like an additional source of noise.
We imposed a cut at $\ell_\mathrm{min} = 30$, and we took $\ell_\mathrm{max} = 5000$ for all spectra.
A model for {\it Planck\/} temperature data was included for $\ell < 30$ on $f_\mathrm{sky} = 0.8$.

We assumed a cosmology described by $\Lambda$CDM+$\Neff+\sum m_\nu$, assuming BBN consistency to fix the primordial helium density $Y_\mathrm{p}$.
The $TT$, $TE$, $EE$, and $\phi\phi$ spectra were included, with the lensing reconstruction noise calculated using the minimum variance combination of quadratic estimators \cite{Hu:2001kj,Okamoto:2003zw}, including the improvement from iterative $EB$ reconstruction \cite{Hirata:2003ka,Smith:2010gu}.
The small improvements which come from delensing $T$ and $E$ spectra \cite{Green:2016} were also included.

\begin{figure}[t]
\begin{center}
\includegraphics[width=0.5\columnwidth]{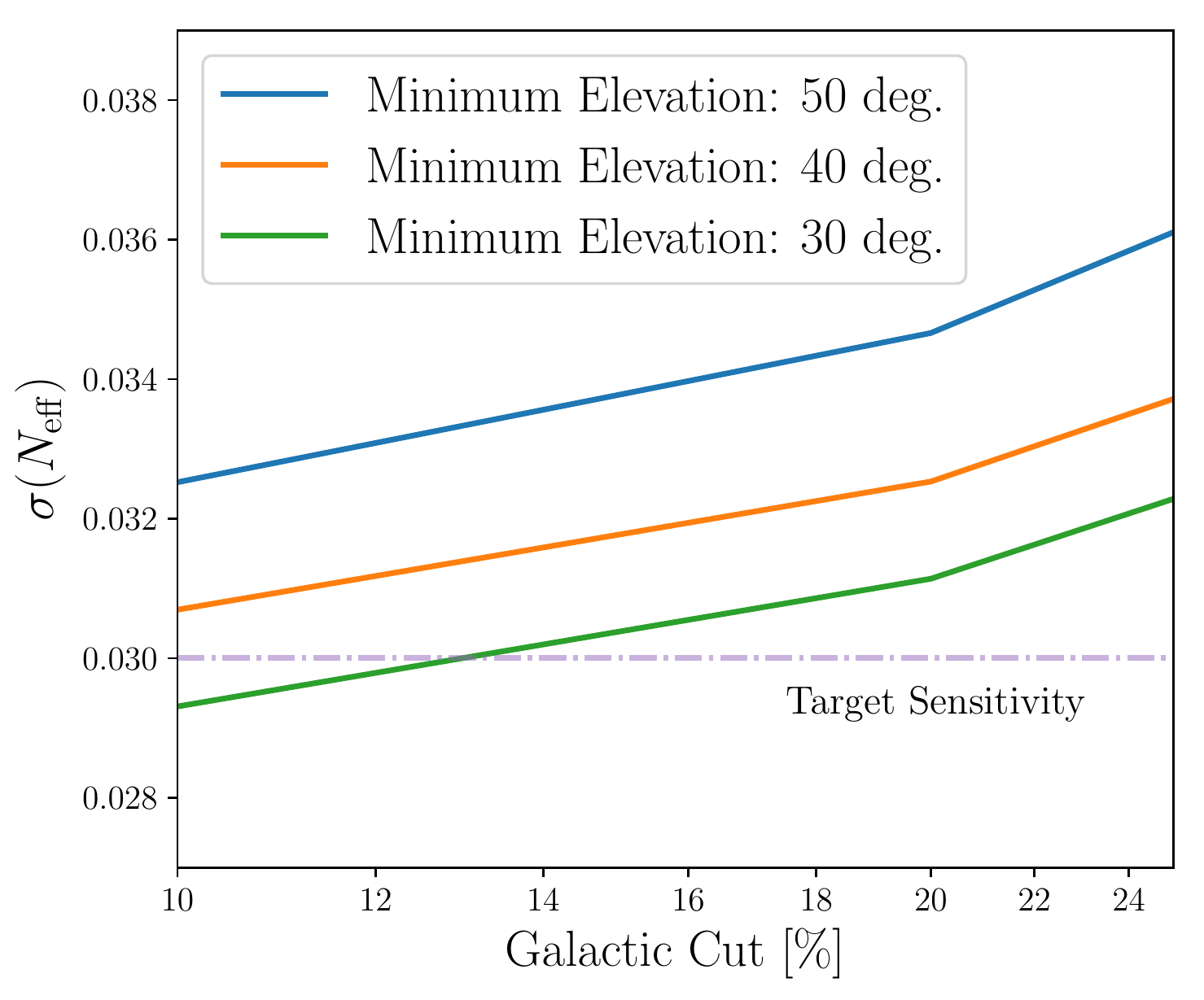}
\end{center}
\caption{Forecasted 1$\sigma$ constraints on $N_\mathrm{eff}$ for different choices of the minimum observing elevation as a function of the size of the galactic mask shown for the nominal scan strategy discussed in the text.}
\label{fig:Neff_WAFTT}
\end{figure}

The resulting noise curves were then used in a Fisher forecast to compute the errors on $\Neff$.
The sky fractions for CMB-S4 observations were calculated for each choice of minimum elevation, galactic cut, and scan strategy as shown in Table~\ref{tab:fsky_WAFTT}.
{\it Planck\/} observations were assumed to cover the rest of the sky which is not observed by CMB-S4 and also lies outside the galactic mask.
As can be seen in Table~\ref{tab:fsky_WAFTT}, the elevation penalty reduces the usable effective sky fraction which leads to the slightly weaker constraints compared to the nominal scan strategy without elevation penalty.
The results for the forecasts for the nominal schedule are shown in Fig.~\ref{fig:Neff_WAFTT}.

\section{Mapping baryons: angular resolution and allocation of detectors across frequency for the LATs}
\label{sec:ClustersFlowdown}

Mapping the pressure, temperature, and distribution of baryons requires high-resolution, multifrequency data in order to separate dust emission, the thermal Sunyaev-Zeldovich (tSZ) effect, and pure blackbody fluctuations, as expected from both the lensed primary CMB and the kinematic Sunyaev-Zeldovich (kSZ) effect.

\subsection{Frequency allocation}

A broad investigation of possible allocations of detectors across frequency was performed. With three extragalactic components known to be present on small angular scales, in addition to possible low-frequency radio emission, and many Galactic components on large and moderate angular scales, at least three well-separated frequencies with high signal-to-noise are required.  In addition, extragalactic dust emission---the cosmic infrared background (CIB)---is already known to exhibit frequency decorrelation due to the different redshift kernels of the CIB signal at different observational frequencies \citep{2014A&A...571A..30P}.  Thus, multiple channels may be needed in order to fully clean this component from the tSZ, kSZ, and lensing signals.

To assess the options for the distribution of detectors, we employed an end-to-end simulation-based optimization framework based on that used in Ref.~\citep{2018arXiv180807445T}.  We focused on temperature-based observables as metrics for optimization: the tSZ power spectrum, kSZ power spectrum, reconstructed CMB lensing power spectrum via the $TT$ quadratic estimator (as a proxy for CMB ``halo lensing,'' which is $TT$-dominated), and the CMB $TT$ power spectrum (this is already well-measured, but included for completeness).  Due to the current lack of knowledge regarding small-scale polarized foregrounds, and the expected stronger need for multifrequency coverage for tSZ and kSZ observables, we did not consider the reconstructed CMB lensing power spectrum from polarization data in this optimization.

We used the CMB-S4 LAT noise calculator described earlier to forecast the S/N of these observables for a large number of experimental configurations.  
We varied the number of optics tubes of each type (LF=27/39\,GHz, MF=93/145\,GHz, UHF=225/280\,GHz), considering here also the possibility of XHF tubes with channels at 281 and 350\,GHz, assumed to be located at a high Chilean site with excellent atmospheric properties.  We assumed two identical LAT copies.  
Several thousand configurations were considered. 
{\it Planck\/} data from 30 to 353\,GHz were also assumed in all forecasts; these channels are useful on large angular scales where the CMB-S4 atmospheric noise is large.

We modeled the temperature sky at all frequencies from 27 to 353\,GHz using the simulated sky maps described in section~2 of \citep{2018arXiv180807445T}.  These maps include models for essentially all Galactic and extragalactic foregrounds (and signals). Simple Galactic-emission-thresholded sky masks that leave the cleanest amount of sky that is visible from Chile were employed to self-consistently include the effect of these sky cuts on the Galactic foreground levels. While several cut levels were explored, it was found that
the sky fraction did not strongly affect the frequency allocation, with a general
trend of higher S/N for larger sky area surveyed. 

For configuration option, we used a harmonic-space internal linear combination (ILC) code to obtain post-component-separation noise power spectra for the blackbody CMB temperature and tSZ fields, using the modeled sky power spectra and the per-frequency noise power spectra computed using the CMB-S4 calculator (as well as {\it Planck\/} noise, assumed to be white). We considered the option of explicitly ``deprojecting'' some contaminants using a constrained ILC, which is a robust way to conservatively assess the frequency coverage that may be needed to sufficiently remove these foregrounds. We considered three deprojection options: no deprojection, deprojection of tSZ (for CMB reconstruction) or CMB (for tSZ reconstruction), and deprojection of a fiducial CIB spectrum (for CMB and tSZ reconstruction). The total number of sky fraction/configuration/deprojection options is 10260.  We then flowed down to determine the experimental setup by optimizing the S/N of the observables described above, for each deprojection choice.

For each observable and each deprojection choice, the optimal tube configuration was slightly different, but general trends were clear.  The optimization preferred a broad frequency coverage, always including at least one tube of each type.  Typical optimal configurations included 1 or 2 LF tubes, 7--14 MF tubes, and 3--11 UHF tubes.  Deprojecting tSZ or CIB foregrounds generally put a stronger demand on the need for UHF tubes.  We also found that inclusion of XHF tubes at even higher frequencies could lead to 10--20\% gains in S/N for some observables, particularly when deprojecting tSZ or CIB foregrounds.  This option may be worthy of further study in the future.

The most important conclusion from this optimization is that the CMB-S4 LAT reference configuration (2 LF tubes, 12 MF tubes, 5 UHF tubes) used throughout this work is sufficiently near-optimal to serve as an excellent choice.  In nearly all cases, it performed within 5--10\% of the maximum S/N found in the optimization; the only exceptions to this were in measurements of the kSZ power spectrum with tSZ or CIB deprojection, where the reference configuration was within 15--20\% of the maximal S/N found in the optimization.  While future refinements of the detector allocation across frequency may be performed, particularly with updated CIB modeling to inform the high-frequency optimization, the flowdown presented here justifies the reference distribution that has been used throughout this document.

\begin{figure}
\begin{center}
\includegraphics[width=6in]{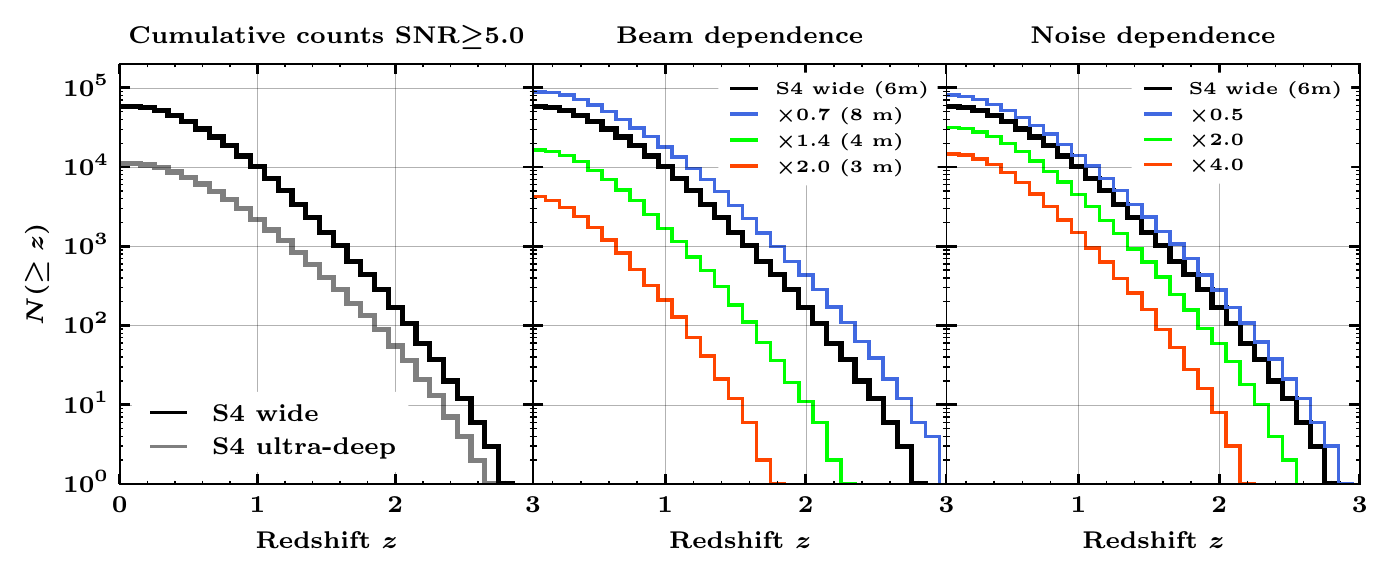}
\end{center}
\caption{Panel 1 shows the expected number of $SNR \ge 5$ clusters as a function of redshift for the deep and wide (black) and ultra-deep (gray) fields with a 6-m aperture corresponding to $1.^{\prime}5$ at 150\,GHz.
Panels 2, 3 show the dependence on telescope size and noise levels for the deep and wide survey. 
To get a sample of clusters at z$\gtrsim 2$, given our current understanding of high-redshift clusters, an aperture size $\ge6$\,m is required. 
The sky coverage has been assumed as $f_{\rm sky} = 0.4$ and $f_{\rm sky} = 0.03$ for the deep and wide and ultra-deep fields respectively. 
\label{fig:cluster_counts}
}
\end{figure}

\subsection{Angular resolution and sensitivity}

In terms of angular resolution, higher resolution will always be better for more precisely constraining the astrophysics of galaxy formation; the minimum resolution is set by the typical angular scale of high-redshift galaxy clusters.
Section~\ref{subsec:GF} shows that CMB-S4 provides new views on the
thermodynamics of galaxy clusters on the scale of the virial region, while
other observations of galaxy feedback are typically on smaller scales.
As resolution is improved it will be easier to make direct comparisons
with optical/IR observations.

\begin{figure}
\begin{center}
\includegraphics[width=4in]{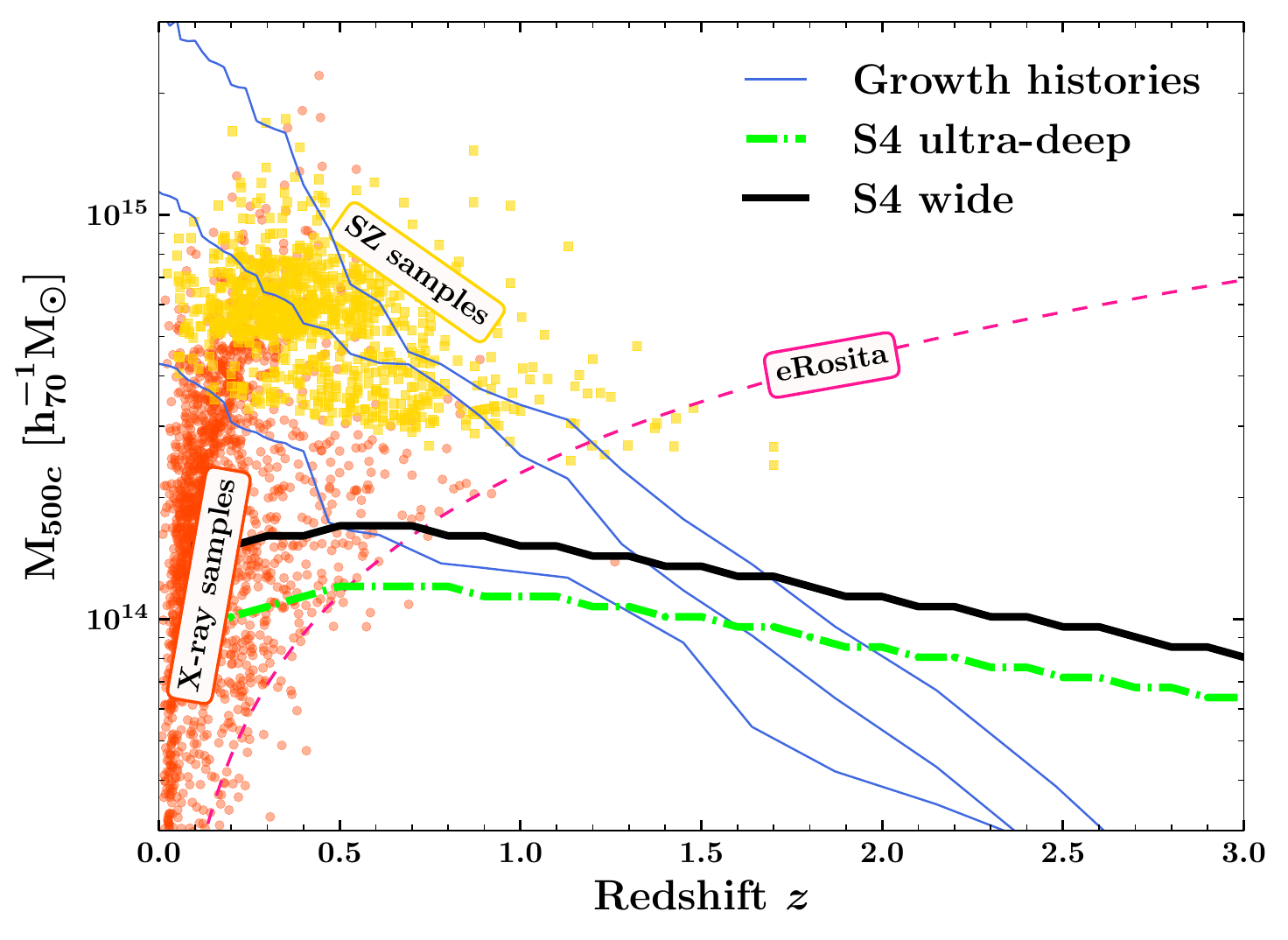}
\end{center}
\caption{Limiting mass as a function of redshift for CMB-S4 galaxy cluster
surveys. Also shown are existing catalogs of clusters selected by
either SZ or X-ray, as well as the projected e-Rosita mass limit.}
\label{fig:m_z_fid}
\end{figure}

A science goal is to be able to detect the progenitors of massive clusters today at the epoch when they were forming the bulk of their stars ($z\approx2$--3).
Fig.~\ref{fig:m_z_fid} shows that CMB-S4 will be able to detect the
typical massive low-redshift massive clusters at $z\approx2$; clusters
will be detected out to $z\approx3$ with reference design sensitivity, assuming
that the thermal pressure is well-described by the model used to
simulate the signal \citep{Arnaud2010}.
For a typical progenitor cluster ($10^{14} M_\odot$ at $z=3$), the typical cluster diameter (2$R_{200}$) is 2$^\prime$. This puts a lower limit on resolution of 2$^\prime$ at 90\,GHz. Similarly, Fig.~\ref{fig:cluster_counts} shows that detecting clusters at $z=3$ at CMB-S4 sensitivity requires a beam size no larger than FWHM=1.5$^\prime$ at 150\,GHz \citep{Madhavacheril:2017}.

\begin{figure}
\begin{center}
\includegraphics[width=3in]{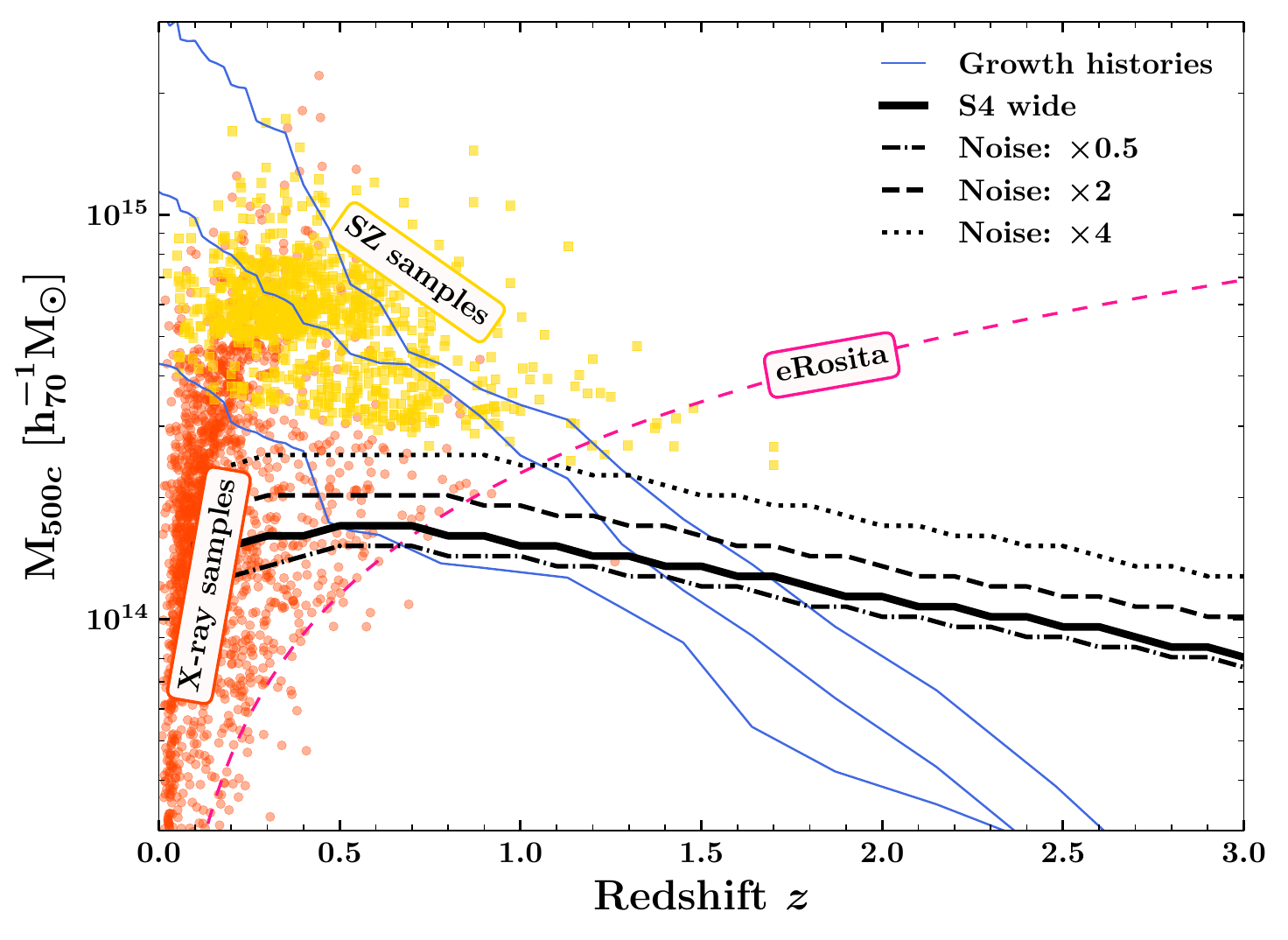}
\includegraphics[width=3in]{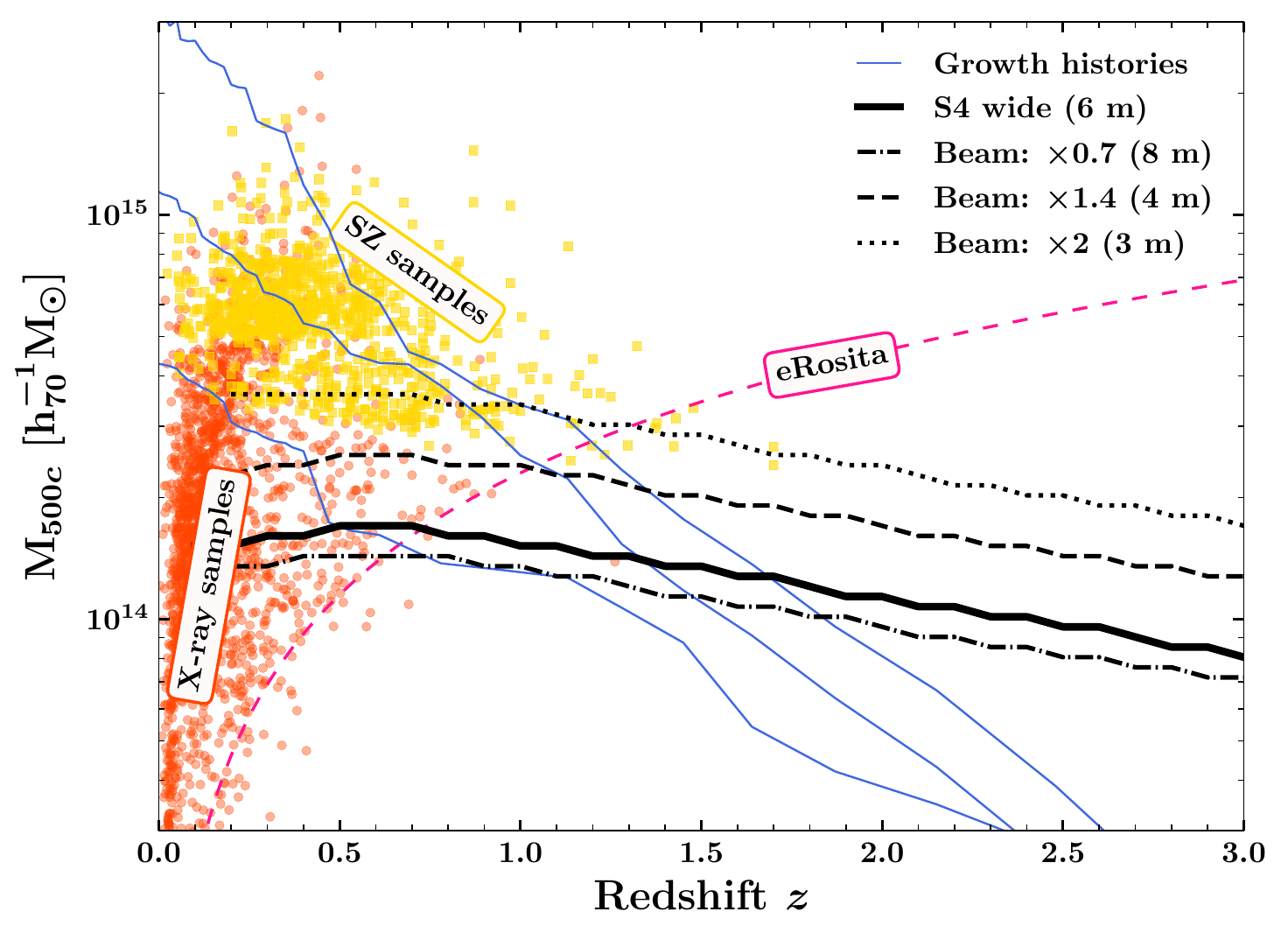}
\end{center}
\caption{Effect on limiting mass as a function of redshift by
varying noise levels (left) and beam size (right). Solid lines (red, green, blue) show
typical progenitor masses as a function of redshift for three different low-redshift
cluster groups.}
\label{fig:m_z_beam_noise}
\end{figure}

For detecting the progenitors of massive clusters at $z\approx2$ and
higher, Fig.~\ref{fig:m_z_beam_noise} shows that either a larger beam
size or higher noise level leads to an inability to access these clusters.

\section{Gamma-ray bursts, sensitivity, angular resolution, and cadence}
\label{sec:GRBflowdown}

Time-variable sources will be spatially unresolved in CMB-S4. Using difference imaging, source confusion
will not be a problem, so the sensitivity to time-variable sources will be determined solely by the
instrument noise and aperture size.

\begin{figure}
\begin{center}
\includegraphics[width=4in]{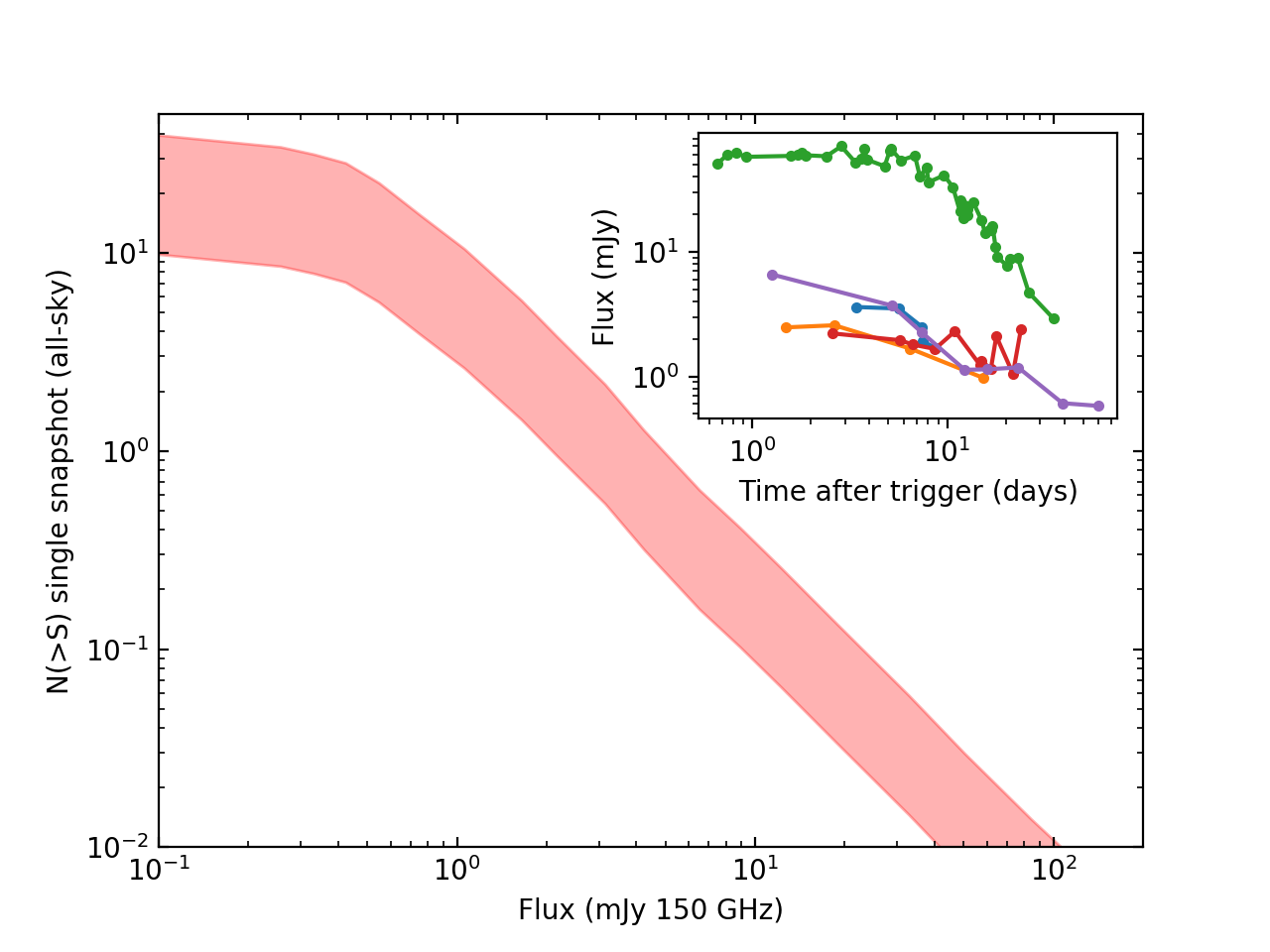}
\end{center}
\caption{Projections for 150-GHz transient source counts from Metzger et al.\ (2015)
\cite{Metzger2015}. Shown are the total
number of on-axis long gamma-ray bursts that are expected to be visible
in the entire sky at any one time as a function of flux at 150\,GHz.
The width of the band schematically represents the uncertainty in
this estimate. The inset shows some mm-wave follow-up observations
of long gamma-ray bursts \citep{deUgarte:2012}, showing typical
variations on time scales of several days.}
\label{fig:transient_counts}
\end{figure}

Some gamma-ray bursts have been measured to be visible at CMB-S4 wavelengths, with measured fluxes
typically ranging from 1--10\,mJy around 90\,GHz in the first week. As shown in
Fig.~\ref{fig:transient_counts}, it is not uncommon for the light curves,
once they start to decline, to
decline by a factor of two within just a few days, motivating a fast cadence.

The detailed number counts are
currently unknown, as no systematic surveys have been performed.
Theoretical calculations have been performed \citep{Metzger2015} that are
broadly consistent with the targeted follow-up.
These theoretical predictions for GRBs are shown in
Fig.~\ref{fig:transient_counts}.
At fluxes of several mJy the
expectation is that there should be a handful of long GRBs visible
somewhere on the sky at any moment in time, with each GRB afterglow
lasting roughly one week.
These calculations are consistent with the
dedicated mm-wave follow-up successfully detecting roughly 1/4 of the GRBs that
were targeted \citep{deUgarte:2012}.

The scaling with flux is uncertain at this point, but
a reasonable expectation when seeing the bright end of a
population, borne out by calculations
\citep{Metzger2015}, would be a power-law with $N(>S) \propto S^{-3/2}$, where a factor of
two improvement in flux sensitivity gains a factor of 2.8 more sources.
A minimum sensitivity
of 5 mJy every two days will ensure that there will be a substantial catalog of gamma-ray bursts.  The ultra-deep survey, covering only a few percent
of the sky but at substantially higher sensitivity, will also be useful
for GRBs.

\begin{figure}
\begin{center}
\includegraphics[width=3.0in]{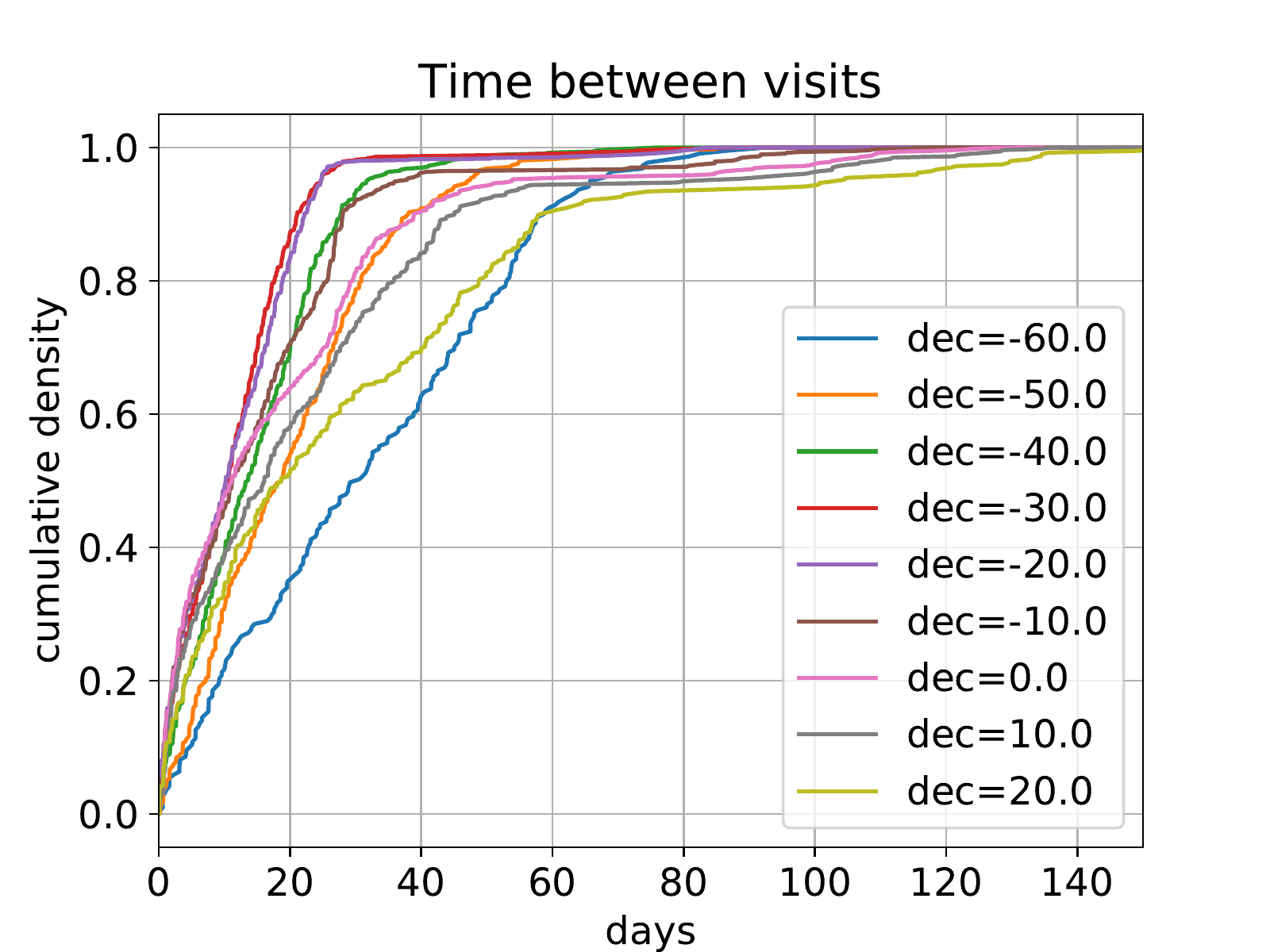}
\includegraphics[width=3.0in]{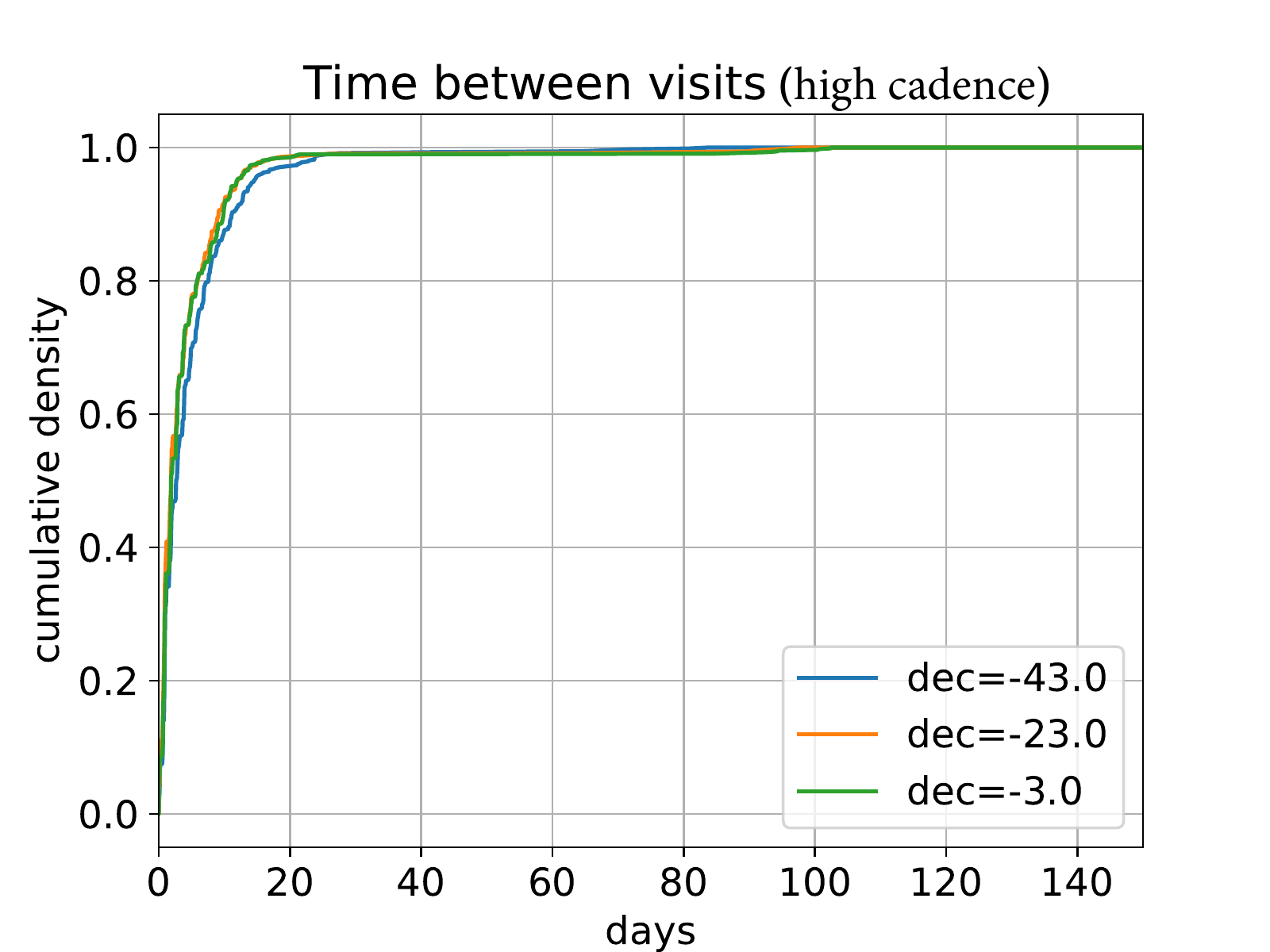}
\end{center}
\caption{Cumulative fraction of time between visits for the deep and wide field, 
split by declination.
Left panel shows the default scan strategy assumed for calculating noise curves
for the reference design: right panel shows results for a scan strategy 
with a high cadence that results in comparable overall survey performance. 
The ultra-deep field is expected to revisit every location
daily in the reference design.}
\label{fig:cadence}
\end{figure}

Observing most of the CMB-S4 area every two days (i.e., the entire survey area while avoiding the Sun and Moon) requires careful survey design.
The ``nominal schedule'' scan strategy discussed in the Light Relics section above has not been optimized for cadence, 
but still leads to good sensitivity to transients. Figure~\ref{fig:cadence}
shows that, with the exception of the lowest declinations, 1/3 of the
survey area is covered weekly in the reference design. From preliminary exploration of alternative observing
schedules we have found we can do significantly better on cadence without degradation to performance
on light relics. Specifically, we have found an observing strategy that achieves the cadence shown in the right
panel of Fig.~\ref{fig:cadence}, with uniform coverage every two days over the same area as that covered by the nominal
schedule. In addition, the ultra-deep
field (for delensing) will have daily coverage of that entire area
(a few percent of the sky).

Some GRBs have
shown evidence for a possible ``reverse shock,'' which will be luminous at mm-wavelengths in the first
few hours \citep{Laskar:2018}. Therefore, it is important to have a large area of the sky covered with a cadence close to
daily (at least every other day), which would provide a high probability of catching some events in the
first 12 hours. The ultra-deep field will do this over a few percent of
the sky; with a nearly daily GRB rate there should be several GRBs per year
in this patch.

\begin{table}[]
\centering
\scriptsize
\begin{tabular}{|l||c c c c c c|}
\hline
CMB-S4 Survey  & \multicolumn{6}{c|}{Point-source rms noise (mJy)} \\
 & 30\,GHz & 40\,GHz & 95\,GHz & 145\,GHz & 220\,GHz & 270\,GHz \\
\hline
\hline
Full depth (wide) &  0.8 & 0.6 & 0.2 & 0.2 & 0.6 & 1.1 \\
1 week (wide)  & 15 & 11 & 3 & 4 & 11 & 21 \\
\hline
Full depth (ultra-deep) & 0.1 & 0.1 & 0.04 & 0.07 & 0.4 & 0.5 \\
1 week (ultra-deep)  & 2 & 3 & 0.8 & 1 &  7 & 9 \\
\hline
\end{tabular}
\caption{Rms noise levels (in mJy) for point sources at different frequencies
for aspects of the CMB-S4 survey, either full depth after coadding the entire
	survey or the characteristic depth from a single week of data}
\label{source_depths}
\end{table}

Better angular resolution and sensitivity
and more frequency coverage are helpful.
As these sources will be spatially unresolved, a key quantity is
point-source sensitivity.
Sensitivity to unresolved sources is improved by having better angular resolution, with an approximate
scaling of signal-to-noise scaling as the inverse of the beam size.
It is necessary to have at least two well-measured frequencies to be able to determine an approximate
spectral index for follow-up at other wavelengths.

\eject

\renewcommand{\floatpagefraction}{.8}%
\chapter{Options \prelim{ ({\it S.~Padin})}}
\label{app:options}

\section{Re-use of existing telescopes and site infrastructure}

Instead of building two new 6-m telescopes in Chile, it may be possible to re-use CCAT-prime and/or the SO-LAT, and their associated site infrastructure. Re-use of existing telescopes could reduce the overall cost of CMB-S4. Another option would be to simply add the existing telescopes to CMB-S4 to increase sensitivity. Re-use would likely involve a fee, upgrades to monitoring and control systems, and some additional testing and alignment to meet CMB-S4 requirements. The details are very uncertain, so the reference design assumes all-new construction.

There are two critical decision points for the re-use option: (i) we could agree to re-use telescopes before awarding a design contract, in which case there would be no design or fabrication costs for new telescopes in Chile; or (ii) we could agree to re-use telescopes before awarding a fabrication contract, in which case there would be no fabrication costs for the telescopes, but we would still pay for the design.
\section{Readout}
\subsection{\fmux{}}

Frequency-division multiplexing (\fmux{}) is an alternative TES readout that has been implemented in several Stage-3 experiments. In an \fmux{} readout, each TES is wired in series with an inductor and a capacitor to create a resonant filter, and many of these TES-filter segments are connected in parallel. A comb of AC carriers is applied to the network to AC voltage-bias the TESs, with each TES-filter segment selecting one of the carriers. The currents flowing through the TESs are summed together and amplified by a SQUID. Since each network of TES-filter segments has just one SQUID, the cold electronics are simple, but the SQUIDs must handle a large signal range. Alternative TES readout schemes all have one SQUID per TES, which relaxes the signal range requirements at the expense of more complicated electronics.

The key advantage of the \fmux{} readout is that the amplitude of the AC voltage bias for each TES can be optimized individually to compensate for variations in TES saturation power across a detector wafer. Some variation in saturation power is inevitable because of fabrication tolerances and variations in optical loading. The ability to adjust each TES bias relaxes fabrication tolerances and improves the yield of working detectors.

A second important advantage of the \fmux{} readout is that there are only two wires for each network of many TES-filter segments, so thermal loading on the camera's sub-kelvin stage is small. As an example from an existing \fmux{} implementation, the thermal load for a 5-cm long niobium-titanium stripline connecting an array of TES-filters at 250\,mK to a thermal intercept at 350\,mK is only $\approx4\,$nW. The filter components at 250\,mK are all superconducting, so they do not dissipate power onto the 250-mK stage.

Several CMB experiments have deployed \fmux{} readout, including SPT, POLARBEAR, and EBEX.  SPT-POL and POLARBEAR used an \fmux{} configuration with 8--12 bolometers multiplexed together, POLARBEAR2 will soon deploy a readout with a MUX factor of 40, and SPT-3G is currently observing with a camera that has a readout MUX factor of 68. SPT-3G provides a concrete demonstration of background-limited performance with an \fmux{} readout at the scale of 16~k detectors.

The main issue for the \fmux{} readout is noise from the readout electronics. To achieve a constant voltage bias, the TES resistance must be larger than the parasitic resistance in the bias filter, but higher TES resistance increases the NEP contribution from SQUID input current noise. The readout NEP in deployed, ground-based \fmux{} systems is a few tens of $\textrm{aW}\sqrt{\text{Hz}}$, which is comparable with the photon noise. Lower readout noise can be achieved if parasitic resistances are carefully controlled. SRON has demonstrated low readout noise in the laboratory, and ANL and LBNL are working on \fmux{} designs that have the SQUIDs at 250\,mK, so there are fewer connections and smaller parasitics. The ANL and LBNL work is supported by LDRD funds.

To be a viable option for CMB-S4, the \fmux{} readout must show integrated performance appropriate for CMB-S4 at the scale of a detector wafer in the laboratory, and preferably on the sky, before the start of the final CMB-S4 design.

\subsection{\umux{}}

Microwave-multiplexed SQUID readout (\umux{}) is an evolution of the DC voltage-biased TES readouts that are used in the small-aperture CMB experiments at the South Pole and some of the large-aperture experiments in Atacama. The \umux{} readout borrows heavily from the DC-SQUID TDM scheme, replacing the TES current sensor (a DC-SQUID) with an equally low-noise extremely high-MUX-factor current sensor that is frequency-division multiplexed. The new current sensor is an RF-SQUID flux-coupled to a GHz-frequency superconducting resonator that can be read out much like an MKID. The signal from each TES modulates the magnetic flux through its SQUID, which in turn modulates the resonant frequency of its resonator.

An important advantage of the \umux{} readout is that the design of the TES bolometer and current sensor can be optimized independently of each other, as in a TDM readout, enabling large safety factors/margins in both bolometer and readout parameters and thus noise. For instance, \umux{} can use low-resistance DC-biased TESs ($\approx10\,{\rm m}\Omega$), which significantly reduces the NEP contribution from the readout. Readout NEP levels of 5--7\,aW/$\sqrt{\text{Hz}}$ (well below the photon noise for a typical ground-based CMB experiment) have been achieved in laboratory tests, using bolometers similar to those that would be deployed in CMB-S4. Additionally, \umux{} provides much higher multiplexing factors than either TDM or \fmux{}, with MUX factors exceeding 400 having been achieved in laboratory demonstrations and MUX factors of $\gtrsim$\,2000 are realistic. A high MUX factor may significantly reduce integration complexity and readout cost per channel, but must be balanced against the risk of a single-point failure.

A drawback of the \umux{} readout is that large numbers of TESs are connected in series to share the same bias, so it is not possible to optimize the operating point of each bolometer individually. TDM readouts have the same issue, and Stage-3 experiments have developed strategies for optimizing the performance of groups of bolometers that share a common bias. As bolometer uniformity achieved at fabrication has improved, the impact of this drawback has declined.

The \umux{} readout uses a phase modulation scheme that greatly reduces sensitivity to low-frequency amplitude-modulation noise (the same technique makes FM/XM radio channels sound less noisy than AM radio channels).  In addition, the readout electronics under development at SLAC use a fully digital approach that is largely insensitive to low-frequency noise induced by temperature fluctuations and phase drifts. The readout uses a ``tone-tracking'' feedback loop in which the microwave carriers that excite the GHz-frequency resonators are continuously maintained at each channel's resonant frequency. Feedback increases the dynamic range and linearity while reducing sensitivity to a variety of noise sources including gain variations and crosstalk.

The \umux{} development effort is supported by LDRD funds. A 64 MUX factor \umux{} readout was deployed on the Green Bank Telescope (GBT) to read out the 215-pixel MUSTANG2 array, prototypes of microwave SQUIDs suitable for CMB observations have been produced, and a full prototype readout has been demonstrated with a MUX factor of 528. SLAC and FNAL are working on warm electronics capable of $\geq 2000$ MUX factor, nearly an order of magnitude higher than \fmux{} and TDM.  In conjunction with this warm readout development, NIST has developed cold \umux{} multiplexers achieving a resonator density which will enable MUX factors of 2000. Both BICEP Array and Simons Observatory have baselined \umux{} in their instruments. 

To be a viable option for CMB-S4, the \umux{} readout must show integrated performance appropriate for CMB-S4 at the scale of a detector wafer in the laboratory, and preferably on the sky, before the start of the final CMB-S4 design.

\section{Kinetic-inductance detectors}

KIDs are superconducting thin-film micro resonators that detect radiation through a change in inductance when the superconductor absorbs photons. Many resonators, each at a slightly different frequency, can be coupled to a single readout line to achieve a high multiplexing factor. Multiplexing on the detector wafer is the key advantage of KIDs, because it allows many thousands of detectors on a wafer, with just a few connections.  In contrast, TESes read with \umux{} require an active squid per TES as well as a pair of wire bonds per TES, requiring high wiring density at the wafer edge.  This challenge is particularly acute for bands with small pixels, e.g., CMB-S4 220/270\,GHz. KIDs have other advantages: a much larger dynamic range than TESs; no SQUID amplifiers in the readout; simple fabrication; and compatibility with warm \umux{} hardware. TKIDs are an alternative that uses the inductor as the thermal sensor of a released bolometer island.  This introduces several additional design parameters that have been used to fine tune the detector noise to 20\,aW/$\sqrt{{\rm Hz}}$ stable to 100\,mHz, measured under loading representative of atmospheric loading at the South Pole.  Such performance would be background limited in observing bands at 90\,GHz and higher.  Their absorptive nature makes them drop-in compatible with antenna-coupling schemes used for multicolor pixels or observing bands below the superconducting band-gap.

Several millimeter-wavelength KID instruments have been deployed on telescopes (e.g., MUSIC, NIKA/NIKA2, MAKO, DESHIMA, A-MKID), but not for CMB observations. To be a viable option for CMB-S4 requires an on-sky demonstration of background limited performance, in a CMB polarization experiment, before CMB-S4 final design starts. Several groups are working towards such a demonstration (e.g., Columbia++ antenna-coupled KIDs, JPL TKIDs, Chicago CMB KIDs). The KID development effort is supported by NSF and NASA funds.

\section{Coupling to detectors}

\subsection{Lenslets} 

Several Stage-3 experiments have deployed lenslet-coupled detectors, which have planar antennas on the detector wafer and an array of lenslets to couple the antennas to the telescope. The principal advantage of this approach is wide bandwidth, which allows more detectors to be squeezed into the focal plane, e.g., each pixel in SPT-3G has 90-, 150-, and 220-GHz detectors. Lenslet-coupled detector arrays are compact, there are no issues with differential thermal contraction, because the detector and lenslet wafers are both made of silicon, and it may be possible to fabricate the lenslets commercially; the POLARBEAR group is pursuing commercial fabrication of monolithic silicon lenslet arrays. The main issue with lenslet-coupled detectors is fairly large frequency-dependent polarization errors in wide-band planar antennas. If the technology is to be a viable option for CMB-S4, we must demonstrate correction of the polarization errors leading to residual systematic errors well below the map noise in Stage-3 experiments. Both SPT-3G and POLARBEAR are working towards such a demonstration.

\subsection{Planar antennas}

Antenna-coupled transition-edge sensor bolometers have been developed for a wide range of CMB polarimetry experiments, including BICEP2, BICEP3, Keck Array, and SPIDER. These photo-lithographed planar antenna arrays synthesize symmetric co-aligned beams, with controlled side-lobes and low cross-polarized response (typically ~0.5\% on boresight, consistent with cross-talk in the multiplexed readout system). End-to-end optical efficiencies in these cameras are routinely 35\% or higher, within well-controlled 25\% wide spectral bands at 95, 150, 220 and 270\,GHz. Thanks to the frequency scalability of this design, more than 90 science wafers have been deployed on the sky to date, achieving a sensitivity of 50\,nK-deg at 95 and 150\,GHz.  Planar antennas show tight control of systematic errors, demonstrated to be sub-dominant to statistical noise at these levels in mapping degree scales with small aperture telescopes\cite{Ade:2018gkx,Ade:2015fpw}.  With this demonstrated performance and high fabrication throughput, planar antennas represent a low-risk implementation for CMB-S4.

Planar antenna technology offers a unique detector packing advantage compared to, e.g., single-band hex-packed spline feedhorns, due to the intrinsically steeper profile of a flat-top vs Gaussian illumination. Holding the edge taper on the aperture stop to be the same, square planar antennas pack more densely than spline feedhorns (in the large pixel / low edge taper regime), including gaps for detectors and wiring.  Switching to a hexagonal planar antenna may confer some additional efficiencies for filling out focal planes.  Planar antennas are currently being developed and tested at 30 and 40\,GHz for upcoming observations. Finally, a dual-band planar antenna that offers further improvements in packing density has been recently developed and will undergo extensive testing..


\section{Commercial detector fabrication}

Detectors for Stage-3 CMB experiments have been fabricated by national laboratories and universities, but CMB-S4 will need far more detectors, so fabrication throughput will be a challenge. Moving at least some of the detector fabrication to industry would increase the fabrication rate without an expensive in-house investment in facilities and staff. The overall cost could be lower because an industrial partner may be able to draw staff and fabrication tools from a larger pool that supports many projects. Shift work is generally also easier to implement in an industrial setting. The main issue with moving detector fabrication to industry is controlling the superconducting properties of the transition edge sensor films, which typically requires dedicated tooling and stringent quality control. Most of the other fabrication steps are standard industry processes.

If commercial detector fabrication is to be a viable option, performance appropriate for CMB-S4 must be demonstrated in the laboratory, and preferably in the field, along with significant fabrication throughput, before CMB-S4 final design starts. LBNL and UCB are working with STARCryo and HYPRES to commercialize the POLARBEAR detector technology, and many process steps have already been demonstrated. The work is supported by SBIR and LDRD awards, and by a DOE Early Career award.

\section{Other developments}

The options described above are supported by LDRD and other funds, with CMB-S4 providing at least some coordination, but there are several other developments that CMB-S4 is watching, and may choose to take advantage of the following.

\begin{enumerate}
\item Monolithic versus segmented mirrors for the large telescopes. Eliminating the gaps between mirror panels would reduce ground pickup, which would allow the large telescopes to recover larger angular scales. Ground pickup is a serious concern for all CMB experiments, so a demonstration of a large monolithic mirror would almost certainly result in CMB-S4 adopting the technology.
\item Full boresight rotation for the large telescopes. Small CMB telescopes use boresight rotation to measure and correct polarization errors, but the technique has not yet been attempted on a large telescope, primarily because it requires an additional axis. The Simons Observatory and CCAT-prime large telescope designs offer partial boresight rotation through receiver rotation, which will enable improved correction of polarization errors compared to existing large telescopes. 
\end{enumerate}

\eject

\newlist{abbrv}{itemize}{1}
\setlist[abbrv,1]{label=,labelwidth=1in,align=parleft,itemsep=0.1\baselineskip,leftmargin=!}
 
\chapter{List of Abbreviations}
\label{chap:abbrvlist}

\vspace{12pt}

\begin{abbrv}

\item[$\Lambda$CDM] 
$\Lambda$ cold dark matter

\item[$\mu$mux] 
Microwave multiplexing

\item[AAAC] 
Astronomy and Astrophysics Advisory Committee

\item[ABS] 
Atacama B-mode Search 

\item[AC] 
Alternating current

\item[ACT] 
Atacama Cosmology Telescope

\item[ADC]
Analog-to-digital converter

\item[ADM] 
Arnowitt-Deser-Misner

\item[ADMX] 
Axion Dark Matter Experiment 

\item[AGN] 
Active galactic nucleus

\item[ALCF]
Argonne Leadership Class Facility

\item[ALMA] 
Atacama Large Millimeter/submillimeter Array

\item[ALP] 
Axion-like particle

\item[AME] 
Anomalous microwave emission

\item[AR] 
Anti-reflection

\item[ASAS-SN]
All-Sky Automated Survey for Supernovae

\item[ASKAP] 
Australian Square Kilometre Array Pathfinder

\item[ATLAS]
A Toroidal LHC ApparatuS detector at the LHC

\item[AU] 
Astronomical unit

\item[BAO] 
Baryon acoustic oscillations

\item[BBN] 
Big--bang nucleosynthesis

\item[BDI]
Baryon density isocurvature

\item[BOE]
Basis of estimate

\item[BOSS]
Baryon Oscillation Spectroscopic Survey

\item[BSM] 
Beyond the Standard Model

\item[CAM] 
Cost account manager

\item[CASPEr] 
Cosmic Axion Spin Precession Experiment

\item[CCAT] 
Cerro Chajnantor Atacama Telescope

\item[CD] 
Critical decision or Crossed Dragone

\item[CDM] 
Cold dark matter

\item[CDI]
CDM density isocurvature

\item[CDT] 
Concept Definition Task force

\item[CFRP] 
Carbon-fiber-reinforced polymer

\item[CGM] 
Circumgalactic medium

\item[CHIME] 
Canadian Hydrogen Intensity Mapping Experiment

\item[CIB] 
Cosmic infrared background

\item[CIP] 
Compensated isocurvature perturbation

\item[CL] 
Critical line or confidence limit

\item[CLASS] 
Cosmology Large Angular Scale Surveyor 

\item[CMB] 
Cosmic microwave background

\item[CMS]
Compact Muon Solenoid detector at the Large Hadron Collider 

\item[CPU]
Central processing unit

\item[CRS]
Cold readout stack

\item[CTE]
Coefficient of thermal expansion

\item[D\&R]
Detetctors and readout

\item[DAC]
Digital-to-analog converter

\item[DAQ]
Data acquisition

\item[DE]
Dark energy

\item[DESI]
Dark Energy Spectroscopic Instrument

\item[DM]
Dark matter

\item[DOE] 
Department of Energy

\item[DR]
Dark radiation or Dilution refrigerator

\item[DSFG]
Dusty star-forming galaxy

\item[DSL]
Dark Sector Laboratory

\item[DSR]
Decadal Survey Report

\item[DUNE]
Deep Underground Neutrino Experiment

\item[EH\&S]
Environment, health, and safety

\item[EMI]
Electromagnetic interference

\item[EMU]
Evolutionary Map of the Universe

\item[ESA]
European Space Agency

\item[FACA] 
Federal Advisory Committee Act

\item[FDM] 
Frequency-division multiplexing

\item[FET] 
Field-effect transistor

\item[FLRW] 
Friedmann-Lema\^{\i}tre-Robertson-Walker

\item[fmux] 
Frequency-division multiplexing

\item[FNAL] 
Fermilab National Accelerator Laboratory

\item[FoM] 
Figure of merit

\item[FOV] 
Field of view

\item[FP] 
Focal plane

\item[FPGA] 
Field-programmable gate array

\item[FRB] 
Fast radio burst

\item[FTE] 
Full-time equivalent

\item[FTS] 
Fourier-transform spectrometer

\item[FWHM] 
Full width at half maximum

\item[GBT] 
Green Bank Telescope

\item[GC] 
Galaxy cluster

\item[GPS] 
Global positioning system

\item[GRB] 
Gamma-ray burst

\item[GSR] 
Generalized slow-roll

\item[GUT] 
Grand unified theory

\item[HDPE] 
High-density polyethylene 

\item[HEP] 
High-energy physics

\item[HPC] 
High-performance computing

\item[HTC] 
High-throughput computing

\item[HWFE] 
Half wavefront error

\item[HWP] 
Half-wave plate

\item[I\&C] 
Integration and commissioning

\item[I\&T] 
Integration and testing

\item[ICM] 
Intracluster medium

\item[ILC] 
Internal linear combination

\item[IPO] 
Interim project office

\item[IPSC]
Integrated Project Steering Committee

\item[IR] 
Infrared

\item[IRIG]
Inter-range instrumentation group 

\item[ISM] 
Interstellar medium

\item[JCMT] 
James Clerk Maxwell Telescope

\item[JPL] 
Jet Propulsion Laboratory

\item[JWST] 
James Webb Space Telescope

\item[KATRIN] 
KArlsruhe TRItium Neutrino (experiment)

\item[KID] 
Kinetic-inductance detector

\item[kSZ] 
Kinematic Sunyaev Zeldovich

\item[L1]
Level 1

\item[LAMBDA]
Legacy Archive for Microwave Background Data

\item[LAT]
Large-aperture telescope

\item[LATR]
Large-aperture telescope receiver

\item[LBNL]
Lawrence Berkeley National Laboratory

\item[LDRD]
Laboratory Directed Research and Development

\item[LHC]
Large Hadron Collider

\item[LIGO]
Laser Interferometer Gravitational-Wave Observatory

\item[LISA]
Laser Interferometer Space Antenna

\item[LR]
Light relics

\item[LRG]
Luminous red galaxy

\item[LSS]
Large-scale structure

\item[LSST]
Large Synoptic Survey Telescope

\item[M\&O]
Management and operations

\item[M\&S]
Materials and services

\item[MAPO]
Martin Pomerantz Observatory

\item[MCMC]
Markov-chain Monte-Carlo

\item[MHD] 
Magnetohydrodynamic

\item[MIE]
Major Item of Equipment

\item[MoU]
Memorandum of understanding

\item[MREFC] 
Major Research Equipment and Facilities Construction

\item[MKID] 
Microwave kinetic-inductance detector

\item[MUX] 
Multiplexing

\item[NASA] 
National Aeronautics and Space Administration

\item[NDI]
Neutrino density isocurvature

\item[NEP] 
Noise-equivalent power

\item[NERSC] 
National Energy Research Scientific Computing Center

\item[NET] 
Noise-equivalent temperature

\item[NIST] 
National Institute of Standards and Technology 

\item[NLDBD]
Neutrino-less double-beta decay 

\item[NSF] 
National Science Foundation

\item[NLDBD] 
Neutrinoless double-beta decay

\item[NLS1] 
Narrow-line Seyfert 1

\item[NTP] 
Network time protocol

\item[OCS] 
Observatory control system

\item[OFHC] 
Oxygen-free high-conductivity

\item[OMT] 
Orthomode transducer

\item[OPC]
Other project costs

\item[P5] 
Particle Physics Project Prioritization Panel

\item[PCI-E]
Peripheral Component Interconnect Express

\item[PD] 
Project director

\item[PDR] 
Preliminary design review

\item[PGW] 
Primordial gravitational waves

\item[PID]
Proportional–integral–derivative 

\item[PMNS] 
Pontecorvo-Maki-Nakagawa-Sakata

\item[pPDG] 
pre-Project Development Group

\item[pPEP] 
preliminary Project Execution Plan

\item[PPS]
Pulse per second

\item[PTFE] 
Polytetrafluoroethylene

\item[PTP]
Precision Time Protocol

\item[PySM] 
Python Sky Model

\item[QA] 
Quality assurance

\item[QCD] 
Quantum chromodynamics

\item[R\&D] 
Research and development

\item[RF] 
Radio frequency

\item[RMS] 
Root mean square

\item[RSD] 
Redshift-space distortion

\item [RT-MLI]
Radio-transparent multi-layer insulation

\item[SAT]
Small-aperture telescope

\item[SBIR]
Small Business Innovation Research

\item[SCUBA]
Submillimeter Common-User Bolometer Array

\item[SDSS] 
Sloan Digital Sky Survey

\item[SED] 
Spectral energy distribution

\item[SKA] 
Square Kilometre Array

\item[SLAC] 
Stanford Linear Accelerator Center

\item[SM] 
Standard Model (of particle physics)

\item[SMuRF]
SLAC Microresonator Radio Frequency

\item[SO] 
Simons Observatory

\item[SOFIA]
Stratospheric Observatory for Infrared Astronomy,

\item[SoW]
Statement of work

\item[SPT] 
South Pole Telescope

\item[SQUID] 
Superconducting quantum-interference device

\item[SRON]
Netherlands Institute for Space Research

\item[SSA] 
Series SQUID array

\item[SWF]
Salaries, wages, and fringe benefits

\item[TDM] 
Time-division multiplexing

\item[TEC]
Total estimated cost

\item[TES] 
Transition-edge sensor

\item[TPC]
Total project cost

\item[TKID] 
Thermal kinetic inductance detector

\item[TLS]
Two-level system

\item[tSZ] 
Thermal Sunyaev Zeldovich

\item[UFM] 
Universal focal-plane module

\item[UHMWPE] 
Ultra-high-molecular-weight polyethylene

\item[ULA] 
Ultralight axion

\item[UV] 
Ultraviolet

\item[VLA] 
Very Large Array

\item[VLASS] 
VLA Sky Survey

\item[WBS]
Work breakdown structure

\item[WFIRST]
Wide Field Infrared Survey Telescope 

\item[WIMP] 
Weakly interacting massive particle

\item[WISE]
Wide-field Infrared Survey Explorer

\item[WMAP]
Wilkinson Microwave Anisotropy Probe

\item[XSEDE]
Extreme Science and Engineering Discovery Environment

\end{abbrv}

\eject

\clearchapterpage
 \addcontentsline{toc}{chapter}{\bibname}
\markboth{\bibname}{}
\markright{\bibname}
\bibliography{cmbs4}

\end{document}